\documentclass[12pt]{book}
\usepackage{here,amsfonts,amssymb,pstricks,pst-node,
pst-plot,pst-coil,epsfig,makeidx}
\usepackage{amsmath}

\newtheorem{thm}{Theorem}[section]
\newtheorem{lemma}[thm]{Lemma}
\newtheorem{cor}[thm]{Corollary}
\newtheorem{prop}[thm]{Proposition}

\newtheorem{defin}[thm]{Definition}

\newbox\sample
\newif\ifproofmode
\newif\ifsymindex
\global\symindexfalse
\newwrite\inx
\def\indsyma#1#2{\ifproofmode\marginpar{$\scriptstyle#1$}\fi%
\ifx#2\empty\write\inx{$\noexpand#1$,\space\thepage}%
\write\inx{\string\newline}\else%
\write\inx{$\noexpand#1$,\space#2,\space\thepage}%
\write\inx{\string\newline}\fi\ignorespaces}%
\def\indsym#1#2{\ifsymindex%
\ifproofmode\marginpar{$\scriptstyle#1$}\fi%
\ifx#2\empty\write\inx{\string\item \space$\noexpand#1$,\space\thepage}%
\else%
\write\inx{\string\item \space$\noexpand#1$,\space#2,\space\thepage}%
\fi\ignorespaces\fi}%
\newskip\dangerskipb
\newskip\dangerskip
\def\hang{\hangindent\dangerskip}

\def\s#1{{\cal #1}}

\def\lag{\left\langle}
\def\rag{\right\rangle}

\def\pairt#1#2{\lag #1, #2\rag}

\def\proof{\noindent{\it Proof\/}.\enspace}

\def\remark{\bigskip\noindent{\bf Remark:}\enspace}

\def\remarks{\bigskip\noindent{\bf Remarks:}\enspace}

\font\manual=manfnt at 12pt
\def\danbend{{\manual\char127}}
\def\datanger{\medbreak\begingroup\clubpenalty=10000
 \def\par{\endgraf\endgroup\medbreak} \noindent\hang\hangafter=-2
 \hbox to0pt{\hskip-3.5pc\danbend\hfill}}
\outer\def\danger{\datanger}%
\def\ddatanger{\medbreak\begingroup\clubpenalty=10000
 \def\par{\endgraf\endgroup\medbreak} \noindent\hang\hangafter=-2
 \hbox to0pt{\hskip-3.5pc\danbend\kern1pt%
\danbend\hfill}}

\def\dobdownarrow{\mathop{\vbox{\kern2pt \hbox{$\Big\downarrow$}\kern-16.5pt
                          \nointerlineskip\hbox{$\Big\downarrow$}}}}

\def\lrightarrow{\hbox to 25pt{\rightarrowfill}}

\def\supexp{exp(m,n,p)=m^{m^{m^{\cdot^{\cdot^{\cdot^{m^{p}}}}}}}
\vbox{\hbox{$\Big\}\scriptstyle n$}\kern0pt}}

\def\supexpo#1#2#3{#1^{#1^{\cdot^{\cdot^{\cdot^{#1^{#2}}}}}}
\vbox{\hbox{$\Big\}\scriptstyle #3$}\kern0pt}}

\def\sqr#1#2{{\vcenter{\hrule height .#2pt
         \hbox{\vrule width.#2pt height#1pt \kern#1pt
             \vrule width.#2pt}
         \hrule height.#2pt}}}

\def\bigsquare{\mathchoice\sqr76\sqr76\sqr{2.1}3\sqr{1.5}3}

\def\lag{\langle}
\def\rag{\rangle}

\def\co{\colon}

\newskip\bogcentering \bogcentering= 0pt plus 1000pt minus 1000pt 

\def\matth{\mathsurround=0pt}
\def\fakrightarrowfill{$\matth \mathord- \mkern-6mu
  \cleaders\hbox{$\mkern-2mu \mathord- \mkern-2mu$}\hfill
 \mkern-6mu \mathord\rightarrow$}

\def\fakoverrightarrow#1{\vbox{\ialign{##\crcr
  \fakrightarrowfill\crcr\noalign{\kern-1pt\nointerlineskip}
 $\hfil\displaystyle{#1}\hfil$\crcr}}}

\def\cases#1{\left\{\,\vcenter{\normalbaselines\matth
  \ialign{$##\hfil$&\quad##\hfil\crcr#1\crcr}}\right.}

\newif\ifdtatp

\def\displaty{%
\global \dtatptrue \openup \jot \matth \everycr{\noalign{\ifdtatp \global 
\dtatpfalse \vskip -\lineskiplimit \vskip \normallineskiplimit \else 
\penalty \interdisplaylinepenalty \fi }}}

\def\displaylignes#1{\displaty
   \halign{\hbox to\displaywidth{$\displaystyle##$}\crcr
   #1\crcr}}
\def\leqaligneno#1{\displaty \tabskip=\bogcentering
 \halign to\displaywidth{\hfil$\displaystyle{##}$\tabskip=0pt
 &$\displaystyle{{}##}$\hfil\tabskip=\bogcentering
 &\kern-\displaywidth\rlap{$##$}\tabskip=\displaywidthpt\crcr
 #1\crcr}}
\def\ligne{\hbox to\hsize}
\newdimen\nouvpagewidth
\newdimen\offwidth
\newdimen\lawidthoui
\nouvpagewidth=\lawidthoui
\def\kboxit#1{\vbox{\hrule\hbox{\vrule\kern3pt
              \vbox{\kern3pt#1\kern3pt}\kern3pt\vrule}\hrule}}

\def\kboxitb#1{\vbox{\hrule\hbox{\vrule\kern3pt
              \vbox{\kern3pt#1\kern3pt}\kern3pt\vrule}\hrule}}

\def\laboxaround#1{
\aboxaround{\hbox to\hsize{\hfill\box2\hfill}}{#1}
}

\def\boxar#1#2{
\aboxaround{\hbox to\hsize{\hfill#1\hfill}}{#2}
}

\def\aboxaround#1#2{
\setbox4=\vbox{\hsize #2\noindent\strut#1\strut}
\kboxitb{\box4}}

\def\kframeit#1{\vbox{\hrule\hbox{\vrule\kern5pt
              \vbox{\kern5pt#1\kern5pt}\kern5pt\vrule}\hrule}}

\newskip\savnormalbaselineskip
\newskip\savnormallineskip
\newdimen\savnormallineskiplimit

\def\ligne{\hbox to\hsize}

\newdimen\tempa
\newdimen\tempb
\newdimen\dimenlastbox
\def\step#1{
       \setbox1\lastbox
       \setbox2\hbox{#1}
       \ifdim \the\dimenlastbox > 1\wd2
       \tempb \the\dimenlastbox \else
       \tempb 1\wd2\fi
       \global\dimenlastbox 1\wd2
       \ifdim 1\wd1 > 1\wd2
       \tempa 1\wd1
       \setbox2\hbox to \the\tempa{\hfill\box2\hfill}
       \else
       \tempa 1\wd2
       \setbox1\hbox to \the\tempa{\hfill\box1\hfill}\fi
       \vbox{\baselineskip 0pt\box1\vskip 3pt%
\hbox to \the\tempa{\hss\vbox{\hrule width \the\tempb}\hss}%
\vskip 3pt%
\box2}}
\def\matth{\mathsurround=0pt}
\def\brokenlinefill{$\matth \mathord-\mkern-5mu%
\cleaders\hbox{$\mathord-$}\hfill\mkern-5mu\mathord-$}
\def\brokstep#1{
       \setbox1\lastbox
       \setbox2\hbox{#1}
       \ifdim \the\dimenlastbox > 1\wd2
       \tempb \the\dimenlastbox \else
       \tempb 1\wd2\fi
       \global\dimenlastbox 1\wd2
       \ifdim 1\wd1 > 1\wd2
       \tempa 1\wd1
       \setbox2\hbox to \the\tempa{\hfill\box2\hfill}
       \else
       \tempa 1\wd2
       \setbox1\hbox to \the\tempa{\hfill\box1\hfill}\fi
       \vbox{\baselineskip 0pt\box1
\hbox to \the\tempa{\hss\vbox{\hbox to \the\tempb{\brokenlinefill}}\hss}%
\box2}}
\def\jstep#1#2{
       \setbox1\lastbox
       \setbox2\hbox{#1}
       \ifdim \the\dimenlastbox > 1\wd2
       \tempb \the\dimenlastbox \else
       \tempb 1\wd2\fi
       \global\dimenlastbox 1\wd2
       \ifdim 1\wd1 > 1\wd2
       \tempa 1\wd1
       \setbox2\hbox to \the\tempa{\hfill\box2\hfill}
       \else
       \tempa 1\wd2
       \setbox1\hbox to \the\tempa{\hfill\box1\hfill}\fi
       \vbox{\baselineskip 0pt\box1%
\vskip-4pt%
\hbox{\hbox to \the\tempa{\hss\vbox{\hrule width \the\tempb}
$\vcenter{\hbox to 0pt{\strut\quad#2\hss}}$%
\hss}}%
\box2}}
\def\istep#1{
       \setbox1\lastbox
       \setbox2\hbox{#1}
       \ifdim \the\dimenlastbox > 1\wd2
       \tempb \the\dimenlastbox \else
       \tempb 1\wd2\fi
       \global\dimenlastbox 1\wd2
       \ifdim 1\wd1 > 1\wd2
       \tempa 1\wd1
       \setbox2\hbox to \the\tempa{\hfill\box2\hfill}
       \else
       \tempa 1\wd2
       \setbox1\hbox to \the\tempa{\hfill\box1\hfill}\fi
       \vbox{\baselineskip 0pt\box1\vskip 6pt\box2}}
\def\strstep#1{
       \setbox1\lastbox
       \setbox2\hbox{#1}
       \ifdim \the\dimenlastbox > 1\wd2
       \tempb \the\dimenlastbox \else
       \tempb 1\wd2\fi
       \global\dimenlastbox 1\wd2
       \ifdim 1\wd1 > 1\wd2
       \tempa 1\wd1
       \setbox2\hbox to \the\tempa{\hfill\box2\hfill}
       \else
       \tempa 1\wd2
       \setbox1\hbox to \the\tempa{\hfill\box1\hfill}\fi
       \vbox{\baselineskip 0pt\box1\vskip3pt%
\hbox{\hbox to \the\tempa{\hss\vbox{\hrule width \the\tempb\vskip3pt%
\hrule width \the\tempb}\hss}}\vskip 3pt\box2}}
\def\settree#1#2{\setbox#1\vbox{#2}}
\def\starttree#1{\setbox1 \hbox{#1}\dimenlastbox 1\wd1\box1}
\def\jointrees#1#2{
    \starttree{\box#1\hskip60pt\box#2}}
\def\njointrees#1#2{
    \starttree{\box#1\hskip25pt\box#2}}

\def\tjointrees#1#2#3{
    \starttree{\box#1\hskip35pt\box#2\hskip35pt\box#3}}
\def\tnjointrees#1#2#3{
    \starttree{\box#1\hskip20pt\box#2\hskip20pt\box#3}}

\def\sequent#1#2{\hbox{{$#1 \rightarrow #2$}}}

\input xy
\xyoption{all}

\newbox\sample

\def\mfrac#1{{\mathfrak{#1}}}
\def\id{\mathrm{id}}
\def\ideal#1{\mfrac{#1}}
\def\reals{\mathbb{R}}
\def\complex{\mathbb{C}}
\def\integs{\mathbb{Z}}
\def\natnums{\mathbb{N}}
\def\rats{\mathbb{Q}}

\def\mapdef#1#2#3{#1\co #2\rightarrow #3}

\def\Ker{\mathrm{Ker}\,}

\def\interio#1{\buildrel \circ\over #1}

\def\transpos#1{#1^{\top}}

\def\res{\upharpoonright}
\def\impl{\Rightarrow}
\def\impliesintr{{\it $\impl$-intro\/}}
\def\implieselim{{\it $\impl$-elim\/}}
\def\landintr{{\it $\land$-intro\/}}
\def\landelim{{\it $\land$-elim\/}}
\def\lorintr{{\it $\lor$-intro\/}}
\def\lorelim{{\it $\lor$-elim\/}}
\def\perpelim{{\it $\perp$-elim\/}}
\def\bycontra{{\it by-contra\/}}
\def\forallintr{{\it $\forall$-intro\/}}
\def\forallelim{{\it $\forall$-elim\/}}
\def\existsintr{{\it $\exists$-intro\/}}
\def\existselim{{\it $\exists$-elim\/}}

\def\rulea#1#2#3{\dfrac{#1}{#2}\quad{\hbox{#3}}}
\def\ruleb#1#2#3#4{\dfrac{#1\quad#2}{#3}\quad{\hbox{#4}}}
\def\rulec#1#2#3#4#5{\dfrac{#1\quad#2\quad#3}{#4}\quad{\hbox{#5}}}

\title{\huge
Discrete Mathematics for\\
Computer Science\\
Some Notes}
\author{Jean Gallier\\
Department of Computer and Information Science\\
University of Pennsylvania\\
Philadelphia, PA 19104, USA\\
e-mail: {\tt jean@cis.upenn.edu}\\
\ \\
\copyright\ Jean Gallier\\
{\bf Please, do not} reproduce {\bf without permission} of the author}
\begin{document}
\maketitle
\ \vfill\eject
\begin{center}
{\large \bf
Discrete Mathematics for Computer Science\\
Some Notes
}

\vspace{1cm}
Jean Gallier
\end{center}

\vspace{2cm}

\noindent
{\bf Abstract:}
These are notes on discrete mathematics for computer scientists.
The presentation is somewhat unconventional. Indeed I begin
with a discussion of the basic rules of mathematical reasoning
and of the notion of proof formalized in a natural deduction
system ``a la Prawitz''. The rest of the material is more or less
traditional but I emphasize partial functions more than usual
(after all, programs may not terminate for all input) and I 
provide a fairly complete account of the basic concepts of
graph theory.

\tableofcontents
\vfill\eject
\chapter*{Preface}
\label{chap0}
The curriculum of most undergraduate programs in computer science 
includes a course untitled {\it Discrete Mathematics\/}.
These days, given that many students who graduate with
a degree in computer science  end up with jobs
where mathematical skills seem basically of no use,%
\footnote{In fact, some people would even argue
that such skills constitute a handicap!}
one may ask why these students should take such a course.
And if they do, what are the most basic notions that they
should learn?

\medskip
As to the first question, I strongly believe that
{\it all\/} computer science students should take such a
course and I will try  justifying this assertion below.

\medskip
The main reason is that, based on my experience of more
than twenty five years of teaching, I have found that the
majority of the students find it very difficult to 
present an argument in a rigorous fashion. The notion of
a proof is something very fuzzy for most students and 
even the need for the rigorous justification of a claim
is not so clear to most of them. Yet, they will all
write complex computer programs and it seems rather crucial
that they should understand the basic issues of program correctness.
It also seems rather crucial
that they should possess some basic mathematical skills to
analyse, even in a crude way, the complexity of the programs they will 
write. Don Knuth has argued these points more eloquently that I can in his
beautiful book, {\sl Concrete Mathematics}, and I will not elaborate
on this anymore.

\medskip
On a scholarly level, I will argue that some basic
mathematical knowledge should be part of the
scientific {\it culture\/} of any computer science
student and more broadly, of any engineering student.

\medskip
Now, if we believe that computer science students should
have some basic mathematical knowledge, what 
should it be?

\medskip
There no simple answer. Indeed, students with an interest
in algorithms  and complexity will need some discrete
mathematics such as combinatorics  and graph theory
but students interested in computer graphics or computer vision
will need some geometry and some continuous mathematics.
Students interested in data bases will need to know
some mathematical logic and students interested in computer
architecture will need yet a different brand of mathematics.
So, what's the common core?

\medskip
As I said earlier, most students have a very fuzzy idea
of what a proof is. This is actually true of most people!
The reason is simple: It is quite difficult
to define precisely what a proof is. To do this, one has to define
precisely  what are the ``rules of mathematical reasoning'' and this
is a lot harder than it looks.  Of course, defining
and analyzing the notion of proof is a major goal of
mathematical logic.

\medskip
Having attempted some twenty years ago to ``demystify''
logic for computer scientists and being an incorrigible
optimist, I still believe that there is great value
in attempting to teach people the basic principles
of mathematical reasoning in a precise but not
overly formal manner.  In these notes, I define the notion of
proof as a certain kind of tree whose inner nodes
respect certain proof rules presented in the style of a
natural deduction system ``a la Prawitz''.
Of course, this has been done before
(for example, in van Dalen \cite{vanDalen}) but 
our presentation has more of a ``computer science'' flavor
which should make it more easily digestible by our 
intended audience. Using such a proof system, it is easy to
describe very clearly what is a proof by contradiction
and to introduce the subtle notion of ``constructive proof''.
We even  question the ``supremacy'' of classical logic,
making our students aware of the fact that there isn't just
one logic, but different systems of logic, which often comes as a 
shock to them.

\medskip
Having provided a firm foundation for the notion
of proof, we proceed with a quick and informal
review of the first seven axioms of Zermelo-Frankel set theory.
Students are usually surprised to hear that axioms are needed to ensure
such a thing as the existence of the union of two sets
and I respond by stressing that one should always keep a healthy dose
of skepticism in life!

\medskip
What next? Again, my experience has been that most students
do not have a clear idea of what a function is, even less of
a partial function. Yet, computer programs may not terminate
for all input, so the notion of partial function is crucial.
Thus, we define carefully relations, functions and partial functions
and investigate some of their properties
(being injective, surjective, bijective). 

\medskip
One of the major stumbling blocks for students is the notion
of proof by induction and its cousin, the definition of
functions by recursion. We spend quite a bit of time clarifying these
concepts and we give a proof of the validity of the induction principle
from the fact that the natural numbers are well-ordered.
We also discuss the pigeonhole principle and some basic
facts about equinumerosity, without introducing cardinal numbers.

\medskip
We introduce some elementary concepts of combinatorics
in terms of counting problems. We introduce the binomial 
and multinomial coefficients and study some of their
properties   and we  conclude with the
Inclusion-Exclusion Principle.

\medskip
Next, we introduce partial orders, well-founded sets and complete
induction. This way, students become aware of the fact that
the induction principle applies to sets with an ordering
far more complex that the ordering on 
the natural numbers. 
As an application, we prove the unique prime factorization
in $\integs$ and discuss GCD's.

\medskip
Another extremely important concept is that of 
an equivalence relation and the related notion of a partition.

\medskip
We have included some material on lattices, Tarski's fixed point 
Theorem, distributive lattices, boolean algebras and
Heyting algebras.  These topics are somewhat more advanced and can
be omitted from the ``core''.

\medskip
The last topic that we consider crucial is graph theory.
We give a fairly complete presentation of the basic
concepts of graph theory: directed and undirected graphs,
paths, cycles, spanning trees, cocycles, cotrees, flows
and tensions, Eulerian and Hamiltonian cycles, matchings, coverings,
and planar graphs. We also discuss the network flow problem
and prove the Max-Flow Min-Cut Theorem in an original way due to M. 
Sakarovitch. 

\medskip
These notes grew out of lectures I gave in 2005 while
teaching CSE260. There is more material than can be covered in one
semester and some choices have to made as to what to omit.
Unfortunately, when I taught this course, I was unable to cover
any graph theory. I also did not cover
lattices and boolean algebras.

\medskip
My unconventional approach of starting with logic may not work
for everybody, as some individuals find such material
too abstract. It is possible to skip the chapter on logic
and proceed directly with sets functions, {\it etc.\/}
I admit that I have raised the bar perhaps higher than the average
compared to other books on discrete maths. However, my experience
when teaching CSE260 was that 70\% of the students enjoyed 
the logic material, as it reminded them of programming.
I hope that these notes will inspire and will be useful to motivated students.

\chapter{Mathematical Reasoning, Proof Principles and Logic}
\label{chap1}
\section{Introduction}
\label{sec1}
Mathematicians write proof; most of us write proofs.
This leads to the question: Which principles of reasoning do we use
when we write proofs?

\medskip
The goal of this Chapter is to try answering this question.
We do so by formalizing the basic rules of reasoning that we use,
most of the time unconsciously, in a certain kind of formalism
known as a {\it natural deduction system\/}. We give a (very) quick
introduction to {\it mathematical logic\/}, with a very deliberate
{\it proof-theoretic\/} bent, that is, neglecting almost
completely all semantic notions, except at a very intuitive level.
We still feel that this approach is  fruitful because
the mechanical and rules-of-the-game flavor of proof systems
is much more easily grasped than semantic concepts. In this approach, we follow
Peter Andrew's motto \cite{Andrews}:

\medskip
``To truth through proof''.

\medskip
We present various natural deduction systems due to Prawitz and Gentzen
(in more modern notation), both in their intuitionistic and classical version.
The adoption of natural deduction systems as proof systems makes
it easy to question the validity of some of
the inference rules, such as
the {\it principle of proof by contradiction\/}. In brief, we try to explain 
to our readers the difference between {\it constructive\/} and
{\it classical\/}  (i.e., not necessarily constructive) proofs.
In this respect, we plant the seed that there is a deep
relationship between {\it constructive proofs\/} and the 
notion of {\it computation\/}
(the ``Curry-Howard isomorphism'' or
``formulae--as--types principle'', see Section \ref{sec5}).

\section[Inference Rules, Deductions, The Proof Systems 
{$\s{N}^{\impl}_{m}$} and
{$\s{NG}^{\impl}_{m}$}]
{Inference Rules, Deductions, The Proof Systems 
$\s{N}^{\impl}_{m}$ and
$\s{NG}^{\impl}_{m}$}
\label{sec2}
In this section, we review some basic proof
principles and attempt to clarify, at least informally,
what constitutes a mathematical proof.

\medskip
In order to define the notion of proof rigorously,
we would have to define a formal language in which
to express statements very precisely and we would have to set up
a proof system  in terms of axioms   and 
proof rules (also called inference rules).
We will not go into this; this would take too much time
and besides, this belongs to a logic course, which
is not what CSE260 is!
Instead, we will content ourselves with an intuitive
idea of what a statement is and focus on stating
as precisely as possible the rules of logic that
are used in constructing proofs.
Readers who really want to see a thorough (and rigorous) introduction
to logic are referred to Gallier \cite{Gall86}
van Dalen \cite{vanDalen} or Huth and Ryan \cite{HuthRyan},
a nice text with a  Computer Science flavor. 
A beautiful exposition of logic (from a proof-theoretic point of view)
is also given in Troelstra and Schwichtenberg \cite{TroelstraSchwi},
but at a more advanced level.
You should also be aware of CSE482, a very exciting course about logic and its
applications in Computer Science. 
By the way, my book has been out
of print for some time but you can get it free (as pdf files) 
from my logic web site

\medskip\noindent
http://www.cis.upenn.edu/$\tilde{\;}$jean/gbooks/logic.html

\medskip
In mathematics, we {\bf prove statements.}
Statements may be {\it atomic\/} or 
{\it compound\/}, that is, built up from
simpler statements using {\it logical connectives\/}, such
as, {\it implication\/} (if--then), {\it conjunction\/} (and),
{\it disjunction\/} (or), {\it negation\/} (not) and
(existential or universal) {\it quantifiers\/}.  

\medskip
As examples of  atomic statements, we have:

\begin{enumerate}
\item
``a student is eager to learn''.
\item
``a students wants an A''.
\item
``an odd integer is never $0$''
\item
``the product of two odd integers is odd''
\end{enumerate}

Atomic statements may also contain ``variables''
(standing for abitrary objects). For example

\begin{enumerate}
\item
$\hbox{human}(x)$: ``$x$ is a human''
\item
$\hbox{needs-to-drink}(x)$: ``$x$'' needs to drink
\end{enumerate}

An example of a compound statement is
\[
\hbox{human}(x) \impl \hbox{needs-to-drink}(x).
\]
In the above statement,  $\impl$ is the symbol used for
logical implication.
If we want to assert that every human needs to drink, we can write
\[
\forall x(\hbox{human}(x) \impl \hbox{needs-to-drink}(x));
\]
This is read: ``for every $x$, if $x$ is a human then $x$ needs to drink''.

\medskip
If we want to assert that some human needs to drink we write
\[
\exists x(\hbox{human}(x) \impl \hbox{needs-to-drink}(x));
\]
This is read: ``for some $x$, if $x$ is a human then $x$ needs to drink''.

\medskip
We often denote statements (also called {\it propositions\/}
or {\it (logical) formulae\/}) using letters,
such as $A, B, P, Q$, etc., typically upper-case letters
(but sometimes greek letters, $\varphi$, $\psi$, etc.).

\medskip
If $P$ and $Q$ are statements, then their {\it conjunction\/}
is denoted $P \land Q$ (say: $P$ and $Q$), their {\it disjunction\/} denoted
$P \lor Q$ (say: $P$ or $Q$), 
their {\it implication\/} $P \impl Q$  or
$P \supset Q$ (say: if $P$ then $Q$). 
Some authors use the symbol $\rightarrow$ and write an implication
as $P \rightarrow Q$. We do not like to use this notation
because the symbol $\rightarrow$ is already used in the notation for functions
($\mapdef{f}{A}{B}$). We will mostly use the symbol $\impl$.

\medskip
We also have the atomic statements $\perp$ ({\it falsity\/}), which
corresponds to {\bf false} (think of it as the
statement which is false no matter what), and
the atomic statement $\top$ ({\it truth\/}), which
corresponds to {\bf true} (think of it as the
statement which is always true).
The constant $\perp$ is also called {\it falsum\/}
or {\it absurdum\/}.
Then, it is convenient to define the {\it negation\/}
of $P$ as $P\impl \perp$ and to abbreviate it as
$\neg P$ (or sometimes $\sim P$). Thus, $\neg P$
(say: not $P$)
is just a shorthand for $P \impl \perp$.

\medskip
Whenever necessary to avoid ambiguities, we
add matching parentheses: $(P \land Q)$, $(P \lor Q)$,
$(P \impl Q)$. For example, $P \lor Q \land R$ is
ambigous; it means either
$(P \lor (Q \land R))$ or $((P \lor Q) \land R)$. 

\medskip
Another important logical operator is 
{\it equivalence\/}. If $P$ and $Q$ are statements, then
their {\it equivalence\/}, denoted
$P \equiv Q$ (or $P \Longleftrightarrow Q$), is an abbreviation
for $(P \impl Q) \land (Q\impl P)$. 
We often say ``$P$ if and only if $Q$''
or even ``$P$ iff $Q$'' for  $P\equiv Q$.
As we will see shortly, to prove a logical equivalence,
$P\equiv Q$, we have to prove {\bf both} implications
$P \impl Q$ and $Q\impl P$.

\medskip
An implication $P \impl Q$ should be understood as an if--then
statement, that is, if $P$ is true then $Q$ is also true.
So, the meaning of negation is that if $\neg P$
holds then $P$ must be false. Otherwise,
as $\neg P$ is really $P \impl \perp$, if $P$
were true, then $\perp$ would have to be true, but this is
absurd. 

\medskip
Of course, there are  problems with the above paragraph.
What does truth have to do with all this?
What do we mean when we say ``$P$ is true''?
What is the relationship between truth and provability?

\medskip
These are actually deep (and tricky!) questions whose answers
are not so obvious. One of the major roles of logic
is to clarify the notion of truth and its
relationship to provability. 
We will avoid  these fundamental
issues by dealing exclusively with the notion
of proof.
So, the big question is: What is a proof?

\medskip
Typically, the statements that we prove depend on
some set of {\it hypotheses\/}, also called {\it premises\/}
(or {\it assumptions\/}).
As we shall see shortly, this amounts to proving
implications of the form
\[
(P_1\land P_2\land \cdots \land P_n ) \impl Q.
\]
However, there are certain advantages in defining the notion of
{\it proof\/} (or {\it deduction\/}) of a proposition
from a set of premises.
Sets of premises are usually denoted using upper-case
greek letters such as $\Gamma$ or $\Delta$.

\medskip
Roughly speaking, a {\it deduction\/} of a proposition $Q$ from a
set of premises $\Gamma$ is a finite  labeled tree whose
root is labeled with $Q$ (the {\it conclusion\/}), whose leaves
are labeled with premises from $\Gamma$ (possibly
with multiple occurrences), and such that
every interior node corresponds to a given set of
{\it proof rules\/} (or {\it inference rules\/}).
Certain simple deduction trees are declared as
obvious proofs, also called {\it axioms\/}.

\medskip
There are many kinds of proofs systems: Hilbert-style systems,
Natural-deduction systems, Gentzen sequents systems, etc.
We describe a so-called {\it natural-deduction system\/} invented by
G. Gentzen in the early 1930's
(and thoroughly investigated by D. Prawitz in the mid
1960's).
The major advantage of this system is that it 
captures quite nicely the ``natural'' rules of reasoning that 
one uses when proving mathematical statements.
This does not mean that it is easy to find proofs in such a system
or that this system is indeed very intuitive!
We begin with the inference rules for implication.

\medskip
In the definition below, the expression $\Gamma, P$ stands for the union
of $\Gamma$ and $P$. So, $P$ may already belong to $\Gamma$.
A picture such as

\settree1{
     \starttree{$\Delta$}
     \step{$P$}
}

\bigskip
\ligne{\hfill\box1\hfill}
\noindent
represents a deduction tree whose root is labeled
with $P$ and whose leaves are labeled with propositions
from $\Delta$ (possibly with multiples occurrences).
Some of the  propositions in $\Delta$ may be tagged be variables.
The list of untagged propositions in $\Delta$ is the list of {\it premises\/}
of the deduction tree.
For example, in the deduction tree below, 

\settree1{
     \starttree{$P \impl (R \impl S)$}
}
\settree2{
     \starttree{$P$}
}
\settree1{
     \njointrees{1}{2}
     \step{$R \impl S$}
}
\settree2{
     \starttree{$Q \impl R$}
}
\settree3{
     \starttree{$P \impl Q$}
}
\settree4{
     \starttree{$P$}
}
\settree3{
     \njointrees{3}{4}
     \step{$Q$}
}
\settree2{
     \njointrees{2}{3}
     \step{$R$}
}
\settree1{
     \njointrees{1}{2}
     \step{$S$}
}

\bigskip
\ligne{\hfill\box1\hfill}

\medskip\noindent
no leaf is tagged, so the premises  form 
the set
\[
\Delta = \{P \impl (R \impl S), P,  Q \impl R, P \impl Q\},
\]
with two occurrences of $P$, and the conclusion is $S$.

\medskip
Certain inferences rules have the effect
that some of the original premises may be
discarded; the traditional jargon is 
that some premises may be {\it discharged\/} (or {\it closed\/}).
This this the case for the inference rule whose
conclusion in an implication. When  one or several occurrences
of some proposition, $P$, are discharged by an inference rule,
these occurrences (which label some leaves) are tagged with 
some new variable not already appearing in the deduction tree.
If  $x$ is a new tag, the tagged occurrences of $P$ are denoted $P^x$
and we indicate the fact that premises were discharged by that inference 
by writing $x$ immediately to the right of the
inference bar. For example,

\settree1{
     \starttree{$P^{x}, Q$}
     \step{$Q$}
     \jstep{$P \impl Q$}{${\scriptstyle x}$}
}

\bigskip
\ligne{\hfill\box1\hfill}

\medskip\noindent
is a deduction tree in which the premise $P$ is discharged  by
the inference  rule. This deduction tree only has $Q$ as a premise,
since $P$ is discharged.

\medskip
What is the meaning of the horizontal bars?
Actually, nothing really! Here, we are victims of an old habit in logic.
Observe that there is always a single proposition immediately under
a bar but there may be several propositions immediately above a bar.
The intended meaning of the bar is that the proposition below it
is obtained as the result of applying an inference rule to the
propositions above it. For example, in

\settree1{
     \starttree{$Q \impl R$}
}
\settree2{
     \starttree{$Q$}
}
\settree1{
     \njointrees{1}{2}
     \step{$R$}
}

\bigskip
\ligne{\hfill\box1\hfill}

\medskip\noindent
the proposition $R$ is the result of applying the $\impl$-elimination
rule (see Definition \ref{implog1} below) to the two premises
$Q \impl R$ and $Q$. Thus, the use of the bar is just a convention
used by logicians going back at least to the 1900's.
Removing the bar everywhere would not change anything to our trees, except
perhaps reduce their readability!
Since most logic books draw proof trees using bars to indicate
inferences, we also use bars in depicting our proof trees.

\medskip
Since propositions do not arise from the vacuum but instead
are built up from a set of atomic propositions using
logical connectives (here, $\impl$), we assume the
existence of an ``official set of atomic propositions'', 
$\mathbf{PS} = \{\mathbf{P}_1,  \mathbf{P}_2, \mathbf{P}_3, \cdots\}$.
So, for example, $\mathbf{P}_1 \impl \mathbf{P}_2$
and $\mathbf{P}_1 \impl (\mathbf{P}_2 \impl \mathbf{P}_1)$
are propositions. Typically, we will use upper-case letters such as
$P, Q, R, S, A, B, C$, etc., to denote arbitrary propositions
formed using atoms from $\mathbf{PS}$.

\begin{defin}
\label{implog1}
{\em 
The axioms and inference rules for {\it implicational logic\/}
are:

\settree1{
     \starttree{$\Gamma, P$}
     \step{$P$}
}

\bigskip
\ligne{\hfill\box1\hfill}

\bigskip
The above is a concise way of denoting a tree whose leaves
are labeled with $P$ and the propositions
in $\Gamma$, each of these proposition (including $P$) 
having possibly multiple occurrences but at least one, 
and whose root is labeled with $P$. 
A more explicit form is

\settree1{
     \starttree{$\overbrace{P_1, \cdots, P_{1}}^{k_1}, \cdots,
\overbrace{P_i, \cdots, P_{i}}^{k_i}, \cdots, 
\overbrace{P_n, \cdots, P_{n}}^{k_n}$}
     \step{$P_i$}
}

\bigskip
\ligne{\hfill\box1\hfill}

\bigskip\noindent
where $k_1, \ldots,  k_n \geq 0$,  $n \geq 1$ and 
$k_i \geq 1$ for some $i$ with $1\leq i \leq n$.
This axiom says that we always have a deduction of $P_i$ from
any set of premises including $P_i$.

\medskip
The {\it $\impl$-introduction rule\/}:

\settree1{
     \starttree{$\Gamma, P^{x}$}
     \step{$Q$}
     \jstep{$P \impl Q$}{${\scriptstyle x}$}
}

\bigskip
\ligne{\hfill\box1\hfill}

\medskip
This inference rule says that if there is a deduction
of $Q$ from the premises in $\Gamma$ and from the premise $P$, then
there is a deduction of $P \impl Q$ from $\Gamma$.
Note that this inference rule has the additional
effect of discharging some occurrences of the premise $P$.
These occurrences are tagged with a new variable, $x$,
and the tag $x$ is also placed immediately to the right
of the inference bar.
This is a reminder that the deduction tree whose conclusion is
$P\impl Q$ no longer has the occurrences of $P$
labeled with  $x$ as premises.

\medskip
The {\it $\impl$-elimination rule\/}:

\settree1{
     \starttree{$\Gamma$}
     \step{$P \impl Q$}
}
\settree2{
     \starttree{$\Delta$}
     \step{$P$}
}
\settree1{
     \njointrees{1}{2}
     \step{$Q$}
}

\bigskip
\ligne{\hfill\box1\hfill}

\medskip
This rule is also known as {\it modus ponens\/}.

\medskip
In the above axioms and rules, $\Gamma$ or $\Delta$ may be empty 
and $P, Q$ denote arbitrary propositions built up from the atoms
in $\mathbf{PS}$.
A {\it deduction tree\/} is a tree whose interior nodes
correspond to applications of the above inference rules.
A {\it proof tree\/} is a deduction tree such that
{\it all its premises are discharged\/}.
The above proof system is denoted $\s{N}^{\impl}_m$
(here, the subscript $m$ stands for {\it minimal\/}, referring to the fact
that this a bare-bone  logical system).
}
\end{defin}

\medskip
In words, the $\impl$-introduction rule says that in order to prove
an implication $P\impl Q$ from a set of premises $\Gamma$, we
assume that $P$ has already been proved, add $P$ to the premises
in $\Gamma$ and then prove $Q$ from $\Gamma$ and $P$.
Once this is done, the premise $P$ is deleted.
This rule formalizes the kind of reasoning that we all perform
whenever we prove an implication statement. In that sense,
it is a natural and familiar rule, except that we perhaps never
stopped to think about what we are really doing.
However, the business about discharging the premise $P$
when we are through with our argument is a bit puzzling. 
Most people probably never carry out this ``discharge step''
consciously, but such a process does takes place implicitely. 

\medskip
It might help to view the action of proving an implication $P\impl Q$
as the construction of a program that converts a proof of
$P$ into a proof of $Q$. Then, if we supply a proof of $P$
as input to this program (the proof of $P\impl Q$), it will
output a proof of $Q$. So, if we don't give the right kind
of input to this program, for example, a ``wrong proof'' of $P$, 
we should not expect that the program retun
a proof of $Q$. However, this does not say that
the program is incorrect; the program was designed to do the right
thing only if it is given the right kind of input. 
From this functional point of view
(also called, constructive), if we take the simplistic
view that $P$ and $Q$  assume the truth values {\bf true} and {\bf false},
we should not be shocked that if we give as input the value {\bf false} (for $P$), 
then the truth value of the whole implication
$P\impl Q$ is {\bf true}. The program $P\impl Q$ is designed to produce
the output value {\bf true} (for $Q$) if it is given the input value {\bf true}
(for $P$). So, this program only goes wrong when,  given the input {\bf true} (for $P$),
it returns the value {\bf false} (for $Q$). In this erroneous case,
$P \impl Q$ should indeed receive the value {\bf false}. However, in all other cases,
the program works correctly, even if it is given the wrong input ({\bf false} for $P$).

\danger

\begin{enumerate}
\item
Only the leaves of a deduction tree may be discharged. Interior nodes, including
the root, are {\it never\/} discharged.
\item
Once a set of leaves labeled with some premise $P$ marked with the label $x$
has been discharged,  none of these leaves can be discharged again.
So, each label (say $x$) can only be used once. This corresponds to the
fact that some leaves of our deduction trees get ``killed off'' (discharged).
\item
A proof is deduction tree whose leaves are {\it all discharged\/} ($\Gamma$ is empty).
This corresponds to the philosophy that  if a proposition has been proved,
then the validity of the proof should not depend on any assumptions 
that are still active. We may think of a deduction tree as an 
unfinished proof tree. 
\item
When constructing a proof tree, we have to be careful not to
include (accidently) extra premises that end up  not beeing discharged.
If this happens, we probably made a mistake and the redundant 
premises should be deleted.
On the other hand, if we have a proof tree, we can always add
extra premises to the leaves and create a new proof tree from the
previous one by discharging all the new premises. 
\item
Beware, when we deduce that an implication $P\impl Q$ is
provable, we {\bf do not} prove that $P$ {\bf and} $Q$
are provable; we only prove that {\bf if} $P$ is provable {\bf then}
$Q$ is provable. 
\end{enumerate}

\medskip
The  $\impl$-elimination rule formalizes the use of {\it auxiliary
lemmas\/}, a mechanism that we use all the time in making
mathematical proofs. Think of $P\impl Q$ as a lemma 
that has already been established and belongs to some
data base of (useful) lemmas. This lemma says if I can prove $P$ then I 
can prove $Q$. Now, suppose that we manage to give a proof of $P$.
It follows from the $\impl$-elimination rule that $Q$ is also provable.

\medskip
Observe that in an introduction rule, the conclusion
contains the logical connective associated
with the rule, in this case, $\impl$; this
jutifies the terminology ``introduction''.
On the other hand, in an elimination rule, the 
logical connective associated
with the rule is gone (although it may still appear in $Q$).
The other inference rules for $\land$, $\lor$, etc.,
will follow this pattern of introduction and elimination.

\medskip
{\bf Examples of proof trees}.

\medskip
(a)

\settree1{
     \starttree{$P^{x}$}
     \step{$P$}
     \jstep{$P\impl P$}{$\scriptstyle x$}
}

\bigskip
\ligne{\hfill\box1\hfill}

\bigskip
So, $P\impl P$ is provable; this is the least we should expect 
from our proof system!

\medskip
(b)

\settree1{
     \starttree{$(P\impl Q)^{z}$}
}
\settree2{
     \starttree{$P^{x}$}
}

\settree3{
     \starttree{$(Q\impl R)^{y}$}
}
\settree1{
     \njointrees{1}{2}
     \step{$Q$}
}
\settree1{
     \njointrees{3}{1}
     \step{$R$}
     \jstep{$P \impl R$}{$\scriptstyle x$}
     \jstep{$(Q\impl R)\impl (P \impl R)$}{$\scriptstyle y$}
     \jstep{$(P\impl Q) \impl ((Q\impl R)\impl (P \impl R))$}
{$\scriptstyle z$}
}

\bigskip
\ligne{\hfill\box1\hfill}

\bigskip
In order to better appreciate the difference between a deduction tree and a proof
tree, consider the following two examples:

\medskip
1. 
The tree below is a deduction tree, since two its leaves are labeled
with the premises $P\impl Q$ and $Q\impl R$, that have not been
discharged yet. So, this tree represents a deduction of $P \impl R$
from the set of premises $\Gamma = \{P\impl Q, Q\impl R\}$
but it is {\it not a proof tree\/} since $\Gamma\not= \emptyset$.
However, observe that the original premise, $P$, labeled $x$, has been
discharged. 

\settree1{
     \starttree{$P\impl Q$}
}
\settree2{
     \starttree{$P^{x}$}
}

\settree3{
     \starttree{$Q\impl R$}
}
\settree1{
     \njointrees{1}{2}
     \step{$Q$}
}
\settree1{
     \njointrees{3}{1}
     \step{$R$}
     \jstep{$P \impl R$}{$\scriptstyle x$}
}

\bigskip
\ligne{\hfill\box1\hfill}

\bigskip
2. The next tree was obtained from  the previous one by
applying the $\impl$-introduction rule which
triggered the discharge of the premise $Q\impl R$ labeled $y$,
which is no longer active. However, the premise $P\impl Q$ is
still active (has not been discharged, yet), so
the tree below is a deduction tree of $(Q\impl R)\impl (P \impl R)$
from the set of premises $\Gamma = \{P\impl Q\}$.
It is not yet a proof tree since $\Gamma\not= \emptyset$.
 
\settree1{
     \starttree{$P\impl Q$}
}
\settree2{
     \starttree{$P^{x}$}
}

\settree3{
     \starttree{$(Q\impl R)^{y}$}
}
\settree1{
     \njointrees{1}{2}
     \step{$Q$}
}
\settree1{
     \njointrees{3}{1}
     \step{$R$}
     \jstep{$P \impl R$}{$\scriptstyle x$}
     \jstep{$(Q\impl R)\impl (P \impl R)$}{$\scriptstyle y$}
}

\bigskip
\ligne{\hfill\box1\hfill}

\bigskip
Finally, one more application of the $\impl$-introduction rule 
will discharged the premise $P\impl Q$, at last, yielding
the proof tree in (b). 

\bigskip
(c)
In the next example, the two occurrences of $A$ labeled
$x$ are discharged simultaneously.

\settree1{
     \starttree{$(A\impl (B\impl C))^{z}$}
}
\settree2{
     \starttree{$(A\impl B)^{y}$}
}
\settree3{
     \starttree{$A^{x}$}
}
\settree4{
     \starttree{$A^{x}$}
}
\settree1{
     \njointrees{1}{3}
     \step{$B\impl C$}
}
\settree2{
     \njointrees{2}{4}
     \step{$B$}
}
\settree1{
     \njointrees{1}{2}
     \step{$C$}
     \jstep{$A\impl C$}{${\scriptstyle x}$}
     \jstep{$(A\impl B)\impl (A\impl C)$}{${\scriptstyle y}$}
     \jstep{$\bigl(A\impl (B\impl C)\bigr)\impl
                \bigl((A\impl B)\impl (A\impl C)\bigr)$}
                      {${\scriptstyle z}$}
}

\bigskip
\ligne{\hfill\box1\hfill}

\bigskip
(d) 
In contrast to Example (c), in the proof
tree below the two occurrences
of $A$ are discharded separately. To this effect, they
are labeled differently.

\settree1{
     \starttree{$(A\impl (B\impl C))^{z}$}
}
\settree2{
     \starttree{$(A\impl B)^{y}$}
}
\settree3{
     \starttree{$A^{x}$}
}
\settree4{
     \starttree{$A^{t}$}
}
\settree1{
     \njointrees{1}{3}
     \step{$B\impl C$}
}
\settree2{
     \njointrees{2}{4}
     \step{$B$}
}
\settree1{
     \njointrees{1}{2}
     \step{$C$}
     \jstep{$A\impl C$}{${\scriptstyle x}$}
     \jstep{$(A\impl B)\impl (A\impl C)$}{${\scriptstyle y}$}
     \jstep{$\bigl(A\impl (B\impl C)\bigr)\impl
                \bigl((A\impl B)\impl (A\impl C)\bigr)$}
                      {${\scriptstyle z}$}
     \jstep{$A\impl \Bigl(\bigl(A\impl (B\impl C)\bigr)\impl
                \bigl((A\impl B)\impl (A\impl C)\bigr)\Bigr)$}
                      {${\scriptstyle t}$}
}

\bigskip
\ligne{\hfill\box1\hfill}

\bigskip
\remark
How do we find these proof trees? Well, we could try to enumerate all
possible proof trees systematically and see if a proof of the desired conclusion
turns up. Obviously, this is a very inefficient procedure and moreover,
how do we know that all possible proof trees will be generated and how do 
we know that such a method will terminate after a finite number of steps
(what if the proposition proposed as a conclusion of a proof is  not provable)?
This is a very difficult problem and, in general, it can be shown that
there is {\bf no} procedure that will give an answer in all cases and 
terminate in a finite number of steps for all possible input propositions.
We will come back to this point in Section \ref{sec5}. However, for the
system $\s{N}^{\impl}_m$, such a procedure exists, but it is not easy to prove
that it terminates in all cases and in fact, it can take a very long time.

\medskip
What we did, and we strongly advise our readers to try it when they
attempt to construct proof trees, is to
construct the proof tree from the bottom-up,  starting from the
proposition labeling the root, rather than top-down, i.e.,  starting from the leaves.
During this process, whenever we are trying to prove a proposition
$P \impl Q$, we use the $\impl$-introduction rule backward, i.e.,
we add $P$ to the set of active premises and we try to prove
$Q$ from this new set of premises.  At some point, we get stuck with an atomic
proposition, say $Q$. Call the resulting deduction $\s{D}_{bu}$; note
that $Q$ is the only active (undischarged) premises of $\s{D}_{bu}$
and the node labeled $Q$ immediately below it plays a special role; we will
call it the special node of $\s{D}_{bu}$.
The trick is to now switch strategy and start building
a proof tree top-down, starting from the  leaves, using the
$\impl$-elimination rule. If everything works out well, we get a deduction
with root $Q$, say $\s{D}_{td}$, and then we glue this deduction 
$\s{D}_{td}$ to the deduction $\s{D}_{bu}$ in such a way that
the root of $\s{D}_{td}$ is identified with the special node of  $\s{D}_{bu}$
labeled $Q$. We also have to make sure that all the discharged premises
are linked to the correct instance of the $\impl$-introduction rule
that caused them to be discharged.  One of the difficulties is that
during the bottom-up process, we don't know how many copies of a premise
need to be discharged in a single step. We only find out how many
copies of a premise need to be discharged during the top-down process.

\medskip
Here is an illustration of this method for our third example.
At the end of the bottom-up process, we get the deduction tree $\s{D}_{bu}$:

\settree1{
     \starttree{$(A\impl (B\impl C))^{z}$}
}
\settree3{
     \starttree{$(A\impl B)^{y}$}
}
\settree2{
     \starttree{$A^{x}$}
}
\settree4{
     \starttree{$C$}
}
\settree1{
     \njointrees{1}{3}
}
\settree1{
     \njointrees{1}{2}
}
\settree1{
     \njointrees{1}{4}
     \step{$C$}
     \jstep{$A\impl C$}{${\scriptstyle x}$}
     \jstep{$(A\impl B)\impl (A\impl C)$}{${\scriptstyle y}$}
     \jstep{$\bigl(A\impl (B\impl C)\bigr)\impl
                \bigl((A\impl B)\impl (A\impl C)\bigr)$}
                      {${\scriptstyle z}$}
}

\bigskip
\ligne{\hfill\box1\hfill}

\bigskip
At the end of the top-down process, we get the deduction tree $\s{D}_{td}$:

\settree1{
     \starttree{$A\impl (B\impl C)$}
}
\settree2{
     \starttree{$A\impl B$}
}
\settree3{
     \starttree{$A$}
}
\settree4{
     \starttree{$A$}
}
\settree1{
     \njointrees{1}{3}
     \step{$B\impl C$}
}
\settree2{
     \njointrees{2}{4}
     \step{$B$}
}
\settree1{
     \njointrees{1}{2}
     \step{$C$}
}

\bigskip
\ligne{\hfill\box1\hfill}

\bigskip
Finally, after glueing $\s{D}_{td}$ on top of $\s{D}_{bu}$ (which has the correct
number of premises to be discharged), we get our proof tree:

\settree1{
     \starttree{$(A\impl (B\impl C))^{z}$}
}
\settree2{
     \starttree{$(A\impl B)^{y}$}
}
\settree3{
     \starttree{$A^{x}$}
}
\settree4{
     \starttree{$A^{x}$}
}
\settree1{
     \njointrees{1}{3}
     \step{$B\impl C$}
}
\settree2{
     \njointrees{2}{4}
     \step{$B$}
}
\settree1{
     \njointrees{1}{2}
     \step{$C$}
     \jstep{$A\impl C$}{${\scriptstyle x}$}
     \jstep{$(A\impl B)\impl (A\impl C)$}{${\scriptstyle y}$}
     \jstep{$\bigl(A\impl (B\impl C)\bigr)\impl
                \bigl((A\impl B)\impl (A\impl C)\bigr)$}
                      {${\scriptstyle z}$}
}

\bigskip
\ligne{\hfill\box1\hfill}

\bigskip

\medskip
Let us return to
the functional interpretation of implication by giving an example.
The proposition $P \impl ((P \impl Q)\impl Q)$ has the following proof:

\settree1{
     \starttree{$(P \impl Q)^x$}
}
\settree2{
     \starttree{$P^y$}
}
\settree1{
     \njointrees{1}{2}
     \step{$Q$}
     \jstep{$(P\impl Q) \impl Q$}{$\scriptstyle x$}
     \jstep{$P \impl ((P\impl Q) \impl Q)$}{$\scriptstyle y$}
}

\bigskip
\ligne{\hfill\box1\hfill}

\bigskip
Now, say $P$ is the proposition $R\impl R$, which has the proof

\settree1{
     \starttree{$R^z$}
     \step{$R$}
     \jstep{$R\impl R$}{$\scriptstyle z$}
}

\bigskip
\ligne{\hfill\box1\hfill}

\bigskip
Using $\impl$-elimination, 
we obtain a proof of $((R\impl R)\impl Q)\impl Q$
from the proof of 
$(R\impl R) \impl (((R\impl R)\impl Q) \impl Q)$
and the proof of $R\impl R$:

\settree1{
     \starttree{$((R\impl R) \impl Q)^x$}
}
\settree2{
     \starttree{$(R\impl R)^y$}
}
\settree1{
     \njointrees{1}{2}
     \step{$Q$}
     \jstep{$((R\impl R)\impl Q) \impl Q$}{$\scriptstyle x$}
     \jstep{$(R\impl R) \impl (((R\impl R)\impl Q) \impl Q)$}{$\scriptstyle y$}
}
\settree2{
     \starttree{$R^z$}
     \step{$R$}
     \jstep{$R\impl R$}{$\scriptstyle z$}
}
\settree1{
     \jointrees{1}{2}
     \step{$((R\impl R)\impl Q) \impl Q$}
}

\bigskip
\ligne{\hfill\box1\hfill}

\bigskip
Note that the above proof is redundant. A more direct proof
can be obtained as follows:
Undo the last $\impl$-introduction
in the proof of $(R\impl R) \impl (((R\impl R)\impl Q) \impl Q)$:

\settree1{
     \starttree{$((R\impl R) \impl Q)^x$}
}
\settree2{
     \starttree{$R\impl R$}
}
\settree1{
     \njointrees{1}{2}
     \step{$Q$}
     \jstep{$((R\impl R)\impl Q) \impl Q$}{$\scriptstyle x$}
}

\bigskip
\ligne{\hfill\box1\hfill}

\bigskip\noindent
and then glue the proof of  $R\impl R$ on top of the leaf $R\impl R$, obtaining the
desired proof of $((R\impl R)\impl Q) \impl Q$:

\settree1{
     \starttree{$((R\impl R) \impl Q)^x$}
}
\settree2{
     \starttree{$R^z$}
     \step{$R$}
     \jstep{$R\impl R$}{$\scriptstyle z$}
}
\settree1{
     \njointrees{1}{2}
     \step{$Q$}
     \jstep{$((R\impl R)\impl Q) \impl Q$}{$\scriptstyle x$}
}

\bigskip
\ligne{\hfill\box1\hfill}

\bigskip\noindent
In general, one has to exercise care with the label variables. 
It may be necessary to rename some of these variables to avoid clashes.
What we have above is an example of {\it proof substitution\/} also
called {\it proof normalization\/}. We will come back to this topic in
Section \ref{sec5}.

\medskip
The process of discharging premises when constructing a deduction
is admittedly a bit confusing. Part of the problem is that
a deduction tree really represents the last of
a sequence of stages (corresponding to the application
of inference rules) during 
which the current set of ``active'' premises, that is, those premises that
have not yet been discharged (closed, cancelled) evolves (in fact, shrinks).
Some mechanism is needed to keep track of which premises are no longer
active and this is what this business of labeling premises with variables
achieves. Historically, this is the first mechanism that was invented.
However, Gentzen (in the 1930's) came up with an alternative solution which
is mathematically easier to handle. Moreover, it turns out that this notation
is also better suited to computer implementations,
if one wishes to implement an automated theorem prover.

\medskip
The point is to keep a record of all undischarged
assumptions at every stage of the deduction. Thus, a deduction
is now a tree whose nodes are labeled with expressions
of the form $\sequent{\Gamma}{P}$, called {\it sequents\/}, 
where $P$ is a proposition, and $\Gamma$ is a record 
of all undischarged assumptions at the stage of the deduction
associated with this node. 

\medskip
During the construction of a deduction tree, it is necessary to discharge
packets of assumptions consisting of one or more occurrences of the same proposition.
To this effect, it is convenient to tag packets of assumptions with labels, 
in order to discharge
the propositions in these packets in a single step. 
We use variables for the labels, and a packet labeled with $x$ consisting
of occurrences of the proposition $P$  is written as $x\co P$.
Thus, in a sequent $\sequent{\Gamma}{P}$, the expression
$\Gamma$ is any finite set 
of the form $x_1\co P_1,\ldots,x_m\co P_m$, where the $x_i$ are pairwise
distinct (but the $P_i$ need not be distinct).
Given  $\Gamma = x_1\co P_1,\ldots,x_m\co P_m$,
the notation $\Gamma, x\co P$ is only well defined when
$x\not= x_i$ for all $i$, $1\leq i\leq m$, in which case
it denotes the set $x_1\co P_1,\ldots,x_m\co P_m, x\co P$.

\medskip
Using sequents, the axioms and rules of Definition  \ref{implog2} 
are now expressed as follows:

\begin{defin}
\label{implog2} 
{\em
The axioms and inference rules of the system $\s{NG}^{\impl}_{m}$
({\it implicational logic, Gentzen-sequent style (the $\s{G}$ in 
$\s{NG}$ stands for Gentzen)\/})
are listed below:
$$\sequent{\Gamma, x\co P}{P}$$

$$\rulea{\sequent{\Gamma, x\co P}{Q}}
        {\sequent{\Gamma}{P\impl Q}}
        {(\impliesintr)}$$

$$\ruleb{\sequent{\Gamma}{P\impl Q}}
        {\sequent{\Gamma}{P}}
        {\sequent{\Gamma}{Q}}
        {(\implieselim)}$$
}
\end{defin}

\medskip
In an application of the rule (\impliesintr), 
observe that in the lower sequent,
the proposition $P$ (labeled $x$) 
is deleted from the list of premises occurring on the left-hand side
of the arrow in the upper sequent.
We say that the
proposition $P$ which appears as a hypothesis of the deduction
is {\it discharged\/} (or {\it closed\/}).
It is important to note that the ability to label packets consisting of
occurrences of the same proposition with different labels is essential, 
in order
to be able to have control over which groups of packets of assumptions
are discharged simultaneously. Equivalently, we could avoid
tagging packets of assumptions with variables if we assumed
that in a sequent $\sequent{\Gamma}{C}$, the expression $\Gamma$,
also called a {\it context\/}, is a {\it multiset\/} of propositions.

\medskip
Below we show  a proof of the third example given above in our new system.
Let
$$\Gamma = x\co A\impl (B\impl C), y\co A\impl B, z\co A.$$

\settree1{
     \starttree{$\sequent{\Gamma}
                {A\impl (B\impl C)}$}
}
\settree2{
     \starttree{$\sequent{\Gamma}
                {A\impl B}$}
}
\settree3{
     \starttree{$\sequent{\Gamma}
                {A}$}
}
\settree4{
     \starttree{$\sequent{\Gamma}
                {A}$}
}
\settree1{
     \njointrees{1}{3}
     \step{$\sequent{\Gamma}
                {B\impl C}$}
}
\settree2{
     \njointrees{2}{4}
     \step{$\sequent{\Gamma}
                {B}$}
}
\settree1{
     \njointrees{1}{2}
     \step{$\sequent{x\co A\impl (B\impl C), y\co A\impl B, z\co A}%
                {C}$}
     \step{$\sequent{x\co A\impl (B\impl C), y\co A\impl B}%
                {A\impl C}$}
     \step{$\sequent{x\co A\impl (B\impl C)}%
                {(A\impl B)\impl (A\impl C)}$}
     \step{$\sequent{}{\bigl(A\impl (B\impl C)\bigr)\impl
                \bigl((A\impl B)\impl (A\impl C)\bigr)}$}
}

\bigskip
\ligne{\hfill\box1\hfill}

\bigskip
In principle, it does not matter which of the two systems  $\s{N}^{\impl}_{m}$
or  $\s{NG}^{\impl}_{m}$ we use to construct deductions; it is a matter of taste.
My experience is that I make fewer mistakes with the Gentzen-sequent style
system  $\s{NG}^{\impl}_{m}$.

\medskip
We now describe the inference rules dealing with the connectives
$\land$, $\lor$ and $\perp$.

\section[Adding $\land$, $\lor$, $\perp$; The Proof Systems 
{$\s{N}^{\impl, \land, \lor, \perp}_c$} and 
{$\s{NG}^{\impl, \land, \lor, \perp}_c$}]
{Adding $\land$, $\lor$, $\perp$; The Proof Systems 
{$\s{N}^{\impl, \land, \lor, \perp}_c$} and 
{$\s{NG}^{\impl, \land, \lor, \perp}_c$}}
\label{sec3}
Recall that $\neg P$ is an abbreviation for $P \impl \perp$.

\begin{defin}
\label{propsys1}
{\em 
The axioms and inference rules for {\it (propositional) classical logic\/}
are:

\medskip
Axioms:

\settree1{
     \starttree{$\Gamma, P$}
     \step{$P$}
}

\bigskip
\ligne{\hfill\box1\hfill}

\medskip
The {\it $\impl$-introduction rule\/}:

\settree1{
     \starttree{$\Gamma, P^{x}$}
     \step{$Q$}
     \jstep{$P \impl Q$}{${\scriptstyle x}$}
}

\bigskip
\ligne{\hfill\box1\hfill}

\medskip
The {\it $\impl$-elimination rule\/}:

\settree1{
     \starttree{$\Gamma$}
     \step{$P \impl Q$}
}
\settree2{
     \starttree{$\Delta$}
     \step{$P$}
}
\settree1{
     \njointrees{1}{2}
     \step{$Q$}
}

\bigskip
\ligne{\hfill\box1\hfill}

\medskip
The {\it $\land$-introduction rule\/}:

\settree1{
     \starttree{$\Gamma$}
     \step{$P$}
}
\settree2{
     \starttree{$\Delta$}
     \step{$Q$}
}
\settree1{
     \njointrees{1}{2}
     \step{$P\land Q$}
}

\bigskip
\ligne{\hfill\box1\hfill}

\medskip
The {\it $\land$-elimination rule\/}:

\settree1{
     \starttree{$\Gamma$}
     \step{$P\land Q$}
     \step{$P$}
}
\settree2{
     \starttree{$\Gamma$}
     \step{$P\land Q$}
     \step{$Q$}
}

\bigskip
\ligne{\hfill\box1\qquad\qquad \box2\hfill}

\medskip
The {\it $\lor$-introduction rule\/}:

\settree1{
     \starttree{$\Gamma$}
     \step{$P$}
     \step{$P\lor Q$}
}
\settree2{
     \starttree{$\Gamma$}
     \step{$Q$}
     \step{$P\lor Q$}
}

\bigskip
\ligne{\hfill\box1\qquad\qquad \box2\hfill}

\medskip
The {\it $\lor$-elimination rule\/}:

\settree1{
     \starttree{$\Gamma$}
     \step{$P\lor Q$}
}
\settree2{
     \starttree{$\Delta, P^{x}$}
     \step{$R$}
}
\settree3{
     \starttree{$\Lambda, Q^{y}$}
     \step{$R$}
}

\settree1{
     \tnjointrees{1}{2}{3}
     \jstep{$R$}{${\scriptstyle x, y}$}
}

\bigskip
\ligne{\hfill\box1\hfill}

\medskip
The {\it $\perp$-elimination rule\/}:

\settree1{
     \starttree{$\Gamma$}
     \step{$\perp$}
     \step{$P$}
}

\bigskip
\ligne{\hfill\box1\hfill}

\medskip
The {\it proof-by-contradiction rule\/} (also known as
{\it reductio ad absurdum rule\/}, for short {\it RAA\/}):

\settree1{
     \starttree{$\Gamma, \neg P^{x}$}
     \step{$\perp$}
     \jstep{$P$}{${\scriptstyle x}$}
}

\bigskip
\ligne{\hfill\box1\hfill}

\bigskip
Since $\neg P$ is an abbreviation
for $P \impl \perp$, the  $\neg$-introduction rule is a special case
of the $\impl$-introduction rule (with $Q = \perp$).
However, it is worth stating it  explicitly:

\medskip
The {\it $\neg$-introduction rule\/}:

\settree1{
     \starttree{$\Gamma, P^{x}$}
     \step{$\perp$}
     \jstep{$\neg P$}{${\scriptstyle x}$}
}

\bigskip
\ligne{\hfill\box1\hfill}

\bigskip
Similarly, the $\neg$-elimination rule 
is a special case of  $\impl$-elimination
applied to  \\
$\neg P \> (= P \impl \perp)$ and $P$: 

\medskip
The {\it $\neg$-elimination rule\/}:

\settree1{
     \starttree{$\Gamma$}
     \step{$\neg P$}
}
\settree2{
     \starttree{$\Delta$}
     \step{$P$}
}
\settree1{
     \njointrees{1}{2}
     \step{$\perp$}
}

\bigskip
\ligne{\hfill\box1\hfill}

\bigskip
In the above axioms and rules, $\Gamma, \Delta$ or $\Lambda$ may be empty,
$P, Q, R$ denote arbitrary propositions built up from the atoms in $\mathbf{PS}$
and all the premises labeled $x$ are discharged.
A {\it deduction tree\/} is a tree whose interior nodes
correspond to applications of the above inference rules.
A {\it proof tree\/} is a deduction tree such that
{\it all its premises\/} are discharged.
The above proof system is denoted $\s{N}^{\impl, \land, \lor, \perp}_c$
(here, the subscript $c$ stands for {\it classical\/}).

\medskip
The system obtained by removing the proof-by-contradiction (RAA) rule
is called {\it (propositional) intuitionistic logic\/} and is denoted
$\s{N}^{\impl, \land, \lor, \perp}_i$.
The system obtained by deleting both the
$\perp$-elimination rule and the proof-by-contradiction rule is
called {\it (propositional) minimal logic\/} and 
is denoted $\s{N}^{\impl, \land, \lor, \perp}_m$.
}
\end{defin}

\medskip
The version of  $\s{N}^{\impl, \land, \lor, \perp}_c$ in terms of
Gentzen sequents is the following:

\begin{defin}
\label{propsys2}
{\em 
The axioms and inference rules of the system
$\s{NG}^{\impl, \land, \lor, \perp}_{i}$
(of {\it propositional classical logic, Gentzen-sequent style\/})
are listed below:
$$\sequent{\Gamma, x\co P}{P}$$
$$\rulea{\sequent{\Gamma, x\co P}{Q}}
        {\sequent{\Gamma}{P\impl Q}}
        {(\impliesintr)}$$

$$\ruleb{\sequent{\Gamma}{P\impl Q}}
        {\sequent{\Gamma}{P}}
        {\sequent{\Gamma}{Q}}
        {(\implieselim)}$$

$$\ruleb{\sequent{\Gamma}{P}}
        {\sequent{\Gamma}{Q}}
        {\sequent{\Gamma}{P\land Q}}
        {(\landintr)}$$

$$\rulea{\sequent{\Gamma}{P\land Q}}
        {\sequent{\Gamma}{P}}
        {(\landelim)}\qquad
  \rulea{\sequent{\Gamma}{P\land Q}}
        {\sequent{\Gamma}{Q}}
        {(\landelim)}
$$

$$\rulea{\sequent{\Gamma}{P}}
        {\sequent{\Gamma}{P\lor Q}}
        {(\lorintr)}\qquad
  \rulea{\sequent{\Gamma}{Q}}
        {\sequent{\Gamma}{P\lor Q}}
        {(\lorintr)}
$$

$$\rulec{\sequent{\Gamma}{P\lor Q}}
        {\sequent{\Gamma, x\co P}{R}}
        {\sequent{\Gamma, y\co Q}{R}}
        {\sequent{\Gamma}{R}}
        {(\lorelim)}$$

$$\rulea{\sequent{\Gamma}{\perp}}
        {\sequent{\Gamma}{P}}
        {(\perpelim)}$$

$$\rulea{\sequent{\Gamma, x\co \neg P}{\perp}}
        {\sequent{\Gamma}{P}}
        {(\bycontra)}$$

$$\rulea{\sequent{\Gamma, x\co P}{\perp}}
        {\sequent{\Gamma}{\neg P}}
        {($\neg$-introduction)}$$

$$\ruleb{\sequent{\Gamma}{\neg P}}
        {\sequent{\Gamma}{P}}
        {\sequent{\Gamma}{\perp}}
        {($\neg$-elimination)}$$

\medskip
Since the rule (\perpelim) is trivial (does nothing) when $P = \perp$,
from now on, we will assume that $P\not= \perp$.
{\it Propositional minimal  logic\/}, denoted
$\s{NG}^{\impl, \land, \lor, \perp}_{m}$,
is obtained by dropping the (\perpelim) and (\bycontra)  rules.
{\it Propositional intuitionistic logic\/}, 
denoted $\s{NG}^{\impl, \land, \lor, \perp}_{i}$, 
is obtained by dropping the (\bycontra) rule.
}
\end{defin}

\medskip
When we say that a proposition, $P$, is {\it provable from $\Gamma$\/},
we  mean that we can construct a proof tree whose
conclusion is $P$ and whose set of premises is $\Gamma$, in one of
the systems $\s{N}^{\impl, \land, \lor, \perp}_{c}$ or
$\s{NG}^{\impl, \land, \lor, \perp}_{c}$. 
Therefore, when we use the
word ``provable'' unqualified, we mean provable in {\it classical logic\/}.
If $P$ is provable from $\Gamma$ in one of the intuitionistic systems
$\s{N}^{\impl, \land, \lor, \perp}_{i}$ or
$\s{NG}^{\impl, \land, \lor, \perp}_{i}$, then we say 
{\it intuitionistically provable\/} (and similarly, if 
$P$ is provable from $\Gamma$ in one of the systems
$\s{N}^{\impl, \land, \lor, \perp}_{m}$ or
$\s{NG}^{\impl, \land, \lor, \perp}_{m}$, then we say 
{\it  provable in minimal logic\/}). When $P$ is provable from 
$\Gamma$, most people write
$\Gamma \vdash  P$, or $\vdash \sequent{\Gamma}{P}$, sometimes
with the name of the corresponding proof system tagged as a subscript on
the sign $\vdash$ if necessary to avoid ambiguities.
When $\Gamma$ is empty, we just say
$P$ is provable (provable in intuitionistic logic, etc.)
and write $\vdash P$.

\medskip
We treat {\it logical equivalence\/} as a derived connective, that is,
we view $P \equiv Q$ as an abbreviation for 
$(P\impl Q) \land (Q \impl P)$. In view of the inference rules for $\land$,
we see that to prove a logical equivalence $P\equiv Q$, we just have to prove
both implications $P\impl Q$ and $Q\impl P$.
 
\medskip
In view of the $\perp$-elimination rule, the best way to interpret the provability
of a negation, $\neg P$, is as ``$P$ is not provable''.
Indeed, if $\neg P$ and $P$ were both provable, then $\perp$ would be provable.
So, $P$ should not be provable if $\neg P$ is. This is not the usual
interpretation of negation in terms of truth values, but it turns out to be
the most fruitful. Beware that if $P$ is not provable, then $\neg P$ is 
{\bf not} provable in general! There are plenty of propositions such
that neither $P$ nor $\neg P$ is provable (for instance, $P$,
with $P$ an atomic proposition).

\medskip
Let us now make some (much-needed) comments about the
above inference rules. There is no need to repeat our comments
regarding the $\impl$-rules.

\medskip
The $\land$-introduction rule says that in order to prove
a conjunction $P \land Q$ from some premises $\Gamma$,
all we have to do is to prove {\it both\/} that
$P$ is provable from $\Gamma$ {\it and\/} 
that $Q$ is provable from $\Gamma$.
The $\land$-elimination rule says that once
we have proved $P\land Q$ from $\Gamma$, then $P$ (and $Q$)
is also provable from $\Gamma$. This makes sense intuitively
as $P\land Q$ is ``stronger'' than $P$ and $Q$ separately
($P\land Q$ is true iff both $P$ and $Q$ are true).

\medskip
The $\lor$-introduction rule says that if 
$P$ (or $Q$) has been proved from $\Gamma$, then $P\lor Q$ is
also provable from $\Gamma$. Again, this makes sense
intuitively as $P\lor Q$ is ``weaker'' than $P$ and $Q$.
The $\lor$-elimination rule formalizes the {\it proof-by-cases\/}
method.  It is a more subtle rule. The idea is that
if we know that in the case where $P$ is already 
assumed to be provable and similarly in the case where $Q$ is already 
assumed to be provable that we can prove $R$ (also 
using  premises in $\Gamma$), then if $P\lor Q$ is also
provable from $\Gamma$, as we have ``covered both cases'',
it should be possible to prove $R$ from $\Gamma$ only
(i.e., the premises $P$ and $Q$ are discarded).

\medskip
The $\perp$-elimination rule formalizes the principle
that once a false statement has been established, then
anything should be provable.

\medskip
The proof-by-contradiction rule formalizes the 
method of proof by contradiction! That is, in order
to prove that $P$ can be deduced from some premises $\Gamma$,
one may assume the negation, $\neg P$, of $P$ (intuitively, assume that
$P$ is false) and then derive a contradiction from $\Gamma$
and $\neg P$ (i.e., derive falsity). Then, $P$
actually follows from $\Gamma$ {\it without using
$\neg P$ as a premise\/}, i.e., $\neg P$ is discharged.

\medskip
Most people, I believe, will be comfortable with the rules
of minimal logic and will agree that they constitute
a ``reasonable'' formalization of the rules of 
reasoning involving $\impl$, $\land$ and $\lor$.
Indeed, these rules seem to
express the intuitive meaning of the connectives
$\impl$, $\land$ and $\lor$. However, 
some may question the two rules $\perp$-elimination
and proof-by-contradiction. Indeed, their meaning is not as
clear and, certainly, the proof-by-contradiction rule
introduces a form of indirect reasoning that is somewhat
worrisome.

\medskip
The problem has to do with the meaning of disjunction and negation 
and more
generally, with the notion of {\it constructivity\/} in mathematics.
In fact, in the early 1900's, some mathematicians,
especially L. Brouwer (1881-1966), questioned the validity
of the   proof-by-contradiction rule, among other principles.
Two specific cases illustrate the problem,
namely, the propositions
\[
P\lor \neg P\quad\hbox{and}\quad \neg \neg P \impl  P.
\]
As we will see shortly, the above propositions
are both provable in classical logic. 
Now,  Brouwer and some mathematicians belonging to his school
of thoughts (the so-called ``intuitionsists'' or
``constructivists'') advocate that in order to prove
a disjunction, $P \lor Q$ (from some premises $\Gamma$)
one has to either exhibit a proof of $P$ or a proof or $Q$
(from $\Gamma$). However, it can be shown that this
fails for $P \lor \neg P$. The fact that $P\lor \neg P$
is provable (in classical logic) {\bf does not\/} imply that either $P$
is provable or that $\neg P$ is provable!
That $P\lor \neg P$ is provable
is sometimes called the {\it principle of the excluded middle\/}!
In intuitionistic logic, $P\lor \neg P$ is {\bf not} provable.
Of course, if one gives up the proof-by-contradiction rule,
then fewer propositions become provable. On the other hand,
one may claim that the propositions that remain provable
have more constructive proofs and thus, feels on safer grounds.

\medskip
A similar controversy arises with  $\neg \neg P \impl  P$.
If we give up the proof-by-contradiction rule, then
this formula is no longer provable, i.e., 
$\neg\neg P$ is no longer equivalent to $P$.
Perhaps this relates to the fact that if one says

\medskip
`` I don't have no money''

\medskip\noindent
then this does not mean that this person has money!
(Similarly with ``I don't get no satisfaction'', ... ).
However, note that one can still prove $P \impl \neg\neg P$
in minimal logic (try doing it!).
Even stranger, $\neg\neg\neg P \impl \neg P$ is provable in 
intuitionistic (and minimal) logic, so  $\neg\neg\neg P$ and
$\neg P$ are equivalent intuitionistically!

\remark
Suppose we have a deduction

\settree1{
     \starttree{$\Gamma, \neg P$}
     \step{$\perp$}
}

\bigskip
\ligne{\hfill\box1\hfill}

\bigskip\noindent
as in the proof by contradiction rule.
Then, by $\neg$-introduction, we get a deduction of $\neg\neg P$
from $\Gamma$:

\settree1{
     \starttree{$\Gamma, \neg P^{x}$}
     \step{$\perp$}
     \jstep{$\neg\neg P$}{${\scriptstyle x}$}
}

\bigskip
\ligne{\hfill\box1\hfill}

\bigskip
So, if we knew that $\neg\neg P$ was equivalent to $P$ (actually,
if we knew that $\neg\neg P \impl P$ is provable)
then  the proof by contradiction rule would be justified 
as a valid rule (it follows from modus ponens).
We can view the proof by contradiction rule as
a sort of act of faith that consists in saying that if we can derive
an inconsistency (i.e., chaos) by assuming the falsity of a 
statement $P$, then $P$ has to hold in the first place.
It not so clear that such an act of faith is justified
and the intuitionists refuse to take it!

\medskip
Constructivity in mathematics is a fascinating subject but 
it is a topic that is really outside the scope of this course.
What we hope is that our brief and very incomplete
discussion of constructivity issues made the reader aware
that the rules of logic are not cast in stone and that,
in particular, there isn't {\bf only one\/} logic.

\medskip
We feel safe in  saying that most mathematicians
work with classical logic and only few of them
have  reservations about using the proof-by-contradiction
rule. Nevertherless, intuitionistic logic has its
advantages, especially when it comes to
proving the correctess of programs (a branch of computer science!).
We will come back to this point several times in this course.

\medskip
In the rest of this section, we make further useful remarks
about (classical) logic and give some explicit
examples of proofs illustrating the inference rules
of classical logic.
We begin by proving that $P \lor \neg P$ is provable
in classical logic.

\begin{prop}
\label{proplog1}
The proposition $P \lor \neg P$ is provable
in classical logic.
\end{prop}

\proof
We  prove that $P \lor (P \impl \perp)$ is provable
by using the proof-by-contradiction rule as shown below:
\settree2{
     \starttree{$((P \lor (P\impl \perp))\impl \perp)^{y}$}
}
\settree3{
     \starttree{$P^{x}$}
     \step{$P \lor (P\impl \perp)$}
}
\settree1{
     \starttree{$((P \lor (P\impl \perp))\impl \perp)^{y}$}
}
\settree2{
     \njointrees{2}{3}
     \step{$\perp$}
     \jstep{$P\impl \perp$}{${\scriptstyle x}$}
     \step{$P \lor (P\impl \perp)$}
}
\settree1{
     \njointrees{1}{2}
     \step{$\perp$}
     \jstep{$P \lor (P\impl \perp)$}{${\scriptstyle y}\>$ (by-contra)}
}

\bigskip
\ligne{\hfill\box1\hfill}
$\bigsquare$

\bigskip
Next, we consider the equivalence of $P$  and $\neg \neg P$.

\begin{prop}
\label{proplog2}
The proposition $P \impl \neg\neg P$ is provable
in minimal logic. The proposition $\neg \neg P \impl P$
is provable in classical logic. Therefore, 
in classical logic, $P$ is equivalent to $\neg\neg P$.
\end{prop}

\proof
We leave that $P \impl \neg\neg P$ is provable
in minimal logic as an exercise. Below is a proof of
$\neg\neg P\impl P$ 
using the proof-by-contradiction rule:

\settree1{
     \starttree{$((P \impl \perp)\impl \perp)^{y}$}
}
\settree2{
     \starttree{$(P \impl \perp)^{x}$}
}

\settree1{
     \njointrees{1}{2}
     \step{$\perp$}
     \jstep{$P$}{${\scriptstyle x}\>$ (by-contra)}
     \jstep{$((P \impl \perp)\impl \perp)\impl P$}{${\scriptstyle y}\>$}
}

\bigskip
\ligne{\hfill\box1\hfill}
$\bigsquare$

\medskip
The next proposition shows why $\perp$
can be viewed as the ``ultimate'' contradiction.

\begin{prop}
\label{proplog4}
In intuitionistic logic, the propositions
$\perp$ and $P \land \neg P$ are
equivalent for all $P$. Thus,
$\perp$ and $P \land \neg P$ are also equivalent
in classical propositional logic
\end{prop}

\proof
We need to show that both $\perp \impl (P \land \neg P)$
and $(P \land \neg P)\impl \perp$ are provable
in intuitionistic logic.
The provability of $\perp \impl (P \land \neg  P)$
is an immediate consequence or $\perp$-elimination,
with $\Gamma = \emptyset$.
For $(P \land \neg P)\impl \perp$, we have the following
proof:

\settree2{
     \starttree{$(P \land \neg P)^x$}
     \step{$P$}
}
\settree1{
     \starttree{$(P \land \neg P)^x$}
     \step{$\neg P$}
}

\settree1{
     \njointrees{1}{2}
     \step{$\perp$}
     \jstep{$(P \land \neg P)\impl \perp$}{${\scriptstyle x}$}
}

\bigskip
\ligne{\hfill\box1\hfill$\bigsquare$}

\medskip
So, in intuitionistic  logic (and also in classical logic),
$\perp$ is equivalent to $P \land \neg P$ for all $P$.
This means that $\perp$ is the ``ultimate'' contradiction,
it corresponds to total inconsistency.
By the way, we could have the bad luck that
the system $\s{N}^{\impl, \land, \lor, \perp}_c$
(or   $\s{N}^{\impl, \land, \lor, \perp}_i$
or even  $\s{N}^{\impl, \land, \lor, \perp}_m$)
is {\it inconsistent\/}, that is, that
$\perp$ is provable! Fortunately, this is not the case,
although this hard to prove.
(It is also the case that $P \lor \neg P$
and $\neg \neg P \impl P$ are {\bf not} provable
in intuitionistic logic, but this too is hard to prove!)

\section[Clearing Up  Differences Between 
 Rules Involving $\perp$]
{Clearing Up Differences Between \\
{$\neg$-introduction},
{$\perp$-elimination} and RAA}
\label{sec3b}
The differences between the rules, $\neg$-introduction,
$\perp$-elimination and the proof by contradiction rule (RAA)
are often unclear to the uninitiated reader and this
tends to cause confusion. In this section, we will try to
clear up some common misconceptions about these rules.

\medskip
{\bf Confusion 1}.
Why is RAA not a special case of $\neg$-introduction?

\settree1{
     \starttree{$\Gamma, P^{x}$}
     \step{$\perp$}
     \jstep{$\neg P$}{${\scriptstyle x\>}$($\neg$-intro)}
}

\settree2{
     \starttree{$\Gamma, \neg P^{x}$}
     \step{$\perp$}
     \jstep{$P$}{${\scriptstyle x}\>$(RAA)}
}

\bigskip
\ligne{\hfill\box1\qquad\qquad\qquad\qquad\qquad\qquad\qquad\box2\hfill}

\bigskip\noindent
The only apparent difference between $\neg$-introduction (on the left)
and RAA (on the right) is that in RAA, the premise $P$ is negated
but  the conclusion is not, 
whereas in $\neg$-introduction the premise $P$ is not negated
but the conslusion is.

\medskip
The important difference is that the conclusion of
RAA is {\bf not} negated. If  we had applied  $\neg$-introduction
instead of RAA on the right, we would have obtained

\settree1{
     \starttree{$\Gamma, \neg P^{x}$}
     \step{$\perp$}
     \jstep{$\neg\neg P$}{${\scriptstyle x\>}$($\neg$-intro)}
}

\bigskip
\ligne{\hfill\box1\hfill}

\bigskip\noindent
where the conclusion would have been $\neg\neg P$ as opposed to $P$.
However, as we already said earlier, $\neg\neg P \impl P$
is {\bf not} provable intuitionistically.
Consequenly, RAA {\bf is not} a special case of
$\neg$-introduction. 

\medskip
{\bf Confusion 2\/}.
Is there any  difference between  $\perp$-elimination and RAA?

\settree1{
     \starttree{$\Gamma$}
     \step{$\perp$}
     \jstep{$P$}{($\perp$-elim)}
}

\settree2{
     \starttree{$\Gamma, \neg P^{x}$}
     \step{$\perp$}
     \jstep{$P$}{${\scriptstyle x}\>$(RAA)}
}

\bigskip
\ligne{\hfill\box1\qquad\qquad\qquad\qquad\qquad\qquad\qquad\box2\hfill}

\bigskip
The difference is that  $\perp$-elimination does not discharge any of its premises.
In fact, RAA is a stronger rule
which implies $\perp$-elimination as we now demonstate.

\medskip
{\bf RAA implies $\perp$-elimination}.

\medskip
Suppose we have a deduction

\settree1{
     \starttree{$\Gamma$}
     \step{$\perp$}
}

\bigskip
\ligne{\hfill\box1\hfill}

\bigskip\noindent
Then, for any proposition $P$, we can add the premise $\neg P$ to every leaf of 
the above deduction tree and we get the deduction tree

\settree1{
     \starttree{$\Gamma, \neg P$}
     \step{$\perp$}
}

\bigskip
\ligne{\hfill\box1\hfill}

\bigskip\noindent
We can now apply RAA to get the following deduction tree of $P$ from
$\Gamma$ (since $\neg P$ is discharged), and this is just the result
of $\perp$-elimination:

\settree1{
     \starttree{$\Gamma, \neg P^{x}$}
     \step{$\perp$}
     \jstep{$P$}{${\scriptstyle x}\>$(RAA)}
}

\bigskip
\ligne{\hfill\box1\hfill}

\bigskip
The above considerations also show that
RAA is obtained from $\perp$-elimination
by adding the new rule of {\it $\neg\neg$-elimination\/}:

\settree1{
     \starttree{$\Gamma$}
     \step{$\neg\neg P$}
     \jstep{$P$}{($\neg\neg$-elimination)}
}

\bigskip
\ligne{\hfill\box1\hfill}

\bigskip
Some authors prefer adding the   $\neg\neg$-elimination rule
to intuitionistic logic instead of RAA 
in order to obtain classical logic. As we just demonstrated,
the two additions are equivalent: by adding either
RAA or $\neg\neg$-elimination to intuitionistic logic,
we get classical logic.

\medskip
There is another way to obtain RAA from the rules
of intuitionistic logic, this time,  using the propositions of the form
$P\lor \neg P$. 
We saw in Proposition \ref{proplog1} that all formulae of the form
$P\lor \neg P$ are provable in classical logic (using RAA).

\medskip
{\bf Confusion 3\/}.
Are propositions of the form $P \lor \neg P$ provable
in intuitionistic logic?

\medskip
The answer is {\bf no}, which may be disturbing to some readers.
In fact, it is quite difficult to prove that propositions
of the form  $P \lor \neg P$ are not provable in intuitionistic logic.
One method consists in using the fact that intuitionistic proofs
can be normalized (see Section \ref{sec5} for more on normalization of proofs).
Another method uses Kripke models (see van Dalen \cite{vanDalen}).

\medskip
Part of the difficulty in understanding at some intuitive level 
why propositions of the form  $P \lor \neg P$ are not provable 
in intuitionistic logic is that the notion of truth based on 
the truth values {\bf true} and {\bf false} is deeply rooted
in all of us. In this frame of mind, it seems ridiculous to
question the provability of $P\lor \neg P$, since its truth
value is {\bf true} whether $P$ is assigned the value {\bf true} or
{\bf false}. Classical two-valued truth value semantics is
too crude  for intuitionistic logic.

\medskip
Another difficulty is that it is tempting to equate the notion
of truth and the notion of provability. Unfortunately,
because classical truth value semantics is too crude
for intuitionistic logic, there are propositions that are
universally true (i.e., they evaluate to {\bf true} for all 
possible truth assignments of the atomic letters in them)
and yet they are {\bf not} provable intuitionistically.
The propositions $P\lor \neg P$ and $\neg\neg P \impl P$
are such examples.

\medskip
One of the major motivations for advocating intuitionistic logic
is that it yields proofs that are more constructive than classical proofs.
For example, in classical logic, when we prove a disjunction $P\lor Q$,
we generally can't conclude that either $P$ or $Q$ is provable,
as examplified by $P \lor \neg P$.
A more interesting example involving  a non-constructive
proof of a disjunction will be given in Section \ref{sec3c}.
But, in intuitionistic logic, from a proof of $P\lor Q$, it is possible to extract
either a proof of $P$ or a proof or $Q$ (and similarly for
existential statements, see Section \ref{sec4}).
This property is not easy to prove.
It is a consequence of the normal form for intuitionistic proofs
(see Section \ref{sec5}).

\medskip
In brief, besides being a fun intellectual game,
intuitionistic logic is only an interesting alternative
to classical logic if we care about the constructive nature
of our proofs. But then, we are forced to abandon the
classical two-valued truth value semantics and adopt
other semantics such as Kripke semantics.
If we do not care about the constructive nature of our proofs
and if we want to stick to  two-valued truth value semantics,
then we should stick to classical logic. 
Most people do that, so don't feel bad if you
are not comfortable with intuitionistic logic!

\medskip
One way to gauge how intuitionisic logic differs
from classical logic is to ask what kind of propositions
need to be added to intuitionisic logic in order to get
classical logic.
It turns out that if all the  propositions of the form
$P\lor \neg P$ are considered to be
axioms, 
then RAA follows from some of the rules of intuitionistic logic.

\medskip
{\bf RAA holds in Intuitionistic logic $+$ all axioms $P\lor \neg P$}.

\medskip
The proof involves a subtle use of the  $\perp$-elimination and
$\lor$-elimination rules  which may be a bit puzzling.
Assume, as we do when when use the proof by contradiction rule (RAA) that
we have a deduction

\settree1{
     \starttree{$\Gamma, \neg P$}
     \step{$\perp$}
}
 
\bigskip
\ligne{\hfill\box1\hfill}

\bigskip\noindent
Here is the deduction tree demonstrating that RAA is a derived rule:

\settree1{
     \starttree{$P \lor \neg P$}
}
\settree2{
     \starttree{$P^x$}
     \step{$P$}
}
\settree3{
     \starttree{$\Gamma, \neg P^y$}
     \step{$\perp$}
     \jstep{$P$}{($\perp$-elim)}
}
\settree1{
     \tjointrees{1}{2}{3}
     \jstep{$P$}{${\scriptstyle x, y\>}$ ($\lor$-elim)}     
}

\bigskip
\ligne{\hfill\box1\hfill}

\bigskip
At first glance, the rightmost subtree

\settree1{
     \starttree{$\Gamma, \neg P^y$}
     \step{$\perp$}
     \jstep{$P$}{($\perp$-elim)}
}
 
\bigskip
\ligne{\hfill\box1\hfill}

\bigskip\noindent
appears to use RAA and
our argument looks circular! But this is not so because
the premise $\neg P$ labeled $y$ is {\it not} discharged in the
step that yields $P$ as conclusion; the step that  yields $P$
is a $\perp$-elimination step.
The premise $\neg P$ labeled $y$ is actually discharged by the
$\lor$-elimination rule (and so is the premise $P$ labeled $x$).
So, our argument establishing RAA is not circular after all!

\medskip
In conclusion, intuitionistic logic is obtained
from classical logic by {\it taking away the proof by contradiction rule (RAA)\/}.
In this more restrictive proof system, we obtain more constructive
proofs. In that sense, the situation is better than in classical logic.
The major drawback is that we can't think in terms of classical truth value
semantics anymore. 

\medskip
Conversely, classical logic is obtained
from intuitionistic logic in at least three ways:
\begin{enumerate}
\item
Add the proof by contradiction rule (RAA).
\item
Add the $\neg\neg$-elimination rule.
\item
Add all propositions of the form $P \lor \neg P$ as axioms.
\end{enumerate}

\section{Other Rules of Classical Logic and Examples of Proofs}
\label{sec3c}
In classical logic, we have the de Morgan laws:

\begin{prop}
\label{proplog3}
The following equivalences (de Morgan laws) are provable
in classical logic:
\begin{align*}
\neg(P \land Q) & \equiv \neg P \lor \neg Q \\
\neg(P \lor Q) & \equiv \neg P \land \neg Q.
\end{align*}
In fact, $\neg(P \lor Q)  \equiv \neg P \land \neg Q$ and
$(\neg P \lor \neg Q)\impl \neg(P\land Q)$ are
provable in intuitionistic logic.
The proposition $(P \land \neg Q) \impl \neg(P \impl Q)$ is provable
in intuitionistic logic and 
$\neg(P \impl Q) \impl (P \land \neg Q)$ is provable in 
classical logic. Therefore, $\neg(P \impl Q)$ and $P \land \neg Q$
are equivalent in classical logic. Furthermore,
$P \impl Q$ and $\neg P \lor Q$
are equivalent in classical logic and
$(\neg P \lor Q) \impl (P \impl Q)$ is provable in intuitionistic logic.

\end{prop}

\proof
Here is an intuitionistic proof of  $(\neg P \lor Q) \impl (P \impl Q)$:

\settree1{
     \starttree{$(\neg P \lor Q)^w$}
}
\settree2{
     \starttree{$\neg P^{z}$}
}
\settree3{
     \starttree{$P^{x}$}
}
\settree2{
     \njointrees{2}{3}
     \step{$\perp$}
     \step{$Q$}
     \jstep{$P\impl Q$}{${\scriptstyle x}$}
}
\settree3{
     \starttree{$P^{y}$}
}
\settree4{
     \starttree{$Q^{t}$}
}
\settree3{
     \njointrees{3}{4}
     \step{$Q$}
     \jstep{$P\impl Q$}{${\scriptstyle y}$}
}
\settree1{
     \tjointrees{1}{2}{3}
     \jstep{$P \impl Q$}{${\scriptstyle z, t}$}
     \jstep{$(\neg P \lor Q) \impl (P \impl Q)$}{${\scriptstyle w}$}
}

\bigskip
\ligne{\hfill\box1\hfill}

\bigskip
Here is a classical proof of $(P \impl Q) \impl (\neg P \lor Q)$:

\settree1{
     \starttree{$(\neg (\neg P \lor Q))^y$}
}
\settree3{
     \starttree{$(\neg (\neg P \lor Q))^y$}
}
\settree4{
     \starttree{$\neg P^{x}$}
     \step{$\neg P \lor Q$}
}
\settree3{
     \njointrees{3}{4}
     \step{$\perp$}
     \jstep{$P$}{${\scriptstyle x}\>$ RAA}
}
\settree2{
     \starttree{$(P \impl Q)^z$}
}
\settree2{
     \njointrees{2}{3}
     \step{$Q$}
     \step{$\neg P \lor Q$}
}
\settree1{
     \njointrees{1}{2}
     \step{$\perp$}
     \jstep{$\neg P \lor Q$}{${\scriptstyle y}\>$ RAA}
     \jstep{$(P \impl Q)\impl (\neg P \lor Q)$}{${\scriptstyle z}$}
}

\bigskip
\ligne{\hfill\box1\hfill}

\bigskip
The other proofs are left as exercises.
$\bigsquare$

\medskip
Propositions \ref{proplog2} and \ref{proplog3} show a property
that is very specific to classical logic, namely, that the logical
connectives $\impl, \land, \lor, \neg$ are not independent.
For example, we have \\
$P \land Q \equiv \neg(\neg P \lor \neg Q)$, which shows that
$\land$ can be expressed in terms of $\lor$ and $\neg$.
In intuitionistic logic, $\land$ and $\lor$ cannot be expressed in terms
of each other via negation. 

\medskip
The fact that  the logical
connectives $\impl, \land, \lor, \neg$ are not independent in
classical logic suggests the following question: Are there propositions,
written in terms of $\impl$ only, that are provable classically but
not provable intuitionistically?

\medskip
The answer is yes! For instance, the proposition 
$((P \impl Q)\impl P) \impl P$ (known as {\it Pierce's law\/})
is provable classically (do it) but it can be shown that it is not
provable intuitionistically.

\medskip
In addition to the proof by cases method and the 
proof by contradiction method, we also have the
proof by contrapositive method valid in classical logic:

\medskip
{\it Proof by contrapositive rule\/}:

\settree1{
     \starttree{$\Gamma, \neg Q^{x}$}
     \step{$\neg P$}
     \jstep{$P \impl Q$}{${\scriptstyle x}$}
}

\bigskip
\ligne{\hfill\box1\hfill}

\bigskip
This rule says that in order to prove an implication
$P \impl Q$ (from $\Gamma$), one may assume $\neg Q$
as proved, and then deduce that $\neg P$ is provable
from $\Gamma$ and $\neg Q$. This inference rule is valid
in classical logic because we can construct the following proof:

\settree1{
     \starttree{$\Gamma, \neg Q^{x}$}
     \step{$\neg P$}
}
\settree2{
     \starttree{$P^y$}
}

\settree1{
     \njointrees{1}{2}
     \step{$\perp$}
     \jstep{$Q$}{${\scriptstyle x}\>$ (by-contra)}
     \jstep{$P \impl Q$}{${\scriptstyle y}\>$}
}

\bigskip
\ligne{\hfill\box1\hfill}

\bigskip
We will now give some explicit examples of proofs illustrating
the proof principles that we just discussed.

\medskip
Recall that the {\it set of integers\/} is the set
\[
\integs = \{\cdots, -2, -1, 0, 1, 2, \cdots\}
\]
and that the {\it set of natural numbers\/} is the set
\[
\natnums = \{0, 1, 2, \cdots\}.
\]
(Some authors exclude $0$ from $\natnums$. We don't
like this discrimination against zero.)
An integer is {\it even\/} if it is divisible by $2$, that is,
if it can be written as $2k$, where $k\in \integs$.
An integer is {\it odd\/} if it is not divisible by $2$, that is,
if it can be written as $2k + 1$, where $k\in \integs$.
The following facts are essentially obvious:

\begin{enumerate}
\item[(a)]
The sum of even integers is even.
\item[(b)]
The sum of an even integer and of an odd integer is odd.
\item[(c)]
The sum of two odd integers is even.
\item[(d)]
The product of odd integers is odd.
\item[(e)]
The product of an even integer with any integer is even.
\end{enumerate}
 
\medskip
Now, we prove the following fact using
the proof by cases method.

\begin{prop}
\label{proplog5}
Let $a, b , c$ be odd integers. For any
integers $p$ and $q$, if $p$ and $q$ are not  both even, then
\[
ap^2 + bpq + cq^2
\]
is odd.
\end{prop}

\proof
We consider the three cases:
\begin{enumerate}
\item
$p$ and $q$ are odd. In this case
as $a, b$ and $c$ are odd, by (d) all the  products
$ap^2, bpq$ and $cq^2$ are odd. By (c), $ap^2 + bpq$ is even and by
(b), $ap^2 + bpq + cq^2$ is odd.

\item
$p$ is even and $q$ is odd. In this case, by (e), both
$ap^2$ and $bpq$ are even and by (d), $cq^2$ is odd.
But then, by (a),   $ap^2 + bpq$ is even and by
(b), $ap^2 + bpq + cq^2$ is odd.

\item
$p$ is odd and $q$ is even. This case is analogous
to the previous case, except that $p$ and $q$ are interchanged.
The reader should have no trouble filling in the details.
\end{enumerate}

Since all three cases exhaust all possibilities
for $p$ and $q$ not to be both even, the proof
is complete by the   $\lor$-elimination rule (applied twice).
$\bigsquare$

\medskip
The set of rational numbers $\rats$ consists of all fractions
$p/q$, where $p, q\in \integs$, with $q\not= 0$.
We now use Proposition \ref{proplog5} and the
proof by contradiction method to prove

\begin{prop}
\label{proplog6}
Let $a, b , c$ be odd integers. Then, the equation
\[
aX^2 + bX + c = 0
\]
has no rational solution $X$.
\end{prop}

\proof
We proceed by contradiction (by this, we mean that 
we use the proof by contradiction rule).
So, assume that there is a rational solution $X = p/q$.
We may assume that $p$ and $q$ have no common divisor,
which implies that $p$ and $q$ are not both even.
As $q\not= 0$, if $aX^2 + bX + c = 0$, then by
multiplying by $q^2$, we get
\[
ap^2 + bpq + cq^2 = 0.
\]
However, as $p$ and $q$ are not both even and $a, b , c$ are odd,
we know from Proposition \ref{proplog5} that 
$ap^2 + bpq + cq^2$ is odd, that is, at least $1$.
This contradicts the fact that $p^2 + bpq + cq^2 = 0$
and thus, finishes the proof.
$\bigsquare$

\medskip
As as example of the proof by contrapositive method, we prove that
if an integer $n^2$ is even, then $n$ must be even.

\medskip
Observe that if an integer is not even then it is odd (and vice-versa).
Thus, the contrapositive of our statement is:
If $n$ is odd, then $n^2$ is odd. But, to say that $n$ is odd is to say that
$n = 2k + 1$ and then, $n^2 = (2k + 1)^2 = 4k^2 + 4k + 1 = 2(2k^2 + 2k) + 1$,
which shows that $n^2$ is odd.

\medskip
A real number $a\in \reals$ is said to be {\it irrational\/} if it cannot be 
expressed as a number in $\rats$ (a fraction). The reader
should prove that $\sqrt{2}$ is irrational by adapting the arguments
used in the two previous  propositions.

\remark
Let us return briefly to the issue of constructivity 
in classical logic, in particular when it comes to disjunctions.
Consider the question: are there 
two irrational real numbers $a$ and $b$ such that $a^b$ is
rational?  Here is a way to prove that this indeed the case.
Consider the number $\sqrt{2}^{\sqrt{2}}$. If this number
is rational, then $a = \sqrt{2}$ and $b =\sqrt{2}$ is an answer
to our question (since we already know that   $\sqrt{2}$ is
irrational). Now, observe that
\[
(\sqrt{2}^{\sqrt{2}})^{\sqrt{2}} = \sqrt{2}^{\sqrt{2}\times \sqrt{2}}
= \sqrt{2}^{2} = 2\quad\hbox{is rational!}
\]
Thus,  if $\sqrt{2}^{\sqrt{2}}$ is irrational, 
then $a = \sqrt{2}^{\sqrt{2}}$ and $b = \sqrt{2}$
is an answer to our question. 
So, we proved that

\medskip
$(\sqrt{2}$ is irrational and $\sqrt{2}^{\sqrt{2}}$ is rational$)$ or 

\smallskip
$(\sqrt{2}^{\sqrt{2}}$ and $\sqrt{2}$ 
are irrational and $(\sqrt{2}^{\sqrt{2}})^{\sqrt{2}}$  is rational$)$.

\medskip
However, the above proof does not tell us 
whether $\sqrt{2}^{\sqrt{2}}$ is rational or not!

\medskip
We see one of the shortcomings of classical reasoning:
certain statements (in particular, disjunctive or existential)
are provable but their proof does provide an explicit answer.
It is in that sense that classical logic
is not constructive.

\medskip
Many more examples of non-constructive arguments in classical logic
can be given. 

\medskip
We now add quantifiers to our language and give the
corresponding inference rules.

\section[Adding Quantifiers; The Proof Systems 
{$\s{N}^{\impl, \land, \lor, \forall, \exists, \perp}_c$}, 
{$\s{NG}^{\impl, \land, \lor, \forall, \exists, \perp}_c$}]
{Adding Quantifiers; The Proof Systems 
{$\s{N}^{\impl, \land, \lor, \forall, \exists, \perp}_c$}, 
{$\s{NG}^{\impl, \land, \lor, \forall, \exists, \perp}_c$}}
\label{sec4}
As we mentioned in Section \ref{sec1}, atomic propositions
may contain variables. The intention is that such variables
correspond to arbitrary  objects. An example is
\[
\hbox{human}(x) \impl \hbox{needs-to-drink}(x).
\]
Now, in mathematics, we usually prove universal 
statements, that is statement that hold
for all possible ``objects'', or existential statement,  that is,
statement asserting the existence of some object satisfying 
a given property. As we saw earlier, we assert that
every human needs to drink by writing the proposition
\[
\forall x(\hbox{human}(x) \impl \hbox{needs-to-drink}(x)).
\]
Observe that once the quantifier $\forall$ (pronounced 
``for all'' or ``for every'') is applied to the variable
$x$, the variable $x$ becomes a place-holder and
replacing $x$ by $y$ or any other variable does not change
anything. What matters is the locations to which
the outer $x$ points to in the inner proposition.
We say that $x$ is a {\it bound variable\/} (sometimes
a ``dummy variable''). 

\medskip
If we want to assert that some human needs to drink we write
\[
\exists x(\hbox{human}(x) \impl \hbox{needs-to-drink}(x));
\]
Again, once the quantifier $\exists$ (pronounced 
``there exists'') is applied to the variable
$x$, the variable $x$ becomes a place-holder. However, the intended
meaning of the second proposition is very different and weaker
than the first. It only asserts the existence of some
object satisfying the statement 
\[
\hbox{human}(x) \impl \hbox{needs-to-drink}(x).
\]

\medskip
Statements may contain variables that are not bound
by quantifiers. For example,  in
\[
\forall y\, \hbox{parent}(x, y)
\]
the variable $y$ is bound but the variable $x$ is not. Here,
the intended meaning of $\hbox{parent}(x, y)$ is that $x$ is
a parent of $y$.
Variables that are not bound are called {\it free\/}.
The proposition
\[
\forall y \exists x\, \hbox{parent}(x, y),
\]
which contains only bound variables in meant to
assert that every $y$ has some parent $x$.
Typically, in mathematics, we only prove statements
without free variables. However, statements with free
variables may occur during intermediate stages of a proof.

\medskip
The intuitive meaning of the statement $\forall x P$ is that
$P$ holds for all possible objects $x$ and the 
intuitive meaning of the statement $\exists x P$ is that
$P$ holds for some object $x$. Thus, we see that it would be useful
to use symbols to denote various objects. 
For example, if we want to assert some facts about the
``parent'' predicate, we may want to introduce some {\it constant
symbols\/} (for short, constants)
such as ``Jean'', ``Mia'', etc. and write
\[
\hbox{parent}(\mathrm{Jean}, \mathrm{Mia})
\]
to assert that Jean is a parent of Mia.
Often, we also have to use {\it function symbols\/} (or {\it operators,
constructors\/}), for instance, to write statement about numbers:
$+$, $*$, etc. Using constant symbols, function symbols and variables,
we can form {\it terms\/}, such as
\[
(x^2 + 1)(3*y + 2).
\]
In addition to function symbols, we also use {\it predicate symbols\/},
which are names for atomic properties. We have already seen
several examples of predicate symbols: ``human'', ``parent''.
So, in general, when we try to prove properties of certain
classes of objects (people, numbers, strings, graphs, etc.),
we assume that we have a certain {\it alphabet\/} consisting
of constant symbols, function symbols and predicate symbols.
Using these symbols and an infinite supply of variables
(assumed distinct from the variables which we use to label
premises) we can form {\it terms and predicate terms\/}.
We say that we have a {\it (logical) language\/}.
Using this language, we can write compound statements.

\medskip
Let us be a little more precise. In a {\it first-order language\/}, $\mathbf{L}$,
in addition to the logical connectives, $\impl, \land, \lor, \neg, \perp$,
$\forall$ and $\exists$, we have  a set, $\mathbf{L}$, of
{\it nonlogical symbols\/} consisting of
\begin{enumerate}
\item[(i)]
A set $\mathbf{CS}$ of constant symbols, $c_1, c_2, \ldots, $.
\item[(ii)]
A set $\mathbf{FS}$ of function symbols, $f_1, f_2, \ldots, $.
Each function symbol, $f$, has a {\it rank\/}, $n_f\geq 1$, which is the 
number of arguments of $f$.
\item[(iii)]
A set $\mathbf{PS}$ of predicate symbols, $P_1, P_2, \ldots, $.
Each predicate symbol, $P$, has a {\it rank\/}, $n_P\geq 0$, which is the 
number of arguments of $P$. Predicate symbols of rank $0$
are propositional letters, as in earlier sections.
\item[(iv)]
The equality predicate, $=$, is added to our language
when we want to deal with equations.
\item[(v)]
First-order variables, $t_1, t_2, \ldots, $ used to form quantified formulae.
\end{enumerate}

\medskip
The difference between function symbols and predicate symbols
is that function symbols are interpreted as functions defined
on a structure (for example, addition, $+$, on $\natnums$),
whereas predicate symbols are interpreted as properties of 
objects, that is, they take the value {\bf true} or {\bf false}.
An example is the language of {\it Peano arithmetic\/},
$\mathbf{L} = \{0, S, +, *, =\}$.  Here, the intended structure
is $\natnums$,  $0$ is of course zero,
$S$ is interpreted as the function $S(n) = n + 1$, the symbol
$+$ is addition, $*$ is multiplication and $=$ is equality.

\medskip
Using a first-order language, $\mathbf{L}$, we can form terms, predicate 
terms and formulae. The {\it terms over $\mathbf{L}$\/} are the following
expressions:
\begin{enumerate}
\item[(i)]
Every variable, $t$, is a term;
\item[(ii)]
Every constant symbol, $c\in \mathbf{CS}$, is a term;
\item[(iii)]
If $f\in \mathbf{FS}$ is a function symbol taking $n$ arguments
and $\tau_1, \ldots, \tau_n$ are terms already constructed, then
$f(\tau_1, \ldots, \tau_n)$ is a term.
\end{enumerate}

The {\it predicate terms over $\mathbf{L}$\/} are the following
expressions:
\begin{enumerate}
\item[(i)]
If $P\in \mathbf{PS}$ is a predicate symbol taking $n$ arguments
and $\tau_1, \ldots, \tau_n$ are terms already constructed, then
$P(\tau_1, \ldots, \tau_n)$ is a predicate term.
When $n = 0$, the predicate symbol, $P$, is a predicate term
called a propositional letter. 
\item[(ii)]
When we allow the equality predicate, for any two terms
$\tau_1$ and $\tau_2$, the expression $\tau_1 = \tau_2$
is a predicate term. It is usually called an {\it equation\/}. 
\end{enumerate}

The {\it (first-order) formulae over $\mathbf{L}$\/} are the following
expressions:
\begin{enumerate}
\item[(i)]
Every predicate term, $P(\tau_1, \ldots, \tau_n)$, is an atomic formula.
This includes all propositional letters. We also view $\perp$
(and sometimes $\top$) as an atomic formula.
\item[(ii)]
When we allow the equality predicate, every equation,
$\tau_1 = \tau_2$, is an atomic formula.
\item[(iii)]
If $P$ and $Q$ are formulae already constructed, then
$P \impl Q$, $P\land Q$, $P\lor Q$, $\neg P$ are
compound formulae. We treat $P \equiv Q$ as an abbreviation
for $(P \impl Q) \land (Q\impl P)$, as before.
\item[(iv)]
If $P$ is a formula already constructed and $t$ is any variable, then
$\forall t P$ and $\exists t P$ are compound formulae.
\end{enumerate}

\medskip
All this can be made very precise but this is quite tedious.
Our primary goal is to explain the basic rules of
logic and not to teach a full-fledged logic course.
We hope that our intuitive explanations will suffice
and we now come to the heart of the matter, the inference
rules for the quantifiers. Once again, for a complete treatment,
readers are referred to
Gallier \cite{Gall86}
van Dalen \cite{vanDalen} or Huth and Ryan \cite{HuthRyan}.

\medskip
Unlike the rules for $\impl, \lor, \land$ and $\perp$,
which are rather straightforward, the rules for quantifiers
are more subtle due the presence of variables (occurring
in terms and predicates). We have to be careful to forbid
inferences that would yield ``wrong'' results and for this
we have to be very precise about the way we use free variables.
More specifically, we have to exercise care when we
make {\it substitutions\/} of terms for variables in
propositions. For example, say we have the predicate
``odd'', intended to express that a number is odd.
Now, we can substitute the term $(2y + 1)^2$ for $x$
in $\mathrm{odd}(x)$ and obtain
\[
\mathrm{odd}((2y + 1)^2).
\]
More generally, if $P(t_1, t_2, \ldots, t_n)$ is a statement
containing the free variables $t_1, \ldots, t_n$
and if $\tau_1, \ldots, \tau_n$ are terms, we can form the new statement
\[
P[\tau_1/t_1, \ldots, \tau_n/t_n]
\]
obtained by substituting the term $\tau_i$ for all free occurrences
of the variable $t_i$, for $i = 1, \ldots, n$.
By the way, we denote terms by the greek letter $\tau$
because we use the letter $t$ for a variable and using $t$
for both variables and terms would be confusing; sorry!

\medskip
However, if $P(t_1, t_2, \ldots, t_n)$ contains quantifiers,
some bad things can happen, namely, some of the variables
occurring in some term $\tau_i$ may become  quantified when
$\tau_i$ is substituted for $t_i$. For example, consider
\[
\forall x\exists y\, P(x, y, z)
\]
which contains the free variable $z$
and substitute the term $x + y$ for $z$: we get
\[
\forall x\exists y\, P(x, y, x + y).
\]
We see that the variables $x$ and $y$ occurring in the term
$x + y$ become bound variables after substitution. 
We say that there is a ``capture of variables''. 

\medskip
This is not what
we intended to happen! To fix this problem, we recall that
bound  variables are really place holders, so they can be renamed
without changing anything. Therefore, we can rename the
bound variables $x$ and $y$ in $\forall x\exists y\, P(x, y, z)$
to $u$ and $v$, getting the statement
$\forall u\exists v\, P(u, v, z)$ and now, the result of the substitution is
\[
\forall u\exists v\, P(u, v, x + y).
\]
Again, all this needs to be explained very carefuly but this can be done!

\medskip
Finally, here are the inference rules for the quantifiers,
first stated in a natural deduction style and then in
sequent style.
It is assumed that we use two disjoint sets of variables
for labeling premises $(x, y, \cdots)$ and free
variables $(t, u, v, \cdots$). As we will see, the
$\forall$-introduction rule and the $\exists$-elimination rule
involve a crucial restriction on the occurrences of certain variables.
Remember, {\it variables are terms\/}!

\begin{defin}
\label{quantrules}
{\em
The {\it inference rules for the quantifiers\/} are

\medskip
{\it $\forall$-introduction\/}:

\settree1{
     \starttree{$\Gamma$}
     \step{$P[u/t]$}
     \step{$\forall t P$}
}

\bigskip
\ligne{\hfill\box1\hfill}

\bigskip
Here, $u$ must be a variable that does not occur
free in any of the propositions in $\Gamma$ or in $\forall t P$;
the notation $P[u/t]$ stands for the result of
substituting $u$ for all free occurrences of $t$ in $P$.

\medskip
{\it $\forall$-elimination\/}:

\settree1{
     \starttree{$\Gamma$}
     \step{$\forall t P$}
     \step{$P[\tau/t]$}
}

\bigskip
\ligne{\hfill\box1\hfill}

\bigskip\noindent
Here $\tau$ is an arbitrary term and it is assumed that
bound variables in $P$ have been renamed
so that none of the variables in $\tau$ are captured
after substitution.

\bigskip
{\it $\exists$-introduction\/}:

\settree1{
     \starttree{$\Gamma$}
     \step{$P[\tau/t]$}
     \step{$\exists t P$}
}

\bigskip
\ligne{\hfill\box1\hfill}

\medskip
As in $\forall$-elimination, $\tau$ is an arbitrary term
and the same proviso on bound variables in $P$ applies.

\medskip
{\it $\exists$-elimination\/}:

\settree1{
     \starttree{$\Gamma$}
     \step{$\exists t P$}
}
\settree2{
     \starttree{$\Delta, P[u/t]^x$}
     \step{$C$}
}

\settree1{
     \njointrees{1}{2}
     \jstep{$C$}{${\scriptstyle x}$}
}

\bigskip
\ligne{\hfill\box1\hfill}

\medskip
Here, $u$ must be a variable that does not occur
free in any of the propositions in $\Delta$, $\exists t P$, or $C$,
and all premises $P[u/t]$ labeled $x$ are discharged.

\medskip
In the above  rules, $\Gamma$ or $\Delta$ may be empty, 
$P, C$ denote arbitrary propositions constructed from
a first-order language, $\mathbf{L}$ and $t$ is {\it any\/} variable.
The system of {\it first-order classical  logic\/},
$\s{N}^{\impl, \lor, \land, \perp, \forall, \exists}_c$ is
obtained by adding the above rules to the system
of propositional classical logic $\s{N}^{\impl, \lor, \land, \perp}_c$.
The system of {\it first-order  intuitionistic logic\/},
$\s{N}^{\impl, \lor, \land, \perp, \forall, \exists}_i$ is
obtained by adding the above rules to the system
of propositional intuitionistic  logic $\s{N}^{\impl, \lor, \land, \perp}_i$.
}
\end{defin}

\medskip
Using sequents, the quantifier rules in first-order logic
are expressed as follows:

\begin{defin}
\label{quantrules2}
{\em
The {\it inference rules for the quantifiers in Gentzen-sequent style\/} are
$$\rulea{\sequent{\Gamma}{P[u/t]}}
        {\sequent{\Gamma}{\forall t P}}
        {(\forallintr)}\qquad
\rulea{\sequent{\Gamma}{\forall t P}}
        {\sequent{\Gamma}{P[\tau/t]}}
        {(\forallelim)}$$
where in (\forallintr),
$u$ does not occur free in $\Gamma$ or $\forall t P$;

$$\rulea{\sequent{\Gamma}{P[\tau/t]}}
        {\sequent{\Gamma}{\exists t P}}
        {(\existsintr)}\qquad
\ruleb{\sequent{\Gamma}{\exists t P}}
        {\sequent{z\co P[u/t], \Gamma}{C}}
        {\sequent{\Gamma}{C}}
        {(\existselim)}$$
where in (\existselim),
$u$ does not occur free in $\Gamma$, $\exists t P$,  or $C$.
Again, $t$ is {\it any\/} variable.

\medskip
The variable $u$ is called the {\it eigenvariable\/} of the inference.
The systems 
$\s{NG}^{\impl, \lor, \land, \perp, \forall, \exists}_c$  and
$\s{NG}^{\impl, \lor, \land, \perp, \forall, \exists}_i$  are defined
from the systems $\s{NG}^{\impl, \lor, \land, \perp}_c$  and
$\s{NG}^{\impl, \lor, \land, \perp}_i$, respectively,  
by adding the above rules.
}
\end{defin}

\medskip
When we say that a proposition, $P$, is {\it provable from $\Gamma$\/},
we  mean that we can construct a proof tree whose
conclusion is $P$ and whose set of premises is $\Gamma$, in one of
the systems $\s{N}^{\impl, \land, \lor, \perp, \forall, \exists}_{c}$ or
$\s{NG}^{\impl, \land, \lor, \perp, \forall, \exists}_{c}$. 
Therefore, as in propositional logic, when we use the
word ``provable'' unqualified, we mean provable in {\it classical logic\/}.
Otherwise, we say {\it intuitionistically provable \/}.

\medskip
A first look at the above rules shows that universal formulae,
$\forall t P$, behave somewhat like infinite conjunctions and
that existential formulae, $\exists t P$, behave somewhat
like infinite disjunctions. 

\medskip
The  $\forall$-introduction rule looks a little strange
but the idea behind it is actually very simple:
Since $u$ is totally unconstrained, if $P[u/t]$
is provable (from $\Gamma$), then intuitively  $P[u/t]$
holds of any arbitrary object, and so,
the statement $\forall t P$ should also be provable (from $\Gamma$).

\medskip
The meaning of the $\forall$-elimination is that
if $\forall t P$ is provable (from $\Gamma$), then
$P$ holds for all objects and so, in particular for
the object denoted by the term $\tau$, i.e.,
$P[\tau/t]$ should be provable (from $\Gamma$).

\medskip
The  $\exists$-introduction rule is dual to the $\forall$-elimination rule.
If $P[\tau/t]$ is provable (from $\Gamma$), this means that
the object denoted by $\tau$ satisfies $P$, so $\exists t P$
should be provable (this latter formula asserts the existence
of some object satisfying $P$, and $\tau$ is such an object).

\medskip
The $\exists$-elimination rule is reminiscent of the
$\lor$-elimination rule and is a little more tricky.
It goes as follows: Suppose that we proved $\exists t P$
(from $\Gamma$). Moreover, suppose that for every
possible case, $P[u/t]$, we were able to prove $C$ (from $\Gamma$).
Then, as we have ``exhausted'' all possible cases and
as we know from the provability of $\exists t P$ that
some case must hold, we can conclude that $C$ is provable (from $\Gamma$)
without using $P[u/t]$ as a premise.

\medskip
Like the $\lor$-elimination rule, the $\exists$-elimination
rule is not very constructive. It allows making a conclusion
($C$) by considering alternatives without knowing which
actually occurs.

\remark
Anagolously to disjunction, in (first-order) intuitionistic logic, if
an existential statement $\exists t P$ is provable (from $\Gamma$), then 
from any proof of  $\exists t P$, some term, $\tau$, can be extracted
so that $P[\tau/t]$ is provable from $\Gamma$. 
Such a term, $\tau$, is called a {\it witness\/}. The witness property
is not easy to prove. It follows from the fact that intuitionistic 
proofs have a normal form (see Section \ref{sec5}).
However, no such property holds in classical logic
(for instance, see the $a^b$ rational with $a, b$ irrational example 
revisited below).

\medskip
Here is an example of a proof 
in the system
$\s{N}^{\impl, \lor, \land, \perp, \forall, \exists}_c$ 
(actually, in $\s{N}^{\impl, \lor, \land, \perp, \forall, \exists}_i$)
of the formula
$\forall t(P\land Q) \impl \forall t P \land \forall t Q$.

\settree1{
     \starttree{$\forall t (P\land Q)^{x}$}
     \step{$P[u/t]\land Q[u/t]$}
     \step{$P[u/t]$}
     \step{$\forall t P$}
}

\settree2{
     \starttree{$\forall t (P\land Q)^{x}$}
     \step{$P[u/t]\land Q[u/t]$}
     \step{$Q[u/t]$}
     \step{$\forall t Q$}
}

\settree1{
     \njointrees{1}{2}
     \step{$\forall t P \land \forall t Q$}
     \jstep{$\forall t(P\land Q) \impl \forall t P \land \forall t Q$}
{${\scriptstyle x}$}
}

\bigskip
\ligne{\hfill\box1\hfill}

\medskip\noindent
In the above proof, $u$ is a new variable, i.e., a variable that
does not occur free in $P$ or $Q$.

\medskip
The reader should show that 
$\forall t P \land \forall t Q \impl \forall t(P\land Q)$ is also
provable in $\s{N}^{\impl, \lor, \land, \perp, \forall, \exists}_i$.
However, in general, one can't just replace $\forall$ by $\exists$ (or
$\land$ by $\lor$)  and still obtain  provable statements.
For example,
$\exists t P \land \exists t Q \impl \exists t(P\land Q)$
is not provable at all!

\medskip
Here are some useful equivalences involving quantifiers.
The first two are analogous to the de Morgan laws for 
$\land $ and $\lor$.

\begin{prop}
\label{proplog7}
The following equivalences are provable in
classical first-order logic:
\begin{align*}
\neg \forall t P &\equiv \exists t \neg P \\
\neg \exists t P &\equiv \forall t \neg P \\
\forall t (P\land Q) &\equiv \forall t P \land \forall t Q\\
\exists t (P\lor Q) &\equiv \exists t P \lor \exists t Q.
\end{align*}
In fact, the last three and $\exists t \neg P \impl \neg \forall t P$
are provable intuitionistically.
Moreover, the propositions
$\exists t (P\land Q) \impl \exists t P \land \exists t Q$
and $\forall t P \lor \forall t Q\impl  \forall t (P\lor Q)$ are
provable in intuitionistic first-order logic (and thus, also in 
classical first-order logic).
\end{prop}

\proof
Left as an exercise to the reader.
$\bigsquare$

\remark
We can illustrate, again, the fact that classical
logic allows for non-constructive proofs by reexamining
the example at the end of Section \ref{sec3}.
There, we proved that if  $\sqrt{2}^{\sqrt{2}}$ is rational, 
then $a = \sqrt{2}$ and $b = \sqrt{2}$ are both
irrational numbers such that $a^b$ is rational and if
$\sqrt{2}^{\sqrt{2}}$ is irrational then 
$a = \sqrt{2}^{\sqrt{2}}$ and $b = \sqrt{2}$  are both
irrational numbers such that $a^b$ is rational.
By $\exists$-introduction, we deduce that if
$\sqrt{2}^{\sqrt{2}}$ is rational then there exist 
some irrational numbers $a, b$ so that
$a^b$ is rational and if $\sqrt{2}^{\sqrt{2}}$ is 
irrational then  there exist
some irrational numbers $a, b$ so that
$a^b$ is rational. In classical logic, as $P\lor \neg P$ is
provable, by $\lor$-elimination, we just proved that
there exist some irrational numbers $a$ and $b$ so that $a^b$ is rational.

\medskip
However, this argument does not give us explicitely numbers $a$ and $b$
with the required properties! It only tells us that such
numbers must exist. Now, it turns out that $\sqrt{2}^{\sqrt{2}}$
is indeed irrational (this follows from the Gel'fond-Schneider Theorem,
a hard theorem in number theory). Furthermore,  there are also
simpler explicit solutions such as $a = \sqrt{2}$ and
$b = \log_2 9$, as the reader should check!

\medskip
We conclude this section by giving an example of a ``wrong proof''.
Here is an example in which the $\forall$-introduction rule
is applied illegally, and thus, yields a statement which
is actually false (not provable). In the incorrect ``proof'' below,
$P$ is an atomic predicate
symbol taking two arguments (for example, ``parent'') and 
$0$ is a constant denoting zero:

\settree1{
     \starttree{$P(t, 0)^{x}$}
     \jstep{$\forall t P(t, 0)$}{illegal step!}
     \jstep{$P(t, 0) \impl \forall t P(t, 0)$}{${\scriptstyle x}$}
     \step{$\forall t(P(t, 0) \impl \forall t P(t, 0))$}
     \step{$P(0, 0) \impl \forall t P(t, 0)$}
}

\bigskip
\ligne{\hfill\box1\hfill}

\bigskip
The problem is that the variable $t$ occurs free in the premise
$P[t/t, 0] = P(t, 0)$ and therefore, 
the application of the $\forall$-introduction
rule in the first step is illegal. However, note that
this premise is discharged in the second step and so, the 
application of the  $\forall$-introduction rule in the third
step is legal. The (false) conclusion of this faulty proof is that
$P(0, 0) \impl \forall t P(t, 0)$ is provable. Indeed, there are plenty
of properties such that the fact that the single instance, $P(0, 0)$, holds
does not imply that $P(t, 0)$ holds for all $t$.

\remark
The above example shows why it is desirable to have premises
that are universally quantified. A premise of the form
$\forall t P$ can be instantiated to $P[u/t]$, using
$\forall$-elimination,  where $u$ is a brand new
variable. Later on, it may be possible to use
$\forall$-introduction without running into trouble with
free occurrences of $u$ in the premises. But we still have to be very careful 
when we use $\forall$-introduction or $\exists$-elimination.

\medskip
Before concluding this section, let us give a few more examples of 
proofs using the rules for the quantifiers.
First, let us prove that
\[
\forall t P \equiv \forall u P[u/t],
\]
where $u$ is any variable not free in $\forall t P$ and
such that $u$ is not captured during the substitution.
This rule allows us to rename bound variables (under very mild
conditions).
We have the proofs

\settree1{
     \starttree{$(\forall t P)^{\alpha}$}
     \step{$P[u/t]$}
     \step{$\forall u P[u/t]$}
     \jstep{$\forall t P \impl \forall u P[u/t]$}{${\scriptstyle \alpha}$}
}

\bigskip
\ligne{\hfill\box1\hfill}

\bigskip\noindent
and

\settree1{
     \starttree{$(\forall u P[u/t])^{\alpha}$}
     \step{$P[u/t]$}
     \step{$\forall t P$}
     \jstep{$\forall u P[u/t] \impl \forall t P$}{${\scriptstyle \alpha}$}
}

\bigskip
\ligne{\hfill\box1\hfill}

\bigskip
Now, we give a proof (intuitionistic) of
\[
\exists t(P \impl Q) \impl (\forall t P \impl Q),
\]
where $t$ does not occur (free or bound) in $Q$.

\settree1{
     \starttree{$(\exists t (P \impl Q))^{z}$}
}

\settree2{
     \starttree{$(P[u/t] \impl Q)^{x}$}
}
\settree3{
     \starttree{$(\forall t P)^{y}$}
     \step{$P[u/t]$}
}
\settree2{
     \njointrees{2}{3}
     \step{$Q$}
}
\settree1{
     \njointrees{1}{2}
     \jstep{$Q$}{${\scriptstyle x}$}
     \jstep{$\forall t P \impl Q$}{${\scriptstyle y}$}
     \jstep{$\exists t(P \impl Q)\impl (\forall t P \impl Q)$}{${\scriptstyle z}$}
}

\bigskip
\ligne{\hfill\box1\hfill}

\bigskip
In the above proof, $u$ is a new variable that does not occur in
$Q$, $\forall t P$, or $\exists t(P \impl Q)$
The converse requires (RAA) and is a bit more complicated.
To conclude, we give a proof (intuitionistic) of
\[
(\forall t P \lor Q)\impl \forall t(P \lor Q),
\]
where $t$ does not occur (free or bound) in $Q$.

\settree1{
     \starttree{$(\forall t P \lor Q)^{z}$}
}

\settree2{
     \starttree{$(\forall t P)^{x}$}
     \step{$P[u/t]$}
     \step{$P[u/t]\lor Q$}
     \step{$\forall t(P\lor Q)$}
}
\settree3{
     \starttree{$Q^{y}$}
     \step{$P[u/t]\lor Q$}
     \step{$\forall t(P\lor Q)$}
}
\settree1{
     \tjointrees{1}{2}{3}
     \jstep{$\forall t(P\lor Q)$}{${\scriptstyle x, y}$}
     \jstep{$(\forall t P \lor Q)\impl \forall t(P \lor Q)$}{${\scriptstyle z}$}
}

\bigskip
\ligne{\hfill\box1\hfill}

\bigskip
In the above proof, $u$ is a new variable that does not occur in
$\forall t P$ or $Q$.
The converse requires (RAA).

\medskip
Several times in this Chapter, we have claimed that certain propositions
are not provable in some logical system. What kind of reasoning do we
use to validate such claims? In the next section, we briefly
address this question as well as related ones.

\section[Decision Procedures, Proof Normalization, etc.]
{Decision Procedures, Proof Normalization, \\ Counter-Examples, Theories, etc.}
\label{sec5}
In the previous sections, we saw how the rules of mathematical reasoning
can be formalized in various natural deduction systems and we
defined a precise notion of proof. We observed that finding a proof
for a given proposition was not a simple matter, nor was it to acertain that
a  proposition is unprovable. Thus, it is natural to ask the following
question:

\medskip
{\it The Decision Problem\/}: Is there a general procedure which takes
any arbitrary proposition, $P$,  as input, always terminates in a finite number
of steps, and tells us whether $P$ is provable or not.

\medskip
Clearly, it would be very nice if such a procedure existed, especially
if it also produced a proof of $P$ when $P$ is provable.

\medskip
Unfortunately, for rich enough languages, such as first-order logic,
it is impossible to find such a procedure. This deep result known
as the {\it undecidability of the decision problem\/} or
{\it Church's Theorem\/}  was proved
by A. Church in 1936 (Actually, Church proved the undecidability of the
validity problem, but by G\"odel's completeness Theorem,
validity and provability are equivalent). 

\medskip
Proving Church's Theorem is hard and a lot of work.
One needs to develop a good deal of what is called the {\it theory of computation\/}.
This involves defining models of computation such as {\it Turing machines\/} and
proving other deeps results such as the {\it undecidability of the halting problem\/}
and the {\it undecidability of the Post Correspondence Problem\/}, among other
things. Some of this material is covered in CSE262, so be patient and your curiosity
will be satisfied!

\medskip
So, our hopes to find a ``universal theorem prover'' are crushed.
However, if we restrict ourselves to propositional logic,
classical or intuitionistic, it turns out that procedures solving the 
decision problem do exist and they even produce a proof of 
the input proposition when that proposition is provable. 

\medskip
Unfortunately, proving that such procedures exist 
and are correct in the propositional case
is rather difficult, especially for intuitionistic logic.
The difficulties have a lot to do with our choice of a natural deduction system.
Indeed, even for the system $\s{N}^{\impl}_m$ (or $\s{NG}^{\impl}_m$),
provable propositions may have  infinitely many proofs.
This makes the search process impossible; when do we know how to stop, especially
if a proposition is not provable!
The problem is that proofs may contain redundancies (Gentzen said ``detours'').
A typical example of redundancy is an
elimination immediately follows an introduction, as in the following example
in which $\s{D}_1$ denotes a deduction with conclusion
$\sequent{\Gamma, x\co A}{B}$ and $\s{D}_2$ denotes a deduction with conclusion
$\sequent{\Gamma}{A}$.
\settree1{
     \starttree{$\s{D}_1$}
     \istep{$\sequent{\Gamma, x\co A}{B}$}
     \step{$\sequent{\Gamma}{A\impl B}$}
}
\settree2{
     \starttree{$\s{D}_2$}
     \istep{$\sequent{\Gamma}{A}$}
}
\settree1{
     \jointrees{1}{2}
     \step{$\sequent{\Gamma}{B}$}
}

\bigskip
\ligne{\hfill\box1\hfill}

\bigskip
Intuitively, it should be possible to construct a deduction for
$\sequent{\Gamma}{B}$ from the two deductions $\s{D}_1$ and
$\s{D}_2$ without  using at all the hypothesis $x\co A$.
This is indeed the case. If we look closely at the deduction
$\s{D}_1$, from the shape of the inference rules,
assumptions are never created, and the leaves must be labeled
with expressions of the form $\sequent{\Gamma', \Delta, x\co A, y\co C}{C}$
or $\sequent{\Gamma, \Delta,  x\co A}{A}$, where $y\not= x$
and either $\Gamma = \Gamma'$ or $\Gamma = \Gamma', y\co C$.
We can form a new deduction for $\sequent{\Gamma}{B}$ as follows:
in $\s{D}_1$, wherever a leaf of the form
$\sequent{\Gamma, \Delta, x\co A}{A}$ occurs, replace it by
the deduction obtained from $\s{D}_2$ by adding $\Delta$
to the premise of each sequent in $\s{D}_2$. Actually,
one should be careful to first make a fresh copy of $\s{D}_2$
by renaming all the variables so that clashes with
variables in $\s{D}_1$ are avoided. 
Finally,  delete the assumption $x\co A$
from the premise of every sequent in the resulting proof.
The resulting deduction
is obtained by a kind of substitution and may be denoted
as $\s{D}_1[\s{D}_2/x]$, with some minor abuse of notation.
Note that the assumptions $x\co A$ occurring in the leaves of the form
$\sequent{\Gamma', \Delta, x\co A, y\co C}{C}$ were never used anyway.
The step which consists in transforming the above redundant
proof figure into the deduction $\s{D}_1[\s{D}_2/x]$ is
called a {\it reduction step\/} or {\it normalization step\/}. 

\medskip
The idea of {\it proof normalization\/} 
goes back to Gentzen (\cite{Gentzen69}, 1935). Gentzen noted
that (formal) proofs can contain redundancies, or ``detours'',
and that most complications in the analysis of proofs are due to
these redundancies.
Thus, Gentzen had the idea that the analysis of proofs 
would be simplified if it was possible to show that
every proof can be converted to an equivalent irredundant
proof, a proof in normal form. 
Gentzen proved a technical result to that effect, the
``cut-elimination theorem'', for a sequent-calculus formulation
of first-order logic \cite{Gentzen69}.
Cut-free proofs are direct, in the sense that they never use
auxiliary lemmas via the cut rule. 

\remark
It is important to 
note that Gentzen's result gives a particular algorithm
to produce a proof in normal form. Thus, we know that every proof can
be reduced to some normal form using a specific strategy, 
but there may be more than one normal form, and certain
normalization strategies may not terminate.

\medskip
About thirty years later, Prawitz (\cite{Praw65}, 1965)
reconsidered the issue of
proof normalization, but in the framework of natural deduction
rather than the framework of sequent calculi.%
\footnote{This is somewhat ironical, since Gentzen began
his investigations using a natural deduction system, but decided to switch
to sequent calculi (known as Gentzen systems!) for
technical reasons.}
Prawitz explained very clearly what redundancies are
in systems of natural deduction, and he proved that every proof
can be reduced to a normal form. Furthermore, this normal
form is unique. A few years later, Prawitz (\cite{Praw71}, 1971) showed
that in fact, every reduction sequence terminates, a property
also called {\it strong normalization\/}.

\medskip
A remarkable connection between proof normalization and
the notion of computation must also be mentioned.
Curry  (1958) made the remarkably insightful observation that 
certain typed combinators can be viewed as representations
of proofs (in a Hilbert system) of certain propositions
(See in Curry and Feys 
\cite{Curry} (1958), Chapter 9E, pages 312-315.)
Building up on this observation, Howard (\cite{Howard69}, 1969) described a
general correspondence between propositions and types,
proofs in natural deduction and certain typed $\lambda$-terms,
and proof normalization and $\beta$-reduction. (The simply-typed-$\lambda$-calculus
was invented by Church, 1940). This correspondence,
usually referred to as the {\it Curry/Howard isomorphism\/} or
{\it formulae--as--types principle\/}, is fundamental and very fruitful.

\medskip
The Curry/Howard isomorphism establishes a deep correspondence
between the notion of proof and the notion of computation. 
Furthermore, and this is the deepest
aspect of the Curry/Howard isomorphism, proof normalization corresponds
to term reduction in the $\lambda$-calculus associated with the
proof system. 
To make the story short, the correspondence between proofs 
in intuitionistic logic and typed
$\lambda$-terms  on one-hand and between proof normalization
and $\beta$-conversion on the other hand can be used to translate results about
typed $\lambda$-terms into results about proofs 
in intuitionistic logic. By the way, some aspects of the 
Curry/Howard isomorphism are covered in CIS500. 

\medskip
In summary, using either some suitable intuitionistic
sequent calculi and
Gentzen's cut elimination theorem or some suitable typed $\lambda$-calculi
and (strong) normalization results about them, it is possible to
prove that there is a decision procedure for propositional
intuitionistic logic. 
However, it can also be shown that the time-complexity of
any such procedure is very high. Here, we are alluding to 
{\it complexity theory\/}, another active area of computer science.
You will learn about some basic and fundamental aspects of this theory
in CSE262 when you learn about the two problems {\it P} and {\it NP}.

\medskip
Readers who wish to learn more about these topics can read
my two survey papers Gallier \cite{Galliercahiers} 
(on the Correspondence Between Proofs  and $\lambda$-Terms)
and Gallier \cite{Gall93} (A Tutorial
on Proof Systems and Typed $\lambda$-Calculi), both available on the web site

\medskip\noindent
http://www.cis.upenn.edu/$\tilde{\;}$jean/gbooks/logic.html

\medskip\noindent
and the excellent introduction to proof theory by
Troelstra and Schwichtenberg \cite{TroelstraSchwi}.

\medskip
Anybody who  really wants to understand logic should of course 
take a look at Kleene \cite{Kleene52} (the famous ``I.M.''), but 
this is not recommended to beginners!

\medskip
Let us return to the question of deciding whether a proposition is
not provable. To simplify the discussion, let us restrict our attention to
propositional classical logic. So far, we have presented a very {\it proof-theoretic\/}
view of logic, that is, a view based on the notion of provability as
opposed to a more {\it semantic\/} view of based on the notions of truth and models.
A possible excuse for our bias is that, as Peter Andrews (from CMU) puts it,
``truth is elusive''. Therefore, it is simpler
to understand what truth is in terms of
the more ``mechanical'' notion of provability. (Peter Andrews even gave the subtitle

\medskip
{\sl To Truth Through Proof}

\medskip\noindent
to his logic book Andrews \cite{Andrews}!)

\medskip
However, mathematicians are not mechanical theorem provers (even if they
prove lots of stuff)! Indeed, mathematicians almost always think
of the objects they deal with (functions, curves, surfaces, groups, rings, etc.)
as rather concrete objects (even if they may not seem concrete to the uninitiated)
and not as abstract entities soleley characterized by arcane axioms.

\medskip
It is indeed natural and fruitful to try to interpret formal statements
semantically. For propositional classical logic, this can be done quite easily
if we interpret atomic propositional letters using  the truth values
{\bf true} and {\bf false}. Then, the crucial point that {\it every provable
proposition (say in $\s{NG}^{\impl, \lor, \land, \perp}_c$) has the value
{\bf true} no matter how we assign truth values to the letters in 
our proposition\/}. In this case, we say that $P$ is {\it valid}.

\medskip 
The fact that provable implies valid 
is called {\it soundness\/} or {\it consistency\/}
of the proof system. The soundness of the proof system 
$\s{NG}^{\impl, \lor, \land, \perp}_c$
is easy to prove. For this, given any sequent, $\sequent{\Gamma}{P}$,
we prove that whenever all the propositions in $\Gamma$ are assigned
the value {\bf true}, then $P$ evaluates to {\bf true}. This is easy to do:
check that this holds for the axioms and that whenever it holds for the
premise(s) of an inference rule then it holds for the conclusion.

\medskip
We now have a method to show that a proposition, $P$, is not provable:
Find some truth assignment that makes $P$ {\bf false}.

\medskip
Such an assignment falsifying $P$ is called a {\it counter-example\/}.
If $P$ has a counter-example, then it can't be provable because if it were, then
by soundness it would be {\bf true} for all possible truth assigments.

\medskip
But now, another question comes up: If a proposition is not provable, can we always 
find a counter-example for it. Equivalently, {\it is every valid proposition provable\/}?
If every valid proposition is provable, we say that our proof system is {\it complete\/}
(this is the {\it completeness\/} of our system).

\medskip
The system $\s{NG}^{\impl, \lor, \land, \perp}_c$ is indeed complete. 
In fact, {\it all\/} the classical systems that we have discussed
are sound and complete. Completeness is usually a lot harder to prove
than soundness. For first-order classical logic, this is known
as {\it G\"odel's completeness Theorem\/} (1929). Again, we refer our readers
to Gallier \cite{Gall86}  van Dalen \cite{vanDalen} or
or Huth and Ryan \cite{HuthRyan} for a thorough
discussion of these matters. In the first-order case, one has to define
{\it first-order structures\/} (or {\it first-order models\/}).

\medskip
What about intuitionistic logic?

\medskip
Well, one has to come up with a richer notion of semantics because
it is no longer true that if a proposition is valid (in the sense
of our two-valued semantics using {\bf true}, {\bf false}), then it is provable.
Several semantics have been given for intuitionistic logic. In our opinion, the most
natural is the notion of  {\it Kripke model\/}. Then, again,
soundness and completeness holds for intuitionistic proof systems,
even in the first-order case
(see   van Dalen \cite{vanDalen}).

\medskip
In summary, semantic models can be use to provide {\it counter-examples\/}
of unprovable propositions. This is a quick method to establish
that a proposition is not provable.

\medskip
The way we presented  deduction trees and proof trees may have given 
our readers the impression that the set of premises, $\Gamma$, was
just an auxiliary notion. Indeed, in all of our examples, $\Gamma$
ends up being empty! However, nonempty $\Gamma$'s are crucially needed if
we want to develop theories about various kinds of
structures and objects, such as the natural numbers, groups, rings, fields, trees,
graphs, sets, etc. Indeed, we need to make definitions about the objects
we want to study and we need to state some axioms asserting the main
properties of these objects. We do this by putting these definitions
and axioms in $\Gamma$. Actually, we have to allow $\Gamma$ to be infinite
but we still require that our deduction trees are finite; they can only
use finitely many of the propositions in $\Gamma$. We are then interested in
all propositions, $P$,  such that $\sequent{\Delta}{P}$ is provable,
where $\Delta$ is any finite subset of $\Gamma$; the set of all such $P$'s
is called a {\it theory\/}. Of course we have the usual problem of
consistency: If we are not careful, our theory may be inconsistent, i.e.,
it may consist of all propositions.

\medskip
Let us give two examples of theories.

\medskip
Our first example is the {\it theory of equality\/}. 
Indeed, our readers may have noticed that
we have avoided to deal with the equality relation. In practice, we
can't do that.  

\medskip
Given a language, $\mathbf{L}$, with a given supply of constant,  function
and predicate symbols,
the theory of equality consists of the following
formulae taken as axioms:
\begin{align*}
& \forall (x = x) \\
& \forall x_1 \cdots\forall x_n \forall y_1 \cdots\forall y_n
[(x_1 = y_1\land\cdots\land x_n = y_n) \impl 
f(x_1, \ldots, x_n) = f(y_1, \ldots, y_n)] \\
& \forall x_1 \cdots\forall x_n \forall y_1 \cdots\forall y_n
[(x_1 = y_1\land\cdots \land x_n = y_n) \land
P(x_1, \ldots, x_n) \impl P(y_1, \ldots, y_n)], 
\end{align*}
for all function symbols (of $n$ arguments) and all predicate symbols
(of $n$ arguments), including the equality predicate, $=$, itself.

\medskip
It is not immediately clear from the above
axioms that $=$ is reflexive and transitive but this
can shown easily.

\medskip
Our second example is the first-order theory of the natural numbers
known as {\it Peano's arithmetic\/}.

\medskip
Here, we have the constant $0$ (zero), the unary function symbol $S$ (for successor
function; the intended meaning is $S(n) = n + 1$) and the binary 
function symbols $+$ (for addition) and $*$ (for multiplication). In addition
to the axioms for the theory of equality we have the following axioms:
\begin{align*}
& \forall x\neg(S(x) = 0)\\
& \forall x\forall y (S(x) = S(y) \impl x = y) \\
& \forall x\forall y(x + 0 = x) \\
& \forall x\forall y(x + S(y) = S(x + y)) \\
& \forall x\forall y(x * 0  = 0) \\
& \forall x\forall y(x * S(y) = x * y + x) \\
& [A(0) \land \forall x(A(x) \impl A(S(x)))] \impl \forall n A(n),
\end{align*}
where $A$ is any first-order formula with one free variable.
This last axiom is the {\it induction axiom\/}. Observe how $+$ and $*$ are
defined recursively in terms of $0$ and $S$ and that there are
infinitely many induction axioms (countably many).

\medskip
Many properties that hold for the natural numbers (i.e., are true 
when the symbols $0, S, +, *$ have their usual interpretation and all variables
range over the natural numbers) 
can be proved in this theory
(Peano's arithmetic), but not all! This is another very famous
result of G\"odel known as {\it  G\"odel's incompleteness Theorem\/} (1931).
However, the topic of incompleteness is definitely oustside the scope of this course, 
so we will not say anymore about it.
Another very interesting theory is {\it set theory\/}.
There are a number of axiomatizations of set theory and we will
discuss one of them (ZF) very briefly in the next section.

\medskip
We close this section by repeating something we said ealier: There isn't
just one logic but instead, {\it many\/} logics. In addition to classical
and intuitionistic logic (propositional and first-order), there
are: modal logics, higher-order logics and 
{\it linear logic\/}, a logic due to Jean-Yves Girard, 
attempting to unify classical and intuitionistic logic (among other goals).
An excellent introduction to these
logics can be found in Troelstra and Schwichtenberg \cite{TroelstraSchwi}.
We warn our readers that most presentations of linear logic are
(very) difficult to follow. This is definitely true of
Girard's seminal paper \cite{Girlin87}. A more approachable version
can be found in Girard, Lafont and Taylor \cite{Gir89}, but 
most readers will still wonder what hit them when they attempt to read it.

\medskip
In computer science, there is also {\it dynamic logic\/}, used
to prove properties of programs and {\it temporal logic\/}
and its variants (originally invented
by A. Pnueli), to prove properties of real-time systems.
So, logic is alive and well! Also, take a look at CSE482!

\section{Basics Concepts of Set Theory}
\label{sec5b}
Having learned some fundamental notions of logic, it is now
a good place before proceeding to more interesting things, such as
functions and relations, to go through a very quick review
of some basic concepts of set theory. This section will take 
the very ``naive'' point of view that a set is a collection of objects,
the collection being regarded as a single object.
Having first-order logic at our disposal, we could
formalize set theory very rigorously in terms of axioms.
This was done by Zermelo first (1908) and in a more satisfactory
form by Zermelo and  Frankel  in 1921, in a theory known as 
the ``Zermelo-Frankel'' (ZF) axioms.
Another axiomatization was given by John von Neumann in 1925
and later improved by Bernays in 1937.
A modification of Bernay's axioms was used by Kurt G\"odel in 1940.
This approach is now known as ``von Neumann-Bernays'' (VNB) or
``G\"odel-Bernays'' (GB) set theory. 
There are many books that give an axiomatic presentation
of set theory. Among them, we recommend Enderton
\cite{Endertonset}, which we find remarkably clear and elegant,
Suppes \cite{Suppes} (a little more advanced) and Halmos
\cite{Halmos}, a classic (at a more elementary level). 

\medskip
However, it must be said that set theory 
was first created by Georg Cantor (1845-1918) between 1871 and 1879.
However, Cantor's work was not unanimously well received by all mathematicians.
Cantor regarded infinite objects as objects to be treated in much
the same way as finite sets, a point of view that was shocking
to a number of very prominent mathematicians who bitterly attacked him
(among them, the powerful Kronecker).
Also, it turns out that some paradoxes in set theory popped up
in the early 1900, in particular, Russell's paradox.
Russell's paradox (found by Russell in 1902) has to to with the 

\medskip
``set of all sets that are not members of themselves''

\medskip\noindent
which we denote by
\[
R = \{x \mid x\notin x\}.
\]
(In general, the notation $\{x \mid P\}$ stand for 
the set of all objects  satisfying the property $P$.)

\medskip
Now, classically, either $R \in R$ or $R\notin R$.
However, if $R\in R$, then the definition of $R$ says that
$R\notin R$; if $R\notin R$, then again, the definition
of $R$ says that $R\in R$! 

\medskip
So, we have a contradiction and
the existence of such a set is a paradox. 
The problem is that  we are allowing a property (here,
$P(x) = x\notin x$), which is ``too wild'' and circular in nature.
As we will see, the way out, as found by Zermelo,
is to place a restriction on the property $P$ and to also
make sure that $P$ picks out elements from some already given set
(see the Subset Axioms below).

\medskip
The apparition of these paradoxes prompted mathematicians,
with Hilbert among its leaders, to put set theory on firmer grounds.
This was achieved by  Zermelo, Frankel, von Neumann, Bernays
and G\"odel, to only name the major players.

\medskip
In what follows, we are assuming that we are working in classical logic.
We will introduce various operations on sets using defintion
involving the logical connectives $\land$, $\lor$, $\neg$, $\forall$ and
$\exists$.
In order to ensure the existence of some of these sets requires
some of the axioms of set theory, but we will be rather casual about that.

\medskip
Given a set, $A$, we write that some object, $a$, is an element of (belongs to)
the set $A$ as
\[
a\in A
\]
and that $a$ is not an element of $A$ (does not belong to $A$) as
\[
a\notin A.
\]

\medskip
When are two sets $A$ and $B$ equal? 
This corresponds to the first axiom of set theory, called

\medskip\noindent
{\bf Extensionality Axiom}

\medskip
Two sets $A$ and $B$ are equal iff they have exactly the same elements, that is
\[
\forall x (x\in A \impl x\in B) \land \forall x(x\in B \impl x\in A).
\]
The above says: 
Every element of $A$ is an element of $B$ and conversely.

\medskip
There is a special set having no elements at all, the {\it empty set\/},
denoted $\emptyset$. This is the 

\medskip\noindent
{\bf Empty Set Axiom}

\medskip
There is a set having no members. This set is denoted $\emptyset$.

\remark
Beginners often wonder whether there is more than one empty set.
For example, is the empty set of professors distinct from the
empty set of potatoes?

\medskip
The answer is, by the extensionality axiom, there is only
{\it one\/} empty set!

\medskip
Given any two objects $a$ and $b$, we can form the set $\{a, b\}$
containing exactly these two objects. Amazingly enough, this 
must also be an axiom:

\medskip\noindent
{\bf Pairing Axiom} 

\medskip
Given any two objects $a$ and $b$ (think sets), there is a set,
$\{a, b\}$, having as members just $a$ and $b$.

\medskip
Observe that if $a$ and $b$ are identical, then we have the
set $\{a, a\}$, which is denoted by $\{a\}$ and is called a
{\it singleton set\/} (this set has $a$ as its only element).

\medskip
To form bigger sets, we use the union operation. This too requires an axiom.

\medskip\noindent
{\bf Union Axiom (Version 1)}

\medskip
For any two sets $A$ and $B$, there is a set, $A\cup B$, called
the {\it union of $A$ and $B$\/} defined by
\[
x\in A\cup B \quad\hbox{iff}\quad (x\in A) \lor (x\in B).
\] 
This reads, $x$ is a member of $A\cup B$ if either $x$ belongs
to $A$ or $x$ belongs to $B$ (or both).
We also write
\[
A\cup B = \{x \mid x\in A\quad\hbox{or}\quad x\in B\}.
\]
Using the union operation, we can form bigger sets by taking unions
with singletons. For example, we can form
\[
\{a, b, c\} = \{a, b\}\cup \{c\}.
\]

\remark
We can systematically construct bigger and bigger sets by the 
following method: Given any set, $A$, let
\[
A^+ = A\cup \{A\}.
\]
If we start from the empty set, we obtain
sets that can be used to define the natural numbers
and the $+$ operation corresponds to the successor function
on the natural numbers, i.e., $n \mapsto n + 1$.

\medskip
Another operation is the power set formation. It is indeed a ``powerful''
operation, in the sense that it allows us to form very big sets.
For this, it is helpful to
define the notion of inclusion between sets. Given any two sets,
$A$ and $B$, we say that {\it $A$ is a subset of $B$\/} 
(or that {\it $A$ is included in $B$\/}), denoted $A\subseteq B$,
iff every element of $A$ is also an element of $B$, i.e.
\[
\forall x(x\in A \impl x \in B).
\] 
We say that {\it $A$ is a proper subset of $B$\/} iff
$A\subseteq B$ and $A\not= B$. This implies that
that there is some $b\in B$ with $b\notin A$.
We usually write $A \subset B$.

\medskip
Observe that the equality of two sets can be expressed by
\[
A = B \quad\hbox{iff}\quad A\subseteq B
\quad\hbox{and}\quad B \subseteq A.
\]

\medskip\noindent
{\bf Power Set Axiom}

\medskip
Given any set, $A$, there is a set, $\s{P}(A)$, (also denoted $2^A$)
called the {\it power set of $A$ \/}
whose members are exactly the subsets of $A$, i.e.,
\[
X\in \s{P}(A)\quad\hbox{iff}\quad X\subseteq A.
\]
 
\medskip
For example, if $A = \{a, b, c\}$, then
\[
\s{P}(A) = \{\emptyset, \{a\}, \{b\}, \{c\}, \{a, b\}, \{a, c\}, 
\{b, c\}, \{a, b, c\}\},
\]
a set containing $8$ elements. Note that the empty set and $A$ itself are always
members of $\s{P}(A)$.

\remark
If $A$ has $n$ elements, it is not hard to show that $\s{P}(A)$ has $2^n$
elements. For this reason, many people, including me, prefer the notation
$2^A$ for the power set of $A$.

\medskip
At this stage, we would like to define intersection and complementation.
For this, given any set, $A$, and given a property, $P$, (specified
by a first-order formula)
we need to be able to define the subset of $A$ consisting of those
elements satisfying $P$. This subset is denoted by
\[
\{x\in A\mid P\}.
\]
Unfortunately, there are problems with this construction.
If the formula, $P$, is somehow a circular definition and
refers to the subset that we are trying to define, 
then some paradoxes may arise!

\medskip
The way out is to place a restriction on the formula used
to define our subsets, and this leads to the subset axioms,
first formulated by Zermelo. These axioms are also
called {\it comprehension axioms\/} or {\it axioms of separation\/}.

\medskip\noindent
{\bf Subset Axioms}

\medskip
For every first-order formula, $P$,  we have the axiom:
\[
\forall A\exists X\forall x(x\in X 
\quad\hbox{iff}\quad (x\in A) \land P),
\]
where $P$ does {\it not\/} contain $X$ as a free variable.
(However, $P$ may contain $x$ free.)

\medskip
The subset axiom says that for every set, $A$, 
there is a set, $X$, consisting exactly of those elements of $A$
so that $P$ holds. For short, we usually write
\[
X = \{x\in A \mid P\}.
\]

\medskip
As an example, consider the formula
\[
P(B, x) = x\in B. 
\]
Then, the subset axiom  says
\[
\forall A\exists X\forall x(x\in A \land x\in B),
\]
which means that $X$ is the set of elements that belong both to $A$ and $B$.
This is called the 
{\it intersection of $A$ and $B$\/}, denoted by $A\cap B$.
Note that
\[
A\cap B = \{x\mid x\in A\quad\hbox{and}\quad x\in B\}.
\]

\medskip
We can also define the {\it relative complement of $B$ in $A$\/},
denoted $A - B$, given by the formula 
$P(x, B) = x\notin B$, so that
\[
A - B = \{x \mid x\in A\quad\hbox{and}\quad x\notin B\}.
\]
In particular, if $A$ is any given set and
$B$ is any subset of $A$, the set $A - B$ is also denoted
$\overline{B}$ and is called the {\it complement of $B$\/}.
Because $\land, \lor$ and $\neg$ satisfy the de Morgan laws
(remember, we are dealing with classical logic),
for any set $X$, 
the operations of union, intersection and complementation
on subsets of $X$ satisfy various identities, in particular
the de Morgan laws
\begin{align*}
\overline{A\cap B} & =  \overline{A} \cup \overline{B}\\
\overline{A\cup B} & =  \overline{A} \cap \overline{B}\\
\overline{\overline{A}} & =  A,
\end{align*}
and various associativity, commutativity and distributivity laws.

\medskip
So far, the union axiom only applies to two sets
but later on we will need to form infinite unions. Thus, it is 
necessary to generalize our union axiom as follows:

\medskip\noindent
{\bf Union Axiom (Final Version)}

\medskip
Given any set $X$ (think of $X$ as a set of sets),
there is a set, $\bigcup X$, defined so that
\[
x\in \bigcup X\quad\hbox{iff}\quad \exists B(B\in X \land x\in B).
\]
This says that $\bigcup X$ consists of all elements
that belong to some member of $X$.

\medskip
If we take $X = \{A, B\}$, where $A$ and $B$ are two sets, we see
that
\[
\bigcup \{A, B\} = A\cup B,
\]
and so, our final version of the union axiom subsumes
our previous union axiom which we now discard in favor
of the more general version.

\medskip
Observe that
\[
\bigcup \{A\} = A, \quad
\bigcup \{A_1, \ldots, A_n\} = A_1\cup \cdots\cup A_n.
\]
and in particular, $\bigcup \emptyset = \emptyset$.

\medskip
Using the subset axiom, we can also define infinite intersections.
For every nonempty set, $X$, there is a set, $\bigcap X$, defined
by
\[
x\in \bigcap X \quad\hbox{iff}\quad \forall B (B\in X  \impl x\in B).
\]

\medskip
The existence of $\bigcap X$ is justified as follows: Since
$X$ is nonempty, it contains some set, $A$; let
\[
P(X, x) = \forall B (B\in X \impl x\in B).
\]
Then, the subset axiom 
asserts the existence of a set $Y$ so that
for every $x$,
\[
x\in Y\quad\hbox{iff}\quad x\in A\quad\hbox{and}\quad P(X, x)
\]
which is equivalent to
\[
x\in Y\quad\hbox{iff}\quad  P(X, x).
\]
Therefore, the set $Y$ is our desired set, $\bigcap X$.

\medskip
Observe that
\[
\bigcap \{A, B\} = A\cap B,\quad
\bigcap \{A_1, \ldots, A_n\} = A_1 \cap \cdots \cap A_n.
\]
Note that $\bigcap \emptyset$ is not defined. 
Intuitively, it would have to be the set of all sets,
but such a set does not exist, as we now show. This
is basically a version of Russell's paradox.

\begin{thm} (Russell)
There is no set of all sets, i.e., there is no set to which
every other set belongs.
\end{thm}

\proof
Let $A$ be any set. We construct a set, $B$, that does
not belong to $A$. If the set of all sets existed,
then we could produce a set that does not belong to 
it, a contradiction.
Let
\[
B = \{a\in A \mid a\notin a\}.
\]
We claim that $B\notin A$. We proceed by contradiction,
so assume $B\in A$. However, by the definition of $B$, we have
\[
B\in B \quad\hbox{iff}\quad B\in A 
\quad\hbox{and}\quad B\notin B.
\]
Since $B\in A$, the above is equivalent to 
\[
B\in B \quad\hbox{iff}\quad  B\notin B,
\]
which is a contradiction. Therefore,  $B\notin A$
and we deduce that there is no set of all sets.
$\bigsquare$

\remarks
\begin{enumerate}
\item[(1)]
We should justify why the equivalence
$B\in B$ iff $B\notin B$ is a contradiction.
What we mean by ``a contradiction'' is 
that if the above equivalence holds,
then we can derive $\perp$ (falsity) and thus, all propositions
become provable. This is because we can show
that for any proposition, $P$, if
$P\equiv \neg P$ is provable, then 
$\neg (P\equiv \neg P)$ is also provable.
We leave the proof of this fact as an easy exercise for the reader.
By the way, this  holds classically as well as intuitionistically.
\item[(2)]
We said that in  the subset axiom, the variable $X$
is not allowed to occur free in $P$. 
A slight modification of Russell's paradox shows
that allowing $X$ to be free in $P$ lead to  
paradoxical sets. For example, pick $A$ to be any nonempty set and 
set $P(X, x) = x\notin X$. Then, look at the (alleged) set
\[
X = \{x\in A \mid x\notin X\}.
\]
As an exercise, the reader should show that
$X$ is empty iff $X$ is nonempty! 
\end{enumerate}

\medskip
This is as far as we can go with the elementary notions
of set theory that we have introduced so far. In order to 
proceed further, we need to define relations and functions,
which is the object of the next Chapter.

\medskip
The reader may also wonder why we have not yet 
discussed infinite sets. This is because we don't 
know how to  show that they exist!
Again, perhaps surprinsingly, this takes another axiom,
the {\it axiom of infinity\/}. We also have to define
when a set is infinite.  
However, we will not go into this right now. 
Instead, we will accept that the set of natural numbers,
$\natnums$, exists and is infinite. 
Once, we have the notion of a function, we will be able to
show that other sets are infinite by comparing their
``size'' with that of $\natnums$ (This is the purpose
of {\it cardinal numbers\/}, but this would lead us
too far afield).

\remark
In an axiomatic presentation of set theory, the natural numbers
can be defined from the empty set using the operation 
$A \mapsto A^+ = A \cup \{A\}$
introduced just after the union axiom.
The idea due to von Neumann is that
\begin{eqnarray*}
0 & = & \emptyset \\
1 & = & 0^+ = \{\emptyset\} = \{0\}\\
2 & = & 1^+ = \{\emptyset, \{\emptyset\}\} = \{0, 1\} \\
3 & = & 2^+ = \{\emptyset, \{\emptyset\}, \{\emptyset, \{\emptyset\}\}\} = \{0, 1, 2\} \\
  & \vdots & \\
n+1 & = & n^+ = \{0, 1, 2, \ldots, n\} \\
  & \vdots &  
\end{eqnarray*}

\medskip
However, the above subsumes induction! Thus, we have to proceed
in a different way to avoid circularities.

\begin{defin}
\label{inducdef}
{\em
We say that a set, $X$, is {\it inductive\/} iff
\begin{enumerate}
\item[(1)]
$\emptyset \in X$;
\item[(2)]
For every $A\in X$, we have $A^+\in X$.
\end{enumerate}
}
\end{defin}

\medskip\noindent
{\bf Axiom of Infinity}

\medskip
There is some inductive set.

\medskip
Having done this, we make the
\begin{defin}
\label{natnumdef}
{\em
A {\it natural number\/} is a set that belongs to every
inductive set. 
}
\end{defin}

\medskip
Using the subset axioms, we can  
show that there is a set whose members are exactly the natural numbers.
The argument is very similar to the one used to prove that
arbitrary intersections exist. By the Axiom of infinity, there is some 
inductive set, say $A$. Now consider the property, $P(x)$, which asserts
that $x$ belongs to every inductive set. By the subset axioms applied to $P$, 
there is a set, $\natnums$, such that
\[
x\in \natnums\quad\hbox{iff}\quad x\in A\quad\hbox{and}\quad P(x)
\]
and since $A$ is inductive and $P$ says that $x$ belongs to every inductive set,
the above is equivalent to
\[
x\in \natnums\quad\hbox{iff}\quad P(x),
\]
that is, $x\in \natnums$ iff $x$ belongs to every inductive set.
Therefore, the set of all natural numbers, $\natnums$, does exist.
The set $\natnums$ is  also denoted $\omega$.
We can now easily show 

\begin{thm}
\label{natinduc}
The set $\natnums$ is inductive  and it is a subset of every inductive set.
\end{thm}

\proof
Recall that $\emptyset$ belongs to
every inductive set; so,  $\emptyset$ is a natural number ($0$). 
As $\natnums$ is the set of natural numbers, $\emptyset\> (= 0)$ belongs to 
$\natnums$.
Secondly, if $n\in \natnums$, 
this means that $n$ belongs to every inductive set ($n$ is a natural number),
which implies that $n^+ = n + 1$ belongs to every inductive set,
which means that $n + 1$ is a natural number, i.e., $n + 1\in \natnums$.
Since $\natnums$ is the set of natural numbers and since every natural number
belongs to every inductive set, we conclude that $\natnums$ is a subset 
of every  inductive set.
$\bigsquare$

\danger
It would be tempting to view $\natnums$ as the intersection of the family
of inductive sets, but unfortunately this family is not a set;
it is too ``big'' to be a set.

\medskip
As a consequence of the above fact,
we obtain the 

\medskip\noindent
{\bf Induction Principle for $\natnums$\/}:
Any inductive subset of $\natnums$ is equal to $\natnums$ itself.

\medskip
Now, in our setting, $0 = \emptyset$ and $n^+ = n + 1$,
so the above principle can be restated as follows:

\medskip\noindent
{\bf Induction Principle for $\natnums$ (Version 2)\/}:
For any subset,  $S\subseteq \natnums$, if $0\in S$ and
$n + 1\in S$ whenever $n\in S$, then $S = \natnums$.

\medskip
We will see how to rephrase this induction principle
a little more conveniently
in terms of the notion of function in the next chapter.

\remarks
\begin{enumerate}
\item
We still don't know what an infinite set is or, for that matter,
that $\natnums$ is infinite! This will be shown in the next Chapter
(see Corollary \ref{pigenhole2}).
\item
Zermelo-Frankel set theory ($+$ Choice) has three more axioms that we did not
discuss: The {\it Axiom of Choice\/}, the {\it Replacement Axioms\/} and
the {\it Regularity Axiom\/}. For our purposes, only the Axiom of Choice
will be needed and we will introduce it in Chapter \ref{chap2}.
Let us just say that  the Replacement Axioms are needed to deal
with ordinals and cardinals and that the Regularity Axiom
is needed to show that every set is grounded. For more about these axioms,
see Enderton \cite{Endertonset}, Chapter 7. The Regularity Axiom
also implies that no set can be a member of itself, an eventuality
that is not ruled out by our current set of axioms!
 
\end{enumerate}

\chapter{Relations, Functions, Partial Functions}
\label{chap2}
\section{What is a Function?}
\label{sec6}
We use functions all the time in Mathematics and in 
Computer Science. But, what exactly is a function?

\medskip
Roughly speaking, a function, $f$, is a rule or mechanism, which
takes input values in some {\it input domain\/}, say $X$, and
produces output values in some {\it output domain\/},
say $Y$, in such a way that to each input $x\in X$
corresponds a {\it unique\/} output value $y \in Y$,  denoted
$f(x)$. We usually write $y = f(x)$, or better, $x \mapsto f(x)$.

\medskip
Often, functions are defined by some sort of closed expression
(a formula), but not always.
For example, the formula
\[
y = 2x 
\]
defines a function. Here, we can take both the input and output
domain to be $\reals$, the set of real numbers. Instead, we could
have taken $\natnums$, the set of natural numbers; this gives
us a different function. In the above example, $2x$ makes sense
for all input $x$, whether the input domain is $\natnums$ or $\reals$,
so our formula yields a function defined
for all of its input values. 

\medskip
Now, look at the function defined by the formula
\[
y = \frac{x}{2}. 
\]
If the input and output domains are both $\reals$, again
this function is well-defined. However, what if we
assume that the input and output domains are both $\natnums$?
This time, we have a problem when $x$ is odd. For example,
$\frac{3}{2}$ is not an integer, so our function is not defined
for all of its input values. It is a {\it partial function\/}.
Observe that this function is defined for the set of 
even natural numbers (sometimes denoted $2\natnums$)
and this set is called the {\it domain\/} (of definition) of $f$.
If we enlarge the output domain to be $\rats$, the set of rational 
numbers, then our function is defined for all inputs.

\medskip
Another example of a partial function is given by
\[
y = \frac{x + 1}{x^2 -3x + 2},
\]
assuming that both the input and output domains are $\reals$.
Observe that for $x = 1$ and $x = 2$, the denominator vanishes,
so we get the undefined fractions $\frac{2}{0}$ and
$\frac{3}{0}$. The function ``blows up'' for $x = 1$ and $x = 2$,
its value is ``infinity'' $(= \infty$), which is not an element of $\reals$.
So, the domain of $f$ is $\reals - \{1, 2\}$.

\medskip
In summary, functions need not be defined for all of their input
values and we need to pay close attention to
both the input and the ouput domain of our functions.

\medskip
The following example illustrates another difficulty:
Consider the function given by
\[
y = \sqrt{x}.
\]
If we assume that the input domain is $\reals$
and that the output domain is  $\reals^+ = \{x\in \reals\mid x\geq 0\}$, 
then this function is not defined for negative values of $x$.
To fix this problem,
we can extend the output domain to be $\complex$, the complex
numbers. Then we can make sense of $\sqrt{x}$ when
$x < 0$. 
However, a new problem comes up: Every negative number, $x$, has two
complex square roots, $-i\sqrt{-x}$ and $+i\sqrt{-x}$
(where $i$ is ``the'' square root of $-1$).
Which of the two should we pick?

\medskip
In this case, we could systematically pick $+i\sqrt{-x}$
but what if we extend the input domain to be $\complex$.
Then, it is not clear which of the two complex roots
should be picked, as there is no obvious total order on $\complex$.
We can treat $f$ as a {\it multi-valued function\/},
that is, a function that may return several possible outputs
for a given input value.

\medskip
Experience shows that it is akward to deal with multi-valued functions
and that it is best to treat them as relations (or to
change the output domain to be a power set, which is equivalent
to view the function as a relation).

\medskip
Let us give one more example showing that it is not always easy
to make sure that a formula is a proper definition of a function.
Consider the function from $\reals$ to $\reals$ given by
\[
f(x) = 1 + \sum_{n = 1}^{\infty} \frac{x^n}{n!}.
\]
Here, $n!$ is the function {\it factorial\/},
defined by
\[
n! = n\cdot (n-1) \cdots 2\cdot 1.
\]
How do we make  sense of this infinite expression?
Well, that's where analysis comes in, with the notion of limit
of a series, etc. It turns out that $f(x)$ is the exponential
function $f(x) = e^x$.
Actually, $e^x$ is even defined when $x$ is a complex number
or even a square matrix (with real or complex entries)!
Don't panic, we will not use such functions in this course.

\medskip
Another issue comes up, that is,  the notion of {\it computability\/}.
In all of our examples, and for most functions we will ever need
to compute, it is clear that it is possible to give
a mechanical procedure, i.e., a computer program which computes
our functions (even if it hard to write such a program or
if such a program takes a very long time
to compute the output from the input).

\medskip
Unfortunately, there are functions which, although
well-defined mathematically, are not computable!
For an example, let us go back to first-order logic and the notion
of provable proposition. Given a finite (or countably infinite)
alphabet of function, predicate, constant symbols, and a countable
supply of variables, it is quite clear
that the set $\s{F}$ of all propositions built up from these symbols
and variables can be enumerated systematically. We can define
the function, $\mathrm{Prov}$, with input domain $\s{F}$ and
output domain $\{0, 1\}$, so that, for every
proposition $P\in \s{F}$, 
\[
\mathrm{Prov}(P) = \cases{
1 & if $P$ is provable (classically) \cr
0 & if $P$ is not provable (classically).\cr
}
\]
Mathematically, for every proposition, $P\in \s{F}$, either
$P$ is provable or it is not, so this function makes sense.
However, by Church's Theorem (see Section \ref{sec5}), we know
that there is {\bf no} computer program that will terminate
for all input propositions and give an answer in  a finite number
of steps! So, although the function $\mathrm{Prov}$ makes
sense as an abstract function, it is not computable.
Is this a paradox? No, if we are careful when defining a function
not to incorporate in the definition any notion of computability
and instead to take a more abstract and, in some some sense naive
view of a function as some kind of input/output process given by  pairs
$\lag$input value, output value$\rag$
(without worrying about the way the output is ``computed'' from the input).
A rigorous way to proceed is to use the notion of
ordered pair and of graph of a function.
Before we do so, let us point out some facts about functions
that were revealed by our examples:
\begin{enumerate}
\item
In order to define a function, in addition to defining its
input/output behavior, it is also important to specify what
is its {\it input domain\/} and its {\it output domain\/}. 
\item
Some functions may not be defined for all of their input values;
a function can be a {\it partial function\/}.
\item
The input/output behavior of a function can be defined
by a set of ordered pairs. As we will see next, this is
the {\it graph\/} of the function.
\end{enumerate}

\medskip
We are now going to formalize the notion of function (possibly partial)
using the concept of ordered pair.

\section[Ordered Pairs, Cartesian Products, Relations, etc.]
{Ordered Pairs, Cartesian Products, Relations, \\
Functions, Partial Functions}
\label{sec7}
Given two sets, $A$ and $B$, one of the basic constructions
of set theory is the formation of an {\it ordered pair\/},
$\lag a, b\rag$, where $a\in A$ and $b\in B$.
Sometimes, we also write $(a, b)$ for an ordered pair.
The main property of ordered pairs is that if
$\lag a_1, b_1\rag$ and $\lag a_2, b_2\rag$ are ordered pairs,
where $a_1, a_2\in A$ and $b_1, b_2\in B$, then
\[
\lag a_1, b_1\rag = \lag a_2, b_2\rag
\quad\hbox{iff}\quad
a_1 = a_2 \quad\hbox{and}\quad b_1 = b_2.
\]
Observe that  this property implies that,
\[
\lag a, b\rag \not= \lag b, a\rag,
\]
unless $a = b$. Thus, the ordered pair, $\lag a, b\rag$, is
not a notational variant for the set $\{a, b\}$; 
implicit to the notion of ordered pair is the fact that
there is an order (even though we have not yet defined
this notion yet!) among the elements of the pair. Indeed, in
$\lag a, b\rag$, the element $a$ comes first and $b$ comes
second. Accordingly, given an ordered pair, $p = \lag a, b\rag$,
we will denote $a$ by $pr_1(p)$ and $b$ by $pr_2(p)$
({\it first an second projection\/} or
{\it first and second coordinate\/}).

\remark
Readers who like set theory will be happy to hear that an ordered
pair, $\lag a, b\rag$, can be defined as the set $\{\{a\}, \{a, b\}\}$.
This definition  is due to Kuratowski, 1921.
An earlier (more complicated) definition given by N. Wiener in 1914 is
$\{\{\{a\}, \emptyset\}, \{\{b\}\}\}$.

\medskip
Now, from set theory, it can be shown that given two sets,
$A$ and $B$, the set of all ordered pairs $\lag a, b\rag$,
with $a\in A$ and $b\in B$, is a set denoted $A\times B$ and
called the {\it Cartesian product of $A$ and $B$\/} (in that order).
By convention, we agree that $\emptyset\times B = A\times \emptyset
= \emptyset$.  To simplify the terminology, we often say
{\it pair\/} for {\it ordered pair\/}, with the understanding
that pairs are always ordered (otherwise, we should say set).

\medskip
Of course, given three sets, $A, B, C$, we can form
$(A\times B)\times C$ and we call its elements (ordered) {\it triples\/}
(or {\it triplets\/}). To simplify the notation, we write
$\lag a, b, c\rag$ instead of $\lag\lag a, b\rag, c\rag$.
More generally, given $n$ sets $A_1, \ldots, A_n$ ($n\geq 2$),
we define the set of {\it $n$-tuples\/}, \\
$A_1\times A_2\times\cdots\times A_n$, as
$(\cdots((A_1\times A_2)\times A_3)\times  \cdots )\times A_n$. 
An element of $A_1\times A_2\times\cdots\times A_n$ is denoted
by $\lag a_1, \ldots, a_n\rag$ (an $n$-tuple).
We agree that when $n = 1$, we just have $A_1$ and
a $1$-tuple is just an element of $A_1$.

\medskip
We now have all we need to define relations.

\begin{defin}
\label{reldef}
{\em
Given two sets, $A$ and $B$, a (binary) {\it relation, $R$,
between $A$ and $B$\/} is any subset 
$R \subseteq A\times B$ of ordered pairs from $A\times B$.
When $\lag a, b\rag\in R$, we also write $a R b$ and we say
that {\it $a$ and $b$ are related by $R$\/}.
The set
\[
\mathit{dom}(R) = \{a\in A\mid \exists b\in B,\> \lag a, b\rag \in R\}
\]
is called the {\it domain of $R$\/} and the set
\[
\mathit{range}(R) = \{b\in B\mid \exists a\in A,\> \lag a, b\rag \in R\}
\]
is called the {\it range of $R$\/}.
Note that $\mathit{dom}(R)\subseteq A$ and $\mathit{range}(R) \subseteq B$.
When $A = B$, we often say that 
{\it $R$ is a (binary) relation over $A$\/}.
}
\end{defin}

\medskip
Among all relations between $A$ and $B$, we mention three relations
that play a special role:
\begin{enumerate}
\item
$R = \emptyset$, the {\it empty relation\/}.
Note that $\mathit{dom}(\emptyset) = \mathit{range}(\emptyset) = \emptyset$.
This is not a very exciting relation!
\item
When $A = B$, we have the {\it identity relation\/},
\[
\id_A = \{\lag a, a\rag \mid a\in A\}.
\]
The identity relation relates every element to itself, 
and that's it!
Note that  \\
$\mathit{dom}(\id_A) = \mathit{range}(\id_A) = A$.
\item
The relation $A\times B$ itself. This relation relates every
element of $A$ to every element of $B$.
Note that  $\mathit{dom}(A\times B) = A$ and
$\mathit{range}(A\times B) = B$.
\end{enumerate}

\medskip
Relations can be represented graphically by pictures
often called graphs. (Beware, the term ``graph'' is
very much overloaded. Later on, we will define
what a graph is.)
We depict the elements of both sets $A$ and $B$ as points
(perhaps with different colors) and we indicate that
$a\in A$ and $b\in B$ are related (i.e., $\lag a, b\rag\in R$)
by drawing an oriented edge (an arrow) starting from $a$ (its source)
and ending in  $b$ (its target). Here is an example:

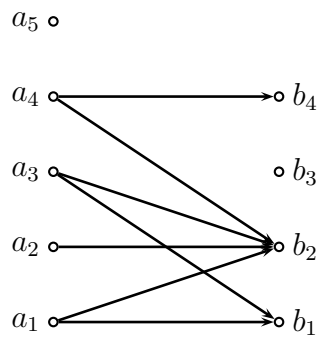
\begin{figure}[H]
 \begin{center}
    \begin{pspicture}(0,0)(3,3.8)
    \cnode(0,0){2pt}{u1}
    \cnode(0,1){2pt}{u2}
    \cnode(0,2){2pt}{u3}
    \cnode(0,3){2pt}{u4}
    \cnode(0,4){2pt}{u5}
    \cnode(3,0){2pt}{v1}
    \cnode(3,1){2pt}{v2}
    \cnode(3,2){2pt}{v3}
    \cnode(3,3){2pt}{v4}
    \ncline[linewidth=1pt]{->}{u1}{v1}
    \ncline[linewidth=1pt]{->}{u1}{v2}
    \ncline[linewidth=1pt]{->}{u2}{v2}
    \ncline[linewidth=1pt]{->}{u3}{v2}
    \ncline[linewidth=1pt]{->}{u3}{v1}
    \ncline[linewidth=1pt]{->}{u4}{v4}
    \ncline[linewidth=1pt]{->}{u4}{v2}
    \uput[180](0,0){$a_1$}
    \uput[180](0,1){$a_2$}
    \uput[180](0,2){$a_3$}
    \uput[180](0,3){$a_4$}
    \uput[180](0,4){$a_5$}
    \uput[0](3,0){$b_1$}
    \uput[0](3,1){$b_2$}
    \uput[0](3,2){$b_3$}
    \uput[0](3,3){$b_4$}
    \end{pspicture}
  \end{center}
  \caption{A binary relation, $R$}
  \label{fig1}
\end{figure}

In Figure \ref{fig1}, $A = \{a_1, a_2, a_3, a_4, a_5\}$ and
$B = \{b_1, b_2, b_3, b_4\}$. Observe that $a_5$ is not related to 
any element of $B$, $b_3$ is not related to any element of $A$  
and that some elements of $A$, namely,  $a_1, a_3, a_4$,
are related some several elements of $B$.

\medskip
Now, given a relation, $R \subseteq A\times B$, some element
$a\in A$ may be related to several distinct elements $b\in B$.
If so, $R$ does not correspond to our notion of a function,
because we want our functions to be single-valued.
So, we impose a natural condition on relations to get
relations that correspond to functions.

\begin{defin}
\label{funcrel}
{\em
We say that a relation, $R$, between two sets $A$ and $B$ is
{\it functional\/} if for every $a\in A$, there is {\it at most one\/}
$b\in B$ so that $\lag a, b\rag \in R$.
Equivalently, $R$ is functional if for all $a\in B$ and all
$b_1, b_2\in B$, if $\lag a, b_1\rag\in R$ and
$\lag a, b_2\rag\in R$, then $b_1 = b_2$.
}
\end{defin}

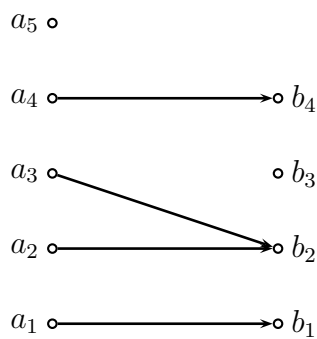
\begin{figure}[H]
 \begin{center}
    \begin{pspicture}(0,0)(3,3.8)
    \cnode(0,0){2pt}{u1}
    \cnode(0,1){2pt}{u2}
    \cnode(0,2){2pt}{u3}
    \cnode(0,3){2pt}{u4}
    \cnode(0,4){2pt}{u5}
    \cnode(3,0){2pt}{v1}
    \cnode(3,1){2pt}{v2}
    \cnode(3,2){2pt}{v3}
    \cnode(3,3){2pt}{v4}
    \ncline[linewidth=1pt]{->}{u1}{v1}
    \ncline[linewidth=1pt]{->}{u2}{v2}
    \ncline[linewidth=1pt]{->}{u3}{v2}
    \ncline[linewidth=1pt]{->}{u4}{v4}
    \uput[180](0,0){$a_1$}
    \uput[180](0,1){$a_2$}
    \uput[180](0,2){$a_3$}
    \uput[180](0,3){$a_4$}
    \uput[180](0,4){$a_5$}
    \uput[0](3,0){$b_1$}
    \uput[0](3,1){$b_2$}
    \uput[0](3,2){$b_3$}
    \uput[0](3,3){$b_4$}
    \end{pspicture}
  \end{center}
  \caption{A functional relation $G$}
  \label{fig2}
\end{figure}

\medskip
The picture in Figure \ref{fig2} shows  an example of a functional relation.

\medskip
Using Definition \ref{funcrel}, we can give a rigorous definition of
a function (partial or not).

\begin{defin}
\label{fundef}
{\em
A {\it partial function, $f$\/}, is a triple,
$f = \lag A, G, B\rag$, where $A$ is a set called
the {\it input domain of $f$\/}, $B$ is a set called
the {\it output domain of $f$\/} (sometimes {\it codomain of $f$\/})
and $G\subseteq A\times B$ is a functional relation
called the {\it graph of $f$\/}; we let
$\mathit{graph}(f) = G$. We write $\mapdef{f}{A}{B}$
to indicate that $A$ is the input domain of $f$ and that
$B$ is the codomain of $f$ and we let
$\mathit{dom}(f) = \mathit{dom}(G)$ and
$\mathit{range}(f) = \mathit{range}(G)$.
For every $a\in \mathit{dom}(f)$,
the unique element, $b\in B$, so that $\lag a, b\rag\in \mathit{graph}(f)$
is denoted by $f(a)$ (so, $b = f(a)$). Often, we say that
$b =f(a)$ is the {\it image of $a$ by $f$\/}.
The range of $\mathit{f}$ is also called the {\it image of $f$\/} and
is denoted $\Im(f)$.  
If $\mathit{dom}(f) = A$,
we say that $f$ is a {\it total function\/}, for short, a
{\it function with domain $A$\/}. 
}
\end{defin}

\remarks
\begin{enumerate}
\item
If $f =  \lag A, G, B\rag$ is a partial function
and $b = f(a)$ for some $a\in \mathit{dom}(f)$, we say that
{\it $f$ maps $a$ to $b$\/}; we may write $f\co a \mapsto b$.
For any $b\in B$, the set
\[
\{a\in A \mid f(a) = b\}
\]
is denoted $f^{-1}(b)$ and called the {\it inverse image\/}
or {\it preimage of $b$ by $f$\/}. (It is also called
the {\it fibre of $f$ above $b$\/}. We will explain this
peculiar language later on.)
Note that $f^{-1}(b)\not= \emptyset$ iff $b$ is in the image (range) of $f$.
Often, a function, 
partial or not, is called a {\it map\/}.
\item 
Note that Definition \ref{fundef} allows $A = \emptyset$.
In this case, we must have $G =\emptyset$ and, technically,
$\lag \emptyset, \emptyset, B\rag$ is  total function!
It is the {\it empty function from $\emptyset$ to $B$\/}.
\item
When a partial function is a total function, we don't
call it a ``partial total function'', but simply
a ``function''. The usual pratice is that the term ``function''
refers to a total function. However, sometimes, we say
``total function'' to stress that a function is indeed
defined on all of its input domain.
\item
Note that if a partial
function $f = \lag A, G, B\rag$ is not a total function,
then $\mathit{dom}(f) \not= A$ and
for all $a\in A - \mathit{dom}(f)$, there is {\bf no}
$b\in B$ so that $\lag a, b\rag\in \mathit{graph}(f)$.
This corresponds to the intuitive fact that $f$ does
not produce any output for any value not in its domain
of definition. We can imagine that $f$ ``blows up'' 
for this input (as in the situation where the denominator of a fraction
is $0$) or that the program computing $f$ loops indefinitely 
for that input.
\item
If $f = \lag A, G, B\rag$ is a total function and $A\not= \emptyset$,
then $B\not= \emptyset$.
\item
For any set, $A$, the identity relation, $\id_A$, is actually
a function $\mapdef{\id_A}{A}{A}$.
\item
Given any two sets, $A$ and $B$, the rules
$\lag a, b\rag \mapsto a = pr_1(\lag a, b\rag)$
and $\lag a, b\rag \mapsto b = pr_2(\lag a, b\rag)$
make $pr_1$ and $pr_2$ into functions
$\mapdef{pr_1}{A\times B}{A}$ and $\mapdef{pr_2}{A\times B}{B}$ 
called the {\it first and second projections\/}.
\item
A function, $\mapdef{f}{A}{B}$, is sometimes denoted
$A \stackrel{f}{\longrightarrow} B$. Some authors
use a different kind of arrow to indicate that $f$ is partial, 
for example, a dotted or dashed arrow. We will not go that far!
\item
The set of all functions, $\mapdef{f}{A}{B}$, is denoted by $B^A$.
If $A$ and $B$ are finite, $A$ has $m$ elements and $B$ has $n$ elements,
it is easy to prove that $B^A$ has $n^m$ elements.
\end{enumerate}

\medskip
The reader might wonder why, in the definition of a 
(total) function, $\mapdef{f}{A}{B}$, we do not require
$B = \Im f$, since we require that $\mathrm{dom}(f) = A$.

\medskip
The reason has to do with experience and convenience. It turns out that
in most cases, we know what the domain of a function is,
but it may be very hard to determine exactly what its
image is. Thus, it is more convenient to be flexible about
the codomain. As long as we know that $f$ maps into
$B$, we are satisfied. 

\medskip
For example, consider functions, 
$\mapdef{f}{\reals}{\reals^2}$, 
from the real line into the plane.
The image of such a function is
a {\it curve\/} in the plane $\reals^2$.
Actually, to really get ``decent'' curves we need to impose some
reasonable conditions on $f$, for example, to be differentiable.
Even continuity may yield very strange curves 
(see Section \ref{sec12b}). But even for a very well behaved function, 
$f$, it may be very hard to figure out what the image of  $f$ is.
Consider the function, $t \mapsto (x(t), y(y))$,
given by
\begin{eqnarray*}
x(t) & = &\frac{t(1 + t^2)}{1 + t^4} \\
y(t) & = &\frac{t(1 - t^2)}{1 + t^4}.\\
\end{eqnarray*}

\begin{figure}
\centerline{
\psfig{figure=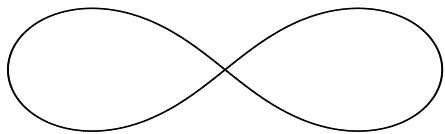,height=2truein,width=2.5truein}}
\caption{Lemniscate of Bernoulli}
\label{lemnisfig}
\end{figure}

The  curve which is the image of this function,
shown in Figure \ref{lemnisfig},
is called the 
``lemniscate of Bernoulli''.  

Observe that this curve 
has a self-intersection at the origin, which is not so obvious
at first glance.

\section{Induction Principle on $\natnums$}
\label{sec7b}
Now that we have the notion of function, we can restate
the induction principle (Version 2) stated at the send of Section \ref{sec5b}
to make it more flexible. We define a {\it property 
of the natural numbers\/}  as any function,
$\mapdef{P}{\natnums}{\{\mathbf{true}, \mathbf{false}\}}$.
The idea is that $P(n)$ holds iff $P(n) = \mathbf{true}$, 
else $P(n) = \mathbf{false}$. Then, we have the following principle:

\medskip\noindent
{\bf Principle of Induction for $\natnums$ (Version 3)}.

\medskip
Let $P$ be any property of the natural numbers. In order to prove
that $P(n)$ holds for all $n\in \natnums$, it is enough to prove that
\begin{enumerate}
\item[(1)]
$P(0)$ holds and
\item[(2)]
For every $n\in \natnums$, the implication $P(n) \impl P(n+1)$ holds.
\end{enumerate}
As a formula, (1) and (2) can be written
\[
[P(0) \land (\forall n\in \natnums)(P(n) \impl P(n+1))] 
\impl (\forall n\in \natnums)P(n).
\]

\medskip
Step (1) is usually called the {\it basis\/} or {\it base step\/}
of the induction and step (2) is called the {\it induction step\/}.
In step (2), $P(n)$ is called the {\it induction hypothesis\/}.
That the above induction principle is valid is given by the

\begin{prop}
\label{inducp1}
The Principle of Induction stated above is valid.
\end{prop}

\proof
Let 
\[
S = \{n\in \natnums \mid P(n) =\mathbf{true}\}.
\]
By the  induction principle (Version 2) stated at the send of Section \ref{sec5b},
it is enough to prove that $S$ is inductive, because then
$S = \natnums$ and we are done. 

\medskip
Since $P(0)$ hold, we have $0\in S$.
Now, if $n\in S$, i.e., if $P(n)$ holds, since
$P(n) \impl P(n+1)$ holds for every $n$, we deduce that
$P(n+1)$ holds, that is, $n+1\in S$. Therefore, $S$ is inductive as claimed
and this finishes the proof.
$\bigsquare$

\medskip
Induction is a very valuable tool for proving properties
of the natural numbers and we will make  extensive use of it.
We will also see other more powerful induction principles. 
Let us give just one example illustrating how it is used.

\medskip
{\it Claim\/}: 
\[
1 + 3 + 5 + \cdots + 2n + 1 = (n+1)^2,
\]
where $n\in \natnums$.

\medskip
For the basis of the induction, where $n = 0$, we get
$1 = 1^2$,  so the base step holds.

\medskip
For the induction step, for any $n\in \natnums$, assume that
\[
1 + 3 + 5 + \cdots + 2n + 1 = (n+1)^2.
\]
Consider $1 + 3 + 5 + \cdots + 2n + 1 + 2(n+1) + 1 = 
1 + 3 + 5 + \cdots + 2n + 1 + 2n + 3$. Then, using the induction hypotesis,
we have
\begin{eqnarray*} 
1 + 3 + 5 + \cdots + 2n + 1 + 2n + 3 & = & (n+1)^2 + 2n + 3 \\
& = & n^2 + 2n + 1 + 2n + 3 = n^2 + 4n + 4 \\
& = & (n + 2)^2. 
\end{eqnarray*} 
Therefore, the induction step holds and this completes the proof
by induction.
$\bigsquare$

\medskip
A useful way to produce new relations or functions is 
to compose them.

\section{Composition of Relations and Functions}
\label{sec8}
We begin with the definition of the composition of relations.

\begin{defin}
\label{comprel}
{\em
Given two relations, $R\subseteq A\times B$ and $S \subseteq B\times C$,
the {\it composition of $R$ and $S$\/}, denoted $R\circ S$, is
the relation between $A$ and $C$ defined by
\[
R\circ S = \{\lag a, c\rag \in A\times C \mid \exists b\in B, \>
\lag a, b\rag \in R\quad\hbox{and}\quad \lag b, c\rag \in S\}.
\]
}
\end{defin}

\medskip
One should check that for any relation $R\subseteq A\times B$,
we have $\id_A\circ R = R$ and $R\circ \id_B = R$.
If $R$ and $S$ are the graphs of functions, possibly partial,
is $R\circ S$ the graph of some function? The answer is yes,
as shown in the following

\begin{prop}
\label{compos1}
Let  $R\subseteq A\times B$ and $S \subseteq B\times C$ be two relations.
\begin{enumerate}
\item[(a)]
If $R$ and $S$ are both functional relations, then $R\circ S$ is
also a functional relation. Consequently,  $R\circ S$ is the graph
of some partial function.
\item[(b)]
If $\mathit{dom}(R) = A$ and $\mathit{dom}(S) = B$, then
$\mathit{dom}(R\circ S) = A$.
\item[(c)]
If $R$ is the graph of a (total) function from $A$ to $B$
and $S$ is the graph of a (total) function from $B$ to $C$, then
$R\circ S$ is the graph of a (total) function from $A$ to $C$.
\end{enumerate}
\end{prop}

\proof
(a)
Assume that $\lag a, c_1\rag\in R\circ S$ and
$\lag a, c_2\rag\in R\circ S$. By definition of $R\circ S$, there
exist $b_1, b_2\in B$ so that
\begin{align*}
& \lag a, b_1\rag \in R, \quad \lag b_1, c_1\rag \in S, \\
& \lag a, b_2\rag \in R, \quad \lag b_2, c_2\rag \in S.
\end{align*}
As $R$ is functional, $\lag a, b_1\rag \in R$ 
and $\lag a, b_2\rag \in R$ implies $b_1 = b_2$. Let $b = b_1 = b_2$, 
so that $\lag b_1, c_1\rag = \lag b, c_1\rag$ and
$\lag b_2, c_2\rag = \lag b, c_2\rag$. But, $S$ is also functional,
so  $\lag b, c_1\rag\in S$ and $\lag b, c_2\rag\in S$ implies
that $c_1 = c_2$, which proves that $R\circ S$ is functional.

\medskip 
(b) Pick any $a\in A$. 
The fact that $\mathit{dom}(R) = A$ means that  
there is some $b\in B$ so that $\lag a, b\rag\in R$.
As $S$ is also functional, there is some $c\in C$ so that 
$\lag b, c\rag\in S$. Then, by the definition of $R\circ S$, we see that
$\lag a, c\rag\in R\circ S$. Since the argument holds for any $a\in A$,
we deduce that $\mathit{dom}(R\circ S) = A$.

\medskip
(c) If $R$ and $S$ are the graphs of partial functions, 
then this means that they are functional and (a) implies that
$R\circ S$ is also functional. This shows that 
$R\circ S$ is the graph of the partial function
$\lag A, R\circ S, C\rag$.
If $R$ and $S$ are the graphs of total functions, then 
$\mathit{dom}(R) = A$ and $\mathit{dom}(S) = B$. By (b), 
we deduce that $\mathit{dom}(R\circ S) = A$. By the first part of
(c), $R\circ S$ is the graph of the partial function
$\lag A, R\circ S, C\rag$, which is a total function,
since  $\mathit{dom}(R\circ S) = A$.
$\bigsquare$

\medskip
Proposition \ref{compos1} shows that it is legitimate to define
the composition of functions, possibly partial. Thus, we make
the following 

\begin{defin}
\label{compfun}
{\em 
Given two functions, $\mapdef{f}{A}{B}$ and $\mapdef{g}{B}{C}$,
possibly partial, the {\it composition of $f$ and $g$\/},
denoted $g\circ f$, is the function (possibly partial)
\[
g\circ f = \lag A, \mathit{graph}(f)\circ  \mathit{graph}(g), C\rag.
\]
}
\end{defin}

\medskip
The reader must have noticed that the composition of
two functions  $\mapdef{f}{A}{B}$ and $\mapdef{g}{B}{C}$ is denoted
$g\circ f$, whereas the graph of $g\circ f$ is denoted
$\mathit{graph}(f)\circ  \mathit{graph}(g)$.
This ``reversal'' of the order in which function composition
and relation composition are written  is unfortunate and somewhat confusing.

\medskip
Once again, we are victim of tradition. The main reason
for writing function composition as $g\circ f$ is that
traditionally, the result of applying a function $f$ to an 
argument $x$ is written $f(x)$. Then, $(g\circ f)(x) = g(f(x))$,
which makes sense. Some people, in  particular
algebraists, write function composition as
$f\circ g$, but then, they write  the result of applying a function $f$ to an 
argument $x$ as  $x f$. With this convention,
$x(f\circ g) = (x f) g$, which also makes sense.

\medskip
We prefer to stick to the convention where we write $f(x)$
for the result of applying a function $f$ to an argument $x$ and, 
consequently, we use the notation $g\circ f$ for the composition of
$f$ with $g$, even though it is the opposite of the convention for
writing the composition of relations.

\medskip
Given any three relations, $R\subseteq A\times B$,
$S\subseteq B\times C$ and $T\subseteq C\times D$, the reader should
verify that
\[
(R\circ S)\circ T = R\circ (S\circ T).
\]
We say that composition is {\it associative\/}.
Similarly, for any three functions (possibly partial),
$\mapdef{f}{A}{B}$, $\mapdef{g}{B}{C}$ and $\mapdef{h}{C}{D}$,
we have (associativity of function composition)
\[
(h\circ g)\circ f = h\circ (g\circ f).
\]  

\section{Recursion on $\natnums$}
\label{sec8b}
The following situation often occurs: We have some set, $A$, 
some fixed element, $a\in A$,
some function, $\mapdef{g}{A}{A}$, and we wish to define a new
function, $\mapdef{h}{\natnums}{A}$, so that
\begin{eqnarray*}
h(0) & = & a, \\
h(n+1) & = & g(h(n))\qquad\hbox{for all}\quad n \in \natnums.
\end{eqnarray*}

\medskip
This way of defining $h$ is called a {\it recursive definition\/} 
(or a definition by {\it primitive recursion\/}).
I would be surprised if any computer scientist had any trouble
with this ``definition'' of $h$ but how can we justify
rigorously that such a function exists and is unique?

\medskip
Indeed, the existence (and uniqueness) of $h$ requires  proof.  
The proof, although not
really hard, is surprisingly involved and, in fact quite subtle.
For those reasons, we will not give a proof of the following theorem
but instead the main idea of the proof. The reader 
will find a complete proof in
Enderton \cite{Endertonset} (Chapter 4).

\begin{thm}
\label{recnat} (Recursion Theorem on $\natnums$)
Given any set, $A$, any fixed element, $a\in A$,
and any  function, $\mapdef{g}{A}{A}$, there is a unique
function, $\mapdef{h}{\natnums}{A}$, so that
\begin{eqnarray*}
h(0) & = & a, \\
h(n+1) & = & g(h(n))\qquad\hbox{for all}\quad n \in \natnums.
\end{eqnarray*}
\end{thm}

\proof
The idea is to approximate $h$. To do this, define a 
function, $f$, to be {\it acceptable\/} iff
\begin{enumerate}
\item
$\mathit{dom}(f) \subseteq \natnums$ and $\mathit{range}(f) \subseteq A$;
\item
If $0 \in \mathit{dom}(f)$, then $f(0) = a$;
\item
If $n + 1\in \mathit{dom}(f)$, then $n\in \mathit{dom}(f)$ and
$f(n+1) = g(f(n))$.
\end{enumerate}
 
Let $\s{F}$ be the collection of all acceptable functions and set
\[
h = \bigcup \s{F}.
\]
All we can say, so far,  is that $h$ is a relation.
We claim that $h$ is the desired function. For this, four things need to be proved:

\begin{enumerate}
\item
The relation $h$ is function.
\item
The function $h$ is acceptable.
\item
The function $h$ has domain $\natnums$.
\item
The function $h$ is unique.
\end{enumerate}

As expected, we make heavy use of induction in proving
(1), (2), (3) and (4). For complete details, see Enderton \cite{Endertonset} (Chapter 4).
$\bigsquare$

\medskip
Theorem \ref{recnat} is very important. Indeed,  experience shows that
it is used almost as much as induction!
As an example, we show how to define addition on $\natnums$.
Indeed, at the moment, we know what the natural numbers are but we don't
know what are the arithmetic operations such as $+$ or $*$! (at least,
not in our axiomatic treatment; of course, nobody needs an axiomatic treatment to
know how to add or multiply).

\medskip
How do we define $m + n$, where $m,n \in \natnums$?

\medskip
If we try to use Theorem \ref{recnat} directly, we seem to have a problem, because
addition is a function of two arguments, but $h$ and $g$ in the theorem
only take one argument. We can overcome this problem in two ways:

\begin{enumerate}
\item[(1)] 
We prove a generalization of Theorem \ref{recnat} involving functions
of several arguments, but with recursion only in a {\it single\/} argument.
This can be done quite easily but we have to be a little careful.
\item[(2)] 
For any fixed $m$, we define $add_m(n)$ as $add_m(n) = m + n$, that is,
we define addition of a {\it fixed\/} $m$ to any $n$.
Then, we let $m + n = add_m(n)$. 
\end{enumerate}

\medskip
Since solution (2) involves much less work, we follow it.
Let $S$ denote the successor function on $\natnums$, that is, the function
given by
\[
S(n) = n^+ = n + 1.
\]
Then, using Theorem \ref{recnat} with $a = m$ and $g = S$, we get a function,
$add_m$, such that

\begin{eqnarray*}
add_m(0) & = & m, \\
add_m(n+1) & = & S(add_m(n)) = add_m(n) + 1\qquad\hbox{for all}\quad n \in \natnums.
\end{eqnarray*}

\medskip
Finally, for all $m, n\in \natnums$, we define $m + n$ by
\[
m + n = add_m(n).
\]
Now, we have our addition function on $\natnums$. 
But this is not the end of the story because we don't
know yet that the above definition yields 
a function having the usual properties of addition, such as
\begin{eqnarray*}
m + 0 & = & m\\
m + n & = & n + m\\
(m + n) + p & = & m + (n + p).
\end{eqnarray*}
To prove these properties, of course, we use induction!

\medskip
We can also define multiplication. Mimicking what we did for
addition, define $mult_m(n)$ by recursion as follows;

\begin{eqnarray*}
mult_m(0) & = & 0, \\
mult_m(n+1) & = & mult_m(n) + m \qquad\hbox{for all}\quad n \in \natnums.
\end{eqnarray*}
Then, we set
\[
m\cdot n = mult_m(n).
\]
Note how the recursive definition of $mult_m$ uses
the adddition function, $+$, previously defined.
Again, to prove the usual properties of multiplication as well
as the distributivity of $\cdot $ over $+$, we use induction.
Using recursion, we can define many more arithmetic functions.
For example, the reader should try defining exponentiation, $m^n$.

\medskip
We still haven't defined the usual ordering on the natural numbers
but we will do so in the next chapter. Of course,
we all know what it is and we will not refrain from using it.
Still, it is interesting to give such a definition in 
our axiomatic framework.

\section{Inverses of Functions and Relations}
\label{sec9}
Given a function, $\mapdef{f}{A}{B}$ (possibly partial),
with $A\not= \emptyset$,
suppose there is some function,
$\mapdef{g}{B}{A}$ (possibly partial),
called a {\it left inverse of $f$\/}, such that
\[
g\circ f =\id_A.
\]
If such a $g$ exists, we see that $f$  must be total but more is true.
Indeed, assume that $f(a) = f(b)$. Then, by applying $g$,
we get
\[
(g\circ f)(a) = g(f(a)) = g(f(b)) = (g\circ f)(b).
\]
However, since $g\circ f =\id_A$, we have $(g\circ f)(a) =\id_A(a) = a$
and $(g\circ f)(b) =\id_A(a) = b$, so we deduce that
\[
a = b.
\]
Therefore, we showed that if a function, $f$, with nonempty domain,
has a left inverse,
then $f$ is total and has the property that for all $a, b\in A$, 
$f(a) = f(b)$ implies that $a = b$, or equivalently
$a\not= b$ implies that $f(a) \not= f(b)$.
We say that $f$ is {\it injective\/}. As we will see later,
injectivity is a very desirable property of functions.

\remark
If $A = \emptyset$, then $f$ is still considered to be injective.
In this case, $g$ is the empty partial function
(and when $B = \emptyset$, both $f$ and $g$ are the empty function
from $\emptyset$ to itself).

\medskip
Now, suppose there is some function,
$\mapdef{h}{B}{A}$ (possibly partial),
with $B\not= \emptyset$,
called a {\it right inverse of $f$\/}, 
but this time, we have
\[
f\circ h =\id_B.
\]
If such an $h$ exists, we see that it must be total but more is true.
Indeed, for any $b\in B$, as $f\circ h =\id_B$, we have
\[
f(h(b)) = (f\circ h)(b) = \id_B(b) = b.
\]
Therefore, we showed that if a function, $f$, with nonempty codomain
has a right inverse, $h$, then $h$ is total  and
$f$ has the property that for all $b\in B$, 
there is some $a\in A$, namely, $a = h(b)$, so that
$f(a) = b$. In other words, $\Im(f) = B$ or equivalently,
every element in $B$ is the image by $f$ of some element of $A$. 
We say that $f$ is {\it surjective\/}. Again,
surjectivity is a very desirable property of functions.

\remark
If $B = \emptyset$, then $f$ is still considered to be
surjective but $h$ is not total unless
$A = \emptyset$, in which case $f$ is the empty function
from $\emptyset$ to itself.

\danger
If a function has a left inverse (respectively a right inverse),
then it may have more than one left inverse
(respectively right inverse).

\medskip
If a function (possibly partial), $\mapdef{f}{A}{B}$,
with $A, B\not= \emptyset$,
happens to have both a left inverse, 
$\mapdef{g}{B}{A}$, and a right inverse, $\mapdef{h}{B}{A}$,
then we know that $f$ and $h$ are total. We 
claim that $g = h$, so that $g$  is total 
and moreover $g$ is uniquely determined by $f$.

\begin{lemma}
\label{invlem}
Let $\mapdef{f}{A}{B}$  be any function and suppose
that $f$ has a left inverse, \\
$\mapdef{g}{B}{A}$, and a right inverse, $\mapdef{h}{B}{A}$.
Then, $g = h$ and moreover, $g$ is unique, which means
that if $\mapdef{g'}{B}{A}$ is any function which is both a left and 
a right inverse of $f$, then $g' = g$.
\end{lemma}

\proof
Assume that
\[
g\circ f = \id_A
\quad\hbox{and}\quad
f\circ h = \id_B.
\]
Then, we have
\[
g = g\circ \id_B = g \circ (f\circ h) = (g \circ f) \circ h =
\id_A \circ h = h.
\]
Therefore, $g = h$. Now, if $g'$ is any other left inverse of $f$
and $h'$ is any other right inverse of $f$, 
the above reasoning applied to $g$ and $h'$
shows that $g = h'$ and the same reasoning applied to $g'$ and $h'$
shows that $g' = h'$.  Therefore, $g' = h' = g = h$,
that is, $g$ is uniquely determined by $f$.
$\bigsquare$

\medskip
This leads to the following definition.

\begin{defin}
\label{funinv}
{\em 
A function, $\mapdef{f}{A}{B}$, is said to be {\it invertible\/}
iff there is a function, $\mapdef{g}{B}{A}$, which
is both a left inverse and a right inverse, that is,
\[
g\circ f = \id_A
\quad\hbox{and}\quad
f\circ g = \id_B.
\]
In this case, we know that $g$ is unique and it is denoted $f^{-1}$.
}
\end{defin}

\medskip
From the above discussion, if a function is invertible, then
it is both injective and surjective. 
This shows that a function {\it generally does not  have an inverse\/}.
In order to have an inverse a function needs to be injective and surjective,
but this fails to be true for many functions. 
It turns out that if a function is injective and surjective
then it has an inverse. We will prove this in the next section.

\medskip
The notion of inverse can also be defined for relations, but it is
a somewhat weaker notion.

\begin{defin}
\label{relinv}
{\em 
Given any relation, $R\subseteq A\times B$, the {\it converse\/}
or {\it inverse\/} of $R$ is the relation,
$R^{-1} \subseteq B\times A$, defined by
\[
R^{-1} = \{\lag b, a\rag \in B\times A \mid \lag a, b\rag \in R\}.
\]
}
\end{defin}

\medskip
In other words, $R^{-1}$ is obtained by swapping $A$ and $B$
and reversing the orientation of the arrows. Figure \ref{fig3}
below shows the inverse of the relation of Figure \ref{fig1}:

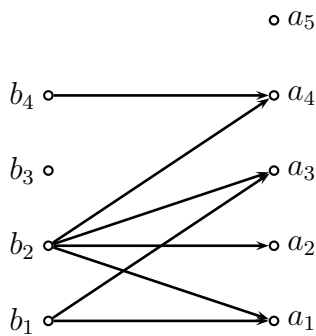
\begin{figure}[H]
 \begin{center}
    \begin{pspicture}(0,0)(3,3.8)
    \cnode(3,0){2pt}{u1}
    \cnode(3,1){2pt}{u2}
    \cnode(3,2){2pt}{u3}
    \cnode(3,3){2pt}{u4}
    \cnode(3,4){2pt}{u5}
    \cnode(0,0){2pt}{v1}
    \cnode(0,1){2pt}{v2}
    \cnode(0,2){2pt}{v3}
    \cnode(0,3){2pt}{v4}
    \ncline[linewidth=1pt]{<-}{u1}{v1}
    \ncline[linewidth=1pt]{<-}{u1}{v2}
    \ncline[linewidth=1pt]{<-}{u2}{v2}
    \ncline[linewidth=1pt]{<-}{u3}{v2}
    \ncline[linewidth=1pt]{<-}{u3}{v1}
    \ncline[linewidth=1pt]{<-}{u4}{v4}
    \ncline[linewidth=1pt]{<-}{u4}{v2}
    \uput[0](3,0){$a_1$}
    \uput[0](3,1){$a_2$}
    \uput[0](3,2){$a_3$}
    \uput[0](3,3){$a_4$}
    \uput[0](3,4){$a_5$}
    \uput[180](0,0){$b_1$}
    \uput[180](0,1){$b_2$}
    \uput[180](0,2){$b_3$}
    \uput[180](0,3){$b_4$}
    \end{pspicture}
  \end{center}
  \caption{The inverse of the relation, $R$, from Figure \ref{fig1}}
  \label{fig3}
\end{figure}

\medskip
Now, if $R$ is the graph of a (partial) function, $f$, 
beware that $R^{-1}$ is generally {\it not\/} the graph of a function
at all, because $R^{-1}$ may not be functional.
For example, the inverse of the graph $G$ in Figure \ref{fig2}
is {\it not\/} functional, see below:

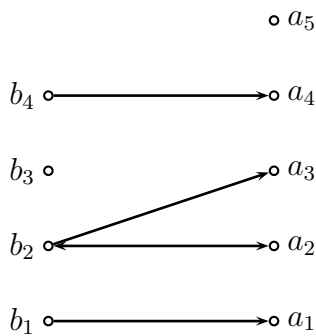
\begin{figure}[H]
 \begin{center}
    \begin{pspicture}(0,0)(3,3.8)
    \cnode(3,0){2pt}{u1}
    \cnode(3,1){2pt}{u2}
    \cnode(3,2){2pt}{u3}
    \cnode(3,3){2pt}{u4}
    \cnode(3,4){2pt}{u5}
    \cnode(0,0){2pt}{v1}
    \cnode(0,1){2pt}{v2}
    \cnode(0,2){2pt}{v3}
    \cnode(0,3){2pt}{v4}
    \ncline[linewidth=1pt]{<-}{u1}{v1}
    \ncline[linewidth=1pt]{<->}{u2}{v2}
    \ncline[linewidth=1pt]{<-}{u3}{v2}
    \ncline[linewidth=1pt]{<-}{u4}{v4}
    \uput[0](3,0){$a_1$}
    \uput[0](3,1){$a_2$}
    \uput[0](3,2){$a_3$}
    \uput[0](3,3){$a_4$}
    \uput[0](3,4){$a_5$}
    \uput[180](0,0){$b_1$}
    \uput[180](0,1){$b_2$}
    \uput[180](0,2){$b_3$}
    \uput[180](0,3){$b_4$}
    \end{pspicture}
  \end{center}
  \caption{The inverse, $G^{-1}$, of the graph of Figure \ref{fig2}}
  \label{fig4}
\end{figure}

\medskip
The above example shows that one has to be careful
not to view a function as a relation in
order to take its inverse. In general, this process does not produce
a function. This only works if the function is invertible.

\medskip
Given any two relations, $R\subseteq A\times B$ and
$S\subseteq B\times C$, the reader should prove that
\[
(R\circ S)^{-1} = S^{-1}\circ R^{-1}.
\]
(Note the switch in the order of composition on the right hand side.)
Similarly, if $\mapdef{f}{A}{B}$ and $\mapdef{g}{B}{C}$
are any two invertible functions, then
$g\circ f$ is invertible and
\[
(g\circ f)^{-1} = f^{-1}\circ g^{-1}.
\]

\section{Injections, Surjections, Bijections, Permutations}
\label{sec10}
We encountered injectivity and surjectivity in Section 
\ref{sec9}. For the record, let us give 

\begin{defin}
\label{injecsurjec}
{\em
Given any function, $\mapdef{f}{A}{B}$, we say that {\it $f$ is
injective\/} (or {\it one-to-one\/}) iff
for all $a, b\in A$, if $f(a) = f(b)$, then $a = b$, or
equivalently, if $a\not= b$, then $f(a) \not= f(b)$. 
We say that {\it $f$ is surjective\/} (or {\it onto\/}) iff
for every $b\in B$, there is some $a\in A$ so that $b = f(a)$,
or equivalently if $\Im(f) = B$. The function $f$ is {\it bijective\/}
iff it is both injective and surjective.
When $A = B$, a bijection $\mapdef{f}{A}{A}$ is called
a {\it permutation of $A$\/}.
}
\end{defin}

\remarks
\begin{enumerate}
\item
If $A = \emptyset$, then any function, $\mapdef{f}{\emptyset}{B}$
is (trivially) injective. 
\item
If $B = \emptyset$, then
$f$ is the empty function from $\emptyset$ to itself and
it is (trivially) surjective.
\item
A function, $\mapdef{f}{A}{B}$, is {\bf not injective} iff
{\bf there exist} $a, b\in A$ with $a\not= b$ and {\bf yet} $f(a) = f(b)$.
\item
A function, $\mapdef{f}{A}{B}$, is {\bf not surjective} iff
{\bf for some} $b\in B$, {\bf there is no} $a\in A$ with $b = f(a)$.
\item
Since $\Im f = \{b\in B \mid (\exists a\in A)(b = f(a))\}$,
a function $\mapdef{f}{A}{B}$ is always surjective onto its image.
\item
The notation $f\co A \hookrightarrow B$ is often used to indicate
that a function, $\mapdef{f}{A}{B}$, is an injection.
\item
Observe that if $A\not= \emptyset$, a function $f$
is surjective iff $f^{-1}(b) \not= \emptyset$ for all $b\in B$.
\item
When $A$ is the finite set $A = \{1, \ldots, n\}$, also denoted
$[n]$, it is not hard to show that there are $n!$ permutations
of $[n]$.
\end{enumerate}

\medskip
The function, $\mapdef{f_1}{\integs}{\integs}$, given by
$f_1(x) = x + 1$ is injective and surjective.
However, the function,  $\mapdef{f_2}{\integs}{\integs}$, given by
$f_2(x) = x^2$ is neither injective nor surjective (why?).
The function, $\mapdef{f_3}{\integs}{\integs}$, given by
$f_3(x) = 2x$ is injective but not surjective.
The function,  $\mapdef{f_4}{\integs}{\integs}$, given by
\[
f_4(x) = \cases{
k & if $x = 2k$\cr
k & if $x = 2k + 1$\cr
}
\]
is surjective but not injective.

\remark
The reader should prove that
if $A$ and $B$ are finite sets, $A$ has $m$ elements and $B$ has
$n$ elements (so, $m \leq n$)
then the set of injections from $A$ to $B$ has
\[
\frac{n!}{(n - m)!}
\]
elements.
The following Theorem relates the notions of injectivity
and surjectivity to the existence of left and right inverses.

\begin{thm}
\label{invers1}
Let $\mapdef{f}{A}{B}$ be any function and assume $A\not= \emptyset$.
\begin{enumerate}
\item[(a)]
The function $f$ is injective iff it has a left inverse, $g$
(i.e., a function $\mapdef{g}{B}{A}$ so that $g\circ f = \id_A$).
\item[(b)]
The function $f$ is surjective iff it has a right inverse, $h$
(i.e., a function $\mapdef{h}{B}{A}$ so that $f\circ h = \id_B$).
\item[(c)] The function $f$ is invertible iff
it is injective and surjective.
\end{enumerate}
\end{thm}

\proof
(a) We already proved in Section \ref{sec9} that the existence
of a left inverse implies injectivity.
Now, assume $f$ is injective. Then, for every
$b\in \mathit{range}(f)$, there is a unique $a_b\in A$
so that $f(a_b) = b$. Since $A\not= \emptyset$, we may 
pick some $a$ in $A$. We define $\mapdef{g}{B}{A}$ by
\[
g(b) = \cases{
a_b & if $b\in \mathit{range}(f)$\cr
a & if $b\in B - \mathit{range}(f)$.\cr
}
\]
Then, $g(f(a)) = a$, since $f(a)\in \mathit{range}(f)$ and
$a$ is the only element of $A$ so that $f(a) = f(a)$!
This shows that $g \circ f = \id_A$, as required.

\medskip
(b) We already proved in Section \ref{sec9} that the existence
of a right inverse implies surjectivity. For the converse, 
assume that $f$ is  surjective. As $A\not= \emptyset$ and
$f$ is a function (i.e., $f$ is total), $B\not= \emptyset$. 
So, for every $b\in B$,
the preimage $f^{-1}(b) = \{a\in A\mid f(a) = b\}$ is nonempty.
We make a function, $\mapdef{h}{B}{A}$, as follows:
For each $b\in B$, pick some element $a_b\in f^{-1}(b)$ (which
is nonempty) and let $h(b) = a_b$. By definition of $f^{-1}(b)$, 
we have $f(a_b) = b$ and so,
\[
f(h(b)) = f(a_b) = b, \quad\hbox{for all}\> b\in B.
\]
This shows that $f \circ h = \id_B$, as required.

\medskip
(c) If $f$ is invertible, we proved in Section \ref{sec9}
that $f$ is injective and surjective. Conversely, if $f$ is both
injective and surjective, by (a), the function $f$ has
a left inverse $g$ and by (b) it has a right inverse $h$.
However, by Lemma \ref{invlem}, $g = h$, which shows that
$f$ is invertible.
$\bigsquare$

\medskip
The alert reader may have noticed a ``fast turn'' in
the proof of the converse in (b). Indeed, we constructed
the function $h$ by choosing, for each $b\in B$, some
element in $f^{-1}(b)$. How do we justify this procedure
from the axioms of set theory?

\medskip
Well, we can't! For this, we need another (historically
somewhat controversial) axiom, the {\it Axiom of Choice\/}.
This axiom has many equivalent forms. We state
the following form which is intuitively quite plausible:

\vskip 1cm
{\bf Axiom of Choice (Graph Version)}.

\bigskip
For every relation, $R\subseteq A\times B$, there is a function,
$\mapdef{f}{A}{B}$, with $\mathit{graph}(f) \subseteq R$ and
$\mathit{dom}(f) = \mathit{dom}(R)$.

\bigskip
We see immediately that the Axiom of choice justifies the existence
of the function $g$ in part (b) of Theorem \ref{invers1}.

\begin{figure}[htbp]
\begin{center}
    \begin{pspicture}(0,0)(6,4)
    \psline[linewidth=1.5pt](0,0)(6,0)
    \psline[linewidth=1.5pt](0,2)(6,2)
    \psline[linewidth=1.5pt](0,2)(0,4)
    \psline[linewidth=1.5pt](0,4)(6,4)
    \psline[linewidth=1.5pt](6,2)(6,4)
    \psline[linewidth=1.5pt](2,2)(2,4)
    \psline[linewidth=1.5pt](4,2)(4,4)
    \psline[linewidth=1pt]{->}(2,1.5)(2,0.5)
    \psline[linewidth=1pt]{->}(4,0.5)(4,1.5)
    \psdots[dotstyle=o,dotscale=2](4,0)
    \psdots[dotstyle=o,dotscale=2](4,3.1)
    \psdots[dotstyle=o,dotscale=2](2,0)
    \psbezier[linewidth=1pt,showpoints=false](0,2.2)(3.5,5)(4,2)(6,3)
    \uput[180](2,1){$f$}
    \uput[0](1.9,2.85){$f^{-1}(b_1)$}
    \uput[0](4,1){$h$}
    \uput[180](0,0){$B$}
    \uput[180](0,3){$A$}
    \uput[-90](2,0){$b_1$}
    \uput[-90](4,0){$b_2$}
    \uput[0](4,3.2){$h(b_2)$}
    \end{pspicture}
\end{center}
\caption{A section, $h$, of a surjective function, $f$.}
\label{Figsection}
\end{figure}
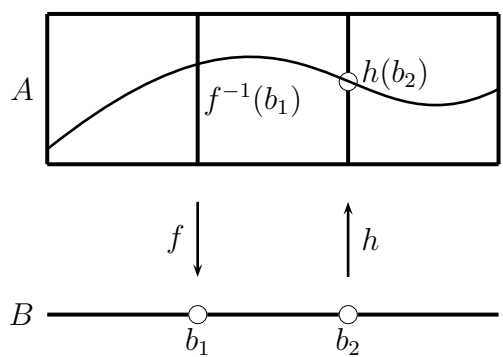

\remarks
\begin{enumerate}
\item
Let $\mapdef{f}{A}{B}$ and $\mapdef{g}{B}{A}$ be any two functions
and assume that
\[
g\circ f = \id_A.
\]
Thus, $f$ is a right inverse of $g$ and $g$ is a
left inverse of $f$. So, by Theorem \ref{invers1} (a) and (b),
we deduce that $f$ is injective and $g$ is surjective.
In particular, this shows that any left inverse of
an injection is a surjection and that any right inverse
of a surjection is an injection.
\item
Any right inverse, $h$, of a surjection,  $\mapdef{f}{A}{B}$, is
called a {\it section\/} of $f$ (which is an abbreviation
for {\it cross-section\/}).
This terminology can be 
better understood as follows: Since $f$ is surjective,
the preimage, $f^{-1}(b) = \{a\in A\mid f(b)\}$
of any element $b\in B$ is nonempty. Moreover,
$f^{-1}(b_1) \cap f^{-1}(b_2) = \emptyset$
whenever $b_1 \not= b_2$. Therefore, the pairwise
disjoint and nonempty subsets, $f^{-1}(b)$, where $b\in B$,
partition $A$. We can think of $A$ as a big ``blob''
consisting of the union of the sets  $f^{-1}(b)$
(called fibres) and lying over $B$. The function $f$
maps each fibre, $f^{-1}(b)$ onto the element, $b\in B$.
Then, any right inverse, $\mapdef{h}{B}{A}$, of $f$ picks
out some element in each fibre, $f^{-1}(b)$, 
forming a sort of horizontal section of $A$ shown as a curve
in Figure \ref{Figsection}.
\item
Any left inverse, $g$, of an injection,  $\mapdef{f}{A}{B}$, is
called a {\it retraction\/} of $f$.
The terminology reflects the fact that intuitively,
as $f$ is injective (thus, $g$ is surjective),
$B$ is bigger than $A$ and since $g\circ f = \id_A$, 
the function $g$ ``squeezes'' $B$ onto $A$ in such a way
that each point $b = f(a)$ in $\Im f$ is mapped back to its
ancestor $a\in A$. So, $B$ is ``retracted'' onto $A$ by $g$.
\end{enumerate}

\medskip
Before discussing direct and inverse images, we define the
notion of restriction and extension of functions.

\begin{defin}
\label{restrictdef}
Given two functions, $\mapdef{f}{A}{C}$ and
$\mapdef{g}{B}{C}$, with $A\subseteq B$,
we say that {\it $f$ is the restriction
of $g$ to $A$\/} if $\mathit{graph}(f) \subseteq \mathit{graph}(g)$;
we write  $f = g\res A$. In this case, we also say that 
{\it $g$ is an extension of $f$ to $B$\/}.
\end{defin}

\section{Direct Image and Inverse Image}
\label{sec11}
A function, $\mapdef{f}{X}{Y}$, induces a function from
$2^X$ to $2^Y$ also denoted $f$ and a
function from $2^Y$ to $2^X$, as shown in the following definition:

\begin{defin}
\label{dirinvimag}
{\em 
Given any function,  $\mapdef{f}{X}{Y}$, we define the
function $\mapdef{f}{2^X}{2^Y}$ so that, for every subset
$A$ of $X$,
\[
f(A) = \{y\in Y \mid \exists x \in A, \> y = f(x)\}.
\]
The subset, $f(A)$, of $Y$ is called the {\it direct image
of $A$ under $f$\/},
for short, the {\it image of $A$ under $f$\/}.
We also define the function  
$\mapdef{f^{-1}}{2^Y}{2^X}$ so that, for every subset
$B$ of $Y$,
\[
f^{-1}(B) = \{x\in X \mid \exists y \in B, \> y = f(x)\}.
\]
The subset, $f^{-1}(B)$, of $X$ is called the {\it inverse image
of $A$ under $f$\/} or
the {\it preimage of $A$ under $f$\/}. 
}
\end{defin}

\remarks
\begin{enumerate}
\item
The overloading of notation where $f$ is used both for denoting
the original function  $\mapdef{f}{X}{Y}$ and the new function
$\mapdef{f}{2^X}{2^Y}$ may be slightly confusing. 
If we observe that $f(\{x\}) =\{f(x)\}$, for all $x\in X$,
we see that the new $f$ is a natural extension of the old $f$ to
the subsets of $X$ and so, using the same symbol $f$ for both functions is
quite natural after all. To avoid any confusion, some authors
(including Enderton)
use a different notation for $f(A)$, for example, $f[\![A]\!]$.
We prefer not to introduce  more notation and we hope that
the context will make it clear which $f$ we are dealing with.
\item
The use of the notation $f^{-1}$ for the function
$\mapdef{f^{-1}}{2^Y}{2^X}$ may  even be more confusing,
because  we know that $f^{-1}$ is generally not a function
from $Y$ to $X$. However, it {\it is\/} a function from 
$2^Y$ to $2^X$. Again, some authors use a different notation
for $f^{-1}(B)$, for example,  $f^{-1}[\![A]\!]$.
Again, we will stick to $f^{-1}(B)$.
\item
The set $f(A)$ is sometimes called the {\it push-forward of $A$
along  $f$\/} and $f^{-1}(B)$ is sometimes called 
the {\it pullback of $B$ along $f$\/}.
\item
Observe that $f^{-1}(y) = f^{-1}(\{y\})$, 
where $f^{-1}(y)$ is the preimage defined just after Definition
\ref{fundef}.
\item
Although this may seem counter-intuitive, the function
$f^{-1}$ has a better behavior than $f$ with respect to union, 
intersection and complementation.
\end{enumerate}

\medskip
Some useful properties of $\mapdef{f}{2^X}{2^Y}$ and
$\mapdef{f^{-1}}{2^Y}{2^X}$ are now stated without proof.
The proofs are easy and left as exercises.

\begin{prop}
\label{invdirp1}
Given any function, $\mapdef{f}{X}{Y}$, the following properties
hold:
\begin{enumerate}
\item[(1)]
For any $B\subseteq Y$, we have
\[
f(f^{-1}(B)) \subseteq B.
\]
\item[(2)]
If $\mapdef{f}{X}{Y}$ is surjective, then
\[
f(f^{-1}(B)) = B.
\]
\item[(3)]
For any $A\subseteq X$, we have
\[
A \subseteq f^{-1}(f(A)).
\]
\item[(4)]
If $\mapdef{f}{X}{Y}$ is injective, then
\[
A = f^{-1}(f(A)).
\]
\end{enumerate}
\end{prop}

\medskip
The next proposition deals with the behavior of  $\mapdef{f}{2^X}{2^Y}$ and
$\mapdef{f^{-1}}{2^Y}{2^X}$  with respect to union, intersection and
complementation.

\begin{prop}
\label{invdirp2}
Given any function, $\mapdef{f}{X}{Y}$, the following properties
hold:
\begin{enumerate}
\item[(1)]
For all $A, B\subseteq X$, we have
\[
f(A\cup B) = f(A) \cup f(B).
\]
\item[(2)]
\[
f(A\cap B)\subseteq f(A) \cap f(B).
\]
Equality holds if $\mapdef{f}{X}{Y}$ is injective.
\item[(3)]
\[
f(A) - f(B) \subseteq f(A - B).
\]
Equality holds if $\mapdef{f}{X}{Y}$ is injective.
\item[(4)]
For all $C, D\subseteq Y$, we have
\[
f^{-1}(C\cup D) = f^{-1}(C)\cup f^{-1}(D).
\]
\item[(5)]
\[
f^{-1}(C\cap D) = f^{-1}(C)\cap f^{-1}(D).
\]
\item[(6)]
\[
f^{-1}(C - D) = f^{-1}(C) - f^{-1}(D).
\]
\end{enumerate}
\end{prop}

\medskip
As we can see from Proposition \ref{invdirp2}, 
the function $\mapdef{f^{-1}}{2^Y}{2^X}$
has a better behavior than  $\mapdef{f}{2^X}{2^Y}$
with respect to union, intersection and
complementation.

\section[Equinumerosity; Pigeonhole Principle;
Schr\" oder--Bernstein]
{Equinumerosity; The Pigeonhole Principle and the
Schr\" oder--Bernstein Theorem}
\label{sec12}
The notion of size of a set is fairly intuitive for finite
sets but what does it mean for infinite sets?
How do we give a precise meaning to the questions:
\begin{enumerate}
\item[(a)]
Do $X$ and $Y$ have the same size?
\item[(b)]
Does $X$ have more elements than $Y$?
\end{enumerate}

\medskip
For finite sets, we can rely on the natural numbers. We count the elements
in the two sets  and compare the resulting numbers.  
If one of the two sets is finite and the other is infinite,
it seems fair to say that the infinite set has more elements than 
the finite one. 

\medskip
But what is both sets are infinite?

\remark
A critical reader should object that we have
not yet  defined what a finite set is
(or what an infinite set is). Indeed, we have not!
This can be done in terms of the natural numbers,
but for the time being, we will rely on intuition.
We should also point out that when it comes to infinite sets,
experience shows that
our intuition fails us miserably. So, we should be very
careful.

\medskip
Let us return to the case where we have two infinite sets.
For example, consider $\natnums$ and the set of even natural numbers,
$2\natnums = \{0, 2, 4, 6, \ldots\}$. Clearly, the second set
is properly contained in the first. Does that make $\natnums$ bigger?
On the other hand, the function $n \mapsto 2n$ is
a bijection between the two sets, which seems to indicate that
they have the same number of elements.
Similarly, the set of squares of natural numbers,
$\mathrm{Squares} = \{0, 1, 4, 9, 16, 25, \ldots\}$
is properly contained in $\natnums$, yet many natural numbers
are missing from $\mathrm{Squares}$.
But, the map $n \mapsto n^2$ is a bijection between $\natnums$
and $\mathrm{Squares}$, which seems to indicate that they have
the same number of elements. 

\medskip
A more extreme example is provided
by $\natnums\times\natnums$ and $\natnums$. Intuitively,
$\natnums\times\natnums$ is two-dimensional and  $\natnums$ is
one-dimensional, so  $\natnums$ seems much smaller than 
$\natnums\times\natnums$. However, it is possible to construct
bijections between  $\natnums\times\natnums$ and $\natnums$
(try to find one!). In fact, such a function,
$J$, has the graph partially showed below:
\[
\begin{array}{lllllllll}
\vdots  & &    &     &     &         &   &         &           \\
3 & 6 & \ldots  &     &         &   &         &   &        \\
  &  &\searrow &     &         &   &         &   &        \\
2 &3 &         & 7   & \ldots  &   &         &   &        \\
  &  &\searrow &     &\searrow &   &         &   &        \\
1 & 1 &         & 4   &         & 8 & \ldots  &   &        \\
  &  &\searrow &     &\searrow &   &\searrow &   &        \\
0 & 0 &         & 2   &         & 5 &         & 9 & \\  
  & 0 &         & 1   &         & 2 &         & 3 & \ldots
\end{array}
\]

\medskip
The function $J$ corresponds to a certain way of enumerating
pairs of integers. 
Note that the value of $m + n$
is constant along each diagonal, and consequently, we have
\begin{eqnarray*}
J(m, n) & = & 1 + 2 + \cdots + (m + n) + m,\\
        & = & ((m + n)(m + n + 1) + 2m)/2,\\
        & = & ((m + n)^2 + 3m + n)/2.
\end{eqnarray*}
For example, $J(2, 1) = ((2 + 1)^2 + 3\cdot 2 + 1)/2 = (9 + 6 + 1)/2
= 16/2 = 8$.
The function
\[
J(m, n) = \frac{1}{2}((m + n)^2 + 3m + n)
\]
is a bijection but that's not so easy to prove!

\medskip
Perhaps even more surprising, there are bijections between
$\natnums$ and $\rats$. What about between $\reals\times \reals$
and $\reals$? Again, the answer is yes, but that's a lot harder
to prove.

\medskip
These examples suggest that the notion of bijection can be used
to define rigorously when two sets have the same size.
This leads to the concept of equinumerosity.

\begin{defin}
\label{equinum}
{\em 
A set $A$ is {\it equinumerous\/} to a set $B$, written
$A \approx B$, iff there is a bijection
$\mapdef{f}{A}{B}$. We say that $A$ is {\it dominated\/}
by $B$, written $A \preceq B$,
iff there is an injection from $A$ to $B$.
Finally, we say that $A$ is {\it strictly dominated\/} by $B$,
written $A\prec B$, iff $A \preceq B$ and
$A \not\approx B$.
}
\end{defin}

\medskip
Using the above concepts, we can give a precise definition
of finiteness. Firstly, recall that for any $n\in \natnums$,
we defined $[n]$ as the set
$[n] = \{1, 2, \ldots, n\}$, with $[0] = \emptyset$.

\begin{defin}
\label{finitedef}
{\em 
A set, $A$, is {\it finite\/} if it is equinumerous 
to a set of the form $[n]$, for some $n\in \natnums$.
We say that $A$ is {\it countable\/} (or {\it denumerable\/}) iff
$A$ is dominated by $\natnums$.
}
\end{defin}

\medskip
Two pretty results due to Cantor (1873) are given in the next Theorem.
These are among the earliest results of set theory.
We assume that the reader is familiar with the fact that
every number, $x\in \reals$, can be expressed in decimal expansion
(possibly infinite).
For example,
\[
\pi = 3.14159265358979 \cdots
\]

\begin{thm} (Cantor)
\label{cantor1}
(a) The set $\natnums$ is not equinumerous to the set $\reals$
of real numbers.

\medskip
(b) No set, $A$, is equinumerous to its power set, $2^A$. 
\end{thm}

\medskip
(a) We use a famous proof method due to Cantor and known
as a {\it diagonal argument\/}. We will prove that if we assume
that there is a bijection, $\mapdef{f}{\natnums}{\reals}$, then
there is a real number $z$ not belonging to the image of $f$,
contradicting the surjectivity of $f$.
Now, if $f$ exists, we can form a bi-infinite array
\begin{align*}
f(0) &= k_0.d_{0\, 1}d_{0\, 2}d_{0\, 3}d_{0\, 4}\cdots , \\
f(1) &= k_1.d_{1\, 1}d_{1\, 2}d_{1\, 3}d_{1\, 4}\cdots , \\
f(2) &= k_2.d_{2\, 1}d_{2\, 2}d_{2\, 3}d_{2\, 4}\cdots , \\
     & \vdots \\
f(n) &= k_n.d_{n\, 1}d_{n\, 2}\cdots d_{n\, n+1}\cdots ,\\
     & \vdots 
\end{align*}
where $k_n$ is the integer part of $f(n)$ and
the $d_{n\, i}$ are the decimals of $f(n)$, with $i\geq 1$.

The number 
\[ 
z = 0.d_1d_2d_3\cdots d_{n+1}\cdots
\]
is defined as follows:
$d_{n+1} = 1$ if $d_{n\, n+1}\not= 1$, else 
$d_{n+1} = 2$ if $d_{n\, n+1}= 1$, for 
every $n \geq 0$,
The definition of $z$ shows that 
\[
d_{n+1} \not= d_{n\, n+1},
\quad\hbox{for all}\quad n \geq 0,
\]
which implies that $z$ is not in the above array, i.e.,
$z\notin \Im\, f$.

\medskip
(b)
The proof is a variant of Russell's paradox.
Assume that there is a bijection \\
$\mapdef{g}{A}{2^A}$; we construct 
a set $B \subseteq A$ that is not in the image of $g$, a contradiction.
Consider the set
\[
B = \{a\in A \mid a\notin g(a)\}.
\]
Obviously, $B \subseteq A$. However, for every $a\in A$,
\[
a\in B \quad\hbox{iff}\quad a\notin g(a),
\]
which shows that $B \not= g(a)$ for all $a\in A$,
i.e.,  $B$ is not in the image of $g$.
$\bigsquare$

\medskip
As there is an obvious injection of $\natnums$ into $\reals$,
Theorem \ref{cantor1} shows that $\natnums$ is strictly
dominated by $\reals$. Also, as we have the injection
$a \mapsto \{a\}$ from $A$ into $2^A$, we see that every set
is strictly dominated by its power set. 
So, we can form
sets as big as we want by repeatedly using the power set operation.

\remarks
\begin{enumerate}
\item
The proof of part (b) of Theorem \ref{cantor1}
only requires $g$ to be a surjection.
\item
In fact, $\reals$ is equinumerous to $2^{\natnums}$, but we
will not prove this here.
\end{enumerate}

\medskip
The following proposition shows an interesting connection between 
the notion of power set and certain sets of functions.
To state this proposition, we need the concept of 
characteristic function of a subset.

\medskip
Given any set, $X$, for any subset, $A$, of $X$, define
the {\it characteristic function of $A$\/}, denoted
$\chi_A$, as the function, $\mapdef{\chi_A}{X}{\{0, 1\}}$, given 
by
\[
\chi_A(x) = \cases{
1 & if $x\in A$\cr
0 & if $x\notin A$.\cr
}
\]
In other words, $\chi_A$ tests membership in $A$: For any $x\in X$,
$\chi_A(x) = 1$ iff $x\in A$. Observe that we obtain a function,
$\mapdef{\chi}{2^X}{\{0, 1\}^X}$, from the power set of $X$ to the set 
of characteristic functions from $X$ to $\{0, 1\}$, given by
\[
\chi(A) = \chi_A.
\]
We also have the function, $\mapdef{\s{S}}{\{0, 1\}^X}{2^X}$,
mapping any characteristic function to the set that it defines
and given by
\[
\s{S}(f) = \{x\in X \mid f(x) = 1\},
\]
for every  characteristic function, $f\in \{0, 1\}^X$.

\begin{prop}
\label{powesetp1}
For any set, $X$, the function $\mapdef{\chi}{2^X}{\{0, 1\}^X}$ 
from the power set of $X$ to the set of characteristic functions on $X$
is a bijection whose inverse is
$\mapdef{\s{S}}{\{0, 1\}^X}{2^X}$.
\end{prop}

\proof
Simply check that $\chi\circ \s{S} = \id$
and $\s{S}\circ \chi = \id$, which is straightforward.
$\bigsquare$

\medskip
In view of Proposition \ref{powesetp1}, there is a bijection between
the power set $2^X$ and the set of functions in $\{0, 1\}^X$.
If we write $2 = \{0, 1\}$, then we see that the two sets looks the same!
This is the reason why the notation $2^X$ is often used for the power set
(but others prefer $\s{P}(X)$).

\medskip
There are many other interesting results about  equinumerosity.
We only mention four more, all very important.

\begin{thm}
\label{pigenhole} (Pigeonhole Principle)
No set of the form $[n]$ is equinumerous to a proper
subset of itself, where $n\in\natnums$, 
\end{thm}

\proof
Although the Pigeonhole Principle seems obvious,
the proof is not. In fact, the proof requires induction.
We advice the reader to skip this proof
and come back to it later after we have given more examples
of proof by induction.

\medskip
Suppose  we can prove the following Claim:

\medskip
{\it Claim\/}. Whenever a function,
$\mapdef{f}{[n]}{[n]}$, is an injection, then it is
a surjection onto $[n]$ (and thus, a bijection).

\medskip
Observe that the above Claim implies
the Pigeonhole Principle.
This is proved by contradiction. So, assume there is
a function, $\mapdef{f}{[n]}[n]$, 
such that $f$ is injective and $\Im f = A \subseteq [n]$ 
with $A\not= [n]$, i.e., $f$ is a bijection
between $[n]$ and $A$, a proper subset of $[n]$.
Since $\mapdef{f}{[n]}{[n]}$ is injective, by the Claim, 
we deduce that $\mapdef{f}{[n]}{[n]}$ is surjective, 
i.e., $\Im f = [n]$, contradicting the fact that
$\Im f = A\not= [n]$.

\medskip
It remains to prove by induction on $n\in\natnums$ that
if $\mapdef{f}{[n]}{[n]}$ is an injection, then it is
a surjection (and thus, a bijection).
For $n = 0$, $f$ must be the empty function, which
is a bijection.

\medskip
Assume that the induction hypothesis holds for any $n\geq 0$
and consider any injection, $\mapdef{f}{[n+1]}{[n+1]}$.
Observe that the restriction of $f$ to $[n]$ is injective.

\medskip
{\it Case\/} 1. 
The subset $[n]$ is closed under $f$, i.e.,
$f([n]) \subseteq [n]$. Then, we know that $f\res [n]$ is injective
and by the induction hypothesis, $f([n]) = [n]$.
Since $f$ is injective, we must have $f(n+1) = n+1$.
Hence, $f$ is surjective, as claimed.

\medskip
{\it Case\/} 2. 
The subset $[n]$ is not closed under $f$, i.e., there is
some $p \leq n$ such that $f(p) = n+1$.
We can create a new injection, $\widehat{f}$, from $[n+1]$ to itself
with the same image as $f$ by interchanging two values
of $f$ so that  $[n]$ closed under  $\widehat{f}$.
Define $\widehat{f}$ by
\begin{align*}
\widehat{f}(p) & = f(n+1) \\
\widehat{f}(n+1) & = f(p) = n+1 \\
\widehat{f}(i) & =  f(i),\qquad 1 \leq i \leq n, \> i \not= p.
\end{align*}
Then,  $\widehat{f}$ is an injection from $[n+1]$ to itself
and $[n]$ is closed under  $\widehat{f}$. By Case 1,
$\widehat{f}$ is surjective, and as 
$\Im\, f = \Im \widehat{f}$, we conclude that $f$ is also
surjective.
$\bigsquare$

\begin{cor}
\label{pigenhole1} (Pigeonhole Principle for finite sets)
No finite set is equinumerous to a proper subset of
itself.
\end{cor}

\proof
To say that a set,  $A$, is finite is to say that there is
a bijection, $\mapdef{g}{A}{[n]}$, for some $n\in \natnums$.
Assume that there is a bijection, $f$, between $A$ and some proper subset
of $A$. Then, consider the function $g\circ f\circ g^{-1}$,
from $[n]$ to itself. The rest of proof consists in showing that
$[n]$ would be equinumerous to a proper subset of itself,
contradicting Theorem \ref{pigenhole}.
We leave the details as an exercise.
$\bigsquare$

\medskip
The pigeonhole principle is often used in  the following way:
If we have $m$ distinct slots and $n > m$ distinct objects
(the pigeons), then when we put all $n$ objects into the $m$ slots,
two objects must end up in the same slot.
This fact was apparently first stated explicitly by Dirichlet in 1834.
As such, it is also known as {\it Dirichlet's box principle\/}.

\medskip
Let $A$ be a finite set. Then, by definition, there is a bijection,
$\mapdef{f}{A}{[n]}$, for some $n\in\natnums$.
We claim that such an $n$ is unique. Otherwise, there
would be another bijection, $\mapdef{g}{A}{[p]}$, for some $p\in\natnums$
with $n \not= p$. But now, we would have a bijection
$g\circ f^{-1}$ between $[n]$ and $[p]$ with $n \not= p$.
This would imply that there is either an injection
from $[n]$ to a proper subset of itself or
an injection from $[p]$ to a proper subset of itself,%
\footnote{Recall that
$n + 1 = \{0, 1 ,\ldots, n\} = [n]\cup \{0\}$.
Here in our argument, we are using the fact that
for any two natural numbers $n, p$, either $n \subseteq p$
or $p \subseteq n$. This fact is indeed true but requires
a proof. The proof uses induction and some special properties
of the natural numbers implied by the definition of a natural number
as a set that belongs to every inductive set. For details,
see Enderton \cite{Endertonset}, Chapter 4.}
contradicting the Pigeonhole Principle.

\medskip
If $A$ is a finite set, the unique natural number, $n\in \natnums$,
such that $A\approx [n]$ is called the
{\it cardinality of $n$\/}
and we write $|A| = n$ (or sometimes, $\mathrm{card}(A) = n$). 

\remark
The notion of cardinality also makes sense for infinite sets.
What happens is that every set is equinumerous to a special
kind of set (an initial ordinal) called a {\it cardinal number\/}
but this topic is beyond the scope of this course.
Let us simply mention that the cardinal number of $\natnums$
is denoted $\aleph_0$ (say ``aleph'' $0$).

\begin{cor}
\label{pigenhole2}
(a)
Any set equinumerous to a proper subset of itself is infinite.

\medskip
(b) The set $\natnums$ is infinite.
\end{cor}

\proof
Left as an exercise to the reader.
$\bigsquare$

\medskip
Let us give another application of the pigeonhole principle
involving sequences of integers.
Given a finite sequence, $S$, of integers, $a_1, \ldots, a_n$,
a {\it subsequence of $S$\/} is a sequence,
$b_1, \ldots, b_m$, obtained by deleting elements from the
original sequence and keeping the remaining elements in the 
same order as they originally appeared.
More precisely, $b_1, \ldots, b_m$ is a subsequence of
$a_1, \ldots, a_n$ if there is an injection,
$\mapdef{g}{\{1, \ldots, m\}}{\{1, \ldots, n\}}$, such that
$b_i = a_{g(i)}$ for all $i\in \{1, \ldots, m\}$ and
$i \leq j$ implies $g(i) \leq g(j)$ for all 
$i, j\in \{1, \ldots, m\}$.
For example, the sequence
\[
1 \quad \mathbf{9}\quad 10 \quad \mathbf{8}\quad 3\quad 7\quad 5 \quad 2
\quad \mathbf{6}\quad \mathbf{4}
\]
contains the subsequence
\[
9\quad 8\quad 6\quad 4.
\]
An {\it increasing subsequence\/} is a subsequence whose
elements are in strictly increasing order and
a {\it decreasing subsequence\/} is a subsequence whose
elements are in strictly decreasing order.
For example, $9\> 8\> 6 \>4$ is a decreasing subsequence
of our original sequence.
We now prove the following beautiful result due to
Erd\"os and Szekeres:

\begin{thm} (Erd\"os and Szekeres)
\label{Erdos1}
Let $n$ be any nonzero natural number. Every sequence of
$n^2 + 1$ pairwise distinct natural numbers must contain
either an increasing subsequence or a decreasing subsequence of
length $n + 1$.
\end{thm}

\proof
The proof proceeds by contradiction. So, assume there is a sequence, $S$,
of $n^2 + 1$ pairwise distinct natural numbers so that
all increasing  or decreasing subsequences of $S$ have length at most $n$.
We assign to every element, $s$, of the sequence, $S$, a
pair of natural numbers, $(u_s, d_s)$, called a {\it label\/},
where $u_s$, is the length of a longest increasing subsequence of $S$
that starts at $s$ and where $d_s$ 
is the length of a longest decreasing subsequence of $S$
that starts at $s$. 

\medskip
Since there are no increasing or descreasing subsequences of
length $n + 1$ in $S$, observe that $1 \leq u_s, d_s \leq n$
for all $s\in S$. Therefore, 

\medskip
{\it Claim\/} 1: 
There are at most $n^2$ distinct labels
$(u_s, d_s)$, where $s\in S$.

\medskip
We also assert

\medskip
{\it Claim\/} 2: 
If $s$ and $t$ are any two distinct elements of $S$, then
$(u_s, d_s) \not= (u_t, d_t)$.

\medskip
We may assume that $s$ precedes $t$ in $S$ since otherwise, we 
interchange $s$ and $t$ in the following argument.
Since $s\not= t$, there are two cases:
\begin{enumerate}
\item[(a)]
$s < t$. In this case, we know that there is an increasing subsequence
of length $u_t$ starting with $t$. If we insert $s$ in front of this
subsequence, we get an increasing subsequence of $u_t + 1$ elements
starting at $s$. Then, as $u_s$ is the maximal length of all 
increasing subsequences starting with $s$, we must have
$u_t + 1 \leq u_s$, i.e.,
\[
u_s > u_t,
\]
which implies $(u_s, d_s) \not= (u_t, d_t)$.
\item[(b)]
$s > t$. This case is similar to case (a), except that
we consider a decreasing subsequence of length $d_t$ starting with
$t$. We conclude that 
\[
d_s > d_t
\]
which implies $(u_s, d_s) \not= (u_t, d_t)$.
\end{enumerate}
Therefore, in all cases, we proved that $s$ and $t$ have distinct labels.

\medskip
Now, by Claim 1, there are only $n^2$ distinct labels and $S$ has $n^2 + 1$
elements so, by the Pigeonhole Principle, two elements of $S$ must have the
same label. But, this contradicts Claim 2, which
says that distinct elements of $S$ have distinct labels.
Therefore, $S$ must have either an increasing subsequence or
a decreasing subsequence of length $n  + 1$, as originally claimed.
$\bigsquare$

\remark
Note that this proof is not constructive in the sense that it does
not produce the desired subsequence; it merely asserts that
such a sequence exists.

\medskip
Our next theorem is the historically famous
Schr\"oder-Bernstein Theorem, sometimes  called the
``Cantor-Bernstein Theorem.'' 
Cantor proved the theorem in 1897 but his proof
used a principle equivalent to the axiom of choice.
Schr\"oder announced the theorem in an 1896 abstract. His proof,
published in 1898, had problems and he published a correction in
1911. The first fully satisfactory proof was given by Felix Bernstein 
and was published in 1898 in a book by Emile Borel.
A shorter  proof was given later by Tarski (1955)
as a consequence of his fixed point theorem.
We postpone giving this proof until the section on lattices
(see Section \ref{sec15}).

\begin{thm} (Schr\"oder-Bernstein Theorem)
\label{CantorBernstein}
Given any two sets, $A$ and $B$, if there is an injection from
$A$ to $B$ and an injection from $B$ to $A$, then there is
a bijection between $A$ and $B$. 
Equivalently, if $A\preceq B$ and $B\preceq A$, then
$A \approx B$.
\end{thm}

\medskip
The Schr\"oder-Bernstein Theorem is quite a remarkable result
and it is a main tool to develop cardinal arithmetic,
a subject beyond the scope of this course.

\medskip
Our third theorem is perhaps the one that is the
more surprising from an intuitive point of view. If nothing else, 
it shows that our intuition about infinity is rather poor.

\begin{thm}
\label{cardprod}
If $A$ is any infinite set, then $A\times A$ is equinumerous
to $A$.
\end{thm}

\proof
The proof is more involved than any of the proofs given so far
and it makes use of the axiom of choice in the form known as
{\it Zorn's Lemma\/} (see Theorem \ref{Zornlem}). 
For these reasons, we omit the proof and instead
refer the reader to  Enderton \cite{Endertonset} (Chapter 6).
$\bigsquare$

\medskip
In particular, Theorem \ref{cardprod} implies that
$\reals\times\reals$ is in bijection with $\reals$.
But, geometrically, $\reals\times\reals$ is a plane and 
$\reals$ is a line and, intuitively it is
surprising that a plane and a line would have ``the same number
of points.'' Nevertheless, that's what mathematics tells us!

\medskip
Our fourth theorem also plays an 
important role in the theory of cardinal numbers.

\begin{thm} (Cardinal comparability)
\label{cardcomp}
Given any two sets, $A$ and $B$, either there is an injection from
$A$ to $B$ or there is an injection from $B$ to $A$
(that is, either $A\preceq B$ or $B\preceq A$).
\end{thm}

\proof
The proof requires the axiom of choice in a form known
as the {\it Well-Ordering Theorem\/}, which is also equivalent  
to  Zorn's lemma. For details, see Enderton 
\cite{Endertonset} (Chapters 6 and 7).
$\bigsquare$

\medskip
Theorem \ref{cardprod} implies that there is a bijection
between the closed line segment
\[
[0, 1] = \{x\in \reals \mid 0 \leq x \leq 1\}
\]
and the closed unit square
\[
[0, 1]\times [0, 1] = \{(x, y) \in \reals^2 \mid
 0 \leq x , y\leq 1\}
\]
As an interlude, in the next section, we describe 
a famous space-filling function due to Hilbert.
Such a function is obtained as the limit of 
a sequence of curves that can be defined recursively.

\section[An Amazing Surjection: Hilbert's Space Filling Curve]
{An Amazing Surjection: Hilbert's Space Filling Curve}
\label{sec12b}
In the years 1890-1891, Giuseppe Peano and David Hilbert
discovered examples of {\it space filling functions\/} (also called
{\it space filling curves\/}). These are surjective
functions from the line segment, $[0, 1]$ onto the unit square
and thus, their image is the whole unit square!
Such functions defy intuition since they seem to 
contradict our intuition about the notion of
dimension, a line segment is one-dimensional, yet the
unit square is two-dimensional. They also seem to 
contradict our intuitive notion of area.
Nevertheless, such functions do exist, 
even continuous ones, although
to justify their existence rigouroulsy requires
some tools from mathematical analysis. Similar curves were found by others,
among which we mention Sierpinski, Moore and Gosper.

\begin{figure}
\centerline{
\psfig{figure=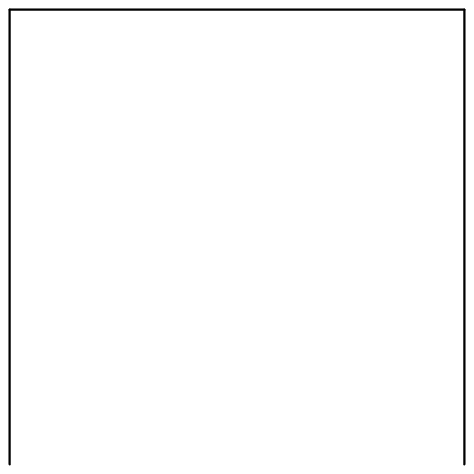,height=1.6truein,width=1.6truein}
\psfig{figure=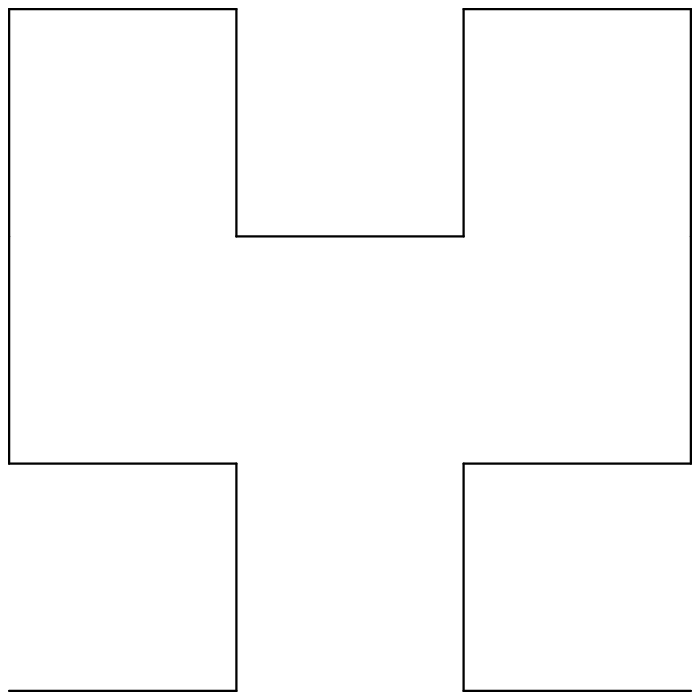,height=1.6truein,width=1.6truein}
\psfig{figure=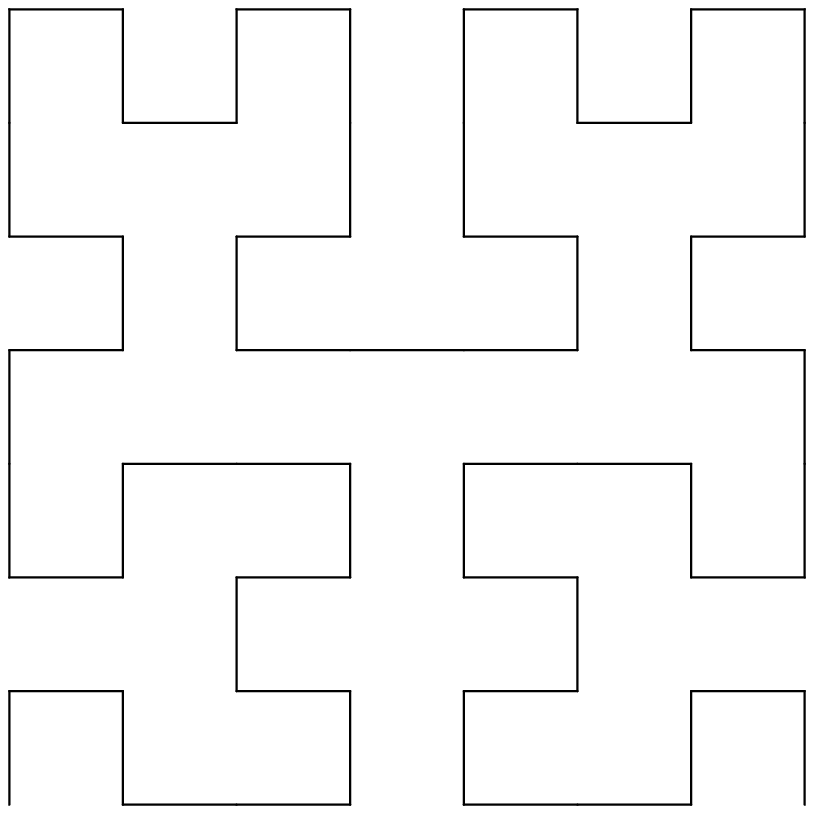,height=1.5truein,width=1.5truein}
}
\caption{A sequence of Hilbert curves $h_0, h_1, h_2$}
\label{hilcurfig1}
\end{figure}

\medskip
We will describe Hilbert's scheme for constructing
such a square-filling curve. 
We define  a sequence, $(h_n)$, of polygonal lines, 
$\mapdef{h_n}{[0, 1]}{[0, 1]\times [0, 1]}$, starting from
the simple pattern $h_0$ (a ``square cap'' $\sqcap$)
shown on the left in Figure \ref{hilcurfig1}.

\medskip
The curve $h_{n+1}$ is obtained by scaling down $h_n$
by a factor of $\frac{1}{2}$, and connecting the four copies of
this scaled--down version of $h_n$ obtained by
rotating by $\pi/2$ (left lower part), rotating by $-\pi/2$
and translating right
(right lower part), translating up (left upper part),
and translating diagonally (right upper part),
as illustrated in Figure  \ref{hilcurfig1}.

\begin{figure}
\centerline{
\psfig{figure=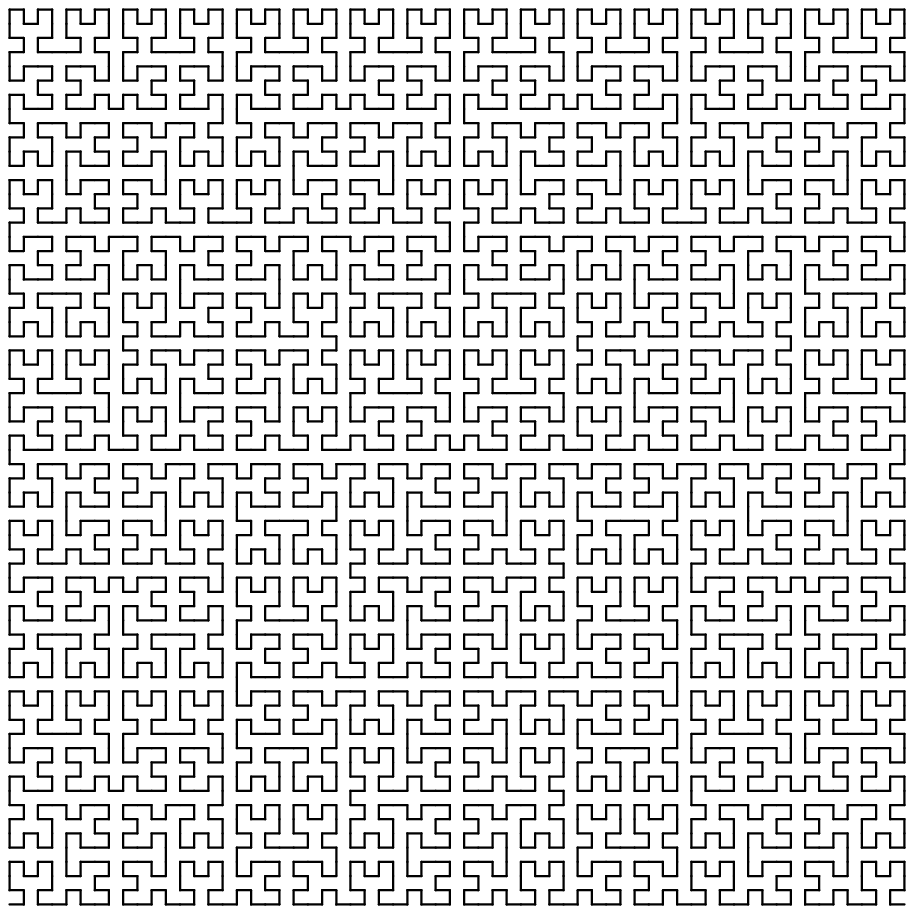,height=3truein,width=3truein}
}
\caption{The Hilbert curve $h_5$}
\label{hilcurfig2}
\end{figure}

\medskip
It can be shown that the sequence $(h_n)$ converges
(uniformly)  to a continuous curve
$\mapdef{h}{[0, 1]}{[0, 1]\times [0, 1]}$
whose trace is the entire square $[0, 1]\times [0, 1]$.
The Hilbert curve $h$ is surjective, continuous,
and nowhere differentiable.
It also has infinite length!
The curve $h_5$ is shown in Figure \ref{hilcurfig2}.
%
You should try writing a computer program to plot 
these curves! By the way, it can be shown that
no continuous square-filling function can be injective.
It is also possible to define cube-filling curves
and even higher-dimensional cube-filling curves!
(see some of the web page links in the home page for 
CSE260)

\medskip
Before we close this chapter and move on to special kinds
of relations, namely, partial orders and equivalence relations,
we illustrate how the notion of function can be used
to define strings, multisets and indexed families rigorously.

\section{Strings, Multisets, Indexed Families}
\label{sec13}
Strings play an important role in computer science and linguistics
because they are the basic tokens  that languages are made of.
In fact, formal language theory takes the (somewhat crude) view that
a language is a set of strings (you will study some formal
language theory in CSE262).
A string is a finite sequence of letters,
for example ``Jean'', ``Val'', ``Mia'', ``math'', ``gaga'', ``abab''.
Usually, we have some alphabet in mind and we form strings
using letters from this alphabet. Strings are not sets, the order of the letters
matters: ``abab'' and ``baba'' are different strings.
What matters is the position of every letter. In the string ``aba'',
the leftmost ``a'' is in position 1, ``b'' is in position 2 and the rightmost ``b''
is in position 3. All this suggests defining strings as certain kinds
of functions whose domains are the sets $[n] = \{1, 2, \ldots, n\}$
(with $[0] = \emptyset$) encountered earlier. Here is the very beginning 
of the theory of formal languages.

\begin{defin}
\label{alphadef}
{\em 
An {\it alphabet\/}, $\Sigma$, is any {\bf finite} set.
}
\end{defin}

We often write $\Sigma = \{a_1,\ldots,a_k\}$.
The $a_i$ are called the {\it symbols\/} of the alphabet.

\remark
There will be a few occasions where we will
allow infinite alphabets but normally an alphabet is
assumed to be finite.

\medskip
{\it Examples\/}:

\medskip
$\Sigma = \{a\}$

\medskip
$\Sigma = \{a, b, c\}$

\medskip
$\Sigma = \{0, 1\}$

\medskip
A string is a finite sequence of symbols.
Technically, it is convenient to define strings
as functions. 

\begin{defin}
\label{stringdef}
{\em 
Given an alphabet, $\Sigma$, a {\it string over $\Sigma$ (or 
simply a string) of length $n$\/} is any function
$$\mapdef{u}{[n]}{\Sigma}.$$
The integer $n$ is the {\it length\/} of the string, $u$, 
and it is denoted by $|u|$.
When $n = 0$, the special string, $\mapdef{u}{[0]}{\Sigma}$,
of length $0$ is called the {\it empty string, or null string\/},
and is denoted by $\epsilon$.
}
\end{defin}

\medskip
Given a string, $\mapdef{u}{[n]}{\Sigma}$, of length
$n\geq 1$, $u(i)$ is the $i$-th letter in the string $u$.
For simplicity of notation, we denote the string
$u$ as
$$u = u_1u_2 \ldots u_n,$$
with each $u_i \in \Sigma$.

\medskip
For example, if $\Sigma = \{a, b\}$ and
$\mapdef{u}{[3]}{\Sigma}$ is defined such that
$u(1) = a$, $u(2) = b$, and $u(3) = a$, we write
$$u = aba.$$
Strings of length $1$ are functions
$\mapdef{u}{[1]}{\Sigma}$ simply picking some element
$u(1) = a_i$ in $\Sigma$. Thus, we will identify every
symbol $a_i\in\Sigma$ with the corresponding string
of length $1$.

\medskip
The set of all strings over an alphabet $\Sigma$,
including the empty string, is denoted as $\Sigma^*$.
Observe that when $\Sigma = \emptyset$, then
$$\emptyset^* = \{\epsilon\}.$$
When $\Sigma \not= \emptyset$, the set $\Sigma^*$ is
countably infinite. Later on, we will see ways of ordering
and enumerating strings.

\medskip
Strings can be juxtaposed, or concatenated.

\begin{defin}
\label{concatdef}
{\em 
Given an alphabet, $\Sigma$, given two strings,
$\mapdef{u}{[m]}{\Sigma}$ and $\mapdef{v}{[n]}{\Sigma}$, the
{\it concatenation, $u\cdot v$, (also written  $uv$)
of $u$ and $v$\/} is the string
\hfill\break
$\mapdef{uv}{[m + n]}{\Sigma}$, defined such that
$$uv(i) = \cases{u(i) & if $1\leq i\leq m$,\cr
                 v(i - m) & if $m + 1\leq i\leq m + n$.\cr
}$$
In particular, $u\epsilon = \epsilon u = u$.
}
\end{defin}

\medskip
It is immediately verified that
$$u(vw) = (uv)w.$$
Thus, concatenation is a binary operation on $\Sigma^*$
which is associative and has $\epsilon$ as an identity.
Note that generally, $uv \not= vu$, for example
for $u = a$ and $v= b$.

\begin{defin}
\label{pssfix}
{\em 
Given an alphabet $\Sigma$, given any two strings
$u, v\in\Sigma^*$
we define the following notions as follows:

\medskip 
{\it $u$ is a prefix of $v$\/} iff
there is some $y\in\Sigma^*$ such that
$$v = uy.$$

\medskip
{\it $u$ is a suffix of $v$\/} iff
there is some $x\in\Sigma^*$ such that
$$v = xu.$$

\medskip
{\it $u$ is a substring of $v$\/} iff
there are some $x, y\in\Sigma^*$ such that
$$v = xuy.$$

\medskip
We say that {\it $u$ is a proper prefix (suffix, substring) of $v$\/}
iff   $u$ is a  prefix (suffix, substring) of $v$ and $u\not= v$.
}
\end{defin}

\medskip
For example, $ga$ is a prefix of $gallier$, the string
$lier$ is a suffix of $gallier$ and $all$ is a substring of $gallier$

Finally, languages are defined as follows.

\begin{defin}
\label{langdef}
{\em 
Given an alphabet $\Sigma$, a {\it language over $\Sigma$ (or
simply a language)\/} is any subset, $L$, of $\Sigma^*$.
}
\end{defin}

\medskip
The next step would be to introduce various formalisms to define
languages, such as automata or grammars but you'll have to take CSE262
to learn about these things!

\medskip
We now consider multisets. We already encountered multisets
in Section \ref{sec2} when we defined the axioms of 
propositional logic. As for sets, in a multiset, the order of
elements does not matter, but as in strings, multiple
occurrences of elements matter. 
For example, 
\[
\{a, a, b, c, c, c\}
\]
is a multiset with two occurrences of $a$, one occurrence of $b$
and three occurrences of $c$. This suggests defining a multiset
as a function with range $\natnums$, to specify the multiplicity
of each element.

\begin{defin}
\label{multisetdef}
{\em
Given any set, $S$, a
{\it multiset, $M$, over $S$\/} is any function,
$\mapdef{M}{S}{\natnums}$.
A {\it finite multiset, $M$, over $S$\/} is any function,
$\mapdef{M}{S}{\natnums}$, such that $M(a) \not= 0$
only for finitely many $a\in S$.
If $M(a) = k > 0$, we say that $a$ {\it appears with
mutiplicity $k$ in $M$\/}.
}
\end{defin}

\medskip
For example, if $S = \{a, b, c\}$,
we may use the notation
$\{a, a, a, b, c, c\}$ for the multiset where
$a$ has multiplicity $3$, $b$ has multiplicity $1$,
and $c$ has multiplicity $2$.

\medskip
The empty multiset is the function having the constant
value $0$.
The {\it cardinality $|M|$\/} of a (finite) multiset is
the number
\[|M| = \sum_{a\in S} M(a).\]
Note that this is well-defined since $M(a) = 0$
for all but  finitely many $a\in S$.
For example
\[|\{a, a, a, b, c, c\}| = 6.\]
We can define the {\it union\/} of multisets as follows:
If $M_1$ and $M_2$ are two multisets, then $M_1\cup M_2$
is the multiset given by
\[
(M_1\cup M_2)(a) = M_1(a) + M_2(a),
\quad\hbox{for all}\quad a\in S.
\]
A multiset, $M_1$, is a {\it submultiset\/} of a multiset, $M_2$, if
$M_1(a) \leq M_2(a)$, for all $a\in S$. The {\it difference of $M_1$ and $M_2$\/}
is the multiset, $M_1 - M_2$, given by
\[
(M_1 -  M_2)(a) = \cases{
M_1(a) -  M_2(a) & if $M_1(a) \geq M_2(a)$ \cr
0 & if $M_1(a) < M_2(a)$. \cr
}
\]

\medskip
Intersection of multisets can also be defined
but we will leave this as an exercise.

\medskip
Let us now discuss indexed families.
The Cartesian product construct, 
$A_1\times A_2\times \cdots \times A_n$, allows
us to form  finite indexed sequences,
$\lag a_1, \ldots, a_n\rag$, but there are situations
where we need to have infinite indexed sequences.
Typically, we want to be able to consider families
of elements indexed by some index set of our choice, say $I$.
We can do this as follows:

\begin{defin}
\label{indexseq}
{\em
Given any, $X$, and any other set, $I$, called the {\it index set\/},
the set of {\it $I$-indexed families (or sequences) 
of elements from $X$\/} is the set
of all functions, $\mapdef{A}{I}{X}$; such functions are
usually denoted $A = (A_i)_{i\in I}$.
When $X$ is a set of sets, each $A_i$ is some set in $X$ 
and we call  $(A_i)_{i\in I}$ a {\it family of sets (indexed by $I$)\/}.
}
\end{defin}

\medskip
Observe that if $I = [n] = \{1, \ldots, n\}$, then
an $I$-indexed family is just a string over $X$.
When $I = \natnums$, an $\natnums$-indexed family is
called an {\it infinite sequence\/} or often just
a {\it sequence\/}. In this case, we usually write
$(x_n)$ for such a sequence ($(x_n)_{n\in \natnums}$
is we want to be more precise).
Also, note that although the notion of indexed family
may seem less general than the notion of arbitrary 
collection of sets, this is
an illusion. Indeed, given any collection of sets, $X$, we may choose
the set index set $I$ to be $X$ itself, in wich case
$X$ appears as the range of the identity function, $\mapdef{\id}{X}{X}$.

\medskip
The point of indexed families is that the operations of union
and intersection can be generalized in an interesting way.
We can also form infinite Cartesian products, which are very
useful in algebra and geometry.

\medskip
Given any indexed family of sets,  $(A_i)_{i\in I}$, 
the {\it union of the family  $(A_i)_{i\in I}$\/}, denoted
$\bigcup_{i\in I} A_i$, is simply the union of the range
of $A$, that is,
\[
\bigcup_{i\in I} A_i = \bigcup \mathit{range}(A) =
 \{a \mid (\exists i\in I),\> a \in A_i\}.
\]
Observe that when $I = \emptyset$, the union 
of the family is the empty set.
When $I\not= \emptyset$, we say that we have a 
{\it nonempty family\/} (even though some of the $A_i$ may be empty).

\medskip
Similarly, if $I\not= \emptyset$, then
the {\it intersection of the family,  $(A_i)_{i\in I}$\/}, denoted
$\bigcap_{i\in I} A_i$, is simply the intersection of the range
of $A$, that is,
\[
\bigcap_{i\in I} A_i = \bigcap \mathit{range}(A) =
\{a \mid (\forall  i\in I),\> a \in A_i\}.
\]
Unlike the situation for union, when $I = \emptyset$, 
the intersection of the family does not exist.
It would be the set of all sets, which does not exist.

\medskip
It is easy to see that the laws for union, intersection
and complementation generalize to families but we will leave this to
the exercises. 

\medskip
An important construct generalizing the notion of finite
Cartesian product is the product of families.

\begin{defin}
\label{prodef}
{\em
Given any  family  of sets,
$(A_i)_{i\in I}$, the {\it product of the family  $(A_i)_{i\in I}$\/},
denoted $\prod_{i\in I} A_i$, is the set
\[
\prod_{i\in I} A_i = \{\mapdef{a}{I}{\bigcup_{i\in I} A_i} \mid
(\forall i\in I),\> a(i) \in A_i\}. 
\]
}
\end{defin}

\medskip
Definition \ref{prodef} says that the elements of the product
$\prod_{i\in I} A_i$ are the functions, \\
$\mapdef{a}{I}{\bigcup_{i\in I} A_i}$,
such that $a(i) \in A_i$ for every $i\in I$. We denote the members
of $\prod_{i\in I} A_i$ by $(a_i)_{i\in I}$  and we usually call
them {\it $I$-tuples\/}.
When $I = \{1, \ldots, n\} = [n]$, the members of
$\prod_{i\in [n]} A_i$ are the functions whose graph
consists of the sets of pairs 
\[
\{\lag 1, a_1\rag, \lag 2, a_2\rag, \ldots, \lag n, a_n\rag\},
\quad a_i \in A_i,\> 1\leq i \leq n,
\]
and we see that the function
\[
\{\lag 1, a_1\rag, \lag 2, a_2\rag, \ldots, \lag n, a_n\rag\}
\mapsto \lag a_1, \ldots, a_n\rag
\]
yields a bijection between $\prod_{i\in [n]} A_i$ and the Cartesian product
$A_1\times \cdots\times A_n$.
Thus, if each $A_i$ is nonempty, the product $\prod_{i\in [n]} A_i$
is nonempty. But what if $I$ is infinite?

\medskip
If $I$ is infinite, we smell choice functions.
That is, an element of $\prod_{i\in I} A_i$ is obtained by
choosing for every $i\in I$ some $a_i\in A_i$.
Indeed, the axiom of choice is needed to ensure that
$\prod_{i\in I} A_i \not= \emptyset$ if $A_i\not= \emptyset$
for all $i\in I$! For the record, we state this version (among many!)
of the axiom of choice:

\medskip\noindent
{\bf Axiom of Choice (Product Version)}

\medskip
For any family of sets, $(A_i)_{i\in I}$, if $I \not= \emptyset$
and $A_i\not= \emptyset$ for all $i\in I$, then
$\prod_{i\in I} A_i \not= \emptyset$.

\medskip
Given the product of a family of sets, $\prod_{i\in I} A_i$, 
for each $i\in I$, we have the function
$\mapdef{pr_i}{\prod_{i\in I} A_i}{A_i}$, called the
{\it $i$th projection function\/}, defined by
\[
pr_i((a_i)_{i\in I}) = a_i.
\]

\chapter[Some Counting Problems; Binomial Coefficients]
{Some Counting Problems; Binomial Coefficients}
\label{chap3}
\section[Counting Permutations and Functions]
{Counting Permutations and Functions}
\label{sec13b}
In this short section, we consider some simple counting
problems. Let us begin with permutations.
Recall that a {\it permutation\/} of a set, $A$, is any bijection between
$A$ and itself. If $A$ is a finite set with $n$ elements,
we mentioned earlier (without proof) that $A$
has $n!$ permutations, where the {\it factorial function\/},
$n \mapsto n!\>$ ($n\in \natnums$), is given recursively by:
\begin{eqnarray*}
0! & = & 1 \\
(n + 1)! & = & (n+1) n!.
\end{eqnarray*}
The reader should check that the existence of the
function, $n \mapsto n!$, can be justified using
the Recursion Theorem
(Theorem \ref{recnat}).

\begin{prop}
\label{permp1}
The number of permutations of a set of $n$ elements is $n!$.
\end{prop}

\proof
We prove that if $A$ and $B$ are any two finite sets
of the same cardinality, $n$, then the number of bijections
between $A$ and $B$ is $n!$. Now, in the special case 
where $B = A$, we get our theorem.

\medskip
The proof is by induction on $n$.
For $n = 0$, the empty set has one
bijection (the empty function). So,
there are $0! = 1$ permutations, as desired.

\medskip
Assume inductively that if $A$ and $B$ are any two finite sets
of the same cardinality, $n$, then the number of bijections
between $A$ and $B$ is $n!$.
If $A$ and $B$ are sets with $n + 1$
elements, then pick any element, $a\in A$, and write
$A = A' \cup \{a\}$, where $A' = A - \{a\}$ has $n$ elements.
Now, any bijection, $\mapdef{f}{A}{B}$, must assign
some element of $B$ to $a$ and then  $f\res A'$ is a bijection
between $A'$ and $B' = B - \{f(a)\}$. By the induction hypothesis,
there are $n!$ bijections between $A'$ and $B'$. Since there
are $n + 1$ ways of picking $f(a)$ in $B$, the total number of
bijections between $A$ and $B$ is $(n + 1)n! = (n + 1)!$,
establishing the induction hypothesis.
$\bigsquare$

\medskip
Let us also count the number of functions between two 
finite sets.

\begin{prop}
\label{numfunp1}
If $A$ and $B$ are finite sets with $|A| = m$ and $|B| = n$,
then the set of function, $B^A$, from $A$ to $B$ has $n^m$ elements.
\end{prop}

\proof
We proceed by induction on $m$. For $m = 0$, we have $A = \emptyset$,
and the only function is the empty function. In this case,
$n^0 = 1$ and the base base holds.

\medskip
Assume the induction hypothesis holds for $m$ and assume
$|A| = m + 1$. Pick any element, $a\in A$, and let $A' = A - \{a\}$,
a set with $m$ elements. Any function, $\mapdef{f}{A}{B}$,
assigns an element, $f(a)\in B$, to $a$ and $f\res A'$ is
a function from $A'$ to $B$. By the induction hypothesis, there
are $n^m$ functions from $A'$ to $B$. Since there are $n$ ways
of assigning $f(a)\in B$ to $a$, there are 
$n\cdot n^m = n^{m+1}$ functions from $A$ to $B$,
establishing the induction hypothesis.  
$\bigsquare$

\medskip
As a corollary, we determine the cardinality of a finite power set.
\begin{cor}
\label{numfunp2}
For any finite set, $A$, if $|A| = n$, then $|2^A| = 2^n$.
\end{cor}

\proof
By proposition \ref{powesetp1}, there is a bijection between
$2^A$ and the set of functions $\{0, 1\}^A$.
Since $|\{0, 1\}| = 2$, we get $|2^A| = |\{0, 1\}^A| = 2^n$,
by Proposition \ref{numfunp1}.
$\bigsquare$

\medskip
Computing the value of the factorial function
for a few inputs, say $n = 1, 2\ldots, 10$,
shows that it  grows very fast. For example, 
\[
10! = 3,628,800.
\]
It is possible to quantify how fast factorial grows
compared to other functions, say $n^n$ or $e^n$? 
Remarkably, the answer is yes. A beautiful formula due
to James Stirling (1692-1770) tells us that
\[
n! \cong \sqrt{2\pi n}\left(\frac{n}{e}\right)^n,
\]
which means that
\[
\lim_{n \rightarrow \infty} \frac{n!}{\sqrt{2\pi n}\left(\frac{n}{e}\right)^n} = 
1.
\]
Here, of course,
\[
e = 1 + \frac{1}{1!} + \frac{1}{2!} + \frac{1}{3!} + \cdots + 
\frac{1}{n!} + \cdots
\]  
the base of the natural logarithm.
It is even possible to estimate the error. It turns out that
\[
n! = \sqrt{2\pi n}\left(\frac{n}{e}\right)^n e^{\lambda_n},
\]
where 
\[
\frac{1}{12n + 1} < \lambda_n < \frac{1}{12n},
\]
a formula due to Jacques Binet (1786-1856).

\medskip
Let us introduce some notation  used for comparing
the rate of growth of functions. We begin with the ``Big oh'' notation.

\medskip
Given any two functions, 
$\mapdef{f}{\natnums}{\reals}$ and $\mapdef{g}{\natnums}{\reals}$,
we say that {\it $f$ is $O(g)$ (or $f(n)$ is $O(g(n))$)\/} iff there
is some $N > 0$ and a constant $c > 0$ such that
\[
|f(n)| \leq c |g(n)|,
\quad\hbox{for all}\quad n \geq N.
\]
In other words, for $n$ large enough, 
$|f(n)|$ is bounded by $c|g(n)|$.
We sometimes write $n >> 0$ to indicate that $n$ is ``large.''
 
\medskip
For example $\lambda_n$ is $O(\frac{1}{12n})$.
By abuse of notation, we often write $f(n) = O(g(n))$
even though this does not make sense. 

\medskip
The ``Big omega'' notation means the following:
{\it $f$ is $\Omega(g)$ (or $f(n)$ is $\Omega(g(n))$)\/} iff there
is some $N > 0$ and a constant $c > 0$ such that
\[
|f(n)| \geq c |g(n)|,
\quad\hbox{for all}\quad n \geq N.
\]

\medskip
The reader should check that  $f(n)$ is $O(g(n))$ iff
$g(n)$ is $\Omega(f(n))$.

\medskip
We can combine $O$ and $\Omega$ to get the
``Big theta'' notation:
{\it $f$ is $\Theta(g)$ (or $f(n)$ is $\Theta(g(n))$)\/} iff there
is some $N > 0$ and some constants $c_1 > 0$ and $c_2> 0$ such that
\[
c_1 |g(n)| \leq  |f(n)| \leq c_2 |g(n)|,
\quad\hbox{for all}\quad n \geq N.
\]

\medskip
Finally, the ``Little oh'' notation expresses the fact that
a function, $f$, has much slower growth than a function $g$.
We say that {\it $f$ is $o(g)$ (or $f(n)$ is $o(g(n))$)\/} iff
\[
\lim_{n\rightarrow \infty} \frac{f(n)}{g(n)} = 0.
\]
For example, $\sqrt{n}$ is $o(n)$.

\section[Counting Subsets of Size $k$; Binomial Coefficients]
{Counting Subsets of Size $k$; Binomial Coefficients}
\label{sec13c}
Let us now count the number of subsets of cardinality $k$ of
a set of cardinality $n$, with $0 \leq k \leq n$.
Denote this number by $\binom{n}{k}$ (say ``$n$ choose $k$'').
Actually, in the proposition below,
it will be more convenient to assume that $k\in \integs$.

\begin{prop}
\label{countp1}
For all $n\in \natnums$ and all $k\in \integs$,
if $\binom{n}{k}$ denotes the number of subsets of cardinality $k$ of
a set of cardinality $n$, then
\begin{eqnarray*}
\binom{0}{0} & = &  1\\
\binom{n}{k} & = & 0
\quad\hbox{if}\quad k \notin \{0, 1, \ldots, n\} \\
\binom{n}{k} & = & \binom{n-1}{k} + \binom{n-1}{k-1}
\quad (n \geq 1). 
\end{eqnarray*}
\end{prop}

\proof
We proceed by induction on $n \geq 0$. 
Clearly, we may assume that  our set is \\
$[n] = \{1, \ldots, n\}$ ($[0] = \emptyset$).
The base case $n = 0$
is trivial since the empty set is the only subset
of size $0$. When $n\geq 1$,  there are two kinds of subsets
of $\{1,\ldots,n\}$  having
$k$ elements: those containing $1$, and those not containing $1$.
Now, there are as many  subsets of $k$ elements from $\{1,\ldots,n\}$
containing $1$ 
as there are subsets of $k-1$ elements from  $\{2,\ldots,n\}$,
namely $\binom{n-1}{k-1}$, and there are as many subsets of $k$
elements from $\{1,\ldots,n\}$
not containing $1$ as there are subsets of $k$ elements
from  $\{2,\ldots,n\}$, namely $\binom{n-1}{k}$.
Thus, the number of subsets of $\{1,\ldots,n\}$
consisting of $k$ elements is $\binom{n-1}{k} + \binom{n-1}{k-1}$,
which is equal to $\binom{n}{k}$.
$\bigsquare$

\medskip
The numbers $\binom{n}{k}$ are also called {\it binomial coefficients\/},
because they arise in the expansion of the binomial
expression $(a + b)^n$, as we will see shortly.
The binomial coefficients can be computed inductively
using the formula
\[
\binom{n}{k}  =  \binom{n-1}{k} + \binom{n-1}{k-1}
\]
(sometimes known as Pascal's recurrence formula)
by forming what is usually called {\it Pascal's triangle\/},
which is based on the recurrence for $\binom{n}{k}$:

\medskip
\[
\begin{array}{llllllllll}
n & \binom{n}{0} & \binom{n}{1} & \binom{n}{2} & \binom{n}{3} &
\binom{n}{4} & \binom{n}{5} & \binom{n}{6} & \binom{n}{7} &\ldots\\ 
  &    &   &    &   &     &    &   &   & \\
0 &  1 &   &    &   &     &    &   &   & \\
1 &  1 & 1 &    &   &     &    &   &   & \\
2 &  1 & 2 & 1  &   &     &    &   &   & \\
3 &  1 & 3 & 3  & 1 &     &    &   &   & \\
4 &  1 & 4 & 6  & 4  & 1  &    &   &   & \\
5 &  1 & 5 & 10 & 10 & 5  & 1  &   &   & \\
6 &  1 & 6 & 15 & 20 & 15 & 6  & 1 &   & \\
7 &  1 & 7 & 21 & 35 & 35 & 21 & 7 & 1 & \\
\vdots&\vdots&\vdots&\vdots&\vdots&\vdots&\vdots&\vdots&\vdots&\vdots
\end{array}
\]

\medskip
We can also give the following explicit formula
for $\binom{n}{k}$ in terms of the factorial function:

\begin{prop}
\label{countp2}
For all $n, k\in \natnums$, with $0 \leq k \leq n$, we have
\[
\binom{n}{k} = \frac{n!}{k!(n - k)!}.
\]
\end{prop}

\proof
Left as an exercise to the reader
(use induction on $n$ and Pascal's recurrence formula).
$\bigsquare$ 

\medskip
Then, it  is very easy to see that
\[
\binom{n}{k} = \binom{n}{n-k}.
\]

\remark
The binomial coefficients were already known in the
twelfth century by the  Indian Scholar Bhaskra.
Pascal's triangle was taught back in 1265 by the Persian philosopher,
Nasir-Ad-Din. 

\medskip
We now prove the ``binomial formula'' (also called ``binomial theorem'').

\begin{prop} (Binomial Formula)
\label{binomp1}
For any two reals $a, b\in \reals$ (or more generally, any two commuting
variables $a, b$, i.e. satisfying $ab = ba$), we have the
formula:
\[
(a + b)^n = a^n + \binom{n}{1}a^{n - 1}b + \binom{n}{2}a^{n - 2}b^2 
+ \cdots + \binom{n}{k}a^{n - k}b^k + \cdots + 
\binom{n}{n - 1}ab^{n - 1} + b^n.
\]
The above can be written concisely as
\[
(a + b)^n = \sum_{k = 0}^n \binom{n}{k} a^{n - k}b^k.
\]   
\end{prop}

\proof
We proceed by induction on $n$.
For $n = 0$, we have $(a + b)^0 = 1$ and
the sum on the righthand side is also $1$, since
$\binom{0}{0} = 1$.

\medskip
Assume inductively that the formula holds for $n$.
Since
\[
(a + b)^{n + 1} = (a + b)^n(a + b),
\]
using the induction hypothesis, we get
\begin{eqnarray*}
(a + b)^{n + 1} & = & (a + b)^n(a + b) \\
& = & \left(\sum_{k = 0}^n \binom{n}{k} a^{n - k}b^k  \right)(a + b) \\
& = & \sum_{k = 0}^n \binom{n}{k} a^{n + 1 - k}b^k  +
\sum_{k = 0}^n \binom{n}{k} a^{n - k}b^{k+1}  \\
& = & a^{n + 1} + \sum_{k = 1}^n \binom{n}{k} a^{n + 1 - k}b^k  +
\sum_{k = 0}^{n-1} \binom{n}{k} a^{n - k}b^{k+1} + b^{n+1} \\
& = & a^{n + 1} + \sum_{k = 1}^n \binom{n}{k} a^{n + 1 - k}b^k  +
\sum_{k = 1}^{n} \binom{n}{k-1} a^{n + 1- k}b^{k} + b^{n+1} \\
& = &  a^{n + 1} + \sum_{k = 1}^{n} \left(\binom{n}{k}  + \binom{n}{k-1}\right)  
a^{n + 1 - k}b^k  +  b^{n+1}\\
& = &   \sum_{k = 0}^{n+1}\binom{n+1}{k} a^{n + 1 - k}b^k,
\end{eqnarray*}
where we used Proposition \ref{countp1} to go from the
next to the last line to the last line.
This establishes the induction step and
thus, proves the binomial formula. 
$\bigsquare$

\medskip
We also stated earlier that the number of injections between a
set with $m$ elements and a set with $n$ elements, where $m \leq n$, is given
by $\frac{n!}{(n - m)!}$ and we now prove it.

\begin{prop}
\label{countp3}
The number of injections between a
set, $A$, with $m$ elements and a set, $B$, with $n$ elements, 
where $m \leq n$, is given
by $\frac{n!}{(n - m)!} = n(n - 1)\cdots (n - m + 1)$.
\end{prop}

\proof
We proceed by induction on $m\leq n$. If $m = 0$, then
$A = \emptyset$ and there is only one injection,
namely the empty function from $\emptyset$ to $B$.
Since $\frac{n!}{(n - 0)!} = \frac{n!}{n!} = 1$, the base case holds.

\medskip
Assume the induction hypothesis holds for $m$ and consider
a set, $A$, with $m + 1$ elements, where $m + 1 \leq n$.
Pick any element $a\in A$ and let $A' = A - \{a\}$, a set
with $m$ elements. Any injection, $\mapdef{f}{A}{B}$,
assigns some element, $f(a)\in B$, to $a$ and then
$f\res A'$ is an injection from $A'$ to $B' = B - \{f(a)\}$,
a set with $n - 1$ elements. By the induction hypothesis, there are
\[
\frac{(n - 1)!}{(n - 1 - m)!}
\]
injections from $A'$ to $B'$. Since there are $n$
ways of picking $f(a)$ in $B$, the number of injections from $A$ to $B$ is
\[
n \frac{(n - 1)!}{(n - 1 - m)!} = \frac{n!}{(n - (m + 1))!},
\]
establishing the induction hypothesis.
$\bigsquare$

\medskip
Counting the number of surjections between a
set with $n$ elements and a set with $p$ elements, where $n \geq  p$,
is harder. We state the following formula without proof, leaving
the proof as an interesting exercise.

\begin{prop}
\label{countp4}
The number of surjections, $S_{n\, p}$,  between a
set, $A$, with $n$ elements and a set, $B$, with $p$ elements, 
where $n \geq p$, is given by 
\[
S_{n\, p} = p^n - \binom{p}{1}(p - 1)^n + \binom{p}{2}(p - 2)^n
+ \cdots + (-1)^{p - 1} \binom{p}{p - 1}.
\]
\end{prop}

\def\stirltwo#1#2{\genfrac{\{}{\}}{0pt}{}{#1}{#2}}

\remarks
\begin{enumerate}
\item
It can be shown that $S_{n\, p}$ satisfies the following peculiar
version of Pascal's identity:
\[
S_{n\, p} = p(S_{n-1\, p} + S_{n-1\, p-1}).
\]
\item
The numbers,  $S_{n\, p}$, are intimately related to the so-called
{\it Strirling numbers of the second kind\/}, denoted
$\stirltwo{n}{p}$,  $S(n, p)$, or $S_n^{(p)}$, which
count the number of partitions of a set of $n$ elements into
$p$ nonempty pairwise disjoint blocks.
In fact,
\[
S_{n\, p} = p! \stirltwo{n}{p}.
\]
\end{enumerate}

\medskip
The binomial coefficients can be generalized as follows. 
For all $n, m, k_1, \ldots, k_m\in \natnums$,
with $k_1 +\cdots + k_m = n$
and $m \geq 2$, we have the {\it multinomial coefficient\/},
\[
\binom{n}{k_1 \cdots k_m},
\]
which counts the number of ways of splitting
a set of $n$ elements into $m$ disjoint subsets, 
the $i$th subset having $k_i$ elements. Note that
when $m = 2$, the number of ways splitting a set of $n$ elements 
into two disjoint subsets where one of the two subsets
has $k_1$ elements and the other subset has $k_2 = n - k_1$
elements is precisely the number of subsets of size $k_1$
of a set of $n$ elements, that is
\[
\binom{n}{k_1\,  k_2} = \binom{n}{k_1}.
\]

\begin{prop}
\label{multinomp1}
For all $n, m, k_1, \ldots, k_m\in \natnums$,
with $k_1 +\cdots + k_m = n$
and $m \geq 2$, we have
\[
\binom{n}{k_1 \cdots k_m} = \frac{n!}{k_1! \cdots k_m!}.
\]
\end{prop}

\proof
There are $\binom{n}{k_1}$ ways of forming a subset of
$k_1$ elements from the set of $n$ elements; 
there are $\binom{n - k_1}{k_2}$ ways of forming a subset of
$k_2$ elements from the remaining $n - k_1$ elements;
there are $\binom{n - k_1 - k_2}{k_3}$ ways of forming a subset of
$k_3$ elements from the remaining $n - k_1 - k_2$ elements and so on;
finally, there are $\binom{n - k_1 - \cdots - k_{m - 2}}{k_{m - 1}}$ 
ways of forming a subset of
$k_{m - 1}$ elements from the remaining $n - k_1 - \cdots - k_{m - 2}$ elements
and there remains a set of $n - k_1 - \cdots - k_{m - 1} = k_m$
elements. This shows that
\[
\binom{n}{k_1 \cdots k_m} = \binom{n}{k_1}\binom{n - k_1}{k_2}\cdots
 \binom{n - k_1 - \cdots - k_{m - 2}}{k_{m - 1}}.
\]
But then, using the fact that $k_m = n - k_1 - \cdots - k_{m-1}$, we get
\begin{eqnarray*}
\binom{n}{k_1 \cdots k_m} & = & \frac{n!}{k_1!(n - k_1)!}
\frac{(n - k_1)!}{k_2!(n - k_1 - k_2)!} \cdots
\frac{(n - k_1 - \cdots - k_{m - 2})!}{k_{m - 1}!(n - k_1 - \cdots - k_{m - 1})!} \\
& = & \frac{n!}{k_1! \cdots k_m!},
\end{eqnarray*}
as claimed.

\medskip
As in the binomial case, it is convenient to set
\[
\binom{n}{k_1 \cdots k_m} = 0
\]
if $k_i < 0$ or $k_i> n$, for any $i$, with $1 \leq i \leq m$.
Then, Proposition \ref{countp1} is generalized as follows:

\begin{prop}
\label{multinomp2}
For all $n, m, k_1, \ldots, k_m\in \natnums$,
with $k_1 +\cdots + k_m = n$, $n \geq 1$
and $m \geq 2$, we have
\[
\binom{n}{k_1 \cdots k_m} = 
\sum_{i = 1}^m \binom{n-1}{k_1 \cdots (k_i - 1) \cdots k_m}. 
\]
\end{prop}

\proof
Note that we have $k_i - 1 = -1$ when $k_i = 0$.
If we observe that
\[
k_i\binom{n}{k_1 \cdots k_m} = n\binom{n-1}{k_1 \cdots (k_i - 1) \cdots k_m}
\]
even if $k_i = 0$, then we have
\begin{eqnarray*}
\sum_{i = 1}^m 
\binom{n-1}{k_1 \cdots (k_i - 1) \cdots k_m}  & = & 
\left(\frac{k_1}{n} + \cdots + \frac{k_m}{n}\right)\binom{n}{k_1 \cdots k_m} \\
& = & \binom{n}{k_1 \cdots k_m},
\end{eqnarray*}
since $k_1 + \cdots + k_m = n$.
$\bigsquare$

\remark
Proposition \ref{multinomp2} shows that Pascal's triangle
generalizes to ``higher dimensions'', that is, to $m \geq 3$.
Indeed, it is possible to give a geometric interpretation
of Proposition \ref{multinomp2} in which the multinomial
coefficients corresponding to those $k_1, \ldots, k_m$
with $k_1 + \cdots + k_m = n$ lie on the hyperplane
of equation $x_1 + \cdots + x_m = n$ in $\reals^m$,
and all the multinomial coefficients for which
$n \leq N$, for any fixed $N$, lie in a generalized tetrahedron called
a {\it simplex\/}. When $m = 3$, the multinomial coefficients 
for which $n \leq N$ lie in a tetrahedron whose faces are the planes
of equations, $x = 0$; $y = 0$; $z = 0$; and
$x + y + z = N$. 

\medskip
We have also the following generalization of  
Proposition \ref{binomp1}:

\begin{prop} (Multinomial Formula)
\label{multinomp3}
For all $n, m\in \natnums$ with  $m \geq 2$,
for all pairwise commuting variables $a_1, \ldots, a_m$, we have
\[
(a_1 + \cdots + a_m)^n = 
\sum_{\begin{subarray}{c}
k_1, \ldots, k_m \geq 0 \\
k_1 + \cdots + k_m = n 
\end{subarray}}
\binom{n}{k_1 \cdots k_m} a_1^{k_1} \cdots a_m^{k_m}.
\]
\end{prop}

\proof
We proceed by induction on $n$
and use Proposition \ref{multinomp2}.
The case $n = 0$ is trivially true.

\medskip
Assume the induction hypothesis holds for $n\geq 0$, then we have
\begin{eqnarray*}
(a_1 + \cdots + a_m)^{n+1} & = & (a_1 + \cdots + a_m)^n(a_1 + \cdots + a_m) \\
& = & \left(\sum_{\begin{subarray}{c}
k_1, \ldots, k_m \geq 0 \\
k_1 + \cdots + k_m = n 
\end{subarray}} 
\binom{n}{k_1 \cdots k_m} a_1^{k_1} \cdots a_m^{k_m} \right)
(a_1 + \cdots + a_m) \\
& = & \sum_{i = 1}^m 
\sum_{\begin{subarray}{c}
k_1, \ldots, k_m \geq 0 \\
k_1 + \cdots + k_m = n 
\end{subarray}} 
\binom{n}{k_1 \cdots k_i \cdots k_m} 
a_1^{k_1} \cdots a_i^{k_i + 1} \cdots a_m^{k_m} \\
& = & \sum_{i = 1}^m 
\sum_{\begin{subarray}{c}
k_1, \ldots, k_m \geq 0,\, k_i \geq 1 \\
k_1 + \cdots + k_m = n + 1 
\end{subarray}} 
\binom{n}{k_1 \cdots (k_i - 1) \cdots k_m} 
a_1^{k_1} \cdots a_i^{k_i} \cdots a_m^{k_m}. 
\end{eqnarray*}
We seem to hit a snag, namely, that $k_i \geq 1$, but recall that
\[
\binom{n}{k_1 \cdots - 1 \cdots k_m} = 0,
\]
so we have
\begin{eqnarray*}
(a_1 + \cdots + a_m)^{n+1} & = & 
\sum_{i = 1}^m 
\sum_{\begin{subarray}{c}
k_1, \ldots, k_m \geq 0,\, k_i \geq 1 \\
k_1 + \cdots + k_m = n + 1 
\end{subarray}} 
\binom{n}{k_1 \cdots (k_i - 1) \cdots k_m} 
a_1^{k_1} \cdots a_i^{k_i} \cdots a_m^{k_m} \\
& = & \sum_{i = 1}^m 
\sum_{\begin{subarray}{c}
k_1, \ldots, k_m \geq 0, \\
k_1 + \cdots + k_m = n + 1 
\end{subarray}} 
\binom{n}{k_1 \cdots (k_i - 1) \cdots k_m} 
a_1^{k_1} \cdots a_i^{k_i} \cdots a_m^{k_m} \\
& = & \sum_{\begin{subarray}{c}
k_1, \ldots, k_m \geq 0, \\
k_1 + \cdots + k_m = n + 1 
\end{subarray}} 
\left( \sum_{i = 1}^m 
\binom{n}{k_1 \cdots (k_i - 1) \cdots k_m} 
\right)
 a_1^{k_1} \cdots a_i^{k_i} \cdots a_m^{k_m} \\
& = & \sum_{\begin{subarray}{c}
k_1, \ldots, k_m \geq 0, \\
k_1 + \cdots + k_m = n + 1 
\end{subarray}} 
\binom{n+1}{k_1 \cdots k_i \cdots k_m} 
 a_1^{k_1} \cdots a_i^{k_i} \cdots a_m^{k_m},
\end{eqnarray*}
where we used Proposition \ref{multinomp2}
to justify the last equation. Therefore, the induction step
is proved and so is our proposition.
$\bigsquare$

\medskip
How many terms occur on the right-hand side of the
multinomial formula? After a moment of reflexion,
we see that this is the number of finite multisets of size
$n$ whose elements are drawn from a set of $m$ elements,
which is also equal to the number of $m$-tuples,
$k_1, \ldots, k_m$, with $k_i\in \natnums$ and
\[
k_1 + \cdots + k_m = n.
\]
The following proposition is left an exercise:

\begin{prop}
\label{multcount}
The number of finite multisets of size $n \geq 0$ whose
elements come from a set of size $m\geq 1$ is
\[
\binom{m + n - 1}{n}.
\]
\end{prop}

\section[The Inclusion-Exclusion Principle]
{The Inclusion-Exclusion Principle}
\label{sec13d}
We close this chapter with the proof of a poweful formula
for determining the cardinality of the union of a finite
number of (finite) sets in terms of the cardinalities of the
various intersections of these sets. This identity
variously attributed Nicholas Bernoulli, de Moivre, 
Sylvester and Poincar\'e
has many applications to counting problems
and to probability theory. We begin with the ``baby case'' of
two finite sets.

\begin{prop}
\label{inclexcl1}
Given any two finite sets, $A$, and $B$, we have
\[
|A \cup B| = |A| + |B| - |A\cap B|.
\]
\end{prop}

\proof
This formula is intuitively obvious because if
some element, $a\in A\cup B$, belongs to both $A$ and $B$ then
it is counted twice in $|A| + |B|$ and so we need to subtract
its contribution to $A\cap B$. Nevertherless,
it is worth giving a rigorous proof by induction on
$n = |A \cup B|$.

\medskip
If $n = 0$, then $A = B = \emptyset$ and the formula is clear:
$0 = 0 - 0$.

\medskip
For the induction step, assume that $A\cup B$ has $n + 1$
elements and pick any element, $a\in A\cup B$. There
are three cases:
\begin{enumerate}
\item
$a\in A$ and $a\notin B$. Since $a\notin B$, $(A - \{a\})\cap B = A\cap B$.
By the induction hypothesis,
\[
|(A - \{a\}) \cup B| = |A - \{a\}| + |B| + |A \cap B|.
\]
Then, adding $a$ to $A - \{a\}$  adds $1$ to $|(A - \{a\}) \cup B|$
and to $|A - \{a\}|$, so we get
\[
|A\cup B| = |A| + |B| - |A\cap B|,
\] 
proving the induction step.
\item
$a\notin A$ and $a\in B$. This case is analogous
to the previous one except that the roles of $A$ and $B$
are swapped.
\item
$a\in A$ and $a\in B$, i.e., $a\in A\cap B$.
In this case, by the induction hypothesis, we have
\begin{equation}
 |(A - \{a\}) \cup (B - \{a\})| = 
 |A - \{a\}| + |B - \{a\}| + |(A - \{a\})\cap (B - \{a\})|.
\tag{$*$}
\end{equation}
Adding $a$ to $A - \{a\}$ and $B - \{a\}$ adds $1$ to 
$|(A - \{a\}) \cup (B - \{a\})|$; it also adds $1$ to both
$|A - \{a\}|$ and $|B - \{a\}|$ and adds
$1$ to $|(A - \{a\}) \cap (B - \{a\})|$. 
So, the contribution of $a$ to the righthand side of $(*)$
is $2 - 1 = 1$ and we get
\[
|A\cup B| = |A| + |B| - |A\cap B|,
\] 
proving the induction step. $\bigsquare$
\end{enumerate}

\medskip
We would like to generalize the formula of Proposition \ref{inclexcl1}
to any finite collection of finite sets, $A_1, \ldots, A_n$.
A moment of reflexion shows that when $n = 3$, we have
\[
|A\cup B\cup C | = |A| + |B| + |C|
- |A\cap B| - |A\cap C| - |B\cap C|
+ |A\cap B\cap C|.
\] 
One of the obstacles in generalizing the above formula to $n$ sets
is purely notational: We need a way of
denoting arbitrary intersections of sets belonging to 
a family of sets indexed by  $\{1, \ldots, n\}$.
We can do this by using indices ranging
over subsets of $\{1, \ldots, n\}$, 
as opposed to indices ranging over integers. So, for example,
for any nonempty subset, $I \subseteq \{1, \ldots, n\}$, the expression
$\bigcap_{i \in I} A_i$ denotes the intersection of all the
subsets whose index, $i$, belongs to $I$. 

\begin{thm} (Inclusion-Exclusion Principle)
\label{inclexcl2}
For any finite sequence, $A_1, \ldots, A_n$, of  \\
$n \geq 2$ 
subsets of a finite set, $X$, we have
\[
\left|\bigcup_{k = 1}^n A_k\right| =
\sum_{\begin{subarray}{c}
 I \subseteq \{1, \ldots, n\} \\
 I \not= \emptyset
      \end{subarray}}
(-1)^{(|I| - 1)}
 \left|\bigcap_{i\in I} A_i\right|.
\]
\end{thm}

\proof
We proceed by induction on $n \geq 2$.
The base case, $n = 2$, is exactly Proposition \ref{inclexcl1}.
Let us now consider the induction step. We can write
\[
\bigcup_{k = 1}^{n+1} A_k = \left(\bigcup_{k = 1}^{n} A_k\right) 
\cup \{A_{n+1}\}
\]
and so, by Proposition \ref{inclexcl1}, we have
\begin{eqnarray*}
\left| \bigcup_{k = 1}^{n+1} A_k \right| & = & 
\left| \left(\bigcup_{k = 1}^{n} A_k\right) \cup \{A_{n+1}\} \right| \\
& = & \left| \bigcup_{k = 1}^{n} A_k\right| + |A_{n+1}| - 
\left|\left(\bigcup_{k = 1}^{n} A_k\right) \cap \{A_{n+1}\} \right|.
\end{eqnarray*}
We can  apply the induction hypothesis to the first term and we get
\[
\left|\bigcup_{k = 1}^{n} A_k\right| =
\sum_{\begin{subarray}{c}
 J \subseteq \{1, \ldots, n\} \\
 J \not= \emptyset
      \end{subarray}}
(-1)^{(|J| - 1)}
 \left|\bigcap_{j\in J} A_j\right|.
\]
Using distributivity of intersection over union, we have
\[
\left(\bigcup_{k = 1}^{n} A_k\right) \cap \{A_{n+1}\} =
\bigcup_{k = 1}^{n} (A_k\cap A_{n+1}). 
\]
Again, we can apply the induction hypothesis and obtain
\begin{eqnarray*}
- \left|\bigcup_{k = 1}^{n} (A_k\cap A_{n+1}) \right| 
& = &  - \sum_{\begin{subarray}{c}
 J \subseteq \{1, \ldots, n\} \\
 J \not= \emptyset
      \end{subarray}}
(-1)^{(|J| - 1)}
 \left|\bigcap_{j\in J} (A_j\cap A_{n+1})\right| \\
& = &   \sum_{\begin{subarray}{c}
 J \subseteq \{1, \ldots, n\} \\
 J \not= \emptyset 
      \end{subarray}}
(-1)^{|J|}
 \left|\bigcap_{j\in J\cup \{n + 1\}} A_j\right| \\
& = &   \sum_{\begin{subarray}{c}
 J \subseteq \{1, \ldots, n\} \\
 J \not= \emptyset 
      \end{subarray}}
(-1)^{(|J \cup \{n + 1\}| - 1)}
 \left|\bigcap_{j\in J\cup \{n + 1\}} A_j\right|.
\end{eqnarray*}
Putting all this together, we get
\begin{eqnarray*}
\left| \bigcup_{k = 1}^{n+1} A_k \right|  & = &
\sum_{\begin{subarray}{c}
 J \subseteq \{1, \ldots, n\} \\
 J \not= \emptyset
      \end{subarray}}
(-1)^{(|J| - 1)}
 \left|\bigcap_{j\in J} A_j\right| + |A_{n+1}| 
+ 
 \sum_{\begin{subarray}{c}
 J \subseteq \{1, \ldots, n\} \\
 J \not= \emptyset 
      \end{subarray}}
(-1)^{(|J \cup \{n + 1\}| - 1)}
 \left|\bigcap_{j\in J\cup \{n + 1\}} A_j\right| \\
& = & 
\sum_{\begin{subarray}{c}
 J \subseteq \{1, \ldots, n + 1\} \\
 J \not= \emptyset,\> n + 1\notin J
      \end{subarray}}
(-1)^{(|J| - 1)}
 \left|\bigcap_{j\in J} A_j\right| 
+ 
 \sum_{\begin{subarray}{c}
 J \subseteq \{1, \ldots, n + 1\} \\
 n + 1 \in J  
      \end{subarray}}
(-1)^{(|J| - 1)}
 \left|\bigcap_{j\in J} A_j\right| \\
& = & \sum_{\begin{subarray}{c}
 I \subseteq \{1, \ldots, n+1\} \\
 I \not= \emptyset
      \end{subarray}}
(-1)^{(|I| - 1)}
 \left|\bigcap_{i\in I} A_i\right|,
\end{eqnarray*}
establishing the induction hypothesis and finishing the proof.
$\bigsquare$

\medskip
As an application of the Inclusion-Exclusion Principle, let us prove the
formula for counting the number of surjections from 
$\{1, \ldots, n\}$ to  $\{1, \ldots, p\}$, with $p \leq n$, given in
Proposition \ref{countp4}.

\medskip
Recall that the total number of functions from 
$\{1, \ldots, n\}$ to  $\{1, \ldots, p\}$ is
$p^n$. The trick is to count the number
of functions that are {\it not\/} surjective. 
Any such function has the property that its image
misses one element from $\{1, \ldots, p\}$. So, if we let
\[
A_i = \{\mapdef{f}{\{1, \ldots, n\}}{\{1, \ldots, p\}} \mid
i \notin \Im(f)\},
\]
we need to count
$|A_1 \cup \cdots \cup A_p|$. But, we can easily do this using
the Inclusion-Exclusion Principle. Indeed, for any nonempty
subset, $I$, of $\{1, \ldots, p\}$, with $|I| = k$,
the functions in $\bigcap_{i\in I} A_i$ are exactly the functions
whose range misses $I$. But, these are exactly the functions
from $\{1, \ldots, n\}$ to  $\{1, \ldots, p\} - I$
and there are $(p - k)^n$ such functions. Thus,
\[
\left| \bigcap_{i\in I} A_i\right| = (p - k)^n.
\]
As there are
$\binom{p}{k}$ subsets, $I \subseteq  \{1, \ldots, p\}$, with $|I| = k$,
the contribution of all $k$-fold intersections to the
Inclusion-Exclusion Principle is
\[
\binom{p}{k} (p - k)^n.
\]
Note that $A_1\cap \cdots \cap  A_p = \emptyset$, since  
functions have a nonempty image.
Therefore, the Inclusion-Exclusion Principle yields
\[
|A_1 \cup \cdots \cup A_p| = 
\sum_{k = 1}^{p-1} (-1)^{k-1} \binom{p}{k} (p - k)^n ,
\]
and so, the number of surjections, $S_{n\, p}$, is
\begin{eqnarray*}
S_{n\, p} & = & p^n - |A_1\cup \cdots \cup A_p| =
 p^n - \sum_{k = 1}^{p-1} (-1)^{k-1} \binom{p}{k} (p - k)^n \\
&  = & \sum_{k = 0}^{p-1} (-1)^{k} \binom{p}{k} (p - k)^n \\
& = & p^n - \binom{p}{1}(p - 1)^n + \binom{p}{2}(p - 2)^n
+ \cdots + (-1)^{p - 1} \binom{p}{p - 1},
\end{eqnarray*}
which is indeed the formula of Proposition \ref{countp4}.

\medskip
Another amusing application of the 
Inclusion-Exclusion Principle is the formula
giving the number, $p_n$, of permutations
of $\{1, \ldots, n\}$ that leave no element fixed
(i.e., $f(i) \not = i$, for all $i\in \{1, \ldots, n\}$).
Such permutations are often called {\it derangements\/}.
We get
\begin{eqnarray*}
p_n  & = & n!\left(1 - \frac{1}{1!} + \frac{1}{2!} + \cdots + 
\frac{(-1)^k}{k!} + \cdots +  \frac{(-1)^n}{n!} \right) \\
 & = & n! - \binom{n}{1}(n - 1)! + \binom{n}{2}(n - 2)!
+ \cdots + (-1)^{n} .
\end{eqnarray*}

\remark
We know (using the series expansion for $e^x$ in which we set $x = -1$)
that
\[
\frac{1}{e} = 1 - \frac{1}{1!} + \frac{1}{2!} + \cdots + 
\frac{(-1)^k}{k!} + \cdots . 
\]
Consequently, the factor of $n!$ in the above formula for $p_n$
is the sum of the first $n + 1$ terms of $\frac{1}{e}$ and so,
\[
\lim_{n \rightarrow \infty} \frac{p_n}{n!} = \frac{1}{e}.
\]

It turns out that the series for $\frac{1}{e}$ converges very
rapidly, so $p_n \approx \frac{1}{e} n!$.
The ratio $p_n/n!$ has an interesting interpretation in terms of 
probabilities. Assume $n$ persons go to a restaurant (or to the theatre, etc.)
and that they all check their coats. Unfortunately, the cleck loses all 
the coat tags. Then, $p_n/n!$ is the probability that
nobody will get her or his own coat back! As we just explained,
this probability is roughly $\frac{1}{e} \approx \frac{1}{3}$,
a surprisingly large number.

\medskip
The Inclusion-Exclusion Principle can be easily generalized in 
a useful way as follows: Given a finite set, $X$, let $m$ be any
given function, $\mapdef{m}{X}{\reals}_+$, 
and for any nonempty subset,
$A\subseteq X$, set
\[
m(A) = \sum_{a\in A} m(a),
\]
with the convention that $m(\emptyset) = 0$
(Recall that $\reals_+ = \{x\in \reals \mid x\geq 0\}$).
For any $x\in X$, the number $m(x)$ is called the
{\it weight\/} (or {\it measure\/}) of $x$ and
the quantity $m(A)$ is often called the {\it measure of the set $A$\/}.
For example, if $m(x) = 1$ for all $x\in A$, then
$m(A) = |A|$, the cardinality of $A$, which is the special case
that we have been considering. For any two subsets, $A, B\subseteq X$,
it is obvious that
\begin{eqnarray*}
m(A\cup B) & = & m(A) + m(B) \\
m(X - A) & = & m(X) - m (A) \\
m(\overline{A\cup B}) & = & m(\overline{A} \cap \overline{B}) \\
m(\overline{A\cap B}) & = & m(\overline{A} \cup \overline{B}), 
\end{eqnarray*}
where $\overline{A} = X - A$.
Then, we have the following version of Theorem \ref{inclexcl2}:

\begin{thm} (Inclusion-Exclusion Principle, Version 2)
\label{inclexcl3}
Given any measure function, $\mapdef{m}{X}{\reals}_+$, 
for any finite sequence, $A_1, \ldots, A_n$, of  $n \geq 2$ 
subsets of a finite set, $X$, we have
\[
m\left(\bigcup_{k = 1}^n A_k\right) =
\sum_{\begin{subarray}{c}
 I \subseteq \{1, \ldots, n\} \\
 I \not= \emptyset
      \end{subarray}}
(-1)^{(|I| - 1)} \,
m \left(\bigcap_{i\in I} A_i\right).
\]
\end{thm}

\proof
The proof is obtained from the proof of Theorem \ref{inclexcl2}
by changing everywhere any expression of the form
$|B|$ to $m(B)$.
$\bigsquare$

\medskip
A useful corollary of Theorem \ref{inclexcl3} often
known as Sylvester's formula is:

\begin{thm} (Sylvester's Formula)
\label{inclexcl4}
Given any measure, $\mapdef{m}{X}{\reals}_+$, 
for any finite sequence, $A_1, \ldots, A_n$, of  $n \geq 2$ 
subsets of a finite set, $X$, the measure of the set of elements of $X$ 
that do not belong to any of the sets $A_i$ is given by
\[
m\left(\bigcap_{k = 1}^n \overline{A}_k\right) = m(X) +
\sum_{\begin{subarray}{c}
 I \subseteq \{1, \ldots, n\} \\
 I \not= \emptyset
      \end{subarray}}
(-1)^{|I|} \,
m \left(\bigcap_{i\in I} A_i\right).
\]
\end{thm}

\proof
Observe that
\[
\bigcap_{k = 1}^n \overline{A}_k = X - \bigcup_{k = 1}^n A_k.
\]
Consequently, using  Theorem \ref{inclexcl3}, we get
\begin{eqnarray*}
m\left(\bigcap_{k = 1}^n \overline{A}_k\right) & = & 
m\left(X - \bigcup_{k = 1}^n A_k\right) \\
& = & m(X) - m\left(\bigcup_{k = 1}^n A_k\right) \\
& = & m(X) -  \sum_{\begin{subarray}{c}
 I \subseteq \{1, \ldots, n\} \\
 I \not= \emptyset
      \end{subarray}}
(-1)^{(|I| - 1)} \,
m \left(\bigcap_{i\in I} A_i\right) \\
& = & m(X) +
\sum_{\begin{subarray}{c}
 I \subseteq \{1, \ldots, n\} \\
 I \not= \emptyset
      \end{subarray}}
(-1)^{|I|} \,
m \left(\bigcap_{i\in I} A_i\right),
\end{eqnarray*}
establishing Sylvester's formula.
$\bigsquare$

\medskip
Note that if we use the convention that when the index set, $I$, is empty
then
\[
\bigcap_{i\in \emptyset} A_i = X,
\]
then the term $m(X)$ can be included in the above sum by removing the
condition that $I \not= \emptyset$. Sometimes, it is also
convenient to regroup terms involving subsets, $I$, having the
same cardinality and another way to state Sylvester's formula is as
follows:
\begin{equation}
m\left(\bigcap_{k = 1}^n \overline{A}_k\right) =
\sum_{k = 0}^n (-1)^k 
\sum_{\begin{subarray}{c} 
 I \subseteq \{1, \ldots, n\} \\
 |I| = k
      \end{subarray}}
m \left(\bigcap_{i\in I} A_i\right).
\tag{Sylvester's Formula}
\end{equation}

\medskip
Finally, Sylvester's formula can be generalized to a formula
usually known as the ``Sieve Formula'':

\begin{thm} (Sieve Formula)
\label{inclexcl5}
Given any measure, $\mapdef{m}{X}{\reals}_+$, 
for any finite sequence, $A_1, \ldots, A_n$, of  $n \geq 2$ 
subsets of a finite set, $X$, the measure of the set of elements of $X$ 
that belong to exactly $p$  of the sets $A_i$ ($0 \leq p \leq n$)
is given by
\[
T^p_n = 
\sum_{k = p}^n (-1)^{k- p}\binom{k}{p}  
\sum_{\begin{subarray}{c} 
 I \subseteq \{1, \ldots, n\} \\
 |I| = k
      \end{subarray}}
m \left(\bigcap_{i\in I} A_i\right).
\]
\end{thm}

\proof
For any subset, $I \subseteq \{1, \ldots, n\}$, apply
Sylvester's formula to 
$X = \bigcap_{i\in I} A_i$ and to the subsets
$A_j \cap  \bigcap_{i\in I} A_i$. We get
\[
m\left(\bigcap_{i\in I} A_i \cap \bigcap_{j\notin I} \overline{A}_j \right)
=
\sum_{\begin{subarray}{c} 
 J \subseteq \{1, \ldots, n\} \\
 I \subseteq J
      \end{subarray}}
 (-1)^{|J| - |I|}\, m \left(\bigcap_{j\in J} A_j\right).
\]
Hence,
\begin{eqnarray*}
T_n^p & = & \sum_{\begin{subarray}{c} 
 I \subseteq \{1, \ldots, n\} \\
 |I| = p
      \end{subarray}}
m\left(\bigcap_{i\in I} A_i \cap \bigcap_{j\notin I} \overline{A}_j \right) \\
& = &
\sum_{\begin{subarray}{c} 
 I \subseteq \{1, \ldots, n\} \\
 |I| = p
      \end{subarray}}
\sum_{\begin{subarray}{c} 
 J \subseteq \{1, \ldots, n\} \\
 I \subseteq J
      \end{subarray}}
(-1)^{|J| - |I|}\, m \left(\bigcap_{j\in J} A_j\right) \\
& = &
\sum_{\begin{subarray}{c} 
 J \subseteq \{1, \ldots, n\} \\
 |J| \geq p
      \end{subarray}}
\sum_{\begin{subarray}{c} 
 I \subseteq J \\
 |I| = p
      \end{subarray}}
 (-1)^{|J| - |I|}\, m \left(\bigcap_{j\in J} A_j\right) \\
& = & 
\sum_{k = p}^n (-1)^{k- p}\binom{k}{p}  
\sum_{\begin{subarray}{c} 
 J \subseteq \{1, \ldots, n\} \\
 |J| = k
      \end{subarray}}
m \left(\bigcap_{j\in J} A_j\right),
\end{eqnarray*}
establishing the Sieve formula.
$\bigsquare$

\medskip
Observe that Sylvester's Formula is the special case of the Sieve Formula
for which $p = 0$.
The Inclusion-Exclusion Principle (and its relatives)
plays an important role in
combinatorics and probablity theory as the reader will
verify by consulting any text on combinatorics.
A classical reference on combinatorics is Berge \cite{Berge71};
a more recent is Cameron \cite{Cameron}; a more recent
and more advanced is Stanley \cite{Stanley1}. 
Another fascinating (but deceptively tough)
reference covering discrete mathematics
and including a lot of combinatorics
is Graham, Knuth and Patashnik \cite{GrahamKnuth}.

\medskip
We are now ready to study special kinds of relations:
Partial orders and equivalence relations.

\chapter[Partial Orders and Equivalence Relations]
{Partial Orders, Complete Induction 
and Equivalence Relations}
\label{chap4}
\section{Partial Orders}
\label{sec14}
There are two main kinds of relations that play a very important role
in mathematics and computer science: 
\begin{enumerate}
\item
Partial orders
\item
Equivalence relations.
\end{enumerate}

\medskip
In this section and the next few ones, we define partial orders
and investigate some of their properties. As we will see, 
the ability to use induction is intimately related to 
a very special property of partial orders known as
well-foundedness. 

\medskip
Intuitively, the notion of order among elements of a set,
$X$, captures the fact some elements are bigger than others,
perhaps more important, or perhaps that they carry
more information. For example, we are all familiar with
the natural ordering, $\leq $, of the integers
\[ 
 \cdots, -3 \leq -2 \leq -1 \leq 0 \leq 1\leq 2 \leq 3 \leq \cdots,
\]
the ordering
of the rationals (where $\frac{p_1}{q_1} \leq \frac{p_2}{q_2}$
iff $\frac{p_2q_1 - p_1q_2}{q_1q_2} \geq 0$, i.e.,
$p_2q_1 - p_1q_2\geq 0$ if $q_1q_2 > 0$
else $p_2q_1 - p_1q_2\leq 0$ if $q_1q_2 < 0$),
and  the ordering of the real numbers.
In all of the above orderings, note that
for any two number $a$ and $b$, either $a \leq b$ or $b \leq a$.
We say that such orderings are {\it total\/} orderings.

\medskip
A natural example of an ordering which is not total is
provided by the subset ordering.
Given a set, $X$, we can order the subsets of $X$
by the subset relation: $A\subseteq B$, where $A, B$ are
any subsets of $X$. For example, if $X = \{a, b, c\}$,
we have $\{a\}\subseteq \{a, b\}$. However, note that
neither $\{a\}$ is a subset of $\{b, c\}$ nor
$\{b, c\}$ is a subset of $\{a\}$. We say that 
$\{a\}$ and $\{b, c\}$ are {\it incomparable\/}.
Now, not all relations are partial orders, so which
properties characterize partial orders?  Our next definition
gives us the answer.

\begin{defin}
\label{posetdef}
{\em
A binary relation, $\leq$, on a set, $X$, is a {\it partial order\/}
(or {\it partial ordering\/}) iff it is
{\it reflexive\/}, {\it transitive\/} and {\it antisymmetric\/}, 
that is:
\begin{enumerate}
\item[(1)] 
({\it Reflexivity\/}):
$a\leq a$, for all $a\in X$;
\item[(2)]
({\it Transitivity\/}):
If $a \leq b$  and $b \leq c$, then $a \leq c$, for all
$a, b, c \in X$. 
\item[(3)]
({\it antisymmetry\/}): If $a\leq b$ and $b \leq a$, then $a = b$,
for all $a, b\in X$.
\end{enumerate}

\medskip
A partial order is a {\it total order (ordering)\/}
(or {\it linear order (ordering)\/})  
iff for all $a, b\in X$, either
$a\leq b$ or $b \leq a$. When neither $a\leq b$ nor
$b\leq a$, we say that {\it $a$ and $b$ are incomparable\/}.
A subset, $C\subseteq X$, is a {\it chain\/} iff
$\leq$ induces a total order on $C$
(so,  for all $a, b\in C$, either
$a\leq b$ or $b \leq a$).
The {\it strict order (ordering), $<$, associated with $\leq$\/} is 
the relation defined by: $a < b$ iff $a\leq b$ and $a\not= b$.
If $\leq$ is a partial order on $X$, we say that the pair
$\lag X, \leq\rag$ is a {\it partially ordered set\/} or for short,
a {\it poset\/}. 
}
\end{defin}

\remark
Observe that if $<$ is the strict order associated with 
a partial order, $\leq$, then $<$ is transitive and 
{\it anti-reflexive\/}, which means that
\begin{enumerate}
\item[(4)]
$a \not< a$, for all $a\in X$.
\end{enumerate}
Conversely, let $<$ be a relation on $X$ and assume
that $<$ is transitive and anti-reflexive. Then, we can define
the relation $\leq$ so that $a \leq b$ iff $a = b$ or $a < b$.
It is easy to check that $\leq$ is a partial order and
that the strict order associated with $\leq$ is our original relation, $<$.

\medskip
Given a poset, $\lag X, \leq\rag$, by abuse of notation, we often refer
to $\lag X, \leq\rag$ as the {\it poset $X$\/}, the partial order $\leq$
being implicit. If confusion may arise, for example when
we are dealing with several posets, we denote the partial
order on $X$ by $\leq_X$.

\medskip
Here are a few examples of partial orders.

\begin{enumerate}
\item 
{\bf  The subset ordering}.
We leave it to the reader to check that 
the subset relation, $\subseteq$,  on a set,  $X$, is indeed a partial order.
For example, if $A \subseteq B$ and $B\subseteq A$,
where $A, B \subseteq X$, then $A = B$, since
these assumptions are exactly those needed by the extensionality
axiom.
\item
{\bf The natural order on $\natnums$}.
Although we all know what is the ordering of the natural numbers,
we should realize that if we stick to our axiomatic
presentation where we defined the natural numbers as sets that belong to every
inductive set (see Definition \ref{natnumdef}), then we haven't yet defined
this ordering. However, this is easy to do since the natural numbers are sets.
For any $m, n\in \natnums$, define $m \leq n$ as $m = n$ or $m \in n$!
Then, it is  not hard check that this relation is a total order
(Actually, some of the details are a bit tedious and require induction,
see Enderton  \cite{Endertonset}, Chapter 4).
\item
{\bf Orderings on strings}.
Let $\Sigma = \{a_1, \ldots, a_n\}$ be an alphabet. 
The prefix, suffix and substring relations
defined in Section \ref{sec13} are easily seen to be partial orders.
However, these orderings are not total. It is sometimes
desirable to have a total order on strings and, fortunately, 
the lexicographic order (also called dictionnary order) 
achieves this goal. In order to define the {\it lexicographic order\/}
we assume that the symbols in $\Sigma$ are totally ordered, 
$a_1 < a_2 < \cdots < a_n$. Then, given any two strings, $u, v\in \Sigma^*$,
we set 
\[
u \preceq v \quad \cases{ & if $v = uy$, for some $y\in\Sigma^*$, or\cr
                    & if $u = xa_iy$, $v = xa_jz$,\cr
                    & and $a_i < a_j$, for some $x, y, z\in\Sigma^*$.\cr
}
\]
In other words, either $u$ is a prefix of $v$ or else
$u$ and $v$ share a common prefix, $x$, and then there is a
differring symbol, $a_i$ in $u$ and $a_j$ in $v$, with $a_i < a_j$.
It is fairly tedious to prove that the lexicographic order
is a partial order. Moreover, the lexicographic order
is a total order.
\item
{\bf The divisibility order on $\natnums$}.
Let us begin by defining divisibility in $\integs$.
Given any two integers, $a, b\in \integs$, with $b\not= 0$, we say that
{\it $b$ divides $a$\/} ({\it $a$ is a multiple of $b$\/}) iff
$a = bq$ for some $q\in \integs$. Such a $q$ is called the {\it quotient
of $a$ and $b$\/}. Most number theory books use the notation
$b \mid a$ to express that $b$ divides $a$. 
We leave  the verification that the
divisibility relation is reflexive and transitive
as an easy exercise. What about
antisymmetry? So, assume that $b \mid a$ and $a \mid b$
(thus, $a, b \not= 0$).
This means that there exist $q_1, q_2\in \integs$ so that
\[
a = bq_1\quad\hbox{and}\quad b = aq_2.
\]
From the above, we deduce that $b = bq_1q_2$, that is
\[
b(1 - q_1q_2) = 0.
\]
As $b\not = 0$, we conclude that 
\[
q_1q_2 = 1.
\]
Now, let us restrict ourselves to $\natnums_+ = \natnums - \{0\}$,
so that $a, b \geq 1$. It follows that $q_1, q_2\in \natnums$
and in this case, $q_1q_2 = 1$ is only possible iff $q_1 = q_2 = 1$.
Therefore,  $a = b$ and the divisibility relation is 
indeed a partial order on $\natnums_+$. 
Why is divisibility not a partial order on $\integs - \{0\}$?
\end{enumerate}

\medskip
Given a poset,  $\lag X \leq\rag$, if $X$ is finite, then
there is a convenient way to describe the partial
order $\leq$ on $X$ using a graph. In preparation for that,
we need a few preliminary notions.

\medskip 
Consider an arbitrary poset, $\lag X \leq\rag$ (not necessarily finite).
Given any element, $a\in X$, the following situations
are of interest:
\begin{enumerate}
\item
For {\bf no} $b\in X$ do we have $b < a$.
We say that $a$ is a {\it minimal element\/} (of $X$).
\item
There is some $b\in X$ so that $b < a$ and
there is {\bf no} $c\in X$ so that $b < c < a$.
We say that $b$ is an {\it immediate predecessor of $a$\/}.
\item
For {\bf no} $b\in X$ do we have $a < b$.
We say that $a$ is a {\it maximal element\/} (of $X$).
\item
There is some $b\in X$ so that $a < b$ and
there is {\bf no} $c\in X$ so that $a < c < b$.
We say that $b$ is an {\it immediate successor of $a$\/}.
\end{enumerate}

\medskip
Note that an element may have more than one immediate predecessor
(or more than one immediate successor).

\medskip
If $X$ is a finite set, then it is easy to see that every element
that is not minimal has an immediate predecessor and any element
that is not maximal has an immediate successor
(why?). But if $X$ is infinite, for
example, $X = \rats$, this may not be the case. Indeed,
given any two distinct rational numbers, $a, b\in \rats$, we have
\[
a < \frac{a + b}{2} < b.
\]

\medskip
Let us now use our notion of immediate predecessor
to draw a diagram representing a finite poset,  $\lag X, \leq\rag$.
The trick is to draw a picture consisting of nodes and oriented edges,
where the nodes are all the elements of $X$ and where we draw an oriented
edge from $a$ to $b$ iff $a$ is an immediate predecessor of $b$.
Such a diagram is called a {\it Hasse diagram\/} for $\lag X, \leq\rag$.
Observe that if $a < c < b$, then the diagram does {\bf not}
have edges corresponding to the relations $a < c$ and $c < b$.
However, such information can be recovered from the diagram by following
paths consisting of one or several consecutive edges (we are a bit informal
here, but we will define directed graphs and paths more rigorously later).
Similarly, the self-loops corresponding to the the reflexive relations
$a\leq a$ are omitted. A Hasse diagram is an economical representation
of a finite poset and it contains the same amount of information as the
partial order, $\leq$.

\medskip
Here is the diagram associated with the partial order on the power set of 
the two element set, $\{a, b\}$:

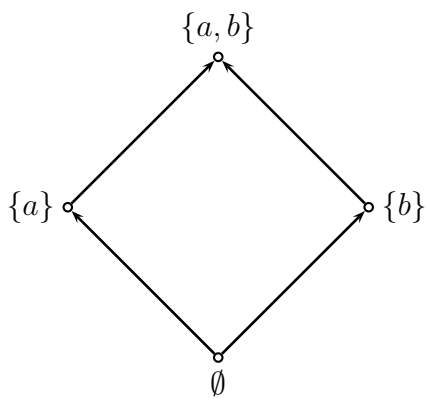
\begin{figure}[H]
 \begin{center}
    \begin{pspicture}(0,0)(4,4.5)
    \cnode(2,0){2pt}{u1}
    \cnode(0,2){2pt}{u2}
    \cnode(4,2){2pt}{u3}
    \cnode(2,4){2pt}{u4}
    \ncline[linewidth=1pt]{->}{u1}{u2}
    \ncline[linewidth=1pt]{->}{u1}{u3}
    \ncline[linewidth=1pt]{->}{u2}{u4}
    \ncline[linewidth=1pt]{->}{u3}{u4}
    \uput[-90](2,0){$\emptyset$}
    \uput[180](0,2){$\{a\}$}
    \uput[0](4,2){$\{b\}$}
    \uput[90](2,4){$\{a,b\}$}
    \end{pspicture}
  \end{center}
  \caption{The partial order of the power set $2^{\{a, b\}}$}
  \label{order1}
\end{figure}

\medskip
Here is the diagram associated with the partial order on the power set of 
the three element set, $\{a, b, c\}$:

\begin{figure}[H]
 \begin{center}
    \begin{pspicture}(0,0)(4,5)
    \cnode(2,0){2pt}{u1}
    \cnode(0,1.5){2pt}{u2}
    \cnode(2,1.5){2pt}{u3}
    \cnode(4,1.5){2pt}{u4}
    \cnode(0,3){2pt}{u5}
    \cnode(2,3){2pt}{u6}
    \cnode(4,3){2pt}{u7}
    \cnode(2,4.5){2pt}{u8}
    \ncline[linewidth=1pt]{->}{u1}{u2}
    \ncline[linewidth=1pt]{->}{u1}{u3}
    \ncline[linewidth=1pt]{->}{u1}{u4}
    \ncline[linewidth=1pt]{->}{u2}{u6}
    \ncline[linewidth=1pt]{->}{u2}{u7}
    \ncline[linewidth=1pt]{->}{u3}{u5}
    \ncline[linewidth=1pt]{->}{u3}{u7}
    \ncline[linewidth=1pt]{->}{u4}{u5}
    \ncline[linewidth=1pt]{->}{u4}{u6}
    \ncline[linewidth=1pt]{->}{u5}{u8}
    \ncline[linewidth=1pt]{->}{u6}{u8}
    \ncline[linewidth=1pt]{->}{u7}{u8}
    \uput[-90](2,0){$\emptyset$}
    \uput[210](0,1.5){$\{a\}$}
    \uput[-30](2,1.5){$\{b\}$}
    \uput[-30](4,1.5){$\{c\}$}
    \uput[150](0,3){$\{b,c\}$}
    \uput[30](2,3){$\{a,c\}$}
    \uput[30](4,3){$\{a,b\}$}
    \uput[90](2,4.5){$\{a, b, c\}$}
    \end{pspicture}
  \end{center}
  \caption{The partial order of the power set $2^{\{a, b, c\}}$}
  \label{order2}
\end{figure}
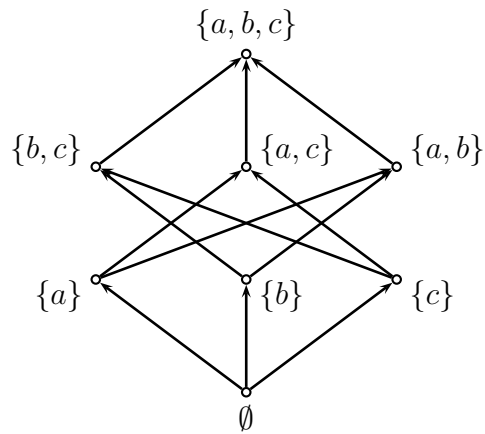

\medskip
Note that $\emptyset$ is a minimal element of the above poset
(in fact, the smallest element) and $\{a, b, c\}$ is a maximal element
(in fact, the greatest element). In the above example, there is a unique
minimal (resp. maximal) element.
A less trivial example with multiple minimal and maximal elements
is obtained by deleting $\emptyset$ and $\{a, b, c\}$: 

\begin{figure}[H]
 \begin{center}
    \begin{pspicture}(0,0)(4,2.2)
    \cnode(0,0){2pt}{u2}
    \cnode(2,0){2pt}{u3}
    \cnode(4,0){2pt}{u4}
    \cnode(0,2){2pt}{u5}
    \cnode(2,2){2pt}{u6}
    \cnode(4,2){2pt}{u7}
    \ncline[linewidth=1pt]{->}{u2}{u6}
    \ncline[linewidth=1pt]{->}{u2}{u7}
    \ncline[linewidth=1pt]{->}{u3}{u5}
    \ncline[linewidth=1pt]{->}{u3}{u7}
    \ncline[linewidth=1pt]{->}{u4}{u5}
    \ncline[linewidth=1pt]{->}{u4}{u6}
    \uput[210](0,0){$\{a\}$}
    \uput[-30](2,0){$\{b\}$}
    \uput[-30](4,0){$\{c\}$}
    \uput[150](0,2){$\{b,c\}$}
    \uput[30](2,2){$\{a,c\}$}
    \uput[30](4,2){$\{a,b\}$}
    \end{pspicture}
  \end{center}
  \caption{Minimal and maximal elements in a poset}
  \label{order3}
\end{figure}
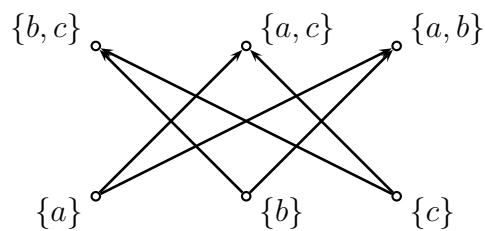

\medskip
Given a poset, $\lag X, \leq\rag$, observe that if there is some
element $m\in X$ so that $m \leq x$ for all $x\in X$, then
$m$ is unique. Indeed, if $m'$ is another element so that
$m' \leq x$ for all $x\in X$, then if we set $x = m'$ in the
first case, we get $m \leq m'$ and if we set $x = m$ in the second
case, we get $m' \leq m$, from which we deduce that $m = m'$,
as claimed. Such an element, $m$, is called the {\it smallest\/} or
the {\it least element\/} of $X$. Similarly, an element, $b\in X$,
so that $x \leq b$ for all $x\in X$ is unique and is called
the {\it greatest element\/} of $X$.

\medskip
We summarize some of our previous definitions and introduce a few more useful
concepts in

\begin{defin}
\label{partialdef2}
{\em
Let $\lag X, \leq\rag$ be a poset and let $A\subseteq X$ be any
subset of $X$. An element, $b\in X$, is a {\it lower bound of $A$\/}
iff $b \leq a$ for all $a\in A$. 
An element, $m\in X$, is an {\it upper bound of $A$\/} 
iff $a \leq m$ for all $a\in A$. 
An element, $b\in X$, is the {\it least element of $A$\/}
iff $b\in A$ and $b \leq a$ for all $a\in A$.
An element, $m\in X$, is the {\it greatest element of $A$\/}
iff $m\in A$ and $a \leq m$ for all $a\in A$. 
An element, $b\in A$, is  {\it minimal in $A$\/}  iff
$a < b$ for no $a\in A$, or equivalently, if for all
$a\in A$, $a\leq b$ implies that $a = b$.
An element, $m\in A$, is  {\it maximal in $A$\/} iff
$m < a$ for no $a\in A$, or equivalently, if for all
$a\in A$, $m\leq a$ implies that $a = m$.
An element, $b\in X$, is the  {\it greatest lower bound of $A$\/} 
iff  the set of lower bounds of $A$ is nonempty and if $b$ is the greatest 
element of this set.
An element, $m\in X$, is the  {\it least upper bound of $A$\/} 
iff  the set of upper bounds of $A$ is nonempty and if $m$ is the least 
element of this set.
}
\end{defin}

\remarks
\begin{enumerate}
\item 
If $b$ is a lower bound of $A$ (or $m$ is an upper bound of $A$),
then $b$ (or $m$) may not belong to $A$.
\item 
The least element of $A$ is a lower bound of $A$ that also belongs to $A$
and the greatest  element of $A$ is an upper bound of $A$ that also belongs to $A$.
When $A = X$, the least element is often denoted $\perp$, sometimes $0$,
and the greatest element is often denoted $\top$, sometimes $1$.
\item
Minimal or maximal elements of $A$ belong to $A$ but they are not necessarily
unique.
\item
The greatest lower bound (or the least upper bound) of $A$
may not belong to $A$. We use the notation $\bigwedge A$ for
the greatest lower bound of $A$ and the notation
$\bigvee A$ for the least upper bound of $A$. 
In computer science,
some people also use $\bigsqcup A$ instead of   $\bigvee A$ and
the symbol $\bigsqcup$  upside down instead of $\bigwedge$. 
When $A = \{a, b\}$,
we write $a\land b$ for $ \bigwedge \{a, b\}$ and $a\lor b$
for $\bigvee  \{a, b\}$. The element $a\land b$ is called the {\it meet
of $a$ and $b$\/} and $a\lor b$ is the {\it join of $a$ and $b$\/}.
(Some computer scientists use $a\sqcap b$ for $a\land b$
and $a\sqcup b$ for $a\lor b$.)
\item
Observe that if it exists, $\bigwedge \emptyset = \top$, the greatest
element of $X$ and if its exists, $\bigvee \emptyset = \>\perp$,
the least element of $X$.
Also, if it exists, $\bigwedge X = \>\perp$ and if it exists,
$\bigvee X = \top$.
\end{enumerate}

\medskip
The reader should look at the posets in Figures \ref{order2} and \ref{order3} 
for examples of the above notions.

\medskip
For the sake of completeness, we state the following fundamental
result known as Zorn's Lemma even though it is unlikely that we will
use it in this course. Zorn's lemma turns out to be equivalent to
the axiom of choice. For details and a proof, the reader is referred to
Suppes \cite{Suppes} or Enderton \cite{Endertonset}.

\begin{thm} (Zorn's Lemma)
\label{Zornlem}
Given a poset, $\lag X, \leq\rag$, if every nonempty chain in $X$ 
has an upper-bound, then
$X$ has some maximal element.
\end{thm}

\medskip
When we deal with posets, it is useful to use functions that
are order-preserving as defined next.

\begin{defin}
\label{monotone}
{\em
Given two posets $\lag X, \leq_X\rag$ and $\lag Y, \leq_Y\rag$,
a function, $\mapdef{f}{X}{Y}$, is {\it monotonic\/} (or 
{\it order-preserving\/}) iff for all $a, b\in X$,
\[
\hbox{if}\quad a \leq_X b\quad\hbox{then}\quad f(a) \leq_Y f(b).
\]
}
\end{defin}

\section{Lattices and Tarski's Fixed Point Theorem}
\label{sec15}
We now take a closer look at posets having the property that every two 
elements have a meet and a join (a greatest lower bound and a least
upper bound). Such posets occur a lot more than we think. 
A typical example is the power set under inclusion, where
meet is intersection and join is union.

\begin{defin}
\label{latticedef}
{\em
A {\it lattice\/} is a poset in which any two elements have
a meet and a join. A {\it complete lattice\/} is a poset 
in which any subset has a greatest lower bound and a least upper
bound.
}
\end{defin}

\medskip
According to part (5) of the remark just before Zorn's Lemma, observe that
a complete lattice must have a least element, $\perp$, and a greatest
element, $\top$.

\remark 
The notion of complete lattice is due to G. Birkhoff (1933).
The notion of a lattice is due to Dedekind (1897) but his
definition used properties (L1)-(L4) listed in Proposition
\ref{latticep1}. The use of meet and join in posets was first studied
by C. S. Peirce (1880).

\begin{figure}[H]
 \begin{center}
    \begin{pspicture}(0,0)(4,5)
    \cnode(2,0){2pt}{u1}
    \cnode(0,1.5){2pt}{u2}
    \cnode(2,1.5){2pt}{u3}
    \cnode(4,1.5){2pt}{u4}
    \cnode(0,3){2pt}{u5}
    \cnode(2,3){2pt}{u6}
    \cnode(4,3){2pt}{u7}
    \cnode(2,4.5){2pt}{u8}
    \ncline[linewidth=1pt]{->}{u1}{u2}
    \ncline[linewidth=1pt]{->}{u1}{u3}
    \ncline[linewidth=1pt]{->}{u1}{u4}
    \ncline[linewidth=1pt]{->}{u2}{u6}
    \ncline[linewidth=1pt]{->}{u2}{u7}
    \ncline[linewidth=1pt]{->}{u3}{u5}
    \ncline[linewidth=1pt]{->}{u3}{u7}
    \ncline[linewidth=1pt]{->}{u4}{u5}
    \ncline[linewidth=1pt]{->}{u4}{u6}
    \ncline[linewidth=1pt]{->}{u5}{u8}
    \ncline[linewidth=1pt]{->}{u6}{u8}
    \ncline[linewidth=1pt]{->}{u7}{u8}
    \uput[-90](2,0){$\emptyset$}
    \uput[210](0,1.5){$\{a\}$}
    \uput[-30](2,1.5){$\{b\}$}
    \uput[-30](4,1.5){$\{c\}$}
    \uput[150](0,3){$\{b,c\}$}
    \uput[30](2,3){$\{a,c\}$}
    \uput[30](4,3){$\{a,b\}$}
    \uput[90](2,4.5){$\{a, b, c\}$}
    \end{pspicture}
  \end{center}
  \caption{The lattice  $2^{\{a, b, c\}}$}
  \label{order4}
\end{figure}
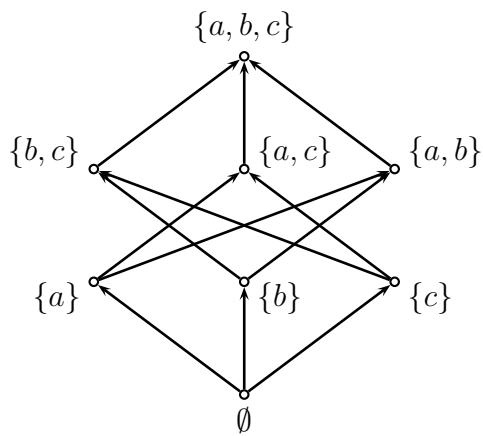

\medskip
Figure \ref{order4} shows the lattice structure of
the power set of $\{a, b, c\}$. It is actually a complete lattice.

\medskip
It is easy to show that any finite lattice is a complete lattice
and that a finite poset is a lattice iff it has a least element and a greatest
element. 

\medskip
The poset $\natnums_+$ under the divisibility ordering is a lattice!
Indeed, it turns out that the meet operation corresponds to 
{\it greatest common divisor\/} and the join operation corresponds to 
{\it least common multiple\/}. However, it is not a complete lattice.
The power set of any set, $X$,  is a complete lattice under the  subset
ordering. Indeed, one will verify immediately that for
any collection, $\s{C}$, of subsets of $X$, the least upper bound
of $\s{C}$ is its union, $\bigcup \s{C}$, and the greatest
lower bound of $\s{C}$ is its intersection, $\bigcap \s{C}$.
The least element of $2^X$ is $\emptyset$ and its greatest element
is $X$ itself.

\medskip
The following proposition gathers some useful properties
of meet and join.

\begin{prop}
\label{latticep1}
If $X$ is a lattice, then the following identities hold for all
$a, b, c\in X$: 
\begin{alignat*}{3}
& L1 \quad &   & a\lor b = b\lor a, & \qquad & a\land b = b\land a   \\
& L2 \quad &   & (a\lor b)\lor c = a\lor (b\lor c), &\qquad &
(a\land b)\land c = a\land (b\land c)\\
& L3 \quad &   & a \lor a = a, & \qquad & a \land a = a \\
& L4 \quad &   &(a\lor b)\land a = a, & \qquad & (a\land b) \lor a = a.
\end{alignat*}
Properties (L1) correspond to {\it commutativity\/}, properties (L2) to
{\it associativity\/}, properties (L3) to {\it idempotence\/} and
properties (L4) to {\it absorption\/}.
Furthermore, for all $a, b\in X$, we have
\[
a \leq b\quad\hbox{iff}\quad a\lor b = b \quad\hbox{iff}\quad
a\land b = a,
\]    
called {\it consistency\/}.
\end{prop}

\proof
The proof is left as an exercise to the reader.
$\bigsquare$

\medskip
Properties (L1)-(L4) are algebraic properties that were found by
Dedekind (1897). A pretty symmetry  reveals itself in these identities:
they all  come in pairs, one involving $\land$, the other involving $\lor$.
A useful consequence of this symmetry is {\it duality\/}, namely, that
each equation derivable from (L1)-(L4) has a dual 
statement obtained by exchanging the symbols $\land $ and $\lor$.
What is even more interesting is that it is possible to 
use these properties to define lattices. Indeed, if $X$ is a set 
together with two operations, $\land$ and $\lor$,
satisfying (L1)-(L4), we can define the relation
$a\leq b$ by $a\lor b = b$ and  then show
that $\leq$  is a partial order such that $\land$ and $\lor$ are
the corresponding meet and join. The first step is to show that
\[
a\lor b = b \quad\hbox{iff}\quad
a\land b = a.
\]
If $a\lor b = b$, then substituting $b$ for $a\lor b$ in (L4), namely
\[
(a\lor b)\land a = a,
\]
we get
\[
b\land a = a,
\]
which, by (L1), yields
\[
a\land b = a,
\]
as desired. Conversely, if $a\land b = a$, then by (L1) we have
$b\land a = a$, and substituting $a$ for
$b\land a$ in the instance of (L4) where $a$ and $b$ are switched, namely
\[
 (b\land a) \lor b = b,
\]
we get
\[
a\lor b = b,
\]
as claimed. Therefore, we can define $a\leq b$ as $a\lor b = b$
or equivalently as $a\land b = a$. 
After a little work, we obtain

\begin{prop}
\label{latticep2}
Let $X$ be a set together with two operations $\land $ and $\lor$
satisfying the axioms (L1)-(L4) of proposition \ref{latticep1}.
If we define the relation $\leq$ by $a\leq b$ iff $a\lor b = b$
(equivalently, $a\land b = a$), then $\leq$ is a partial order
and $(X, \leq)$ is a lattice whose meet and join agree with
the original operations $\land$ and $\lor$.
\end{prop}

\medskip
The following proposition shows that the existence of 
arbitrary least upper bounds (or arbitrary greatest lower bounds)
is already enough ensure that a poset is a complete lattice.

\begin{prop}
\label{complp1}
Let $\lag X, \leq\rag$ be a poset. If $X$ has a greatest element, $\top$,
and if every nonempty subset, $A$, of $X$ has a greatest lower bound,
$\bigwedge A$, then $X$ is a complete lattice. Dually, if $X$ has
a least element, $\perp$, and if every nonempty subset, $A$, 
of $X$ has a least upper bound,
$\bigvee A$, then $X$ is a complete lattice
\end{prop}

\proof
Assume  $X$ has a greatest element, $\top$,
and that every nonempty subset, $A$, of $X$ has a greatest lower bound,
$\bigwedge A$. We need to show that
any subset, $S$, of $X$ has a least upper bound.
As $X$ has a greatest element, $\top$, the set, $U$, of upper bounds
of $S$ is nonempty and so, $m = \bigwedge U$ exists.
We claim that $\bigwedge U = \bigvee S$, i.e, $m$ is the least upper bound of
$S$. First, note that every element of $S$ is a lower bound of $U$ since
$U$ is the set of upper bounds of $S$. As $m = \bigwedge U$ is
the greatest lower bound of $U$, we deduce that $s \leq m$
for all $s\in S$, i.e., $m$ is an upper bound of $S$.
Next, if $b$ is any upper bound for $S$, then
$b\in U$ and as $m$ is a lower bound of $U$ (the greatest one),
we have $m \leq b$, i.e., $m$ is the least upper bound of $S$.
The other statement is proved by duality.
$\bigsquare$

\medskip
We are now going to prove a remarkable result due to A. Tarski (discovered in 1942,
published in 1955).  A special case (for power sets)
was proved by B. Knaster (1928). First, we define fixed points.

\begin{defin}
\label{fixpoint}
{\em
Let $\lag X, \leq \rag$ be a poset and let $\mapdef{f}{X}{X}$ be a
function. An element, $x\in X$, is a {\it fixed point of $f$\/}
(sometimes spelled {\it fixpoint\/}) iff
\[
f(x) = x.
\]
An element, $x\in X$, is a {\it least (resp. greatest) fixed point of $f$\/}
if it is a fixed point of $f$ and if  $x\leq y$ (resp. $y \leq x$)
for every fixed point $y$ of $f$.
}
\end{defin}

\medskip
Fixed points play an important role in certain areas of mathematics
(for example, topology, differential equations)
and also in economics because they tend to capture the notion of
stability or equilibrium.

\medskip
We now prove the following pretty theorem due to Tarski and then
immediately proceed to use it to give a very short proof of
the Schr\"oder-Bernstein Theorem (Theorem \ref{CantorBernstein}).

\begin{thm} (Tarski's Fixed Point Theorem)
\label{Tarskithm}
Let $\lag X, \leq\rag$ be a complete lattice and
let $\mapdef{f}{X}{X}$ be any monotonic function.
Then, the set, $F$, of fixed points of $f$ is a complete lattice.
In particular, $f$ has a least fixed point,
\[
x_{\mathrm{min}} = \bigwedge \{x\in X \mid f(x) \leq x\}
\]
and a greatest fixed point
\[
x_{\mathrm{max}} = \bigvee \{x\in X \mid x \leq f(x)\}.
\]
\end{thm}

\proof
We proceed in three steps.

\medskip
{\it Step 1\/}.
We prove that $x_{\mathrm{max}}$ is the largest fixed point of $f$.

\medskip
Since $x_{\mathrm{max}}$ is an upper bound of
$A = \{x\in X \mid x \leq f(x)\}$ (the smallest one),
we have $x\leq x_{\mathrm{max}}$
for all $x\in A$. By monotonicity of $f$, we get
$f(x) \leq f(x_{\mathrm{max}})$ and since $x\in A$, we deduce
\[
x \leq f(x) \leq   f(x_{\mathrm{max}})
\quad\hbox{for all}\quad x\in A,
\]
which shows that $f(x_{\mathrm{max}})$ is an upper bound of $A$.
As  $x_{\mathrm{max}}$ is the least upper bound of $A$, we get
\begin{equation}
x_{\mathrm{max}} \leq f(x_{\mathrm{max}}).
\tag{$*$}
\end{equation}
Again, by monotonicity, from the above inequality, we get
\[
f(x_{\mathrm{max}}) \leq f(f(x_{\mathrm{max}})),
\]
which shows that $f(x_{\mathrm{max}}) \in A$.
As  $x_{\mathrm{max}}$ is an upper bound of $A$, we deduce that
\begin{equation}
f(x_{\mathrm{max}}) \leq x_{\mathrm{max}}.
\tag{$**$}
\end{equation}
But then, $(*)$ and $(**)$ yield 
\[
f(x_{\mathrm{max}}) = x_{\mathrm{max}},
\]
which shows that $x_{\mathrm{max}}$ is a fixed point of $f$.
If $x$ is any fixed point of $f$, that is, if $f(x) = x$,
we also have $x\leq f(x)$, i.e., $x\in A$. As $x_{\mathrm{max}}$
is the least upper bound of $A$, we have
$x\leq x_{\mathrm{max}}$, which proves that $x_{\mathrm{max}}$
is the greatest fixed point of $f$.

\medskip
{\it Step 2\/}.
We prove that $x_{\mathrm{min}}$ is the least fixed point of $f$.

\medskip
This proof is dual to the proof given in Step 1.

\medskip
{\it Step 3\/}.
We know that the set of fixed points, $F$, of $f$
has a least element and a greatest element, so by
Proposition \ref{complp1}, it is enough to prove that any 
nonempty subset, $S\subseteq F$, has a greatest lower bound.
If we let
\[
I = \{x\in X \mid x \leq s\quad\hbox{for all}\quad s\in S
\quad\hbox{and}\quad x \leq f(x)\},
\]
then we claim that $a = \bigvee I$ is a fixed point of $f$ and that it is
the greatest lower bound of $S$.

\medskip
The proof that  $a = \bigvee I$ is a fixed point of $f$ 
is analogous to the proof used in Step 1.
Since $a$ is an upper bound of $I$, we have
$x\leq a$ for all $x \in I$. By monotonicity of $f$ and the fact
that $x\in I$, we get
\[ 
x \leq f(x) \leq f(a).
\]
Thus, $f(a)$ is an upper bound of $I$ and so, as $a$ is the least upper bound
of $I$, we have
\begin{equation}
a \leq f(a).
\tag{$\dagger$}
\end{equation}
By monotonicity of $f$, we get $f(a) \leq f(f(a))$.
Now, to claim that $f(a)\in I$, 
we need to check that $f(a)$ is a lower bound of $S$.
However, by definition of $I$, every element of $S$ is an upper
bound of $I$ and since $a$ is the least upper bound of $I$,
we must have $a \leq s$ for all $s\in S$ i.e.,
$a$ is a lower bound of $S$.
By monotonicity of $f$ and the fact that $S$ is a set
of fixed points, we get 
\[
f(a) \leq f(s) = s,\quad\hbox{for all}\quad \in S,
\]
which shows that $f(a)$ is a lower bound of $S$
and thus, $f(a) \in I$, as contended.
As $a$ is an upper bound of $I$ and $f(a)\in I$, we must have
\begin{equation}
f(a) \leq a,
\tag{$\dagger\!\dagger$}
\end{equation}
and together with $(\dagger)$, we conclude that $f(a) = a$, i.e., $a$ is a fixed
point of $f$.

\medskip
Since we  already proved that $a$ is a lower bound of $S$ it only remains
to show that if $x$ is any fixed point of $f$ and $x$ is a lower
bound of $S$, then $x \leq a$. But, if $x$ is any fixed point of $f$
then $x \leq f(x)$  and since $x$ is also a lower bound of $S$, then
$x\in I$. As $a$ is an upper bound of $I$, we do get
$x \leq a$.
$\bigsquare$

\medskip
It should be noted that the least upper bounds and the
greatest lower bounds in $F$ do not necessarily agree with those in $X$.
In technical terms, $F$ is generally not a sublattice of $X$.

\medskip
Now, as promised, we use Tarski's Fixed Point Theorem to prove the 
Schr\"oder-Bernstein Theorem.

\bigskip
{\bf Theorem$\>$ \ref{CantorBernstein}} \  
{\it Given any two sets, $A$ and $B$, if there is an injection from
$A$ to $B$ and an injection from $B$ to $A$, then there is
a bijection between $A$ and $B$.\/} 

\medskip\noindent
{\it Proof\/}.
Let $\mapdef{f}{A}{B}$ and $\mapdef{g}{B}{A}$ be two injections.
We define the function, $\mapdef{\varphi}{2^A}{2^A}$, by
\[
\varphi(S) = A - g(B - f(S)),
\]
for any $S\subseteq A$. Because of the two complementations,
it is easy to check that $\varphi$ is monotonic (ckeck it).
As $2^A$ is a complete lattice, by Tarski's fixed point theorem, the
function $\varphi$ has a fixed point, that is, there is some
subset $C\subseteq A$ so that
\[
C = A - g(B - f(C)).
\]
By taking the complement of $C$ in $A$, we get
\[
A - C = g(B - f(C)).
\]
Now, as $f$ and $g$ are injections, the restricted functions
$\mapdef{f\res C}{C}{f(C)}$ and \\
$\mapdef{g\res (B - f(C))}{(B - f(C))}{(A - C)}$  are bijections.
Using these functions, we define the function, $\mapdef{h}{A}{B}$, as follows:
\[
h(a) = \cases{
f(a) & if $a\in C$ \cr
(g\res (B - f(C))^{-1}(a) & if $a \notin C$.\cr
}
\]
The reader will check that $h$ is indeed a bijection.
$\bigsquare$

\medskip
The above proof is probably the shortest known proof
of the Schr\"oder-Bernstein Theorem because it uses Tarski's
fixed point theorem, a  powerful result. If one looks
carefully at the proof, one realizes that there are two crucial 
ingredients:
\begin{enumerate}
\item
The set $C$ is closed under $g\circ f$, that is, 
$g\circ f(C) \subseteq C$.
\item
$A - C \subseteq g(B)$.
\end{enumerate}

\medskip
Using these observations, it is possible to give a proof that circumvents 
the use of Tarski's theorem. Such a proof is given in Enderton 
\cite{Endertonset}, Chapter 6, and we give a sketch of this proof below.

\medskip
Define a sequence of subsets, $C_n$, of $A$ by recursion as follows:
\begin{eqnarray*}
C_0 & = & A - g(B) \\
C_{n+1} & = & (g\circ f)(C_n),
\end{eqnarray*}
and set 
\[
C = \bigcup_{n \geq 0} C_n.
\]
Clearly, $A - C \subseteq g(B)$ and since direct images
preserve unions, $(g\circ f)(C) \subseteq C$.
The definition of $h$ is similar to the one used in  our proof:
\[
h(a) = \cases{
f(a) & if $a\in C$ \cr
(g\res (A - C))^{-1}(a) & if $a \notin C$.\cr
}
\]
When $a\notin C$, i.e., $a\in A - C$, as $A - C \subseteq g(B)$
and $g$ is injective, $g^{-1}(a)$ is indeed well-defined. 
As $f$ and $g$ are injective,
so is $g^{-1}$ on $A - C$. So, to check that $h$ is injective, 
it is enough to prove that
$f(a) = g^{-1}(b)$ with $a\in C$ and $b\notin C$ is impossible.
However, if $f(a) = g^{-1}(b)$, then
$(g\circ f)(a) = b$. Since $(g\circ f)(C) \subseteq C$
and $a\in C$, we get $b = (g\circ f)(a)\in C$, yet
$b\notin C$, a contradiction.  
It is not hard to verify that $h$ is surjective and therefore,
$h$ is a bijection between $A$ and $B$.
$\bigsquare$

\medskip
The classical reference on lattices is Birkhoff
\cite{Birkhoff}. We highly recommend this beautiful book
(but it is not easy reading!).

\medskip
We now turn to special properties of partial orders 
having to do with induction.

\section{Well-Founded Orderings and Complete Induction}
\label{sec16}
Have you ever wondered why induction on $\natnums$ actually ``works''?
The answer, of course, is that $\natnums$ was defined in such a way
that, by Theorem \ref{natinduc}, it is the ``smallest'' inductive set!
But this is not a very illuminating answer. The key point is that
every nonempty subset of $\natnums$ has a least element.
This fact is intuitively clear since if we had  some nonempty
subset of $\natnums$ with no smallest element, then we could
construct an infinite strictly decreasing sequence,
$k_0 > k_1> \cdots > k_n > \cdots$. But this is absurd, as
such a sequence would  eventually run into $0$ and stop.
It turns out that the deep reason why induction ``works''
on a poset is indeed that the poset ordering has a very special property
and this leads us to the following definition:

\begin{defin}
\label{wellord}
{\em
Given a poset, $\lag X, \leq\rag$, we say that $\leq$ is a 
{\it well-order (well ordering)\/} and that $X$ is {\it well-ordered by $\leq$\/}
iff every nonempty subset of $X$ has a least element.
}
\end{defin}

\medskip
When $X$ is nonempty, if we pick any two-element subset,
$\{a, b\}$, of $X$, since the subset $\{a, b\}$ must have
a least element, we see that either $a \leq b$ or $b\leq a$,
i.e., {\it every well-order is a total order\/}.
First, let us confirm that $\natnums$ is indeed well-ordered.

\begin{thm} (Well-Ordering of $\natnums$)
\label{natwellord}
The set of natural numbers, $\natnums$, is well-ordered.
\end{thm}

\proof
Not surprisingly we use induction, but we have to be a little
shrewd. Let $A$ be any nonempty subset of $\natnums$.
We prove by contradiction that $A$ has a least element.
So, suppose $A$ does not have a least element and let $P(m)$
be the predicate
\[
P(m) \equiv (\forall k\in \natnums)(k < m \impl k\notin A),
\]
which says that no natural number strictly smaller than $m$ is in $A$.
We will prove by induction on $m$ that $P(m)$ holds. But then,
the fact that $P(m)$ holds for all $m$ shows that $A = \emptyset$,
a contradiction.

\medskip
Let us now prove $P(m)$ by induction. The base case $P(0)$  holds 
trivially. Next, assume $P(m)$ holds; we want to prove that
$P(m+1)$ holds. Pick any $k < m + 1$. Then, either
\begin{enumerate}
\item[(1)]
$k < m$, in which case, by the induction hypothesis, $k\notin A$;
or
\item[(2)]
$k = m$. By the induction hypothesis, $P(m)$ holds. Now, if $m$ was in $A$, 
as $P(m)$ holds no $k < m$ would belong to $A$ and $m$ would
be the least element of $A$, contradicting the assumption that
$A$ has no least element. Therefore, $m\notin A$.
\end{enumerate}
Thus, in both cases, we proved that if $k < m + 1$, then
$k \notin A$, establishing the induction hypothesis.
This concludes the induction and the proof of Theorem \ref{natwellord}.
$\bigsquare$

\medskip
Theorem \ref{natwellord} yields another induction principle
which is often more flexible that our original induction
principle. This principle, called {\it complete induction\/}
(or sometimes {\it strong induction\/}), is stated below.
Recall that $\natnums_+ = \natnums - \{0\}$.

\medskip\noindent
{\bf Complete Induction Principle on $\natnums$}.

\medskip
In order to prove that a predicate, $P(n)$, holds for all $n\in \natnums$
it is enough to prove that
\begin{enumerate}
\item[(1)] 
$P(0)$ holds (the base case) and
\item[(2)]
for every $m\in \natnums_+$, 
if $(\forall k\in \natnums) (k < m \impl P(k))$  then $P(m)$.
\end{enumerate}
As a formula, complete induction is stated as
\[
P(0) \land (\forall m\in \natnums_+)[(\forall k\in \natnums) 
(k < m \impl P(k)) \impl P(m)] \impl
(\forall n\in \natnums) P(n).
\]

\medskip
The difference between ordinary induction and complete induction is that
in complete induction, the induction hypothesis,
$(\forall k \in \natnums)(k < m \impl P(k))$, assumes that $P(k)$ holds for
all $k < m$ and not just for $m - 1$ (as in ordinary induction),
in order to deduce $P(m)$. This gives us more proving power as
we have more knowledge in  order to prove $P(m)$.

\medskip
We will have many occasions to use complete induction but let us first
check that it is a valid principle. Even though we will
give a more general proof of the validity of complete
induction for a well-ordering, we feel that it will be helpful to give 
the proof in the case of $\natnums$ as a warm-up.

\begin{thm}
\label{natcomplind}
The complete induction principle for $\natnums$ is valid.
\end{thm}

\proof
Let $P(n)$ be a predicate on $\natnums$ and  assume that
$P(n)$ satisfies conditions (1) and (2) of complete induction
as stated above. We proceed by contradiction. So, assume
that $P(n)$ fails for some $n\in \natnums$. If so, the set
\[
F = \{n\in \natnums \mid P(n) = {\bf false}\}
\]
is nonempty. By Theorem \ref{natwellord}, the set $A$ has a least element,
$m$, and thus,  
\[
P(m) = {\bf false}.
\]
Now, we can't have $m = 0$, as we assumed that $P(0)$ holds (by (1))
and since $m$ is the least element for which $P(m) = {\bf false}$,
we must have
\[
P(k) = {\bf true}\quad\hbox{for all}\quad k < m.
\]
But, this is exactly the premise in (2) and as
we assumed that (2) holds,  we deduce that
\[
P(m) = {\bf true},
\]
contradicting the fact that we already know that $P(m) = {\bf false}$.
Therefore, $P(n)$ must hold for all $n\in\natnums$.
$\bigsquare$

\remark
In our statement of the principle of complete induction, we singled out the
base case, (1), and consequently, we stated the induction step (2)
for  every $m\in \natnums_+$,  excluding the case $m = 0$, which
is already covered by the base case. It is also possible to
state  the principle of complete induction in a more concise fashion 
as follows:
\[
 (\forall m\in \natnums)[(\forall k\in \natnums) 
(k < m \impl P(k)) \impl P(m)] \impl
(\forall n\in \natnums) P(n).
\]
In the above formula, observe that when $m = 0$, which is
now allowed, the premise \\
$(\forall k\in \natnums)(k < m \impl P(k))$ of the implication within the
brackets is trivially true and so,
$P(0)$ must still be established.
In the end, exactly the same amount of work is required
but some people prefer the second more concise version
of the principle of complete induction.
We feel that it would be easier for the reader to make  the transition
from ordinary induction to complete induction if we make explicit
the fact that the base case must be established.

\medskip
Let us illustrate the use of the complete induction principle
by proving that every natural number factors as a product of primes.
Recall that for any two natural numbers, $a, b\in \natnums$
with $b\not= 0$, we say that {\it $b$ divides $a$\/} iff
$a = bq$, for some $q\in \natnums$. In this case, we say that {\it $a$ is
divisible by $b$\/} and that {\it $b$ is a  factor of $a$\/}.
Then, we say that a natural 
number, $p\in \natnums$, is a {\it prime number\/} (for short,
a {\it prime\/}) if $p \geq 2$ and if $p$ is only divisible
by itself and by $1$. Any prime number but $2$ must be odd
but the converse is false.
For example, $2, 3, 5, 7, 11, 13, 17$
are prime numbers, but $9$ is not. It can be shown that
there are infinitely many prime numbers but to prove this,
we need the following Theorem:

\begin{thm}
\label{primedecomp}
Every natural number, $n\geq 2$ can be factored as
a product of primes, that is, $n$ can be written as a product, 
$n = p_1^{m_1}\cdots p_k^{m_k}$,
where the $p_i$s  are pairwise distinct  prime numbers
and $m_i \geq 1\>$  ($1\leq i \leq k$).
\end{thm}

\proof
We proceed by complete induction on $n \geq 2$.
The base case, $n = 2$ is trivial, since $2$ is prime.

\medskip
Consider any $n > 2$ and assume that the induction hypothesis
holds, that is, every $m$ with $2 \leq  m < n$
can be factored as a product of primes. There are two cases:
\begin{enumerate}
\item[(a)]
The number $n$ is prime. Then, we are done.
\item[(b)]
The number $n$ is not a prime. In this case, $n$ factors
as $n = n_1n_2$, where $2 \leq n_1, n_2 < n$.
By the induction hypothesis,  $n_1$ has some prime
factorization and so does  $n_2$.
If $\{p_1, \ldots, p_k\}$
is the union of all the primes occurring in these factorizations
of $n_1$ and $n_2$, we can write
\[
n_1 = p_1^{i_1}\cdots p_k^{i_k}
\quad\hbox{and}\quad
n_2 = p_1^{j_1}\cdots p_k^{j_k},
\]
where $i_h, j_h \geq 0$ and, in fact, $i_h + j_h \geq 1$,
for $1 \leq h \leq k$. Consequently,
$n$ factors as the product of primes, 
\[
n =  p_1^{i_1+ j_1}\cdots p_k^{i_k+ j_k},
\]
with  $i_h + j_h \geq 1$,
establishing the induction hypothesis. $\bigsquare$
\end{enumerate}

\remark
It can be shown that the prime factorization 
of a natural number is unique
up to permutation of the primes $p_1, \ldots, p_k$
but this requires the Euclidean Division Lemma.
However, we can prove right away that there are infinitely
primes.

\begin{thm}
\label{primeinf}
Given any natural number, $n \geq 1$, there is a prime
number, $p$, such that $p > n$. Consequently, there are 
infinitely many primes.
\end{thm}

\proof
Let $m = n! + 1$. If $m$ is prime, we are done.
Otherwise, by Theorem \ref{primedecomp}, the number
$m$ has a prime decomposition. We claim  that $p > n$
for every prime in this decomposition. If not,
$2 \leq p \leq n$ and then $p$ would divide both
$n! + 1$ and $n!$, so $p$ would
divide $1$, a contradiction.  
$\bigsquare$

\medskip
As an application of Theorem \ref{natwellord}, we prove the
``Euclidean Division Lemma'' for the integers.

\begin{thm} (Euclidean Division Lemma for $\integs$)
\label{divthm}
Given any two integers, $a, b\in\integs$, with $b\not = 0$, there 
is some unique integer, $q\in \integs$ (the quotient), and 
some unique natural number, $r\in \natnums$ 
(the remainder or residue), so that
\[
a = bq + r
\quad\hbox{with}\quad 0 \leq r <  |b|.
\]
\end{thm}

\proof
First, let us prove the existence of $q$ and $r$ with the required
condition on $r$.
We claim that if we show existence in the
special case where $a, b\in \natnums$ (with $b\not= 0$), then
we can prove existence in the general case. There are four cases:
\begin{enumerate}
\item
If $a, b\in \natnums$, with $b\not= 0$, then we are done.
\item
If $a \geq 0$ and $b < 0$, then $-b > 0$, so we know that
there exist $q, r$ with
\[
a = (-b)q + r
\quad\hbox{with}\quad 0 \leq r \leq  -b - 1.
\]
Then,
\[
a = b(-q) + r
\quad\hbox{with}\quad 0 \leq r \leq  |b| - 1.
\]
\item
If $a < 0$ and $b > 0$, then $-a > 0$, so we 
know that
there exist $q, r$ with
\[
-a = bq + r
\quad\hbox{with}\quad 0 \leq r \leq  b - 1.
\]
Then,
\[
a = b(-q) - r
\quad\hbox{with}\quad 0 \leq r \leq  b - 1.
\]
If $r = 0$, we are done. Otherwise, $1 \leq r \leq b - 1$,
which implies $1 \leq b - r \leq b - 1$, so we get
\[
a = b(-q) - b + b - r = b(-(q+1)) + b - r
\quad\hbox{with}\quad 0 \leq b - r \leq  b - 1.
\]
\item
If $a < 0$ and $b <0$, then $-a > 0$  and $-b > 0$, so we 
know that
there exist $q, r$ with
\[
-a = (-b)q + r
\quad\hbox{with}\quad 0 \leq r \leq  -b - 1.
\]
Then,
\[
a = bq - r
\quad\hbox{with}\quad 0 \leq r \leq  -b - 1.
\]
If $r = 0$, we are done. Otherwise, $1 \leq r \leq -b - 1$,
which implies $1 \leq -b - r \leq -b - 1$, so we get
\[
a = bq + b - b - r = b(q+1) + (-b - r)
\quad\hbox{with}\quad 0 \leq -b - r \leq  |b| - 1.
\]
\end{enumerate}
We are now reduced to proving the existence of $q$  and $r$
when $a, b\in \natnums$ with $b\not= 0$.
Consider the set
\[
R = \{a - bq  \in \natnums \mid q\in \natnums\}.
\]
Note that $a\in R$, by setting $q = 0$, since $a\in \natnums$.
Therefore, $R$ is nonempty. By Theorem \ref{natwellord}, 
the nonempty set, $R$, has a least element, $r$.
We claim that $r \leq b - 1$ (of course, $r \geq 0$ as
$R \subseteq \natnums$). 
If not, then $r\geq b$, and so $r - b \geq 0$.
As $r\in R$, there is some
$q\in \natnums$ with $r = a - bq$. But now, we have
\[
r - b = a - bq - b = a - b(q+1)
\]
and as $r - b \geq 0$, we see that $r - b \in R$
with $r - b < r$ (since $b\not= 0$), contradicting the
minimality of $r$. Therefore, $0 \leq r \leq b -1$,
proving the existence of $q$ and $r$ with the required 
condition on $r$.

\medskip
We now go back to the general case where $a, b\in \integs$
with $b\not= 0$ and we prove uniqueness of $q$  and $r$ (with
the required condition on $r$).
So, assume that
\[
a = bq_1 + r_1 = bq_2 + r_2
\quad\hbox{with}\quad 0 \leq r_1 \leq  |b| - 1
\quad\hbox{and}\quad 0 \leq r_2 \leq  |b| - 1.
\]
Now, as $0 \leq r_1 \leq  |b| - 1$ and $0 \leq r_2 \leq  |b| - 1$,
we have $|r_1 - r_2| < |b|$, and from
$bq_1 + r_1 = bq_2 + r_2$, we get
\[
b(q_2 - q_1) = r_1 - r_2,
\]
which yields
\[
|b||q_2 - q_1| = |r_1 - r_2|.
\]
Since $|r_1 - r_2| < |b|$, we must have $r_1 = r_2$.
Then, from $b(q_2 - q_1) = r_1 - r_2 = 0$, 
as $b\not= 0$, we get $q_1 = q_2$,
which concludes the proof.
$\bigsquare$

\medskip
We will now show that complete induction holds for a very broad
class of partial orders called {\it well-founded orderings\/}
that subsume well-orderings.

\begin{defin}
\label{wellordef}
{\em
Given a poset, $\lag X, \leq \rag$, we say that $\leq$ is a 
{\it well-founded ordering (order)\/} and that  $X$
is {\it well-founded\/} iff $X$ has {\bf no}
infinite strictly decreasing sequence \\
$x_0 > x_1 > x_2 > \cdots > x_n > x_{n+1} > \cdots$.
}
\end{defin}

\medskip
The following property of well-founded sets is fundamental:

\begin{prop}
\label{wellfound1}
A poset, $\lag X, \leq \rag$, is well-founded iff every nonempty
subset of $X$ has a minimal element.
\end{prop}

\proof
First, assume that every nonempty
subset of $X$ has a minimal element. If we had an 
infinite strictly decreasing sequence,
$x_0 > x_1 > x_2 > \cdots > x_n > \cdots$,
then the set $A = \{x_n\}$ would have no minimal element,
a contradiction. Therefore, $X$ is well-founded.

\medskip
Now, assume that $X$ is well-founded. We prove
that $A$ has a least element by contradiction.
So, let $A$ be some nonempty subset of $X$ and suppose $A$ has no
least element. This means that for every $a\in A$, there is some
$b\in A$ with $a > b$. Using the Axiom of Choice (Graph Version),
there is some function, $\mapdef{g}{A}{A}$, with the property that
\[
a > g(a), \quad\hbox{for all}\quad a\in A.
\]
Now, since $A$ is nonempty, we can pick some element, say $a\in A$.
By the recursion Theorem (Theorem \ref{recnat}), there is a unique
function, $\mapdef{f}{\natnums}{A}$, so that
\begin{eqnarray*}
f(0) & = & a, \\
f(n+1) & = & g(f(n))\qquad\hbox{for all}\quad n \in \natnums.
\end{eqnarray*}
But then, $f$ defines an infinite sequence, $\{x_n\}$,
with $x_n = f(n)$, so that $x_n > x_{n+1}$ for all $n\in \natnums$,
contradicting the fact that $X$ is well-founded.
$\bigsquare$

\medskip
So, the seemingly weaker condition that there is  {\bf no}
infinite strictly decreasing sequence in $X$ is equivalent to the fact
that every nonempty subset of $X$ has a minimal element.
If $X$ is a total order, any minimal element  is actually
a least element and so, we get

\begin{cor}
\label{wellfound2}
A poset, $\lag X, \leq \rag$, is well-ordered iff $\leq$ is 
total  and $X$ is well-founded.
\end{cor}

\medskip
Note that the notion of a well-founded set is more general than
that of a well-ordered set, since a well-founded set is
not necessarily totally ordered. 

\remark
Suppose we can prove some property, $P$, by (ordinary) induction on $\natnums$.
Then, I claim that $P$ can also be proved by complete induction on $\natnums$.
To see this, observe first that the base step is identical.  Also,
for all $m\in \natnums_+$, the implication
\[
(\forall k\in \natnums) 
(k < m \impl P(k))  \impl P(m - 1)
\]
holds and since the induction step (in ordinary induction) consists 
in proving for all $m\in \natnums_+$ that
\[
P(m - 1) \impl P(m)
\]
holds, from this  implication and the previous implication we deduce that
for all $m\in \natnums_+$, the implication
\[
(\forall k\in \natnums) 
(k < m \impl P(k)) \impl P(m)
\]
holds, which is exactly the induction step of the
complete induction method.
So, we see that complete induction on $\natnums$
implies ordinary induction  on $\natnums$. The converse is also
true but we leave it as a fun exercise. 
But now, by Theorem \ref{natwellord}, (ordinary) induction on 
$\natnums$ implies that $\natnums$ is well-ordered and by
Theorem \ref{natcomplind}, the fact that  
$\natnums$ is well-ordered implies complete induction on $\natnums$.
Since we just showed that  complete induction on $\natnums$
implies  (ordinary) induction on $\natnums$, we conclude that
all three are equivalent, that is
\begin{center}
(ordinary) induction on $\natnums$ is valid \\
iff \\
 complete induction on $\natnums$ is valid\\
iff\\
$\natnums$ is well-ordered.
\end{center}
These equivalences justify our earlier claim that the ability 
to do induction hinges on some key property of the ordering,
in this case, that it is a well-ordering.

\medskip
We finally come to the principle of {\it complete induction\/}
(also called {\it transfinite induction\/} or {\it structural induction\/)},
which, as we shall prove, is valid for all well-founded sets.
Since every well-ordered set is also well-founded, 
complete induction is a very general induction method.

\medskip
Let $(X, \leq)$ be a well-founded poset and let $P$ be a predicate on $X$
(i.e., a function $\mapdef{P}{X}{\{\mathbf{true}, \mathbf{false}\}}$).

\bigskip\noindent
{\bf Principle of Complete Induction on a Well-Founded Set}.

\medskip 
To prove that a property $P$ holds for all $z\in X$, it suffices to show that,
for every $x\in X$, 
\begin{enumerate}
\item[$(*)$] 
if $x$ is minimal or $P(y)$ holds for all $y<x$, 
\item[$(**)$]
then $P(x)$ holds.
\end{enumerate}

\medskip 
The statement $(*)$ is called the {\it induction hypothesis\/}, and 
the implication  

\medskip 
for all $x$, $(*)$ implies $(**)$ 
is called the {\it induction step\/}.  Formally, the induction principle
can be stated as:
\begin{equation}
(\forall x\in X)[(\forall y\in X)(y<x \impl  P(y)) \impl  P(x)] 
\impl  (\forall z\in X)P(z)
\tag{CI}
\end{equation}
Note that if $x$ is minimal, then there is no $y\in X$ such that
$y<x$, and \\
$(\forall y\in X)(y<x \impl  P(y))$ is true. Hence, we must show
that $P(x)$ holds for every minimal element, $x$.
These cases are called the {\it base cases\/}.

\medskip
Complete induction is not valid for  arbitrary posets (see the problems)
but holds for well-founded sets as shown in the following theorem.

\begin{thm}
\label{wellfthm}
The principle of complete induction 
holds for  every well-founded set.
\end{thm}
 
\proof
We proceed by contradiction. Assume that $(CI)$ is false.
Then, 
\begin{equation}
(\forall x\in X)[(\forall y\in X)(y<x \impl  P(y)) \impl  P(x)]
\tag{1}
\end{equation}
holds and
\begin{equation}
(\forall z\in X)P(z)
\tag{2}
\end{equation}
is false, that is, there is some $z\in X$ so that
\[
P(z)={\bf false}.
\]
Hence, the subset $F$ of $X$ defined by 
\[
F=\{x\in X \mid P(x)={\bf false}\}
\] 
is nonempty. Since $X$ is well founded, by Proposition \ref{wellfound1},
$F$ has some
minimal element, $b$. Since (1) holds for all $x\in X$,
letting $x=b$, we see that
\begin{equation}
[(\forall y\in X)(y<b \impl  P(y)) \impl  P(b)]
\tag{3} 
\end{equation}
holds.
If $b$ is also minimal in $X$, then there is no $y\in X$ such that $y<b$
and so,  
\[
(\forall y\in X)(y<b \impl  P(y))
\]
holds trivially and 
(3) implies that 
$P(b)={\bf true}$, which contradicts the fact that $b\in F$. 
Otherwise, for every $y\in X$ such that $y<b$,
$P(y)={\bf true}$, since otherwise $y$ would belong to $F$ 
and $b$ would not be minimal.
But then,  
\[
(\forall y\in X)(y<b \impl P(y))
\] 
also holds and (3) implies
that $P(b)={\bf true\/}$, contradicting the fact that $b\in F$.
Hence, complete induction is valid for well-founded sets.
$\bigsquare$

\medskip 
As an illustration of well-founded sets, we define the {\it lexicographic
ordering\/} on pairs.
Given a partially ordered set $\lag X,\leq\rag$, the {\it lexicographic
ordering\/}, $<<$, on $X \times X$ induced by $\leq$  is defined a follows:
For all $x,y,x',y'\in X$, 
\[
(x,y)<<(x',y') \quad\hbox{iff either}
\]
\begin{align*}
& x=x'\quad\hbox{and}\quad y=y'\quad \hbox{or}\\
& x < x' \quad \hbox{or}\\
& x=x' \quad\hbox{and}\quad  y<y'.
\end{align*}

\medskip
We leave it as an exercise to check that $<<$ is indeed a partial order on 
$X \times X$.
The following proposition will be useful.

\begin{prop}
\label{lexicop1}
If $\lag X,\leq\rag$ is a well-founded 
set, then the lexicographic ordering
$<<$ on $X \times X$ is also well founded.
\end{prop}

\proof
We proceed by contradiction. Assume that there is an
infinite decreasing  sequence $(\lag x_{i},y_{i}\rag)_i$ in 
$X \times X$. Then,
either,
\begin{enumerate}
\item[(1)] 
There is an infinite number of distinct $x_{i}$, or

\medskip 
\item[(2)] 
There is only a finite number of distinct $x_{i}$.
\end{enumerate}

\medskip 
In case (1), the subsequence consisting of these distinct elements forms
a decreasing sequence in $X$, contradicting the fact that $\leq$  is
well founded. In case (2), there is some $k$ such that 
$x_{i}=x_{i+1}$,  for all $i\geq k$. 
By definition of $<<$, the sequence $(y_{i})_{i\geq k}$ is
a decreasing sequence in $X$, 
contradicting the fact that $\leq$  is well founded. 
Hence, $<<$ is well founded on $X \times X$.
$\bigsquare$

\medskip 
As an illustration of the principle of complete induction, consider
the following example in which it is shown that a function defined recursively
is a total function.

\medskip
{\bf Example} (Ackermann's function) 
The following 
function,  $\mapdef{A}{\natnums\times\natnums}{\natnums}$,
known as {\it Ackermann's function\/} is well known in
recursive function theory for its extraordinary rate of growth. It is
defined recursively as follows:
\begin{align*}
A(x,y)  = &\>\> \mathbf{if}\>  x=0\>  \mathbf{then}\> y+1 \\
& \>\>  \mathbf{else}\>   \mathbf{if}\>  y=0 \>  \mathbf{then}\>   A(x-1,1)\\
&\>\>    \mathbf{else} \> A(x-1,A(x,y-1)). 
\end{align*}

\medskip
We wish to prove that
$A$ is a total function. We proceed by complete induction over the
lexicographic ordering on   $\natnums \times \natnums$.

\begin{enumerate}
\item
The base case is $x=0$, $y=0$. In this case, since $A(0,y)=y+1$,
$A(0,0)$ is defined and equal to 1. 
\item 
The induction hypothesis is that for any $(m,n)$, $A(m',n')$ is defined
for all \\
$(m',n')<<(m,n)$, with $(m,n)\not= (m',n')$. 
\medskip
\item 
For the induction step, we  have three cases:
\begin{enumerate}
\item
If $m=0$, since $A(0,y)=y+1$, $A(0,n)$ is defined and equal to $n+1$.
\item
If $m\not= 0$ and $n=0$, since $(m-1,1)<<(m,0)$ and $(m-1,1)\not= (m,0)$,
by the induction hypothesis, $A(m-1,1)$ is defined, and so $A(m,0)$ is
defined since it is equal to $A(m-1,1)$.
\item
If $m\not= 0$ and $n\not= 0$, since $(m,n-1)<<(m,n)$ and 
$(m,n-1)\not= (m,n)$,
by the induction hypothesis, $A(m,n-1)$ is defined. Since
$(m-1,y)<<(m,z)$ and $(m-1,y)\not= (m,z)$ no matter what $y$ and $z$ are, \\
$(m-1,A(m,n-1))<<(m,n)$ and $(m-1,A(m,n-1))\not= (m,n)$, and by the induction
hypothesis, $A(m-1,A(m,n-1))$ is defined. But this is precisely $A(m,n)$, 
and so
$A(m,n)$ is defined. This concludes the induction step.
 \end{enumerate}
\end{enumerate}
Hence, $A(x,y)$ is defined for all $x,y\geq 0$.
$\bigsquare$

\section{Unique Prime Factorization in $\integs$ and GCD's}
\label{sec16b}
In the previous section, we proved that every natural number,
$n\geq 2$, can be factored as a product of primes numbers. 
In this section, we use the Euclidean Division Lemma to prove that
such a factorization is unique. For this, we need to introduce
greatest common divisors (gcd's) and prove some of their
properties. 

\medskip
In this section, it will be convenient to allow $0$ to be a divisor. 
So, given any two integers, $a, b\in \integs$, we will say that
{\it $b$ divides $a$ and that $a$ is a multiple of $b$\/} iff
$a = bq$, for some $q\in\integs$. Contrary to our previous definition,
$b = 0$ is allowed as a divisor. However, this changes very little because
if $0$ divides $a$, then $a = 0q = 0$, that is, {\it the only
integer divisible by $0$ is $0$\/}.

\medskip
We begin by introducing a very important notion in algebra,
that of an ideal, and prove a fundamental property of 
the ideals of $\integs$.

\begin{defin}
\label{idealdef}
{\em
An {\it ideal of $\integs$\/} is
any nonempty subset, $\ideal{I}$, of $\integs$ satisfying the following two
properties:
\begin{enumerate}
\item[(ID1)]
If $a, b\in \ideal{I}$, then $b - a\in \ideal{I}$.
\item[(ID2)]
If $a\in \ideal{I}$, then $ak\in \ideal{I}$ for every $k\in \integs$.
\end{enumerate}
An ideal, $\ideal{I}$, is a {\it principal ideal\/} if there is some
$a\in \ideal{I}$, {\it called a generator\/}, such that \\
$\ideal{I} = \{ak \mid k\in \integs\}$.
The equality $\ideal{I} = \{ak\mid k\in \integs\}$
is also written as $\ideal{I} = a\integs$ or as $\ideal{I} = (a)$.
The ideal $\ideal{I} = (0) = \{0\}$ is called the {\it null ideal\/}.
}
\end{defin}

Note that if $\ideal{I}$ is an ideal, then $\ideal{I} = \integs$
iff $1\in \ideal{I}$. Since by definition,
an ideal $\ideal{I}$ is nonempty, there
is some $a\in\ideal{I}$, and by (ID1) we get $0 = a-a\in\ideal{I}$.
Then, for every $a\in\ideal{I}$, since $0\in\ideal{I}$, by (ID1) 
we get $-a\in \ideal{I}$.

\begin{thm}
\label{ZPID}
Every ideal, $\ideal{I}$, of $\integs$, is a principal
ideal, i.e., $\ideal{I} = m\integs$ for some unique $m\in \natnums$,
with $m > 0$ iff $\ideal{I} \not= (0)$.
\end{thm}

\proof
Note that $\ideal{I} = (0)$ iff $\ideal{I} = 0\integs$ and
the theorem holds in this case.
So, assume that $\ideal{I}\not = (0)$. Then, our previous observation
that $-a\in \ideal{I}$ for every $a\in \ideal{I}$
implies that some positive integer belongs to $\ideal{I}$ and so,
the set $\ideal{I}\cap \natnums_+$ is nonempty. As
$\natnums$ is well-ordered, this set has a smallest element,
say $m > 0$. We claim that $\ideal{I} = m\integs$.

\medskip
As $m\in \ideal{I}$, by (ID2),  $m\integs\subseteq \ideal{I}$.
Conversely, pick any $n \in \ideal{I}$. By the Euclidean
division Theorem, there are unique $q\in\integs$ and
$r\in\natnums$ so that
$n = mq + r$, with $0\leq r < m$.
If $r > 0$, 
since $m\in \ideal{I}$, by (ID2), $mq\in \ideal{I}$ and by
(ID1), we get $r = n - mq \in \ideal{I}$. Yet
$r < m$, contradicting the minimality of $m$. Therefore,
$r = 0$, so $n = mq\in m\integs$, establishing
that $\ideal{I} \subseteq m\integs$ and thus,
$\ideal{I} = m\integs$, as claimed.
As to uniqueness, clearly $(0) \not= m\integs$ if $m \not= 0$,
so assume $m\integs = m'\integs$, with $m > 0$ and $m' > 0$.
Then, $m$ divides $m'$ and $m'$ divides $m$, but we already proved 
earlier that this implies $m = m'$.
$\bigsquare$

\medskip
Theorem \ref{ZPID} is often phrased: $\integs$ is a
{\it principal ideal domain\/}, for short, a {\it PID\/}.
Note that the natural number $m$ such that $\ideal{I} = m\integs$ 
is a divisor of every element in $\ideal{I}$.

\begin{cor}
\label{ZPID2}
For any two integers, $a, b\in \integs$, 
there is a unique natural number, $d\in \natnums$, 
and some integers,
$u, v\in\integs$, so that $d$ divides both $a$ and $b$ and
\[
ua + vb = d.
\]
(The above is called the Bezout identity.)
Furthermore, $d = 0$ iff $a = 0$ and $b = 0$.
\end{cor}

\proof
It is immediately verified that
\[
\ideal{I} = \{ha + kb \mid h, k\in \integs\}
\]
is an ideal of $\integs$ with  $a, b \in \ideal{I}$.
Therefore, by Theorem \ref{ZPID}, there
is a unique, $d\in \natnums$, so that $\ideal{I} = d\integs$.
We already observed that $d$ divides every number in $\ideal{I}$ 
so, as $a, b\in \ideal{I}$, we see that $d$ divides $a$ and $b$.
If $d = 0$, as $d$ divides $a$ and $b$, we must have $a = b = 0$.
Conversely, if $a = b = 0$, then $d = ua + bv = 0$.
$\bigsquare$

\medskip
The natural number  $d$ of corollary \ref{ZPID2} 
divides both $a$ and $b$.
Moreover,  every divisor of $a$ and $b$ divides $d = ua + vb$.
This motivates the definition:

\begin{defin}
\label{gcddef}
{\em
Given any two  integers, $a, b\in\integs$,
an integer, $d\in \integs$, is a {\it greatest common
divisor of $a$ and $b$\/} (for short, a {\it gcd of $a$ and $b$\/})
if $d$ divides $a$ and $b$ and, for any integer,
$h\in\integs$, if $h$ divides $a$ and $b$, then $h$ divides $d$.
We say that $a$ and $b$ are {\it relatively prime\/} 
if $1$ is a gcd of $a$ and $b$.
}
\end{defin}
\remarks
\begin{enumerate}
\item
Assume $a = b = 0$. Then, any integer, $d\in\integs$, is a divisor of
$0$. In particular, $0$ divides $0$. According to Definition \ref{gcddef}, 
this implies $\mathrm{gcd}(0, 0) = 0$. If $(a, b)\not= (0,0)$,
then $1$ divides $a$ and $b$, so $\mathrm{gcd}(a, b) = d > 0$.
In this case, if $d'$ is any other gcd of $a$ and $b$, then
$d = qd'$ and $d' = q'd$ for some $q, q'\in \integs$.
So, $d = qq'd$ which implies $qq' = 1$ (since $d\not= 0$) and 
thus, $d' = \pm d$.
So, according to the above definition, when $(a, b) \not= (0, 0)$,
gcd's are not unique. However, exactly one of $d$ or $-d$ is positive,
so we will refer to this positive  gcd as ``the'' gcd of $a$ and $b$
and write $d = \mathrm{gcd}(a, b)$.
\item
Observe that $d = \mathrm{gcd}(a, b)$ is indeed the largest
positive common divisor of $a$ and $b$ since every
divisor of $a$ and $b$ must divide $d$.
However, we did not use this property as one of the conditions
for being a gcd because such a condition does not generalize
to other rings where a total order is not available.
Another minor reason is that if we had used in the definition of a gcd
the condition that
$\mathrm{gcd}(a,b)$ should be the largest common divisor of $a$ and $b$, 
as every integer divides $0$,
$\mathrm{gcd}(0,0)$ would be undefined! 
\item
Our definition of the gcd makes sense
even if  we allow $a = b = 0$. In this case,
$\mathrm{gcd}(0, 0) = 0$. If we did not allows $0$ to be a divisor,
the situation would be different. Indeed,
if we had $\mathrm{gcd}(0, 0) = d$
for some $d > 0$, as every other positive integer, $d'$,
divides $0$, every integer $d'$ would have to divide
$d$, which is absurd. 
This is why we relaxed our definition to allow $0$ to be a divisor.
Nevertheless, the cases where $a = 0$ or $b = 0$
are somewhat degenerate cases so we prefer to stick to
the simpler situation where 
we only consider gcd's for two nonzero integers.
\end{enumerate}

\medskip
Let $p\in \natnums$ be a prime number. Then, note that 
for any other integer, $n$, if
$p$ does not divide $n$, then $\mathrm{gcd}(p, n) = 1$,
as the only divisors of $p$ are $1$ and $p$.

\begin{prop}
\label{gcdp1}
Given any two integers,  $a, b\in\integs$,
a natural number, $d\in \natnums$, is the 
greatest common divisor of $a$ and $b$ iff
$d$ divides $a$ and $b$ and if there are some integers,
$u, v\in \integs$, so that
\begin{equation}
ua + vb = d.
\tag{Bezout~Identity}
\end{equation}
In particular, $a$ and $b$ are relatively prime iff
there are some integers,
$u, v\in \integs$, so that
\begin{equation}
ua + vb = 1.
\tag{Bezout~Identity}
\end{equation}
\end{prop}

\proof
We already observed that half of Proposition \ref{gcdp1}
holds, namely if  $d\in \natnums$ divides $a$ and $b$ 
and if there are some integers,
$u, v\in \integs$, so that $ua + vb = d$,
then, $d$ is the  $\mathrm{gcd}$ of $a$ and $b$.
Conversely, assume that $d = \mathrm{gcd}(a, b)$.
If $d = 0$, then $a = b = 0$ and
the proposition holds trivially. So, assume $d > 0$, in which
case $(a, b) \not= (0, 0)$.
By Corollary \ref{ZPID2}, there is a unique $m \in \natnums$
with $m > 0$
that divides $a$ and $b$ and there are some integers,
$u, v\in \integs$, so that
\[
ua + vb = m.
\]
But now, $m$ is also the (positive) gcd of $a$ and $b$, so $d = m$
and our Proposition holds. Now, $a$ and $b$ are relatively
prime iff  $\mathrm{gcd}(a, b) = 1$ in which case
the condition that $d = 1$ divides $a$ and $b$ is trivial.
$\bigsquare$

\remark
The gcd of two nonzero integers can be found using a method
involving Euclidean division and so can the numbers $u$ and $v$.

\medskip
Proposition \ref{gcdp1} implies a very crucial
property of divisibility in any PID.

\begin{prop}
\label{gcdcol2} (Euclid's proposition)
Let $a, b, c\in\integs$ be any integers.
If $a$ divides $bc$ and $a$ is relatively prime to $b$,
then $a$ divides $c$.
\end{prop}

\proof 
From Proposition \ref{gcdp1}, $a$ and $b$ are relatively
prime iff there exist some integers, $u, v\in\integs$
such that 
\[
ua + vb = 1.
\]
Then, we have
\[
uac + vbc = c,
\]
and since $a$ divides $bc$, it divides both
$uac$ and $vbc$ and so, $a$ divides $c$.
$\bigsquare$

\medskip
In particular, if $p$ is a prime number and if
$p$ divides $ab$, where $a, b\in \integs$ are nonzero, then
either $p$ divides $a$ or $p$ divides $b$ since
if $p$ does not divide $a$, by a previous remark,
then $p$ and $a$ are relatively prime, so 
Proposition \ref{gcdcol2} implies that $p$ divides $c$.

\begin{prop}
\label{gcdcol3}
Let $a, b_1,\ldots,b_m\in\integs$
be any integers.
If $a$ and $b_i$ are relatively prime for all $i$, with $1\leq i \leq m$,
then $a$ and $b_1\cdots b_m$ are relatively prime.
\end{prop}

\proof 
We proceed by induction on $m$. The case $m=1$ is trivial.
Let $c = b_2\cdots b_m$. By the induction hypothesis, 
$a$ and $c$ are relatively prime. 
Let $d$ the gcd of $a$ and $b_1c$. 
We claim that $d$ is relatively prime to $b_1$.
Otherwise, $d$ and  $b_1$ would have some gcd $d_1\not= 1$
which would divide both $a$ and $b_1$, contradicting
the fact that $a$ and $b_1$ are relatively prime.
Now, by Proposition \ref{gcdcol2}, since $d$ divides $b_1c$ and $d$ and $b_1$
are relatively prime, $d$ divides $c = b_2\cdots b_m$.
But then, $d$ is a divisor of $a$ and $c$, 
and since $a$ and $c$ are relatively prime, 
$d = 1$, which means that $a$ and $b_1\cdots b_m$ are relatively prime.
$\bigsquare$

\medskip
We can now prove the uniqueness of prime factorizations in 
$\natnums$.
 divisor to be $0$.

\begin{thm} (Unique Prime Factorization in $\natnums$)
\label{uniquefac1}
For every nonzero natural number, $a\geq 2$,
there exists a unique set,
$\{\pairt{p_1}{k_1}, \ldots, \pairt{p_m}{k_m}\}$,
where the $p_i$'s are distinct prime numbers and
the $k_i$'s are (not necessarily distinct) integers, with 
$m\geq 1$, $k_i\geq 1$, and 
\[
a =  p_{1}^{k_{1}}\cdots p_{m}^{k_{m}}.
\]
\end{thm}

\proof 
The existence of such a factorization has already been proved in Theorem
\ref{primedecomp}.

\medskip
Let us now prove uniqueness.
Assume that
\[ 
a =  p_{1}^{k_{1}}\cdots p_{m}^{k_{m}}
\quad\hbox{and}\quad
a =  q_{1}^{h_{1}}\cdots q_{n}^{h_{n}}.
\]
Thus, we have
\[ 
p_{1}^{k_{1}}\cdots p_{m}^{k_{m}}
=  q_{1}^{h_{1}}\cdots q_{n}^{h_{n}}.
\]
We prove that $m = n$, $p_i = q_i$
and $h_i = k_i$, for all $i$, with $1\leq i\leq n$.
The proof proceeds by induction on $h_1 + \cdots + h_n$.

\medskip
If $h_1 + \cdots + h_n = 1$, then $n=1$ and $h_1 = 1$.
Then, 
\[
p_{1}^{k_{1}}\cdots p_{m}^{k_{m}} = q_1,
\]
and since  $q_1$ and the $p_i$ are prime numbers,
we must have $m = 1$ and $p_1 = q_1$
(a prime is only divisible by $1$ or itself).

\medskip
If $h_1 + \cdots + h_n  \geq 2$, 
since $h_1\geq 1$, we have
\[
p_{1}^{k_{1}}\cdots p_{m}^{k_{m}} = q_1q,
\]
with 
\[
q = q_{1}^{h_{1} - 1}\cdots q_{n}^{h_{n}},
\]
where $(h_1 - 1) + \cdots + h_n  \geq 1$
(and $q_{1}^{h_{1} - 1} = 1$  if $h_1 = 1$).
Now, if $q_1$ is not equal to any of the $p_i$, by a previous
remark, $q_1$ and $p_i$ are relatively prime, and
by Proposition \ref{gcdcol3}, $q_1$ and
$p_{1}^{k_{1}}\cdots p_{m}^{k_{m}}$ are relatively prime.
But this contradicts the fact that $q_1$ divides
$p_{1}^{k_{1}}\cdots p_{m}^{k_{m}}$.
Thus, $q_1$ is equal to one of the $p_i$.
Without loss of generality, we can assume that $q_1 = p_1$.
Then, as $q_1\not= 0$, we get
\[
p_{1}^{k_{1} - 1}\cdots p_{m}^{k_{m}} =
q_{1}^{h_{1} - 1}\cdots q_{n}^{h_{n}},
\]
where $p_{1}^{k_{1} - 1} = 1$  if $k_1 = 1$, 
and  $q_{1}^{h_{1} - 1} = 1$  if $h_1 = 1$.
Now, $(h_1 - 1) + \cdots + h_n < h_1 + \cdots + h_n$,
and we can apply the induction hypothesis to conclude that
$m = n$, $p_i = q_i$  and $h_i = k_i$, with $1\leq i\leq n$.
$\bigsquare$

\medskip
Theorem \ref{uniquefac1} is a basic but very important result of
number theory and it has many applications. It also
reveals the importance of the primes as the building
blocks of all numbers. 

\remark
Theorem \ref{uniquefac1} also applies to any nonzero
integer $a\in \integs - \{-1, +1\}$, by adding a suitable sign in front
of the prime factorization.  That is, we have
a unique prime factorization of the form
\[
a =  \pm p_{1}^{k_{1}}\cdots p_{m}^{k_{m}}.
\]
Theorem \ref{uniquefac1} shows that $\integs$ is a 
{\it unique factorization domain\/}, for short, a {\it UFD\/}.
Such rings play an important role because every nonzero element
which is not a unit (i.e., which is not invertible) 
has a unique factorization (up to some unit factor)
into so-called {\it irreducible elements\/} which generalize the primes.

\medskip
We now take a well-deserved break from partial orders and
induction and study equivalence relations, an equally important class
of relations.
 
\section{Equivalence Relations and Partitions}
\label{sec17}
Equivalence relations basically generalize the identity relation.
Technically, the definition of an equivalence relation is obtained
from the definition of a partial order (Definition \ref{posetdef})
by changing the third condition, antisymmetry, to {\it symmetry\/}.

\begin{defin}
\label{equivreldef}
{\em
A binary relation, $R$, on a set, $X$, is an {\it equivalence relation\/}
iff it is
{\it reflexive\/}, {\it transitive\/} and {\it symmetric\/}, 
that is:
\begin{enumerate}
\item[(1)] 
({\it Reflexivity\/}):
$a R a$, for all $a\in X$;
\item[(2)]
({\it Transitivity\/}):
If $a R b$  and $b R c$, then $a R c$, for all
$a, b, c \in X$. 
\item[(3)]
({\it symmetry\/}): If $a R b$,  then $b R a$,
for all $a, b\in X$.
\end{enumerate}
}
\end{defin}

\medskip
Here are some examples of equivalence relations.
\begin{enumerate}
\item
The identity relation, $\id_X$,  on a set $X$ is an equivalence relation.
\item
The relation $X\times X$ is an equivalence relation.
\item
Let $S$ be the set of students in CSE260. 
Define two students to be equivalent iff they were born 
the same year. It is trivial to check that this relation is indeed
an equivalence relation.
\item
Given any natural number, $p \geq 1$, define a relation
on $\integs$ as follows: 
\[
m \equiv n\> (\mathrm{mod}\> p)
\]
iff $p \mid m - n$, i.e., $p$ divides $m - n$.
It is an easy exercise to check that this is indeed
an equivalence relation called {\it congruence modulo $p$\/}.
\item
Equivalence of propositions is the relation defined so that
$P\equiv Q$ iff $P \impl Q$ and $Q\impl P$ are both provable
(say, classically). It is easy to check that logical equivalence
is an equivalence relation.
\item
Suppose $\mapdef{f}{X}{Y}$ is a  function. Then, we define
the relation $\equiv_f$ on $X$ by
\[
x \equiv_f y\quad\hbox{iff}\quad f(x) = f(y).
\]
It is immediately verified that $\equiv_f$ is an equivalence relation.
Actually, we are going to show that every equivalence relation arises
in this way, in terms of (surjective) functions.
\end{enumerate}

\medskip
The crucial property of equivalence relations is that they
{\it partition\/} their domain, $X$, into
pairwise disjoint nonempty blocks. Intuitively, they carve out
$X$ into a bunch of puzzle pieces.

\begin{defin}
\label{equivclass}
{\em
Given an equivalence relation, $R$, on a set, $X$, for any 
$x\in X$, the set
\[
[x]_R = \{y\in X \mid x R y\}
\]
is the {\it equivalence class of $x$\/}.
Each equivalence class, $[x]_R$, is also denoted $\overline{x}_R$
and the subscript $R$ is often omitted when no confusion arises.
The set of equivalence classes of $R$ is denoted by $X/R$.
The set $X/R$ is called the {\it quotient of $X$ by $R$\/}
or {\it quotient of $X$ modulo $R$\/}. The function,
$\mapdef{\pi}{X}{X/R}$, given by
\[
\pi(x) = [x]_R, \quad x\in X,
\]
is called the {\it canonical projection\/} (or {\it projection\/})
of $X$ onto $X/R$.
}
\end{defin}

\medskip
Since every equivalence relation is reflexive, i.e., $x R x$
for every $x\in X$, observe that
$x\in [x]_R$ for any $x\in R$, that is, every equivalence class
is {\it nonempty\/}. It is also clear that the projection,
$\mapdef{\pi}{X}{X/R}$, is surjective.
The main properties of equivalence classes are given by 

\begin{prop}
\label{equivp1}
Let $R$ be an equivalence relation on a set, $X$. For any two elements
$x, y\in X$, we have
\[
x R y
\quad\hbox{iff}\quad
[x] = [y].
\]
Moreover, the equivalences classes of $R$ satisfy the following 
properties:
\begin{enumerate}
\item[(1)]
$[x] \not= \emptyset$, for all $x\in X$;
\item[(2)]
If $[x]\not= [y]$ then $[x]\cap [y] = \emptyset$;
\item[(3)]
$X = \bigcup_{x\in X} [x]$.
\end{enumerate}
\end{prop}

\proof
First, assume that $[x] = [y]$. We observed that by reflexivity, $y\in [y]$.
As $[x] = [y]$, we get $y\in [x]$ and by definition of $[x]$, this means that
$x R y$. 

\medskip
Next, assume that $x R y$. Let us prove that
$[y] \subseteq [x]$. Pick any $z\in [y]$; this means that
$y R z$. By transitivity, we get $x R z$, ie.,
$z\in [x]$, proving that $[y] \subseteq [x]$.
Now, as $R$ is symmetric, $x R y$ implies that $y R x$ and
the previous argument yields $[x] \subseteq [y]$.
Therefore, $[x] = [y]$, as needed.

\medskip
Property (1) follows from the fact that $x\in [x]$
(by reflexivity).

\medskip
Let us prove the contrapositive of (2). So, assume
$[x]\cap [y] \not= \emptyset$. Thus, there is some $z$
so that $z\in [x]$ and $z\in [y]$, i.e.,
\[
x R z\quad\hbox{and}\quad y R z.
\]
By symmetry, we get $z R y$ and by transitivity,
$x R y$. But then, by the first part of the proposition,
we deduce $[x] = [y]$, as claimed.

\medskip
The third property follows again from the fact that $x\in [x]$.
$\bigsquare$

\medskip
A useful way of interpreting 
Proposition \ref{equivp1} is to say that the equivalence classes of 
an equivalence relation form a partition, as defined next.

\begin{defin}
\label{partdef}
{\em
Given a set, $X$, a {\it partition of $X$\/} is any family,
$\Pi = \{X_i\}_{i\in I}$, of subsets of $X$ such that
\begin{enumerate}
\item[(1)]
$X_i \not= \emptyset$, for all $i\in I$
(each $X_i$ is nonempty);
\item[(2)]
If $i\not= j$ then $X_i\cap X_j = \emptyset$
(the $X_i$ are pairwise disjoint);
\item[(3)]
$X = \bigcup_{i\in I} X_i$ (the family is exhaustive).
\end{enumerate}
Each set $X_i$ is called a {\it block\/} of the partition.
}
\end{defin}

\medskip
In the example where equivalence is determined by the same
year of birth, each equivalence class consists of those
students having the same year of birth.
Let us now go back to the example of congruence modulo $p$ (with $p > 0$)
and figure out what are the blocks of the corresponding partition.
Recall that
\[
m \equiv n\> (\mathrm{mod}\> p)
\]
iff $m - n = pk$ for some $k\in\integs$. By the division Theorem
(Theorem \ref{divthm}), 
we know that there exist some unique $q, r$, with
$m = pq + r$ and $0 \leq r \leq p - 1$. Therefore,
for every $m\in \integs$, 
\[
m \equiv r\> (\mathrm{mod}\> p)
\quad\hbox{with}\quad 0 \leq r \leq p  - 1,
\]
which shows that there are $p$ equivalence classes,
$[0], [1], \ldots, [p - 1]$, where the equivalence class,
$[r]$ (with $0 \leq r \leq p - 1$), consists of all integers
of the form $pq + r$, where $q \in \integs$, i.e.,
those integers whose residue modulo $p$ is $r$.

\medskip
Proposition \ref{equivp1} defines a map from the set of equivalence relations
on $X$ to the set of partitions on $X$.
Given any set, $X$, let $\mathrm{Equiv}(X)$ denote the set of equivalence
relations on $X$ and let $\mathrm{Part}(X)$ denote the set
of partitions on $X$. Then, Proposition \ref{equivp1} defines the 
function, $\mapdef{\Pi}{\mathrm{Equiv}(X)}{\mathrm{Part}(X)}$,
given by,
\[
\Pi(R) = X/R = \{[x]_R\mid x\in X\},
\]
where $R$ is any equivalence relation on $X$. We also
write $\Pi_R$ instead of $\Pi(R)$.

\medskip
There is also a function, $\mapdef{\s{R}}{\mathrm{Part}(X)}{\mathrm{Equiv}(X)}$,
that assigns an equivalence relation to a partition a shown 
by the next proposition.

\begin{prop}
\label{equivp2}
For any partition, $\Pi = \{X_i\}_{i\in I}$, on a set, $X$,
the relation, $\s{R}(\Pi)$, defined by
\[
x\s{R}(\Pi) y \quad\hbox{iff}\quad 
(\exists i \in I)(x, y\in X_i),
\]
is an equivalence relation whose equivalence
classes are exactly the blocks $X_i$.
\end{prop}

\proof
We leave this easy proof as an exercise to the reader.
$\bigsquare$

\medskip
Putting Propositions \ref{equivp1} and \ref{equivp2} together we obtain
the useful fact there is a bijection between
$\mathrm{Equiv}(X)$ and  $\mathrm{Part}(X)$. Therefore, in principle,
it is a matter of taste whether we prefer to work with
equivalence relations or partitions. In computer science, it is often preferable 
to work with partitions, but not always.

\begin{prop}
\label{equivp3}
Given any set, $X$, the functions 
$\mapdef{\Pi}{\mathrm{Equiv}(X)}{\mathrm{Part}(X)}$ and \\
$\mapdef{\s{R}}{\mathrm{Part}(X)}{\mathrm{Equiv}(X)}$ are mutual inverses,
that is,
\[
\s{R} \circ \Pi = \id
\quad\hbox{and}\quad 
\Pi\circ \s{R} = \id.
\]
Consequently, there is a bijection between the set,
$\mathrm{Equiv}(X)$, of equivalence relations on $X$
 and the set, $\mathrm{Part}(X)$, of partitions on $X$.
\end{prop}

\proof
This is a routine verication left to the reader.
$\bigsquare$

\medskip
Now, if $\mapdef{f}{X}{Y}$ is a surjective function, we have the
equivalence relation, $\equiv_f$, defined by
\[
x \equiv_f y\quad\hbox{iff}\quad f(x) = f(y).
\]
It is clear that the equivalence class of any $x\in X$ is
the inverse image, $f^{-1}(f(x))$, of $f(x)\in Y$.
Therefore, there is a bijection between $X/\equiv_f$ and $Y$.
Thus, we can identify $f$ and the projection, $\pi$, from $X$ onto $X/\equiv_f$.
If $f$ is not surjective, note that $f$ is surjective onto $f(X)$
and so, we see that $f$ can be written as the composition 
\[
f = i \circ \pi,
\]
where $\mapdef{\pi}{X}{f(X)}$ is the canonical projection and 
$\mapdef{i}{f(X)}{Y}$ is the {\it inclusion function\/}
mapping $f(X)$ into $Y$ (i.e., $i(y) = y$,
for every $y\in f(X)$).

\medskip
Given a set, $X$, the inclusion ordering on $X\times X$ defines
an ordering on binary relations on $X$, namely,
\[
R \leq S
\quad\hbox{iff}\quad
(\forall x, y\in X)(x R y \impl x Sy).
\]
When $R \leq S$, we say that {\it $R$ refines $S$\/}.
If $R$ and $S$ are equivalence relations and $R \leq S$, we 
observe that every equivalence class of $R$ is contained
in some equivalence class of $S$. Actually, in view of
Proposition \ref{equivp1}, we see that {\it every equivalence class
of $S$ is the union of equivalence classes of $R$\/}.
We also note that $\id_X$ is the least equivalence relation on
$X$ and $X\times X$ is the largest equivalence relation on $X$.
This suggests the following question: Is $\mathrm{Equiv}(X)$
a lattice under refinement?

\medskip
The answer is yes. It is easy to see that the meet
of two equivalence relations is $R\cap S$, their intersection.
But beware, their join is not $R\cup S$, because in general,
$R\cup S$ is not transitive.  However,
there is a least equivalence relation containing $R$ and $S$, and
this is the join of $R$  and $S$. This leads us to look
at various closure properties of relations.

\section[Transitive Closure, Reflexive and Transitive Closure]
{Transitive Closure, Reflexive and Transitive Closure, 
Smallest Equivalence Relation}
\label{sec18}
Let $R$ be any relation on a set $X$. Note that $R$ is reflexive
iff $\id_X \subseteq R$.  Consequently, the smallest
reflexive relation containing $R$ is $\id_X \cup R$.
This relation is called the {\it reflexive closure of $R$\/}.

\medskip
Note that $R$ is transitive iff $R\circ R \subseteq R$.
This suggests a way of making the smallest transitive
relation containing $R$ (if $R$ is not already transitive).
Define $R^n$ by induction as follows:
\begin{eqnarray*}
R^0 & = & \id_X \\
R^{n + 1} & = & R^n \circ R.
\end{eqnarray*}

\begin{defin}
\label{transclo}
{\em
Given any relation, $R$, on a set, $X$, the {\it transitive
closure of $R$\/} is the relation, $R^+$, given by
\[
R^+ = \bigcup_{n \geq 1} R^n. 
\]
The {\it reflexive and transitive
closure of $R$\/} is the relation, $R^*$, given by
\[
R^* = \bigcup_{n \geq 0} R^n = \id_X \cup R^+. 
\]
}
\end{defin}

\medskip
The proof of the following proposition is left an an easy
exercise.

\begin{prop}
\label{transclo2}
Given any relation, $R$, on a set, $X$,
the relation $R^+$ is the smallest transitive relation
containing $R$ and $R^*$ is the smallest 
reflexive and transtive relation
containing $R$.  
\end{prop}

\medskip
If $R$ is reflexive, then it is easy to see that
$R \subseteq R^2$
and so, $R^k \subseteq R^{k + 1}$ for all $k \geq 0$.
From this, we can show that if $X$ is a finite set, then
there is a smallest $k$ so that $R^k = R^{k + 1}$.
In this case, $R^k$ is the reflexive and transitive closure 
of $R$. If $X$ has $n$ elements it can be shown that
$k \leq n - 1$.

\medskip
Note that a relation, $R$, is symmetric iff
$R^{-1} = R$. As a consequence, $R\cup R^{-1}$
is the smallest symmetric relation containing $R$.
This relation is called the {\it symmetric closure of $R$\/}.
Finally, given a relation, $R$, what is the smallest
equivalence relation containing $R$?
The answer is given by

\begin{prop}
\label{equivclo}
For any relation, $R$, on a set, $X$, the relation
\[
(R \cup R^{-1})^*
\]
is the smalest equivalence relation containing $R$.
\end{prop}

\section[Distributive Lattices, Boolean Algebras, Heyting Algebras]
{Distributive Lattices, Boolean Algebras, Heyting Algebras}
\label{sec19}
If we go back to one of our favorite  examples of a lattice,
namely, the power set, $2^X$, of some set, $X$, we observe
that it is more than a lattice. For example, if we look at Figure
\ref{order4}, we can check that the two identities 
D1 and D2 stated in the next definition hold.

\begin{defin}
\label{distlatdef}
{\em
We say that a lattice, $X$,  is a {\it distributive lattice\/} if 
(D1) and (D2) hold:
\begin{alignat*}{2}
& D1 \qquad &  & a\land (b\lor c) =  (a\land b)\lor (a\land c) \\
& D2 \qquad &  & a\lor (b\land c) =  (a\lor b)\land (a\lor c). 
\end{alignat*}
}
\end{defin}

\remark
Not every lattice is distributive but many lattices of interest
are distributive.

\medskip
It is a bit surprising that in a lattice, (D1) and (D2) are 
actually equivalent, as we now show. Suppose (D1) holds, then
\begin{align}
(a\lor b)\land (a\lor c) & =  ((a\lor b)\land a)\lor ((a\lor b) \land c) \tag{D1} \\
& =  a \lor ((a\lor b) \land c) \tag{L4} \\
& =  a \lor ((c \land (a\lor b))  \tag{L1} \\
& =  a \lor ((c \land a) \lor (c\land b))  \tag{D1} \\
& =  a \lor ((a \land c) \lor (b\land c))  \tag{L1} \\
& =  (a \lor (a \land c)) \lor (b\land c)  \tag{L2} \\
& =  ((a \land c)\lor a) \lor (b\land c)  \tag{L1} \\
& =  a  \lor (b\land c)  \tag{L4}
\end{align}
which is (D2). Dually, (D2) implies (D1). 

\medskip
The reader should prove that every totally ordered poset is
a distributive lattice.
The lattice $\natnums_+$ under the divisibility ordering also 
turns out to be a distributive lattice.

\medskip
Another useful fact about distributivity which is worth noting is that in any lattice
\[
a\land (b\lor c) \geq (a\land b) \lor (a \land c).
\]
This is because in any lattice, $a\land (b\lor c) \geq a \land b$ and
$a\land (b\lor c) \geq a\land c$.
Therefore, in order to establish associativity in a lattice it suffices to show that
\[
a\land (b\lor c) \leq (a\land b) \lor (a \land c).
\]

\medskip
Another important property of distributive lattices 
is the following:

\begin{prop}
\label{distpro1}
In a distributive lattice, $X$, if
$z\land x = z \land y$ and $z\lor x = z\lor y$, then $x = y$
(for all $x, y, z\in X$).
\end{prop}

\proof
We have
\begin{align}
x  & = (x \lor z)  \land x \tag{L4} \\
   & =  x \land (z \lor x) \tag{L1} \\
   & =  x \land (z \lor y) \notag \\
   & =  (x \land z) \lor (x\land y) \tag{D1} \\
   & =  (z \land x) \lor (x\land y) \tag{L1} \\
   & =  (z \land y) \lor (x\land y) \notag \\
   & =  (y \land z) \lor (y\land x) \tag{L1} \\
   & =  y\land (z \lor x) \tag{D1} \\
   & =  y\land (z \lor y) \notag \\
   & =  (y \lor z)\land y \tag{L1} \\
   & =   y, \tag{L4} 
\end{align}
that is, $x = y$, as claimed.
$\bigsquare$

\medskip
The power set lattice has yet some additional properties
having to do with complementation. First, the power
lattice $2^X$ has a least element $0 = \emptyset$ and a greatest
element, $1 = X$. If a lattice, $X$, has a least element, $0$,
and a greatest element, $1$, the following properties are clear:
For all $a\in X$, we have
\begin{alignat*}{2}
& a \land 0  =  0  & \qquad  & a\lor 0 = a  \\
& a \land 1  =  a  & \qquad  & a\lor 1 = 1.
\end{alignat*}

\medskip
More importantly, for any subset, $A\subseteq X$,
we have the complement, $\overline{A}$, of $A$ in $X$, which
satisfies the identities:
\[
A\cup \overline{A} = X,\qquad A\cap \overline{A} = \emptyset.
\] 
Moreover, we know that the de Morgan identities hold.
The generalization of these properties leads to  what is called
a complemented lattice. 

\begin{defin}
\label{complatdef}
{\em
Let $X$ be a lattice and assume that $X$ has
a least element, $0$, and a greatest element, $1$
(we say that $X$ is a {\it bounded lattice\/}).
For any $a\in X$, a {\it complement of $a$\/} is any element,
$b\in X$, so that
\[
a\lor b = 1\quad\hbox{and}\quad a\land b = 0. 
\]
If every element of $X$ has a complement, we say that $X$ is
a {\it complemented lattice\/}.
}
\end{defin}

\remarks
\begin{enumerate}
\item
When $0 = 1$, the lattice $X$ collapses to the degenerate lattice
consisting of a single element. As this lattice is of little interest,
from now on, we will always assume that $0 \not= 1$. 
\item
In a complemented lattice, complements are generally not unique.
However, as the next proposition shows, this is
the case for distributive lattices.
\end{enumerate}

\begin{prop}
\label{comlatp1}
Let $X$ be a lattice with least element $0$ and greatest element $1$.
If $X$ is distributive, then complements are unique if they
exist. Moreover, if $b$ is the complement of $a$, then
$a$ is the complement of $b$.
\end{prop}

\proof
If $a$ has two complements, $b_1$ and $b_2$, then
$a\land b_1 = 0$, $a\land b_2 = 0$, $a\lor b_1 = 1$, 
and $a\lor b_2 = 1$. By commutativity, if follows that
$b_1\land a = b_2\land a = 0$ and
$b_1\lor a = b_2\lor a = 1$.
By Proposition \ref{distpro1}, we deduce that $b_1 = b_2$, that is,
$a$ has a unique complement.

\medskip
By commutativity, the equations
\[
a\lor b = 1\quad\hbox{and}\quad a\land b = 0 
\]
are equivalent to the equations
\[
b\lor a = 1\quad\hbox{and}\quad b\land a = 0, 
\]
which shows that $a$ is indeed a complement of $b$.
By uniqueness, $a$ is {\it the\/} complement of $b$.
$\bigsquare$

\medskip
In view of Proposition \ref{comlatp1}, if $X$ is a complemented
distributive lattice, we denote the complement of any
element, $a\in X$, by $\overline{a}$. We have the identities
\begin{eqnarray*}
a\lor \overline{a} & = & 1 \\
a\land \overline{a} & = & 0 \\
\overline{\overline{a}} & = & a.
\end{eqnarray*}

\medskip
We also have the following proposition about the
de Morgan laws.
\begin{prop}
\label{boolatp1}
Let $X$ be a lattice with least element $0$ and greatest element $1$.
If $X$ is distributive and complemented, then
the de Morgan laws hold:
\begin{eqnarray*}
\overline{a\lor b} & = &\overline{a} \land \overline{b} \\
\overline{a\land b} & = &\overline{a} \lor \overline{b}.
\end{eqnarray*}
\end{prop}

\proof
We prove that
\[
\overline{a\lor b} = \overline{a} \land \overline{b},
\]
leaving the dual identity as an easy exercise. Using the uniqueness
of complements, it is enough to check that $\overline{a} \land \overline{b}$
works, i.e., satisfies the conditions of Definition \ref{complatdef}.
For the first condition, we have
\begin{eqnarray*}
(a\lor b) \lor (\overline{a} \land \overline{b})
 & = & ((a\lor b) \lor \overline{a})\land ((a\lor b)\lor \overline{b}) \\
 & = & (a\lor (b \lor \overline{a}))\land (a\lor (b\lor \overline{b})) \\
 & = & (a\lor (\overline{a} \lor b))\land (a\lor 1) \\
& = & ((a\lor \overline{a}) \lor b)\land 1 \\
& = & (1 \lor b)\land 1 \\
& = & 1\land 1  = 1. 
\end{eqnarray*}
For the second condition, we have
\begin{eqnarray*}
(a\lor b) \land (\overline{a} \land \overline{b})
 & = & (a \land (\overline{a} \land \overline{b}))\lor 
(b \land (\overline{a} \land \overline{b})) \\
 & = & ((a \land \overline{a}) \land \overline{b})\lor 
(b \land (\overline{b} \land \overline{a})) \\
 & = & (0 \land \overline{b})\lor 
((b \land \overline{b}) \land \overline{a}) \\
 & = & 0 \lor (0 \land \overline{a}) \\
 & = & 0 \lor 0 = 0.
\end{eqnarray*}
$\bigsquare$

\medskip
All this leads to the definition of a boolean lattice

\begin{defin}
\label{boolatdef}
{\em
A {\it Boolean lattice\/} is a lattice
with a least element, $0$, a greatest element, $1$, 
and which is distributive and complemented.
}
\end{defin}

\medskip
Of course, every power set is a boolean lattice, but there are 
boolean lattices that are not power sets. 
Putting together what we have done, we see that a boolean lattice
is a set, $X$, with two special elements, $0$, $1$, and three
operations, $\land$, $\lor$ and $a\mapsto \overline{a}$
satisfying the axioms stated in 

\begin{prop}
\label{boolp1}
If $X$ is a boolean lattice, then the following equations hold
for all \\
$a, b, c \in X$:
\begin{alignat*}{3}
& L1 \quad &   & a\lor b = b\lor a, & \qquad & a\land b = b\land a   \\
& L2 \quad &   & (a\lor b)\lor c = a\lor (b\lor c), &\qquad &
(a\land b)\land c = a\land (b\land c)\\
& L3 \quad &   & a \lor a = a, & \qquad & a \land a = a \\
& L4 \quad &   &(a\lor b)\land a = a, & \qquad & (a\land b) \lor a = a \\
& D1\hbox{-}D2 \quad &   &  a\land (b\lor c) =  (a\land b)\lor (a\land c), 
& \qquad &   a\lor (b\land c) =  (a\lor b)\land (a\lor c) \\
& \mathit{LE} \quad &  & a \lor 0 = a, &\qquad & a\land 0 = 0 \\
& \mathit{GE} \quad &  & a \lor 1 = 1, &\qquad & a\land 1 = a \\
& C \quad &  & a \lor \overline{a} = 1, & \qquad & a\land \overline{a} = 0 \\
& I \quad &  & \overline{\overline{a}} = a & & \\
& \mathit{dM} \quad &  & \overline{a\lor b} = \overline{a}\land \overline{b}, &\qquad &
\overline{a\land b} = \overline{a}\lor \overline{b}. 
\end{alignat*}
Conversely, if $X$ is a set together  with two special elements, $0$, $1$, and three
operations, $\land$, $\lor$ and $a\mapsto \overline{a}$
satisfying the axioms above, then it is a boolean lattice under
the ordering given by $ a\leq b$ iff $a\lor b = b$.
\end{prop}

\medskip
In view of Proposition \ref{boolp1}, we make the definition:

\begin{defin}
\label{boolalg}
{\em
A set, $X$,  together  with two special elements, $0$, $1$, and three
operations, $\land$, $\lor$ and $a\mapsto \overline{a}$
satisfying the axioms of Proposition \ref{boolp1} is called a
{\it Boolean algebra\/}.
}
\end{defin}

\medskip
Proposition \ref{boolp1} shows that the notions of a Boolean lattice
and of a Boolean algebra are equivalent. The first one is order-theoretic and the
second one is algebraic. 

\remarks
\begin{enumerate}
\item
As the name indicates, Boolean algebras were invented by
G. Boole (1854). One of the first comprehensive accounts is due to 
E. Schr\"oder (1890-1895).
\item
The axioms for Boolean algebras given in Proposition \ref{boolp1}
are not independent. There is a set of independent axioms known as
the {\it Huntington axioms\/} (1933).
\end{enumerate}

\medskip
Let $p$ be any integer with $p\geq 2$.
Under the division ordering, it turns out that
the set, $\mathrm{Div}(p)$, of divisors of $p$
is a distributive lattice.
In general not every integer, $k\in\mathrm{Div}(p)$, has a complement 
but when it does, $\overline{k} = p/k$.
It can be shown that $\mathrm{Div}(p)$ is a Boolean algebra iff $p$
is not divisible by any square integer
(an integer of the form $m^2$, with $m > 1$).

\medskip
Classical logic is also a rich source of Boolean algebras.
Indeed, it is easy to show that logical equivalence is an equivalence
relation and, as Homework problems, you have shown (with great pain)
that all the axioms of Proposition \ref{boolp1} are provable
equivalences (where $\lor$ is disjunction, $\land$ is conjunction,
$\overline{P} = \neg P$, i.e., negation, $0 = \> \perp$ and
$1 = \top$).
Furthermore, again, as a Homework problem, you have shown that
logical equivalence is compatible with $\lor, \land, \neg$ in the 
following sense: If $P_1 \equiv Q_1$ and $P_2 \equiv Q_2$, then
\begin{eqnarray*}
(P_1\lor P_2) & \equiv & (Q_1\lor Q_2) \\
(P_1\land P_2) & \equiv & (Q_1\land Q_2) \\
\neg P_1 & \equiv & \neg Q_1.
\end{eqnarray*}
Consequently, for any set, $T$, of propositions
we can define the relation, $\equiv_T$, by 
\[
P \equiv_T Q
\quad\hbox{iff}\quad
T \vdash P \equiv Q,
\]
i.e.,  iff $P \equiv Q$ is provable from $T$
(as explained in Section \ref{sec5}). 
Clearly, $\equiv_T$ is an equivalence relation on propositions and so,
we can define the operations $\lor, \land$ and $\overline{\> }$
on the set of equivalence classes, $\mathbf{B}_T$, of propositions
as follows:
\begin{align*}
 [P]\lor [Q] & =   [P\lor Q] \\
 [P]\land [Q] & =   [P\land Q] \\
\overline{[P]}& =   [\neg P].
\end{align*}
We also let $0 = [\perp]$ and $1 = [\top]$.
Then, we get the Boolean algebra, $\mathbf{B}_T$, called
the {\it Lindenbaum algebra\/} of $T$.

\medskip
It also turns out that Boolean algebras are just what's needed
to give  truth-value semantics to classical logic.
Let $B$ be any Boolean algebra. A {\it truth assignment\/}
is any function, $v$, from the set 
$\mathbf{PS} = \{\mathbf{P}_1, \mathbf{P}_2, \cdots\}$
of propositional symbols to $B$. Then, we can
evaluate recursively the truth value, $P_B[v]$, in $B$ of
any proposition, $P$,
with respect to the truth assigment, $v$, as follows:
\begin{align*}
(\mathbf{P}_i)_B[v] & = v(P) \\
\perp_B[v] & = 0 \\
\top_B[v] & = 1 \\
(P\lor Q)_B[v] & =  P_B[v] \lor P_B[v] \\
(P\land Q)_B[v] & =  P_B[v] \land P_B[v] \\
(\neg P)_B[v] & = \overline{P[v]_B}.
\end{align*} 
In the equations above, on the
right hand side, $\lor$ and $\land$ are the lattice operations
of the Boolean algebra, $B$.
We say that a proposition, $P$, is {\it valid 
in the Boolean algebra $B$ (or $B$-valid)\/} if $P_B[v] = 1$
for all truth assigments, $v$.
We say that $P$ is {\it (classially) valid\/} if $P$ is $B$-valid
in all Boolean algebras, $B$. It can be shown that every
provable proposition is valid. This property is called
{\it soundness\/}. Conversely, if
$P$ is valid, then it is provable. This second property
is called {\it completeness\/}. 
Actually completeness holds in a much stronger sense:
If a proposition is valid in the two element
Boolean algebra, $\{0, 1\}$, then it is provable!

\medskip
One might wonder if there are certain kinds of algebras
similar to Boolean algebras well suited for 
intuitionistic logic. The answer is yes: Such
algebras are called {\it Heyting algebras\/}.

\medskip
In our study of intuitionistic logic, we learned that
negation is not a primary connective but instead it is 
defined in terms of implication by
$\neg P = P \impl \perp$.
This suggests adding to the two lattice operations
$\lor $ and $\land$ a new operation, $\rightarrow$,
that will behave like $\impl$.
The trick is, what kind of axioms should we require on
$\rightarrow$ to ``capture'' the properties of intuitionistic logic?
Now, if $X$ is a lattice with $0$ and $1$, given any two elements,
$a, b\in X$, experience shows that $a\rightarrow b$ should be the
largest element, $c$, such that
$c\land a \leq b$. 
This leads to

\begin{defin}
\label{Heyting1}
{\em
A lattice, $X$, with $0$ and $1$ is a {\it Heyting lattice\/} iff
it has a third binary operation, $\rightarrow$, such that
\[
c\land a \leq b 
\quad\hbox{iff}\quad
c \leq (a\rightarrow b)
\]
for all $a, b, c\in X$.
We define the {\it negation (or pseudo-complement) of $a$\/} as
$\>\overline{a} = (a \rightarrow 0)$.
}
\end{defin}

\medskip
At first glance, it is not clear that a Heyting lattice is 
distributive but in fact, it is. The following proposition
(stated without proof) gives an algebraic characterization
of Heyting lattices which is useful to prove various
properties of Heyting lattices.
 
\begin{prop}
\label{Heytingp1}
Let $X$ be a lattice with $0$ and $1$ and with a binary operation,
$\rightarrow$. Then, $X$ is a Heyting lattice iff
the following equations hold  for all $a, b, c\in X$:
\begin{eqnarray*}
a \rightarrow a & = & 1 \\
a\land (a\rightarrow b) & = & a\land b \\
b\land (a\rightarrow b) & = & b \\
a \rightarrow (b \land c) & = & (a\rightarrow b) \land (a\rightarrow c).
\end{eqnarray*}
\end{prop}

\medskip
A lattice with $0$ and $1$ and with a binary operation,
$\rightarrow$, satisfying the equations of Proposition
\ref{Heytingp1} is called a {\it Heyting algebra\/}.
So, we see that Proposition \ref{Heytingp1} shows that
the notions of Heyting lattice and Heyting algebra are equivalent
(this is analogous to Boolean lattices and Boolean algebras).

\medskip
The reader will notice that these axioms are 
propositions that were shown to be
provable intuitionistically in Homework Problems!
The proof of Proposition \ref{Heytingp1}
is not really difficult but it is a bit 
tedious so we will omit it. 
Let us simply show that the fourth equation implies
that for any fixed $a\in X$, the map
$b \mapsto (a\rightarrow b)$ is monotonic.
So, assume $b \leq c$, i.e., $b\land c = b$.
Then, we get
\[
a \rightarrow b = a \rightarrow (b\land c) = 
(a\rightarrow b) \land (a\rightarrow c),
\]
which means that $(a \rightarrow b) \leq (a\rightarrow c)$, 
as claimed. 

\medskip
The following theorem shows that every Heyting algebra
is distributive, as we claimed earlier.
This theorem also shows ``how close'' to a Boolean algebra
a Heyting algebra is.

\begin{thm}
\label{Heytingp2}
(a) Every Heyting algebra is distributive.

\medskip
(b)
A Heyting algebra, $X$, is a boolean algebra iff
$\>\overline{\overline{a}} = a$ for all $a\in X$.
\end{thm}

\proof
(a) From a previous remark, to show distributivity, it is enough to
show the inequality
\[
a\land (b\lor c) \leq (a\land b) \lor (a \land c).
\]
Observe that from the property characterizing $\rightarrow$, we have
\[
b \leq a \rightarrow (a\land b)
\quad\hbox{iff}\quad
b\land a \leq a \land b
\]
which holds, by commutativity of $\land$. Thus,
$b \leq a \rightarrow (a\land b)$ and similarly,
$c \leq a \rightarrow (a\land c)$. 

\medskip
Recall that for any fixed $a$, the map
$x \mapsto (a\rightarrow x)$ is monotonic.
Since $a\land b \leq (a\land b) \lor (a \land c)$
and $a\land c \leq (a\land b) \lor (a \land c)$, we get
\[
a \rightarrow (a\land b) \leq 
a \rightarrow ((a\land b) \lor (a \land c)) 
\quad\hbox{and}\quad
a \rightarrow (a\land c) \leq 
a \rightarrow ((a\land b) \lor (a \land c)).
\]
These two inequalities imply
$\>(a \rightarrow (a\land b))\lor (a \rightarrow (a\land c)) \leq 
a \rightarrow ((a\land b) \lor (a \land c))$,
and since we also have
$b \leq a \rightarrow (a\land b)$ and
$c \leq a \rightarrow (a\land c)$,
we deduce that
\[
b\lor c \leq a \rightarrow ((a\land b) \lor (a \land c)),
\]
which, using the fact that $(b\lor c)\land a = a \land (b\lor c)$,
means that
\[
a\land (b\lor c) \leq (a\land b) \lor (a \land c),
\]
as desired.

\medskip
(b) We leave this part as an exercise. The trick is to
see that the de Morgan laws hold and to apply one of them
to $a \land \overline{a} = 0$.
$\bigsquare$

\remarks
\begin{enumerate}
\item
Heyting algebras were invented by A. Heyting in 1930.
Heyting algebras are sometimes known as ``Brouwerian lattices''.
\item
Every Boolean algebra is automatically a Heyting algebra:
Set $a \rightarrow b = \overline{a} \lor b$.
\item
It can be shown that every finite distributive lattice is
a Heyting algebra.
\end{enumerate}

\medskip
We conclude this brief exposition of Heyting algebras
by explaining how they provide a truth semantics for
intuitionistic logic analogous to the thuth semantics
that Boolean algebras provide for classical logic.

\medskip
As in the classical case,
it is easy to show that intuitionistic 
logical equivalence is an equivalence
relation and  you have shown (with great pain)
that all the axioms of Heyting algebras are intuitionistically provable
equivalences (where $\lor$ is disjunction, $\land$ is conjunction,
and $\rightarrow$ is $\impl$).
Furthermore, you have also shown that
intuitionistic logical equivalence is compatible with 
$\lor, \land, \impl$ in the 
following sense: If $P_1 \equiv Q_1$ and $P_2 \equiv Q_2$, then
\begin{eqnarray*}
(P_1\lor P_2) & \equiv & (Q_1\lor Q_2) \\
(P_1\land P_2) & \equiv & (Q_1\land Q_2) \\
(P_1\impl P_2) & \equiv & (Q_1\impl Q_2).
\end{eqnarray*}
Consequently, for any set, $T$, of propositions
we can define the relation, $\equiv_T$, by 
\[
P \equiv_T Q
\quad\hbox{iff}\quad
T \vdash P \equiv Q,
\]
i.e.,  iff $P \equiv Q$ is provable intuitionistically from $T$
(as explained in Section \ref{sec5}). 
Clearly, $\equiv_T$ is an equivalence relation on propositions and 
we can define the operations $\lor, \land$ and $\rightarrow$
on the set of equivalence classes, $\mathbf{H}_T$, of propositions 
as follows:
\begin{align*}
 [P]\lor [Q] & =   [P\lor Q] \\
 [P]\land [Q] & =   [P\land Q] \\
 [P]\rightarrow  [Q] & =   [P\impl Q].
\end{align*}
We also let $0 = [\perp]$ and $1 = [\top]$.
Then, we get the Heyting algebra, $\mathbf{H}_T$, called
the {\it Lindenbaum algebra\/} of $T$, as in the classical case.

\medskip
Now, let $H$ be any Heyting algebra.
By analogy with the case of  Boolean algebras,
a {\it truth assignment\/}
is any function, $v$, from the set 
$\mathbf{PS} = \{\mathbf{P}_1, \mathbf{P}_2, \cdots\}$
of propositional symbols to $H$. Then, we can
evaluate recursively the truth value, $P_H[v]$, in $H$ of
any proposition, $P$,
with respect to the truth assigment, $v$, as follows:
\begin{align*}
(\mathbf{P}_i)_H[v] & = v(P) \\
\perp_H[v] & = 0\\
\top_H[v] & = 1\\
(P\lor Q)_H[v] & =  P_H[v] \lor P_H[v] \\
(P\land Q)_H[v] & =  P_H[v] \land P_H[v] \\
(P\impl Q)_H[v] & = (P_H[v]\rightarrow P_H[v]) \\
(\neg P)_H[v] & =  (P_H[v]\rightarrow 0).
\end{align*} 
In the equations above, on the
right hand side, $\lor$,   $\land$ and $\rightarrow$  are the operations
of the Heyting algebra, $H$.
We say that a proposition, $P$, is {\it valid 
in the Heyting algebra $H$ (or $H$-valid)\/} if $P_H[v] = 1$
for all truth assigments, $v$.
We say that $P$ is {\it HA-valid (or intuitionistically valid)\/} if $P$ is $H$-valid
in all Heyting algebras, $H$. As in the classical case, 
it can be shown that every
intuitionistically provable proposition is 
HA-valid. This property is called
{\it soundness\/}. Conversely, if
$P$ is HA-valid, then it is 
intuitionistically provable. This second property
is called {\it completeness\/}. 
A stronger completeness result actually holds:
If a proposition is $H$-valid in all {\it finite\/} 
Heyting algebras, $H$, then it is intuitionistically provable.
As a consequence, if a proposition is {\it not\/} provable
intuitionistically, then it can be falsified in some finite
Heyting algebra.

\remark
If $X$ is any set, a {\it topology on $X$\/} is a family, $\s{O}$,
of subsets of $X$ satisfying the following conditions:
\begin{enumerate}
\item[(1)]
$\emptyset\in \s{O}$ and $X\in \s{O}$;
\item[(2)]
For every family (even infinite), $(U_i)_{i\in I}$,
of sets $U_i\in \s{O}$, we have
$\bigcup_{i\in I} U_i\in \s{O}$.
\item[(3)]
For every {\it finite\/} family,  $(U_i)_{1 \leq i\leq n}$,
of sets $U_i\in \s{O}$, we have
$\bigcap_{1 \leq i\leq n} U_i\in \s{O}$.
\end{enumerate}

\medskip
Every subset in $\s{O}$ is called an {\it open subset\/} of $X$
(in the topology $\s{O}$) .
The pair, $\lag X, \s{O}\rag$, is called a {\it topological space\/}.
Given any subset, $A$, of $X$, the union of all open subsets 
contained in $A$ is the largest open subset of $A$ and is denoted
$\interio{A}$. 

\medskip
Given a topological space,  $\lag X, \s{O}\rag$, 
we claim that $\s{O}$ with the inclusion ordering
is a Heyting algebra with $0 = \emptyset$; $1 = X$;
$\lor = \cup$ (union);
$\land = \cap$ (intersection); and with
\[
(U \rightarrow V ) = \overbrace{(X - U)\cup V}^{\circ}.
\]
(Here, $X - U$ is the complement of $U$ in $X$.)
In this Heyting algebra, we have
\[ 
\overline{U} = \overbrace{X - U}^{\circ}.
\]
Since $X - U$ is usually not open, we generally have
$\overline{\overline{U}} \not= U$.
Therefore, we see that topology yields another 
supply of Heyting algebras.

\chapter[Graphs, Basic Notions]
{Graphs, Basic Notions}
\label{chap5}
\section[Why Graphs? Some Motivations]
{Why Graphs? Some Motivations}
\label{sec20}
Graphs are mathematical structures that have
many applications to computer science,
electrical engineering and
more widely to engineering as a whole, but also to
sciences such as
biology, linguistics, and sociology, among others.
For example, relations among objects can usually be encoded
by graphs. Whenever a system has a notion of state
and state transition function, graph methods may be
applicable. Certain problems are naturally modeled
by undirected graphs whereas  others require
directed graphs. Let us give a concrete example.

\medskip
Suppose a city decides to create a public-transportation system.
It would be desirable if this system allowed
transportation between certain locations considered
important. Now, if this system consists of buses,
the traffic will probably get worse so the city engineers
decide that the traffic will be improved 
by making certain streets one-way streets.
The problem then is, given a map of the city
consisting of the important locations and of the two-way
streets linking them, find an orientation of the
streets so that it is still possible to travel 
between any two locations. The problem requires
finding a directed graph, given an undirected graph.
Figure \ref{graphfig1a} shows the undirected graph 
corresponding to the city map and Figure \ref{graphfig1b}
shows a proposed choice of
one-way streets. Did the engineers do a good job
or are there locations such that it is impossible to travel from 
one to the other while respecting the one-way signs?

\medskip
The answer to this puzzle will be revealed in Section 
\ref{sec22}.
\begin{figure}
  \begin{center}
    \begin{pspicture}(0,0)(10,6.5)
    \cnodeput(0,0){v16}{$16$}
    \cnodeput(2,0){v17}{$17$}
    \cnodeput(6,0){v18}{$18$}
    \cnodeput(10,0){v19}{$19$}
    \cnodeput(0,2){v10}{$10$}
    \cnodeput(2,2){v11}{$11$}
    \cnodeput(4,2){v12}{$12$}
    \cnodeput(6,2){v13}{$13$}
    \cnodeput(8,2){v14}{$14$}
    \cnodeput(10,2){v15}{$15$}
    \cnodeput(0,4){v5}{$5$}
    \cnodeput(4,4){v6}{$6$}
    \cnodeput(6,4){v7}{$7$}
    \cnodeput(8,4){v8}{$8$}
    \cnodeput(10,4){v9}{$9$}
    \cnodeput(0,6){v1}{$1$}
    \cnodeput(6,6){v2}{$2$}
    \cnodeput(8,6){v3}{$3$}
    \cnodeput(10,6){v4}{$4$}
    \ncline[linewidth=1pt]{v11}{v10}
    \ncline[linewidth=1pt]{v12}{v11}
    \ncline[linewidth=1pt]{v13}{v12}
    \ncline[linewidth=1pt]{v13}{v14}
    \ncline[linewidth=1pt]{v14}{v15}
    \ncline[linewidth=1pt]{v12}{v6}
    \ncline[linewidth=1pt]{v14}{v8}
    \ncline[linewidth=1pt]{v6}{v5}
    \ncline[linewidth=1pt]{v7}{v6}
    \ncline[linewidth=1pt]{v8}{v7}
    \ncline[linewidth=1pt]{v8}{v9}
    \ncline[linewidth=1pt]{v7}{v2}
    \ncline[linewidth=1pt]{v8}{v3}
    \ncline[linewidth=1pt]{v11}{v17}
    \ncline[linewidth=1pt]{v13}{v18}
    \ncarc[arcangle=20, linewidth=1pt]{v16}{v17}
    \ncarc[arcangle=20, linewidth=1pt]{v17}{v16}
    \ncarc[arcangle=10, linewidth=1pt]{v17}{v18}
    \ncarc[arcangle=10, linewidth=1pt]{v18}{v17}
    \ncarc[arcangle=10, linewidth=1pt]{v18}{v19}
    \ncarc[arcangle=10, linewidth=1pt]{v19}{v18}
    \ncarc[arcangle=20, linewidth=1pt]{v16}{v10}
    \ncarc[arcangle=20, linewidth=1pt]{v10}{v16}
    \ncarc[arcangle=20, linewidth=1pt]{v10}{v5}
    \ncarc[arcangle=20, linewidth=1pt]{v5}{v10}
    \ncarc[arcangle=20, linewidth=1pt]{v5}{v1}
    \ncarc[arcangle=20, linewidth=1pt]{v1}{v5}
    \ncarc[arcangle=20, linewidth=1pt]{v19}{v15}
    \ncarc[arcangle=20, linewidth=1pt]{v15}{v19}
    \ncarc[arcangle=20, linewidth=1pt]{v15}{v9}
    \ncarc[arcangle=20, linewidth=1pt]{v9}{v15}
    \ncarc[arcangle=20, linewidth=1pt]{v9}{v4}
    \ncarc[arcangle=20, linewidth=1pt]{v4}{v9}
    \ncarc[arcangle=10, linewidth=1pt]{v1}{v2}
    \ncarc[arcangle=10, linewidth=1pt]{v2}{v1}
    \ncarc[arcangle=20, linewidth=1pt]{v2}{v3}
    \ncarc[arcangle=20, linewidth=1pt]{v3}{v2}
    \ncarc[arcangle=20, linewidth=1pt]{v3}{v4}
    \ncarc[arcangle=20, linewidth=1pt]{v4}{v3}
    \ncarc[arcangle=20, linewidth=1pt]{v7}{v13}
    \ncarc[arcangle=-20, linewidth=1pt]{v7}{v13}
    \end{pspicture}
  \end{center}
  \caption{An undirected graph modeling a city map}
\label{graphfig1a}
\end{figure}
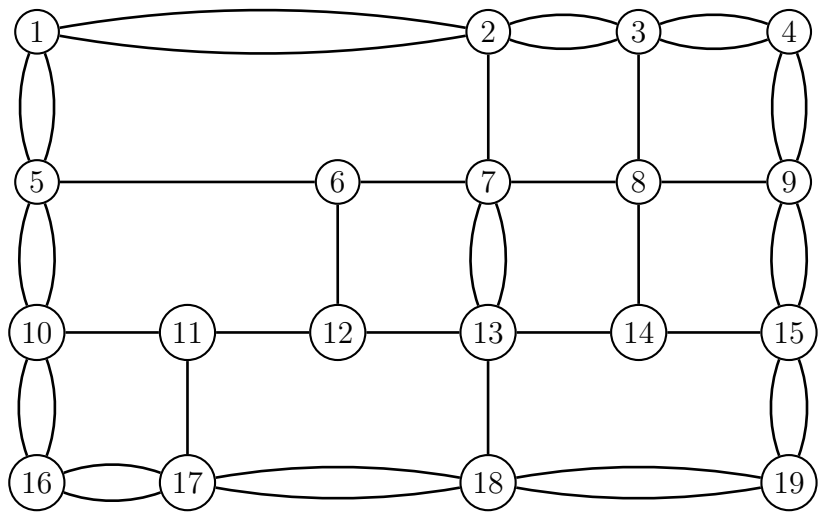
\begin{figure}
  \begin{center}
    \begin{pspicture}(0,0)(10,6.5)
    \cnodeput(0,0){v16}{$16$}
    \cnodeput(2,0){v17}{$17$}
    \cnodeput(6,0){v18}{$18$}
    \cnodeput(10,0){v19}{$19$}
    \cnodeput(0,2){v10}{$10$}
    \cnodeput(2,2){v11}{$11$}
    \cnodeput(4,2){v12}{$12$}
    \cnodeput(6,2){v13}{$13$}
    \cnodeput(8,2){v14}{$14$}
    \cnodeput(10,2){v15}{$15$}
    \cnodeput(0,4){v5}{$5$}
    \cnodeput(4,4){v6}{$6$}
    \cnodeput(6,4){v7}{$7$}
    \cnodeput(8,4){v8}{$8$}
    \cnodeput(10,4){v9}{$9$}
    \cnodeput(0,6){v1}{$1$}
    \cnodeput(6,6){v2}{$2$}
    \cnodeput(8,6){v3}{$3$}
    \cnodeput(10,6){v4}{$4$}
    \ncline[linewidth=1pt]{->}{v11}{v10}
    \ncline[linewidth=1pt]{->}{v12}{v11}
    \ncline[linewidth=1pt]{->}{v13}{v12}
    \ncline[linewidth=1pt]{->}{v13}{v14}
    \ncline[linewidth=1pt]{->}{v14}{v15}
    \ncline[linewidth=1pt]{->}{v12}{v6}
    \ncline[linewidth=1pt]{->}{v14}{v8}
    \ncline[linewidth=1pt]{->}{v6}{v5}
    \ncline[linewidth=1pt]{->}{v7}{v6}
    \ncline[linewidth=1pt]{->}{v8}{v7}
    \ncline[linewidth=1pt]{->}{v8}{v9}
    \ncline[linewidth=1pt]{->}{v7}{v2}
    \ncline[linewidth=1pt]{->}{v8}{v3}
    \ncline[linewidth=1pt]{->}{v11}{v17}
    \ncline[linewidth=1pt]{->}{v13}{v18}
    \ncarc[arcangle=20, linewidth=1pt]{->}{v16}{v17}
    \ncarc[arcangle=20, linewidth=1pt]{->}{v17}{v16}
    \ncarc[arcangle=10, linewidth=1pt]{->}{v17}{v18}
    \ncarc[arcangle=10, linewidth=1pt]{->}{v18}{v17}
    \ncarc[arcangle=10, linewidth=1pt]{->}{v18}{v19}
    \ncarc[arcangle=10, linewidth=1pt]{->}{v19}{v18}
    \ncarc[arcangle=20, linewidth=1pt]{->}{v16}{v10}
    \ncarc[arcangle=20, linewidth=1pt]{->}{v10}{v16}
    \ncarc[arcangle=20, linewidth=1pt]{->}{v10}{v5}
    \ncarc[arcangle=20, linewidth=1pt]{->}{v5}{v10}
    \ncarc[arcangle=20, linewidth=1pt]{->}{v5}{v1}
    \ncarc[arcangle=20, linewidth=1pt]{->}{v1}{v5}
    \ncarc[arcangle=20, linewidth=1pt]{->}{v19}{v15}
    \ncarc[arcangle=20, linewidth=1pt]{->}{v15}{v19}
    \ncarc[arcangle=20, linewidth=1pt]{->}{v15}{v9}
    \ncarc[arcangle=20, linewidth=1pt]{->}{v9}{v15}
    \ncarc[arcangle=20, linewidth=1pt]{->}{v9}{v4}
    \ncarc[arcangle=20, linewidth=1pt]{->}{v4}{v9}
    \ncarc[arcangle=10, linewidth=1pt]{->}{v1}{v2}
    \ncarc[arcangle=10, linewidth=1pt]{->}{v2}{v1}
    \ncarc[arcangle=20, linewidth=1pt]{->}{v2}{v3}
    \ncarc[arcangle=20, linewidth=1pt]{->}{v3}{v2}
    \ncarc[arcangle=20, linewidth=1pt]{->}{v3}{v4}
    \ncarc[arcangle=20, linewidth=1pt]{->}{v4}{v3}
    \ncarc[arcangle=20, linewidth=1pt]{->}{v7}{v13}
    \ncarc[arcangle=-20, linewidth=1pt]{->}{v7}{v13}
    \end{pspicture}
  \end{center}
  \caption{A choice of one-way streets}
\label{graphfig1b}
\end{figure}
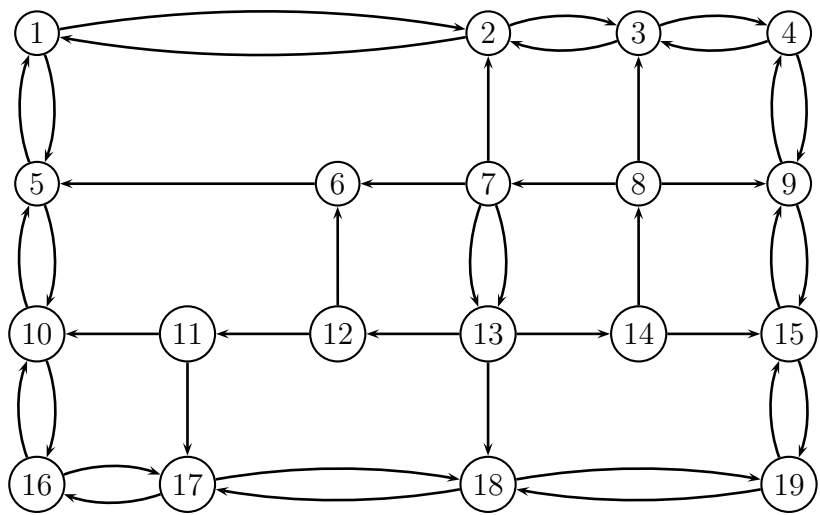

\medskip
There is a peculiar aspect of graph theory having to do
with its terminology. Indeed, unlike most  branches
of mathematics, it appears that the terminology of
graph theory is not  standardized, yet. This can be quite
confusing to the beginner who has to struggle
with many different and often inconsistent terms
denoting the same concept, one of the worse being the
notion of a {\it path\/}. Our attitude has been to
use terms that we feel are as simple as possible.
As a result, we have not followed a single book.
Among the many books on graph theory, we have 
been inspired by the classic texts, Harary \cite{Harary},
Berge \cite{Berge} and Bollobas \cite{Bollobas}.
This chapter on graphs is heavily inspired by
Sakarovitch \cite{Sakarovitch1}, because
we find Sakarovitch's book  extremely clear and because it has more emphasis
on applications than the previous two.
Another more recent (and more advanced) text which
is also excellent is Diestel \cite{Diestel}.

\medskip
Many books begin by discussing undirected graphs and 
introduce directed graph only later on. We disagree with this approach.
Indeed, we feel that the notion of a directed graph is more fundamental
than the notion of an undirected graph. For one thing, 
a unique undirected graph is obtained from a directed graph by
forgetting the direction of the arcs, whereas there are many
ways of orienting an undirected graph. Also, in general, 
we believe that most definitions about directed graphs are cleaner
than the corresponding ones for undirected graphs
(for instance, we claim that
the definition of a directed graph is simpler than the definition
of an undirected graph, and similarly for paths).
Thus, we begin with directed graphs.

\section{Directed Graphs}
\label{sec21}
Informally, a directed graph consists of
a set of nodes together with a set of oriented arcs (also called edges)
between these nodes. Every arc has a single source (or initial point)
and a single target (or endpoint), both of which are nodes.
There are various ways of formalizing what a directed graph is
and some decisions must be made. Two issues must be confronted:
\begin{enumerate}
\item 
Do we allow ``loops,'' that is, arcs whose source and target 
are identical?
\item
Do we allow ``parallel arcs,'' that is distinct arcs having the
same source and target?
\end{enumerate}

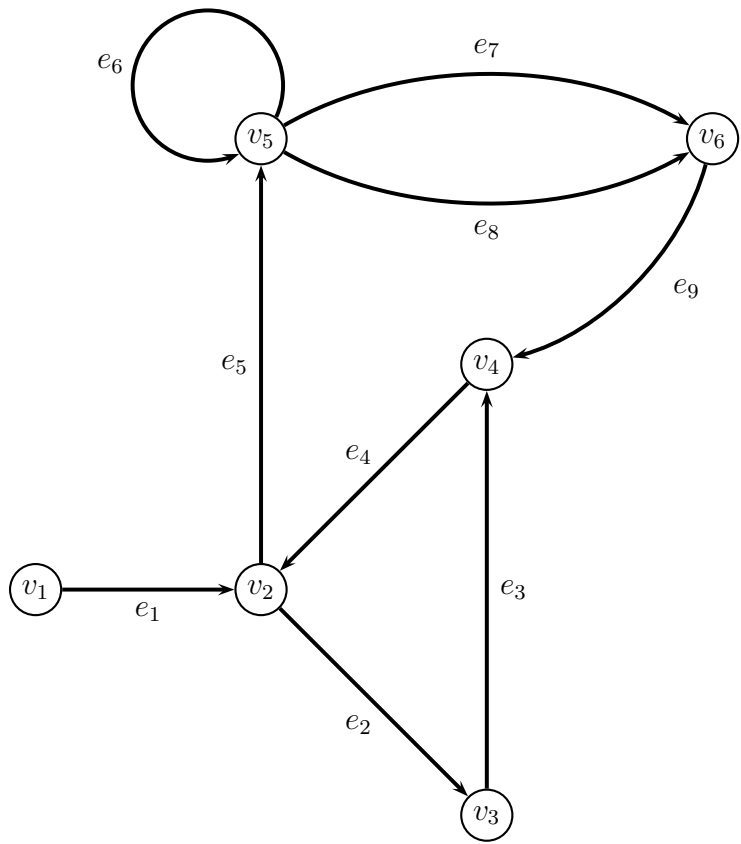
\begin{figure}
  \begin{center}
    \begin{pspicture}(0,-3)(9,7.7)
    \cnodeput(0,0){v1}{$v_1$}
    \cnodeput(3,0){v2}{$v_2$}
    \cnodeput(6,-3){v3}{$v_3$}
    \cnodeput(6,3){v4}{$v_4$}
    \cnodeput(3,6){v5}{$v_5$}
    \cnodeput(9,6){v6}{$v_6$}
    \ncline[linewidth=1.5pt]{->}{v1}{v2}
    \ncline[linewidth=1.5pt]{->}{v2}{v3}
    \ncline[linewidth=1.5pt]{->}{v3}{v4}
    \ncline[linewidth=1.5pt]{->}{v4}{v2}
    \ncline[linewidth=1.5pt]{->}{v2}{v5}
    \ncarc[arcangle=30, linewidth=1.5pt]{->}{v5}{v6}
    \aput{:U}{$e_7$}
    \ncarc[arcangle=-30, linewidth=1.5pt]{->}{v5}{v6}
    \bput{:U}{$e_8$}
    \ncarc[arcangle=30, linewidth=1.5pt]{->}{v6}{v4}
    \nccircle[linewidth=1.5pt,angleA=45]{->}{v5}{1cm}
    \uput[0](8.3,4){$e_9$}
    \uput[-90](1.5,0){$e_1$}
    \uput[-120](4.5,-1.5){$e_2$}
    \uput[0](6,0){$e_3$}
    \uput[120](4.5,1.5){$e_4$}
    \uput[180](3,3){$e_5$}
    \uput[90](1,6.7){$e_6$}
    \end{pspicture}
  \end{center}
  \caption{A directed graph, $G_1$}
\label{graphfig1}
\end{figure}

\medskip
Since every binary relation on a set can be represented as
a directed graph with loops, our definition allows loops.
Since the directed graphs used in automata theory must
accomodate parallel arcs (usually labeled with different
symbols), our definition also allows parallel arcs.
Thus, we choose a more inclusive definition in order to accomodate
as many applications as possible, even though some authors
place restrictions on the definition of a graph, for example,
forbidding loops and parallel arcs (we will call such graphs,
simple graphs).
Before giving a formal definition, let us say that
graphs are usually depicted by drawings (graphs!) where
the nodes are represented by circles containing the node name
and oriented line segments labeled with their arc name
(see Figure \ref{graphfig1}).

\begin{defin}
\label{graph1}
{\em 
A {\it directed graph\/} (or {\it digraph\/}) is a quadruple,
$G = (V, E, s, t)$, where $V$ is a set of {\it nodes or vertices\/},
$E$ is a set of {\it arcs or edges\/} and
$\mapdef{s, t}{E}{V}$ are two functions, $s$ being called the
{\it source function\/}  and $t$ the {\it target function\/}.
Given an edge $e\in E$, we also call $s(e)$ the {\it origin\/}
or {\it source\/} of $e$, and $t(e)$ the {\it endpoint\/}
or {\it target\/} of $e$.
}
\end{defin}

\medskip
If the context makes it clear that we are dealing only with
directed graphs, we usually say simply ``graph'' instead
of ``directed graph''.
A directed graph, $G = (V, E, s, t)$,  is {\it finite\/} 
iff both $V$ and $E$ are finite. In this case, $|V|$,
the number of nodes of $G$ is called the {\it order\/} of $G$.

\medskip
{\bf Example\/}: Let $G_1$ be the directed graph defined such that

\medskip
$E = \{e_1, e_2, e_3, e_4, e_5, e_6, e_7, e_8, e_9\},$

\smallskip
$V = \{v_1, v_2, v_3, v_4, v_5, v_6\},$ and
\begin{align*}
 s(e_1) & = v_1, s(e_2) = v_2, s(e_3) = v_3, s(e_4) = v_4,\\
 s(e_5) & = v_2, s(e_6) = v_5, s(e_7) = v_5, s(e_8) = v_5, s(e_9) =  v_6\\
 t(e_1) & = v_2, t(e_2) = v_3, t(e_3) = v_4, t(e_4) = v_2,\\
 t(e_5) & = v_5, t(e_6) = v_5, t(e_7) = v_6, t(e_8) = v_6,
 t(e_9) =  v_4.
\end{align*}

The graph $G_1$ is represented by the  diagram shown in
Figure \ref{graphfig1}.

\medskip
It should be noted that there are many different ways of ``drawing''
a graph. Obviously, we would like as much as possible to avoid
having too many intersecting arrows but this is not always possible if
we insist in drawing a graph on a sheet of paper (on the plane).

\begin{defin}
\label{graph2}
{\em 
Given a directed graph, $G$,
an edge, $e\in E$, such that $s(e) = t(e)$ is called
a {\it loop\/} (or {\it self-loop\/}).
Two edges, $e, e'\in E$ are said to be {\it parallel edges\/}
iff $s(e) = s(e')$ and $t(e) = t(e')$.
A directed graph is {\it simple\/} iff it has no parallel edges.
}
\end{defin}

\remarks
\begin{enumerate}
\item
The functions $s, t$ need not be injective
or surjective. Thus, we allow ``isolated vertices'',
that is, vertices that are not the  source or the target of any edge.
\item
When $G$ is simple, every edge, $e\in E$, is uniquely determined
by the ordered pair of vertices, $(u, v)$, such that
$u = s(e)$ and $v = t(e)$. In this case, we may denote the
edge $e$ by $(u v)$ 
(some books also use the notation  $u v$). 
Also, a graph without
parallel edges can be defined as a pair,
$(V, E)$, with $E\subseteq V\times V$. In other words,
a simple graph is equivalent to a binary relation
on a set ($E \subseteq V\times V$). This definition is often
the one used to define  directed graphs. 
\item
Given any edge, $e\in E$, the nodes $s(e)$ and $t(e)$ are 
often called the {\it boundaries\/} of $e$
and the expression $t(e) - s(e)$ is called the {\it boundary of $e$\/}.
\item
Given a graph, $G = (V, E, s, t)$, we may also
write $V(G)$ for $V$ and $E(G)$ for $E$.
Sometimes, we even  drop $s$ and $t$ and write simply
$G = (V, E)$ instead of $G = (V, E, s, t)$.
\item
Some authors define a simple graph to be a graph without
loops and without parallel edges.
\end{enumerate}

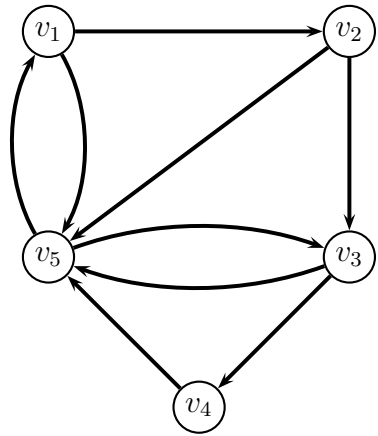
\begin{figure}
  \begin{center}
    \begin{pspicture}(0,0)(4,5.3)
    \cnodeput(2,0){v4}{$v_4$}
    \cnodeput(0,2){v5}{$v_5$}
    \cnodeput(4,2){v3}{$v_3$}
    \cnodeput(0,5){v1}{$v_1$}
    \cnodeput(4,5){v2}{$v_2$}
    \ncline[linewidth=1.5pt]{->}{v1}{v2}
    \ncline[linewidth=1.5pt]{->}{v2}{v3}
    \ncline[linewidth=1.5pt]{->}{v2}{v5}
    \ncline[linewidth=1.5pt]{->}{v3}{v4}
    \ncline[linewidth=1.5pt]{->}{v4}{v5}
    \ncarc[arcangle=30, linewidth=1.5pt]{->}{v1}{v5}
    \ncarc[arcangle=30, linewidth=1.5pt]{->}{v5}{v1}
    \ncarc[arcangle=20, linewidth=1.5pt]{->}{v5}{v3}
    \ncarc[arcangle=20, linewidth=1.5pt]{->}{v3}{v5}
    \end{pspicture}
  \end{center}
  \caption{A directed graph, $G_2$}
\label{graphfig2}
\end{figure}

\medskip
Observe that the graph $G_1$ has the loop $e_6$ and the two
parallel edges $e_7$ and $e_8$.
When we draw pictures of graphs, we often omit the edge names
(sometimes even the node names) as illustrated in Figure 
\ref{graphfig2}.

\begin{defin}
\label{graph3}
{\em 
Given a directed graph, $G$, for any edge $e\in E$,
if $u = s(e)$ and $v = t(e)$, we say that
\begin{enumerate}
\item[(i)]
The nodes $u$ and $v$ are {\it adjacent\/}
\item[(ii)]
The nodes $u$ and $v$ are {\it incident to the arc $e$\/}
\item[(iii)]
The arc $e$ is  {\it incident to the nodes $u$ and $v$\/}
\item[(iv)]
Two edges, $e, e'\in E$ are {\it adjacent\/} if they are incident to
some common node (that is, either $s(e) = s(e')$ or
$t(e) = t(e')$ or $t(e) = s(e')$ or $s(e) = t(e')$).
\end{enumerate}
For any node, $u\in V$, set
\begin{enumerate}
\item[(a)]
$d_G^+(u) = |\{e\in E \mid s(e) = u\}|\>$, the
{\it outer half-degree or outdegree of $u$\/}   
\item[(b)]
$d_G^-(u) = |\{e\in E \mid t(e) = u\}|\>$, the
{\it inner half-degree or indegree of $u$\/}   
\item[(c)]
$d_G(u) = d_G^+(u) + d_G^-(u)\>$, the {\it degree of $u$\/}.
\end{enumerate}
A graph is {\it regular\/} iff
every node has the same degree.
}
\end{defin}

\medskip
Note that $d_G^+$ (respectively $d_G^-(u)$) counts the number
of arcs ``coming out from $u$'', that is, 
whose source is $u$
(resp. counts the number
of arcs ``coming into $u$'', that is, 
whose target is $u$). For example,
in the graph of Figure \ref{graphfig2},
$d_{G_2}^+(v_1) = 2$, $d_{G_2}^-(v_1) = 1$, 
$d_{G_2}^+(v_5) = 2$, $d_{G_2}^-(v_5) = 4$,
$d_{G_2}^+(v_3) = 2$, $d_{G_2}^-(v_3) = 2$.
Neither $G_1$ nor $G_2$ are regular graphs.

\medskip
The first result of graph theory is the following simple
but very useful proposition:

\begin{prop}
\label{graphp1}
For any finite graph, $G = (V, E, s, t)$, we have
\[
\sum_{u\in V} d_G^+(u) = \sum_{u\in V} d_G^-(u).
\]
\end{prop}

\proof
Every arc, $e\in E$, has a single source and a single target
and each side of the above equations simply counts the number
of edges in the graph.
$\bigsquare$

\begin{cor}
\label{graphp2}
For any finite graph, $G = (V, E, s, t)$, we have
\[
\sum_{u\in V} d_G(u) =  2 |E|,
\]
that is, the sum of the degrees of all the nodes is equal to twice
the number of edges.
\end{cor}

\begin{cor}
\label{graphp3}
For any finite graph, $G = (V, E, s, t)$,
there is an even number of nodes with an odd degree.
\end{cor}

\medskip
The notion of homomorphism and isomorphism of graphs
is fundamental.

\begin{defin}
\label{graph4}
{\em 
Given two directed graphs, $G_1 = (V_1, E_1, s_1, t_1)$
and $G_2 = (V_2, E_2, s_2, t_2)$, a {\it homomorphism\/}
(or {\it morphism\/}), $\mapdef{f}{G_1}{G_2}$,
{\it from $G_1$ to $G_2$\/} is a pair, 
$f = (f^v, f^e)$, with $\mapdef{f^v}{V_1}{V_2}$
and $\mapdef{f^e}{E_1}{E_2}$ preserving incidence, that is,
for every edge, $e\in E_1$, we have
\[
s_2(f^e(e)) = f^v (s_1(e))
\quad\hbox{and}\quad
t_2(f^e(e)) = f^v (t_1(e)).
\]
These conditions can also be expressed by saying that
the following two diagrams commute:
\[
\xymatrix{
E_1 \ar[r]^-{f^e} \ar[d]_-{s_1} & \> E_2 \ar[d]^-{s_2} \\
V_1 \ar[r]_-{f^v} & \> V_2
}
\qquad\qquad\qquad
\xymatrix{
E_1 \ar[r]^-{f^e} \ar[d]_-{t_1} & \> E_2 \ar[d]^-{t_2} \\
V_1 \ar[r]_-{f^v} & \> V_2 .
}
\]
}
\end{defin}

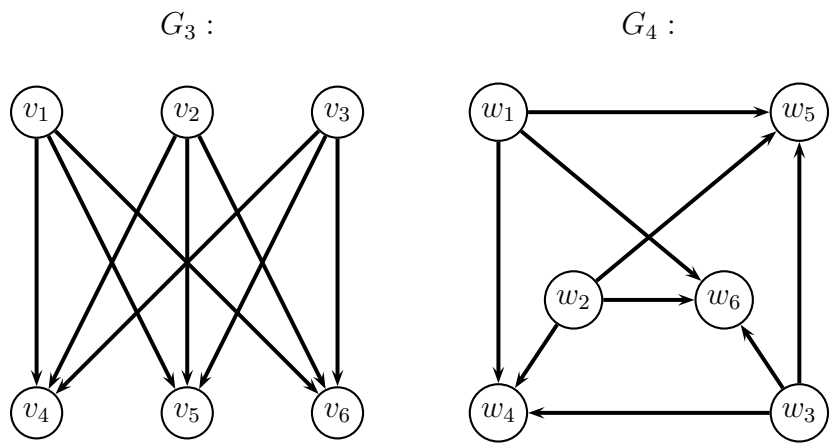
\begin{figure}
  \begin{center}
    \begin{pspicture}(0,0)(4,5.2)
    \cnodeput(0,0){v4}{$v_4$}
    \cnodeput(2,0){v5}{$v_5$}
    \cnodeput(4,0){v6}{$v_6$}
    \cnodeput(0,4){v1}{$v_1$}
    \cnodeput(2,4){v2}{$v_2$}
    \cnodeput(4,4){v3}{$v_3$}
    \ncline[linewidth=1.5pt]{->}{v1}{v4}
    \ncline[linewidth=1.5pt]{->}{v1}{v5}
    \ncline[linewidth=1.5pt]{->}{v1}{v6}
    \ncline[linewidth=1.5pt]{->}{v2}{v4}
    \ncline[linewidth=1.5pt]{->}{v2}{v5}
    \ncline[linewidth=1.5pt]{->}{v2}{v6}
    \ncline[linewidth=1.5pt]{->}{v3}{v4}
    \ncline[linewidth=1.5pt]{->}{v3}{v5}
    \ncline[linewidth=1.5pt]{->}{v3}{v6}
    \uput[90](2,4.8){$G_3:$}
    \end{pspicture}
\hskip 2 cm
    \begin{pspicture}(0,0)(4,5.2)
    \cnodeput(0,0){w4}{$w_4$}
    \cnodeput(4,4){w5}{$w_5$}
    \cnodeput(3,1.5){w6}{$w_6$}
    \cnodeput(0,4){w1}{$w_1$}
    \cnodeput(1,1.5){w2}{$w_2$}
    \cnodeput(4,0){w3}{$w_3$}
    \ncline[linewidth=1.5pt]{->}{w1}{w4}
    \ncline[linewidth=1.5pt]{->}{w1}{w5}
    \ncline[linewidth=1.5pt]{->}{w1}{w6}
    \ncline[linewidth=1.5pt]{->}{w2}{w4}
    \ncline[linewidth=1.5pt]{->}{w2}{w5}
    \ncline[linewidth=1.5pt]{->}{w2}{w6}
    \ncline[linewidth=1.5pt]{->}{w3}{w4}
    \ncline[linewidth=1.5pt]{->}{w3}{w5}
    \ncline[linewidth=1.5pt]{->}{w3}{w6}
    \uput[90](2,4.8){$G_4:$}
    \end{pspicture}
  \end{center}
  \caption{Two isomorphic graphs, $G_3$ and $G_4$}
\label{graphfig3}
\end{figure}

\medskip
Given three graphs, $G_1, G_2, G_3$ and two homomorphisms,
$\mapdef{f}{G_1}{G_2}$ and $\mapdef{g}{G_2}{G_3}$, with
$f = (f^v, f^e)$ and $g = (g^v, g^e)$, it is easily checked
that $(g^v\circ f^v, g^e\circ f^e)$ is a homomorphism
from $G_1$ to $G_3$. The homomorphism  $(g^v\circ f^v, g^e\circ f^e)$ is
denoted $g\circ f$.
Also, for any graph, $G$, the map $\id_G = (\id_V, \id_E)$
is a homomorphism called the {\it identity homomorphism\/}.
Then, a homomorphism, $\mapdef{f}{G_1}{G_2}$, is an
{\it isomorphism\/} iff there is a homomorphism,
$\mapdef{g}{G_2}{G_1}$, such that
\[
g\circ f = \id_{G_1}
\quad\hbox{and}\quad
f\circ g = \id_{G_2}.
\]
In this case, $g$ is unique and it is called the {\it inverse\/} of
$f$ and denoted $f^{-1}$. If $f = (f^v, f^e)$ is
an isomorphism, we see immediately that $f^v$ and $f^e$
are bijections.  Checking whether two finite graphs
are isomorphic is not as easy as it looks. In fact,
no general efficient algorithm for checking graph isomorphism 
is known at this time and determining the exact complexity
of this problem is a major open question in computer science.
For example, the graphs $G_3$ and $G_4$ shown in Figure
\ref{graphfig3} are isomorphic. The bijection $f^v$ is given 
by $f^v(v_i) = w_i$, for $i = 1, \ldots, 6$ and the reader
will easily figure out the bijection on arcs.
As we can see, isomorphic graphs can look quite different.

\section[Path in Digraphs; Strongly Connected Components]
{Paths in Digraphs; Strongly Connected Components}
\label{sec22}
Many problems about graphs
can be formulated  as  path existence problems.
Given a directed graph, $G$, intuitively, a path from a node
$u$ to a node $v$ is a way to travel from $u$ in $v$ 
by following edges of the graph that ``link up correctly''.
Unfortunately, if we look up the definition of a path
in two different graph theory books, we are almost guaranteed
to find different and usually clashing definitions!
This has to do with the fact that for some authors, a path
may not use the same edge more than once and for others,
a path may not pass through the same  node more than once.
Moreover, when parallel edges are present (i.e.. when a graph is not 
simple), a sequence of nodes does not define a path unambiguously!.

\medskip
The terminology that we have chosen may not be standard,
but it is used by a number of authors (some very distinguished,
for example, Fields medalists!) and we believe that
it is less taxing on one's memory (however, this
point is probably the most debatable).

\begin{defin}
\label{graph5}
{\em 
Given any digraph, $G = (V, E, s, t)$, and any two nodes, $u, v\in V$,
a {\it path  from $u$ to $v$\/} is a triple,
$\pi = (u, e_1\cdots e_n, v)$, where $n \geq 1$ and 
$e_1 \cdots e_n$ is a sequence of edges, $e_i \in E$
(i.e., a nonempty string in $E^*$), such that
\[
s(e_1) = u; \> t(e_n) = v; \>
t(e_i) = s(e_{i + 1}),\> 1 \leq i \leq n - 1.
\]
We call $n$ the {\it length of the path $\pi$\/}
and we write $|\pi| = n$. 
When $n = 0$, we have the {\it null path\/},
$(u, \epsilon, u)$, from $u$ to $u$ (recall, $\epsilon$ denotes
the empty string); the null path has length $0$.
If $u = v$, then $\pi$
is called a {\it closed path\/}, else an {\it open path\/}.
The path, $\pi = (u, e_1\cdots e_n, v)$, determines the sequence
of nodes, $\mathrm{nodes}(\pi) = \lag u_0, \ldots, u_n\rag$,
where $u_0 = u$, $u_n = v$ and
$u_i= t(e_i)$, for $1 \leq i \leq n - 1$.
We also set $\mathrm{nodes}((u, \epsilon, u)) = \lag u, u\rag$.
A path, $\pi = (u, e_1\cdots e_n, v)$, 
is {\it simple\/} iff $e_i \not= e_j$ for all $i\not= j$
(i.e., no edge in the path is used twice). 
A path, $\pi$, from $u$ to $v$ is {\it elementary\/}
iff no vertex in $\mathrm{nodes}(\pi)$ occurs twice,
except possibly for $u$ if $\pi$ is closed.
Equivalently, if
$\mathrm{nodes}(\pi) = \lag u_0, \ldots, u_n\rag$,
then $\pi$ is elementary iff either
\begin{enumerate}
\item
$u_i \not = u_j$ for all $i, j$ with $i \not= j$
and $0\leq i, j \leq n$, or $\pi$ is closed  ($u_0 = u_n$) in which case
\item
$u_i\not= u_0$ and $u_i \not= u_n$
for all $i$ with $1 \leq i \leq n -1$,
and $u_i \not = u_j$ for all $i, j$ with $i \not= j$
and $1\leq i, j \leq n - 1$.
\end{enumerate}
The null path, $(u, \epsilon, u)$, is considered simple and elementary. 
}
\end{defin}

\remarks
\begin{enumerate}
\item
Other authors use the term {\it walk\/} for what we call a path.
These authors also use the term {\it trail\/} for what we call
a simple path and the term {\it path\/} for what we call an 
elementary path.
\item
If a digraph is not simple, then 
even if a sequence of nodes is of the form
$\mathrm{nodes}(\pi)$ for some path, that sequence of nodes
does not uniquely determine a path. For example,
in the graph of Figure \ref{graphfig1}, the sequence
$\lag v_2, v_5, v_6\rag$ corresponds to the two distinct paths
$(v_2, e_5e_7, v_6)$ and $(v_2, e_5e_8, v_6)$.
\end{enumerate}

\medskip
In the graph $G_1$ from Figure \ref{graphfig1},
\[
(v_2, e_5e_7e_9e_4e_5e_8, v_6)
\]
is a path from $v_2$ to $v_6$ which is neither simple nor elementary,
\[
(v_2, e_2e_3e_4e_5,  v_5)
\]
is a simple path from $v_2$ to $v_5$ which is not elementary and
\[
(v_2, e_5e_7e_9,  v_4), \qquad
(v_2, e_5e_7e_9e_4,  v_2)
\]
are elementary paths, the first one open and the second one closed.

\medskip
Recall  the notion of subsequence of a sequence
defined just before stating Theorem \ref{Erdos1}.
Then, if  $\pi = (u, e_1 \cdots e_n, v)$
is any path from $u$ to $v$ in a digraph, $G$, a {\it subpath\/} of 
$\pi$ is any path $\pi' = (u, e_1'\cdots e_m', v)$
such that $e_1',\ldots, e_m'$ is a subsequence of
$e_1, \ldots, e_n$.
The following simple proposition is actually very important:

\begin{prop}
\label{graphp4}
Let  $G$ be any digraph. 
(a)
For any two nodes, $u, v$, in $G$,
every non-null path, $\pi$, from $u$ to $v$ contains
an elementary non-null subpath.

\medskip
(b)
If $|V| = n$, then every open elementary path  has length
at most $n - 1$ and every closed elementary path 
has length at most $n$.
\end{prop}

\proof
(a)
Let $\pi$ be any non-null path from $u$ to $v$ in $G$ and let
\[
S = \{k \in \natnums \mid k = |\pi'|,
\quad\hbox{$\pi'$ is a non-null subpath of $\pi$}\}.
\]
The set $S\subseteq \natnums$ is nonempty since $|\pi|\in S$ and as
$\natnums$ is well-ordered, $S$ has a least element, say $m\geq 1$.
We claim that any  subpath of $\pi$ of length $m$ is
elementary. Consider any such path, say 
$\pi' = (u, e_1'\cdots e_m',v)$, let
\[
\mathrm{nodes}(\pi') = \lag v_0, \ldots, v_m\rag,
\]
and assume that  $\pi'$ is not elementary. There are two cases:
\begin{enumerate}
\item[(1)]
$u \not= v$. Then, 
some node occurs twice in $\mathrm{nodes}(\pi')$,
say $v_i = v_j$, with $i < j$.
Then, we can delete the path $(v_i, e_{i+1}'\cdots, e_{j}', v_j)$
from $\pi'$ to obtain a non-null 
(because $u \not= v$) subpath $\pi''$ of $\pi'$
from $u$ to $v$ 
with $|\pi''| = |\pi'| - (j - i)$
and since $i < j$, we see that $|\pi''| < |\pi'|$,
contradicting the minimality of $m$. Therefore,
$\pi'$ is a non-null elementary subpath of $\pi$.
\item[(2)]
$u = v$. In this case, either some node occurs twice in the sequence
$\lag v_0, \ldots, v_{m-1}\rag$
or some node  occurs twice in the sequence
$\lag v_1, \ldots, v_{m}\rag$.
In either case, as in (1), we can strictly shorten the
path from $v_0$ to $v_{m -1}$ or the path from
$v_1$ to $v_m$. Even though the resulting path may be the null path,
as one of two two edges  $e_1'$ or $e_m'$ remains from
the original path $\pi'$, we get a
non-null path from $u$ to $u$ strictly shorter than $\pi'$, 
contradicting the minimality of $\pi'$.
\end{enumerate}

\medskip
(b)
As in (a), let $\pi'$ be an open elementary path from $u$ to $v$ and let
\[
\mathrm{nodes}(\pi') = \lag v_0, \ldots, v_m\rag.
\]
If $m \geq n = |V|$, as the above sequence has $m + 1 > n$
nodes, by the Pigeonhole Principle, some node
must occur twice, contradicting the fact that $\pi'$
is an open elementary path. If $\pi'$ is
a non-null closed path and $m \geq n + 1$,
then by the Pigeonhole Principle, either some node
occurs twice in
$ \lag v_0, \ldots, v_m\rag$
or some node occurs twice in
$ \lag v_1, \ldots, v_{m+1}\rag$.
In either case, this contradicts the fact that
$\pi'$ is a non-null elementary path.
$\bigsquare$

\medskip
Like strings, paths can be concatenated.

\begin{defin}
\label{graph6}
{\em 
Two paths,
$\pi = (u, e_1\cdots e_m, v)$ and
$\pi' = (u', e_1'\cdots e_n', v')$  in a digraph $G$
can be {\it concatenated\/} iff
$v = u'$ in which case their {\it concatenation\/}, $\pi \pi'$, is the path
\[
\pi\pi' = (u, e_1\cdots e_m e_1'\cdots e_n', v').
\]
We also let
\[
(u, \epsilon, u)\pi = \pi = \pi (v, \epsilon, v).
\]  
}
\end{defin}

\medskip
Concatenation of paths is obviously associative and
observe that $|\pi\pi'| = |\pi| + |\pi'|$.

\begin{defin}
\label{graph7}
{\em 
Let $G = (V, E, s, t)$ be a digraph. We define the binary
relation, $\widehat{C}_G$, on $V$ as follows:
For all $u, v\in V$,
\[
u \widehat{C}_G v
\quad\hbox{iff}\quad
\hbox{there is a path from $u$ to $v$ and there
is a path from $v$ to $u$}.
\]
When $u \widehat{C}_G v$, we say that {\it $u$ and $v$
are strongly connected\/}.
}
\end{defin}

\medskip
Oberve that the relation $\widehat{C}_G$ is an equivalence relation.
It is reflexive because we have the null path from
$u$ to $u$, symmetric by definition, and transitive because
paths can be concatenated. The equivalence classes
of the relation $\widehat{C}_G$ are called the
{\it strongly connected components of $G$ (SCC's)\/}.
A graph is {\it strongly connected\/} iff it has
a single strongly connected component.

\medskip
For example, we see that the graph, $G_1$, of
Figure \ref{graphfig1} has two strongly connected components
\[
\{v_1\}, \quad \{v_2, v_3, v_4, v_5, v_6\},
\]
since there is a closed path
\[
(v_4, e_4e_2e_3e_4e_5e_7e_9, v_4).
\]
The graph $G_2$ of Figure \ref{graphfig2} is strongly connected.

\medskip
Let us give a simple algorithm for 
computing the strongly connected components of
a graph since this is often 
the key to solving many problems. The algorithm works as
follows: Given some vertex, $u\in V$, the algorithm
computes the two sets, $X^+(u)$ and $X^-(u)$,
where
\begin{eqnarray*}
X^+(u) & = &\{v\in V \mid \hbox{there exists a path from $u$ to $v$}\}\\
X^-(u) & = & \{v\in V \mid \hbox{there exists a path from $v$ to $u$}\}.
\end{eqnarray*}
Then, it is clear that the connected component, $C(u)$, or $u$,
is given by
$C(u) = C^+(u) \cap X^-(u)$.
For simplicity, 
we assume that $X^+(u),  X^-(u)$ and $C(u)$ are represented
by linear arrays.
In order to make sure that the algorithm makes progress,
we used a simple marking scheme. We use the variable
$\mathit{total}$ to count how many nodes are in
$X^+(u)$ (or in $X^-(u)$) and the variable
$\mathit{marked}$ to keep track of how many
nodes in $X^+(u)$ (or in $X^-(u)$) have been processed so far.
Whenever the algorithm
considers some unprocessed node, the first thing it does is to
increment $\mathit{marked}$ by $1$.
Here is the algorithm in high-level form.

\begin{tabbing}
\quad \= \quad \= \quad \= \quad \= \quad \= \quad \= \quad \\
{\bf function} $\mathit{strcomp}$($G$: graph; $u$: node): set \\
 \> {\bf begin} \\
 \> \>  $X^+(u)[1] : = u$; $X^-(u)[1] : = u$; 
  $\mathit{total} : = 1$; $\mathit{marked} := 0$; \\
 \> \> {\bf while $\mathit{marked} < \mathit{total}$ do} \\
\> \> \> $\mathit{marked} := \mathit{marked} + 1$; 
$v := X^+(u)[marked]$;  \\
 \> \> \> {\bf for each $e\in E$} \\
 \> \> \> \> {\bf if $(s(e) = v) \land (t(e)\notin X^+(u))$ then} \\
\> \> \> \> \>  $\mathit{total} := \mathit{total} + 1$;  
  $X^+(u)[\mathit{total}] := t(e)$   {\bf endif} \\
 \> \> \> {\bf endfor} \\  
 \> \> {\bf endwhile}; \\
 \> \> $\mathit{total} : = 1$; $\mathit{marked} := 0$; \\
 \> \> {\bf while $\mathit{marked} < \mathit{total}$ do} \\
\> \> \> $\mathit{marked} := \mathit{marked} + 1$; 
$v := X^-(u)[marked]$;  \\
 \> \> \> {\bf for each $e\in E$} \\
 \> \> \> \> {\bf if $(t(e) = v) \land (s(e)\notin X^-(u))$ then} \\
\> \> \> \> \>  $\mathit{total} := \mathit{total} + 1$;  
  $X^-(u)[\mathit{total}] := s(e)$  {\bf endif} \\
 \> \> \> {\bf endfor} \\  
 \> \> {\bf endwhile}; \\
 \> \> $C(u) = X^+(u) \cap X^-(u)$; $\mathit{strcomp} := C(u)$ \\ 
 \> {\bf end} 
\end{tabbing}

\medskip
If we want to obtain all the strongly connected components (SCC's) of
a finite graph, $G$, we proceed as follows: Set $V_1 = V$,
pick any node, $v_1$, in $V_1$ and use the above algorithm to compute
the strongly connected component, $C_1$, of $v_1$.
If $V_1 = C_1$, stop. Otherwise, let $V_2 = V_1 - C_1$. 
Again, pick any node, $v_2$ in $V_2$ and determine
the strongly connected component, $C_2$, of $v_2$.
If $V_2 = C_2$, stop. Otherwise, let $V_3 = V_2 - C_2$,
pick $v_3$ in $V_3$, and continue  in the same manner as before.
Ultimately, this process will stop 
and produce all the strongly connected components $C_1, \ldots, C_k$
of $G$.

\medskip
It should be noted that the function $\mathit{strcomp}$
and the simple algorithm that we just described are ``naive''
algorithms that are not particularly efficient.
Their main advantage is their simplicity.
There are more efficient algorithms,
in particular, there is a beautiful algorithm for 
computing the SCC's due to Robert Tarjan.

\medskip
Going back to our city traffic problem from Section 
\ref{sec20}, if we compute the strongly connected components
for the proposed solution shown in Figure \ref{graphfig1b} ,
we find three SCC's:
\[
\{6, 7, 8, 12, 13, 14\}, \quad
\{11\},\quad
\{1, 2, 3, 4, 5,  9, 10, 15, 16, 17, 18, 19\}.
\]
Therefore, the city engineers did not do a good job!
We will show after proving Proposition \ref{graphp6}
how to ``fix'' this faulty solution. 

\medskip
Closed simple paths also play an important role.

\begin{defin}
\label{graph8}
{\em 
Let $G = (V, E, s, t)$ be a digraph. 
A {\it circuit\/} is a closed simple path
(i.e., no edge occurs twice) and an {\it elementary circuit\/}
is an elementary closed path. 
The null path, $(u, \epsilon, u)$, is an elementary circuit.
}
\end{defin}

\remark
A closed path is sometimes called a {\it pseudo-circuit\/}.
In a pseudo-circuit, some edge may occur move than once.

\medskip
The significance of elementary circuits is revealed
by the next proposition.

\begin{prop}
\label{graphp5}
Let  $G$ be any digraph. (a) Every circuit, $\pi$, in $G$
is the concatenation of pairwise edge-disjoint 
elementary circuits.

\medskip
(b)
A circuit is elementary iff it is a minimal circuit, that is,
iff it does not contain any proper circuit.
\end{prop}

\proof
We proceed by induction on the length of $\pi$.
The proposition is trivially true if $\pi$ is the null path.
Next, let $\pi = (u, e_1\cdots e_m, u)$ be any non-null circuit and let
\[
\mathrm{nodes}(\pi) = \lag v_0, \ldots, v_m\rag.
\]
If $\pi$ is an elementary circuit, we are done.
Otherwise,  some node occurs twice in the sequence
$\lag  v_0, \ldots, v_{m - 1}\rag$ or
in the sequence $\lag  v_1, \ldots, v_{m}\rag$.
Let us consider the first case, the second one being similar.
Pick two occurrences of the same, node, say $v_i = v_j$,
with $i < j$, such that  $j - i$ is minimal. Then,
due to the minimality of $j - i$, no node occurs twice
in  $\lag  v_i, \ldots, v_{j - 1}\rag$ or
$\lag  v_{i+1}, \ldots, v_{j}\rag$, which shows that
$\pi_1 =  (v_i, e_{i+1} \cdots e_{j}, v_i)$ is 
an elementary circuit.
Now, we can write
$\pi = \pi' \pi_1 \pi''$, with $|\pi'| < |\pi|$ and 
$|\pi''| < |\pi|$.
Thus, we can apply the induction hypothesis
to both $\pi'$ and $\pi''$, which shows that
$\pi'$  and $\pi''$ are concatenations of elementary
circuits. Then, $\pi$ itself is the concatenation
of elementary circuits. All these elementary circuits
are pairwise edge-disjoint since $\pi$ has no
repeated edges.

\medskip
(b)
This is clear by definition of an elementary circuit.
$\bigsquare$

\remarks
\begin{enumerate}
\item
If $u$ and $v$ are two nodes that belong to a circuit, $\pi$, in $G$,
(i.e., both $u$ and $v$ are incident to some edge in $\pi$),
then $u$ and $v$ are strongly connected.
Indeed, $u$ and $v$ are connected by a portion of the circuit
$\pi$, and $v$ and $u$ are connected by the complementary
portion of the circuit.
\item
If $\pi$ is a pseudo-circuit, the above proof shows that it is 
still possible to decompose $\pi$ into elementary circuits, but
it may not be possible
to write $\pi$ as the concatenation of pairwise
edge-disjoint elementary circuits.
\end{enumerate}

\medskip
Given a graph, $G$, we can form a new and simpler graph
from $G$ by connecting the strongly connected components of $G$
as shown below.

\begin{defin}
\label{graph9}
{\em 
Let $G = (V, E, s, t)$ be a digraph. The {\it reduced graph\/},
$\widehat{G}$, is the simple digraph whose set of nodes,
$\widehat{V} = V/\widehat{C}_G$, is the set
of strongly connected components of $V$ and whose set of edges,
$\widehat{E}$, is defined as follows:
\[
(\widehat{u}, \widehat{v}) \in \widehat{E}
\quad\hbox{iff}\quad
(\exists e\in E)(s(e) \in \widehat{u}
\quad\hbox{and}\quad t(e) \in \widehat{v}),
\]
where we denote the strongly connected component of $u$ by
$\widehat{u}$.
}
\end{defin}

\medskip
That $\widehat{G}$ is ``simpler'' than $G$ is 
the object of the next proposition.

\begin{prop}
\label{graphp6}
Let  $G$ be any digraph. The reduced graph,
$\widehat{G}$, contains no circuits.
\end{prop}

\proof
Suppose that $u$ and $v$ are nodes of $G$ and that
$u$ and $v$ belong to two disjoint strongly connected components
that belong to a circuit, $\widehat{\pi}$, in $\widehat{G}$.
Then, the circuit, $\widehat{\pi}$, yields
a closed sequence of edges $e_1, \ldots, e_n$
between  strongly connected components
and we can arrange the numbering so that
these components are
$C_0, \ldots, C_n$, with $C_n = C_0$, 
with $e_i$  an edge
between $s(e_i)\in C_i$ and $t(e_i)\in C_{i + 1}$
for $0 \leq i \leq n - 1$, 
$e_n$  an edge between between $s(e_n)\in C_n$ 
and $t(e_n)\in C_{0}$, 
$\widehat{u} = C_p$ and $\widehat{v} = C_q$,
for some $p < q$.
Now, we have
$t(e_i)\in C_{i+1}$  and $s(e_{i+1})\in C_{i+1}$
for $0 \leq i \leq n - 1$ and 
$t(e_n)\in C_0$ and $s(e_1)\in C_0$ and as each
$C_i$ is strongly connected, we have elementary paths from
$t(e_i)$ to $s(e_{i+1})$ and from $t(e_n)$ to $s(e_1)$.
Also, as $\widehat{u} = C_p$ and $\widehat{v} = C_q$
for some $p < q$, we have some elementary paths
from $u$ to $s(e_p)$ and from $t(e_{q - 1})$ to $v$.
By concatenating the appropriate paths, we get
a circuit in $G$ containing $u$ and $v$,
showing that $u$  and $v$ are strongly connected,
contradicting that 
$u$ and $v$ belong to two disjoint strongly connected components.
$\bigsquare$

\remark
Digraphs without circuits are called {\it DAG's\/}. 
Such graphs have many nice properties. In particular,
it is easy to see that any finite DAG has
nodes with no incoming edges. Then, it is easy to see that
finite DAG's are basically collections of trees with 
shared nodes.

\medskip
The reduced graph of the graph
shown in Figure \ref{graphfig1b} is showed in Figure
\ref{redgraph1}, where its SCC's are labeled
A, B  and C as shown below:
\[
A = \{6, 7, 8, 12, 13, 14\}, \quad
B = \{11\},\quad
C = \{1, 2, 3, 4, 5,  9, 10, 15, 16, 17, 18, 19\}.
\]
\begin{figure}
  \begin{center}
    \begin{pspicture}(0,0)(3,2.5)
    \cnodeput(1.5,2){v1}{$A$}
    \cnodeput(0,0){v2}{$B$}
    \cnodeput(3,0){v3}{$C$}
    \ncline[linewidth=1pt]{->}{v1}{v2}
    \ncline[linewidth=1pt]{->}{v2}{v3}
    \ncline[linewidth=1pt]{->}{v1}{v3}
    \end{pspicture}
  \end{center}
  \caption{The reduced graph of the graph in Figure \ref{graphfig1b}}
\label{redgraph1}
\end{figure}
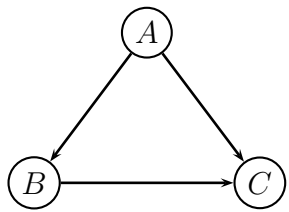
The locations in the component $A$ are inaccessible.
Observe that changing the direction of the street between
13 and 18 yields a solution, that is, a strongly connected graph.
So, the engineers were not too far off after all!
The solution to our traffic problem is shown in Figure \ref{graphfig1c}.
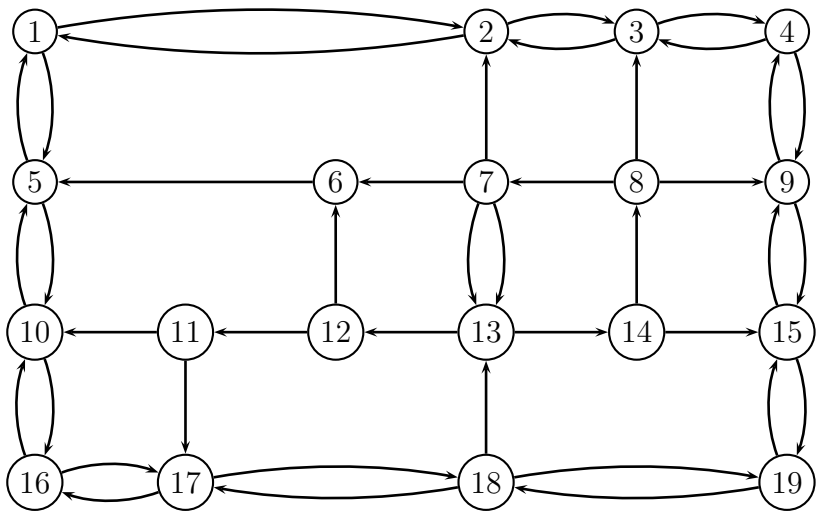
\begin{figure}[H]
  \begin{center}
    \begin{pspicture}(0,0)(10,6.5)
    \cnodeput(0,0){v16}{$16$}
    \cnodeput(2,0){v17}{$17$}
    \cnodeput(6,0){v18}{$18$}
    \cnodeput(10,0){v19}{$19$}
    \cnodeput(0,2){v10}{$10$}
    \cnodeput(2,2){v11}{$11$}
    \cnodeput(4,2){v12}{$12$}
    \cnodeput(6,2){v13}{$13$}
    \cnodeput(8,2){v14}{$14$}
    \cnodeput(10,2){v15}{$15$}
    \cnodeput(0,4){v5}{$5$}
    \cnodeput(4,4){v6}{$6$}
    \cnodeput(6,4){v7}{$7$}
    \cnodeput(8,4){v8}{$8$}
    \cnodeput(10,4){v9}{$9$}
    \cnodeput(0,6){v1}{$1$}
    \cnodeput(6,6){v2}{$2$}
    \cnodeput(8,6){v3}{$3$}
    \cnodeput(10,6){v4}{$4$}
    \ncline[linewidth=1pt]{->}{v11}{v10}
    \ncline[linewidth=1pt]{->}{v12}{v11}
    \ncline[linewidth=1pt]{->}{v13}{v12}
    \ncline[linewidth=1pt]{->}{v13}{v14}
    \ncline[linewidth=1pt]{->}{v14}{v15}
    \ncline[linewidth=1pt]{->}{v12}{v6}
    \ncline[linewidth=1pt]{->}{v14}{v8}
    \ncline[linewidth=1pt]{->}{v6}{v5}
    \ncline[linewidth=1pt]{->}{v7}{v6}
    \ncline[linewidth=1pt]{->}{v8}{v7}
    \ncline[linewidth=1pt]{->}{v8}{v9}
    \ncline[linewidth=1pt]{->}{v7}{v2}
    \ncline[linewidth=1pt]{->}{v8}{v3}
    \ncline[linewidth=1pt]{->}{v11}{v17}
    \ncline[linewidth=1pt]{->}{v18}{v13}
    \ncarc[arcangle=20, linewidth=1pt]{->}{v16}{v17}
    \ncarc[arcangle=20, linewidth=1pt]{->}{v17}{v16}
    \ncarc[arcangle=10, linewidth=1pt]{->}{v17}{v18}
    \ncarc[arcangle=10, linewidth=1pt]{->}{v18}{v17}
    \ncarc[arcangle=10, linewidth=1pt]{->}{v18}{v19}
    \ncarc[arcangle=10, linewidth=1pt]{->}{v19}{v18}
    \ncarc[arcangle=20, linewidth=1pt]{->}{v16}{v10}
    \ncarc[arcangle=20, linewidth=1pt]{->}{v10}{v16}
    \ncarc[arcangle=20, linewidth=1pt]{->}{v10}{v5}
    \ncarc[arcangle=20, linewidth=1pt]{->}{v5}{v10}
    \ncarc[arcangle=20, linewidth=1pt]{->}{v5}{v1}
    \ncarc[arcangle=20, linewidth=1pt]{->}{v1}{v5}
    \ncarc[arcangle=20, linewidth=1pt]{->}{v19}{v15}
    \ncarc[arcangle=20, linewidth=1pt]{->}{v15}{v19}
    \ncarc[arcangle=20, linewidth=1pt]{->}{v15}{v9}
    \ncarc[arcangle=20, linewidth=1pt]{->}{v9}{v15}
    \ncarc[arcangle=20, linewidth=1pt]{->}{v9}{v4}
    \ncarc[arcangle=20, linewidth=1pt]{->}{v4}{v9}
    \ncarc[arcangle=10, linewidth=1pt]{->}{v1}{v2}
    \ncarc[arcangle=10, linewidth=1pt]{->}{v2}{v1}
    \ncarc[arcangle=20, linewidth=1pt]{->}{v2}{v3}
    \ncarc[arcangle=20, linewidth=1pt]{->}{v3}{v2}
    \ncarc[arcangle=20, linewidth=1pt]{->}{v3}{v4}
    \ncarc[arcangle=20, linewidth=1pt]{->}{v4}{v3}
    \ncarc[arcangle=20, linewidth=1pt]{->}{v7}{v13}
    \ncarc[arcangle=-20, linewidth=1pt]{->}{v7}{v13}
    \end{pspicture}
  \end{center}
  \caption{A good choice of one-way streets}
\label{graphfig1c}
\end{figure}
%
Before discussing undirected graphs, let us collect various definitions
having to do with the notion of subgraph.
\begin{defin}
\label{subgraph1}
{\em
Given any two digraphs, $G = (V, E, s, t)$  and  $G' = (V', E', s', t')$,
we say that {\it $G'$ is a subgraph of $G$\/} iff $V' \subseteq V$, $E' \subseteq E$,
$s'$ is the retriction of $s$ to $E'$ and $t'$ is the retriction of $t$ to $E'$.
If $G'$ is a subgraph of $G$ and $V' = V$, we say that $G'$ is a {\it spanning
subgraph of $G$\/}. Given any subset, $V'$, of $V$, the {\it induced subgraph,
$G\langle V'\rangle$, of $G$\/} is the graph whose set of edges is
\[
E_{V'} = \{e\in E \mid s(e) \in V'; t(e)\in V'\}.
\]
(Clearly, $s'$  and $t'$ are the restrictions of $s$ and $t$ to $E_{V'}$,
respectively.)
Given any subset, $E'\subseteq E$, the graph
$G' = (V, E', s', t')$, where $s'$ and $t'$ are the  restrictions of $s$ and $t$ to $E'$,
respectively, is called the {\it partial graph of $G$ generated by $E'$\/}.
The graph, $(V', E'\cap V_{V'}, s', t')$, is a {\it partial subgraph of $G$\/}
(here,  $s'$ and $t'$ are the  restrictions of $s$ and $t$ to $E'\cap V_{V'}$,
respectively.)
}
\end{defin}

\section{Undirected Graphs, Chains, Cycles, Connectivity}
\label{sec23}
The edges of a graph express relationships among its nodes.
Sometimes, these relationships are not symmetric, in which
case it is desirable to use directed arcs, as we have in the
previous sections. However, there is a class of problems
where these relationships are naturally symmetric or where
there is no a priori preferred orientation of the arcs.
For example, if $V$ is the population of individuals
that were students at Penn between 1900 until now and if 
we are interested in the relation where two people $A$ and $B$
are related iff they had the same professor in some course,
then this relation is clearly symmetric. As a consequence,
if we want to find the set of individuals that
are related to a given individual, $A$, it seems
unnatural and, in fact, counter-productive, to model
this relation using a directed graph.

\medskip
As another example suppose we want to investigate
the vulnerabilty of an internet network under two kinds
of attacks: (1) disabling a node; (2) cutting a link.
Again, whether of not a link between two sites is
oriented  is irrelevant. What is important is that
the two sites are either connected or disconnected.

\medskip
These examples suggest that we should consider
an ``unoriented'' version of a graph. How should we proceed?

\medskip
One way to proceed is to still assume that we have a
directed graph but to modify certain notions
such as paths and circuits to account for the fact that
such graphs are really ``unoriented.''
In particular, we should redefine paths to allow
edges to be traversed in the ``wrong direction''.
Such an approach is possible but slightly akward
and ultimately it is really better to define
undirected graphs. However, to show that this approach is feasible,
let us give a new definition of a path that corresponds to the 
notion of path in an undirected graph.

\begin{defin}
\label{graph10}
{\em 
Given any digraph, $G = (V, E, s, t)$,
and any two nodes, $u, v\in V$,
a {\it chain\/} (or {\it walk\/}) {\it from $u$ to $v$\/} is a sequence
$\pi = (u_0, e_1, u_1, e_2, u_2, \ldots, u_{n - 1}, e_n, u_n)$, 
where $n \geq 1$; $u_i \in V$;  $e_i \in E$ and
\[
u_0= u; \> u_n = v 
\quad\hbox{and}\quad
\{s(e_i),  t(e_{i})\} = \{u_{i-1 }, u_{i}\},\quad 1 \leq i \leq n.
\]
We call $n$ the {\it length of the chain $\pi$\/}
and we write $|\pi| = n$. 
When $n = 0$, we have the {\it null chain\/},
$(u, \epsilon, u)$, from $u$ to $u$, a chain of length $0$.
If $u = v$, then $\pi$
is called a {\it closed chain\/}, else an {\it open chain\/}.
The chain, $\pi$, determines the sequence
of nodes, $\mathrm{nodes}(\pi) = \lag u_0, \ldots, u_n\rag$,
with  $\mathrm{nodes}((u, \epsilon, u)) = \lag u, u\rag$.
A chain, $\pi$,
is {\it simple\/} iff $e_i \not= e_j$ for all $i\not= j$
(i.e., no edge in the chain is used twice). 
A chain, $\pi$, from $u$ to $v$ is {\it elementary\/}
iff no vertex in $\mathrm{nodes}(\pi)$ occurs twice,
except possibly for $u$ if $\pi$ is closed.
The null chain, $(u, \epsilon, u)$, is considered simple and elementary.
}
\end{defin}

\medskip
The main difference between Definition \ref{graph10} and
Definition \ref{graph5} is that Definition \ref{graph10}
ignores the orientation: in a chain,  
an edge may be traversed backwards, from
its endpoint back to its source.
This implies that the reverse of a chain
\[
\pi^R = (u_n, e_n, u_{n-1}, , \ldots, u_2, e_2, u_{1}, e_1, u_0) 
\]
is a chain from $v = u_n$ to $u = u_0$. In general, this fails
for paths.
Note, as before, that if $G$ is a simple graph, then
a chain is more simply defined by a sequence of nodes
\[
(u_0, u_1, \ldots, u_n).
\]

\begin{figure}
  \begin{center}
    \begin{pspicture}(0,0)(4.5,4.2)
    \cnodeput(3,0){v4}{$v_4$}
    \cnodeput(4.5,1.5){v5}{$v_5$}
    \cnodeput(0,3){v1}{$v_1$}
    \cnodeput(3,3){v2}{$v_2$}
    \cnodeput(0,0){v3}{$v_3$}
    \ncline[linewidth=1.5pt]{->}{v1}{v3}
    \ncline[linewidth=1.5pt]{->}{v1}{v2}
    \aput{:U}{$a$}
    \ncline[linewidth=1.5pt]{->}{v2}{v5}
    \ncline[linewidth=1.5pt]{->}{v3}{v2}
    \ncline[linewidth=1.5pt]{->}{v3}{v4}
    \bput{:U}{$g$}
    \ncline[linewidth=1.5pt]{->}{v4}{v2}
    \ncline[linewidth=1.5pt]{->}{v5}{v4}
    \uput[180](0,1.6){$b$}
    \uput[45](1,1.4){$c$}
    \uput[0](2.95,1.6){$d$}
    \uput[45](3.8,2.2){$e$}
    \uput[0](3.8,0.7){$f$}
    \end{pspicture}
\hskip 2 cm
    \begin{pspicture}(0,0)(4,4.2)
    \cnodeput(2,0){v8}{$v_8$}
    \cnodeput(0,3){v6}{$v_6$}
    \cnodeput(4,3){v7}{$v_7$}
    \ncline[linewidth=1.5pt]{->}{v6}{v7}
    \aput{:U}{$i$}
    \ncline[linewidth=1.5pt]{->}{v6}{v8}
    \ncline[linewidth=1.5pt]{->}{v7}{v8}
    \uput[180](0.9,1.5){$j$}
    \uput[0](3,1.5){$h$}
    \end{pspicture}
  \end{center}
  \caption{Graph $G_5$}
\label{graphfig4}
\end{figure}
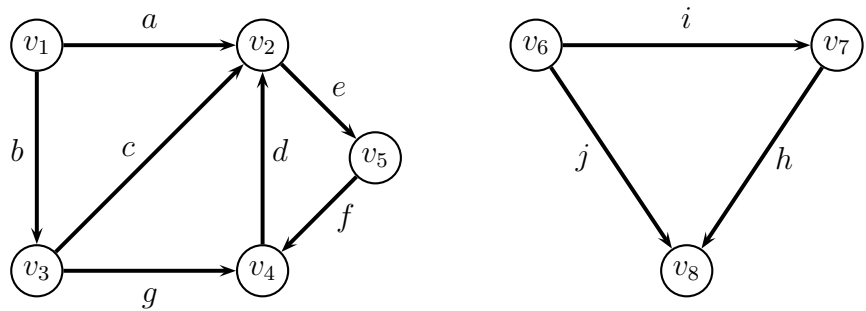

\medskip
For example, in the gaph $G_5$ shown in Figure \ref{graphfig4},
we have the chains
\[
(v_1, a, v_2, d, v_4, f, v_5, e, v_2, d, v_4, g, v_3),
\>
(v_1, a, v_2, d, v_4, f, v_5, e, v_2, c, v_3), 
\>
(v_1, a, v_2, d, v_4, g, v_3)
\]
from $v_1$ to $v_3$, The second chain is simple and
the third is elementary.
Note that none of these chains are paths.

\medskip
Chains are concatenated the same way as paths
and the notion of subchain is analogous to the notion of subpath.
The  undirected version of Proposition \ref{graphp4} also holds.
The proof is obtained by changing the word ``path'' to ``chain''.

\begin{prop}
\label{graphp4b}
Let  $G$ be any digraph. 
(a)
For any two nodes, $u, v$, in $G$,
every non-null chain, $\pi$, from $u$ to $v$ contains
an elementary non-null subchain.

\medskip
(b)
If $|V| = n$, then every open elementary chain has length
at most $n - 1$ and every closed elementary chain 
has length at most $n$.
\end{prop}

\medskip
The undirected version of strong connectivity is the following:

\begin{defin}
\label{graph11}
{\em 
Let $G = (V, E, s, t)$ be a digraph. We define the binary
relation, $\widetilde{C}_G$, on $V$ as follows:
For all $u, v\in V$,
\[
u \widetilde{C}_G v
\quad\hbox{iff}\quad
\hbox{there is a chain from $u$ to $v$}.
\]
When $u \widetilde{C}_G v$, we say that {\it $u$ and $v$
are connected\/}.
}
\end{defin}

\medskip
Oberve that the relation $\widetilde{C}_G$ is an equivalence relation.
It is reflexive because we have the null chain from
$u$ to $u$, symmetric because the reverse of a chain is also a chain
and transitive because
chains can be concatenated. The equivalence classes
of the relation $\widetilde{C}_G$ are called the
{\it  connected components of $G$ (CC's)\/}.
A graph is {\it connected\/} iff it has
a single connected component.

\medskip
Observe that strong connectivity implies connectively but the converse
is false. For example, the graph $G_1$ of Figure  \ref{graphfig1} is connected
but it is not strongly connected. The function $\mathit{strcomp}$
and the method for computing the strongly connected components of a graph can
easily be adapted to compute the connected components of a graph.

\medskip
The undirected version of a circuit is the following:

\begin{defin}
\label{graph12}
{\em 
Let $G = (V, E, s, t)$ be a digraph. 
A {\it cycle\/} is a closed simple chain
(i.e., no edge occurs twice) and an {\it elementary cycle\/}
is an elementary closed chain. 
The null chain, $(u, \epsilon, u)$, is an elementary cycle.
}
\end{defin}

\remark
A closed cycle is sometimes called a {\it pseudo-cycle\/}.
The undirected version of Proposition \ref{graphp5} also holds.
Again, the proof consist in changing the word ``circuit'' to ``cycle''.

\begin{prop}
\label{graphp5b}
Let  $G$ be any digraph. (a) Every cycle, $\pi$, in $G$
is the concatenation of pairwise edge-disjoint 
elementary cycles.

\medskip
(b)
A cycle is elementary iff it is a minimal cycle, that is,
iff it does not contain any proper cycle.
\end{prop}

\medskip
The reader should now be convinced that it is actually possible to
use the notion of a directed graph to model a large class
of problems where the notion of orientation is irrelevant.
However, this is somewhat unnatural and often inconvenient, so
it is desirable to introduce the notion of an undirected graph
as a ``first-class'' object. How should we do that?

\medskip
We could  redefine the set of edges of an undirected graph 
to be of the form $E^+\cup E^-$,
where $E^+ = E$ is the original set of edges of a digraph 
and with
\[
E^- = \{e^- \mid e^+\in E^+,\> s(e^-) = t(e^+),\> t(e^-) = s(e^+)\},
\]
each edge, $e^-$, being the ``anti-edge'' (opposite edge) of $e^+$.
Such an approach is workable but experience shows that it 
not very satisfactory.

\medskip
The solution adopted by most people is to relax the 
condition that every edge, $e\in E$, is assigned an {\it ordered pair\/},
$\lag u, v\rag$, of nodes  (with $u = s(e)$ and $v = t(e)$)
to the condition that  every edge, $e\in E$, is assigned a {\it set\/},
$\{u, v\}$ of nodes (with $u = v$ allowed). 
To this effect, let $[V]^2$ denote the subset of the power set consisting
of all two-element subsets of $V$ (the notation
$\binom{V}{2}$ is sometimes used instead of $[V]^2$) :
\[
[V]^2 = \{\{u, v\}\in 2^V \mid u\not= v\}.
\]

\begin{defin}
\label{ungraph1}
{\em 
A {\it  graph\/} is a triple,
$G = (V, E, st)$, where $V$ is a set of {\it nodes or vertices\/},
$E$ is a set of {\it arcs or edges\/} and
$\mapdef{st}{E}{V\cup [V]^2}$ is a function
that assigns a set of {\it endpoints \/} (or {\it endnodes\/})
to every edge. 
}
\end{defin}

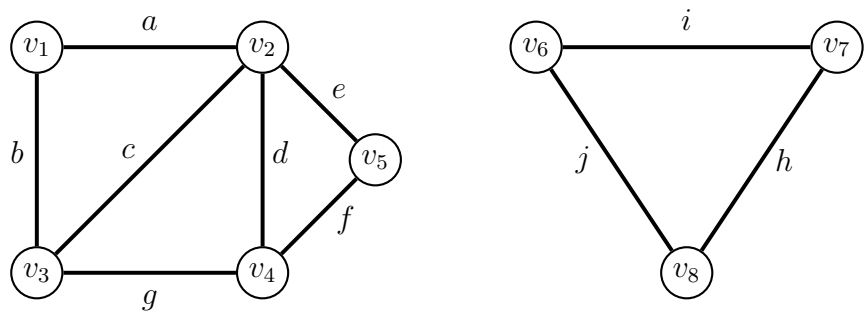
\begin{figure}
  \begin{center}
    \begin{pspicture}(0,0)(4.5,4.2)
    \cnodeput(3,0){v4}{$v_4$}
    \cnodeput(4.5,1.5){v5}{$v_5$}
    \cnodeput(0,3){v1}{$v_1$}
    \cnodeput(3,3){v2}{$v_2$}
    \cnodeput(0,0){v3}{$v_3$}
    \ncline[linewidth=1.5pt]{v1}{v3}
    \ncline[linewidth=1.5pt]{v1}{v2}
    \aput{:U}{$a$}
    \ncline[linewidth=1.5pt]{v2}{v5}
    \ncline[linewidth=1.5pt]{v3}{v2}
    \ncline[linewidth=1.5pt]{v3}{v4}
    \bput{:U}{$g$}
    \ncline[linewidth=1.5pt]{v4}{v2}
    \ncline[linewidth=1.5pt]{v5}{v4}
    \uput[180](0,1.6){$b$}
    \uput[45](1,1.4){$c$}
    \uput[0](2.95,1.6){$d$}
    \uput[45](3.8,2.2){$e$}
    \uput[0](3.8,0.7){$f$}
    \end{pspicture}
\hskip 2 cm
    \begin{pspicture}(0,0)(4,4.2)
    \cnodeput(2,0){v8}{$v_8$}
    \cnodeput(0,3){v6}{$v_6$}
    \cnodeput(4,3){v7}{$v_7$}
    \ncline[linewidth=1.5pt]{v6}{v7}
    \aput{:U}{$i$}
    \ncline[linewidth=1.5pt]{v6}{v8}
    \ncline[linewidth=1.5pt]{v7}{v8}
    \uput[180](0.9,1.5){$j$}
    \uput[0](3,1.5){$h$}
    \end{pspicture}
  \end{center}
  \caption{The Undirected Graph $G_6$}
\label{graphfig5}
\end{figure}

\medskip
When we want to stress that we are dealing with an undirected graph
as opposed to a digraph, we use the locution {\it undirected graph\/}.
When we draw an undirected graph we suppress the tip on the extremity
of an arc. For example, the undirected graph, $G_6$, corresponding
to the directed graph $G_5$ is shown in Figure \ref{graphfig5}.

\begin{defin}
\label{ungraph2}
{\em 
Given a  graph, $G$,
an edge, $e\in E$, such that  $st(e) \in V$ is called
a {\it loop\/} (or {\it self-loop\/}).
Two edges, $e, e'\in E$ are said to be {\it parallel edges\/}
iff $st(e) =  st(e')$.
A graph is {\it simple\/} iff it has no loops and no 
parallel edges.
}
\end{defin}

\remarks
\begin{enumerate}
\item
The functions $st$ need not be injective
or surjective. 
\item
When $G$ is simple, every edge, $e\in E$, is uniquely determined
by the set of  vertices, $\{u, v\}$, such that
$\{u, v\} = st(e)$. In this case, we may denote the
edge $e$ by $\{u, v\}$ 
(some books also use the notation $(u v)$ or even $u v$).
\item
Some authors call a graph with no loops but possibly
parallel edges a {\it multigraph\/} and a graph
with loops and parallel edges a {\it pseudograph\/}.
We prefer to use the term graph for the most
general concept.
\item
Given an undirected graph, $G = (V, E, st)$,
we can form  directed graphs from $G$
by assigning an arbitrary orientation to the edges of $G$.
This means that we assign to every set,
$st(e) = \{u, v\}$, where $u \not= v$, one of the two pairs
$(u, v)$ or $(v, u)$ and define $s$ and $t$
such that $s(e) = u$ and $t(e) = v$ in the first case or
such that $s(e) = v$ and $t(e) = u$ in the second case
(when $u = v$, we have $s(e) = t(e) = u$)).
\item
When a graph is simple,
the function $st$ is often omitted and we simply write
$(V, E)$, with the understanding that $E$ is a 
set of two-elements subsets of $V$.
\item
The concepts or adjacency and incidence transfer immediately
to (undirected) graphs.
\end{enumerate}

\medskip
It is clear that the Definition of chain, connectivity, 
and cycle (Definitions \ref{graph10}, \ref{graph11} and
\ref{graph12}) immediately apply to (undirected) graphs.
However, only the notion of {\it degree\/} (or {\it valency\/}) 
of a node applies to undirected graph
where it is given by
\[
d_G(u) = |\{e\in E \mid u \in st(e)\}|.
\]
We can check immediately that
Corollary \ref{graphp2} and Corollary \ref{graphp3} apply to
undirected graphs.

\remark
When it is clear that we are dealing with undirected graphs,
we will sometimes allow ourselves some abuse of language.
For example, we will occasionally use the term path
instead of chain. 

\medskip
The notion of homomorphism and isomorphism 
also makes sense for undirected graphs.
In order to adapt Definition \ref{graph4}, observe that
any function, $\mapdef{g}{V_1}{V_2}$, can be extended
in a natural  way to a function from $V_1\cup [V_1]^2$ to
$V_2\cup [V_2]^2$, also denoted $g$, so that
\[
g(\{u, v\}) = \{g(u), g(v)\},
\]
for all $\{u, v\}\in [V_1]^2$.

\begin{defin}
\label{ungraph4}
{\em 
Given two graphs, $G_1 = (V_1, E_1, st_1)$
and $G_2 = (V_2, E_2, st_2)$, a {\it homomorphism\/}
(or {\it morphism\/}), $\mapdef{f}{G_1}{G_2}$,
{\it from $G_1$ to $G_2$\/} is a pair, 
$f = (f^v, f^e)$, with $\mapdef{f^v}{V_1}{V_2}$
and $\mapdef{f^e}{E_1}{E_2}$ preserving incidence, that is,
for every edge, $e\in E_1$, we have
\[
st_2(f^e(e)) = f^v(st_1(e)).
\]
These conditions can also be expressed by saying that
the following diagram commute:
\[
\xymatrix{
E_1 \ar[r]^-{f^e} \ar[d]_-{st_1} & \> E_2 \ar[d]^-{st_2} \\
V_1\cup [V_1]^2 \ar[r]_-{f^v} & \> V_2\cup [V_2]^2 .
}
\]
}
\end{defin}

\medskip
As for directed graphs, we can compose homomorphisms of
undirected graphs and the definition of an isomorphism of undirected graphs
is the same as the definition of an isomorphism of digraphs.

\medskip
We are now going to investigate the properties of a very important
subclass of graphs, trees.

\section{Trees and Arborescences}
\label{sec24}
In this section, until further notice, 
we will be dealing with undirected graphs.
Given a graph, $G$, edges having the property that
their deletion increases the number of connected
components of $G$ play an important role and we
would like to characterize such edges.

\begin{defin}
\label{bridge}
{\em
Given any graph, $G = (V, E, st)$, any edge, $e\in E$,
whose deletion increases the number of connected components
of $G$ (i.e., $(V, E - \{e\}, st\res (E - \{e\})$) has more connected
components than $G$) is called a {\it bridge\/}. 
}
\end{defin}

\begin{figure}
  \begin{center}
    \begin{pspicture}(0,0)(6.5,4.1)
    \cnodeput(3,2){v4}{$v_4$}
    \cnodeput(4.5,2){v5}{$v_5$}
    \cnodeput(0,2){v1}{$v_1$}
    \cnodeput(1.5,0){v2}{$v_2$}
    \cnodeput(1.5,4){v3}{$v_3$}
    \cnodeput(6.5,4){v6}{$v_6$}
    \cnodeput(6.5,0){v7}{$v_7$}
    \ncline[linewidth=1pt]{v1}{v2}
    \ncline[linewidth=1pt]{v1}{v3}
    \ncline[linewidth=1pt]{v3}{v4}
    \ncline[linewidth=1pt]{v2}{v4}
    \ncline[linewidth=1.5pt]{v4}{v5}
    \ncline[linewidth=1pt]{v5}{v6}
    \ncline[linewidth=1pt]{v5}{v7}
    \ncline[linewidth=1pt]{v6}{v7}
    \end{pspicture}
  \end{center}
  \caption{A bridge in the graph $G_7$}
\label{graphfig6}
\end{figure}
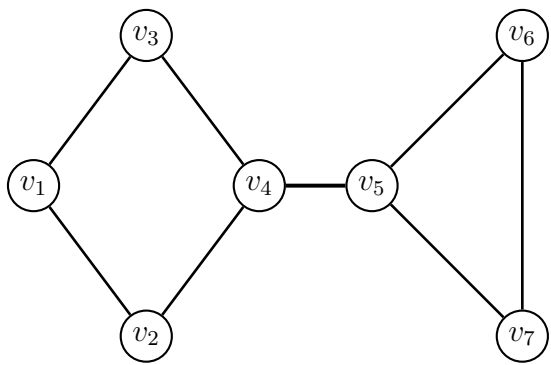

\medskip
For example, the edge $(v_4 v_5)$ in the graph shown in Figure
\ref{graphfig6} is a bridge.

\begin{prop}
\label{bridgep1}
Given any graph, $G = (V, E, st)$, adjunction of a new edge, $e$, 
between  $u$ and  $v$ (this means that $st$ is extended to 
$st_e$, with
$st_e(e) = \{u, v\}$) 
to $G$ has the following effect:
\begin{enumerate}
\item
Either the number of components of $G$ decreases by $1$, in wich
case the edge $e$ does not belong to any cycle of 
$G' = (V, E\cup \{e\}, st_e)$, or
\item
The number of components of $G$ is unchanged, in wich case
the edge $e$ belongs to some cycle of $G' = (V, E\cup \{e\}, st_e)$.
\end{enumerate}
\end{prop}

\proof
Two mutually exclusive cases are possible:
\begin{enumerate}
\item[(a)]
The endpoints $u$ and $v$ (of $e$) belong to two disjoint
connected components of $G$. In $G'$, these components are merged.
The edge $e$ can't belong to a cycle of $G'$ because the chain obtained
by deleting  $e$ from this cycle would connect
$u$ and $v$ in $G$, a contradiction.
\item[(b)]
The endpoints $u$ and $v$ (of $e$) belong to the same
connected component of $G$. Then, $G'$ has the same connected
components as $G$. Since $u$ and $v$ are connected,  there
is an elementary chain from $u$ to $v$ (by Proposition \ref{graphp4b})
and by adding $e$ to this elementary chain, we get a cycle of $G'$
containing $e$. $\bigsquare$
\end{enumerate}

\begin{cor}
\label{bridgep2}
Given any graph, $G = (V, E, st)$, an edge, $e\in E$, is a bridge
iff it does not belong to any cycle of $G$.
\end{cor}

\begin{thm}
\label{bridgep3}
Let $G$ be a finite graph and let $m = |V| \geq 1$.
The following properties hold:
\begin{enumerate}
\item[(i)]
If $G$ is connected, then $|E| \geq m - 1$.
\item[(ii)]
If $G$ has no cycle, then $|E| \leq m - 1$.
\end{enumerate}
\end{thm}

\proof
We can build the graph $G$ progressively by adjoining edges
one at a time starting from the graph $(V, \emptyset)$,
which has $m$ connected components.

\medskip
(i) Every time a new edge is added, the number of connected components
decreases by at most $1$. Therefore, it will take at least $m - 1$
steps to get a connected graph.

\medskip
(ii)
If $G$ has no cycle, then every spannning graph has
no cycle. Therefore, at every step, we are in case (1) of 
Proposition \ref{bridgep1} and the number of connected 
components decreases by exactly $1$. As $G$ has at least 
one connected component, the number of steps (i.e., of edges)
is at most $m - 1$.
$\bigsquare$

\medskip
In view of Theorem \ref{bridgep3}, it makes sense to define
the following kind of graphs:

\begin{defin}
\label{treedef}
{\em
A {\it tree\/} is a graph that is connected and acyclic
(i.e., has no cycles). A {\it forest\/} is a graph whose 
connected components are trees.
}
\end{defin}

\medskip
The picture of a tree is shown in Figure \ref{graphfig7}.

\begin{figure}
  \begin{center}
    \begin{pspicture}(0,0)(6,4.1)
    \cnodeput(2,4){v2}{$v_2$}
    \cnodeput(4,4){v5}{$v_5$}
    \cnodeput(0,4){v1}{$v_1$}
    \cnodeput(0,2){v4}{$v_4$}
    \cnodeput(2,2){v3}{$v_3$}
    \cnodeput(4,2){v6}{$v_6$}
    \cnodeput(1,0){v7}{$v_7$}
    \cnodeput(3,0){v8}{$v_8$}
    \cnodeput(6,2){v9}{$v_9$}
    \ncline[linewidth=1pt]{v1}{v2}
    \ncline[linewidth=1pt]{v1}{v4}
    \ncline[linewidth=1pt]{v2}{v3}
    \ncline[linewidth=1pt]{v3}{v7}
    \ncline[linewidth=1pt]{v3}{v8}
    \ncline[linewidth=1pt]{v3}{v6}
    \ncline[linewidth=1pt]{v6}{v5}
    \ncline[linewidth=1pt]{v6}{v9}
    \end{pspicture}
  \end{center}
  \caption{A Tree, $T_1$}
\label{graphfig7}
\end{figure}

\medskip
Our next theorem gives several equivalent characterizations
of a tree.

\begin{thm}
\label{bridgep4}
Let $G$ be a finite graph with $m = |V|\geq 2$ nodes. The following
properties characterize trees:
\begin{enumerate}
\item[(1)]
$G$ is connected and acyclic.
\item[(2)]
$G$ is connected and minimal for this property (if we delete any edge of 
$G$, then the resulting graph is no longer connected).
\item[(3)]
$G$ is connected and has $m - 1$ edges.
\item[(4)]
$G$ is acyclic  and maximal for this property (if we add any edge to 
$G$, then the resulting graph is no longer acyclic).
\item[(5)]
$G$ is acyclic   and has $m - 1$ edges.
\item[(6)]
Any two nodes of $G$ are joined by a unique chain.
\end{enumerate}
\end{thm}

\proof
The implications
\begin{align*}
(1) & \Longrightarrow (3), (5) \\
(3) & \Longrightarrow (2) \\
(5) & \Longrightarrow (4)
\end{align*}
all follow immediately from Theorem \ref{bridgep3}.

\medskip
$(4)  \Longrightarrow (3)$.
If $G$ was not connected, we could add an edge between
to disjoint connected components without creating any cycle
in $G$, contradicting the maximality of $G$ with respect
to acyclicity. By Theorem \ref{bridgep3}. as $G$ is
connected and acyclic, it must have $m - 1$ edges.

\medskip
$(2)  \Longrightarrow (6)$. 
As $G$ is connected, 
there is a chain joining any two nodes of $G$. If,
for two nodes $u$ and $v$, we had two distinct chains
from $u$  to $v$, deleting any edge from one of these
two chains would not destroy the connectivity of $G$
contradicting the fact that $G$ is minimal with respect to
connectivity.

\medskip
$(6)  \Longrightarrow (1)$. 
If $G$ had a cycle, then there would be at least two distinct
chains joining two nodes in this cycle, a contradiction.

\medskip
The reader should then draw the directed graph of implications
that we just established and check that this graph is strongly
connected! Indeed, we have the cycle of implications
\[
(1) \Longrightarrow (5) \Longrightarrow (4) \Longrightarrow (3)
\Longrightarrow (2) \Longrightarrow (6) \Longrightarrow (1).
\]
$\bigsquare$

\remark
The equivalence of (1) and (6) holds for infinite graphs too.

\begin{cor}
\label{bridgep5}
For any tree, $G$,  adding a new edge, $e$, to $G$ yields
a graph, $G'$,  with a unique cycle.
\end{cor}

\proof
Because $G$ is a tree, all cycles of $G'$ must contain $e$.
If $G'$ had two distinct cycles, there would be two distinct
chains in $G$ joining the endpoints of $e$, contradicting
property (6) of Theorem \ref{bridgep4}.
$\bigsquare$

\begin{cor}
\label{bridgep6}
Every finite connected graph possesses a spanning tree.
\end{cor}

\proof
This is a consequence of property (2) of Theorem \ref{bridgep4}.
Indeed, if there is some edge, $e\in E$, such that
deleting $e$ yields a connected graph, $G_1$, we consider
$G_1$ and repeat this deletion procedure. Eventually, we will get
a minimal connected graph that must be a tree.
$\bigsquare$

\medskip
An example of a spanning tree (shown in thicker lines)
in a graph is shown in Figure 
\ref{graphfig8}.

\begin{figure}
  \begin{center}
    \begin{pspicture}(0,0)(9,6.6)
    \cnodeput(1.5,0){v1}{$1$}
    \cnodeput(4.5,0){v2}{$2$}
    \cnodeput(7.5,0){v3}{$3$}
    \cnodeput(0,2){v4}{$4$}
    \cnodeput(3,2){v5}{$5$}
    \cnodeput(6,2){v6}{$6$}
    \cnodeput(9,2){v7}{$7$}
    \cnodeput(1.5,4){v8}{$8$}
    \cnodeput(4.5,4){v9}{$9$}
    \cnodeput(7.5,4){v10}{$10$}    
    \cnodeput(3,6){v11}{$11$}    
    \cnodeput(6,6){v12}{$12$}
    \ncline[linewidth=2pt]{v1}{v4}
    \ncline[linewidth=2pt]{v1}{v5}
    \ncline[linewidth=1pt]{v1}{v2}
    \ncline[linewidth=1pt]{v2}{v5}
    \ncline[linewidth=2pt]{v2}{v6}
    \ncline[linewidth=1pt]{v2}{v3}
    \ncline[linewidth=2pt]{v3}{v6}
    \ncline[linewidth=2pt]{v3}{v7}
    \ncline[linewidth=1pt]{v4}{v5}
    \ncline[linewidth=1pt]{v4}{v8}
    \ncline[linewidth=2pt]{v5}{v8}
    \ncline[linewidth=2pt]{v5}{v9}
    \ncline[linewidth=1pt]{v8}{v9}
    \ncline[linewidth=1pt]{v6}{v5}
    \ncline[linewidth=2pt]{v6}{v9}
    \ncline[linewidth=1pt]{v6}{v7}
    \ncline[linewidth=1pt]{v6}{v10}
    \ncline[linewidth=1pt]{v7}{v10}
    \ncline[linewidth=1pt]{v8}{v11}
    \ncline[linewidth=2pt]{v9}{v11}
    \ncline[linewidth=1pt]{v9}{v12}
    \ncline[linewidth=1pt]{v9}{v10}
    \ncline[linewidth=2pt]{v10}{v12}
    \ncline[linewidth=2pt]{v11}{v12}
    \end{pspicture}
  \end{center}
  \caption{A Spanning Tree}
\label{graphfig8}
\end{figure}

\medskip
An {\it endpoint\/} or {\it leaf\/} in a graph is a node of degree $1$.

\begin{prop}
\label{bridgep7}
Every finite tree with $m\geq 2$ nodes has at least two endpoints.
\end{prop}

\proof
By Theorem \ref{bridgep4}, our tree has $m - 1$ edges and by
the version of Proposition \ref{graphp2} for undirected graphs,
\[
\sum_{u\in V} d_G(u) = 2(m - 1).
\]
If we had $d_G(u) \geq 2$ except for a single node $u_0$, we would have
\[
\sum_{u\in V} d_G(u) \geq  2m - 1, 
\]
contradicting the above.
$\bigsquare$

\remark
A forest with $m$ nodes  and $p$ connected components has
$m - p$ edges. Indeed, if each connected component has
$m_i$ nodes, then the total number of edges is
\[
(m_1 - 1)  + (m_2 - 1) +  \cdots + (m_p - 1) = m - p.
\] 

\medskip
We now consider briefly directed versions of a tree.

\begin{defin}
\label{rootdef}
{\em
Given a digraph, $G = (V, E, s, t)$, a node, $a\in V$
is a {\it root\/} (resp. {\it anti-root\/}) iff
for every node $u\in V$, there is a path from $a$ to $u$
(resp. there is a path from $u$ to $a$).
A digraph with at least two nodes is an {\it arborescence with
root $a$\/}
iff
\begin{enumerate}
\item
The node $a$ is a root of $G$
\item
$G$ is a tree (as an undirected graph).
\end{enumerate}
A digraph with at least two nodes is an {\it anti-arborescence with
anti-root $a$\/}
iff
\begin{enumerate}
\item
The node $a$ is an  anti-root of $G$
\item
$G$ is a tree (as an undirected graph).
\end{enumerate}
}
\end{defin}

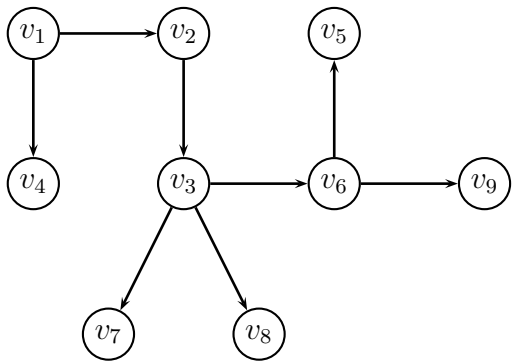
\begin{figure}
  \begin{center}
    \begin{pspicture}(0,0)(6,4.1)
    \cnodeput(2,4){v2}{$v_2$}
    \cnodeput(4,4){v5}{$v_5$}
    \cnodeput(0,4){v1}{$v_1$}
    \cnodeput(0,2){v4}{$v_4$}
    \cnodeput(2,2){v3}{$v_3$}
    \cnodeput(4,2){v6}{$v_6$}
    \cnodeput(1,0){v7}{$v_7$}
    \cnodeput(3,0){v8}{$v_8$}
    \cnodeput(6,2){v9}{$v_9$}
    \ncline[linewidth=1pt]{->}{v1}{v2}
    \ncline[linewidth=1pt]{->}{v1}{v4}
    \ncline[linewidth=1pt]{->}{v2}{v3}
    \ncline[linewidth=1pt]{->}{v3}{v7}
    \ncline[linewidth=1pt]{->}{v3}{v8}
    \ncline[linewidth=1pt]{->}{v3}{v6}
    \ncline[linewidth=1pt]{->}{v6}{v5}
    \ncline[linewidth=1pt]{->}{v6}{v9}
    \end{pspicture}
  \end{center}
  \caption{An Arborescence, $T_2$}
\label{graphfig9}
\end{figure}

\medskip
Note that orienting the edges in a tree does not 
necessarily yield an arborescence (or an anti-arborescence).
Also, if we reverse the orientation of the arcs of
an  arborescence we get  an anti-arborescence.
An  arborescence is shown is Figure \ref{graphfig9}.

\medskip
There is a version of Theorem \ref{bridgep4} 
giving several equivalent characterizations
of an arborescence. The proof of this theorem is
left as an exercise to the reader.

\begin{thm}
\label{bridgep8}
Let $G$ be a finite digraph with $m = |V|\geq 2$ nodes. The following
properties characterize arborescences with root $a$:
\begin{enumerate}
\item[(1)]
$G$ is a tree (as undirected graph) with root $a$.
\item[(2)]
For every $u\in V$, there is a unique path from $a$ to $u$.
\item[(3)]
$G$ has $a$ as a root and is minimal for this property 
(if we delete any edge of
$G$, then $a$ is not a root any longer).
\item[(4)]
$G$ is connected (as undirected graph) and moreover
\[
(*) \cases{
& $d_G^-(a) = 0$ \cr
& $d_G^-(u) = 1$, for all $u\in V, \> u \not= a$.\cr 
}
\]
\item[(5)]
$G$ is acyclic (as undirected graph) 
and the properties $(*)$ are satisfied.
\item[(6)]
$G$ is acyclic (as undirected graph)   and has $a$ as a root.
\item[(7)]
$G$ has $a$ as a root and has $m - 1$ arcs.
\end{enumerate}
\end{thm}

\section{Minimum (or Maximum) Weight Spanning Trees}
\label{sec25}
For a certain class of problems, it is necessary to consider 
undirected graphs (without loops) whose edges are assigned a ``cost''
or ``weight''. 

\begin{defin}
\label{wgraph}
{\em
A {\it weighted graph\/} is a finite graph without loops,
$G = (V, E, st)$, together with a function, $\mapdef{c}{E}{\reals}$,
called a {\it weight function\/} (or {\it cost function\/}).
We will denote  a weighted graph by $(G, c)$.
Given any set of edges, $E' \subseteq E$, we define the 
{\it weight (or cost)\/} of $E'$ by
\[
c(E') = \sum_{e \in E'} c(e).
\]
}
\end{defin}

\medskip
Given a weighted graph, $(G, c)$, an important problem is
to find a spanning tree, $T$ such that $c(T)$ is maximum
(or minimum). This problem is called the 
{\it maximal weight spanning tree\/} (resp.  
{\it minimal weight spanning tree\/}). Actually, it is easy to see that
any algorithm solving any one of the  two problems can be converted
to an algorithm solving the other problem. For example,
if we can solve the maximal weight spanning tree,
we can solve the mimimal weight spanning tree by 
replacing every weight, $c(e)$, by $-c(e)$, and by
looking for a spanning tree, $T$, that is a maximal spanning tree, since
\[
\min_{T\subseteq G} c(T) = - \max_{T\subseteq G} -c(T).
\] 
There are several algorithms for finding 
such spanning trees, including one due to
Kruskal and another one due to Prim. The fastest
known algorithm at the present  is due to Bernard Chazelle (1999).

\medskip
Since every spannning tree of a given graph,
$G = (V, E, st)$, has the same number of edges
(namely, $|V| - 1$), adding the same constant
to the weight of every edge does not affect the maximal
nature a spanning tree, that is, the set of maximal weight 
spanning trees is preserved. Therefore, we may assume that all 
the weights are non-negative.

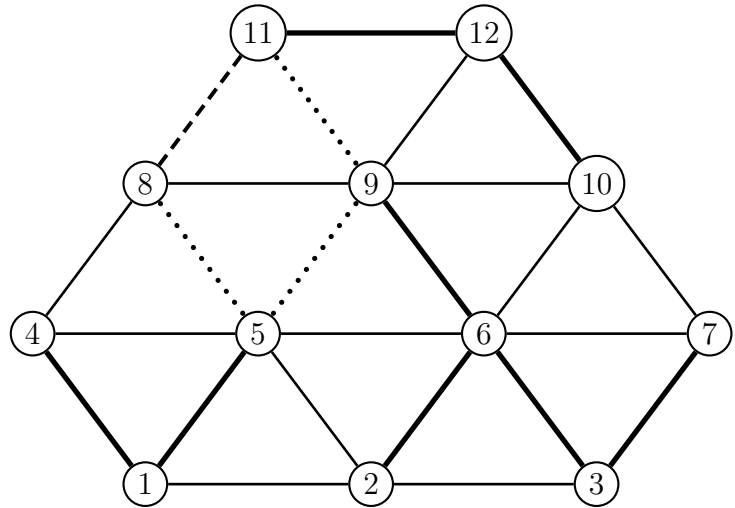
\begin{figure}
  \begin{center}
    \begin{pspicture}(0,0)(9,6.6)
    \cnodeput(1.5,0){v1}{$1$}
    \cnodeput(4.5,0){v2}{$2$}
    \cnodeput(7.5,0){v3}{$3$}
    \cnodeput(0,2){v4}{$4$}
    \cnodeput(3,2){v5}{$5$}
    \cnodeput(6,2){v6}{$6$}
    \cnodeput(9,2){v7}{$7$}
    \cnodeput(1.5,4){v8}{$8$}
    \cnodeput(4.5,4){v9}{$9$}
    \cnodeput(7.5,4){v10}{$10$}    
    \cnodeput(3,6){v11}{$11$}    
    \cnodeput(6,6){v12}{$12$}
    \ncline[linewidth=2pt]{v1}{v4}
    \ncline[linewidth=2pt]{v1}{v5}
    \ncline[linewidth=1pt]{v1}{v2}
    \ncline[linewidth=1pt]{v2}{v5}
    \ncline[linewidth=2pt]{v2}{v6}
    \ncline[linewidth=1pt]{v2}{v3}
    \ncline[linewidth=2pt]{v3}{v6}
    \ncline[linewidth=2pt]{v3}{v7}
    \ncline[linewidth=1pt]{v4}{v5}
    \ncline[linewidth=1pt]{v4}{v8}
    \ncline[linewidth=2pt,linestyle=dotted]{v5}{v8}
    \ncline[linewidth=2pt,linestyle=dotted]{v5}{v9}
    \ncline[linewidth=1pt]{v8}{v9}
    \ncline[linewidth=1pt]{v6}{v5}
    \ncline[linewidth=2pt]{v6}{v9}
    \ncline[linewidth=1pt]{v6}{v7}
    \ncline[linewidth=1pt]{v6}{v10}
    \ncline[linewidth=1pt]{v7}{v10}
    \ncline[linewidth=1.5pt,linestyle=dashed]{v8}{v11}
    \ncline[linewidth=2pt,linestyle=dotted]{v9}{v11}
    \ncline[linewidth=1pt]{v9}{v12}
    \ncline[linewidth=1pt]{v9}{v10}
    \ncline[linewidth=2pt]{v10}{v12}
    \ncline[linewidth=2pt]{v11}{v12}
    \end{pspicture}
  \end{center}
  \caption{The set $C_e$ associated with an edge $e\in G - T$}
\label{graphfig10}
\end{figure}

\medskip
In order to justify the correctness of Kruskal's algorithm,
we need two definitions. Let $G = (V, E, st)$ be any connected weighted 
graph and let $T$ be any spanning tree of $G$. For every edge,
$e\in E - T$, let $C_e$ be the set of edges belonging to the
unique chain joining the endpoints of $e$ (the vertices in
$st(e)$). For example, in the graph shown in Figure \ref{graphfig10},
the set $C_{\{8, 11\}}$ associated with the edge  $\{8, 11\}$ (shown
as a dashed line) corresponds to the following set of edges 
(shown as dotted lines) in $T$:
\[
C_{\{8, 11\}} = \{\{8, 5\}, \{5, 9\}, \{9, 11\}\}.
\]

\begin{figure}
  \begin{center}
    \begin{pspicture}(0,0)(9,6.6)
    \cnodeput(1.5,0){v1}{$1$}
    \cnodeput(4.5,0){v2}{$2$}
    \cnodeput(7.5,0){v3}{$3$}
    \cnodeput(0,2){v4}{$4$}
    \cnodeput(3,2){v5}{$5$}
    \cnodeput(6,2){v6}{$6$}
    \cnodeput(9,2){v7}{$7$}
    \cnodeput(1.5,4){v8}{$8$}
    \cnodeput(4.5,4){v9}{$9$}
    \cnodeput(7.5,4){v10}{$10$}    
    \cnodeput(3,6){v11}{$11$}    
    \cnodeput(6,6){v12}{$12$}
    \ncline[linewidth=2pt]{v1}{v4}
    \ncline[linewidth=2pt]{v1}{v5}
    \ncline[linewidth=1pt,linestyle=dotted]{v1}{v2}
    \ncline[linewidth=1pt,linestyle=dotted]{v2}{v5}
    \ncline[linewidth=2pt]{v2}{v6}
    \ncline[linewidth=1pt]{v2}{v3}
    \ncline[linewidth=2pt]{v3}{v6}
    \ncline[linewidth=2pt]{v3}{v7}
    \ncline[linewidth=1pt]{v4}{v5}
    \ncline[linewidth=1pt]{v4}{v8}
    \ncline[linewidth=2pt]{v5}{v8}
    \ncline[linewidth=1pt,linestyle=dotted]{v8}{v9}
    \ncline[linewidth=1pt,linestyle=dotted]{v6}{v5}
    \ncline[linewidth=2pt]{v6}{v9}
    \ncline[linewidth=1pt]{v6}{v7}
    \ncline[linewidth=1pt]{v6}{v10}
    \ncline[linewidth=1pt]{v7}{v10}
    \ncline[linewidth=1pt,linestyle=dotted]{v8}{v11}
    \ncline[linewidth=2pt]{v9}{v11}
    \ncline[linewidth=1pt]{v9}{v12}
    \ncline[linewidth=1pt]{v9}{v10}
    \ncline[linewidth=2pt]{v10}{v12}
    \ncline[linewidth=2pt]{v11}{v12}
    \end{pspicture}
  \end{center}
  \caption{The set $\Omega_{\{5, 9\}}$ obtained by deleting the edge 
$\{5, 9\}$ from the spanning tree.}
\label{graphfig11}
\end{figure}

Also, given any edge, $e\in T$, observe that
the result of deleting $e$ yields
a graph denoted $T - e$  consisting of
two disjoint subtrees of $T$. We 
let $\Omega_e$ be the set of edges,
$e'\in G - T$, such that if $st(e') = \{u, v\}$, then
$u$ and $v$ belong to the two distinct connected components of
$T - \{e\}$. For example, in Figure \ref{graphfig11},
deleting the edge $\{5, 9\}$ yields  the 
set of edges (shown as dotted lines)
\[
\Omega_{\{5, 9\}} = \{\{1, 2\}, \{5, 2\}, \{5, 6\}, \{8, 9\}, \{8, 11\}\}.
\]

\medskip
Observe that in the first case, deleting any edge from $C_{e}$ and
adding the edge $e\in E - T$ yields a new spanning tree
and in the second case, deleting any edge $e\in T$ and adding any edge
in $\Omega_{e}$ also yields a new spanning tree.
These observations are crucial ingredients in the proof of
the following theorem:

\begin{thm}
\label{Kruskal1}
Let $(G, c)$ be any connected weighted graph and let $T$ be
any spanning tree of $G$. (1) The tree $T$ is a maximal weight
spanning tree iff any of the following (equivalent) conditions hold:
\begin{enumerate}
\item[(i)]
For every $e\in E - T$,
\[
c(e) \leq \min_{e'\in C_e} c(e')
\]
\item[(ii)]
For   every $e\in T$,
\[
c(e) \geq \max_{e'\in \Omega_e} c(e').
\]
\end{enumerate}

\medskip
(2) 
The tree $T$ is a minimal weight
spanning tree iff any of the following (equivalent) conditions hold:
\begin{enumerate}
\item[(i)]
For every $e\in E - T$,
\[
c(e) \geq \max_{e'\in C_e} c(e')
\]
\item[(ii)]
For   every $e\in T$,
\[
c(e) \leq \min_{e'\in \Omega_e} c(e').
\]
\end{enumerate}
\end{thm}

\proof
(1)
First, assume that $T$ is a maximal weight spanning tree.
Observe that
\begin{enumerate}
\item[(a)]
For any $e\in E - T$  and any $e'\in C_e$, the graph
$T' = (V, (T\cup \{e\}) - \{e'\})$ is acyclic and has $|V| - 1$
edges, so it is a spanning tree. Then, (i) must hold, as otherwise we would
have $c(T') > c(T)$, contradicting the maximality of $T$.
\item[(b)]
For any $e\in T$ and any $e'\in \Omega_e$, the graph
$T' = (V, (T\cup \{e'\}) - \{e\})$ is connected  and has $|V| - 1$
edges, so it is a spanning tree. Then, (ii) must hold, as
otherwise we would
have $c(T') > c(T)$, contradicting the maximality of $T$.
\end{enumerate}

\medskip
Let us now assume that (i) holds. We proceed by contradiction.
Let $T$ be a spanning tree satisfying condition (i) and assume
there is another spanning tree, $T'$, with $c(T') > c(T)$.
Since there are only finitely many spanning trees of $G$, we may assume
that $T'$ is maximal. Consider any edge $e\in T' - T$
and let $st(e) = \{u, v\}$. In $T$, there is a unique chain,
$C_e$, joining $u$ and $v$ and this chain must contain some edge, $e'\in T$,
joining the two connected components of $T' - e$, that is,
$e' \in \Omega_e$. As (i) holds, we get
$c(e) \leq c(e')$.
However, as $T'$ is maximal, (ii) holds (as we just proved), so
$c(e) \geq c(e')$.
Therefore, we get
\[
c(e) = c(e').
\]
Consequently, if we form the graph
$T_2 = (T' \cup \{e'\}) - \{e\})$, we see that $T_2$ is a
spanning tree having some edge  from $T$
and $c(T_2) = c(T')$. We can repeat this process of edge
substitution with 
$T_2$ and $T$  and so on. Ultimately,  we will obtain
the tree $T$ with the weight $c(T') > c(T)$, which is absurd.
Therefore, $T$ is indeed maximal.

\medskip
Finally, assume that (ii) holds. The proof is analogous
to the previous proof: We pick some edge 
$e' \in T - T'$ and $e$ is some edge in $\Omega_{e'}$
belonging to the chain joining the endpoints of
$e'$ in $T'$.

\medskip
(2) 
The proof of (2) is analogous to the proof of (1)
but uses 2(a) and 2(b) instead of 1(a) and 1(b).
$\bigsquare$

\medskip
We are now in the position to present a version of Kruskal's
algorithm and to prove its correctness. 
Here is a version of Kruskal's algorithm for finding a minimal
weight spanning tree using criterion 2(a).
Let $n$ be the number of edges of the weighted graph,
$(G, c)$, where $G = (V, E, st)$.

\begin{tabbing}
\quad \= \quad \= \quad \= \quad \= \quad \= \quad \= \quad \\
{\bf function} $\mathit{Kruskal}$($(G, c)$: weighted graph): tree \\
 \> {\bf begin} \\
 \>\> Sort the edges in non-decreasing order of weights: \\
 \> \> $c(e_1) \leq c(e_2) \leq \cdots \leq c(e_n)$; \\
 \> \> $T := \emptyset$; \\
 \> \> {\bf for $i := 1$ to $n$ do} \\
 \> \> \> {\bf if} $(V, T\cup \{e_i\})$ is acyclic {\bf then}
$T := T\cup \{e_i\}$ \\
 \> \> \> {\bf endif} \\ 
 \> \> {\bf endfor}; \\
 \> \> $\mathit{Kruskal} := T$ \\
 \> {\bf end}
\end{tabbing}

\medskip
We admit that
the above description of Kruskal's algorithm is a bit sketchy
as we have not explicitly specified how we check that
adding an edge to a tree preserves acyclicity.
On the other hand, it is quite easy to prove the 
correctness of the above algorithm. It is not difficult
to refine the above ``naive'' algorithm to make it
totally explicit but this involves
a good choice of data structures. We leave these considerations
to an algorithms course.

\medskip
Clearly, the graph $T$ returned by the algorithm is acyclic,
but why is it connected? Well, suppose $T$ is not connected and
consider two of its  connected components, say $T_1$ and $T_2$.
Being acyclic and connected, $T_1$ and $T_2$ are trees.
Now, as $G$ itself it connected, for any node of $T_1$
and nay node of $T_2$, there is  some chain 
connecting these nodes.
Consider such a chain, $C$, of minimal length. Then, as
$T_1$ is a tree, the first edge, $e_j$, of $C$  cannot
belong to $T_1$ since otherwise, we would get an even shorter
chain connecting $T_1$ and $T_2$ by deleting $e_j$.
Furthermore, $e_j$ does not belong to any other connected
component of $T$, as these connected components are pairwise
disjoint. 
But then, $T + e_j$ is acyclic, which means that when
we considered the addition of  edge $e_j$ to the current
graph, $T^{(j)}$, the test should have been positive and
$e_j$ should have been added to  $T^{(j)}$.  Therefore,
$T$ is connected and so, it is a spanning tree.
Now, observe that as the edges are sorted in non-decreasing 
order of weight, condition 2(a) is enforced  and by Theorem \ref{Kruskal1},
$T$ is a minimal weight spanning tree.

\medskip
We can easily design a version of Kruskal's algorithm
based on considion 2(b). This time, we sort the edges
in non-increasing order of weights and, starting with $G$,
we attempt to delete each edge, $e_j$, as long as the remaining
graph is still connected. We leave the design of this algorithm as
an exercise to the reader.

\medskip
Prim's algorithm is based on a rather different observation.
For any node, $v\in V$, let $U_v$ be the
set of edges incident with $v$ that are not loops,
\[
U_v = \{e\in E \mid v\in st(e), \> st(e) \in [V]^2\}.
\]
Choose in $U_v$ some edge of minimum weight which
we will (ambiguously) denote by $e(v)$.

\begin{prop}
\label{prim1}
Let $(G, c)$ be a connected weighted graph with $G = (V, E, st)$.
For every vertex, $v\in V$, there is a minimum weight spanning
tree, $T$, so that $e(v) \in T$.
\end{prop}

\proof
Let $T'$ be a minimum weight spanning tree of $G$ and assume
that $e(v) \notin T'$. Let $C$ be the chain in $T'$ that joins
the endpoints of $e(v)$ and let $e$ the edge of $C$ incident
with $v$. Then, the graph $T'' = (V, (T'\cup \{e(v)\}) - \{e\})$
is a spanning tree of weight less that or equal to the weight
of $T'$ and as $T'$ has minimum  weight, do does $T''$.
By construction, $e(v) \in T''$.
$\bigsquare$

\medskip
Prim's algorithm uses an edge-contraction operation described below:

\begin{defin}
\label{contracdef}
{\em
Let $G = (V, E, st)$ be a graph, and let $e\in E$
be some edge which is not a loop, i.e.,
$st(e) = \{u, v\}$, with $u \not= v$. The graph, $C_e(G)$,
obtained by {\it contracting the edge $e$\/} is the graph
obtained by merging $u$ and $v$ into a single node and deleting $e$.
More precisely, $C_e(G) = ((V - \{u, v\})\cup \{w\}, E - \{e\}, st_e)$,
where $w$ is any new node not in $V$ and where
\begin{enumerate}
\item
$st_e(e') = st(e')$ iff $u\notin st(e')$ and $v\notin st(e')$
\item
$st_e(e') = \{w, z\}$ iff $st(e') = \{u, z\}$, with $z\notin st(e)$
\item
$st_e(e') = \{z, w\}$ iff $st(e') = \{z, v\}$, with $z\notin st(e)$
\item
$st_e(e') = z$ iff $st(e') = \{u, v\}$.
\end{enumerate}  
}
\end{defin}

\begin{prop}
\label{prim2}
Let $G = (V, E, st)$ be a graph. For any edge, $e\in E$,
the graph $G$ is a tree iff $C_e(G)$ is a tree.
\end{prop}

\proof
Proposition \ref{prim2} follows from Theorem \ref{bridgep4}. Observe that
$G$ is connected iff $C_e(G)$ is connected. Moreover, if $G$ is a tree,
the number of nodes of $C_e(G)$ is $n_e = |V| - 1$
and the number of edges of $C_e(G)$ is $m_e = |E| - 1$. Since
$|E| = |V| - 1$, we get $m_e = n_e - 1$ and $C_e(G)$ is a tree.
Conversely,  if $C_e(G)$ is a tree, then $m_e = n_e - 1$, $|V| = n_e + 1$
and $|E| = m_e + 1$, so $m = n - 1$ and $G$ is a tree. 
$\bigsquare$

\medskip
Here is a ``naive'' version of Prim's algorithm.

\begin{tabbing}
\quad \= \quad \= \quad \= \quad \= \quad \= \quad \= \quad \\
{\bf function} $\mathit{Prim}$($(G = (V, E, st), c)$: weighted graph): tree \\
 \> {\bf begin} \\
 \> \> $T := \emptyset$; \\
 \> \> {\bf while $|V| \geq 2$ do} \\
 \> \> \> pick any vertex $v\in V$; \\
 \> \> \> pick any edge (not a loop), $e$, in $U_v$ of minimum weight; \\
 \> \> \> $T := T\cup \{e\}$; $G := C_e(G)$ \\
 \> \> {\bf endwhile}; \\
 \> \> $\mathit{Prim} := T$ \\
 \> {\bf end} 
\end{tabbing}

\medskip
The correctness of Prim's algorithm is an immediate consequence
of Proposition \ref{prim1} and Proposition \ref{prim2}, 
the details are left to the reader. 

\section{$\Gamma$-Cycles, Cocycles, Cotrees, Flows and Tensions}
\label{sec26}
In this section, we take a closer look at the structure of cycles
in a finite graph, $G$. It turns out that there is a dual notion
to that of a cycle, the notion of
a {\it cocycle\/}. Assuming any orientation of our graph, it is possible
to associate a vector space, $\s{F}$, to the set of cycles in $G$,
another vector space, $\s{T}$, to 
the set of cocycles in $G$, and these vector spaces are mutually orthogonal
(for the usual inner product). Furthermore, these vector spaces
do not depend on the orientation chosen, up to isomorphism.
In fact, if $G$ has $m$ nodes, $n$ edges and $p$ connected components,
we will prove that $\mathrm{dim}\, \s{F} = n - m + p$
and  $\mathrm{dim}\, \s{T} = m - p$.
These vector spaces are the {\it flows\/} and the {\it tensions\/} 
of the graph $G$,
and these notions are important in combinatorial optimization
and the study of networks. This chapter assumes some basic knowledge
of linear algebra.

\medskip
Recall that if $G$ is a directed graph, then a cycle, $C$, is
a closed simple chain, which means that $C$ is
a sequence of the form
$C = (u_0, e_1, u_1, e_2, u_2, \ldots, u_{n - 1}, e_n, u_n)$, 
where $n \geq 1$; $u_i \in V$;  $e_i \in E$ and
\[
u_0 = u_n;
\quad
\{s(e_i),  t(e_{i})\} = \{u_{i-1 }, u_{i}\},\quad 1 \leq i \leq n
\quad\hbox{and}\quad
 e_i \not= e_j\quad \hbox{for all $i\not= j$}.
\]
The cycle, $C$, induces the sets $C^+$ and $C^-$
where $C^+$ consists of the edges whose orientation
agrees with the order of traversal induced by $C$
and where $C^-$ consists of the edges whose orientation
is the inverse of the order of traversal induced by $C$.
More precisely,
\[
C^+  = \{e_i\in C \mid s(e_i) = u_{i - 1},\, t(e_i) = u_i\}
\]
and 
\[
C^-  = \{e_i\in C \mid s(e_i) = u_{i},\, t(e_i) = u_{i-1}\}.
\]

\medskip
For the rest of this section, we assume that $G$ is a finite
graph and that its edges are named, $\mathbf{e}_1, \ldots, \mathbf{e}_n$%
\footnote{We use boldface notation for the edges in $E$
in order to avoid confusion with the edges occurring in a cycle or
in a chain; those are denoted in italic.}.

\begin{defin}
\label{vecrep}
{\em 
Given any finite directed graph, $G$, with $n$ edges, to
every cycle, $C$, is associated a {\it representative vector\/},
$\gamma(C)\in \reals^n$, defined so that for every $i$,
with $1 \leq i \leq n$,
\[
\gamma(C)_i = \cases{
+1 & if $e_i \in C^+$ \cr
-1 & if $e_i \in C^-$ \cr
0 & if $e_i \notin C$. \cr
}
\]
}
\end{defin}

\medskip
For example, if $G = G_8$ is the graph of Figure \ref{graphfig12},
the cycle 
\[
C = (v_3, e_7, v_4, e_6, v_5, e_5, v_2, e_1, v_1, e_2, v_3)
\]
corresponds to the vector
\[
\gamma(C) = (-1, 1, 0, 0, -1, -1 ,1).
\]

\begin{figure}
  \begin{center}
    \begin{pspicture}(0,0)(4.5,4.2)
    \cnodeput(3,0){v4}{$v_4$}
    \cnodeput(4.5,1.5){v5}{$v_5$}
    \cnodeput(0,3){v1}{$v_1$}
    \cnodeput(3,3){v2}{$v_2$}
    \cnodeput(0,0){v3}{$v_3$}
    \ncline[linewidth=1.5pt]{->}{v1}{v3}
    \ncline[linewidth=1.5pt]{->}{v1}{v2}
    \aput{:U}{$e_1$}
    \ncline[linewidth=1.5pt]{->}{v2}{v5}
    \ncline[linewidth=1.5pt]{->}{v3}{v2}
    \ncline[linewidth=1.5pt]{->}{v3}{v4}
    \bput{:U}{$e_7$}
    \ncline[linewidth=1.5pt]{->}{v4}{v2}
    \ncline[linewidth=1.5pt]{->}{v5}{v4}
    \uput[180](0,1.6){$e_2$}
    \uput[45](1,1.4){$e_3$}
    \uput[0](2.95,1.6){$e_4$}
    \uput[45](3.8,2.2){$e_5$}
    \uput[0](3.8,0.7){$e_6$}
    \end{pspicture}
  \end{center}
  \caption{Graph $G_8$}
\label{graphfig12}
\end{figure}

\medskip
Observe that distinct cycles may yield the same representative
vector unless they are elementary cycles. For example, the cycles
\[
C_1 = (v_2, e_5, v_5, e_6, v_4, e_4, v_2, e_1, v_1, e_2, v_3, e_3, v_2)
\]
and
\[
C_2 = (v_2, e_1, v_1, e_2, v_3, e_3, v_2, e_5, v_5, e_6, v_4, e_4, v_2)
\]
yield the same representative vector
\[
\gamma = (-1, 1, 1, 1, 1, 1, 0).
\]
In order to obtain a bijection between representative vectors and
``cycles'', we introduce the notion of a ``$\Gamma$-cycle''
(some authors redefine the notion of cycle and call ``cycle''
what we call a $\Gamma$-cycle, but we find this practice confusing).

\begin{defin}
\label{gammacycl}
{\em
Given a finite directed graph, $G = (V, E, s, t)$, 
a {\it $\Gamma$-cycle\/} is
any set of edges, $\Gamma = \Gamma^+\cup \Gamma^-$, such that
there is some cycle, $C$, in $G$ with $\Gamma^+ = C^+$
and $\Gamma^- = C^-$; we say that the  cycle, $C$, {\it induces
the $\Gamma$-cycle, $\Gamma$\/}.
The {\it representative vector\/}, $\gamma(\Gamma)$,
(for short, $\gamma$)
associated with $\Gamma$ is the vector, $\gamma(C)$,
from Definition \ref{vecrep}, where $C$ is
any cycle inducing $\Gamma$.
We say that a $\Gamma$-cycle, $\Gamma$, is a {\it $\Gamma$-circuit\/}
iff either $\Gamma^+ = \emptyset$ or $\Gamma^- = \emptyset$
and that $\Gamma$ is {\it elementary\/} iff $\Gamma$ arises
from an elementary cycle.
}
\end{defin}

\remarks
\begin{enumerate}
\item
Given a $\Gamma$-cycle, $\Gamma = \Gamma^+\cup \Gamma^-$,
we have the subgraphs $G^+ = (V, \Gamma^+, s, t)$ and
$G^- = (V, \Gamma^-, s, t)$. Then, for every $u\in V$, we have
\[
d^+_{G^+}(u)  - d^-_{G^+}(u) - d^+_{G^-}(u) + d^-_{G^-}(u)  = 0.
\]
\item
If $\Gamma$ is an elementary $\Gamma$-cycle, then every vertex of the graph
$(V, \Gamma, s, t)$ has degree $0$ or $2$.
\item
When the context is clear and no confusion may arise, we 
often drop the ``$\Gamma$'' is $\Gamma$-cycle and simply use the
term ``cycle'.
\end{enumerate}

\begin{prop}
\label{Gamcyclp1}
If $G$ is any finite directed graph, then any $\Gamma$-cycle,
$\Gamma$, is the disjoint union of elementary $\Gamma$-cycles.
\end{prop}

\proof
This is an immediate consequence of Proposition \ref{graphp5b}.
$\bigsquare$

\begin{cor}
\label{Gamcyclp2}
If $G$ is any finite directed graph, then any $\Gamma$-cycle, $\Gamma$,
is elementary iff it is minimal, i.e., if there is no
$\Gamma$-cycle, $\Gamma'$, such that $\Gamma' \subseteq \Gamma$
and $\Gamma' \not= \Gamma$.
\end{cor}

\medskip
We now consider a concept which will turn out to be dual to the
notion of $\Gamma$-cycle.

\begin{defin}
\label{cocycldef}
{\em
Let $G$ be a finite directed graph, $G = (V, E, s, t)$, with $n$ edges.
For any subset of nodes, $Y \subseteq V$, define the sets of edges,
$\Omega^+(Y)$ and $\Omega^-(Y)$ by
\begin{eqnarray*}
\Omega^+(Y) & = & \{e\in E \mid s(e) \in Y,\> t(e) \notin Y\} \\
\Omega^-(Y) & = & \{e\in E \mid s(e) \notin Y,\> t(e) \in Y\} \\
\Omega(Y) & = & \Omega^+(Y) \cup \Omega^-(Y).
\end{eqnarray*}
Any set, $\Omega$, of edges of the form
$\Omega = \Omega(Y)$, for some set of nodes,
$Y\subseteq V$, is called a {\it cocycle\/} (or {\it cutset\/}).
To every, cocycle, $\Omega$, we associate the
{\it representative vector\/}, $\omega(\Omega)\in \reals^n$,
defined so that
\[
\omega(\Omega)_i = \cases{
+1 & if $e_i \in \Omega^+$ \cr
-1 & if $e_i \in \Omega^-$ \cr
0 & if $e_i \notin \Omega$, \cr
}
\]
with $1 \leq i \leq n$. We also write $\omega(Y)$ for 
$\omega(\Omega)$ when $\Omega = \Omega(Y)$. 
If either $\Omega^+(Y) = \emptyset$ or $\Omega^-(Y) = \emptyset$,
then $\Omega$ is called a {\it cocircuit\/} and
an {\it elementary cocycle\/} (or {\it bond\/}) is
a minimal cocycle (i.e., there is no cocycle,
$\Omega'$, such that $\Omega' \subseteq \Omega$
and $\Omega' \not= \Omega$).
}
\end{defin}

\medskip
In the graph, $G_8$, of Figure \ref{graphfig12}, 
\[
\Omega = \{e_5\}\cup \{e_1, e_2, e_6\}
\]
is a cocycle induced by the set of nodes, $Y = \{v_2, v_3, v_4\}$
and it corresponds to the vector
\[
\omega(\Omega) = (-1, -1, 0, 0, 1, -1, 0).
\]
This is not an elementary cocycle because 
\[
\Omega' = \{e_5\}\cup \{e_6\}
\]
is also a cocycle (induced by $Y' = \{v_1, v_2, v_3, v_4\}$).
Observe that $\Omega'$ is a minimal cocycle, so it is an elementary
cocycle. Observe that the inner product
\[
\gamma(C_1) \cdot \omega(\Omega) = 
(-1, 1, 1, 1, 1, 1, 0)\cdot (-1, -1, 0, 0, 1, -1, 0) =
1 -1 + 0 + 0 + 1 - 1 + 0 = 0
\]
is zero. This is a general property that we will prove shortly.

\medskip
Observe that a cocycle, $\Omega$, is the set of edges of $G$ that join
the vertices in a set, $Y$, to the vertices in its complement,
$V - Y$. Consequently, deletetion of all the edges in $\Omega$
will increase the number of connected components of $G$.
We say that $\Omega$ is a {\it cutset\/} of $G$.
Generally, a set of edges, $K\subseteq E$, is a
{\it cutset\/} of $G$ if the graph $(V, E - K, s, t)$ has
more connected components than $G$.

\begin{figure}
  \begin{center}
    \begin{pspicture}(0,0)(4,4.3)
    \cnodeput(0,0){v1}{$1$}
    \cnodeput(4,0){v2}{$2$}
    \cnodeput(4,4){v3}{$3$}
    \cnodeput(0,4){v4}{$4$}
    \cnodeput(1,1){v5}{$5$}
    \cnodeput(3,1){v6}{$6$}
    \cnodeput(3,3){v7}{$7$}
    \cnodeput(1,3){v8}{$8$}
    \ncline[linewidth=2pt]{->}{v1}{v2}
    \ncline[linewidth=2pt]{->}{v2}{v3}
    \ncline[linewidth=2pt]{->}{v3}{v4}
    \ncline[linewidth=2pt]{->}{v4}{v1}
    \ncline[linewidth=1pt]{->}{v1}{v5}
    \ncline[linewidth=1pt]{->}{v2}{v6}
    \ncline[linewidth=1pt]{->}{v3}{v7}
    \ncline[linewidth=1pt]{->}{v4}{v8}
    \ncline[linewidth=2pt]{->}{v5}{v6}
    \ncline[linewidth=2pt]{->}{v6}{v7}
    \ncline[linewidth=2pt]{->}{v7}{v8}
    \ncline[linewidth=2pt]{->}{v8}{v5}
    \end{pspicture}
  \end{center}
  \caption{A coycle, $\Omega$, equal to the edge set of a cycle, $\Gamma$}
\label{cocyclecycle}
\end{figure}
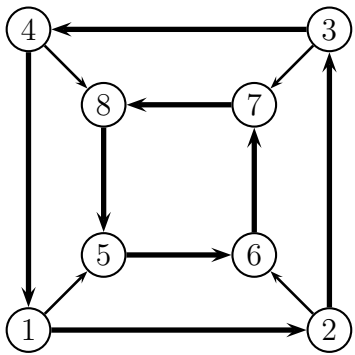

\medskip
It should be noted that a cocycle, $\Omega = \Omega(Y)$, may coincide
with the set of edges of some cycle, $\Gamma$.
For example, in the graph displayed in Figure \ref{cocyclecycle},
the cocycle, $\Omega = \Omega(\{1, 3, 5, 7\})$,
shown in thicker lines, is equal to the set of edges of the cycle,
\[
(1, 2), (2, 3), (3, 4), (4, 1), (5, 6), (6, 7), (7, 8), (8, 5).
\]
If the edges of the graph are listed in the order
\[
(1, 2), (2, 3), (3, 4), (4, 1), (5, 6), (6, 7), (7, 8), 
(8, 5), (1, 5), (2, 6), (3, 7), (4, 8) 
\]
the reader should check that the vectors
\[
\gamma = (1, 1, 1, 1, 1, 1, 1, 1, 0, 0, 0, 0)\in \s{F}
\quad\hbox{and}\quad
\omega = (1, -1, 1, -1, 1, -1, 1, -1, 0, 0, 0, 0, )\in \s{T}
\]
correspond to $\Gamma$ and $\Omega$, respectively.

\medskip
We now give several characterizations of elementary cocycles.

\begin{prop}
\label{cocyclp1}
Given a finite directed graph, $G = (V, E, s, t)$,
a set of edges, $S\subseteq E$, is an elementary cocycle 
iff it is a minimal cutset.
\end{prop}

\proof
We already observed that every cocycle is a cutset. Furthermore,
we claim that every cutset contains a cocyle.
To prove this, it is enough to consider a minimal cutset, $S$,
and to prove the following satement:

\medskip
{\it Claim\/}. Any minimal cutset,  $S$, 
is the set of edges of $G$ that join two 
nonempty sets of vertices, $Y_1$ and $Y_2$, such that
\begin{enumerate}
\item[(i)]
$Y_1\cap Y_2 = \emptyset$;
\item[(ii)]
$Y_1\cup Y_2 = C$, some connected component of $G$;
\item[(iii)]
The subgraphs $G_{Y_1}$ and $G_{Y_2}$, induced by $Y_1$ and $Y_2$
are connected.
\end{enumerate}
Indeed, if $S$ is a minimal cutset, it disconnects a unique
connected component of $G$, say $C$. Let
$C_1, \ldots, C_k$ be the connected components of
the graph, $C - S$, obtained from $C$ by deleting the edges in $S$.
Adding any edge, $e\in S$, to $C - S$, must connect
two components of $C$ since otherwise, $S - \{e\}$
would disconnect $C$, contradicting the minimality of $C$.
Furthermore, $k = 2$, since otherwise, again, $S - \{e\}$
would disconnect $C$. 
Then, if $Y_1$ is the set of nodes of $C_1$ and $Y_2$ is the set of
nodes of $C_2$, it is clear that the Claim holds.

\medskip
Now, if $S$ is a minimal cutset, the above argument shows that
$S$ contains a cocyle and this cocycle must be elementary
(i.e., minimal as a cocycle)  as it
is a cutset. Conversely, if $S$ is an elementary cocycle,
i.e., minimal as a cocycle, it must be a minimal cutset
since otherwise, $S$ would contain a strictly smaller
cutset which would then contain a cocycle strictly contained
in $S$.
$\bigsquare$

\begin{prop}
\label{cocyclp2}
Given a finite directed graph, $G = (V, E, s, t)$,
a set of edges, $S\subseteq E$, is an elementary cocycle 
iff $S$ is the set of edges of $G$ that join two 
nonempty sets of vertices, $Y_1$ and $Y_2$, such that
\begin{enumerate}
\item[(i)]
$Y_1\cap Y_2 = \emptyset$;
\item[(ii)]
$Y_1\cup Y_2 = C$, some connected component of $G$;
\item[(iii)]
The subgraphs $G_{Y_1}$ and $G_{Y_2}$, induced by $Y_1$ and $Y_2$
are connected.
\end{enumerate}
\end{prop}

\proof
It is clear that if $S$ satisfies (i)--(iii), then
$S$ is a minimal cutset and by Proposition \ref{cocyclp2},
it is an elementary cocycle.

\medskip
Let us first assume that $G$ is connected and that $S = \Omega(Y)$
is an elementary cocycle, i.e., is minimal as a cocycle.
If we let $Y_1 = Y$ and $Y_2 = X - Y_1$, it is clear that
(i) and (ii) are satisfied. If $G_{Y_1}$ or $G_{Y_2}$ is not connected,
then if $Z$ is a connected component of one of these two graphs,
we see that $\Omega(Z)$ is a cocycle strictly contained in
$S = \Omega(Y_1)$, a contradiction. Therefore, (iii) also holds.
If $G$ is not connected, as $S$ is a minimal cocycle it is a minimal cutset
and so, it is contained in some connected component, $C$, of $G$
and we apply the above argument to $C$.
$\bigsquare$

\medskip
The following proposition is the analog of Proposition \ref{Gamcyclp1}
for cocycle:

\begin{prop}
\label{cocyclp3}
Given a finite directed graph, $G = (V, E, s, t)$,
every cocycle, $\Omega = \Omega(Y)$, is the disjoint union
of elementary cocycles.
\end{prop}

\proof
We give two proofs. 

\medskip
{\it Proof 1\/}: (Claude Berge) $\>$
Let $Y_1, \ldots, Y_k$ be the connected components of the
subgraph of $G$ induced by $Y$. Then, it is obvious that
\[
\Omega(Y) = \Omega(Y_1) \cup \cdots \cup \Omega(Y_k),
\]
where the $\Omega(Y_i)$ are pairwise disjoint.
So, it is enough to show that each $\Omega(Y_i)$ is the union
of disjoint elementary cycles.

\medskip
Let $C$ be the connected component of $G$ that contains $Y_i$
and let $C_1, \ldots, C_m$ be the connected components of the
subgraph, $C - Y$, obtained from $C$ by deleting the nodes in $Y_i$ and the
edges incident to these nodes. Observe that the set of edges
that are deleted when the nodes in $Y_i$ are deleted is the
union of $\Omega(Y_i)$ and the edges of the connected
subgraph induced by $Y_i$. As a consequence, we see that
\[
\Omega(Y_i) = \Omega(C_1) \cup \cdots \cup \Omega(C_m),
\]
where $\Omega(C_k)$ is the set of edges joining $C_k$ and
nodes from $Y_i$ in the connected subgraph induced by
the nodes in
$Y_i \cup \bigcup_{j\not= k} C_{j}$. By Proposition
\ref{cocyclp3}, the set  $\Omega(C_k)$ is an elementary cocycle
and it is clear that the sets $\Omega(C_k)$ are pairwise
disjoint since the $C_k$ are disjoint.

\medskip
{\it Proof 2\/}: (Michel Sakarovitch) $\>$
Let $\Omega = \Omega(Y)$ be a cocycle in $G$. Now, $\Omega$ is a
cutset and we can pick
some minimal cocycle, $\Omega_1 = \Omega(Z)$,
contained in $\Omega_1$. We proceed by induction on
$|\Omega - \Omega_1|$. If $\Omega = \Omega_1$, we are done.
Otherwise, we claim that 
$E_1 = \Omega - \Omega_1$ is a cutset in $G$.
If not, let $e$ be any edge in $E_1$;
we may assume that $a = s(e)\in Y$ and $b = t(e)\in V - Y$.
As $E_1$ is not a cutset, there is a chain, $C$, from $a$ to $b$
in $(V, E - E_1, s, t)$ and as $\Omega$ is a cutset,
this chain must contain some edge $e'$, in $\Omega$,
so $C = C_1 (x, e', y) C_2$, where $C_1$ is a chain from
$a$ to $x$ and $C_2$ is a chain from $y$ to $b$.
Then, since $C$ has its edges in $E - E_1$ and
$E_1 = \Omega - \Omega_1$, we must have $e'\in \Omega_1$.
We may assume that $x = s(e') \in Z$ and $y = t(e') \in  V - Z$.
But, we have the chain, $C_1^R (a, e, b) C_2^R$,  
joining $x$ and $y$ in $(V, E - \Omega_1)$,
a contradiction. Therefore, $E_1$ is indeed a cutset of $G$.
Now, there is some minimal cocycle, $\Omega_2$, contained
in $E_1$, and if we let $E_2 = E_1 - \Omega_1$,
we can show as we just did that $E_2$ is a cutset of $G$ with
$|E_2| < |E_1$. Thus, we finish the proof by applying the induction
hypothesis to $E_2$.
$\bigsquare$

\medskip
We now prove the key property of orthogonality between
cycles and cocycles.

\begin{prop}
\label{orthop}
Given any finite directed graph, $G = (V, E, s, t)$,
if $\gamma = \gamma(C)$ is the representative vector
of any $\Gamma$-cycle, $\Gamma = \Gamma(C)$, and
$\omega = \omega(Y)$ is the representative vector of
any cocycle, $\Omega = \Omega(Y)$, then 
\[
\gamma\cdot \omega = \sum_{i = 1}^n \gamma_i\omega_i = 0,
\]
i.e., $\gamma$ and $\omega$ are orthogonal.
(Here, $|E| = n$.)
\end{prop}

\proof
Recall that $\Gamma = C^+\cup C^-$, where $C$ is a cycle in $G$, say
\[
C = (u_0, e_1, u_1, \ldots, u_{k-1}, e_k, u_k),
\quad\hbox{with}\quad u_k = u_0.
\]
Then, by definition, we see that
\begin{equation}
\gamma\cdot \omega = |C^+\cap \Omega^+(Y)|- 
|C^+\cap \Omega^-(Y)| - |C^-\cap \Omega^+(Y)| + |C^-\cap \Omega^-(Y)|.
\tag{$*$}
\end{equation}
As we traverse the cycle, $C$, when we traverse the edge
$e_i$ between $u_{i - 1}$ and $u_i$ ($1\leq i \leq k$), we note that
\begin{align*}
& e_i \in (C^+\cap \Omega^+(Y))\cup (C^-\cap \Omega^-(Y))
\quad\hbox{iff}\quad u_{i-1}\in Y,\> u_i\in V - Y \\
& e_i \in (C^+\cap \Omega^-(Y))\cup (C^-\cap \Omega^+(Y))
\quad\hbox{iff}\quad u_{i-1}\in V - Y,\> u_i\in Y. 
\end{align*}
In other words,  every time we traverse an edge 
coming out from $Y$, its contribution to $(*)$ is $+1$ and
every time  we traverse an edge coming into $Y$ its contribution to
$(*)$ is $-1$. After traversing the cycle $C$ entirely, we must have come out
from $Y$ as many times as we came into $Y$, so these contributions
must cancel out.
$\bigsquare$.

\medskip
Note that Proposition \ref{orthop} implies that
$|\Gamma\cap \Omega|$ is even.

\begin{defin}
\label{flowdef}
{\em
Given any finite digraph, $G = (V, E, s, t)$,
where $E = \{\mathbf{e}_1, \ldots, \mathbf{e}_n\}$, 
the subspace, $\s{F}(G)$, of $\reals^n$
spanned by all vectors, $\gamma(\Gamma)$, where
$\Gamma$ is any $\Gamma$-cycle, is called the
{\it cycle space of $G$\/} or {\it flow space of $G$\/}
and the  subspace, $\s{T}(G)$, of $\reals^n$ spanned by all vectors, 
$\omega(\Omega)$, where
$\Omega$ is any cocycle, is called the
{\it cocycle space of $G$\/} or {\it tension space of $G$\/}
(or {\it cut space of $G$\/}).
}
\end{defin}

\medskip
When no confusion is possible, we write $\s{F}$ for $\s{F}(G)$ and
$\s{T}$ for $\s{T}(G)$.
Thus, $\s{F}$ is the space consisting of all linear combinations
$
\sum_{i = 1}^k \alpha_i \gamma_i
$
of representative vectors of $\Gamma$-cycles, $\gamma_i$
and 
$\s{T}$ is the the space consisting of all linear combinations
$
\sum_{i = 1}^k \alpha_i \omega_i
$
of representative vectors of cocycles, $\omega_i$,
with $\alpha_i \in \reals$.
Proposition \ref{orthop} says that the spaces $\s{F}$ and $\s{T}$
are mutually orthogonal.

\remark
The  seemingly odd terminology ``flow space'' and ``tension space'' will be
explained later.

\medskip
Our next goal will be to determine the dimensions of
$\s{F}$ and $\s{T}$ in terms of the number of edges, the number of nodes
and the number of connected components of $G$ and to
give a convenient method for finding bases of $\s{F}$ and $\s{T}$.
For this, we will use spanning trees and their dual, cotrees.
But first, we will need  a crucial theorem that also plays an important
role in the theory of flows in networks.

\begin{thm} (Arc Coloring Lemma; Minty [1960])
\label{Minty}
Let $G = (V, E, s, t)$, be a finite directed graph and assume that
the edges of $G$ are colored either in black, red or green.
Pick any edge, $e$, and color it black. Then, exactly one of two
possibilities may occur:
\begin{enumerate}
\item[(1)]
There is an elementary cycle containing $e$ whose edges are 
only red or black with all the black edges oriented in the same direction;
\item[(2)]
There is an elementary cocycle containing $e$ whose edges are only
green or black with all the black edges oriented in the same direction.
\end{enumerate}
\end{thm}

\proof
Let $a = s(e)$ and $b = t(e)$.
Apply the following procedure for making nodes:

\begin{tabbing}
\quad \= \quad \= \quad \= \quad \= \quad \= \quad \= \quad \\
\> Intitially, only $b$ is marked. \\
\> {\bf while} there is some marked node $x$ and some unmarked node $y$ with \\
\> \> either a black edge, $e'$, with $(x, y) = (s(e'), t(e'))$ or \\
\> \> a red edge, $e'$, with $(x, y) = \{s(e'), t(e')\}$  \\
\> \> {\bf then} mark $y$; $\mathrm{arc}(y) = e'$ \\
\> {\bf endwhile}
\end{tabbing}

\medskip
When the marking algorithm stops, exactly one of the following two cases
occurs:

\begin{enumerate}
\item[(i)]
Node $a$ has been marked. Let $e' = \mathrm{arc}(a)$ be the edge
that caused $a$ to be marked and let $x$ be the other endpoint of $e'$.
If $x = b$, we found an elementary cycle satisfying (i). If not,
let $e'' =  \mathrm{arc}(x)$ and let $y$ be the other endpoint of $e''$
and continue in the same manner. This procedure will stop with $b$
and yields the chain, $C$,  from $b$ to $a$ along which nodes have been marked.
This chain must be elementary because every edge in it was used once to
mark some node (check that the set of edges used for the marking is a tree).
If we add the edge, $e$, to the chain, $C$, we obtain an elementary
cycle, $\Gamma$ whose edges are colored black or red and with all edges
colored black oriented in the same direction due to the marking scheme.
It is impossible to have a cocycle whose edges are colored black
or green containing $e$ because it would have been impossible to
conduct the marking through this cocycle and $a$ would not have been
marked.
\item[(ii)]
Node $a$ has not been marked.
Let $Y$ be the set of unmarked nodes. The set $\Omega(Y)$ is a cocycle
whose edges are colored green or black containing $e$ with all black edges 
in $\Omega^+(Y)$. This cocycle is the disjoint of elementary cocycles
(by Proposition \ref{cocyclp3}) and one of these elementary cocycles
contains $e$. 
If a cycle with black or red edges 
containing $e$ with all black edges oriented in the same direction
existed, then $a$ would have been marked, a contradiction.
$\bigsquare$
\end{enumerate}

\begin{cor}
\label{Minty2}
Every edge of a finite directed graph, $G$, belongs either
to an elementary circuit or to an elementary cocircuit but not both.
\end{cor}

\proof
Color all edges black and apply Theorem \ref{Minty}.
$\bigsquare$

\medskip
Although Minty's Theorem looks more like an amusing fact
than a deep result, it is actually a rather powerful theorem.
For example, we will see in Section \ref{sec29} that 
Minty's Theorem can be used to prove the ``hard part'' of
the Max-flow Min-cut Theorem (Theorem \ref{MaxfMinc}), an
important theorem that has many applications.
Here are a few more applications of Theorem \ref{Minty}.

\begin{prop}
\label{Minty3}
Let $G$ be a finite connected directed graph with at lest one edge. Then,
the following conditions are equivalent:
$G$ is strongly connected iff
\begin{enumerate}
\item[(i)]
$G$ is strongly connected.
\item[(ii)]
Every edge belongs to some circuit.
\item[(iii)]
$G$ has no cocircuit.
\end{enumerate}
\end{prop}

\proof
$(i)\Longrightarrow (ii)$. 
If $x$ and $y$ are the endpoints of any
edge, $e$, in $G$, as $G$ is strongly connected, 
there is an elementary path from $y$ to $x$ and thus,
an elementary circuit through $e$.

\medskip
$(ii)\Longrightarrow (iii)$. 
This follows from Corollary \ref{Minty2}.

\medskip
$(iii)\Longrightarrow (i)$. 
Assume that $G$ is not strongly connected and let $Y'$ and $Y''$
be two strongly connected components linked by some edge, $e$,
and let $a = s(e)$ and $b = t(e)$, with $a\in Y'$ and $b\in Y''$.
The edge $e$ does not belong to any circuit since otherwise,
$a$ and $b$ would belong to the same strongly connected component.
Thus, by Corollary \ref{Minty2}, the edge $e$ should belong
to some cocircuit, a contradiction.
$\bigsquare$

\medskip
In order to determine the dimension of the cycle space, $\s{T}$,
we will use spanning trees. Let us assume that $G$ is connected
since otherwise the same reasoning applies to the connected components
of $G$. If $T$ is any spanning tree of $G$, we know from
Theorem \ref{bridgep4}, part (4), that adding
any edge, $e\in E - T$, (called a {\it chord\/} of $T$) 
will create a (unique) cycle.
We will see shortly that the vectors associated with 
these cycles form a basis of the Cycle space.
We can find a basis of the cocycle space by considering
sets of edges of the form $E - T$, where $T$ is a spanning tree.
Such sets of edges are called {\it cotrees\/}.

\begin{defin}
\label{cotreedef}
{\em
Let $G$ be a finite directed connected graph, $G = (V, E, s, t)$.
A spanning subgraph, $(V, K, s, t)$, is a {\it cotree\/} iff
$(V, E - K, s, t)$ is a spanning tree.
}
\end{defin}

\medskip
Cotrees  are characterized
in the following proposition:

\begin{prop}
\label{cotreep1}
Let $G$ be a finite directed connected graph, $G = (V, E, s, t)$.
If $E$ is partitioned into two subsets, $T$ and $K$,
(i.e., $T\cup K = E$; $T\cap K = \emptyset$; $T, K \not= \emptyset$),
then the following conditions ar equivalent:
\begin{enumerate}
\item[(1)]
$(V, T, s, t)$ is tree.
\item[(2)]
$(V, K, s, t)$ is a cotree.
\item[(3)]
$(V, K, s, t)$ contains no elementary coycles of $G$ and
upon addition of any edge, $e\in T$, it does contain
an elementary cocycle of $G$.
\end{enumerate}
\end{prop}

\proof
By definition of a cotree, (1) and (2) are equivalent, so we will prove
the equivalence of (1) and (3).

\medskip
$(1)\Longrightarrow (3)$.
We claim that $(V, K, s, t)$
contains no elementary coycles of $G$. Otherwise, $K$ would
contain some elementary coycle, $\Gamma(A)$, of $G$ and then
no chain in the tree $(V, T, s, t)$ would connect $A$ and
$V - E$, a contradiction.

\medskip
Next, for any edge, $e\in T$, observe that $(V, T - \{e\}, s, t)$
has two connected components, say $A$ and $B$ and then,
$\Omega(A)$ is an elementary cocycle contained in $(V, K\cup \{e\}, s, t)$
(in fact, it is easy to see that it is the only one).
Therefore, (3) holds
 
\medskip
$(3)\Longrightarrow (1)$.
We need to prove that $(V, T, s, t)$ is tree. First, we show that
$(V, T, s, t)$ has no cycles. Let $e\in T$ be any edge; color
$e$ black; color all edges in $T - \{e\}$ red; color all
edges in $K = E - T$ green. By (3), by adding $e$ to $K$, we find
an elementary cocycle of black or green edges that contains $e$.
Thus, there is no cycle of red or black edges containing $e$.
As $e$ is arbitrary, there are no cycles in $T$.

\medskip
Finally, we prove that $(V, T, s, t)$ is connected.
Pick any edge, $e\in K$, and color it black; color edges in $T$ red; color
edges in $K - \{e\}$ green. Since $G$ has no cocycle of black and green edges
containing $e$, there is a cycle of black or red edges containing $e$.
Therefore, $T \cup \{e\}$ has a cycle, which means that there is a path
from any two nodes in $T$.
$\bigsquare$

\medskip
We are now ready for the main theorem of this section.

\begin{thm}
\label{flowdim1}
Let $G$ be a finite directed graph, $G = (V, E, s, t)$
and assume that $|E| = n$, $|V| = m$ and that $G$ has
$p$ connected components. Then, the cycle space, $\s{F}$,
and the cocycle space, $\s{T}$, are subspaces 
of $\reals^n$ of dimensions $\mathrm{dim}\, \s{F} = n - m + p$
and $\mathrm{dim}\, \s{T} = m - p$ and
$\s{T} = \s{F}^{\perp}$ is the orthogonal complement of $\s{F}$.
Furthermore, if $C_1, \ldots, C_p$ are the connected components
of $G$, bases of $\s{F}$ and $\s{T}$ can be found as follows:
\begin{enumerate}
\item[(1)] 
Let $T_1, \ldots, T_p$, be any spanning  trees in $C_1, \ldots, C_p$.
For each spanning tree, $T_i$, form all the elementary cycles, $\Gamma_{i, e}$,
obtained by adding any chord, $e\in C_i - T_i$,
to $T_i$. Then, the vectors $\gamma_{i, e} = \gamma(\Gamma_{i, e})$
form a basis of $\s{F}$.
\item[(2)]
For any spanning tree, $T_i$, as above, let $K_i = C_i - T_i$
be the corresponding cotree. For every edge, $e\in T_i$
(called a {\it twig\/}), there is a unique elementary
cocycle, $\Omega_{i, e}$, contained in $K_i\cup \{e\}$.
Then, the vectors $\omega_{i, e} = \omega(\Omega_{i, e})$
form a basis of $\s{T}$.
\end{enumerate}
\end{thm}

\proof
We know from Proposition \ref{orthop} that $\s{F}$ and $\s{T}$
are orthogonal. Thus,  
\[
\mathrm{dim}\, \s{F} + \mathrm{dim}\, \s{T}  \leq n.
\] 
Let us follow the procedure specified in (1).
Let $C_i = (E_i, V_i)$, be the $i$-th connected component of $G$ and let
$n_i = |E_i|$ and $|V_i| = m_i$, so that
$n_1 + \cdots + n_p = n$ and $m_1 + \cdots + m_p = m$.  
For any spanning tree, $T_i$, for $C_i$, recall that
$T_i$ has $m_i - 1$ edges and so, $|E_i - T_i| = n_i - m_i + 1$.
If $e_{i, 1}, \ldots, e_{i, n_i - m_i + 1}$ are the edges in $E_i - T_i$,
then the vectors 
\[
\gamma_{i, e_{i, 1}}, \ldots, \gamma_{i, e_{i, m_i}}
\]
must be linearly independent, because 
$\gamma_{i, e_{i, j}} = \gamma(\Gamma_{i, e_{i, j}})$
and the elementary cycle, $\Gamma_{i, e_{i, j}}$, contains the edge,
$e_{i, j}$, that none of the other  $\Gamma_{i, e_{i, k}}$ contain
for $k \not= j$. So, we get
\[
(n_1 - m_1 + 1) + \cdots + (n_p - m_p + 1) = n - m + p \leq 
\mathrm{dim}\, \s{F}.
\]

\medskip
Let us now follow the procedure specified in (2).
For every spanning tree, $T_i$, let
$e_{i, 1}, \ldots, e_{i, m_i - 1}$ be the edges in $T_i$.
We know from proposition \ref{cotreep1} that
adding any edge, $e_{i, j}$ to $C_i - T_i$
determines a unique elementary cocycle, $\Omega_{i, e_{i,j}}$,
containing $e_{i, j}$ and the vectors
\[
\omega_{i, e_{i, 1}}, \ldots, \omega_{i, e_{i, m_i - 1}}
\]
must be linearly independent since the elementary cocycle,
$\Omega_{i, e_{i,j}}$, contains the edge,
$e_{i, j}$, that none of the other  $\Omega_{i, e_{i, k}}$ contain
for $k \not= j$. So, we get
\[
(m_1 - 1) + \cdots + (m_p - 1) = m - p \leq 
\mathrm{dim}\, \s{T}.
\]
But then, 
$n \leq \mathrm{dim}\, \s{F} + \mathrm{dim}\, \s{T}$,
and since we also have
$\mathrm{dim}\, \s{F} + \mathrm{dim}\, \s{T} \leq n$,
we get
\[
\mathrm{dim}\, \s{F} = n - m + p
\quad\hbox{and}\quad 
\mathrm{dim}\, \s{T} = m - p.
\]
Since the vectors produced in (1) and (2) are linearly
independent and in each case, their number is equal to the dimension
of the space to which they belong, they are bases of these spaces.
$\bigsquare$

\medskip
Since $\mathrm{dim}\, \s{F} = n - m + p$ and
$\mathrm{dim}\, \s{T} = m - p$ do not depend on the orientation
of $G$, we conclude that the spaces $\s{F}$ and $\s{T}$
are uniquely determined by $G$, independently of the
orientation of $G$, up to isomorphism.
The number $n - m + p$ is called the {\it cyclomatic number of $G$\/}
and $m - p$ is called the {\it cocyclomatic number of $G$\/}.

\remarks
\begin{enumerate}
\item
Some authors, including  Harary  \cite{Harary} and Diestel
\cite{Diestel}, define the vector spaces $\s{F}$ and $\s{T}$
over the two-element field, $\mathbb{F}_2 = \{0, 1\}$.
The same dimensions are obtained for $\s{F}$ and $\s{T}$ and
$\s{F}$ and $\s{T}$ still orthogonal.
On the other hand, because $1 + 1 = 0$,
some interesting phenomena happen. For example, orientation
is irrelevant, 
the sum of two cycles (or cocycles) is their symmetric difference and
the space $\s{F}\cap \s{T}$ is {\bf not} necessarily reduced to 
the trivial space, $(0)$. The space $\s{F}\cap \s{T}$ is called
the {\it bicycle space\/}. The bicycle space induces a partition
of the edges of a graph called the {\it principal tripartition\/}.
For more on this, Godsil and Royle \cite{Godsil},
Sections 14.15 an 14.16 (and  Chapter 14).
 
\item
For those who know homology, of course,
$p = \mathrm{dim}\, H_0$, the dimension of the zero-th homology group
and $n - m + p = \mathrm{dim}\, H_1$, the dimension of the first homology
group of $G$ 
viewed as a topological space. Usually, the notation used is
$b_0 = \mathrm{dim}\, H_0$ and $b_1 = \mathrm{dim}\, H_1$
(the first two {\it Betti numbers\/}). Then, the above equation 
can be rewritten as
\[
m - n = b_0 - b_1,
\]
which is just the formula for the {\it Euler-Poincar\'e characteristic\/}.
\end{enumerate}

\begin{figure}
  \begin{center}
    \begin{pspicture}(0,0)(4,4.3)
    \cnodeput(0,0){v1}{$1$}
    \cnodeput(4,0){v2}{$2$}
    \cnodeput(4,4){v3}{$3$}
    \cnodeput(0,4){v4}{$4$}
    \cnodeput(1,1){v5}{$5$}
    \cnodeput(3,1){v6}{$6$}
    \cnodeput(3,3){v7}{$7$}
    \cnodeput(1,3){v8}{$8$}
    \ncline[linewidth=2pt]{v1}{v2}
    \ncline[linewidth=2pt]{v2}{v3}
    \ncline[linewidth=2pt]{v3}{v4}
    \ncline[linewidth=2pt]{v4}{v1}
    \ncline[linewidth=1pt]{v1}{v5}
    \ncline[linewidth=1pt]{v2}{v6}
    \ncline[linewidth=1pt]{v3}{v7}
    \ncline[linewidth=1pt]{v4}{v8}
    \ncline[linewidth=2pt]{v5}{v6}
    \ncline[linewidth=2pt]{v6}{v7}
    \ncline[linewidth=2pt]{v7}{v8}
    \ncline[linewidth=2pt]{v8}{v5}
    \end{pspicture}
  \end{center}
  \caption{A bicycle in a graph (a cube)}
\label{bicyclefig}
\end{figure}
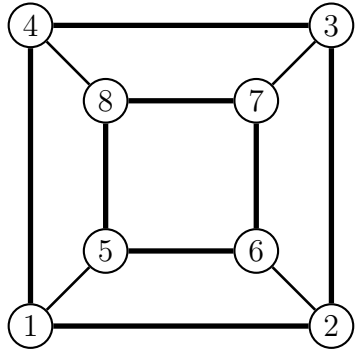

\medskip
Figure \ref{bicyclefig}, shows an unoriented  graph (a cube) and 
a cocycle, $\Omega$, which is also a cycle, $\Gamma$, shown in thick lines
(i.e., a bicycle, over the field $\mathbb{F}_2$).
However, as we saw in the example from Figure \ref{cocyclecycle},
for any orientation of the cube, 
the vectors, $\gamma$ and $\omega$, corresponding to $\Gamma$ and
$\Omega$ are different (and orthogonal). 

\medskip
Let us illustrate the procedures for constructing
bases of $\s{F}$ and $\s{T}$ 
on the graph $G_8$. 
Figure \ref{graphfig13} shows a spanning tree, $T$ and a cotree, $K$
for $G_8$.

\medskip
We have
$n = 7; m = 5$; $p = 1$, and so,
$\mathrm{dim}\, \s{F} = 7 - 5 + 1 = 3$ and 
$\mathrm{dim}\, \s{T} = 5 - 1 = 4$.
If we add successively the edges $e_2$, $e_6$, and $e_7$ to the spanning tree,
$T$, we get the three elementary cycles shown in Figure \ref{graphfig14}
with thicker lines.

\medskip
If we add successively the edges $e_1$, $e_3$, $e_4$ and $e_5$ to the cotree,
$K$, we get the four elementary cocycles shown in Figures \ref{graphfig15}
and \ref{graphfig16}
with thicker lines.

\medskip
Given any node, $v\in V$, in a graph, $G$, for simplicity of
notation let us denote the cocycle $\Omega(\{v\})$ by 
$\Omega(v)$. Similarly, we will write $\Omega^+(v)$ 
for $\Omega^+(\{v\})$;   $\Omega^-(v)$ 
for $\Omega^-(\{v\})$,  and similarly for the
the vectors, $\omega(\{v\})$, etc.
It turns our that vectors of the
form $\omega(v)$ generate the cocycle space and this has
important consequences.

\begin{prop}
\label{genp1}
Given any finite directed graph, $G = (V, E, s, t)$, 
for every cocycle, $\Omega = \Omega(Y)$, we have
\[
\omega(Y) = \sum_{v\in Y} \omega(v).
\]
Consequently, the vectors of the form $\omega(v)$,
with $v\in V$,  generate the cocycle space, $\s{T}$.
\end{prop}

\proof
For any edge, $e\in E$, if $a = s(e)$ and $b = t(e)$, 
observe that
\[
\omega(v)_e = \cases{
+1 & if $v = a$ \cr
-1 & if $v = b$ \cr
0 & if $v \not= a, b$. \cr
}
\]
As a consequently, if we evaluate $\sum_{v\in Y} \omega(v)$, we find that
\[
\left(\sum_{v\in Y} \omega(v)\right)_e = \cases{
+1 & if $a\in Y$ and $b\in V - Y$ \cr
-1 & if $a\in V- Y$ and $b\in Y$\cr
0 & if $a, b\in Y$ or $a, b\in V - Y$, \cr
}
\]
which is exactly $\omega(Y)_v$.
$\bigsquare$

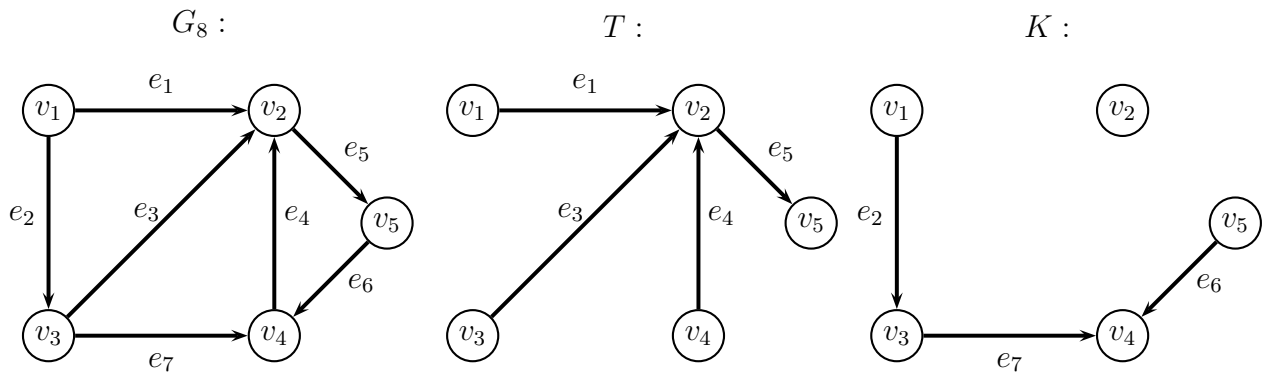
\begin{figure}
  \begin{center}
    \begin{pspicture}(0,0)(4.5,4.2)
    \cnodeput(3,0){v4}{$v_4$}
    \cnodeput(4.5,1.5){v5}{$v_5$}
    \cnodeput(0,3){v1}{$v_1$}
    \cnodeput(3,3){v2}{$v_2$}
    \cnodeput(0,0){v3}{$v_3$}
    \ncline[linewidth=1.5pt]{->}{v1}{v3}
    \ncline[linewidth=1.5pt]{->}{v1}{v2}
    \aput{:U}{$e_1$}
    \ncline[linewidth=1.5pt]{->}{v2}{v5}
    \ncline[linewidth=1.5pt]{->}{v3}{v2}
    \ncline[linewidth=1.5pt]{->}{v3}{v4}
    \bput{:U}{$e_7$}
    \ncline[linewidth=1.5pt]{->}{v4}{v2}
    \ncline[linewidth=1.5pt]{->}{v5}{v4}
    \uput[180](0,1.6){$e_2$}
    \uput[45](1,1.4){$e_3$}
    \uput[0](2.95,1.6){$e_4$}
    \uput[45](3.8,2.2){$e_5$}
    \uput[0](3.8,0.7){$e_6$}
    \uput[90](2,3.8){$G_8:$}
    \end{pspicture}
\hskip 1cm
    \begin{pspicture}(0,0)(4.5,4.2)
    \cnodeput(3,0){v4}{$v_4$}
    \cnodeput(4.5,1.5){v5}{$v_5$}
    \cnodeput(0,3){v1}{$v_1$}
    \cnodeput(3,3){v2}{$v_2$}
    \cnodeput(0,0){v3}{$v_3$}
    \ncline[linewidth=1.5pt]{->}{v1}{v2}
    \aput{:U}{$e_1$}
    \ncline[linewidth=1.5pt]{->}{v2}{v5}
    \ncline[linewidth=1.5pt]{->}{v3}{v2}
    \ncline[linewidth=1.5pt]{->}{v4}{v2}
    \uput[45](1,1.4){$e_3$}
    \uput[0](2.95,1.6){$e_4$}
    \uput[45](3.8,2.2){$e_5$}
    \uput[90](2,3.8){$T:$}
    \end{pspicture}
\hskip 1cm
    \begin{pspicture}(0,0)(4.5,4.2)
    \cnodeput(3,0){v4}{$v_4$}
    \cnodeput(4.5,1.5){v5}{$v_5$}
    \cnodeput(0,3){v1}{$v_1$}
    \cnodeput(3,3){v2}{$v_2$}
    \cnodeput(0,0){v3}{$v_3$}
    \ncline[linewidth=1.5pt]{->}{v1}{v3}
    \ncline[linewidth=1.5pt]{->}{v3}{v4}
    \bput{:U}{$e_7$}
    \ncline[linewidth=1.5pt]{->}{v5}{v4}
    \uput[180](0,1.6){$e_2$}
    \uput[0](3.8,0.7){$e_6$}
    \uput[90](2,3.8){$K:$}
    \end{pspicture}
  \end{center}
  \caption{Graph $G_8$; A Spanning Tree, $T$; A Cotree, $K$}
\label{graphfig13}
\end{figure}
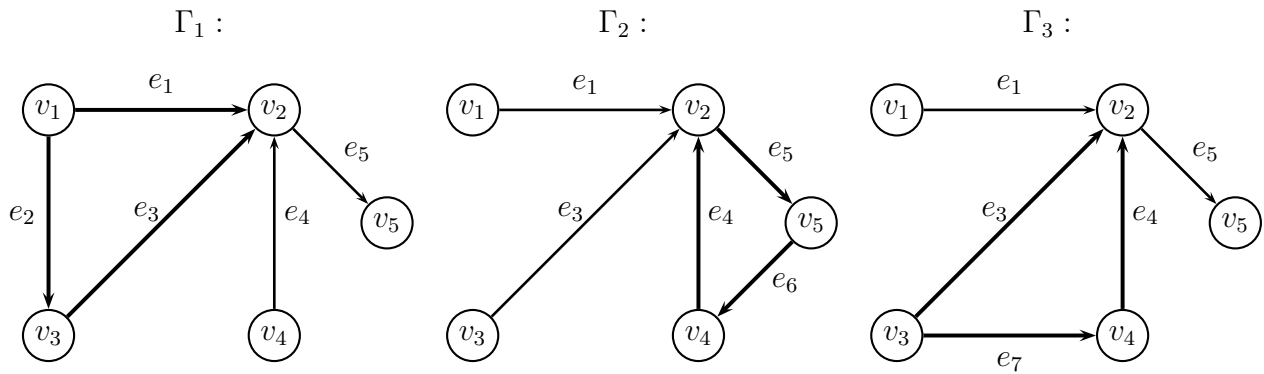
\begin{figure}
  \begin{center}
    \begin{pspicture}(0,0)(4.5,4.2)
    \cnodeput(3,0){v4}{$v_4$}
    \cnodeput(4.5,1.5){v5}{$v_5$}
    \cnodeput(0,3){v1}{$v_1$}
    \cnodeput(3,3){v2}{$v_2$}
    \cnodeput(0,0){v3}{$v_3$}
    \ncline[linewidth=1.5pt]{->}{v1}{v3}
    \ncline[linewidth=1.5pt]{->}{v1}{v2}
    \aput{:U}{$e_1$}
    \ncline[linewidth=1pt]{->}{v2}{v5}
    \ncline[linewidth=1.5pt]{->}{v3}{v2}
    \ncline[linewidth=1pt]{->}{v4}{v2}
    \uput[180](0,1.6){$e_2$}
    \uput[45](1,1.4){$e_3$}
    \uput[0](2.95,1.6){$e_4$}
    \uput[45](3.8,2.2){$e_5$}
    \uput[90](2,3.8){$\Gamma_1:$}
    \end{pspicture}
\hskip 1cm
    \begin{pspicture}(0,0)(4.5,4.2)
    \cnodeput(3,0){v4}{$v_4$}
    \cnodeput(4.5,1.5){v5}{$v_5$}
    \cnodeput(0,3){v1}{$v_1$}
    \cnodeput(3,3){v2}{$v_2$}
    \cnodeput(0,0){v3}{$v_3$}
    \ncline[linewidth=1pt]{->}{v1}{v2}
    \aput{:U}{$e_1$}
    \ncline[linewidth=1.5pt]{->}{v2}{v5}
    \ncline[linewidth=1pt]{->}{v3}{v2}
    \ncline[linewidth=1.5pt]{->}{v4}{v2}
    \ncline[linewidth=1.5pt]{->}{v5}{v4}
    \uput[45](1,1.4){$e_3$}
    \uput[0](2.95,1.6){$e_4$}
    \uput[45](3.8,2.2){$e_5$}
    \uput[0](3.8,0.7){$e_6$}
    \uput[90](2,3.8){$\Gamma_2:$}
    \end{pspicture}
\hskip 1cm
    \begin{pspicture}(0,0)(4.5,4.2)
    \cnodeput(3,0){v4}{$v_4$}
    \cnodeput(4.5,1.5){v5}{$v_5$}
    \cnodeput(0,3){v1}{$v_1$}
    \cnodeput(3,3){v2}{$v_2$}
    \cnodeput(0,0){v3}{$v_3$}
    \ncline[linewidth=1pt]{->}{v1}{v2}
    \aput{:U}{$e_1$}
    \ncline[linewidth=1pt]{->}{v2}{v5}
    \ncline[linewidth=1.5pt]{->}{v3}{v2}
    \ncline[linewidth=1.5pt]{->}{v4}{v2}
    \ncline[linewidth=1.5pt]{->}{v3}{v4}
    \bput{:U}{$e_7$}
    \uput[45](1,1.4){$e_3$}
    \uput[0](2.95,1.6){$e_4$}
    \uput[45](3.8,2.2){$e_5$}
    \uput[90](2,3.8){$\Gamma_3:$}
    \end{pspicture}
  \end{center}
  \caption{A Cycle Basis for $G_8$}
\label{graphfig14}
\end{figure}
\begin{figure}
  \begin{center}
    \begin{pspicture}(0,0)(4.5,4.2)
    \cnodeput(3,0){v4}{$v_4$}
    \cnodeput(4.5,1.5){v5}{$v_5$}
    \cnodeput(0,3){v1}{$v_1$}
    \cnodeput(3,3){v2}{$v_2$}
    \cnodeput(0,0){v3}{$v_3$}
    \ncline[linewidth=1.5pt]{->}{v1}{v2}
    \aput{:U}{$e_1$}
    \ncline[linewidth=1.5pt]{->}{v1}{v3}
    \ncline[linewidth=1pt]{->}{v3}{v4}
    \bput{:U}{$e_7$}
    \ncline[linewidth=1pt]{->}{v5}{v4}
    \uput[180](0,1.6){$e_2$}
    \uput[0](3.8,0.7){$e_6$}
    \uput[90](2,3.8){$\Omega_1:$}
    \end{pspicture}
\hskip 1cm
    \begin{pspicture}(0,0)(4.5,4.2)
    \cnodeput(3,0){v4}{$v_4$}
    \cnodeput(4.5,1.5){v5}{$v_5$}
    \cnodeput(0,3){v1}{$v_1$}
    \cnodeput(3,3){v2}{$v_2$}
    \cnodeput(0,0){v3}{$v_3$}
    \ncline[linewidth=1.5pt]{->}{v3}{v2}
    \ncline[linewidth=1.5pt]{->}{v1}{v3}
    \ncline[linewidth=1.5pt]{->}{v3}{v4}
    \bput{:U}{$e_7$}
    \ncline[linewidth=1pt]{->}{v5}{v4}
    \uput[180](0,1.6){$e_2$}
    \uput[45](1,1.4){$e_3$}
    \uput[0](3.8,0.7){$e_6$}
    \uput[90](2,3.8){$\Omega_2:$}
    \end{pspicture}
\hskip 1cm
    \begin{pspicture}(0,0)(4.5,4.2)
    \cnodeput(3,0){v4}{$v_4$}
    \cnodeput(4.5,1.5){v5}{$v_5$}
    \cnodeput(0,3){v1}{$v_1$}
    \cnodeput(3,3){v2}{$v_2$}
    \cnodeput(0,0){v3}{$v_3$}
    \ncline[linewidth=1pt]{->}{v1}{v3}
    \ncline[linewidth=1.5pt]{->}{v3}{v4}
    \bput{:U}{$e_7$}
    \ncline[linewidth=1.5pt]{->}{v5}{v4}
    \ncline[linewidth=1.5pt]{->}{v4}{v2}
    \uput[180](0,1.6){$e_2$}
    \uput[0](2.95,1.6){$e_4$}
    \uput[0](3.8,0.7){$e_6$}
    \uput[90](2,3.8){$\Omega_3:$}
    \end{pspicture}
  \end{center}
  \caption{A Cocycle Basis for $G_8$}
\label{graphfig15}
\end{figure}
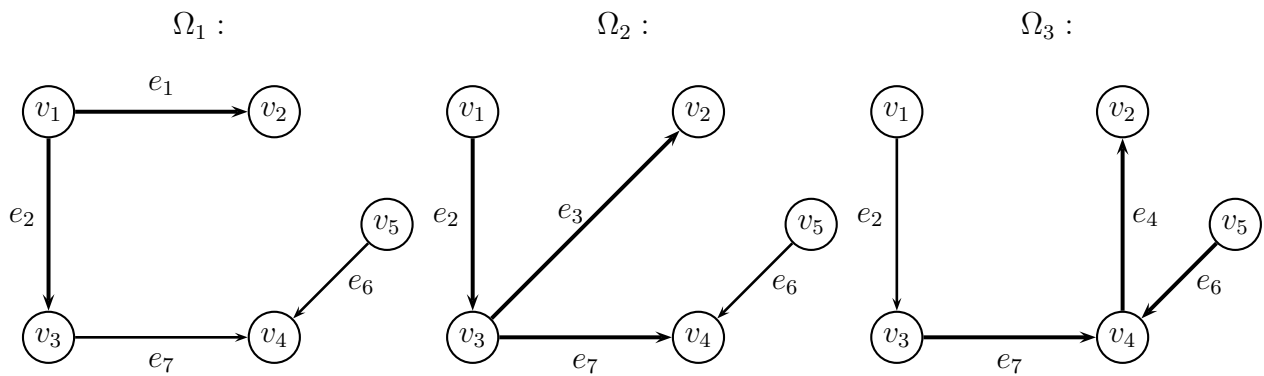
\begin{figure}
  \begin{center}
    \begin{pspicture}(0,0)(4.5,4.2)
    \cnodeput(3,0){v4}{$v_4$}
    \cnodeput(4.5,1.5){v5}{$v_5$}
    \cnodeput(0,3){v1}{$v_1$}
    \cnodeput(3,3){v2}{$v_2$}
    \cnodeput(0,0){v3}{$v_3$}
    \ncline[linewidth=1.5pt]{->}{v2}{v5}
    \ncline[linewidth=1pt]{->}{v1}{v3}
    \ncline[linewidth=1pt]{->}{v3}{v4}
    \bput{:U}{$e_7$}
    \ncline[linewidth=1.5pt]{->}{v5}{v4}
    \uput[180](0,1.6){$e_2$}
    \uput[45](3.8,2.2){$e_5$}
     \uput[0](3.8,0.7){$e_6$}
    \uput[90](2,3.8){$\Omega_4:$}
    \end{pspicture}
  \end{center}
  \caption{A Cocycle Basis for $G_8$ (continued)}
\label{graphfig16}
\end{figure}
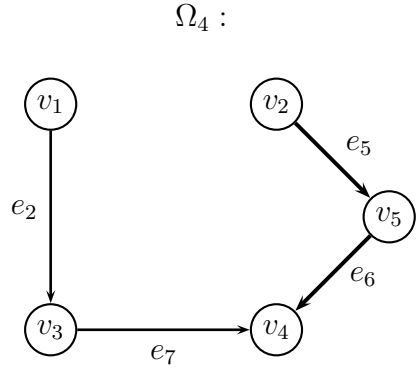

\medskip
Proposition \ref{genp1} allows us to characterize 
flows (the vectors in $\s{F}$) in an interesting way which
also reveals the reason behind the terminology.

\begin{thm}
\label{flowchar1}
Given any finite directed graph, $G = (V, E, s, t)$,
a vector, $f\in \reals^n$, is a flow in $\s{F}$ iff
\begin{equation}
\sum_{e\in \Omega^+(v)}[f(e)]  - \sum_{e\in \Omega^-(v)}[f(e)] = 0,
\quad\hbox{for all}\quad v\in V.  
\tag{$\dagger$}
\end{equation}
\end{thm} 

\proof
By Theorem \ref{flowdim1}, we know that $\s{F}$ is the orthogonal
complement of $\s{T}$. Thus, for any $f\in \reals^n$,
we have $f\in \s{F}$ iff $f\cdot \omega = 0$
for all $\omega\in \s{T}$. Moreover,
Proposition \ref{genp1} says that $\s{T}$ is generated by the vectors
of the form $\omega(v)$, where $v\in V$, so  
 $f\in \s{F}$ iff $f\cdot \omega(v) = 0$ for all $v\in V$.
But, $(\dagger)$ is exactly the assertion that
$f\cdot \omega(v) = 0$ and the theorem is proved.
$\bigsquare$

\medskip
Equation $(\dagger)$ justifies the terminology of ``flow''
for the elements of the space $\s{F}$.
Indeed, a {\it flow\/}, $f$, in a (directed) graph, $G = (V, E, s, t)$,
is defined as a function, $\mapdef{f}{E}{\reals}$, and we say that
a  flow is {\it conservative\/}
(Kirchhoff's first law) iff for every node, $v\in V$, the total flow, 
$\sum_{e\in \Omega^+(v)}[f(e)]$, coming into the vertex, $v$, is equal to 
the total flow, $\sum_{e\in \Omega^-(v)}[f(e)]$, coming out of that
vertex. This is exactly what equation $(\dagger)$ says.

\medskip
We can also characterize tensions as follows:

\begin{thm}
\label{tensionchar1}
Given any finite simple directed graph, $G = (V, E, s, t)$,
for any, $t\in \reals^n$, we have:
\begin{enumerate}
\item[(1)]
The vector, $t$, is a tension in $\s{T}$ iff
for every elementary cycle, $\Gamma = \Gamma^+\cup \Gamma^-$, we have
\begin{equation}
\sum_{e\in \Gamma^+}[t(e)]  - \sum_{e\in \Gamma^-}[t(e)] = 0.
\tag{$*$}
\end{equation}
\item[(2)]
If $G$ has no parallel edges (and no loops), then
$t\in \reals^n$ is a tension in $\s{T}$ iff
the following condition holds: 
There is a function, $\mapdef{\pi}{V}{\reals}$, called a 
{\it ``potential function''\/}, such that
\begin{equation}
t(e) = \pi(t(e)) - \pi(s(e)),
\tag{$**$}
\end{equation}
for every $e\in E$.
\end{enumerate}
\end{thm} 

\proof
(1)
The equation, $(*)$, asserts that $\gamma(\Gamma)\cdot t  = 0$
for every elementary cycle, $\Gamma$. Since every cycle is the disjoint 
union of elementary cycles, the vectors of the form $\gamma(\Gamma)$
generate the flow space, $\s{F}$, and by Theorem \ref{flowdim1}, 
the tension space $\s{T}$ is the orthogonal
complement of $\s{F}$, so $t$ is a tension iff $(*)$ holds.

\medskip
(2)
Assume  a potential function, $\mapdef{\pi}{V}{\reals}$, 
exists, let $\Gamma = (v_0, e_1, v_1, \ldots, v_{k-1}, e_k, v_k)$,
with $v_k = v_0$, be an elementary cycle
and and let $\gamma = \gamma(\Gamma)$.
We have
\begin{eqnarray*}
\gamma_1 t(e_1) & = & \pi(v_1) - \pi(v_0) \\
\gamma_2 t(e_2) & = & \pi(v_2) - \pi(v_1) \\
       & \vdots & \\
\gamma_{k - 1} t(e_{k - 1}) & = & \pi(v_{k - 1}) - \pi(v_{k - 2}) \\
\gamma_k t(e_{k}) & = & \pi(v_{0}) - \pi(v_{k - 1}) \\
\end{eqnarray*}
and we see that when we add up both sides of these equations that we get
$(*)$:
\[
\sum_{e\in \Gamma^+}[t(e)]  - \sum_{e\in \Gamma^-}[t(e)] = 0.
\]

Let us now assume that $(*)$ holds for every elementary cycle
and let $t\in \s{T}$ be any tension.
Consider the following procedure for assigning a value, $\pi(v)$,
to every vertex, $v\in V$, so that $(**)$ is satisfied.
Pick any vertex, $v_0$, and assign it the value, $\pi(v_0) = 0$.

\medskip
Now, for every vertex, $v\in V$, that has not yet been assigned a 
value, do the following:
\begin{enumerate}
\item
If there is an edge, $e = (u, v)$, with $\pi(u)$ already determined,
set 
\[
t(v) = t(u) + t(e);
\]

\item
If there is an edge, $e = (v, u)$, with $\pi(u)$ already determined,
set 
\[
t(v) = t(u) - t(e).
\]
\end{enumerate}
At the end of this process, all the nodes in the connected
component of $v_0$ will have received a value and we repeat this process for
all the other connected components. However, we have to check that
each nodes receives a unique value (given the choice of $v_0$).
If some node, $v$, is assigned two different values, $\pi_1(v)$ 
and $\pi_2(v)$, then there exist two chains, $\sigma_1$ and $\sigma_2$,
from $v_0$ to $v$, and if $C$ is the cycle $\sigma_1\sigma_2^R$,
we  have
\[
\gamma(C)\cdot t \not= 0.
\]
However, any cycle is the disjoint union of elementary cycles,
so there would be some elementary cycle, $\Gamma$, with
\[
\gamma(\Gamma)\cdot t \not= 0,
\]
contradicting $(*)$.
Therefore, the function $\pi$ is indeed well-defined and, by construction,
satisfies $(**)$.
$\bigsquare$

\medskip
Some of these results can be improved in various ways.
For example, flows have what is called a ``conformal decomposition''.

\begin{defin}
\label{conformflow}
{\em
Given any finite directed graph, $G = (V, S, s, t)$,
we say that a flow, $f\in \s{F}$, has a {\it conformal decomposition\/},
iff there are some cycles, $\Gamma_1, \ldots, \Gamma_k$, such that
if $\gamma_i = \gamma(\Gamma_i)$, then
\[
f = \alpha_1\gamma_1 + \cdots + \alpha_k\gamma_k,
\]
with
\begin{enumerate}
\item
$\alpha_i \geq 0$, for $i = 1, \ldots, k$;
\item
For any edge, $e\in E$,
if $f(e) > 0$ (resp. $f(e) < 0$) 
and $e\in \Gamma_j$, then $e\in \Gamma_j^+$ (resp. $e\in \Gamma_j^-$).
\end{enumerate}
}
\end{defin}

\begin{prop}
\label{conformflow1}
Given any finite directed graph, $G = (V, S, s, t)$,
every flow, $f\in \s{F}$, has some conformal decomposition.
In particular, if $f(e) \geq 0$ for all $e\in E$, then
all the $\Gamma_j$'s  are circuits. 
\end{prop}

\proof
We proceed by induction on the number on nonzero components of $f$.
First, note that $f = 0$ has a trivial conformal decomposition.
Next, let $f\in \s{F}$ be a flow and assume that
every flow, $f'$, having at least one more zero component than $f$ has
some conformal decomposition. Let $\overline{G}$ be the
graph obtained by reversing the orientation of all edges, $e$,
for which $f(e) <  0$ and deleting all the edges for wich $f(e) = 0$.
Observe that $\overline{G}$
has no cocircuit, as the inner product of any
elementary cocircuit with any nonzero
flow cannot  be zero. Hence, by the corollary to the
Coloring Lemma, $\overline{G}$ has some circuit, $C$, and let $\Gamma$ be
a cycle of $G$ corresponding to $C$. Let
\[
\alpha = \min\{\min_{e\in \Gamma^+}\, f(e),\, 
\min_{e\in \Gamma^-}\, -f(e)\}\geq 0.  
\]
Then, the flow
\[
f' = f - \alpha \gamma(\Gamma)
\]
has at least one more zero component than $f$. Thus, $f'$ has 
some conformal decomposition and, by construction,
$f = f' + \alpha\gamma(\Gamma)$
is a conformal decomposition of $f$.
$\bigsquare$

\medskip
We now take a quick look at various matrices associated with a graph.

\section{Incidence and Adjacency Matrices of a Graph}
\label{sec27}
In this section, we are assuming that our graphs
are finite, directed, without loops and without parallel edges.

\begin{defin}
\label{incidef}
{\em 
Let $G = (V, E)$ be a graph with $V = \{\mathbf{v}_1, \ldots, \mathbf{v}_m\}$
and $E = \{\mathbf{e}_1, \ldots, \mathbf{e}_n\}$. The 
{\it incidence matrix, $D(G)$, of $G$\/}, is the
$m\times n$-matrix whose entries, $d_{i\, j}$, are
\[
d_{i\, j} = \cases{
+1 & if $\mathbf{v}_i = s(\mathbf{e}_j)$ \cr
-1 & if $\mathbf{v}_i = t(\mathbf{e}_j)$ \cr
0 & otherwise. \cr
}
\]
}
\end{defin}

\remark
The incidence matrix actually makes sense for
a graph, $G$, with parallel edges but without loops.

\medskip
For simplicity of notation and when no confusion is possible, we write
$D$ instead of $D(G)$.

\medskip
Since we assumed that $G$ has no loops,
observe that every column of $D$ contains
exactly two nonzero entries, $+1$ and $-1$. Also, the
$i$th row of  $D$ is the vector, $\omega(\mathbf{v}_i)$,
representing the cocycle, $\Omega(\mathbf{v}_i)$.
For example, here is the incidence matrix of the
graph $G_8$ shown again in Figure \ref{graphfig17}.

\begin{figure}
  \begin{center}
    \begin{pspicture}(0,0)(4.5,4.2)
    \cnodeput(3,0){v4}{$v_4$}
    \cnodeput(4.5,1.5){v5}{$v_5$}
    \cnodeput(0,3){v1}{$v_1$}
    \cnodeput(3,3){v2}{$v_2$}
    \cnodeput(0,0){v3}{$v_3$}
    \ncline[linewidth=1.5pt]{->}{v1}{v3}
    \ncline[linewidth=1.5pt]{->}{v1}{v2}
    \aput{:U}{$e_1$}
    \ncline[linewidth=1.5pt]{->}{v2}{v5}
    \ncline[linewidth=1.5pt]{->}{v3}{v2}
    \ncline[linewidth=1.5pt]{->}{v3}{v4}
    \bput{:U}{$e_7$}
    \ncline[linewidth=1.5pt]{->}{v4}{v2}
    \ncline[linewidth=1.5pt]{->}{v5}{v4}
    \uput[180](0,1.6){$e_2$}
    \uput[45](1,1.4){$e_3$}
    \uput[0](2.95,1.6){$e_4$}
    \uput[45](3.8,2.2){$e_5$}
    \uput[0](3.8,0.7){$e_6$}
    \end{pspicture}
  \end{center}
  \caption{Graph $G_8$}
\label{graphfig17}
\end{figure}

\[
D = 
\begin{pmatrix}
1  & 1  & 0  & 0  & 0  & 0  & 0  \\
-1 & 0  & -1 & -1 & 1  & 0  & 0  \\ 
0  & -1 & 1  & 0  & 0  & 0  & 1  \\
0  & 0  & 0  & 1  & 0  & -1 & -1 \\
0  & 0  & 0  & 0  & -1 & 1  & 0
\end{pmatrix}.
\]

\medskip
The incidence matrix,  $D$, of a graph, $G$, represents a linear map
from $\reals^n$ to $\reals^m$ called the 
{\it incidence map\/} (or {\it boundary map\/} and denoted by $D$
(or $\partial$).
For every $e\in E$, we have
\[
D(\mathbf{e_j}) = s(\mathbf{e_j}) - t(\mathbf{e_j}).
\]   

\remark
Sometimes, it is convenient to consider the vector space, 
$C_1(G) = \reals^E$,
of all functions, $\mapdef{f}{E}{\reals}$,
called the {\it edge space of $G$\/} and
the vector space, $C_0(G) = \reals^V$,
of all functions, $\mapdef{g}{V}{\reals}$,
called the {\it vertex space of $G$\/}. 
Obviously, $C_1(G)$ is isomorphic to $\reals^n$ and
$C_0(G)$ is isomorphic to $\reals^m$.
The transpose, $\transpos{D}$, of $D$, is a linear map
from $C_0(G)$ to $C_1(G)$ also called the
{\it coboundary map\/} and often denoted by $\delta$.
Observe that $\delta(Y) = \Omega(Y)$ (viewing
the subset, $Y\subseteq V$, as a vector in $C_0(G)$).

\medskip
The spaces of flows and tensions can be recovered from
the incidence matrix.

\begin{thm}
\label{incidp1}
Given any finite graph, $G$, if $D$ is the incidence
matrix of $G$ and $\s{F}$ and $\s{T}$ are the spaces of flows
and tensions on $G$, then
\begin{enumerate}
\item[(1)]
$\s{F} = \Ker\, D$;
\item[(2)]
$\s{T} = \Im\, \transpos{D}$.
\end{enumerate}
Futhermore, if $G$ has $p$ connected components and
$m$ nodes, then
\[
\mathrm{rank}\, D = m - p.
\]
\end{thm}
  
\proof
We already observed that the $i$th row of $D$ is
the vector $\omega(\mathbf{v}_i)$ and we know from
Theorem \ref{flowchar1} that $\s{F}$ is exactly the
set of vectors orthogonal to all  vectors of the form
$\omega(\mathbf{v}_i)$. Now, for any $f\in \reals^n$,
\[
\mathbf{A} f = 
\begin{pmatrix}
\omega(\mathbf{v}_1)\cdot f \\
\vdots \\
\omega(\mathbf{v}_m)\cdot f, 
\end{pmatrix}
\]
and so, $\s{F} = \Ker\, D$.
Since the vectors  $\omega(\mathbf{v}_i)$ generate $\s{T}$,
the rows of $D$ generate $\s{T}$, i.e.,
$\s{T} = \Im\, \transpos{D}$.

\medskip
From Theorem \ref{flowdim1}, we know that
\[
\mathrm{dim}\, \s{T} = m - p
\]
and since we just proved that $\s{T} = \Im\, \transpos{D}$, 
we get
\[
\mathrm{rank}\, D = \mathrm{rank}\, \transpos{D} =
m - p, 
\]
which proves the last part of our theorem.
$\bigsquare$

\begin{cor}
\label{incidp2}
For any graph, $G = (V, E, s, t)$, if
$|V| = m$, $|E| = n$ and $G$ has $p$ connected components,
then the incidence matrix, $D$, of $G$ has
rank $n$ (i.e., the columns of $D$ are linearly
independent) iff $\s{F} = (0)$ iff $n = m - p$.
\end{cor}

\proof
By Theorem \ref{incidp2}, we have
$\mathrm{rank}\, D = m - p$. So, 
$\mathrm{rank}\, D = n$ iff $n = m - p$
iff $n - m + p = 0$ iff $\s{F}= (0)$
(since $\mathrm{dim}\, \s{F} = n - m + p$).
$\bigsquare$

\medskip
The incidence matrix of a graph has another interesting property
observed by Poincar\'e. First, let us define a variant
of triangular matrices.

\begin{defin}
\label{pseudotrig}
{\em
An $n\times n$ (real or complex) matrix, $A = (a_{i\, j})$, is
said to be {\it pseudo-triangular and non-singular\/}
iff either 
\begin{enumerate}
\item[(i)] 
$n = 1$  and $a_{1\, 1} \not= 0$, or
\item[(ii)]
$n \geq 2$ and $A$ has some row, say  $k$, with a unique nonzero
entry, $a_{h,\, k}$, such that the submatrix, $B$, obtained
by deleting the $h$-th row and the $k$-th column from $A$ is also
pseudo-triangular and non-singular.
\end{enumerate}
}
\end{defin}

\medskip
It is easy to see a matrix defined as in Definition \ref{pseudotrig} 
can be transformed into a usual triangular matrix by permutation of its 
columns.

\begin{prop} (Poincar\'e, 1901)
\label{pseudotrigp1}
If $D$ is the incidence matrix of a graph, then
every square $k\times k$  nonsingular submatrix%
\footnote{
Given any $m\times n$ matrix, $A = (a_{i\, j})$, if
$1 \leq h \leq m$ and $1\leq k\leq n$, then a $h\times k$-submatrix,
$B$, of $A$ is obtained by picking any $k$ columns of $A$
and then any $h$ rows of this new matrix.
},
$B$, of  $D$ is pseudo-triangular.
Consequently, $\det(B) = +1, -1$, or $0$, for
any  square $k\times k$  submatrix, $B$, of  $D$.
\end{prop}

\proof
We proceed by induction on $k$. The result is obvious for $k = 1$.

\medskip
Next,  let $B$ be a square $k\times k$-submatrix
of $D$ which is nonsingular, not pseudo-triangular
and yet, every nonsingular $h\times h$-submatrix of $B$ is 
pseudo-triangular if $h < k$.
We know that every column of $B$ has at most two
nonzero entries (since every column of $D$
contains two nonzero entries: $+1$ and $-1$).
Also, as $B$ is not pseudo-triangular (but nonsingular)
every row of $B$ contains at least two nonzero elements.
But then, no row of $B$ may contain three of more elements, because
the number of nonzero slots in all columns is at most $2k$
and by the pigeonhole principle, we could fit $2k + 1$
objects in $2k$ slots, which is impossible. Therefore, every
row of $B$ contains exactly two nonzero entries. Again,
the pigeonhole principle implies that every column
also contains exactly two nonzero entries. 
But now, the nonzero entries in each column are $+1$ and $-1$,
so if we add all the rows of $B$, we get the zero vector,
which shows that $B$ is singular, a contradiction.
Therefore, $B$ is pseudo-triangular.

\medskip
Since the entries in $D$ are $+1, -1, 0$,
the above immediately implies that \\
$\det(B) = +1, -1$, or $0$, for
any  square $k\times k$  submatrix, $B$, of  $D$.
$\bigsquare$

\medskip
A square matrix such, $A$, such that
$\det(B) = +1, -1$, or $0$, for
any  square $k\times k$  submatrix, $B$, of  $A$
is said to be {\it totally unimodular\/}. This is a very strong 
property of incidence matrices that has
far reaching implications in the study of optimization problems
for networks.

\medskip
Another important matrix associated to a graph is its
adjacency matrix.

\begin{defin}
\label{adjamatdef}
{\em
Let $G = (V, E)$ be a graph with $V = \{\mathbf{v}_1, \ldots, \mathbf{v}_m\}$.
The {\it ajacency matrix, $A(G)$, of $G$\/}, is the
$m\times m$-matrix whose entries, $a_{i\, j}$, are
\[
a_{i\, j} = \cases{
1 & if ($\exists e\in E) (\{s(e), t(e)\} = \{\mathbf{v}_i, \mathbf{v}_j\}$) \cr
0 & otherwise. \cr
}
\]
}
\end{defin}

\medskip
When no confusion is possible, we write $A$ for
$A(G)$. Note that the matrix  $A$ is symmetric
and $a_{i\, i} = 0$. Here is the adjacency matrix of the graph $G_8$ shown
in Figure \ref{graphfig17}:

\[
A = 
\begin{pmatrix}
0 & 1 & 1 & 0 & 0 \\
1 & 0 & 1 & 1 & 1 \\
1 & 1 & 0 & 1 & 0 \\
0 & 1 & 1 & 0 & 1 \\
0 & 1 & 0 & 1 & 0
\end{pmatrix}.
\]

\medskip
We have the following useful relationship between
the incidence matrix and the adjacency matrix of a graph:

\begin{prop}
\label{adjp2}
Given any graph, $G$, if $D$ is the incidence
matrix of $G$, $A$ is the adjacency matrix of $G$
and $\Delta$ is the diagonal matrix such
that $\Delta_{i\, i} = d(\mathbf{v}_i)$,
the degree of node $\mathbf{v}_i$, then
\[
D\transpos{D} = \Delta - A.
\]
Consequently, $D\transpos{D}$ is 
independent of the orientation of $G$ and
$\Delta - A$ is symmetric
positive, semi-definite, i.e., the eigenvalues of
$\Delta - A$ are real and non-negative.
\end{prop}

\proof
It is well-known that 
$D\transpos{D}_{i\, j}$ is the inner
product of the $i$th row, $d_i$  and the $j$th row, $d_j$, of $D$.
If $i = j$, then as 
\[
d_{i\, k} = \cases{
+1 & if $s(\mathbf{e}_k) = \mathbf{v}_i$ \cr
-1 & if $t(\mathbf{e}_k) = \mathbf{v}_i$ \cr
0 & otherwise,\cr
}
\]
we see that $d_i\cdot d_i =  d(\mathbf{v}_i)$.
If $i \not= j$, then  $d_i\cdot d_j \not= 0$ iff
there is some edge, $\mathbf{e}_k$ with 
$s(\mathbf{e}_k) = \mathbf{v}_i$ and $t(\mathbf{e}_k) = \mathbf{v}_i$,
in which case, $d_i\cdot d_j = -1$.
Therefore,
\[
D\transpos{D} = \Delta - A,
\]
as claimed.
Now, $D\transpos{D}$ is obviously symmetric
and it is well kown that its eigenvalues are non-negative
(for example, see Gallier \cite{Gallbook2}, Chapter 12). 
$\bigsquare$

\medskip
\remarks
\begin{enumerate}
\item
The matrix, $L = D\transpos{D} = \Delta - A$,
is known as the {\it Laplacian (matrix) \/} of the graph, $G$.
Another common notation for the matrix $D\transpos{D}$ is $Q$.
Since the colums of $D$ contain
exactly the two nonzero entries, $+1$ and $-1$, we see that
the vector, $\mathbf{1}$, defined such that
$\mathbf{1}_i = 1$, is an eigenvector for the eigenvalue $0$.
\item
If $G$ is connected, then $D$ has rank $m - 1$,
so the rank of  $D\transpos{D}$ is also $m - 1$
and the other eigenvalues of  $D\transpos{D}$
besides $0$
are strictly positive. The smallest positive eigenvalue of
$L = D\transpos{D}$ has some remarkable
properties. There is an area of graph theory
overlapping (linear) algebra, called {\it spectral
graph theory\/}  that investigates the properties of graphs
in terms of the eigenvalues of its Laplacian matrix
but this is beyond the scope of these notes.
Some good references for algebraic graph theory
include Biggs \cite{Biggs}, 
Godsil and Royle \cite{Godsil} and  Chung \cite{Chung},
for spectral graph theory.

\medskip
One of the classical and surprising results in algebraic graph theory 
is a formula that gives the number of spanning trees,
$\tau(G)$, of a connected graph, $G$,  in terms of its Laplacian,
$L = D\transpos{D}$.
If $J$ denotes the square matrix whose entries are all $1$'s
and if $\mathrm{adj}\, L$ denotes the adjoint matrix of $L$
(the transpose of the matrix of cofactors of $L$), i.e.
the matrix given  by
\[
(\mathrm{adj}\, L)_{i\, j} = (-1)^{i + j}  \det L(j, i),
\]
where $L(j, i)$ is the matrix obtained by deleting the
$j$th row and the $i$-column of $L$, then we have
\[
\mathrm{adj}\, L = \tau(G) J. 
\] 
We also have
\[
\tau(G) = m^{-2}\det(J + L),
\]
where $m$ is the number of nodes of $G$.
\item
As we already observed,
the incidence matrix also makes sense for graphs
with parallel edges and no loops.
But now, in order for the equation
$D\transpos{D} = \Delta - A$ to hold,
we need to define $A$ differently. 
We still have the same definition as before for the adjacency 
matrix but we can define the new matrix, $\s{A}$, such that
\[
\s{A}_{i\, j} = |\{e\in E \mid s(e) = \mathbf{v}_i,\,  t(e) = \mathbf{v}_j\}|, 
\]
i.e., $\s{A}_{i\, j}$ is the number of parallel edges between
$\mathbf{v}_i$ and $\mathbf{v}_j$. Then, we can check that
\[
D\transpos{D} = \Delta - \s{A}.
\]
\item
There are also versions of the adjacency matrix and of the 
incidence matrix for undirected graphs. In this case,
$D$ is no longer totally unimodular. 
\end{enumerate}

\section{Eulerian and Hamiltonian Cycles}
\label{sec28}
In this short section, we discuss two classical problems
that go back to the very beginning of graph theory. These problems
have to do with the existence of certain kinds of cycles in graphs.
These problems come in two flavors depending whether the graphs are directed
or not but there are only minor differences between the two versions
and traditionally the focus is on undirected graphs.

\medskip
The first problems goes back to Euler and is usually known
as the {\it K\"onigsberg bridge problem\/}.
In 1736, the town of K\"onigsberg had seven bridges
joining four areas of land. 
Euler was asked whether it is possible to find  a cycle that
crosses every bridge exactly once (and returns to the starting
point). 

\begin{figure}
  \begin{center}
    \begin{pspicture}(0,0)(3,4.3)
    \cnode(0,0){2pt}{B}
    \cnode(0,2){2pt}{A}
    \cnode(0,4){2pt}{C}
    \cnode(3,2){2pt}{D}
    \ncline[linewidth=1pt]{D}{C}
    \ncline[linewidth=1pt]{D}{A}
    \ncline[linewidth=1pt]{D}{B}
    \ncarc[arcangle=-30, linewidth=1pt]{C}{A}
    \ncarc[arcangle=-30, linewidth=1pt]{A}{C}
    \ncarc[arcangle=30, linewidth=1pt]{A}{B}
    \ncarc[arcangle=30, linewidth=1pt]{B}{A}
    \uput[-90](0,0){$B$}
    \uput[180](0,2){$A$}
    \uput[90](0,4){$C$}
    \uput[0](3,2){$D$}
    \end{pspicture}
  \end{center}
  \caption{A graph modeling the K\"onigsberg bridge problem}
  \label{eul1}
\end{figure}
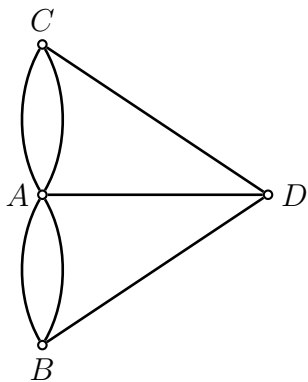

\medskip
The graph shown in Figure \ref{eul1} models  the 
K\"onigsberg bridge problem.
The nodes $A, B, C, D$ correspond to four areas of land in
K\"onigsberg and the edges to the seven bridges joining
these areas of land. 

\medskip
In fact, the problem is unsolvable,
as shown by Euler,  because some nodes do not have 
an even degree. We will now define the problem precisely and 
give a complete solution.

\begin{defin}
\label{Eulerdef}
{\em
Given a finite undirected graph, $G = (V, E)$, 
(resp., a directed graph, $G = (V, E, s, t)$), an {\it Euler cycle\/}
(or {\it Euler tour\/}) (resp. an {\it Euler circuit\/})
is a cycle in $G$ that passes through every node and every edge 
(exactly once)  (resp. a circuit in $G$ that passes through every 
node and every edge (exactly once)).
The 
{\it Eulerian cycle (resp. circuit) Problem\/} 
is the problem: Given a graph $G$, is there
an Eulerian cycle (resp. circuit) in $G$?
}
\end{defin}

\begin{thm}
\label{eulerthm}
(1)
An undirected graph, $G = (V, E)$, has an Eulerian cycle iff
the following properties hold:
\begin{enumerate}
\item[(a1)]
The graph $G$ is connected.
\item[(b1)]
Every node has even degree.
\end{enumerate}

(2)
A directed graph, $G = (V, E, s, t)$, has an Eulerian circuit iff
the following properties hold:
\begin{enumerate}
\item[(a2)]
The graph $G$ is strongly connected.
\item[(b2)]
Every node has the same number of incoming
and outgoing edges, i.e., 
$d^+(v) = d^-(v)$, for all $v\in $V.
\end{enumerate}
\end{thm}

\proof
We prove (1) leaving (2) as an easy exercise
(the proof of (2) is very similar to the proof of (1)).
Clearly, if a Euler cycle exists, $G$ is connected
and since every edge is traversed exactly once, every node
is entered as many times as it is exited so the degree
of every node is even.

\medskip
For the converse, observe that $G$ must contain a cycle
as otherwise, being connected, $G$ would be a tree
but we proved earlier that every tree has some 
node of degree $1$. (If $G$ is directed and strongly
connected, then we know that every edge belongs to a circuit.)
Let $\Gamma$ be any cycle in $G$.
We proceed by induction on the
number of edges in $G$. If $G$ has a single edge, clearly
$\Gamma = G$ and we are done. 
If $G$ has no loops and $G$ has two edges, again
$\Gamma = G$ and we are done. If $G$ has no loops and
no parallel edges and if $G$ has three edges, then
again, $\Gamma = G$.  
Now, consider the induction step.
Assume $\Gamma \not = G$ and consider the graph
$G ' = (V, E - \Gamma)$. Let $G_1, \ldots, G_p$
be the connected components of $G'$.
Pick any connected component,
$G_i$, of $G'$. Now, all nodes in $G_i$ have even degree,
$G_i$ is connected and $G_i$ has strictly fewer
edges than $G$ so, by the induction hypothesis, $G_i$ contains
an Euler cycle, $\Gamma_i$. But then, $\Gamma$ and each $\Gamma_i$
share some vertex and we can combine $\Gamma$ and the $\Gamma_i$'s 
to form an Euler cycle in $G$.
$\bigsquare$

\medskip
There are iterative algorithms that will find
an Euler cycle if one exists. It should also be noted that
testing whether or not a graph has an Euler cycle is
computationally quite an easy problem. This is not so for
the Hamiltonian cycle problem described next.

\medskip
A game invented by Sir William Hamilton in 1859 uses a
regular solid dodecahedron whose twenty vertices are labeled
with the names of famous cities. The player is challenged to
``travel around the world'' by finding a 
circuit along the edges of the dodecahedron which passes 
through every city exactly once.

\begin{figure}
  \begin{center}
    \begin{pspicture}(0,0)(8,8)
    \cnode(1,0){2pt}{v18}
    \cnode(7,0){2pt}{v17}
    \cnode(2,1.5){2pt}{v11}
    \cnode(4,2){2pt}{v12}
    \cnode(6,1.5){2pt}{v13}
    \cnode(2,3){2pt}{v10}
    \cnode(3,3.5){2pt}{v6}
    \cnode(4,3){2pt}{v5}
    \cnode(5,3.5){2pt}{v4}
    \cnode(6,3){2pt}{v14}
    \cnode(0,5.5){2pt}{v19}
    \cnode(1.3,4.7){2pt}{v9}
    \cnode(3,5.5){2pt}{v8}
    \cnode(3.5,4.5){2pt}{v7}
    \cnode(4.5,4.5){2pt}{v3}
    \cnode(5,5.5){2pt}{v2}
    \cnode(6.7,4.7){2pt}{v15}
    \cnode(8,5.5){2pt}{v16}
    \cnode(4,6.5){2pt}{v1}
    \cnode(4,8){2pt}{v20}
    \ncline[linewidth=1pt]{->}{v1}{v2}
    \ncline[linewidth=1pt]{->}{v2}{v3}
    \ncline[linewidth=1pt]{->}{v3}{v4}
    \ncline[linewidth=1pt]{->}{v4}{v5}
    \ncline[linewidth=1pt]{->}{v5}{v6}
    \ncline[linewidth=1pt]{->}{v6}{v7}
    \ncline[linewidth=1pt]{->}{v7}{v8}
    \ncline[linewidth=1pt]{->}{v8}{v9}
    \ncline[linewidth=1pt]{->}{v9}{v10}
    \ncline[linewidth=1pt]{->}{v10}{v11}
    \ncline[linewidth=1pt]{->}{v11}{v12}
    \ncline[linewidth=1pt]{->}{v12}{v13}
    \ncline[linewidth=1pt]{->}{v13}{v14}
    \ncline[linewidth=1pt]{->}{v14}{v15}
    \ncline[linewidth=1pt]{->}{v15}{v16}
    \ncline[linewidth=1pt]{->}{v16}{v17}
    \ncline[linewidth=1pt]{->}{v17}{v18}
    \ncline[linewidth=1pt]{->}{v18}{v19}
    \ncline[linewidth=1pt]{->}{v19}{v20}
    \ncline[linewidth=1pt]{->}{v16}{v20}
    \ncline[linewidth=1pt]{->}{v1}{v8}
    \ncline[linewidth=1pt]{->}{v15}{v2}
    \ncline[linewidth=1pt]{->}{v1}{v8}
    \ncline[linewidth=1pt]{->}{v11}{v18}
    \ncline[linewidth=1pt]{->}{v13}{v17}
    \ncline[linewidth=1pt]{->}{v20}{v1}
    \ncline[linewidth=1pt]{->}{v9}{v19}
    \ncline[linewidth=1pt]{->}{v7}{v3}
    \ncline[linewidth=1pt]{->}{v6}{v10}
    \ncline[linewidth=1pt]{->}{v5}{v12}
    \ncline[linewidth=1pt]{->}{v4}{v14}
    \end{pspicture}
  \end{center}
  \caption{A tour ``around the world.''}
  \label{ham1}
\end{figure}
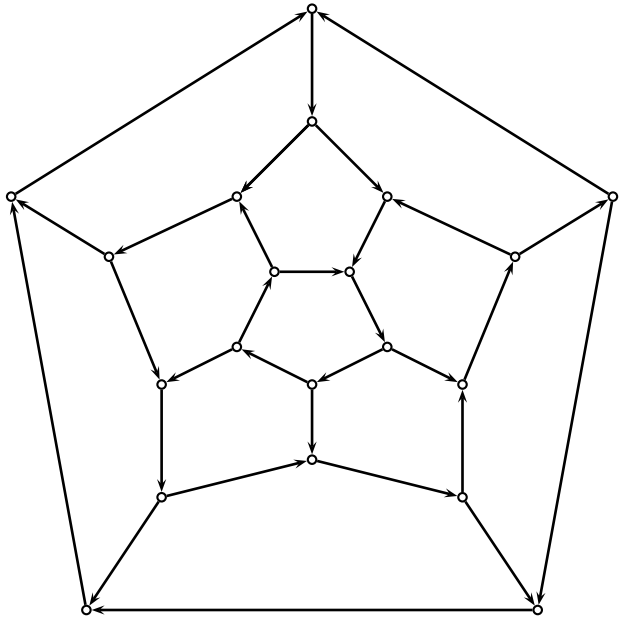

\medskip
In graphical terms, assuming an orientation of the edges
between cities, the graph $D$ shown in Figure \ref{ham1}
is a plane projection of a regular dodecahedron and
we want to know if there is a Hamiltonian cycle in this directed graph
(this is a directed version of the problem).

\medskip
Finding a Hamiltonian cycle in this graph
does not appear to be so easy!
A solution is shown in Figure \ref{ham2} below:

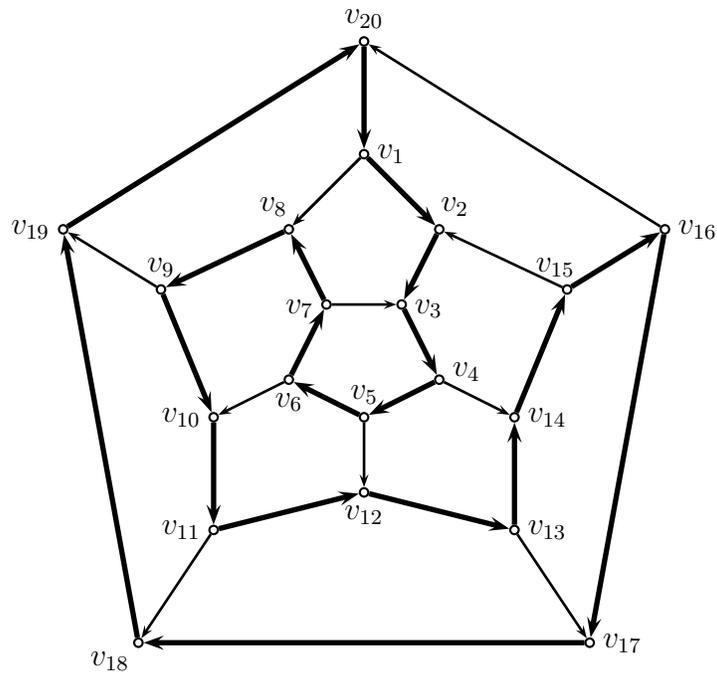
\begin{figure}
  \begin{center}
    \begin{pspicture}(0,0)(8,8)
    \cnode(1,0){2pt}{v18}
    \cnode(7,0){2pt}{v17}
    \cnode(2,1.5){2pt}{v11}
    \cnode(4,2){2pt}{v12}
    \cnode(6,1.5){2pt}{v13}
    \cnode(2,3){2pt}{v10}
    \cnode(3,3.5){2pt}{v6}
    \cnode(4,3){2pt}{v5}
    \cnode(5,3.5){2pt}{v4}
    \cnode(6,3){2pt}{v14}
    \cnode(0,5.5){2pt}{v19}
    \cnode(1.3,4.7){2pt}{v9}
    \cnode(3,5.5){2pt}{v8}
    \cnode(3.5,4.5){2pt}{v7}
    \cnode(4.5,4.5){2pt}{v3}
    \cnode(5,5.5){2pt}{v2}
    \cnode(6.7,4.7){2pt}{v15}
    \cnode(8,5.5){2pt}{v16}
    \cnode(4,6.5){2pt}{v1}
    \cnode(4,8){2pt}{v20}
    \ncline[linewidth=2pt]{->}{v1}{v2}
    \ncline[linewidth=2pt]{->}{v2}{v3}
    \ncline[linewidth=2pt]{->}{v3}{v4}
    \ncline[linewidth=2pt]{->}{v4}{v5}
    \ncline[linewidth=2pt]{->}{v5}{v6}
    \ncline[linewidth=2pt]{->}{v6}{v7}
    \ncline[linewidth=2pt]{->}{v7}{v8}
    \ncline[linewidth=2pt]{->}{v8}{v9}
    \ncline[linewidth=2pt]{->}{v9}{v10}
    \ncline[linewidth=2pt]{->}{v10}{v11}
    \ncline[linewidth=2pt]{->}{v11}{v12}
    \ncline[linewidth=2pt]{->}{v12}{v13}
    \ncline[linewidth=2pt]{->}{v13}{v14}
    \ncline[linewidth=2pt]{->}{v14}{v15}
    \ncline[linewidth=2pt]{->}{v15}{v16}
    \ncline[linewidth=2pt]{->}{v16}{v17}
    \ncline[linewidth=2pt]{->}{v17}{v18}
    \ncline[linewidth=2pt]{->}{v18}{v19}
    \ncline[linewidth=2pt]{->}{v19}{v20}
    \ncline[linewidth=1pt]{->}{v16}{v20}
    \ncline[linewidth=2pt]{->}{v20}{v1}
    \ncline[linewidth=1pt]{->}{v1}{v8}
    \ncline[linewidth=1pt]{->}{v15}{v2}
    \ncline[linewidth=1pt]{->}{v1}{v8}
    \ncline[linewidth=1pt]{->}{v11}{v18}
    \ncline[linewidth=1pt]{->}{v13}{v17}
    \ncline[linewidth=1pt]{->}{v9}{v19}
    \ncline[linewidth=1pt]{->}{v7}{v3}
    \ncline[linewidth=1pt]{->}{v6}{v10}
    \ncline[linewidth=1pt]{->}{v5}{v12}
    \ncline[linewidth=1pt]{->}{v4}{v14}
    \uput[-135](1,0){$v_{18}$}
    \uput[0](7,0){$v_{17}$}
    \uput[180](2,1.5){$v_{11}$}
    \uput[-90](4,2){$v_{12}$}
    \uput[0](6,1.5){$v_{13}$}
    \uput[180](2,3){$v_{10}$}
    \uput[-90](3,3.5){$v_{6}$}
    \uput[90](4,3){$v_{5}$}
    \uput[20](5,3.5){$v_{4}$}
    \uput[0](6,3){$v_{14}$}
    \uput[180](0,5.5){$v_{19}$}
    \uput[90](1.3,4.7){$v_{9}$}
    \uput[120](3,5.5){$v_{8}$}
    \uput[190](3.5,4.5){$v_{7}$}
    \uput[-10](4.5,4.5){$v_{3}$}
    \uput[60](5,5.5){$v_{2}$}
    \uput[110](6.7,4.7){$v_{15}$}
    \uput[0](8,5.5){$v_{16}$}
    \uput[0](4,6.5){$v_{1}$}
    \uput[90](4,8){$v_{20}$}
    \end{pspicture}
  \end{center}
  \caption{A Hamiltonian cycle in $D$}
  \label{ham2}
\end{figure}

\begin{defin}
\label{Hamiltondef}
{\em
Given any undirected graph, $G$, (resp. directed graph, $G$)
a {\it Hamiltonian cycle\/} in $G$ (resp. {\it Hamiltonian circuit\/}
in $G$) is a cycle that passes though every vertex of $G$
exactly once (resp. circuit that passes though every vertex of $G$
exactly once).
The {\it Hamiltonian cycle (resp. circuit) problem\/} is
to decide whether a graph, $G$ has a Hamiltonian cycle
(resp. Hamiltonian circuit). 
}
\end{defin}

\medskip
Unfortunately, no theorem analogous to Theorem \ref{eulerthm}
is known for Hamiltonian cycles. In fact, the 
Hamiltonian cycle problem is known to be NP-complete
and so far, appears to be a computationally hard problem
(of exponential time complexity).
Here is a proposition that may be used to prove that
certain graphs are not Hamiltonian. However, there are graphs
satisfying the condition of that proposition that are not 
Hamiltonian!

\begin{prop}
\label{Hamilp1}
If a graph, $G = (V, E)$, possesses a Hamiltonian cycle then, for every
nonempty set, $S$, of nodes, if $G\langle V - S\rangle$ is
the induced subgraph of $G$ generated by $V - S$
and if $c(G\langle V - S\rangle)$ is the number of
connected components of $G\langle V - S\rangle$, then
\[
c(G\langle V - S\rangle) \leq |S|.
\]
\end{prop}

\proof
Let $\Gamma$ be a Hamiltonian cycle in $G$ and let $\widetilde{G}$
be the graph $\widetilde{G} = (V, \Gamma)$. If we delete $k$ vertices
we can't cut a cycle into more than $k$ pieces and so
\[
c(\widetilde{G}\langle V - S\rangle) \leq |S|.
\]
However, we also have
\[
c(\widetilde{G}\langle V - S\rangle) \leq
c(G\langle V - S\rangle), 
\]
which proves the proposition.
$\bigsquare$

\section{Network Flow Problems; The Max-Flow Min-Cut Theorem}
\label{sec29}
The network flow problem is a perfect example of a problem 
which is important practically but  also
theoretically because in both cases it has unexpected applications.
In this section, we solve the network flow problem
using some of the notions from Section \ref{sec26}.
First, let us describe the kinds of graphs that we are dealing with,
usually called networks (or transportation networks or flow networks).

\begin{defin}
\label{networkdef}
{\em
A {\it network\/} (or {\it flow network\/}) is a
quadruple, $N = (G, c, v_s, s_t)$, where $G$ is a finite
diagraph, $G = (V, E, s, t)$, without loops,
$\mapdef{c}{E}{\reals_+}$, is a function 
called {\it capacity function\/} assigning a {\it capacity\/}, $c(e) > 0$,
(or {\it cost\/} or {\it weight\/}),  to every
edge, $e\in E$, and $v_s, v_t\in V$ are two (distinct) distinguished
nodes.%
\footnote{Most books use the notation $s$ and $t$ for $v_s$ and $v_t$.
Sorry, $s$ and $t$ are already used in the definition of a digraph!}
Moreover, we assume that there are no edges incoming into $v_s$
($d_G^-(v_s) = 0$), which is called the {\it source\/} 
and that there are no edges outgoing from $v_t$
($d_G^+(v_t) = 0$), which is called 
the  {\it terminal\/} (or {\it sink\/}).
}
\end{defin}

\medskip
An example of a network is showed in Figure \ref{net1}
with the capacity of each edge showed within parentheses.

\begin{figure}
  \begin{center}
    \begin{pspicture}(0,0)(5,4.3)
    \cnodeput(0,2){s}{$v_s$}
    \cnodeput(2.5,4){a}{$v_1$}
    \cnodeput(2.5,0){b}{$v_2$}
    \cnodeput(5,2){t}{$v_t$}
    \ncline[linewidth=1pt]{->}{s}{a}
    \ncline[linewidth=1pt]{->}{s}{b}
    \ncline[linewidth=1pt]{->}{a}{b}
    \ncline[linewidth=1pt]{->}{a}{t}
    \ncline[linewidth=1pt]{->}{b}{t}
    \uput[135](1.5,3){$(1)$}
    \uput[-135](1.5,1){$(4)$}
    \uput[45](3.5,3){$(5)$}
    \uput[-45](3.5,1){$(3)$}
    \uput[0](2.5,2){$(2)$}
    \end{pspicture}
  \end{center}
  \caption{A network, $N$}
  \label{net1}
\end{figure}

\medskip
Intuitively, we can think of the edges of a network as
conduits for fluid, or wires for electricity, or highways
for vehicle, {\it etc.\/}, and the capacity of each edge is the
maximum amount of ``flow'' that can pass through that edge.
The purpose of a network is to carry ``flow'', defined as follows:

\begin{defin}
\label{netflowdef}
{\em
Given a network, $N = (G, c, v_s, v_t)$, a {\it flow in $N$\/} is a 
function, $\mapdef{f}{E}{\reals}$, such that
the following  conditions hold:
\begin{enumerate}
\item[(1)] (Conservation of flow)
\[
\sum_{t(e) = v} f(e) = \sum_{s(e) = v} f(e),
\quad\hbox{for all $v \in V - \{v_s, v_t\}$}
\] 
\item[(2)] (Admissibility of flow)
\[
0 \leq f(e) \leq c(e),
\quad\hbox{for all $e\in E$}
\]
\end{enumerate}
Given any two sets of nodes, $S, T\subseteq V$, let
\[
f(S, T) = \sum_{
\begin{subarray}{c}
e\in E \\
s(e)\in S,\> t(e) \in T
\end{subarray}
} 
f(e) 
\quad\hbox{and}\quad
c(S, T) = \sum_{
\begin{subarray}{c}
e\in E \\
s(e)\in S,\> t(e) \in T
\end{subarray}
} 
c(e).
\]
When $S = \{u\}$ or $T = \{v\}$, we write $f(u, T)$ for
$f(\{u\}, T)$ and $f(S, v)$ for $f(S, \{v\})$ (similarly,
we write $c(u, T)$ for $c(\{u\}, T)$
and $c(S, v)$ for $c(S, \{v\})$).
The {\it net flow out of $S$\/} is
defined as $f(S, \overline{S}) - f(\overline{S}, S)$
(where $\overline{S} = V - S$).
The {\it value, $|f|$ (or $v(f)$) of the flow $f$\/} is the quantity
\[
|f| = f(v_s, V - \{v_s\}).
\]
}
\end{defin}

\medskip
We can now state the 

\medskip\noindent
{\bf Network Flow Problem\/}:
Find a flow, $f$, in $N$, for which
the value, $|f|$, is maximum (we call such a flow
a {\it maximum flow\/}).

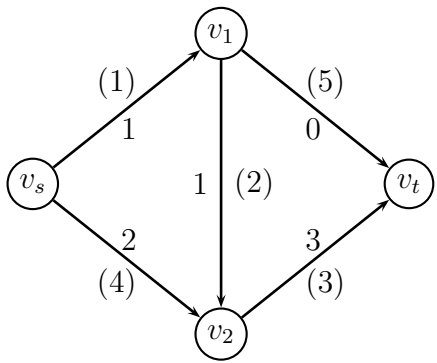
\begin{figure}
  \begin{center}
    \begin{pspicture}(0,0)(5,4.3)
    \cnodeput(0,2){s}{$v_s$}
    \cnodeput(2.5,4){a}{$v_1$}
    \cnodeput(2.5,0){b}{$v_2$}
    \cnodeput(5,2){t}{$v_t$}
    \ncline[linewidth=1pt]{->}{s}{a}
    \ncline[linewidth=1pt]{->}{s}{b}
    \ncline[linewidth=1pt]{->}{a}{b}
    \ncline[linewidth=1pt]{->}{a}{t}
    \ncline[linewidth=1pt]{->}{b}{t}
    \uput[135](1.5,3){$(1)$}
    \uput[-135](1.5,1){$(4)$}
    \uput[45](3.5,3){$(5)$}
    \uput[-45](3.5,1){$(3)$}
    \uput[0](2.5,2){$(2)$}
    \uput[-135](1.5,3){$1$}
    \uput[135](1.5,1){$2$}
    \uput[-45](3.5,3){$0$}
    \uput[45](3.5,1){$3$}
    \uput[180](2.5,2){$1$}
    \end{pspicture}
  \end{center}
  \caption{A flow in the network, $N$}
  \label{net2}
\end{figure}

\medskip
Figure \ref{net2} shows a flow in the network $N$, with 
value, $|f| = 3$. This is not a maximum flow, 
as the reader should check (the maximum flow value is $4$).

\remarks
\begin{enumerate}
\item
For any set of edges,  $\s{E}\subseteq E$, let
\begin{eqnarray*}
f(\s{E}) & = & \sum_{e\in \s{S}} f(e) \\
c(\s{E}) & = & \sum_{e\in \s{S}} c(e).
\end{eqnarray*}
Then, note that the net flow out of $S$ can also be expressed as
\[
f(\Omega^+(S)) -  f(\Omega^-(S)) =  
f(S, \overline{S}) - f(\overline{S}, S).
\]
Now, recall that $\Omega(S) = \Omega^+(S)\cup \Omega^-(S))$ is 
a cocycle (see Definition \ref{cocycldef}). So if we define
the value, $f(\Omega(S))$,  of the cocycle,  $\Omega(S)$, to be
\[
f(\Omega(S)) =  f(\Omega^+(S)) -  f(\Omega^-(S)), 
\]
the net flow through $S$ is the value of the cocycle, $\Omega(S)$.
\item
By definition, $c(S, \overline{S}) = c(\Omega^+(S))$.
\item
Since $G$ has no loops, there are no edges from $u$ to itself, so
\[
f(u, V - \{u\}) = f(u, V)
\]
and similarly,
\[
f(V - \{v\}, v) = f(V, v).
\]
\item
Some authors (for example, Wilf \cite{Wilf})
do not require the distinguished node, $v_s$, to be a source
and the distinguished node, $v_t$, to be a sink.
This makes essentially no difference but if so, the value of the
flow $f$ must be defined as
\[
|f| = f(v_s, V - \{v_s\}) - f(V - \{v_s\}, v_s) = 
f(v_s, V) - f(V, v_s).
\]
\end{enumerate}

\medskip
Intuitively, because  flow conservation holds for every node
except $v_s$ and $v_t$, the net flow, $f(V, v_t)$, into
the sink should be equal to the  net flow, $f(v_s, V)$
our of the source, $v_s$. This is indeed true and follows
from the following proposition:

\begin{prop}
\label{netflowp1}
Given a network, $N = (G, c, v_s, v_t)$, for any flow, $f$, in $N$
and for any subset, $S\subseteq V$, if $v_s\in S$ and $v_t\notin S$, then
the net flow through $S$ has the same value, namely $|f|$, that is
\[
|f| = f(\Omega(S)) = f(S, \overline{S}) - f(\overline{S}, S) 
\leq c(S, \overline{S}) = c(\Omega^+(S)).
\]
In particular,
\[
|f| = f(v_s, V) = f(V, v_t).
\]
\end{prop}

\proof
Recall that $|f| = f(v_s, V)$. Now, for any node,
$v \in S - \{v_s\}$, since $v \not = v_t$, the equation
\[
\sum_{t(e) = v} f(e) = \sum_{s(e) = v} f(e)
\] 
holds and we see that
\[
|f| =  f(v_s, V) = \sum_{v\in S} (\sum_{s(e) = v} f(e) - \sum_{t(e) = v} f(e)) =
\sum_{v\in S} (f(v, V) - f(V, v)) = f(S, V) - f(V, S).
\]
However, $V = S \cup \overline{S}$, so
\begin{eqnarray*}
|f| & = &  f(S, V) - f(V, S) \\
 & = & f(S, S \cup \overline{S}) - f(S \cup \overline{S}, S) \\
& = & f(S, S) + f(S, \overline{S}) - f(S, \overline{S}) - f(S, S) \\
& = & f(S, \overline{S}) - f(S, \overline{S}),
\end{eqnarray*}
as claimed.
Since the capacity of every edge is non-negative, it is obvious that
\[
|f| = f(S, \overline{S}) - f(S, \overline{S}) \leq f(S, \overline{S})
\leq c(S, \overline{S}) = c(\Omega^+(S)),
\]
since a flow is admissible.
Finally, if we set $S = V - \{v_t\}$, we get
\[
f(S, \overline{S}) - f(S, \overline{S}) = f(V, v_t)
\]
and so, $|f| = f(v_s, V) = f(V, v_t)$.
$\bigsquare$

\medskip
Proposition \ref{netflowp1} shows that the sets of edges,
$\Omega^+(S)$, with $v_s\in S$ and $v_t\notin S$, play a very special role.
Indeed, as a corollary of Proposition \ref{netflowp1}, we see
that the value any flow in $N$ is bounded by the capacity, 
$c(\Omega^+(S))$, of the set $\Omega^+(S)$,
for any $S$ with $v_s\in S$ and $v_t\notin S$.
This suggests the following definition:

\begin{defin}
\label{cutdef}
{\em
Given a network,  $N = (G, c, v_s, v_t)$, 
a {\it cut separating $v_s$  and $v_t$, for short a 
$v_s$-$v_t$-cut\/}, is any subset of edges,
$\s{C} = \Omega^+(W)$, where 
$W$ is a subset of $V$ with $v_s\in W$ and $v_t\notin W$.
The {\it capacity of a $v_s$-$v_t$-cut, $\s{C}$,\/} is
\[
c(\s{C}) = c(\Omega^+(W)) = \sum_{e\in \Omega^+(W)} c(e).
\]
}
\end{defin}

\remark
Some authors, including Papadimitriou and Steiglitz
\cite{PapadimitriouSteiglitz} and Wilf \cite{Wilf}, 
 define a $v_s$-$v_t$-cut as a pair
$(W, \overline{W})$, where $W$ is a subset of $V$ with
with $v_s\in W$ and $v_t\notin W$. 
This definition is clearly equivalent to our
definition above, which is due to Sakarovitch \cite{Sakarovitch2}.
We have a slight prerefence for Definition \ref{cutdef} 
because it places the emphasis on edges as opposed to nodes.
Indeed, the intuition behind  $v_s$-$v_t$-cuts is that
any flow from $v_s$ to $v_t$ must pass through some
edge of any  $v_s$-$v_t$-cut. Thus, it is not surprising
that the capacity of $v_s$-$v_t$-cuts places a restriction on how 
much flow can be sent from  $v_s$ to $v_t$.

\medskip
We can rephrase Proposition \ref{netflowp1} as follows:

\begin{prop}
\label{MaxfMincp1}
The maximum value of any flow, $f$,  in $N$ is 
bounded by the minimum  capacity, $c(\s{C})$,  of
any $v_s$-$v_t$-cut, $\s{C}$, in $N$, i.e.,
\[
\max |f| \leq \min c(\s{C}).
\]
\end{prop}

\medskip
Proposition \ref{MaxfMincp1} is half of the so-called 
{\it Max-flow Min-cut Theorem\/}.
The other half of this theorem says that
the above inequality is indeed an equality. That is, there
is actually some $v_s$-$v_t$-cut, $\s{C}$, whose capacity, 
$c(\s{C})$, is the maximum value of the flow in $N$.

\medskip
A $v_s$-$v_t$-cut of minimum capacity is called a 
{\it minimum $v_s$-$v_t$-cut\/}, for short, 
a {\it minimum cut\/}.

\begin{figure}
  \begin{center}
    \begin{pspicture}(0,0)(5,4.3)
    \cnodeput(0,2){s}{$v_s$}
    \cnodeput(2.5,4){a}{$v_1$}
    \cnodeput(2.5,0){b}{$v_2$}
    \cnodeput(5,2){t}{$v_t$}
    \ncline[linewidth=2pt]{->}{s}{a}
    \ncline[linewidth=1pt]{->}{s}{b}
    \ncline[linewidth=1pt]{->}{a}{b}
    \ncline[linewidth=1pt]{->}{a}{t}
    \ncline[linewidth=2pt]{->}{b}{t}
    \uput[135](1.5,3){$(1)$}
    \uput[-135](1.5,1){$(4)$}
    \uput[45](3.5,3){$(5)$}
    \uput[-45](3.5,1){$(3)$}
    \uput[0](2.5,2){$(2)$}
    \uput[-135](1.5,3){$1$}
    \uput[135](1.5,1){$3$}
    \uput[-45](3.5,3){$1$}
    \uput[45](3.5,1){$3$}
    \uput[180](2.5,2){$0$}
    \end{pspicture}
  \end{center}
  \caption{A maximum flow and a minimum cut in the network, $N$}
  \label{net3}
\end{figure}
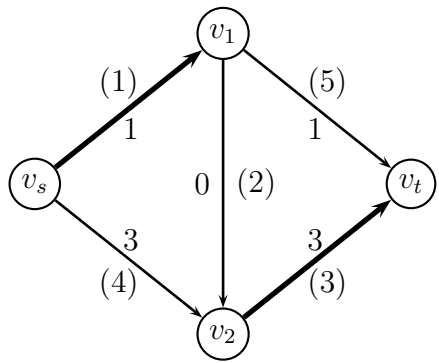

\medskip
An example of a minimum cut is shown in Figure \ref{net3},
where 
\[
\s{C} = \Omega^+(\{v_s, v_2\}) = \{(v_s v_1), (v_2 v_t)\},
\]
these two edges being shown as thicker lines. The capacity of
this cut is $4$ and a maximum flow is also shown in Figure \ref{net3}.

\medskip
What we intend to do next is to prove the celebrated
``Max-flow, Min-cut Theorem'' (due to Ford and Fulkerson, 1957)
and then to give an algorithm (also due to Ford and Fulkerson)
for finding a maximum flow, provided some reasonable assumptions
on the capacity function.  
In preparation for this, we present a handy trick (found
both in Berge \cite{Berge} and  Sakarovitch \cite{Sakarovitch2}),
the {\it return edge\/}. 

\medskip
Recall that one of the consequences of Proposition \ref{netflowp1}
is that the net flow out from $v_s$ is equal to the net
flow into $v_t$. Thus, if we add a  new edge, $e_{r}$, 
called the {\it return edge\/}, to $G$, obtaining the graph $\widetilde{G}$
(and the network $\widetilde{N})$, we see that any flow, $f$, in
$N$ satisfying condition (1) of
Definition \ref{netflowdef} yields a genuine flow, $\widetilde{f}$, 
in $\widetilde{N}$
(a flow according to Definition \ref{flowdef}, by Theorem \ref{flowchar1}),
such that $f(e) = \widetilde{f}(e)$ for every edge of $G$ and
$\widetilde{f}(e_r) = |f|$. Consequently, the Network flow problem
is equivalent to find a (genuine) flow in $\widetilde{N}$ such that
$\widetilde{f}(e_r)$ is maximum. Another advantage of this formulation is that
all the results on flows from Section \ref{sec26} can be 
applied directly to $\widetilde{N}$. To simplify the notation,
as $\widetilde{f}$ extends $f$, let us also use thw notation $f$
for  $\widetilde{f}$. Now, if $D$ is the indicence matrix of
$\widetilde{G}$ (again, we use the simpler notation, $D$, instead
of $\widetilde{D}$), we know that $f$ is a flow iff
\[
Df = 0.
\]
Therefore, the network flow problem can be stated as 
a {\it linear programing problem\/} as follows:

\medskip
\[
\mathrm{Maximize}\> \> z = f(e_r) 
\]
subject to the linear constraints
\begin{eqnarray*}
Df & = & 0 \\
0 & \leq & f \\
f & \leq & c,
\end{eqnarray*}
where we view $f$ as a vector in $\reals^{n + 1}$,
with $n = |E(G)|$.

\medskip
Consequently, we obtain the existence of maximal flows,
a fact which is not immediately obvious.

\begin{prop}
\label{maxflowexist}
Given any network, $N = (G, c, v_s, v_t)$, 
there is some flow, $f$, of maximum value.
\end{prop}

\proof
If we go back to the formulation of the Max-flow problem
as a linear program, we see that the set
\[
C = \{x\in \reals^{n + 1} \mid 0 \leq x \leq c\} \cap \Ker\, D
\]
is compact, as the intersection of a compact subset 
and a closed subset of $\reals^{n + 1}$ (in fact,
$C$ is also convex) and nonempty, as $0$ (the zero vector) is a flow.
But then, the projection, 
$\pi\co  x\mapsto x(e_r)$, is a continuous function,
$\mapdef{\pi}{C}{\reals}$, on a nonempty compact, so it 
achieves its maximum value for some $f\in C$.
Such an $f$ is a flow on $\widetilde{N}$
with maximal value.
$\bigsquare$

\medskip
Now that we know that maximum flows exist, it remains to prove that
a maximal flow is realized by some minimal cut to
complete the Max-flow, Min-cut Theorem of Ford and Fulkerson.
This can be done in various ways usually using some version
of an algorithm due to  Ford and Fulkerson. Such proofs can be found
in Papadimitriou and Steiglitz
\cite{PapadimitriouSteiglitz}, Wilf \cite{Wilf},
Cameron \cite{Cameron} and Sakarovitch \cite{Sakarovitch2}.

\medskip
Sakarovitch makes the interesting observation 
(given as an exercise) that
the Arc Coloring Lemma due to  Minty (Theorem \ref{Minty})
yields a simple proof of the part of the Max-flow, Min-cut Theorem 
that we seek to establish (See  \cite{Sakarovitch2}, Chapter 4,
Exercise 1, page 105).
Therefore, we choose to present such a proof
since it is rather original and quite elegant.

\begin{thm} (Max-Flow, Min-Cut Theorem (Ford and Fulkerson))
\label{MaxfMinc}
For any network, $N = (G, c, v_s, v_t)$, the maximum
value,  $|f|$,  of any flow, $f$, in $N$ is
equal to the minimum capacity, $c(\s{C})$, of
any $v_s$-$v_t$-cut, $\s{C}$, in $N$.
\end{thm}
 
\proof
By Proposition \ref{MaxfMincp1}, we already have half of our theorem.
By Proposition \ref{maxflowexist}, we know that
some maximum flow, say $f$, exists. It remains to show that
there is some $v_s$-$v_t$-cut, $\s{C}$, such that
$|f| = c(\s{C})$.

\medskip
We proceed as follows:

\medskip
Form the graph, 
$\widetilde{G} = (V, E\cup \{e_r\}, s, t)$ from 
$G = (V, E, s, t)$, 
with $s(e_r) = v_t$ and $t(e_r) = v_s$.
Then,  form the graph, 
$\widehat{G} = (V, \widehat{E}, \widehat{s}, \widehat{t})$, 
whose edges are defined as follows:
\begin{enumerate}
\item[(a)]
$e_r\in \widehat{E}$; $\widehat{s}(e_r) = s(e_r)$, $\widehat{t}(e_r) = t(e_r)$; 
\item[(b)]
If $e\in E$ and $0 < f(e) < c(e)$, then
$e\in \widehat{E}$; $\widehat{s}(e) = s(e)$, $\widehat{t}(e) = t(e)$; 
\item[(c)]
If $e\in E$ and $f(e) = 0$, then $e\in  \widehat{E}$; 
$\widehat{s}(e) = s(e)$, $\widehat{t}(e) = t(e)$; 
\item[(d)]
If $e\in E$ and $f(e) = c(e)$, then
$e\in  \widehat{E}$, with $\widehat{s}(e) = t(e)$
and $\widehat{t}(e) = s(e)$. 
\end{enumerate}

\medskip
In order to apply Minty's Theorem, we color all edges
constructed in (a), (c) and (d) in black and all edges
constructed in (b) in red and we pick $e_r$ as the
distinguished edge. Now, apply Minty's Lemma.
We have two possibilities:
\begin{enumerate}
\item 
There is an elementary cycle, $\Gamma$, in $\widehat{G}$,
with all black edges
oriented the same way. Since $e_r$ is incoming into $v_s$,
the direction of  the cycle is from from $v_s$ to $v_t$,
so $e_r \in \Gamma^+$.
This implies that all edges of type (d), $e\in \widehat{E}$, 
have an orientation
consistent with the direction of the cycle.
Now, $\Gamma$ is also a cycle in
$\widetilde{G}$ and, in $\widetilde{G}$,
each edge, $e\in E$, with $f(e) = c(e)$ is oriented 
in  the inverse direction of
the cycle, i.e, $e\in \Gamma^{-1}$ in $\widetilde{G}$.
Also, all edges of type (c), $e\in \widehat{E}$, with
$f(e) = 0$, are oriented in the direction of the cycle,
i.e., $e\in \Gamma^{+}$ in  $\widetilde{G}$.
We also have $e_r\in \Gamma^+$  in  $\widetilde{G}$.

\medskip
We show that the value of the flow, $|f|$, can be increased.
Since $0 < f(e) < c(e)$ for every red edge,
$f(e) = 0$ for every edge of type (c) in $\Gamma^+$,
$f(e) = c(e)$ for every  edge of type (d) in $\Gamma^-$,
and since all capacities are strictly positive, if we let
\begin{eqnarray*}
\delta_1 & = & \min_{e\in \Gamma^+} \{c(e) - f(e)\} \\
\delta_2 & = & \min_{e\in \Gamma^-} \{f(e)\} 
\end{eqnarray*}
and
\[  
\delta = \min\{\delta_1, \delta_2\},
\]
then $\delta > 0$.
We can increase the flow, $f$, in $\widetilde{N}$,
by adding $\delta$ to $f(e)$ for every
edge $e\in \Gamma^+$ (including edges of type (c) for which $f(e) = 0$)
and subtracting $\delta$ from $f(e)$ for every edge $e\in \Gamma^-$ 
(including edges of type (d) for which $f(e) = c(e)$) obtaining a flow, $f'$
such that 
\[
|f'| =  f(e_r)  + \delta = |f| + \delta > |f|,
\]
as $e_r\in \Gamma^+$, 
contradicting the maximality of $f$.
Therefore, we conclude that alternative (1) is impossible and we must
have the second alternative:
\item
There is an elementary cocycle, $\Omega_{\widehat{G}}(W)$, in
$\widehat{G}$ with all edges  black and oriented in the same direction
(there are no green edges).
Since $e_r\in \Omega_{\widehat{G}}(W)$, either $v_s\in W$ or $v_t\in W$
(but not both).
In the second case ($v_t\in W$), we have $e_r\in \Omega_{\widehat{G}}^+(W)$
and $v_s\in \overline{W}$. Then, consider
$\Omega_{\widehat{G}}^+(\overline{W}) = 
\Omega_{\widehat{G}}^-(W)$, with  $v_s\in \overline{W}$.
Thus, we are reduced to the
case where $v_s \in W$.

\medskip
If  $v_s \in W$, then $e_r\in \Omega_{\widehat{G}}^-(W)$ and since all edges
are black,  $\Omega_{\widehat{G}}(W) = 
\Omega_{\widehat{G}}^-(W)$, in $\widehat{G}$.
However, as every edge, $e\in \widehat{E}$, of type (d) 
corresponds to an inverse edge, $e\in E$,
we see that $\Omega_{\widehat{G}}(W)$ defines a cocycle,
$\Omega_{\widetilde{G}}(W) = 
\Omega_{\widetilde{G}}^+(W) \cup \Omega_{\widetilde{G}}^-(W)$, with
\begin{eqnarray*}
\Omega_{\widetilde{G}}^+(W) & = & \{e\in E \mid s(e) \in W\} \\
\Omega_{\widetilde{G}}^-(W) & = & \{e\in E \mid t(e) \in W\}.
\end{eqnarray*}
Moreover, by construction, $f(e) = c(e)$ for all 
$e\in \Omega_{\widetilde{G}}^+(W)$, $f(e) = 0$
for all \\
$e\in \Omega_{\widetilde{G}}^-(W) - \{e_r\}$, and
$f(e_r) = |f|$. We say that the edges of the
cocycle $\Omega_{\widetilde{G}}(W)$ are {\it saturated\/}.
Consequently, $\s{C} = \Omega_{\widetilde{G}}^+(W)$
is a $v_s$-$v_t$-cut in $N$ with
\[
c(\s{C}) = f(e_r) = |f|,
\]
establishing our theorem. $\bigsquare$
\end{enumerate}

\medskip
It is interesting that the proof in part (1) of 
Theorem \ref{MaxfMinc} contains the main idea behind the algorithm
of Ford and Fulkerson that we now describe.

\medskip
The main idea is to look for an (elementary) chain from
$v_s$ to $v_t$ so that together with the return edge, $e_r$,
we obtain a cycle, $\Gamma$, such that the edges in
$\Gamma$ satisfy the following properties:
\begin{enumerate}
\item[(1)]
$\delta_1 = \min_{e\in \Gamma^+} \{c(e) - f(e)\} > 0$;
\item[(2)]
$\delta_2 = \min_{e\in \Gamma^-} \{f(e)\} > 0$.
\end{enumerate}
Such a chain is called a {\it flow augmenting chain\/}.
Then, if we let $\delta = \min\{\delta_1, \delta_2\}$,
we can increase the value of the flow by
adding $\delta$ to $f(e)$ for every edge $e\in \Gamma^+$
(including the edge, $e_r$, which belongs to $\Gamma^+$) 
and subtracting $\delta$ from $f(e)$ for all
edges $e\in \Gamma^-$. This way, we get a new flow, $f'$,
whose value is $|f'| = |f| + \delta$. Indeed,
$f' = f + \delta\gamma(\Gamma)$, where $\gamma(\Gamma)$ is the
flow associated with the cycle, $\Gamma$.
The algorithms goes through rounds each consisting of two phases:
%
During phase 1, a flow augmenting chain is found by the procedure
{\it findchain};
During phase 2, the flow along the edges
of the augmenting chain is increased using the
function {\it changeflow}. 

\medskip
During phase 1, the nodes of the
augmenting chain are saved in the (set) variable, $Y$, 
and the edges of this chain are saved in the (set) variable, $\s{E}$.
We assign the special capacity value $\infty$ to $e_r$,
with the convention that $\infty \pm \alpha = \alpha$
and that $\alpha < \infty$ for all $\alpha\in \reals$.

\begin{tabbing}
\quad \= \quad \= \quad \= \quad \= \quad \= \quad \= \quad \\
{\bf procedure} $\mathit{findchain}$($N$: network; $e_r$: edge; 
$Y$: node set; $\s{E}$: edge set; $\delta$; real; $f$; flow)  \\
 \> {\bf begin} \\
 \> \>  $\delta := \delta(v_s) := \infty$; $Y := \{v_s\}$; \\
 \> \> {\bf while} $(v_t\notin Y)\land (\delta > 0)$ {\bf do} \\
 \> \> \> {\bf if} there is an edge $e$ with $s(e)\in Y$, $t(e)\notin Y$
and $f(e) < c(e)$ {\bf then} \\
 \> \> \> \>  $Y := Y \cup \{t(e)\}$; $\s{E}(t(e)) := e$; 
$\delta(t(e)) :=  \min\{\delta(s(e)), c(e) - f(e)\}$ \\
 \> \> \>  {\bf else} \\
 \> \> \> \>  {\bf if} there is an edge $e$ with $t(e)\in Y$, $s(e)\notin Y$
and $f(e) > 0$ {\bf then} \\
 \> \> \> \> \> $Y := Y \cup \{s(e)\}$; $\s{E}(s(e)) := e$; 
$\delta(s(e)) :=  \min\{\delta(t(e)), f(e)\}$ \\
 \> \> \> \>  {\bf else} $\delta := 0$  (no new arc can be traversed) \\
 \> \> \> \>  {\bf endif} \\
 \> \> \>  {\bf endif} \\
 \> \> {\bf endwhile}; \\
\> \> {\bf if} $v_t\in Y$ {\bf then} $\delta := \delta(v_t)$ {\bf endif} \\ 
\> {\bf end}
\end{tabbing}

\medskip
Here is now the procedure to update the flow:

\begin{tabbing}
\quad \= \quad \= \quad \= \quad \= \quad \= \quad \= \quad \\
{\bf procedure} $\mathit{changeflow}$($N$: network; $e_r$: edge; 
 $\s{E}$: edge set; $\delta$: real; $f$; flow)  \\
 \> {\bf begin} \\
\> \> $u := v_t$; $f(e_r) := f(e_r) + \delta$; \\
\> \> {\bf while} $u\not= v_s$ {\bf do} $e := \s{E}(u)$; \\
\> \> \> {\bf if} $u = t(e)$ {\bf then} $f(e) := f(e) + \delta$; 
$u := s(e)$; \\  
\> \> \> {\bf else} $f(e) := f(e) - \delta$; $u = t(e)$ \\
\> \> \> {\bf endif} \\
\> \> {\bf endwhile} \\
\> {\bf end}
\end{tabbing}

\medskip
Finally, the algorithm {\it maxflow} is given below:

\begin{tabbing}
\quad \= \quad \= \quad \= \quad \= \quad \= \quad \= \quad \\
{\bf procedure} $\mathit{maxflow}$($N$: network; $e_r$: edge; 
$Y$: set of nodes; $\s{E}$: set of edges;  $f$; flow)  \\
 \> {\bf begin} \\
\> \> {\bf for each} $e\in E$ {\bf do} $f(e) := 0$ {\bf enfdor}; \\ 
\> \> {\bf repeat until} $\delta = 0$ \\ 
\> \> \> $\mathit{findchain}(N, e_r, Y, \s{E}, \delta, f)$; \\
\> \> \> {\bf if} $\delta > 0$ {\bf then} \\
\>\> \> \> $\mathit{changeflow}(N, e_r, \s{E}, \delta, f)$ \\
\> \> \> {\bf endif} \\
\> \> {\bf endrepeat} \\
\> {\bf end}
\end{tabbing}

\medskip
The reader should run the algorithm {\it maxflow} on the
network of Figure \ref{net1} to verify that the maximum flow
shown in Figure \ref{net3} is indeed found, with
$Y = \{v_s, v_2\}$ when the algorithm stops.

\medskip
The correctness of the algorithm {\it maxflow} is easy to prove.

\begin{thm}
\label{maxflowthm1}
If the algorithm, {\it maxflow}, terminates and during the last
round through {\it findchain} the node $v_t$ is not marked,
then the flow, $f$, returned by the algorithm is a maximum flow.
\end{thm}

\proof
Observe that if $Y$ is the set of nodes returned when {\it maxflow}
halts, then $v_s\in Y$, $v_t\notin Y$ and
\begin{enumerate}
\item
If $e\in \Omega^+(Y)$, then $f(e) = c(e)$, as otherwise,
procedure {\it findchain} would have added $t(e)$ to $Y$;
\item
If $e\in \Omega^-(Y)$, then $f(e) = 0$, as otherwise,
procedure {\it findchain} would have added $s(e)$ to $Y$.
\end{enumerate}
But then, as in the end of the proof of Theorem \ref{MaxfMinc},
we see that the edges of the coycle $\Omega(Y)$ are saturated
and we know that $\Omega^+(Y)$ is a minimal cut and that
$|f| = c(\Omega^+(Y))$ is maximal.
$\bigsquare$

\medskip
We still have to show that the algorithm terminates but
there is a catch. Indeed, the version of
the Ford and Fulkerson algorithm that we just presented may
not terminate if the capacities are irrational!
Moreover, in the limit, the flow found by the
algorithm may not be maximum!
An example of this bad behavior
due to Ford and Fulkerson is reproduced in
Wilf \cite{Wilf} (Chapter 3, Section 5). However, we
can prove the following termination result which,
for all practical purposes, is good enough, since 
only rational numbers can be stored by a computer.

\begin{thm}
\label{maxflowthm2}
Given a network, $N$, if all the capacities are multiple of
some number, $\lambda$, then the algorithm, {\it maxflow},
always terminates. In particular, the algorithm {\it maxflow}
always terminates if the capacites are rational (or integral).
\end{thm}

\proof
The number $\delta$ will always be a multiple of $\lambda$,
so $f(e_r)$ will increase by at least $\lambda$ during each
iteration. Thus, eventually, the value of a minimal cut,
which is a multiple of $\lambda$, will be reached.
$\bigsquare$

\medskip
If all the capacities are integers, an easy induction
yields the following useful and non-trivial proposition:

\begin{prop}
\label{maxflowthm3}
Given a network, $N$, if all the capacities are integers, then
the algorithm {\it maxflow} outputs a maximum flow,
$\mapdef{f}{E}{\natnums}$, such that  the flow in every
edge is an integer.
\end{prop}

\remark
Proposition \ref{maxflowthm3} only asserts that some
maximum flow is of the form \\
$\mapdef{f}{E}{\natnums}$.
In general, there is more than one
maximum flow and other maximum flows may not
have integer values on all edges.

\medskip
Theorem \ref{maxflowthm2} is good news but it is also bad news
from the point of view of complexity. Indeed, the present version of
the Ford and Fulkerson algorithm has a running time that depends
on  capacities and so, it can be very bad.

\medskip
There are various ways of getting around this difficulty
to find algorithms that do not depend on capacities and 
quite a few researchers have studied this problem.
An excellent discussion of the progress in network flow
algorithms can be found in Wilf (Chapter 3). 

\medskip
A fairly simple modification of the
Ford and Fulkerson algorithm consists in looking
for flow augmenting chains of shortest length. 
To explain this algorithm we need  the
concept of {\it residual network\/}, which is
a useful tool in any case.
Given a network, $N = (G, c, s, t)$ and given any flow, $f$,
the {\it residual network\/}, $N_f = (G_f, c_f, v_f, v_t)$
is defined as follows:

\begin{enumerate}
\item
$V_f = V$;
\item
For every edge, $e\in E$, if $f(e) < c(e)$, then
$e^+\in E_f$, $s_f(e^+) = s(e)$, $t_f(e^+) = t(e)$
and $c_f(e^+) = c(e) - f(e)$; 
the edge $e^+$  is called a {\it forward edge\/};
\item
For every edge, $e\in E$, if $f(e) > 0$, then
$e^-\in E_f$, $s_f(e^-) = t(e)$, $t_f(e^-) = s(e)$
and $c_f(e^-) = f(e)$; the edge $e^-$ is 
called a {\it backward edge\/} because
it has the inverse orientation of the original edge, $e\in E$;
\end{enumerate}

The capacity, $c_f(e^{\epsilon})$, of an edge $e^{\epsilon}\in E_f$ 
(with $\epsilon = \pm$) is usually called the
{\it residual capacity\/} of $e^{\epsilon}$. Observe that
the same edge, $e$, in $G$, will give rise to two edges
$e^+$ and $e^-$
(with the same set of endpoints but with opposite orientations)
in $G_f$ if $0 < f(e) < c(e)$. Thus, $G_f$ has at most twice
as many edges as $G$. Also, note that every edge, $e\in E$,
which is {\it saturated\/}, i.e., for which $f(e) = c(e)$,
does not survive in $G_f$. 

\medskip
Observe that there is a one-to-one correspondence between
(elementary)  flow augmenting chains in the original graph, $G$,
and (elementary) flow augmenting paths in $G_f$. Furthermore,
in order to check that an elementary path, $\pi$, 
from $v_s$ to $v_t$ in $G_f$
is a flow augmenting path, all we have to do is to 
compute
\[
c_f(\pi) = \min_{e^{\epsilon}\in \pi} \{c_f(e^{\epsilon})\},
\]
the {\it bottleneck\/} of the path, $\pi$.
Then, as before, we can update the flow, $f$ in 
$N$, to get the new flow, $f'$, by setting
\begin{alignat*}{2}
f'(e) & =  f(e) + c_f(\pi), & \qquad & \hbox{if}\quad e^+\in \pi\\
f'(e) & =  f(e) - c_f(\pi)  & \qquad & \hbox{if}\quad e^-\in \pi, \\
f'(e) & =  f(e)   & \qquad & \hbox{if}\quad e\in E\quad\hbox{and}\quad 
e^{\epsilon}\notin\pi,
\end{alignat*}
for every edge $e\in E$.
Note that the function, $\mapdef{f_{\pi}}{E}{\reals}$, defined by
\begin{alignat*}{2}
f_{\pi}(e) & =  c_f(\pi), & \qquad & \hbox{if}\quad e^+\in \pi\\
f_{\pi}(e) & =   - c_f(\pi)  & \qquad & \hbox{if}\quad e^-\in \pi,\\
f_{\pi}(e) & =  0   & \qquad & \hbox{if}\quad e\in E\quad\hbox{and}\quad 
e^{\epsilon}\notin\pi,
\end{alignat*}
is a flow in $N$ with $|f_{\pi}| = c_f(\pi)$ 
and $f' = f + f_{N, \pi}$ is a flow in $N$, with
$|f'| = |f| + c_f(\pi)$ (same reasoning as before).
Now, we can repeat this process: Compute the new residual graph,
$N_{f'}$ from $N$ and $f'$, update the flow $f'$ to get the new
flow $f''$ in $N$, etc.

\medskip
The same reasoning as before shows that
if we obtain a residual graph with no flow augmenting
path from $v_s$ to $v_t$, then a maximum flow has been found.

\begin{figure}
  \begin{center}
    \begin{pspicture}(0,0)(5,4.3)
    \cnodeput(0,2){s}{$v_s$}
    \cnodeput(2.5,4){a}{$v_1$}
    \cnodeput(2.5,0){b}{$v_2$}
    \cnodeput(5,2){t}{$v_t$}
    \ncline[linewidth=1pt]{->}{s}{a}
    \ncline[linewidth=1pt]{->}{s}{b}
    \ncline[linewidth=1pt]{->}{a}{b}
    \ncline[linewidth=1pt]{->}{a}{t}
    \ncline[linewidth=1pt]{->}{b}{t}
    \uput[135](1.5,3){$(100)$}
    \uput[-135](1.5,1){$(100)$}
    \uput[45](3.5,3){$(100)$}
    \uput[-45](3.5,1){$(100)$}
    \uput[0](2.5,2){$(1)$}
    \uput[-135](1.5,3){$0$}
    \uput[135](1.5,1){$0$}
    \uput[-45](3.5,3){$0$}
    \uput[45](3.5,1){$0$}
    \uput[180](2.5,2){$0$}
    \end{pspicture}
  \end{center}
  \caption{A poor choice of augmenting paths yields a slow method}
  \label{badnet}
\end{figure}
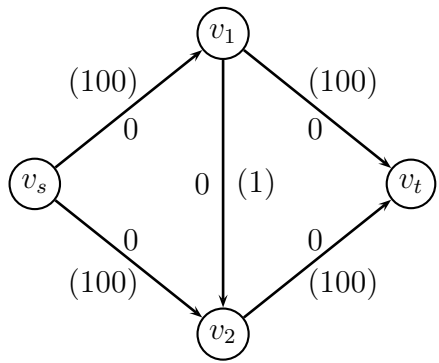

\medskip
It should be noted that a poor choice of augmenting paths
may cause the algorithm to perform a lot more steps than necessary.
For example, if we consider the network shown in Figure 
\ref{badnet}, and if we pick the flow augmenting paths in
the residual graphs to be alternatively
$(v_s, v_1, v_2, v_t)$ and $(v_s, v_2, v_1, v_t)$, at each step,
we only increase the flow by $1$, so it will take $200$ steps to find
a maximum flow!

\medskip
One of the main advantages of using residual graphs is that they make
is convenient to look for better strategies for picking
flow augmenting paths. For example, we can choose 
an elementary flow augmenting path
shortest length (for example, using breadth-first search).
Then, it can be shown that this revised algorithm terminates
in $O(|V|\cdot |E|)$ steps
(see Cormen, Leiserson, Rivest and Stein \cite{Cormen}, Section 26.2,
and Sakarovitch \cite{Sakarovitch2}, Chapter 4, Exercise 5).
Edmonds and Karp designed an algorithm running in time
$O(|E|\cdot |V|^2)$ based on this idea (1972), see
\cite{Cormen}, Section 26.2.
Another way of selecting ``good'' augmenting paths,
the {\it scaling Max-Flow algorithm\/}, 
is described in Kleinberg and Tardos \cite{Kleinberg} 
(see Section 7.3).

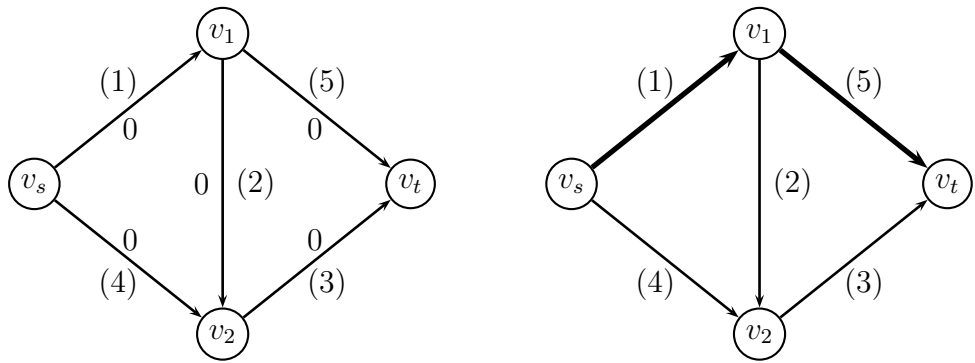
\begin{figure}
  \begin{center}
    \begin{pspicture}(0,0)(5,4.3)
    \cnodeput(0,2){s}{$v_s$}
    \cnodeput(2.5,4){a}{$v_1$}
    \cnodeput(2.5,0){b}{$v_2$}
    \cnodeput(5,2){t}{$v_t$}
    \ncline[linewidth=1pt]{->}{s}{a}
    \ncline[linewidth=1pt]{->}{s}{b}
    \ncline[linewidth=1pt]{->}{a}{b}
    \ncline[linewidth=1pt]{->}{a}{t}
    \ncline[linewidth=1pt]{->}{b}{t}
    \uput[135](1.5,3){$(1)$}
    \uput[-135](1.5,1){$(4)$}
    \uput[45](3.5,3){$(5)$}
    \uput[-45](3.5,1){$(3)$}
    \uput[0](2.5,2){$(2)$}
    \uput[-135](1.5,3){$0$}
    \uput[135](1.5,1){$0$}
    \uput[-45](3.5,3){$0$}
    \uput[45](3.5,1){$0$}
    \uput[180](2.5,2){$0$}
    \end{pspicture}
\hskip 2cm
    \begin{pspicture}(0,0)(5,4.3)
    \cnodeput(0,2){s}{$v_s$}
    \cnodeput(2.5,4){a}{$v_1$}
    \cnodeput(2.5,0){b}{$v_2$}
    \cnodeput(5,2){t}{$v_t$}
    \ncline[linewidth=2pt]{->}{s}{a}
    \ncline[linewidth=1pt]{->}{s}{b}
    \ncline[linewidth=1pt]{->}{a}{b}
    \ncline[linewidth=2pt]{->}{a}{t}
    \ncline[linewidth=1pt]{->}{b}{t}
    \uput[135](1.5,3){$(1)$}
    \uput[-135](1.5,1){$(4)$}
    \uput[45](3.5,3){$(5)$}
    \uput[-45](3.5,1){$(3)$}
    \uput[0](2.5,2){$(2)$}
    \end{pspicture}
  \end{center}
  \caption{Construction of the residual graph, $N_f$, from $N$, round 1}
  \label{augflow1}
\end{figure}

\medskip
Here is an illustration of this faster algorithm,
starting with the network, $N$, shown in Figure \ref{net1}.
The sequence of residual network construction and flow
augmentation steps is shown in Figures \ref{augflow1},
\ref{augflow2} and \ref{augflow3}. During the first two rounds,
the augmented path chosen is shown in thicker lines.
In the third and final round, there is no path
from $v_s$ to $v_t$ in the residual graph, indicating that
a maximum flow has been found. 

\begin{figure}
  \begin{center}
    \begin{pspicture}(0,0)(5,4.3)
    \cnodeput(0,2){s}{$v_s$}
    \cnodeput(2.5,4){a}{$v_1$}
    \cnodeput(2.5,0){b}{$v_2$}
    \cnodeput(5,2){t}{$v_t$}
    \ncline[linewidth=1pt]{->}{s}{a}
    \ncline[linewidth=1pt]{->}{s}{b}
    \ncline[linewidth=1pt]{->}{a}{b}
    \ncline[linewidth=1pt]{->}{a}{t}
    \ncline[linewidth=1pt]{->}{b}{t}
    \uput[135](1.5,3){$(1)$}
    \uput[-135](1.5,1){$(4)$}
    \uput[45](3.5,3){$(5)$}
    \uput[-45](3.5,1){$(3)$}
    \uput[0](2.5,2){$(2)$}
    \uput[-135](1.5,3){$1$}
    \uput[135](1.5,1){$0$}
    \uput[-45](3.5,3){$1$}
    \uput[45](3.5,1){$0$}
    \uput[180](2.5,2){$0$}
    \end{pspicture}
\hskip 2cm
    \begin{pspicture}(0,0)(5,4.3)
    \cnodeput(0,2){s}{$v_s$}
    \cnodeput(2.5,4){a}{$v_1$}
    \cnodeput(2.5,0){b}{$v_2$}
    \cnodeput(5,2){t}{$v_t$}
    \ncline[linewidth=1pt]{->}{a}{s}
    \ncline[linewidth=2pt]{->}{s}{b}
    \ncline[linewidth=1pt]{->}{a}{b}
    \ncline[linewidth=2pt]{->}{b}{t}
    \ncarc[arcangle=20, linewidth=1pt]{->}{a}{t}
    \ncarc[arcangle=20, linewidth=1pt]{->}{t}{a}
    \uput[135](1.5,3){$(1)$}
    \uput[-135](1.5,1){$(4)$}
    \uput[45](4,3.1){$(4)$}
    \uput[45](3,2.1){$(1)$}
    \uput[-45](3.5,1){$(3)$}
    \uput[0](2.4,2){$(2)$}
    \end{pspicture}
  \end{center}
  \caption{Construction of the residual graph, $N_f$, from $N$, round 2}
  \label{augflow2}
\end{figure}

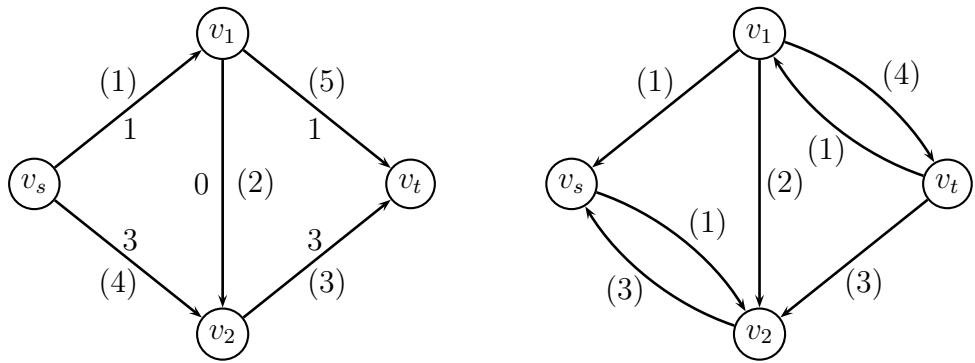
\begin{figure}
  \begin{center}
    \begin{pspicture}(0,0)(5,4.3)
    \cnodeput(0,2){s}{$v_s$}
    \cnodeput(2.5,4){a}{$v_1$}
    \cnodeput(2.5,0){b}{$v_2$}
    \cnodeput(5,2){t}{$v_t$}
    \ncline[linewidth=1pt]{->}{s}{a}
    \ncline[linewidth=1pt]{->}{s}{b}
    \ncline[linewidth=1pt]{->}{a}{b}
    \ncline[linewidth=1pt]{->}{a}{t}
    \ncline[linewidth=1pt]{->}{b}{t}
    \uput[135](1.5,3){$(1)$}
    \uput[-135](1.5,1){$(4)$}
    \uput[45](3.5,3){$(5)$}
    \uput[-45](3.5,1){$(3)$}
    \uput[0](2.5,2){$(2)$}
    \uput[-135](1.5,3){$1$}
    \uput[135](1.5,1){$3$}
    \uput[-45](3.5,3){$1$}
    \uput[45](3.5,1){$3$}
    \uput[180](2.5,2){$0$}
    \end{pspicture}
\hskip 2cm
    \begin{pspicture}(0,0)(5,4.3)
    \cnodeput(0,2){s}{$v_s$}
    \cnodeput(2.5,4){a}{$v_1$}
    \cnodeput(2.5,0){b}{$v_2$}
    \cnodeput(5,2){t}{$v_t$}
    \ncline[linewidth=1pt]{->}{a}{s}
    \ncline[linewidth=1pt]{->}{a}{b}
    \ncline[linewidth=1pt]{->}{t}{b}
    \ncarc[arcangle=20, linewidth=1pt]{->}{a}{t}
    \ncarc[arcangle=20, linewidth=1pt]{->}{t}{a}
    \ncarc[arcangle=20, linewidth=1pt]{->}{s}{b}
    \ncarc[arcangle=20, linewidth=1pt]{->}{b}{s}
    \uput[135](1.5,3){$(1)$}
    \uput[-135](2.2,1.8){$(1)$}
    \uput[-135](1.1,0.9){$(3)$}
    \uput[45](4,3.1){$(4)$}
    \uput[45](3,2.1){$(1)$}
    \uput[-45](3.5,1){$(3)$}
    \uput[0](2.4,2){$(2)$}
    \end{pspicture}
  \end{center}
  \caption{Construction of the residual graph, $N_f$, from $N$, round 3}
  \label{augflow3}
\end{figure}

Another idea originally due to Dinic (1970)
is to use {\it layered networks\/},
see Wilf \cite{Wilf} (Sections 3.6-3.7) and
Papadimitriou and Steiglitz \cite{PapadimitriouSteiglitz}
(Chapter 9). An algorithm using layered networks
running in time $O(V^3)$ is given in the two references above.
There are yet other faster algorithms, for instance
``preflow-push  algorithms'' also called ``preflow-push relabel algorithms'',
originally due to Goldberg.
A {\it preflow\/} is a function, $\mapdef{f}{E}{\reals}$, that
satisfies condition (2) of Definition \ref{netflowdef} but
which, instead of satisfying condition (1), satisfies the inequality
\begin{enumerate}
\item[($1'$)] (Non-negativity of net flow)
\[
 \sum_{s(e) = v} f(e) \geq  \sum_{t(e) = v} f(e)
\quad\hbox{for all $v \in V - \{v_s, v_t\}$},
\] 
\end{enumerate}
that is, the net flow out of $v$ is non-negative.
Now, the principle of all methods using preflows
is to augment a preflow until it becomes
a maximum flow. In order to do this, a labeling algorithm
assigning a {\it height\/}.
Algorithms of this type are discussed
in  Cormen, Leiserson, Rivest and Stein \cite{Cormen}, Sections 26.4
and 26.5 and in Kleinberg and Tardos \cite{Kleinberg}, Section 7.4.

\medskip
The Max-flow, Min-cut Theorem (Theorem \ref{MaxfMinc}) 
is a surprisingly powerful theorem in the sense that it can be
used to prove a number of other results whose original proof
is sometimes quite hard. Among these results, let us mention
the {\it maximum matching problem\/} in a bipartite graph,
discussed in Wilf  \cite{Wilf} (Sections 3.8), 
Cormen, Leiserson, Rivest and Stein \cite{Cormen} (Section 26.3)
Kleinberg and Tardos \cite{Kleinberg} (Section 7.5) and
Cameron \cite{Cameron} (Chapter 11, Section 10), 
finding the {\it edge connectivity\/} of a graph,
discussed in Wilf  \cite{Wilf} (Sections 3.8),
and a beautiful {\it theorem of Menger\/} on edge-disjoint paths 
and {\it Hall's Marriage Theorem}, both discussed
in Cameron  \cite{Cameron} (Chapter 11, Section 10).
More problems that can be solved effectively 
using flow algorithms, including image
segmentation, are discussed in Sections 7.6--7.13 of
Kleinberg and Tardos \cite{Kleinberg}.
We only mention one of Menger's theorems, as it is particularly elegant.

\begin{thm}
\label{Menger1} (Menger)
Given any finite digraph, $G$, for any two nodes, $v_s$ and $v_t$,
the maximum number of pairwise edge-disjoint paths from 
$v_s$ to $v_t$ is equal to the the minimum number of edges
in a $v_s$-$v_t$-separating set. (A a $v_s$-$v_t$-separating set
in $G$ is a set of edges, $C$, such every path from
$v_s$ to $v_t$ uses some edge in $C$.)
\end{thm} 

\medskip
It is also possible to generalize the basic flow problem
in which our flows, $f$,  have the property that
$0 \leq f(e) \leq c(e)$ for every edge, $e\in E$,
to {\it channeled flows\/}. This generalization consists
in adding another capacity function, $\mapdef{b}{E}{\reals}$,
relaxing the condition that $c(e) > 0$ for all $e\in E$,
and in allowing flows such that condition (2) of Definition
\ref{netflowdef} is replaced by
\begin{enumerate}
\item[$(2')$] (Admissibility of flow)
\[
b(e) \leq f(e) \leq c(e),
\quad\hbox{for all $e\in E$}
\]
\end{enumerate}
Now, the ``flow'' $f = 0$  is no longer necessarily admissible
and the channeled flow problem does not always have a solution.
However, it is possible to characterize when it has a solution.

\begin{thm} (Hoffman)
\label{Hoffmanth1}
A network, $N = (G, b, c, v_s, v_t)$, has a channeled flow
iff for every cocycle, $\Omega(Y)$, of $G$, we have
\begin{equation}
\sum_{e\in \Omega^-(Y)} b(e) \leq \sum_{e\in \Omega^+(Y)} c(e).
\tag{$\dagger$}
\end{equation}
\end{thm} 

\medskip
Observe that the necessity of the condition
of Theorem \ref{Hoffmanth1}  is an immediate 
consequence of Proposition \ref{orthop}.
That it is sufficient can be proved by modifying the algorithm
{\it maxflow} or its version using residual networks.
The principle of this method is to start with a flow, $f$, in $N$
that does not necessarily satisfy condition $(2')$ and
to gradually convert it to an admissible flow in $N$ (if one exists)
by applying the method for finding a maximum flow to
a modified version, $\widetilde{N}$, of $N$ in which the capacities have been
adjusted to that $f$ is an admissible flow in $\widetilde{N}$.
Now, if a flow, $f$, in $N$ does not satisfy condition $(2')$, then
there are some {\it offending edges\/}, $e$, for which either
$f(e) < b(e)$ or $f(e) > c(e)$. The new method makes sure that
at the end of every (successful) round through the basic {\it maxflow}
algorithm applied to the modified network, $\widetilde{N}$, 
some offending edge of $N$ is no longer offending.

\medskip
Let $f$ be a flow in $N$ and assume that $\widetilde{e}$ is
an offending edge (i.e.  either $f(e) < b(e)$ or $f(e) > c(e)$).
Then,  we construct the network, $\widetilde{N}(f, \widetilde{e})$,
as follows: The capacity functions, $\widetilde{b}$ and 
$\widetilde{c}$ are given by
\[
\widetilde{b}(e) = \cases{
b(e) & if $b(e) \leq f(e)$ \cr
f(e) & if $f(e) < b(e)$ \cr
}
\]
and
\[
\widetilde{c}(e) = \cases{
c(e) & if $f(e) \leq c(e)$ \cr
f(e) & if $f(e) > c(e)$.\cr
}
\]
We also add one new edge, $\widetilde{e}_r$, to $N$ whose
endpoints and capacities are determined by:
\begin{enumerate}
\item
If $f(\widetilde{e}) > c(\widetilde{e})$, then 
$s(\widetilde{e}_r) = t(\widetilde{e})$, 
$t(\widetilde{e}_r) = s(\widetilde{e})$, 
$\widetilde{b}(\widetilde{e}_r) = 0$ 
and $\widetilde{c}(\widetilde{e}_r) = f(\widetilde{e}) - c(\widetilde{e})$.
\item
If $f(\widetilde{e}) < b(\widetilde{e})$, then 
$s(\widetilde{e}_r) = s(\widetilde{e})$, 
$t(\widetilde{e}_r) = t(\widetilde{e})$, 
$\widetilde{b}(\widetilde{e}_r) = 0$ 
and $\widetilde{c}(\widetilde{e}_r) = b(\widetilde{e}) - f(\widetilde{e})$.
\end{enumerate}

\medskip
Now, observe that the original flow, $f$, in $N$ extended so that 
$f(\widetilde{e}_r) = 0$ is a channeled flow in 
$\widetilde{N}(f, \widetilde{e})$
(i.e., conditions (1) and ($2'$) are satisfied). 
Starting from the new network, $\widetilde{N}(f, \widetilde{e})$,
apply the Max-flow algorithm, say using residual graphs, with the 
following small change in 2:
\begin{enumerate}
\item
For every edge, $e\in \widetilde{E}$, if $f(e) < \widetilde{c}(e)$, then
$e^+\in \widetilde{E}_f$, 
$s_f(e^+) = s(e)$, $t_f(e^+) = t(e)$
and $c_f(e^+) = \widetilde{c}(e) - f(e)$; 
the edge $e^+$  is called a {\it forward edge\/};
\item
For every edge, $e\in \widetilde{E}$, if $f(e) > \widetilde{b}(e)$, then
$e^-\in \widetilde{E}_f$, $s_f(e^-) = t(e)$, $t_f(e^-) = s(e)$
and $c_f(e^-) = f(e) - \widetilde{b}(e)$; the edge $e^-$ is 
called a {\it backward edge\/}.
\end{enumerate}

\medskip
Now, we consider augmenting paths from $t(\widetilde{e}_r)$ to
$s(\widetilde{e}_r)$.
For any such elementary path, $\pi$, in $\widetilde{N}(f, \widetilde{e})_f$, 
as before
we compute
\[
c_f(\pi) = \min_{e^{\epsilon}\in \pi} \{c_f(e^{\epsilon})\},
\]
the {\it bottleneck\/} of the path, $\pi$,
and we say that $\pi$ is a flow augmenting path iff
$c_f(\pi) > 0$.
Then,  we can update the flow, $f$ in 
$\widetilde{N}(f, \widetilde{e})$, to get the new flow, $f'$, by setting
\begin{alignat*}{2}
f'(e) & =  f(e) + c_f(\pi)  & \qquad & \hbox{if}\quad e^-\in \pi, \\
f'(e) & =  f(e) - c_f(\pi)   & \qquad & \hbox{if}\quad e^-\in \pi, \\
f'(e) & =  f(e)   & \qquad & \hbox{if}\quad 
e \in \widetilde{E}\quad\hbox{and}\quad 
e^{\epsilon}\notin\pi,
\end{alignat*}
for every edge $e\in \widetilde{E}$. 

\medskip
We run the flow augmenting path procedure on $\widetilde{N}(f, \widetilde{e})$
and $f$ until it terminates with a maximum flow, $\widetilde{f}$.
If we recall that the offending edge is $\widetilde{e}$, 
then, there are four cases:

\begin{enumerate}
\item
$f(\widetilde{e}) > c(\widetilde{e})$.
\begin{enumerate}
\item 
When the Max-flow algorithm terminates,
$\widetilde{f}(\widetilde{e}_r) = \widetilde{c}(\widetilde{e}_r) =
f(\widetilde{e})  - c(\widetilde{e})$.
If so, define $\widehat{f}$ as follows:
\begin{equation}
\widehat{f}(e) = \cases{
\widetilde{f}(\widetilde{e}) - \widetilde{f}(\widetilde{e}_r) & 
if  $e = \widetilde{e}$ \cr
\widetilde{f}(e) & if  $e \not= \widetilde{e}$. \cr
}
\tag{$*$}
\end{equation} 
It is clear that $\widehat{f}$ is a flow in $N$
and $\widehat{f}(\widetilde{e}) = c(\widetilde{e})$
(there are no elementary paths from 
$t(\widetilde{e})$  to $s(\widetilde{e})$). 
But then, $\widetilde{e}$ is not an offending edge for
$\widehat{f}$, so we repeat the procedure of constructing the
modified network, etc.
\item 
When the Max-flow algorithm terminates,
$\widetilde{f}(\widetilde{e}_r) < \widetilde{c}(\widetilde{e}_r)$.
The flow, $\widehat{f}$, defined in $(*)$ above is
still a flow but the Max-flow algorithm must have terminated
with a residual graph with no flow augmenting path from
$s(\widetilde{e})$  to $t(\widetilde{e})$. 
Then, there is a set of nodes, $Y$ with $s(\widetilde{e})\in Y$
and $t(\widetilde{e})\notin Y$. Moreover, the way the Max-flow
algorithm is designed implies that
\begin{alignat*}{2}
& \widehat{f}(\widetilde{e}) > c(\widetilde{e}) & \qquad & \\
& \widehat{f}(e) = \widetilde{c}(e)\geq c(e) & \qquad & 
\hbox{if}\quad e\in \Omega^+(Y) - \{\widetilde{e}\} \\
& \widehat{f}(e) = \widetilde{b}(e)\leq b(e) & \qquad & 
\hbox{if}\quad e\in \Omega^-(Y).
\end{alignat*}
As $\widehat{f}$ also satisfies $(*)$ above, we conclude that
the cocycle condition $(\dagger)$ of Theorem \ref{Hoffmanth1}
fails for $\Omega(Y)$.
\end{enumerate}
\item
$f(\widetilde{e}) < b(\widetilde{e})$.
\begin{enumerate}
\item 
When the Max-flow algorithm terminates,
$\widetilde{f}(\widetilde{e}_r) = \widetilde{c}(\widetilde{e}_r) =
 b(\widetilde{e}) - f(\widetilde{e})$.
If so, define $\widehat{f}$ as follows:
\begin{equation}
\widehat{f}(e) = \cases{
\widetilde{f}(\widetilde{e}) + \widetilde{f}(\widetilde{e}_r) & 
if  $e = \widetilde{e}$ \cr
\widetilde{f}(e) & if  $e \not= \widetilde{e}$. \cr
}
\tag{$**$}
\end{equation} 
It is clear that $\widehat{f}$ is a flow in $N$
and $\widehat{f}(\widetilde{e}) = b(\widetilde{e})$
(there are no elementary paths from 
$s(\widetilde{e})$  to $t(\widetilde{e})$). 
But then, $\widetilde{e}$ is not an offending edge for
$\widehat{f}$, so we repeat the procedure of constructing the
modified network, etc.
\item 
When the Max-flow algorithm terminates,
$\widetilde{f}(\widetilde{e}_r) < \widetilde{c}(\widetilde{e}_r)$.
The flow, $\widehat{f}$, defined in $(**)$ above is
still a flow but the Max-flow algorithm must have terminated
with a residual graph with no flow augmenting path from
$t(\widetilde{e})$  to $s(\widetilde{e})$. 
Then, as in the case where $f(\widetilde{e}) > c(\widetilde{e})$,
there is a set of nodes, $Y$ with $s(\widetilde{e})\in Y$
and $t(\widetilde{e})\notin Y$ and it is easy to show that
the cocycle condition $(\dagger)$ of Theorem \ref{Hoffmanth1}
fails for $\Omega(Y)$.
\end{enumerate}
\end{enumerate}

\medskip
Therefore, if the algorithm  does not fail during every
round through the Max-flow algorithm applied to the modified
network, $\widetilde{N}$, which, as we observed, is the case if
condition $(\dagger)$ holds, then
a channeled flow, $\widehat{f}$,
will be produced and this flow will be a maximum flow.
This proves the converse of Theorem  \ref{Hoffmanth1}.

\medskip
The Max-flow, Min-cut Theorem can also be generalized to
channeled flows as follows:

\begin{thm}  
\label{MaxfMinc2}
For any network, $N = (G, b, c, v_s, v_t)$, 
if a  flow exists in $N$, then the maximum
value,  $|f|$,  of any flow, $f$, in $N$ is
equal to the minimum capacity, 
$c(\Omega(Y)) = c(\Omega^+(Y)) - b(\Omega^-(Y))$, of
any $v_s$-$v_t$-cocycle in $N$ (this means that
$v_s\in Y$ and $v_r\notin Y$).
\end{thm}

\medskip
If the capacity functions $b$ and $c$ have the property
that $b(e) < 0$ and $c(e) > 0$ for all $e\in E$,
then the condition of Theorem \ref{Hoffmanth1} is trivially 
satisfied. Furthermore, in this case, the flow $f = 0$
is admissible, Proposition \ref{maxflowexist} holds
and  we can apply directly the
construction of the residual network, $N_f$, described above.

\medskip
A variation of our last problem appears in 
Cormen, Leiserson, Rivest and Stein \cite{Cormen} (Chapter 26):
In this version, the underlying graph, $G$, of the network, $N$,
is assumed to have no parallel edges (and no loops), so that
every edge, $e$, can be identified with the pair, $(u, v)$, of
its endpoints (so, $E \subseteq V\times V$).
A flow, $f$, in $N$ is a function,
$\mapdef{f}{V\times V}{\reals}$, where is not necessarily the case
that $f(u, v) \geq 0$ for all $(u, v)$, but there is a capacity
function, $\mapdef{c}{V\times V}{\reals}$, such that
$c(u, v) \geq 0$, for all $(u, v)\in V\times V$ and
it is required that
\begin{eqnarray*}
f(v, u) & = & - f(u, v)\quad\hbox{and} \\
f(u, v) & \leq & c(u, v),
\end{eqnarray*}
for all  $(u, v)\in V\times V$.
Moreover, in view of the skew symmetry condition
($f(v, u) = -f(u, v)$), the equations of conservation of flow  are
written as
\[
\sum_{(u, v)\in E} f(u, v) = 0,
\]
for all $u\not = v_s, v_t$.

\medskip
We can reduce this last version of the flow problem to our previous
setting by noticing that in view of skew symmetry,
the capacity conditions are equivalent to having capacity functions,
$b'$, and $c'$, defined such that
\begin{eqnarray*}
b'(u, v) & = & -c(v, u) \\
c'(u, v) & = & c(u, v), 
\end{eqnarray*}
for every $(u, v)\in E$ and $f$ must satisfy
\[
b'(u, v) \leq f(u, v) \leq c'(u, v)
\]
for all $(u, v)\in E$.
However, we must also have $f(v, u) = -f(u, v)$,
which is an additional constraint in case
$G$ has both edges $(u, v)$ and $(v, u)$.
This point may be a little confusing since 
in our previous setting, $f(u, v)$ and $f(v, u)$
are independent values. However, this new problem is solved
essentially as the previous one. The construction
of the residual graph is identical to the previous
case and so is the flow augmentation procedure along
an elementary path, {\it except that\/}
we force $f_{\pi}(v, u) = f_{\pi}(u, v)$ to hold
during this step. For details, the reader is referred to
Cormen, Leiserson, Rivest and Stein \cite{Cormen}, Chapter 26.

\medskip
More could be said about flow problems but 
we believe that we have covered the basics satisfactorily and
we refer the reader to the various references mentioned in
this section for more on this topic.

\section{Matchings, Coverings, Bipartite Graphs}
\label{sec30}
In this section, we will be dealing with finite unoriented graphs.
Consider the following problem: We have a set of $m$ machines,
$M_1, \ldots, M_m$, and $n$ tasks, $T_1, \ldots, T_n$.
Furthermore,  each machine, $M_i$, is capable of performing
a subset of tasks, $S_i\subseteq \{T_1, \ldots, T_n\}$.  
Then, the problem is to find a set of assignments, 
$\{(M_{i_1}, T_{i_1}), \ldots,  (M_{j_p}, T_{j_p})\}$, 
with \\
$\{i_1, \ldots, i_p\} \subseteq \{1, \ldots, m\}$
and  $\{j_1, \ldots, j_p\} \subseteq \{1, \ldots, n\}$, such that
\begin{enumerate}
\item[(1)]
$T_{j_k} \in S_{i_k}$, \quad $1\leq k \leq p$;
\item[(2)]
$p$ is maximum.
\end{enumerate}

\begin{figure}
  \begin{center}
    \begin{pspicture}(0,0)(3,5.2)
    \cnodeput(0,0.5){x1}{$x_1$}
    \cnodeput(0,1.5){x2}{$x_2$}
    \cnodeput(0,2.5){x3}{$x_3$}
    \cnodeput(0,3.5){x4}{$x_4$}
    \cnodeput(3,0){y1}{$y_1$}
    \cnodeput(3,1){y2}{$y_2$}
    \cnodeput(3,2){y3}{$y_3$}
    \cnodeput(3,3){y4}{$y_4$}
    \cnodeput(3,4){y5}{$y_5$}
    \ncline[linewidth=1pt]{x1}{y1}
    \ncline[linewidth=2pt]{x1}{y2}
    \ncline[linewidth=1pt]{x2}{y2}
    \ncline[linewidth=2pt]{x2}{y3}
    \ncline[linewidth=1pt]{x3}{y3}
    \ncline[linewidth=2pt]{x3}{y4}
    \ncline[linewidth=1pt]{x4}{y4}
    \ncline[linewidth=2pt]{x4}{y5}
    \end{pspicture}
  \end{center}
  \caption{A bipartite graph, $G$,  and a maximum matching in $G$}
  \label{bipart1}.
\end{figure}
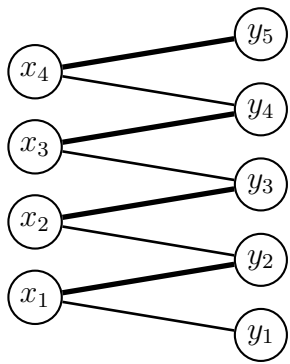

The problem we just described is called a {\it maximum matching problem\/}.
A convenient way to describe this problem is to build a graph, $G$
(undirected), with $m + n$ nodes partitioned into two subsets
$X$ and $Y$, with $X = \{x_1, \ldots, x_m\}$ and
$Y = \{y_1, \ldots, y_n\}$, and with an edge between $x_i$ and $y_j$
iff $T_j \in S_i$, that is, if machine $M_i$ can perform task $T_j$.
Such a graph, $G$, is called a {\it bipartite graph\/}.
An example of a bipartite graph is shown in Figure \ref{bipart1}.
Now, our matching problem is to find an edge set of maximum size, $M$, 
such that no two edges share a common endpoint or, equivalently,
such that every node belongs to at most one edge of $M$.
Such a set of edges is called a {\it maximum matching\/} in $G$.
A maximum matching whose edges are shown as thicker lines is shown in Figure
\ref{bipart1}.

\begin{defin}
\label{bipartdef}
{\em
A graph, $G = (V, E, st)$, is a {\it bipartite graph\/} iff
its set of edges, $V$, can be partitioned into two nonempty
disjoint sets, $V_1, V_2$, 
so that for every edge, $e\in E$,
$|st(e) \cap V_1| = |st(e) \cap V_2| = 1$, i.e., one endpoint of
$e$ belongs to $V_1$ while the other belongs to $V_2$. 
}
\end{defin}

\medskip
Note that in a bipartite graph, there are no edges linking
nodes in $V_1$ (or nodes in $V_2$). Thus, there are no loops.

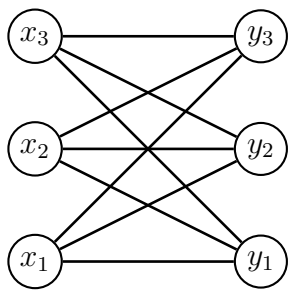
\begin{figure}
  \begin{center}
    \begin{pspicture}(0,0)(3,4)
    \cnodeput(0,0){x1}{$x_1$}
    \cnodeput(0,1.5){x2}{$x_2$}
    \cnodeput(0,3){x3}{$x_3$}
    \cnodeput(3,0){y1}{$y_1$}
    \cnodeput(3,1.5){y2}{$y_2$}
    \cnodeput(3,3){y3}{$y_3$}
    \ncline[linewidth=1pt]{x1}{y1}
    \ncline[linewidth=1pt]{x1}{y2}
    \ncline[linewidth=1pt]{x1}{y3}
    \ncline[linewidth=1pt]{x2}{y1}
    \ncline[linewidth=1pt]{x2}{y2}
    \ncline[linewidth=1pt]{x2}{y3}
    \ncline[linewidth=1pt]{x3}{y1}
    \ncline[linewidth=1pt]{x3}{y2}
    \ncline[linewidth=1pt]{x3}{y3}
    \end{pspicture}
  \end{center}
  \caption{The  bipartite graph $K_{3, 3}$}
  \label{K33}.
\end{figure}

\remark
The {\it complete bipartite graph\/} for which $|V_1| = m$
and $|V_2| = n$ is the bipartite graph that has all edges
$(i, j)$, with $i\in \{1, \ldots, m\}$ and $j\in \{1, \ldots, n\}$.
This graph is denoted $K_{m, n}$. The complete bipartite graph
$K_{3, 3}$ plays a special role, namely, it is not a planar graph,
which means that it is impossible to draw it on a plane without
avoiding that two edges (drawn as continuous simple curves) 
intersect. A picture of $K_{3, 3}$ is shown in Figure \ref{K33}.

\medskip
The maximum matching problem in a bipartite graph can be nicely solved 
using the methods of Section \ref{sec29} for finding Max-flows. Indeed,
our matching problem is equivalent to finding a maximum flow
in the network, $N$, constructed from the bipartite graph $G$ as follows:
\begin{enumerate}
\item
Add a new source, $v_s$ and a new sink, $v_t$;
\item
Add an oriented edge, $(v_s, u)$, for every $u\in V_1$;
\item
Add an oriented edge, $(v, v_t)$, for every $v\in V_2$;
\item
Orient every edge, $e\in E$, from $V_1$ to $V_2$;
\item
Define the capacity function, $c$, so that $c(e) = 1$, for every edge
of this new graph.
\end{enumerate}

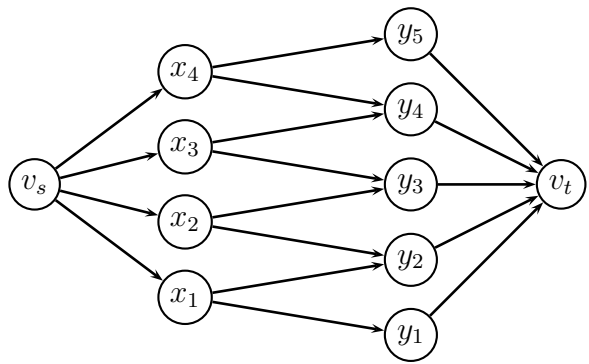
\begin{figure}
  \begin{center}
    \begin{pspicture}(-2,0)(5,5.2)
    \cnodeput(-2,2){s}{$v_s$}
    \cnodeput(5,2){t}{$v_t$}
    \cnodeput(0,0.5){x1}{$x_1$}
    \cnodeput(0,1.5){x2}{$x_2$}
    \cnodeput(0,2.5){x3}{$x_3$}
    \cnodeput(0,3.5){x4}{$x_4$}
    \cnodeput(3,0){y1}{$y_1$}
    \cnodeput(3,1){y2}{$y_2$}
    \cnodeput(3,2){y3}{$y_3$}
    \cnodeput(3,3){y4}{$y_4$}
    \cnodeput(3,4){y5}{$y_5$}
    \ncline[linewidth=1pt]{->}{x1}{y1}
    \ncline[linewidth=1pt]{->}{x1}{y2}
    \ncline[linewidth=1pt]{->}{x2}{y2}
    \ncline[linewidth=1pt]{->}{x2}{y3}
    \ncline[linewidth=1pt]{->}{x3}{y3}
    \ncline[linewidth=1pt]{->}{x3}{y4}
    \ncline[linewidth=1pt]{->}{x4}{y4}
    \ncline[linewidth=1pt]{->}{x4}{y5}
    \ncline[linewidth=1pt]{->}{s}{x1}
    \ncline[linewidth=1pt]{->}{s}{x2}
    \ncline[linewidth=1pt]{->}{s}{x3}
    \ncline[linewidth=1pt]{->}{s}{x4}
    \ncline[linewidth=1pt]{->}{y1}{t}
    \ncline[linewidth=1pt]{->}{y2}{t}
    \ncline[linewidth=1pt]{->}{y3}{t}
    \ncline[linewidth=1pt]{->}{y4}{t}
    \ncline[linewidth=1pt]{->}{y5}{t}
    \end{pspicture}
  \end{center}
  \caption{The network associated with a bipartite graph}
  \label{bipart2}.
\end{figure}

The network corresponding to the bipartite graph of Figure \ref{bipart1}
is shown inFigure \ref{bipart2}.

Now, it is very easy to check that there is a matching, $M$,
containing $p$ edges iff there is a flow of value $p$.
Thus, there is a one-to-one correspondence between maximum matchings
and maximum integral flows. As we know that the algorithm
{\it maxflow\/} (actually, its various versions) produces an integral
solution when ran on the zero flow, this solution yields
a maximum matching.

\medskip
The notion of graph coloring is also important and has bearing
on the notion of bipartite graph.

\begin{defin}
\label{colordef}
{\em
Given a graph, $G = (V, E, st)$, a {\it $k$-coloring\/} of $G$ is
a partition of $V$ into $k$ pairwise
disjoint nonempty subsets, $V_1, \ldots, V_k$,
so that no two vertices in any
subset $V_i$ are adjacent (i.e., the endpoints of every edge, $e\in E$,
must belong to $V_i$ and $V_j$, for some $i\not= j$).
If a graph, $G$, admits a $k$-coloring, we say that that 
{\it $G$ is $k$-colorable\/}. The {\it chromatic number\/}, $\chi(G)$,
of a graph, $G$,
is the minimum $k$ for which $G$ is $k$-colorable.
}
\end{defin}

\remark
Although the notation, $\chi(G)$, for the chromatic number of
a graph is often used in the graph theory literature, it 
is an unfortunate choice because it can be confused
with the Euler characteristic of a graph (see Theorem \ref{Eulerform}).
Other notations for the chromatic number
include $\gamma(G)$, $\nu(G)$ and $\mathrm{chr}(G)$.

\medskip
The following theorem gives some useful characterizations of
bipartite graphs. First, we must define the incidence matrix of
an unoriented graph, $G$. Assume that $G$ has edges
$\mathbf{e}_1, \ldots, \mathbf{e}_n$ and vertices
$\mathbf{v}_1, \ldots, \mathbf{v}_m$. The {\it incidence matrix,
$A$, of $G$\/}, is the $m\times n$ matrix whose entries are given by
\[
a_{i\, j} = 
\cases{
1 & if $\mathbf{v}_i \in st(\mathbf{e}_j)$\cr
0 & otherwise. \cr
}
\]

Note that, unlike the incidence matrix of a directed graph, 
the incidence matrix of an undirected graph only has non-negative entries.
As a consequence, these matrices are not necessarily totally unimodular.
For example, the reader should check that for any elementary cycle, $C$,
of odd length, the incidence matrix, $A$, of $C$ has a
determinant whose value is $\pm 2$. However, the next theorem will 
show that the incidence matrix of a bipartite graph is 
totally unimodular and in fact, this property characterizes
bipartite graphs. 

\medskip
In order to prove part of the next theorem we need the notion of
distance in a graph, an important concept in any case. 
If $G$ is a connected graph, for any two 
nodes $u$ and $v$ of $G$, the length of a chain, $\pi$, from
$u$ to $v$ is the number of edges in $\pi$ and the {\it distance\/},
$d(u, v)$, from   $u$ to $v$ is the minimum length of all 
path from $u$ to $v$. Of course,  $u = v$ iff $d(u, v) = 0$.

\begin{thm}
\label{biparth1}
Given any graph, $G = (V, E, st)$, the following properties are 
equivalent:
\begin{enumerate}
\item[(1)]
$G$ is bipartite.
\item[(2)]
$\gamma(G) = 2$.
\item[(3)]
$G$ has no elementary cycle of odd length.
\item[(4)]
$G$ has no cycle of odd length.
\item[(5)]
The incidence matrix of $G$ is totally unimodular.
\end{enumerate}
\end{thm}

\proof
The equivalence $(1)\Longleftrightarrow (2)$ is clear by definition of
the chromatic number. 

\medskip
$(3)\Longleftrightarrow (4)$ holds because
every cycle is the concatenation of elementary cycles. So,
a cycle of odd length must contain some elementary cycle of odd length.

\medskip
$(1)\Longrightarrow (4)$. 
This is because the vertices of a cycle
belong alternatively to $V_1$ and $V_2$. So, there must be an even number
of them.

\medskip
$(4)\Longrightarrow (2)$.
Clearly, a graph is $k$-colorable iff all its connected
components are $k$-colorable, so we may assume that $G$ is connected.
Pick any node, $v_0$, in $G$ and 
let $V_1$ be the subset of nodes whose distance from $v_0$ is even
and $V_2$ be the subset of nodes whose distance from $v_0$ is odd.
We claim that any two nodes, $u$ and $v$, in $V_1$ (resp. $V_2$) 
are not adjacent.
Otherwise, by going up the chains from $u$ and $v$ back to $v_0$
and by adding the edge from $u$ to $v$, we would obtain a cycle
of odd length, a contradiction. Therefore, $G$, is $2$-colorable.

\medskip
$(1)\Longrightarrow (5)$.
Orient the edges of $G$  so that for every $e\in E$,
$s(e)\in V_1$ and $t(e)\in V_2$. Then, we know from Proposition
\ref{pseudotrigp1} that the incidence matrix, $D$, of the oriented
graph $G$ is totally unimodular. However, because $G$ is 
bipartite, $D$ is obtained from $A$ by multiplying all the
rows corresponding to nodes in $V_2$ by $-1$ and so, $A$ is
also totally unimodular.

\medskip
$(5)\Longrightarrow (3)$. Let us prove the contrapositive.
If $G$ has an elementary cycle, $C$, of odd length, then we
observed that the submatrix of $A$ corresponding to $C$ has determinant 
$\pm 2$. 
$\bigsquare$

\medskip
We now define the general notion of a matching.

\begin{defin}
\label{matchdef}
{\em
Given a graph, $G = (V, E, st)$, a {\it matching, $M$, in $G$\/}
is a subset of edges so that any two distinct edges in $M$ have
no common endpoint (are not adjacent) 
or equivalently, so that every vertex, $v\in E$,
is incident to at most one edge in $M$. A vertex, $v\in V$ is {\it matched\/}
iff it is incident some some edge in $M$ and otherwise it is said to
be {\it unmatched\/}. A matching, $M$, is a {\it perfect matching\/}
iff every node is matched.
}
\end{defin}

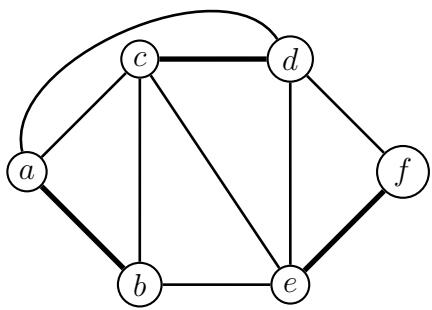
\begin{figure}
  \begin{center}
    \begin{pspicture}(0,0)(5,4)
    \cnodeput(0,1.5){a}{$a$}
    \cnodeput(1.5,3){c}{$c$}
    \cnodeput(1.5,0){b}{$b$}
    \cnodeput(3.5,3){d}{$d$}
    \cnodeput(3.5,0){e}{$e$}
    \cnodeput(5,1.5){f}{$f$}
    \ncline[linewidth=1pt]{a}{c}
    \ncline[linewidth=2pt]{a}{b}
    \ncline[linewidth=1pt]{b}{c}
    \ncline[linewidth=2pt]{c}{d}
    \ncline[linewidth=1pt]{b}{e}
    \ncline[linewidth=1pt]{c}{e}
    \ncline[linewidth=1pt]{d}{e}
    \ncline[linewidth=1pt]{d}{f}
    \ncline[linewidth=2pt]{e}{f}
    \ncarc[arcangleA=-80, arcangleB=-80, linewidth=1pt]{d}{a}
    \end{pspicture}
  \end{center}
  \caption{A perfect matching in a graph}
  \label{match1}
\end{figure}

An example of a perfect matching, $M = \{(ab), (cd), (ef)\}$, 
is shown in Figure \ref{match1},
with the edges of the matching indicated in thicker lines.
The pair $\{(bc), (ed)\}$ is also a matching, in fact, 
a maximal matching (no edge can be added to this matching and still
have a matching).

It is possible to characterize maximum matchings in terms of certain
types of chains called {\it alternating chains\/} defined below:

\begin{defin}
\label{alternatcycle}
{\em
Given a graph, $G = (V, E, st)$, and a matching, $M$, in $G$, 
an elementary chain is an {\it alternating chain w.r.t $M$\/} 
iff the edges in this chain belong alternately to $M$  and $E - M$.
}
\end{defin}

\begin{thm} (Berge)
\label{Bergthm}
Given any graph, $G = (V, E, st)$, a matching, $M$, in $G$ is
a maximum matching iff there are no alternating chains w.r.t. $M$
whose endpoints are unmatched.
\end{thm}

\begin{figure}
  \begin{center}
    \begin{pspicture}(0,0)(2,5)
    \cnodeput(0,0){x1}{$x_1$}
    \cnodeput(0,1){x2}{$x_2$}
    \cnodeput(0,2){x3}{$x_3$}
    \cnodeput(0,3){x4}{$x_4$}
    \cnodeput(0,4){x5}{$x_5$}
    \cnodeput(3,0.5){y1}{$y_1$}
    \cnodeput(3,1.5){y2}{$y_2$}
    \cnodeput(3,2.5){y3}{$y_3$}
    \cnodeput(3,3.5){y4}{$y_4$}
    \cnodeput(3,4.5){y5}{$y_5$}
    \ncline[linewidth=1pt]{x1}{y1}
    \ncline[linewidth=2pt]{x2}{y1}
    \ncline[linewidth=1pt]{x2}{y2}
    \ncline[linewidth=2pt]{x3}{y2}
    \ncline[linewidth=1pt]{x3}{y3}
    \ncline[linewidth=2pt]{x4}{y3}
    \ncline[linewidth=1pt]{x4}{y4}
    \ncline[linewidth=2pt]{x5}{y4}
    \ncline[linewidth=1pt]{x5}{y5}
    \end{pspicture}
  \end{center}
  \caption{An alternating chain in $G$}
  \label{alter1}
\end{figure}
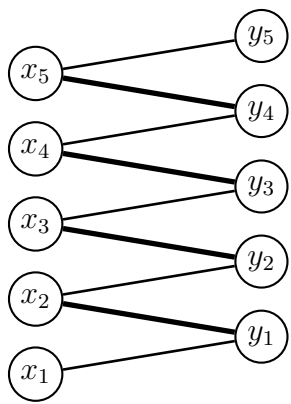

\proof
First, assume that $M$ is a maximum matching and that $C$ is
an alternating chain w.r.t. $M$ whose enpoints, $u$ and $v$ are
unmatched. An an example, consider the alternating chain shown in Figure
\ref{alter1}, where the edges in $C\cap M$ are indicated in
thicker lines.

We can form the set of edges
\[
M' = (M - (C\cap M))\cup (C\cap (E - M)),
\]
which consists in deleting the edges in $M$ from $C$
and adding the edges from $C$ not in $M$. It is immediately verified
that $M'$ is still a matching but $|M'| = |M| + 1$
(see Figure \ref{alter1}),
contradicting the fact that $M$ is a maximum matching.
Therefore, there are no alternating chains w.r.t. $M$ 
whose endpoints are unmatched.

\medskip
Conversely, assume that $G$ has no alternating chains w.r.t. $M$
whose endpoints are unmatched and let $M'$ be another matching
with $|M'| > |M|$ (i.e., $M$ is not a maximum matching).
Consider the spanning subgraph, $H$, of $G$, whose set of edges is
\[
(M - M') \cup (M' - M).
\]
As $M$ and $M'$ are matchings, the connected components of $H$ are either
isolated vertices, or elementary cycles of even length, or
elementary chains, and in these last two cases,
the edges in these cycles or chains  belong alternately to $M$ and $M'$;
this is because $d_H(u) \leq 2$ for every vertex $u\in V$
and if $d_H(u) = 2$, then
$u$ is adjacent to one edge in $M$ and one edge in $M'$.

\medskip
Now, $H$ must possess a connected component that is a chain, $C$, whose
enpoints are in $M'$, as otherwise we would have $|M'| \leq |M|$,
contradicting the assumption  $|M'| > |M|$. However, $C$ is
an alternating chain w.r.t. $M$ whose endpoints are unmatched,
a contradiction.
$\bigsquare$

\medskip
A notion closely related to the concept of a matching 
but, in some sense, dual, is the notion of a {\it line cover\/}.

\begin{defin}
\label{linecov}
{\em
Given any graph, $G = (V, E, st)$, without loops or isolated vertices,
a {\it line cover\/} (or {\it line covering\/}) of $G$ is a set of edges,
$\s{C}\subseteq E$, so that every vertex $u\in V$ is incident to
some edge in $\s{C}$. A {\it minimum line cover\/}, $\s{C}$,
is a line cover of minimum size.
}
\end{defin}

\medskip
The maximum matching, $M$, in the graph of Figure \ref{match1}
is also a minimum line cover. The set
$\{(ab), (bc), (de), (ef)\}$ is also a line cover but it is not
minimum, although minimal.
The relationship between maximum matchings and minimum covers
is given by the following theorem:

\begin{thm}
\label{matchcovth1}
Given any graph, $G = (V, E, st)$, without loops or isolated
vertices, with $|V| = n$, let $M$ be a maximum matching and let
$\s{C}$ be a minimum line cover. Then, the following properties hold:
\begin{enumerate}
\item[(1)]
If we associate to every unmatched vertex of $V$ some edge incident
to this vertex and add all such edges to $M$, then we obtain a
minimum line cover, $\s{C}_M$.
\item[(2)]
Every maximum matching, $M'$, of the spanning subgraph, $(V, \s{C})$,
is a maximum matching of $G$.
\item[(3)]
$|M| + |\s{C}| = n$.
\end{enumerate}
\end{thm} 

\proof
It is clear that $\s{C}_M$ is a line cover. As the number of
vertices umatched by $M$ is $n - 2|M|$ (as
each edge in $M$ matches exactly  two vertices), we have
\begin{equation}
|\s{C}_M| = |M| + n - 2|M| = n - |M|.
\tag{$*$}
\end{equation}
Furthermore, as $\s{C}$ is a minimum line cover, 
the spanning subgraph, $(V, \s{C})$, does not contain any cycle or chain
of length greater than or equal to $2$. Consequently, each edge
$e\in \s{C} - M'$ corresponds to a single vertex unmatched by $M'$.
Thus,
\[
|\s{C}| - |M'| = n - 2|M'|,
\]
that is
\begin{equation}
|\s{C}|  = n - |M'|.
\tag{$**$}
\end{equation}
As $M$ is a maximum matching of $G$,
\[
|M'| \leq |M|
\]
and so, using $(*)$ and $(**)$, we get
\[
|\s{C}_M| = n - |M| \leq n - |M'| = |\s{C}|,
\]
that is, $|\s{C}_M| \leq |\s{C}|$.
However, $\s{C}$ is a minimum matching, so  $|\s{C}| \leq |\s{C}_M|$,
which proves that
\[
|\s{C}| = |\s{C}_M|.
\]
The last equation proves the remaining claims. 
$\bigsquare$

\medskip
There are also notions analogous to matchings and line covers
but applying to vertices instead of edges.

\begin{defin}
\label{indepnodecov}
{\em
Let $G = (V, E, st)$ be any graph. A set, $U\subseteq V$, of nodes
is {\it independent\/}  (or {\it stable\/}) iff no two nodes
in $U$ are adjacent (there is no edge having these nodes as endpoints).
A {\it maximum independent set\/} is an independent set of
maximum size. A set, $\s{U}\subseteq V$, of nodes is a {\it point
cover\/} (or {\it vertex cover\/} or {\it transversal\/}) 
iff every edge of $E$ is incident
to some node in $\s{U}$. A {\it minimum point cover\/}
is a point cover of minimum size.
}
\end{defin}

\medskip
For example, $\{a, b, c, d, f\}$ is point cover of the graph
of Figure \ref{match1}.
The following simple proposition holds:

\begin{prop}
\label{covp1}
Let $G = (V, E, st)$ be any graph, $U$ be any independent set, $\s{C}$ 
be any line cover,  $\s{U}$ be any point cover and  $M$ be
any matching. Then, we have the following inequalities:
\begin{enumerate}
\item[(1)]
$|U| \leq |\s{C}|$;
\item[(2)]
$|M| \leq |\s{U}|$
\item[(3)]
$U$ is an independent set of nodes iff $V - U$ is a point cover.
\end{enumerate}
\end{prop}

\proof
(1) Since $U$ is an independent set of nodes, every edge in $\s{C}$
is incident with at most one vertex in $U$, so $|U| \leq |\s{C}|$.

\medskip
(2) Since $M$ is a matching, every vertex in $\s{U}$ is incident to at most 
one edge in $M$, so $|M| \leq |\s{U}|$. 

\medskip
(3) Clear from the definitions.
$\bigsquare$

\medskip
It should be noted that the inequalities of Proposition \ref{covp1}
can be strict. For example, if $G$ is an elementary cycle with
$2k + 1$ edges, the reader should check that both inequalities
are strict. 

\medskip
We now go back to bipartite graphs and give an algorithm which, 
given a bipartite graph, $G = (V_1\cup V_2, E)$,  will decide
whether a matching, $M$, is a  maximum matching in $G$.
This algorithm, shown in Figure \ref{markingfig}, 
will mark the nodes with the one of the three
tags, $+$, $-$, or $0$.

\begin{figure}
\begin{tabbing}
\quad \= \quad \= \quad \= \quad \= \quad \= \quad \= \quad \\
{\bf procedure} $\mathrm{marking}(G, M, mark)$ \\ 
 \> {\bf begin} \\
 \> \> {\bf for each} $u\in V_1\cup V_2$ {\bf do} $mark(u) := 0$
{\bf endfor}; \\
 \> \> {\bf while} $\exists u\in V_1\cup V_2$ with $mark(u) = 0$ and
$u$ not matched by $M$ {\bf do} \\
\> \> \>  $mark(u) := +$; \\
\> \> \>  {\bf while} $\exists v\in V_1\cup V_2$ with $mark(v) = 0$ and
$v$ adjacent to $w$ with $mark(w) = +$ {\bf do} \\
\> \> \> \> $mark(v) := -$; \\
\> \> \> \> {\bf if} $v$ is not matched by $M$ {\bf then} {\bf exit} 
$(\alpha)$ \\
\> \> \> \> $(*$ an alternating chain has been found $*)$ \\
\> \> \> \>  {\bf else} find $w\in V_1\cup V_2$ so that $(vw)\in M$;
$mark(w) := +$ \\ 
\> \> \> \> {\bf endif} \\
\> \> \>  {\bf endwhile} \\
\> \> {\bf endwhile}; \\
\> \>  {\bf for each} $u\in V_1$ with $mark(u) = 0$ {\bf do} $mark(u) := +$
{\bf endfor}; \\
\> \>  {\bf for each} $u\in V_2$ with $mark(u) = 0$ {\bf do} $mark(u) := -$
{\bf endfor}  $(\beta)$ \\
 \> {\bf end} 
\end{tabbing}
\caption{Procedure {\it marking}}
\label{markingfig}
\end{figure}

\medskip
The following theorem tells us what is the behavior of the procedure 
{\it marking}.

\begin{thm}
\label{markingth}
The procedure {\it marking} always terminates 
in one of the following two (mutually exclusive) situations:
\begin{enumerate}  
\item[(a)]
The algorithm finds an alternating chain w.r.t. $M$ whose endpoints
are unmatched.
\item[(b)]
The algorithm finds a point cover, $\s{U}$, with
$|\s{U}| = |M|$, which shows that $M$ is a maximum matching.
\end{enumerate}  
\end{thm}

\proof
Since nodes keep being marked, the algorithm obviously terminates.
There are no pairs of adjacent nodes bothy marked $+$ since, as soon as
a node is marked $+$, all of its adjacent nodes are labeled $-$.
Consequently, if the algorithm ends in $(\beta)$, those nodes marked $-$
form a point cover.

\medskip
We also claim that the endpoints, $u$ and $v$, of any edge in the matching 
can't both be marked $-$. Otherwise, by following  backward the 
chains that allowed the marking of $u$ and $v$, we would find
an odd cycle, which is impossible in a bipartite graph. Thus, if we
end in $(\beta)$, each node marked $-$ is incident to exactly one
edge in $M$. This shows that  the set, $\s{U}$, of nodes marked $-$
is a point cover with $|\s{U}| = |M|$. By Proposition \ref{covp1},
we see that $\s{U}$ is a minimum point cover and that $M$ is
a maximum matching.

\medskip
If the algorithm ends in $(\alpha)$, by tracing the chain starting
from the unmatched node, $u$, marked $-$ back to the node marked $+$
causing $u$ to marked, and so on, we find an alternating chain w.r.t. $M$ 
whose endpoints are not matched.
$\bigsquare$

\medskip
The following important corollaries follow immediately from Theorem
\ref{markingth}:

\begin{cor}
\label{markingcor2}
In a bipartite graph, the size of a minimum point cover is equal to
the size of maximum matching.
\end{cor}

\begin{cor}
\label{markingcor3}
In a bipartite graph, the size of a maximum independent set is equal to
the size of a minimum line cover.
\end{cor}

\proof
We know from Proposition \ref{covp1} that the complement of
a point cover is an independent set. Consequently, 
by Corollary \ref{markingcor2}, the size of a maximum independent set
is $n - |M|$, where $M$ is a maximum matching and
$n$ is the number of vertices in $G$.
Now, from Theorem \ref{matchcovth1} (3),
for any maximum matching, $M$, and any minimal line cover, $\s{C}$,
we have $|M| + |\s{C}| = n$ and so, the size of a
maximum independent set is equal to the size of a 
minimal line cover. 
$\bigsquare$

\medskip
We can derive more classical theorems from the above results.

\medskip
Given any graph, $G = (V, E, st)$, for any subset of nodes, $U\subseteq V$, 
let
\[
N_G(U) = \{v\in V - U \mid (\exists u\in U)(\exists e\in E)
(st(e) = \{u, v\})\}, 
\]
be the set of {\it neighbours\/} of $U$, i.e.,
the set of vertices {\it not\/} in $U$ and adjacent to vertices in $U$.

\begin{thm} (K\" onig (1931))
\label{Koenig1}
For any bipartite graph, $G = (V_1\cup V_2, E, st)$, the
maximum size of a matching is given by
\[ 
\min_{U \subseteq V_1} (|V_1 - U| + |N_G(U)|).
\]
\end{thm}

\proof
This theorem will follow from Corollary \ref{markingcor2}
if we can show that every minimum point cover is of the form
$(V_1 - U) \cup N_G(U)$, for some subset, $U$, of $V_1$.
However, a moment of reflexion shows that this is indeed the case.
$\bigsquare$

\medskip
Theorem \ref{Koenig1} implies another classical result:

\begin{thm} (K\" onig-Hall)
\label{KoenigHall}
For any bipartite graph, $G = (V_1\cup V_2, E, st)$, 
there is a matching, $M$, such that all nodes in $V_1$ are
matched iff
\[ 
|N_G(U)| \geq |U|
\quad\hbox{for all}\quad U\subseteq V_1.
\]
\end{thm}

\proof
By Theorem \ref{Koenig1}, there is a matching, $M$ in $G$
with $|M| = |V_1|$ iff
\[
|V_1| =  \min_{U \subseteq V_1} (|V_1 - U| + |N_G(U)|)
= \min_{U \subseteq V_1} (|V_1| + |N_G(U)| - |U|),
\]
that is, iff $|N_G(U)| - |U| \geq 0$ for all $U\subseteq V_1$.
$\bigsquare$

\medskip
Now, it is clear that a bipartite graph has a perfect matching
(i.e., a matching such that every vertex is matched), $M$, iff
$|V_1| = |V_2|$ and $M$ matches all nodes in $V_1$. So, as
a corollary of Theorem \ref{KoenigHall}, we see that a
bipartite graph has a perfect matching iff $|V_1| = |V_2|$ and if 
\[ 
|N_G(U)| \geq |U|
\quad\hbox{for all}\quad U\subseteq V_1.
\]
As an exercise, the reader should show the 

\medskip\noindent
{\it Marriage Theorem\/} (Hall, 1935) $\>$
{\it Every $k$-regular
bipartite graph, with $k \geq 1$,
has a perfect matching\/} (a graph is $k$-regular iff
every node has degree $k$).

\medskip
For more on bipartite graphs, matchings, covers, {\it etc}.,
the reader should consult Diestel \cite{Diestel} (Chapter 2),
Berge \cite{Berge} (Chapter 7) and also Harary \cite{Harary}
and Bollobas \cite{Bollobas}.

\section{Planar Graphs}
\label{sec31}
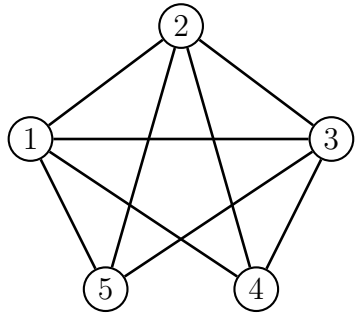
\begin{figure}
  \begin{center}
    \begin{pspicture}(0,0)(4,4)
    \cnodeput(0,2){x1}{$1$}
    \cnodeput(1,0){x5}{$5$}
    \cnodeput(3,0){x4}{$4$}
    \cnodeput(4,2){x3}{$3$}
    \cnodeput(2,3.5){x2}{$2$}
    \ncline[linewidth=1pt]{x1}{x2}
    \ncline[linewidth=1pt]{x1}{x3}
    \ncline[linewidth=1pt]{x1}{x4}
    \ncline[linewidth=1pt]{x1}{x5}
    \ncline[linewidth=1pt]{x2}{x5}
    \ncline[linewidth=1pt]{x2}{x4}
    \ncline[linewidth=1pt]{x2}{x3}
    \ncline[linewidth=1pt]{x3}{x5}
    \ncline[linewidth=1pt]{x3}{x4}
    \end{pspicture}
  \end{center}
  \caption{A Graph, $G$, drawn with intersecting edges}
  \label{planarfig1}
\end{figure}

Suppose we have a graph, $G$, and that we want to draw it
``nicely'' on a piece of paper, which means that we
draw the vertices as points and the edges as line segments
joining some of these points, in such a way that {\it no two edges cross
each other\/}, except possibly at common endpoints. 
We will  have more flexibility
and still have a nice picture if we allow each abstract edge to be 
represented by a continuous simple curve (a curve that has no
self-intersection), that is, a subset of
the plane homeomorphic to the closed interval $[0, 1]$
(in the case of a loop, a subset homeomorphic to the circle, $S^1$).
If a graph can be drawn in such a fashion, it is called a 
{\it planar graph\/}. For example, consider the graph
depicted in Figure \ref{planarfig1}.

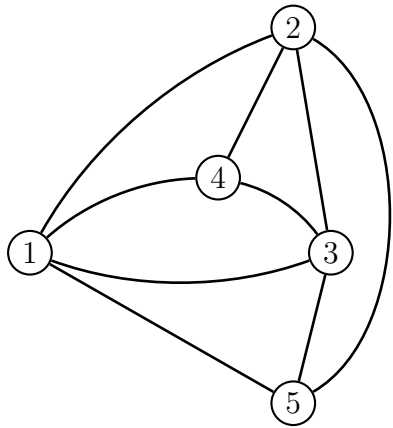
\begin{figure}
  \begin{center}
    \begin{pspicture}(0,0)(5,6)
    \cnodeput(0,2){x1}{$1$}
    \cnodeput(3.5,0){x5}{$5$}
    \cnodeput(2.5,3){x4}{$4$}
    \cnodeput(4,2){x3}{$3$}
    \cnodeput(3.5,5){x2}{$2$}
    \ncline[linewidth=1pt]{x1}{x5}
    \ncline[linewidth=1pt]{x2}{x4}
    \ncline[linewidth=1pt]{x2}{x3}
    \ncline[linewidth=1pt]{x3}{x5}
    \ncarc[arcangleA=-20, arcangleB=-20, linewidth=1pt]{x1}{x3}
    \ncarc[arcangleA=20, arcangleB=20, linewidth=1pt]{x1}{x4}
    \ncarc[arcangleA=20, arcangleB=20, linewidth=1pt]{x1}{x2}
    \ncarc[arcangleA=-20, arcangleB=-20, linewidth=1pt]{x3}{x4}
    \ncarc[arcangleA=60, arcangleB=60, linewidth=1pt]{x2}{x5}
    \end{pspicture}
  \end{center}
  \caption{The  Graph, $G$, drawn as a plane graph}
  \label{planarfig2}
\end{figure}

\medskip
If we look at Figure \ref{planarfig1}, we may believe that
the graph $G$ is not planar, but this is no so. In fact, by moving the
vertices in the plane and by continuously deforming some of the edges,
we can obtain a planar drawing of the same graph, as shown in Figure
\ref{planarfig2}.

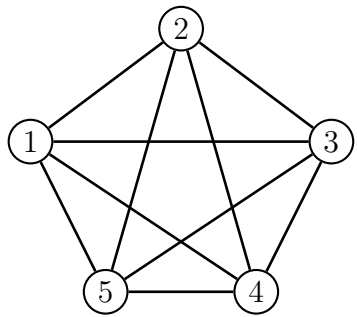
\begin{figure}
  \begin{center}
    \begin{pspicture}(0,0)(4,4)
    \cnodeput(0,2){x1}{$1$}
    \cnodeput(1,0){x5}{$5$}
    \cnodeput(3,0){x4}{$4$}
    \cnodeput(4,2){x3}{$3$}
    \cnodeput(2,3.5){x2}{$2$}
    \ncline[linewidth=1pt]{x1}{x2}
    \ncline[linewidth=1pt]{x1}{x3}
    \ncline[linewidth=1pt]{x1}{x4}
    \ncline[linewidth=1pt]{x1}{x5}
    \ncline[linewidth=1pt]{x2}{x5}
    \ncline[linewidth=1pt]{x2}{x4}
    \ncline[linewidth=1pt]{x2}{x3}
    \ncline[linewidth=1pt]{x3}{x5}
    \ncline[linewidth=1pt]{x3}{x4}
    \ncline[linewidth=1pt]{x5}{x4}
    \end{pspicture}
  \end{center}
  \caption{The complete graph $K_5$, a non-planar graph}
  \label{K5fig}
\end{figure}

\medskip
However, we should not be overly optimistic. Indeed, if we add 
an edge from node $5$ to node $4$, obtaining the graph known
as $K_5$ shown in Figure \ref{K5fig},  
it can be proved that there is no way to
move the nodes around and deform the edge continuously 
to obtain a planar graph (we will prove this a little later
using the Euler formula). 
Another graph that is non-planar is the bipartite grapk $K_{3, 3}$.
The two graphs, $K_5$ and $K_{3, 3}$ play a special role with respect to
planarity. Indeed, a famous theorem of Kuratowski says that
a graph is planar if and only if it does not contain $K_5$ or $K_{3, 3}$
as a minor (we will explain later what a minor is).

\remark
Given $n$ vertices, say $\{1, \ldots, n\}$, the graph whose edges
are all subsets $\{i, j\}$, with $i, j \in \{1, \ldots, n\}$ and
$i \not= j$, is the {\it complete graph on $n$ vertices\/} and is denoted
by $K_n$ (but Diestel uses the notation $K^n$).

\medskip
In order to give a precise definition of a planar graph, let us review
quickly some basic notions about curves. A {\it simple curve\/}
(or {\it Jordan curve\/}) 
is any injective continuous function, $\mapdef{\gamma}{[0, 1]}{\reals^2}$.
Since $[0, 1]$ is compact and $\gamma$ is continuous, it is well-known
that the inverse, $\mapdef{f^{-1}}{\gamma([0, 1])}{[0, 1]}$, of $f$
is also continuous. So,  $\gamma$ is a homeomorphism between
$[0, 1]$ and its image, $\gamma([0, 1])$.
With a slight abuse of language we will also call the image,
$\gamma([0, 1])$, of $\gamma$, a simple curve. This image is a
connected and compact subset of $\reals^2$.
The points $a = \gamma(0)$ and $b = \gamma(1)$ are called the {\it boundaries\/}
or {\it endpoints\/} of $\gamma$ (and $\gamma([0, 1])$).
The open subset $\gamma([0, 1]) - \{\gamma(0), \gamma(1)\}$
is called the {\it interior\/} of $\gamma([0, 1])$ and is denoted
$\interio{\gamma}$. 
A continuous function , $\mapdef{\gamma}{[0, 1]}{\reals^2}$,
such that $\gamma(0) = \gamma(1)$
and $\gamma$ is injective on $[0, 1)$ is called a
{\it simple closed curve\/} or {\it simple loop\/} or {\it closed
Jordan curve\/}. Again, by abuse of language, we call the image,
$\gamma([0, 1])$, of $\gamma$, a simple closed curve {\it etc\/}.
Equivalently, if $S^1 = \{(x, y) \in \reals^2\mid x^2 + y^2 = 1\}$
is  the unit circle in $\reals^2$, a simple closed curve
is any subset of $\reals^2$ homeomorphic to $S^1$. 
In this case, we call $\gamma(0) = \gamma(1)$ 
the {\it boundary\/} or {\it base point\/}
of $\gamma$. The open subset $\gamma([0, 1]) - \{\gamma(0)\}$
is called the {\it interior\/} of $\gamma([0, 1])$ and is also
denoted $\interio{\gamma}$. 

\remark
The notions of simple curve and simple closed curve also make
sense if we replace $\reals^2$ by any topological space, $X$, in particular,
a surface (In this case, a simple (closed) curve is a continuous 
injective function $\mapdef{\gamma}{[0, 1]}{X}$ {\it etc}.).
 
\medskip
We can now define plane graphs as follows:

\begin{defin}
\label{planegraph}
{\em
A {\it plane graph\/} is a pair,
$\s{G} = (V, E)$, where $V$ is a finite set of points in $\reals^2$,
$E$ is a finite set of simple curves and closed simple curves in $\reals^2$ 
called {\it edges\/} and {\it loops\/}, respectively, and
satisfying the following properties:
\begin{enumerate}
\item[(i)]
The endpoints of every edge in $E$ are vertices in $V$ and
the base point of every loop is a vertex in $V$.
\item[(ii)] 
The interior of every edge contains no vertex and
the interiors of any two distinct edges are disjoint.
Equivalently, every edge contains no vertex except for its boundaries
(base point in the case of a loop)  and
any two distinct edges intersect only
at common boundary points.
\end{enumerate}
We say that $G$ is a {\it simple plane graph\/} if it has no
loops and if different edges have different sets of endpoints
}
\end{defin}

\medskip
Obviously, a plane graph, $\s{G} = (V, E)$, defines an ``abstract
graph'', $G = (V, E, st)$, such that
\begin{enumerate}
\item[(a)]
For every simple curve, $\gamma$, 
\[
st(\gamma) = \{\gamma(0), \gamma(1)\}
\]
\item[(b)]
For every simple closed curve, $\gamma$, 
\[
st(\gamma) = \{\gamma(0)\}.
\]
\end{enumerate}

\medskip
For simplicity of notation, we will usually write $\s{G}$
for both the plane graph and the abstract graph associated with $\s{G}$.

\begin{defin}
\label{planardef}
{\em
Given an abstract graph, $G$, we say that {\it $G$ is a planar
graph\/} iff there is some plane graph, $\s{G}$, and an isomorphism,
$\mapdef{\varphi}{G}{\s{G}}$, between $G$ and the abstract graph
associated with $\s{G}$. We call $\varphi$ an {\it embedding
of $G$ in the plane\/} or a {\it planar embedding of $G$\/}. 
}
\end{defin}

\remarks
\begin{enumerate}
\item
If $G$ is a {\it simple\/} planar graph, then by a theorem of Fary, $G$
can be drawn as a plane graph in such a way that the edges are straight
line segments (see Gross and Tucker \cite{GrossTucker}, Section 1.6).
\item
In view of the remark just before Definition \ref{planegraph},
given any topological space, $X$, for instance, a surface,
we can define  a graph on $X$ as a pair, $(V, E)$,
where $V$ is a finite set of points in $X$ and $E$ is a finite
sets of simple (closed) curves on $X$ satisfying the
conditions of Definition \ref{planegraph}. 
\item
Recall the {\it stereographic projection (from the north pole)\/},
$\mapdef{\sigma_N}{(S^2 - \{N\})}{\reals^2}$,  
from the sphere, $S^2 = \{(x, y, x)\in \reals^3 \mid x^2 + y^2 + z^2 = 1\}$
onto the equatorial plane, $z = 0$, with $N = (0, 0, 1)$ (the north pole), 
given by
\[\sigma_N(x, y ,z) = \biggl(\frac{x}{1 - z},\, \frac{y}{1 - z}\biggr).\]
We know that $\sigma_N$ is a homeomorphism, so if $\varphi$
is a planar embedding of a graph $G$ into the plane, then
$\sigma_N^{-1}\circ \varphi$ is an embedding of $G$ into the
sphere. Conversely, if $\psi$ is an embedding of $G$ into the 
sphere, then $\sigma_N\circ \psi$ is a planar embedding of $G$.
Therefore, a graph can be embedded in the plane iff it can be embedded
in the sphere. One of the nice features of  embedding in the
sphere is that the sphere is compact (closed and bounded), so
the faces (see below) of a graph embedded in the sphere are all bounded. 
\item
The ability to embed a graph in a surface other that the sphere
broadens the class of graphs that can be drawn without pairs of
intersecting edges (except at endpoints). For example, it is possible
to embed $K_5$ and $K_{3, 3}$ (which are known {\it not\/} to be planar)
into a torus (try it!). It can be shown that for every (finite) graph, $G$,
there is some surface, $X$, such that $G$ can be embedded in $X$.
Intuitively, whenever two edges cross on a sphere, by lifting one
of the two edges a little bit and adding a ``handle'' 
on which the lifted edge lies we can avoid the crossing.
An excellent reference on the topic of graphs on surfaces is
Gross and Tucker \cite{GrossTucker}.
\end{enumerate}

\medskip
One of the new ingredients of plane graphs is that the notion of a face
makes sense. Given any nonempty open subset, $\Omega$, of the plane
$\reals^2$, we say that two points, $a, b\in \Omega$ are
{\it (arcwise) connected\/}%
\footnote{In topology, a space is connected iff it cannot
be expressed as the union of two nonempty disjoint open subsets.
For {\it open\/} subsets of $\reals^n$, connectedness is equivalent
to arc connectedness. So it is legitimate to use the term connected.} 
iff there is a simple curve, $\gamma$,
such that $\gamma(0) = a$ and $\gamma(1) = b$. Being connected
is an equivalence relation and the equivalence classes of
$\Omega$ w.r.t. connectivity are called the {\it connected components\/}
(or {\it regions\/}) of $\Omega$. Each region is maximally connected
and open. If $R$ is any region of $\Omega$ and if we denote the closure
of $R$ (i.e., the smallest closed set containing $R$) by $\overline{R}$,
then the set $\partial R = \overline{R} - R$ is also a closed set called
the {\it boundary\/} (or {\it frontier\/}) of $R$.

\medskip
Now, given a plane graph, $\s{G}$, if we let $|\s{G}|$ be the
the subset of $\reals^2$ consisting of the union of all the vertices
and edges of $\s{G}$, then this is a closed set and it complement,
$\Omega = \reals^2 - |\s{G}|$, is an open subset of $\reals^2$.

\begin{defin}
\label{facedef}
{\em
Given any plane graph, $\s{G}$, 
the regions of $\Omega = \reals^2 - |\s{G}|$ 
are called the {\it faces\/} of $\s{G}$.
}
\end{defin}

\begin{figure}
  \begin{center}
    \begin{pspicture}(0,0)(6,7)
    \cnodeput(0,0){x1}{$1$}
    \cnodeput(1,5){x2}{$2$}
    \cnodeput(3,4){x3}{$3$}
    \cnodeput(3,2.5){x4}{$4$}
    \cnodeput(2,1.5){x5}{$5$}
    \cnodeput(4,1.5){x6}{$6$}
    \cnodeput(5,5){x7}{$7$}
    \cnodeput(6,0){x8}{$8$}
    \ncline[linewidth=1pt]{x1}{x2}
    \ncline[linewidth=1pt]{x2}{x3}
    \ncline[linewidth=1pt]{x3}{x4}
    \ncline[linewidth=1pt]{x4}{x5}
    \ncline[linewidth=1pt]{x5}{x6}
    \ncline[linewidth=1pt]{x4}{x6}
    \ncline[linewidth=1pt]{x3}{x7}
    \ncline[linewidth=1pt]{x1}{x8}
    \ncarc[arcangle=30, linewidth=1pt]{x7}{x8}
    \ncarc[arcangle=-30, linewidth=1pt]{x7}{x8}
    \nccircle[linewidth=1pt,angleA=30]{x2}{1cm}
    \uput[90](3,0.5){$A$}
    \uput[90](3,1.5){$B$}
    \uput[90](3,5.5){$C$}
    \uput[45](5.2,2.3){$D$}
    \uput[45](0.2,5.5){$E$}
    \end{pspicture}
  \end{center}
  \caption{A planar graph and its faces}
  \label{facefig}
\end{figure}
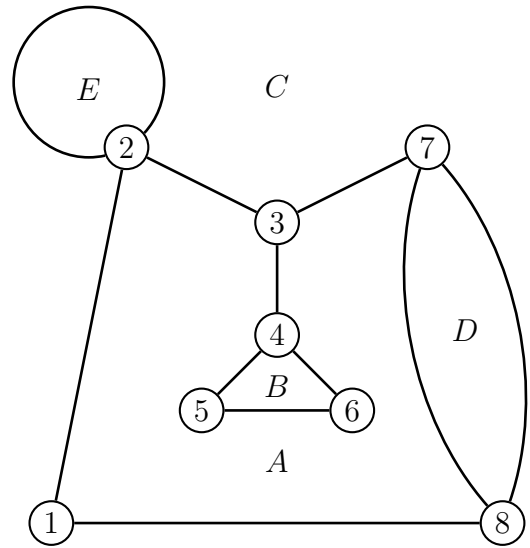

\medskip
As expected, for every face, $F$, of $\s{G}$, the boundary, $\partial F$,
of $F$ is the subset, $|\s{H}|$, associated with some subgraph, $\s{H}$,
of $\s{G}$. However, one should observe that the boundary of a face
may be disconnected and may have several ``holes''. The reader
should draw lots of planar graphs to understand this phenomenon.
Also, since we are considering finite graphs, the set $|\s{G}|$
is bounded and thus, every plane graph has exactly one
unbounded face. Figure \ref{facefig} shows a planar graph and its
faces. Observe that there are five faces, where $A$ is bounded by
the entire graph, $B$ is bounded by the triangle $(4, 5, 6)$
the outside face, $C$, is bounded  by the two edges from $8$ to $2$,
the loop around node $2$, the two edges from $2$ to $7$ and the outer edge from 
$7$ to $8$, $D$ is bounded by the two edges between $7$ and $8$,
and  $E$ is bounded by the loop around node $2$.

\remarks
\begin{enumerate}
\item
Using (inverse) stereographic projection, we see that all the faces
of a graph embedded in the sphere are bounded.
\item
If a graph, $G$, is embedded in a surface, $S$, then
the notion of face still makes sense. Indeed, the faces of $G$
are the regions of the open set $\Omega = S - |G|$.
\end{enumerate}

\medskip
Actually, one should be careful (as usual) not to rely too much on intuition
when dealing with planar graphs. Although certain facts seem obvious,
they may  turn out to be false after closer scrutiny and when
they are true, they may be quite hard to prove. One of the best
examples of an ``obvious'' statement whose proof is much less
trivial than one might expect is the Jordan curve theorem which
is actually needed to justify certain ``obvious'' facts about
faces of plane graphs.

\begin{thm} (Jordan Curve Theorem)
\label{Jordan}
Given any closed simple curve, $\gamma$, in $\reals$,
the complement, $\reals^2 - \gamma([0, 1])$, of $\gamma([0, 1])$,
consist of  exactly two regions both having $\gamma([0, 1])$ as
boundary.
\end{thm}

\proof
There are several proofs all using machinery (such as homology
or differential topology) beyond the scope of 
these notes. A proof using the notion of winding number is given in Guillemin
and Pollack \cite{Guillemin} (Chapter 2, Section 5) and another proof
using homology can be found in Munkres \cite{Munkresalg}
(Chapter 4, Section 36).
$\bigsquare$

\medskip
Using Theorem \ref{Jordan}, the following properties can be proved:

\begin{prop}
\label{planp1}
Let $\s{G} = (V, E)$ be any plane graph and let $e\in E$ be any edge
of $\s{G}$. Then the following properties hold:
\begin{enumerate}
\item[(1)]
For any face, $F$ of $\s{G}$, either $e \subseteq \partial f$ or
$\partial F\> \cap  \interio{e}\> = \emptyset$.
\item[(2)]
If $e$ lies on a cycle, $C$,  of $\s{G}$, then $e$ lies on the
boundary of exactly two faces of $G$ and these are contained in
distinct faces of $C$.
\item[(3)]
If $e$ lies on no cycle, then $e$ lies on the boundary of exactly one
face of $\s{G}$. 
\end{enumerate}
\end{prop}

\proof
See Diestel \cite{Diestel}, Section 4.2.

\medskip
As a corollaries, we also have

\begin{prop}
\label{planp2}
Let $\s{G} = (V, E)$ be any plane graph and let $F$ be any face of
$\s{G}$. Then, the boundary, $\partial F$, of $F$ is a subgraph of 
$\s{G}$ (more accurately, $\partial = |\s{H}|$, for some subgraph, 
$\s{H}$, of $\s{G}$).
\end{prop}

\begin{prop}
\label{planp3}
Every plane forest has a single face.
\end{prop}

\medskip
One of the main theorems about plane graphs is the so-called
{\it Euler formula\/}.

\begin{thm}
\label{Eulerform}
Let $G$ be any connected planar graph with $n_0$ vertices, $n_1$
edges and $n_2$ faces. Then, we have
\[
n_0 - n_1 + n_2 = 2.
\]
\end{thm}

\proof
We proceed by induction on $n_1$. If $n_1 = 0$, the formula
is trivially true, as $n_0 = n_2 = 1$.
Assume  the theorem holds for any $n_1 < n$ and let $G$
be a connected planar graph with $n$ edges. If $G$ has no cycle, then
as it is connected, it is a tree, $n_0 = n + 1$ and $n_2 = 1$,
so $n_0 - n_1 + n_2 = n + 1 - n + 1 = 2$, as desired.
Otherwise, let $e$ be some edge of $G$ belonging to a cycle.
Consider the graph $G' = (V, E - \{e\})$, it is still a connected
planar graph. Therefore, by the induction hypothesis, 
\[
n_0 - (n_1 - 1) + n_2' = 2.
\]
However, by Proposition \ref{planp1}, as $e$ lies on exactly two faces
of $G$, we deduce that \\
$n_2 = n_2' + 1$. Consequently
\[
2 = n_0 - (n_1 - 1) + n_2' = n_0 - n_1 + 1 + n_2 - 1 = n_0 - n_1 + n_2,
\]
establishing the induction hypothesis.
$\bigsquare$

\remarks
\begin{enumerate}
\item
Euler's formula was already known to Descartes in 1640
but the first proof by given by Euler in 1752. Poincar\'e 
generalized it to higher-dimensional polytopes. 
\item
The numbers $n_0$, $n_1$, and $n_2$ are often denoted by $n_v$,
$n_e$ and $n_f$ ($v$ for {\it vertex\/}, $e$ for {\it edge\/} and $f$ for 
{\it face\/}).
\item
The quantity $n_0 - n_1 + n_2$ is called the {\it Euler characteristic\/}
of the graph $G$ and it is usually denoted by $\chi_G$.
\item
If a connected graph, $G$, is embedded in a surface (orientable), $S$, then
we still have an Euler formula of the form
\[
n_0 - n_1 + n_2 = \chi(X) = 2 - 2g,
\]
where $\chi(S)$ is a number depending only on the surface, $S$,
called the {\it Euler characteristic\/} of the surface and
$g$ is called the {\it genus\/} of the surface. It turns out that
$g \geq 0$ is the number of ``handles'' that need to be glued
to the surface of a sphere to get a homeomorphic copy of the
surface $S$. For on this fascinating subject, see Gross and Tucker
\cite{GrossTucker}.
\end{enumerate}

\medskip
It is really remarkable that the quantity $n_0 - n_1 + n_2$
is independent of the way a planar graph is drawn on a sphere
(or in the plane). A neat application of Euler's formula is the proof
that there are only five regular convex polyhedra (the so-called
{\it platonic solids\/}). Such a proof can be found in many places,
for instance Berger \cite{Berger90} and Cromwell \cite{Cromwell}. 
It is easy to generalize Euler's formula to planar graphs
that are not necessarily connected.

\begin{thm}
\label{Eulerform2}
Let $G$ be any planar graph with $n_0$ vertices, $n_1$
edges, $n_2$ faces and $c$ connected components. Then, we have
\[
n_0 - n_1 + n_2 = c + 1.
\]
\end{thm}

\proof
Reduce the proof of Theorem \ref{Eulerform2} to the proof
of Theorem \ref{Eulerform} by adding vertices and edges
between connected components to make $G$ connected.
Details are left as an exercise.
$\bigsquare$

\medskip
Using the Euler formula we can now prove rigorously that
$K_5$ and $K_{3, 3}$ are not planar graphs. For this, we will need the
following fact:

\begin{prop}
\label{planp4}
If $G$ is any simple, connected, plane
graph with $n_1\geq 3$ edges and $n_2$ faces, then
\[
2n_1 \geq 3n_2.
\]
\end{prop}

\proof
Let $F(G)$ be the set of faces of $G$.
Since $G$ is connected, by Proposition \ref{planp1} (2),
every edge belongs to exactly two faces. Thus, if $s_F$ is the number
of sides of a face, $F$, of $G$, we have
\[
\sum_{F\in F(G)} s_F = 2n_1.
\]
Furthermore, as $G$ has no loops, no parallel edges
and $n_0 \geq 3$, every
face has at least three sides, i.e., $s_F \geq 3$. It follows
that
\[
2n_1 = \sum_{F\in F(G)} s_F \geq 3 n_2, 
\]
as claimed.
$\bigsquare$

\medskip
The proof of Proposition \ref{planp4} shows that 
the crucial constant  on the right-hand the inequality is the
the minimum length of all cycles in $G$. This number
is called the {\it girth\/} of the graph $G$. The girth of a
graph with a loop is $1$ and the girth of a graph with parallel
edges is $2$. The girth of a tree is undefined (or infinite).
Therefore, we actually proved:

\begin{prop}
\label{planp4b}
If $G$ is any connected, plane
graph with $n_1$ edges and $n_2$ faces and $G$ is not a tree, then
\[
2n_1 \geq \mathrm{girth}(G)\, n_2.
\]
\end{prop}

\begin{cor}
\label{planp4c}
If $G$ is any simple, connected, plane graph with $n\geq 3$ nodes
then $G$ has at most $3n - 6$ edges and $2n - 4$ faces.
\end{cor}

\proof
By Proposition \ref{planp4}, we have
$2n_1 \geq 3 n_2$, where $n_1$ is the number of edges and
$n_2$ is the number of faces. So, $n_2 \leq \frac{2}{3}n_1$
and by Euler's formula
\[
n - n_1 + n_2 = 2,
\]
we get
\[ 
n - n_1 + \frac{2}{3}n_1 \geq 2,
\]
that is,
\[
n - \frac{1}{3}n_1 \geq 2,
\]
namely $n_1 \leq 3n - 6$. Using $n_2 \leq \frac{2}{3}n_1$, we get
$n_2 \leq 2n - 4$.
$\bigsquare$

\begin{cor}
\label{planp5}
The graphs $K_5$ and $K_{3, 3}$ are not planar.
\end{cor}

\proof
We proceed by contradiction. First, consider $K_5$. We have
$n_0 = 5$ and $K_5$ has $n_1 = 10$ edges.
On the other hand, by Corollary \ref{planp4c}, 
$K_5$ should have at most $3\times 5 - 6 = 15 - 6 = 9$ edges, 
which is absurd.

\medskip
Next, consider $K_{3, 3}$. We have $n_0 = 6$ and $K_{3, 3}$ has 
$n_1 = 9$ edges.
By the Euler formula, we should have
\[
n_2 = 9 - 6 + 2 = 5.
\]
Now, as $K_{3, 3}$ is bipartite, it does not contain any cycle of
odd length, and so each face has at least {\it four\/} sides,
which implies that
\[
2n_1 \geq 4n_2
\]
(because the girth of $K_{3, 3}$ is $4$.)
So, we should have
\[
18 = 2\cdot 9 \geq 4\cdot 5 = 20,
\]
which is absurd. 
$\bigsquare$ 

\medskip
Another important property of simple planar graph is the following:

\begin{prop}
\label{planp6}
If $G$ is any simple, planar graph, then there is a vertex, $u$,
such that $d_G(u) \leq 5$.
\end{prop}

\proof
If the property holds for any connected component of $G$, then
it holds for $G$, so 
we may assume that $G$ is connected. We already know from 
Proposition \ref{planp4} that
$2n_1 \geq 3n_2$. i.e.
\begin{equation}
n_2 \leq \frac{2}{3}n_1. 
\tag{$*$}
\end{equation}
If $d_G(u) \geq 6$ for every vertex, $u$, 
as $\sum x_{u\in V} d_G(u) = 2n_1$, then $6n_0 \leq 2n_1$,
i.e., $n_0 \leq n_1/3$.
By Euler's formula, we would have
\[
n_2 = n_1 - n_0 + 2 \geq n_1 - \frac{1}{3}n_1 + 2 > \frac{2}{3}n_1,
\]
contradicting $(*)$.
$\bigsquare$

\medskip
Remarkably, Proposition \ref{planp6} is the key ingredient in
the proof that every planar graph is $5$-colorable.

\begin{thm}
\label{5colorhm}
Every planar graph, $G$, is $5$-colorable.
\end{thm}

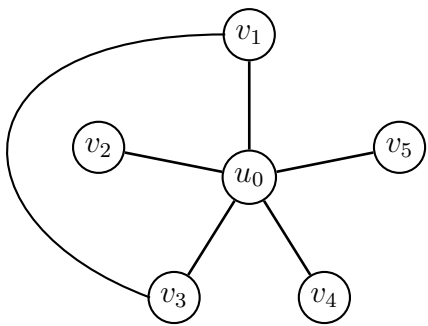
\begin{figure}
  \begin{center}
    \begin{pspicture}(-2,0)(4,4)
    \cnodeput(0,2){v1}{$v_2$}
    \cnodeput(1,0){v5}{$v_3$}
    \cnodeput(3,0){v4}{$v_4$}
    \cnodeput(4,2){v3}{$v_5$}
    \cnodeput(2,3.5){v2}{$v_1$}
    \cnodeput(2,1.6){u0}{$u_0$}
    \ncline[linewidth=1pt]{u0}{v1}
    \ncline[linewidth=1pt]{u0}{v2}
    \ncline[linewidth=1pt]{u0}{v3}
    \ncline[linewidth=1pt]{u0}{v4}
    \ncline[linewidth=1pt]{u0}{v5}
    \psbezier(1.68,3.5)(-2,3.5)(-2,1)(0.68,0)
    \end{pspicture}
  \end{center}
  \caption{The $5$ nodes adjacent to $u_0$}
  \label{5colfig}
\end{figure}

\proof
Clearly, parallel edges and loop play no role in finding
a coloring of the vertices of $G$, so we may assume that
$G$ is a simple graph. Also, the property is clear for graphs
with less than $5$ vertices. We will proceed by induction on the
number of vertices, $m$. By Proposition \ref{planp6}, the graph
$G$ has some vertex, $u_0$, with $d_G(u) \leq 5$. By the induction
hypothesis, we can color the subgraph, $G'$, induced by $V - \{u_0\}$
with $5$ colors. If $d(u_0) < 5$, we can color $u_0$
with one of the colors not used to color the nodes
adjacent to $u_0$ (at most $4$) and we are done.
So, assume $d_G(u_0) = 5$ and let $v_1, \ldots, v_5$ be the nodes
adjacent to $u_0$ and encountered in this order when we rotate
counter-clockwise around $u_0$ (see Figure \ref{5colfig}).
If $v_1, \ldots, v_5$ are not colored with different colors, again,
we are done. 

\medskip
Otherwise, by the induction hypothesis, let $\{X_1, \ldots, X_5\}$
be a coloring of $G'$ and, by renaming the $X_i$'s if necessary,
assume that $v_i\in X_i$, for $i = 1, \ldots, 5$. There are two cases:
\begin{enumerate}
\item[(1)]
There is no chain from $v_1$ to $v_3$ whose nodes belong alternately to
$X_1$ and $X_2$. If so, $v_1$ and $v_3$ must belong to different connected 
components of the subgraph, $H'$, of $G'$ induced by $X_1\cup X_2$. 
Then, we can permute the colors $1$ and $3$ in the connected component of $H'$
that contains $v_3$ and color $u_0$ with color $3$.
\item[(2)]
There is a chain from $v_1$ to $v_3$ whose nodes  belong alternately to
$X_1$ and $X_2$. In this case, as $G$ is a planar graph, there can't
be any chain from $v_2$ to $v_4$ whose nodes belong alternately
to $X_2$ and $X_4$. So, $v_2$ and $v_4$ do not belong to the same
connected component of the subgraph, $H''$, of $G'$ induced by
$X_2\cup X_4$. But then, we can permute the colors $2$ and $4$ in 
the connected component of $H''$
that contains $v_4$ and color $u_0$ with color $4$.
$\bigsquare$
\end{enumerate}

\medskip
Theorem \ref{5colorhm} raises a very famous problem known as the
{\it four color problem\/}: Can every planar graph be colored with four 
colors?

\medskip
This question was apparently first raised by Francis Guthrie in 1850,
communicated to De Morgan by Guthrie's brother Frederick in 1852
and brought to the attention to a wider public by Cayley in 1878. 
In the next hunded years, several incorrect proofs were proposed
and this problem became know as the {\it four color conjecture\/}.
Finally, in 1977, Appel and Haken gave the first ``proof''
of the four color conjecture. However, this proof was somewhat
controversial for various reasons, one of the reasons being
that it relies on a computer program for
checking a large number of unavoidable configurations.
Appel and Haken subsequently published a 741 page paper correcting a
number of errors and addressing various criticisms. More recently (1997) 
a much shorter proof, still relying on a computer program, but
a lot easier to check (including the computer part of it)  has been
given by Robertson, Sanders, Seymour and Thomas.
For more on the four color problem, see Diestel \cite{Diestel},
Chapter 5, and the references given there. 

\medskip
let us now go back to  Kuratowski's criterion for non-planarity.
For this, it is useful to introduce the notion of 
edge contraction in a graph.

\begin{defin}
\label{edgecont}
{\em
Let $G = (V, E, st)$ be any graph and let $e$ be any edge of $G$.
The graph obtained by {\it contracting the edge $e$
into a new vertex, $v_e$\/}, is the graph, $G/e = (V', E', st')$, 
with $V' = (V - st(e))\cup \{v_e\}$ where $v_e$ is a new node
($v_e\notin V$); $E' = E - \{e\}$; and with
\[
st'(e') = \cases{
st(e') & if $st(e') \cap st(e) = \emptyset$ \cr
\{v_e\} & if $st(e') = st(e)$ \cr
\{u, v_e\} & if  $st(e') \cap st(e) = \{z\}$ and $st(e') = \{u, z\}$
with  $u\not= z$\cr
\{v_e\} & if  $st(e') = \{x\}$  or $st(e') = \{y\}$ with
 $st(e) = \{x, y\}$. \cr
}
\]
If $G$ is a simple graph, then we need to eliminate parallel edges
and loops. In, this case, $e = \{x, y\}$ and 
$G/e = (V', E', st)$ is defined so that
$V' = (V - \{x, y\})\cup \{v_e\}$ where $v_e$ is a new node and
\begin{align*}
E' & =  \{\{u, v\} \mid \{u, v\}\cap \{x, y\} = \emptyset\} \\
   &   \quad\> \cup \{\{u, v_e\} \mid \{u, x\} \in E - \{e\}\quad\hbox{or}\quad
\{u, y\}\in E - \{e\}\}.
\end{align*}
}
\end{defin}

\begin{figure}
  \begin{center}
    \begin{pspicture}(0,0)(6.3,4)
    \cnodeput(0,1){v1}{$1$}
    \cnodeput(2.5,1){v2}{$2$}
    \cnodeput(1.25,3){v3}{$3$}
    \cnodeput(5,1){v4}{$4$}
    \cnodeput(3.75,3){v5}{$5$}
    \cnodeput(6.25,3){v6}{$6$}
    \ncline[linewidth=1pt]{v1}{v2}
    \ncline[linewidth=1pt]{v1}{v3}
    \ncline[linewidth=1pt]{v2}{v3}
    \ncline[linewidth=1pt]{v2}{v5}
    \ncline[linewidth=1pt]{v4}{v5}
    \ncline[linewidth=1pt]{v6}{v4}
    \ncarc[arcangleA=30, arcangleB=30, linewidth=2pt]{v2}{v4}
    \ncarc[arcangleA=-30, arcangleB=-30, linewidth=1pt]{v2}{v4}
    \end{pspicture}
  \hskip 2cm
    \begin{pspicture}(0,0)(6,4)
    \cnodeput(0,1){v1}{$1$}
    \cnodeput(2.5,1){v2}{$7$}
    \cnodeput(1.25,3){v3}{$3$}
    \cnodeput(3.25,3){v5}{$5$}
    \cnodeput(5.5,3){v6}{$6$}
    \ncline[linewidth=1pt]{v1}{v2}
    \ncline[linewidth=1pt]{v1}{v3}
    \ncline[linewidth=1pt]{v2}{v3}
    \ncline[linewidth=1pt]{v6}{v2}
    \ncarc[arcangleA=30, arcangleB=30, linewidth=1pt]{v2}{v5}
    \ncarc[arcangleA=-30, arcangleB=-30, linewidth=1pt]{v2}{v5}
    \nccircle[linewidth=1pt,angleA=180]{v2}{0.7cm}
    \end{pspicture}
  \end{center}
  \caption{Edge Contraction in a graph}
  \label{edgecontracfig1}
\end{figure}
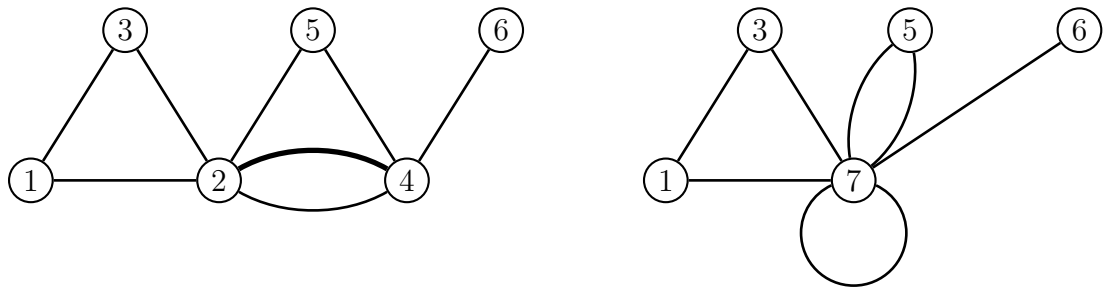

\medskip
Figure \ref{edgecontracfig1} shows the result of contracting 
the upper edge $\{2, 4\}$ (shown as a thicker line) in the graph 
shown on the left, which is not a simple graph.
Observe how the lower edge  $\{2, 4\}$ becomes a loop 
around $7$ and the
two edges $\{5, 2\}$ and $\{5, 4\}$  become parallel edges
between $5$ and $7$.

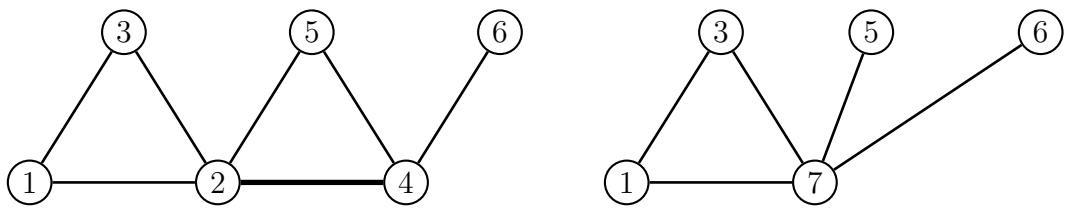
\begin{figure}
  \begin{center}
    \begin{pspicture}(0,0)(6.3,3)
    \cnodeput(0,0){v1}{$1$}
    \cnodeput(2.5,0){v2}{$2$}
    \cnodeput(1.25,2){v3}{$3$}
    \cnodeput(5,0){v4}{$4$}
    \cnodeput(3.75,2){v5}{$5$}
    \cnodeput(6.25,2){v6}{$6$}
    \ncline[linewidth=1pt]{v1}{v2}
    \ncline[linewidth=1pt]{v1}{v3}
    \ncline[linewidth=1pt]{v2}{v3}
    \ncline[linewidth=2pt]{v2}{v4}
    \ncline[linewidth=1pt]{v2}{v5}
    \ncline[linewidth=1pt]{v4}{v5}
    \ncline[linewidth=1pt]{v6}{v4}
    \end{pspicture}
  \hskip 1.5cm
    \begin{pspicture}(0,0)(6,3)
    \cnodeput(0,0){v1}{$1$}
    \cnodeput(2.5,0){v2}{$7$}
    \cnodeput(1.25,2){v3}{$3$}
    \cnodeput(3.25,2){v5}{$5$}
    \cnodeput(5.5,2){v6}{$6$}
    \ncline[linewidth=1pt]{v1}{v2}
    \ncline[linewidth=1pt]{v1}{v3}
    \ncline[linewidth=1pt]{v2}{v3}
    \ncline[linewidth=1pt]{v2}{v5}
    \ncline[linewidth=1pt]{v6}{v2}
    \end{pspicture}
  \end{center}
  \caption{Edge Contraction in a simple graph}
  \label{edgecontracfig2}
\end{figure}

\medskip
Figure \ref{edgecontracfig2} shows the result of contracting 
edge $\{2, 4\}$ (shown as a thicker line) in the simple graph 
shown on the left.
This time, the two edges $\{5, 2\}$ and $\{5, 4\}$ become a single edge
and there is no loop around $7$ as the contracted edge is deleted.

\medskip
Now, given a graph, $G$, we can repeatedly contract edges.
We can also take a subgraph of a graph $G$ and then perform
some edge contractions. We obtain what is known as a minor of $G$.

\begin{defin}
\label{minordef}
{\em
Given any graph, $G$, a graph, $H$, is a {\it minor\/} of $G$ is
there is a sequence of graphs, $H_0, H_1, \ldots, H_n$ $(n \geq 1)$,
such that
\begin{enumerate}
\item[(1)]
$H_0 = G$; $H_n = H$;
\item[(2)]
Either $H_{i+1}$ is obtained from $H_i$ by deleting some 
edge or some node of $H_i$ and all the
edges incident with this node, or
\item[(3)]
$H_{i+1}$ is obtained from $H_i$ by edge contraction,
\end{enumerate}
with $0 \leq i \leq n - 1$.
If $G$ is a simple graph, we require that edge contractions
be of the second type described in Definition \ref{edgecont},
so that $H$ is a a simple graph.
}
\end{defin}

\medskip
It is easily shown that the minor relation is a partial order on graphs
(and simple graphs). Now, the following 
remarkable theorem originally due to
Kuratowski characterizes planarity in terms of the notion of minor:

\begin{thm} (Kuratowski, 1930)
\label{Kuratow1}
For any graph, $G$, the following assertions are equivalent: 
\begin{enumerate}
\item[(1)]
$G$ is planar;
\item[(2)]
$G$ contains neither $K_5$ nor $K_{3, 3}$ as a minor.
\end{enumerate}
\end{thm}

\proof
The proof is quite involved. The first step is to prove the theorem 
for $3$-connected graphs. (A graph $G = (V, E)$ is $k$-connected iff
$|V| > k$ and iff every graph obtained by deleting any set, $S\subseteq V$, 
of nodes with $|S| < k$ and the edges incident to these node 
is still connected. 
So, a $1$-connected graph is just a connected graph.)
We refer the reader to Diestel \cite{Diestel},
Section 4.4, for a complete proof.
$\bigsquare$

\medskip
Another way to state Kuratowski's theorem involves edge
subdivision, an operation of independent interest.
Given a graph, $G = (V, E, st)$, possibly with
loops and parallel edges, the result of subdividing an edge, $e$,
consists in creating a new vertex, $v_e$, deleting the edge $e$,
and adding two new edges from $v_e$ to the old endpoints of $e$
(possibly the same point). Formally, we have the following definition:

\begin{defin}
\label{edgesubdiv}
{\em
Given any graph, $G = (V, E, st)$, for any edge, $e\in E$,
the result of {\it subdividing  the edge $e$\/} is the graph,
$G' = (V \cup \{v_e\}, (E - \{e\}) \cup\{e^1, e^2\}, st')$, where
$v_e$ is a new vertex and $e^1, e^2$ are new edges, 
$st'(e') = st(e')$ for all $e'\in E - \{e\}$ 
and if $st(e) = \{u, v\}$
($u = v$ is possible), then
$st'(e^1) = \{v_e, u\}$ and $st'(e^2) = \{v_e, v\}$.
If a graph, $G'$, is obtained from a graph, $G$, by a sequence
of edge subdivisions, we say that $G'$ is a {\it subdivision\/} of $G$.
}
\end{defin}

\begin{figure}
  \begin{center}
    \begin{pspicture}(0,0)(6.5,3.1)
    \cnode(0,1.5){2pt}{v1}
    \cnode(2,0){2pt}{v2}
    \cnode(2,1.5){2pt}{v3}
    \cnode(2,3){2pt}{v4}
    \cnode(4,1.5){2pt}{v5}
    \cnode(6,1.5){2pt}{v6}
    \ncline[linewidth=1pt]{v1}{v2}
    \ncline[linewidth=1pt]{v1}{v4}
    \ncline[linewidth=1pt]{v2}{v3}
    \ncline[linewidth=1pt]{v3}{v4}
    \ncline[linewidth=1pt]{v2}{v5}
    \ncline[linewidth=1pt]{v4}{v5}
    \ncline[linewidth=1pt]{v5}{v6}
    \end{pspicture}
  \hskip 1.5cm
    \begin{pspicture}(0,0)(7,3.1)
    \cnode(0,1.5){2pt}{v1}
    \cnode(2,0){2pt}{v2}
    \cnode(2,3){2pt}{v3}
    \cnode(4,1.5){2pt}{v4}
    \cnode(5.5,1.5){2pt}{v5}
    \cnode(7,1.5){2pt}{v6}
    \ncline[linewidth=1pt]{v1}{v2}
    \ncline[linewidth=1pt]{v1}{v3}
    \ncline[linewidth=1pt]{v2}{v3}
    \ncline[linewidth=1pt]{v2}{v4}
    \ncline[linewidth=1pt]{v3}{v4}
    \ncline[linewidth=1pt]{v4}{v5}
    \ncline[linewidth=1pt]{v5}{v6}
    \end{pspicture}
  \end{center}
  \caption{Two homeomorphic graphs}
  \label{homeofig}
\end{figure}
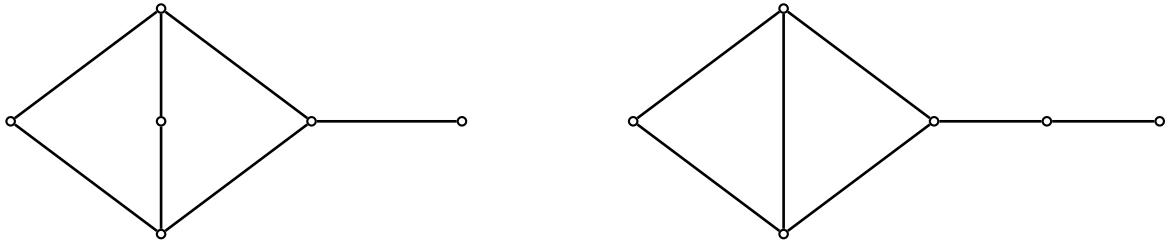

\medskip
Observe that by repeatedly subdividing edges, any graph can be transformed
into a simple graph. Given two graphs, $G$ and $H$, we say that
$G$ and $H$ are {\it homeomorphic\/} iff they have respective
subdivisions $G'$ and $H'$ that are isomorphic graphs.
The idea is that homeomorphic graphs ``look the same'', viewed as
topological spaces. Figure \ref{homeofig} shows
an example of two homeomorphic graphs.
A graph, $H$, that has a subdivision, $H'$,
which is a subgraph of some graph, $G$, is called a {\it topological
minor\/} of $G$.
Then, it is not hard to show 
(see Diestel \cite{Diestel}, Chapter 4, or
Gross and Tucker \cite{GrossTucker}, Chapter 1)
that Kuratowski's Theorem is equivalent to the statement

\medskip
{\it A graph, $G$, is planar iff it does not contain any subgraph
homeomorphic to either $K_5$ or $K_{3, 3}$ or,
equivalently, if it has has neither $K_5$ nor $K_{3, 3}$ 
as a topological minor\/}.

\medskip
Another somewhat surprising characterization of planarity
involving the concept of cycle space over $\mathbb{F}_2$
(see Definition \ref{flowdef}
and the Remarks after Theorem \ref{flowdim1}) 
and due to MacLane is the following: 

\begin{thm}
\label{MacLane1} (MacLane, 1937)
A graph, $G$ is planar iff its cycle space, $\s{F}$, over $\mathbb{F}_2$
has a basis such that every edge of $G$ belongs to at most two cycles 
of this basis.
\end{thm}

\proof
See Diestel \cite{Diestel}, Section 4.4.
$\bigsquare$

\medskip
We conclude this section on planarity with a brief discussion of
the dual graph of a plane graph, a notion originally due to Poincar\'e.
Duality can be generalized
to simplicial complexes and relates Voronoi diagrams and
Delaunay triangulations, two very important tools in computational
geometry.

\medskip
Given a plane graph, $G = (V, E)$, let $F(G)$ be the set of faces
of $G$. The crucial point is that every edge of $G$ is part of the
boundary of at most two faces.
A dual graph, $G^* = (V^*, E^*)$, of $G$ is a graph whose
nodes are in one-to-one correspondence with the faces of $G$,
whose faces are in  one-to-one correspondence with the nodes of $G$
and whose edges are also are in  one-to-one correspondence with the 
the egdes of $G$. For any edge, $e\in E$, a dual edge, $e^*$, links
the two nodes $v_{F_1}$ and $v_{F_2}$ associated with the faces 
$F_1$ and $F_2$ adjacent to $e$ or, $e^*$ is a loop from $v_{F}$ to itself
if $e$ is ajacent to a single face. Here is the precise definition:

\begin{defin}
\label{dualdef}
{\em
Let $G = (V, E)$ be a plane graph and let $F(G)$ be its set of faces.
A {\it dual graph\/} of $G$ is a graph, $G^* = (V^*, E^*)$, where
\begin{enumerate}
\item[(1)]
$V^* = \{v_{F} \mid F\in F(G)\}$, where $v_F$ is a point chosen in
the (open) face, $F$, of $G$;
\item[(2)]
$E^* = \{e^* \mid e\in E\}$, where $e^*$ is a simple curve from
$v_{F_1}$ to $v_{F_2}$ crossing $e$, if $e$ is part of the boundary of 
two faces $F_1$ and $F_2$ or else, a closed simple curve 
crossing $e$ from $v_F$ to itself, if $e$ is part of the boundary of 
exactly one face, $F$.
\item[(3)]
For each $e\in E$,  we have $e^* \cap G = e\cap G^* =\>
\interio{e}\, \cap \, \interio{{e^*}}$, a one point set.
\end{enumerate}
}
\end{defin}

\begin{figure}
  \begin{center}
    \begin{pspicture}(-3,-2.5)(6,7)
    \cnodeput(0,2){v1}{$1$}
    \cnodeput(1.5,4){v2}{$2$}
    \cnodeput(3,6){v3}{$3$}
    \cnodeput(4.5,4){v4}{$4$}
    \cnodeput(3,2){v5}{$5$}
    \cnodeput(4.5,0){v6}{$6$}
    \cnodeput(4.5,0){v6}{$6$}
    \cnodeput(1.5,2.66){u1}{$a$}
    \cnodeput(3,3.33){u2}{$b$}
    \cnodeput(3,4.66){u3}{$c$}
    \cnodeput(6,-2){u4}{$d$}
    \ncline[linewidth=1pt]{v1}{v2}
    \ncline[linewidth=1pt]{v2}{v3}
    \ncline[linewidth=1pt]{v2}{v4}
    \ncline[linewidth=1pt]{v3}{v4}
    \ncline[linewidth=1pt]{v1}{v5}
    \ncline[linewidth=1pt]{v4}{v5}
    \ncline[linewidth=1pt]{v5}{v6}
    \ncline[linewidth=1pt]{v2}{v5}
    \ncline[linewidth=2pt]{u1}{u2}
    \ncline[linewidth=2pt]{u2}{u3}
    \psbezier[linewidth=2pt](5.75,-2)(-6,-5)(-6,8)(2.75,4.66)
    \psbezier[linewidth=2pt](5.75,-2)(-4,-2)(-2,5)(1.25,2.66)
    \psbezier[linewidth=2pt](5.75,-1.9)(-0.2,-0.6)(6,4.5)(6,-1.7)
    \ncarc[arcangleA=45, arcangleB=60, linewidth=2pt]{u4}{u1}
    \ncarc[arcangleA=-60, arcangleB=-60, linewidth=2pt]{u4}{u2}
    \ncarc[arcangleA=-70, arcangleB=-90, linewidth=2pt]{u4}{u3}
    \end{pspicture}
  \end{center}
  \caption{A graph and its dual graph}
  \label{dualfig}
\end{figure}
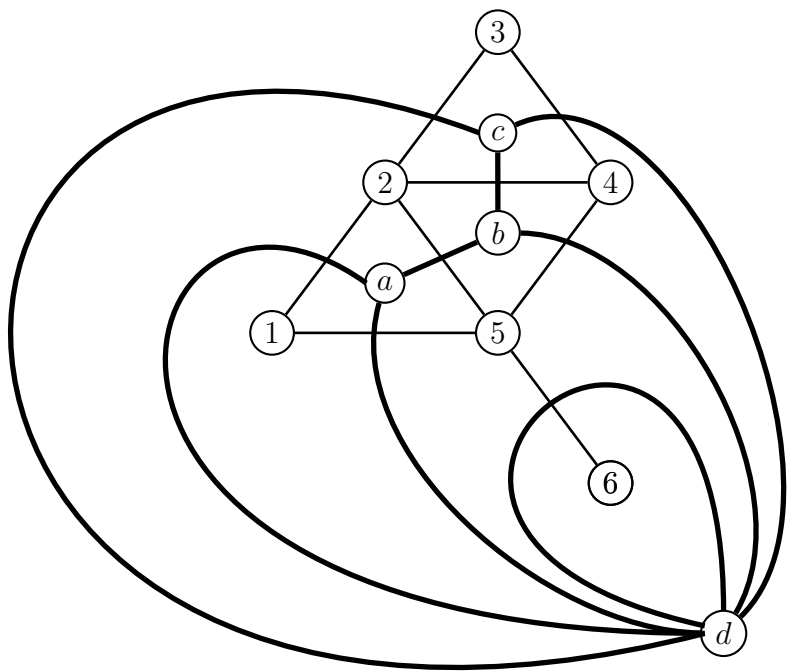

\medskip
An example of a dual graph  is shown in Figure \ref{dualfig}.
The graph $G$ has four faces, $a, b, c, d$ and the dual graph, $G^*$,
has nodes also denoted $a, b, c, d$ enclosed in a small circle,
with the edges of the dual graph shown with thicker lines.

\medskip
Note how the edge $\{5, 6\}$ gives rise to the loop from $d$ to 
itself and that there are parallel edges between $d$ and $a$
and between $d$ and $c$. Thus, even if we start with a simple graph,
a dual graph may have loops and parallel edges.

\medskip
Actually, it is not entirely obvious that a dual of a plane graph is
a plane graph but this is not difficult to prove. 
It is also important to note that a given plane graph, $G$, {\it does not have
a unique dual\/} since the vertices and the edges of a dual graph
can be chosen in infinitely different ways in order to satisfy the
conditions of Definition \ref{dualdef}. However,
given a plane graph, $G$, if $H_1$ and $H_2$
are two dual graphs of $G$, then it is easy to see that 
$H_1$ and $H_2$ are isomorphic. Therefore, with a slight
abuse of language, we may refer to ``the'' dual graph of a plane graph.
Also observe that even if $G$ is not connected, its dual, $G^*$, is 
always connected.

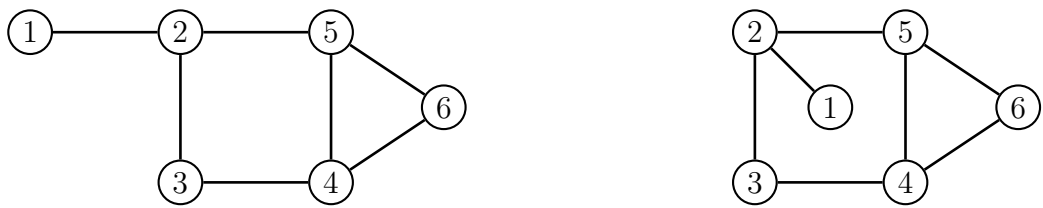
\begin{figure}
  \begin{center}
    \begin{pspicture}(0,0)(6,3)
    \cnodeput(0,2){v1}{$1$}
    \cnodeput(2,2){v2}{$2$}
    \cnodeput(2,0){v3}{$3$}
    \cnodeput(4,0){v4}{$4$}
    \cnodeput(4,2){v5}{$5$}
    \cnodeput(5.5,1){v6}{$6$}
    \ncline[linewidth=1pt]{v1}{v2}
    \ncline[linewidth=1pt]{v2}{v3}
    \ncline[linewidth=1pt]{v2}{v5}
    \ncline[linewidth=1pt]{v3}{v4}
    \ncline[linewidth=1pt]{v4}{v5}
    \ncline[linewidth=1pt]{v4}{v6}
    \ncline[linewidth=1pt]{v5}{v6}
    \end{pspicture}
\hskip 1.5cm
    \begin{pspicture}(0,0)(6,3)
    \cnodeput(3,1){v1}{$1$}
    \cnodeput(2,2){v2}{$2$}
    \cnodeput(2,0){v3}{$3$}
    \cnodeput(4,0){v4}{$4$}
    \cnodeput(4,2){v5}{$5$}
    \cnodeput(5.5,1){v6}{$6$}
    \ncline[linewidth=1pt]{v1}{v2}
    \ncline[linewidth=1pt]{v2}{v3}
    \ncline[linewidth=1pt]{v2}{v5}
    \ncline[linewidth=1pt]{v3}{v4}
    \ncline[linewidth=1pt]{v4}{v5}
    \ncline[linewidth=1pt]{v4}{v6}
    \ncline[linewidth=1pt]{v5}{v6}
    \end{pspicture}
  \end{center}
  \caption{Two isomorphic plane graphs whose dual graphs are not isomorphic}
  \label{dualfig2}
\end{figure}

\danger
The notion of dual graph applies
to a {\it plane\/} graph and {\it not to a planar graph\/}. Indeed, 
the graphs $G_1^*$ and $G_2^*$ associated to two different 
embeddings, $G_1$ and $G_2$, of the same abstract planar graph, $G$,
may {\bf not} be isomorphic, even though $G_1$ and $G_2$
are isomorphic as abstact graphs. For example, the two plane
graphs, $G_1$ and $G_2$, shown in Figure \ref{dualfig2}
are isomorphic but their dual graphs, $G_1^*$ and $G_2^*$,
are not, as the reader should check (one of these two graphs has
a node of degree $7$ but for the other graph all nodes have degree at 
most $6$).

\remark
If a graph, $G$, is embedded in a surface, $S$, then
the notion of dual graph also makes sense. More for on this,
see Gross and Tucker \cite{GrossTucker}.

\medskip
In the following proposition, we summarize some useful
properties of dual graphs.

\begin{prop}
\label{dualp1}
The dual, $G^*$ of any plane graph is connected.
Furthermore, if $G$ is a connected plane graph, then $G^{**}$
is isomorphic to $G$.
\end{prop}

\proof
Left as an exercise.

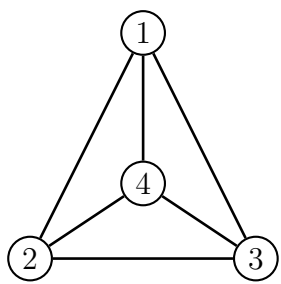
\begin{figure}
  \begin{center}
    \begin{pspicture}(0,0)(3,4)
    \cnodeput(1.5,3){v1}{$1$}
    \cnodeput(0,0){v2}{$2$}
    \cnodeput(3,0){v3}{$3$}
    \cnodeput(1.5,1){v4}{$4$}
    \ncline[linewidth=1pt]{v1}{v2}
    \ncline[linewidth=1pt]{v2}{v3}
    \ncline[linewidth=1pt]{v3}{v1}
    \ncline[linewidth=1pt]{v4}{v1}
    \ncline[linewidth=1pt]{v4}{v2}
    \ncline[linewidth=1pt]{v4}{v3}
    \end{pspicture}
  \end{center}
  \caption{A self-dual graph}
  \label{tetrafig}
\end{figure}

\medskip
We a slight abuse of notation we often write $G^{**} = G$
(when $G$ is connected).
A plane graph, $G$, whose dual, $G^*$, is equal to $G$
(i.e., isomorphic to $G$) is called {\it self-dual\/}. 
For example, the plane graph shown
in Figure \ref{tetrafig} (the projection of a tetrahedron on the plane)
is self dual.

\medskip
The duality of plane graphs is also reflected algebraically
as a duality between their cycle spaces and their cut spaces
(over $\mathbb{F}_2$).

\begin{prop}
\label{dualp2}
If $G$ is any connected plane graph, $G$, then the following
properties hold:
\begin{enumerate}
\item[(1)]
A set of edges, $C\subseteq E$,
is a cycle in $G$ iff $C^* = \{e^* \in E^* \mid e\in C\}$ is a minimal
cutset in $G^*$.
\item[(2)]
If $\s{F}(G)$ and $\s{T}(G^*)$ denote the cycle space of
$G$ over $\mathbb{F}_2$ and the cut space of $G^*$ over $\mathbb{F}_2$,
respectively, then the dual, $\s{F}^*(G)$, of $\s{F}(G)$
(as a vector space)  is equal to the cut space, $\s{T}(G^*)$,  of $G^*$, i.e.,
\[
\s{F}^*(G) = \s{T}(G^*).
\]
\item[(3)]
If $T$ is any spanning tree of $G$, then $(V^*, (E - E(T))^*)$
is a spanning tree of $G^*$ (Here, $E(T)$ is the set of edges of 
the tree, $T$.)
\end{enumerate}
\end{prop}

\proof
See Diestel \cite{Diestel}, Section 4.6.
$\bigsquare$

\medskip
The interesting problem of finding an algorithmic test for planarity
has received quite a bit of attention. Hopcroft and Tarjan
have given an algorithm running in linear time in the
number of vertices. More more on planarity, the reader should
consult Diestel \cite{Diestel}, Chapter 4,
or Harary \cite{Harary}, Chapter 11.

\medskip
Besides the four color ``conjecture'', the other most famous theorem
of graph theory is the {\it graph minor theorem\/}, due to Roberston
and Seymour and we can't resist stating this
beautiful and amazing result. For this, we need to explain what is
a {\it well-quasi order\/}, for short, a {\it w.q.o\/}.

\medskip
Recall that a partial order on a set, $X$, is a binary relation, $\leq$,
which is reflexive, symmetric and anti-symmetric. A {\it quasi-order\/}
(or {\it preorder\/}) is a relation which is reflexive and
transitive (but not necessarily anti-symmetric). A {\it well-quasi-order\/}, 
for short, a {\it w.q.o\/} ,
is a quasi-order with the following property:

\medskip
For every infinite sequence, $(x_n)_{n \geq 1}$, of elements $x_i \in X$,
there exist some indices, $i, j$, with $1 \leq i < j$, so that
$x_i \leq x_j$.

\medskip
Now,  we know that being a minor of another graph is a partial order
and thus, a quasi-order. Here is Robertson and Seymour's theorem:

\begin{thm} (Graph Minor Theorem, Robertson and Seymour, 1985-2004)
\label{RobertsonSeymour1}
The minor relation on finite graphs is a well quasi-order.
\end{thm}

\medskip
Remarkably, the proof of Theorem \ref{RobertsonSeymour1}
is spread  over $20$ Journal papers (under the common title,
{\it Graph Minors\/}) written over nearly $18$ years
and taking well over $500$ pages! Many original 
techniques had to be invented to come up with this proof,
one of which is a careful study of the conditions under
which a graph can be embedded in a surface and a ``Kuratowski-type''
criterion based on a finite family of ``forbidden graphs''.
The interested reader is urged to consult Chapter 12 of Diestel
\cite{Diestel} and the references given there.

\medskip
A precursor of the graph minor theorem is a theorem of Kruskal (1960)
which applies to trees. Although much easier to prove that
the graph minor theorem, the proof fo Kruskal's Theorem is
very ingenious. It turns out that there are also some interesting
connections between Kruskal's Theorem and proof theory, due to 
Harvey Friedman.
A survey on this topic can be found in Gallier \cite{GallierKrus}.

\bibliography{cse260notes}
\bibliographystyle{plain} 
\end{document}